\documentclass[11pt]{report}
\usepackage{epsfig}
\usepackage{amsfonts}
\usepackage{amsmath}
\usepackage{bbm,bm}
\usepackage{makeidx}
\usepackage{cite}
\usepackage{cancel}
\usepackage[hidelinks]{hyperref}
\usepackage{empheq}
\usepackage{scalerel}
\usepackage{verbatim}

\allowdisplaybreaks[1]

\makeindex

\newcommand{\be}{\begin{equation}}
\newcommand{\ee}{\end{equation}}
\newcommand{\ba}{\begin{eqnarray}}
\newcommand{\ea}{\end{eqnarray}}
\newcommand{\bi}{\begin{itemize}}
\newcommand{\ei}{\end{itemize}}
\newcommand{\nn}{\nonumber \\}
\newcommand{\fig}{fig.~}

\newcommand{\eq}{eq.~}
\newcommand{\eqs}{eqs.~}
\newcommand{\ch}{chapter~}

\newcommand{\se}{sec.~}
\newcommand{\ses}{secs.~}
\newcommand{\App}{Appendix~}
\newcommand{\app}{appendix~}
\newcommand{\apps}{appendices~}
\newcommand{\nr}[1]{(\ref{#1})}
\newcommand{\tr}{\mathop{\text{tr}}}
\newcommand{\fr}[2]{{\frac{#1}{#2}\,}}

\newcommand{\tfr}[2]{{\textstyle \frac{#1}{#2}\,}}
\newcommand{\rmi}[1]{{\mbox{\scriptsize #1}}}
\newcommand{\rmii}[1]{{\mbox{\tiny\rm{#1}}}}
\newcommand{\rmiii}[1]{{\mbox{\tiny{$\scriptstyle{\rm#1}$}}}}
\newcommand{\re}{\mathop{\rm {Re}}}
\newcommand{\im}{\mathop{\rm {Im}}}

\renewcommand{\vec}[1]{{\bf #1}}
\newcommand{\mpl}{m_\rmii{pl}} % {m_\rmii{Pl}}

\newcommand{\projP}{\mathbbm{P}}
\newcommand{\projK}{\mathbbm{K}}
\newcommand{\projV}{\mathbbm{V}}

\def\lsi{\raise0.3ex\hbox{$<$\kern-0.75em\raise-1.1ex\hbox{$\sim$}}}
\def\gsi{\raise0.3ex\hbox{$>$\kern-0.75em\raise-1.1ex\hbox{$\sim$}}}
\newcommand{\lsim}{\mathop{\lsi}}
\newcommand{\gsim}{\mathop{\gsi}}

\newcommand{\nB}{n_\rmii{B}}
\newcommand{\Tint}[1]{{\hbox{$\sum$}\!\!\!\!\!\!\!\int\,}_{\!\!\!\!\raise-0.9ex\hbox{$\scriptstyle{#1}$}}}
\newcommand{\Tinti}[1]{{{\Sigma}\!\!\!\!\raise0.3ex\hbox{$\int$}_\rmii{${#1}$}}}
\newcommand{\hide}[1]{ }
\newcommand{\deltabar}{\raise-0.02em\hbox{$\bar{}$}\hspace*{-0.8mm}{\delta}}
\newcommand{\ddeltabar}{\raise-0.18em\hbox{$\bar{}$}\hspace*{-0.8mm}{\delta}}
\renewcommand{\H}{\mathcal{H}}
\newcommand{\I}{\mathcal{I}}
\newcommand{\iI}{\rmii{$I$}}
\renewcommand{\P}{\mathcal{P}}
\newcommand{\K}{\mathcal{K}}
\newcommand{\Q}{\mathcal{Q}}
\newcommand{\iQ}{\rmii{$Q$}}
\newcommand{\R}{\mathcal{R}}
\newcommand{\iR}{\rmii{R}}
\newcommand{\X}{\mathcal{X}}
\newcommand{\Y}{\mathcal{Y}}
\newcommand{\Z}{\mathcal{Z}}
\newcommand{\W}{\mathcal{W}}

\newcommand{\F}{\mathcal{F}}
\newcommand{\E}{\mathcal{S}} % {\rmii{$E$}}

\newcommand{\M}{\mathcal{M}}
\newcommand{\N}{\mathcal{N}}
\newcommand{\T}{\rmii{$T$}}
\newcommand{\der}{,}
\newcommand{\field}{\widehat{\mathcal{Q}}}
\newcommand{\barpPi}{\Pi}
\newcommand{\dd}{{\rm d}}
\newcommand{\Hc}{\mathcal{H}}
\newcommand{\rmO}{{\mathcal{O}}}
\newcommand{\ord}{\mathcal{O}}
\newcommand{\varc}{\mathbbm{c}} 
\newcommand{\varh}{\mathbbm{h}} %   {\mathrm{h}}
\newcommand{\hstrain}{\hspace*{0.3mm}\mathrm{h}}
\newcommand{\fin}{\mbox{\sl f\,}}
\newcommand{\ini}{\mbox{\sl i\,}}
\newcommand{\sfin}{\rmii{\sl f\,}} % {\rmii{f}}
\newcommand{\sini}{\rmii{\sl i\,}}
\newcommand{\bbn}{\rmii{BBN}}
\newcommand{\cmb}{\rmi{dec}}
\newcommand{\icmb}{\rmii{dec}}
\newcommand{\now}{\rmi{0}}
\newcommand{\inow}{\rmii{0}}
\newcommand{\iinow}{\rmiii{0}}
\newcommand{\gw}{\rmi{gw}}
\newcommand{\igw}{\rmii{gw}}
\newcommand{\tensor}{\rmi{t}} % {t} {\rmii{T}}
\newcommand{\itensor}{\rmii{t}} % {t} {\rmii{T}}
\newcommand{\scalar}{\rmi{s}} % {s} 
\newcommand{\iscalar}{\rmii{s}} % {s} 
\newcommand{\vac}{\rmi{vac}}
\newcommand{\dm}{\rmi{dm}}
\newcommand{\idm}{\rmii{dm}}
\newcommand{\tmin}{-\infty}
\newcommand{\tmax}{\infty}
\newcommand{\bit}{\hspace*{0.4mm}}
\newcommand{\ibit}{\hspace*{0.2mm}}
\newcommand{\mbit}{\hspace*{-0.4mm}}
\newcommand{\lift}{\vphantom{ |^b_q }}
\newcommand{\ilift}{\vphantom{ |^b }}
\newcommand{\switch}{s}
\makeatletter
  \@addtoreset{equation}{section} 
  \@addtoreset{figure}{section} 
  \@addtoreset{table}{section} 
  \@addtoreset{enumiv}{section} 
\makeatother

\makeatletter

\renewcommand\thechapter{\@Roman\c@chapter}
\def\@makechapterhead#1{%
  \vspace*{50\p@}%
  {\parindent \z@ \raggedright \normalfont
    \ifnum \c@secnumdepth >\m@ne
        \Large\bfseries \@chapapp\space \thechapter
        \par\nobreak
        \vskip 20\p@
    \fi
    \interlinepenalty\@M
    \Large \bfseries #1\par\nobreak
    \vskip 40\p@
  }}
\renewcommand\section{\@startsection {section}{1}{\z@}%
                                   {-5.5ex \@plus -1ex \@minus -.2ex}% bfr-skip
                                   {2.3ex \@plus.2ex}%
                                   {\normalfont\large\bfseries}}
\renewcommand\subsection{\@startsection{subsection}{2}{\z@}%
                                     {-3.25ex\@plus -1ex \@minus -.2ex}%
                                     {1.5ex \@plus .2ex}%
                                     {\normalfont\normalsize\bfseries}}
\renewcommand\subsubsection{\@startsection{subsubsection}{2}{\z@}%
                                     {-3.25ex\@plus -1ex \@minus -.2ex}%
                                     {1.5ex \@plus .2ex}%
                                     {\normalfont\normalsize\bfseries}}

\renewcommand\thesection {\@arabic\c@section}
\renewcommand\thesubsection   {\thesection.\@arabic\c@subsection}
\renewcommand\thesubsubsection   {\thesection.\@Alph\c@subsubsection}
\renewcommand{\@seccntformat}[1]{%
\csname the#1\endcsname.\hspace{1.0em}}
\makeatother
\makeatletter
\renewcommand\@biblabel[1]{[{\thesection}.#1]}
\makeatother
\begin{document}

\begin{titlepage} 

\begin{flushright}
% v0.012 \\ 
July 2026 
\end{flushright}

\vspace*{0.2cm}

\begin{centering} 

{\Large From inflation to hot big bang --- } \\[2mm] 
{\large a tutorial on cosmological perturbations}\hspace*{0.3mm}\footnote{%
%%%%%%%%%%%%%%%%%%%%%%%%%%%%%%%%%%%%%%%%%%%%%%%%%%%%%%%%%%%%
  A version of these notes is available as an open-access ebook   
  (Springer Lecture Notes in Physics 1047) 
  at
  \href{https://doi.org/10.1007/978-3-032-09893-1}%
  {\tt doi.org/10.1007/978-3-032-09893-1}.
%  An eprint can be found at  
%  \href{https://arxiv.org/abs/...}%
%  {\tt arxiv.org/abs/...}.
  The latest edition, together with the associated computer scripts,  
  is kept up to date at
  \href{https://github.com/laineprocacci/From-inflation-to-hot-big-bang}%
  {\tt github.com/laineprocacci/From-inflation-to-hot-big-bang}.
  }

\vspace*{0.3cm}

Mikko Laine\bit$^\rmi{a}$ 
and 
Simona Procacci\bit$^\rmi{b}{}^{,}_{ }$\hspace*{0.0mm}\footnote{%
 Previously at: 
 D\'epartement de Physique Th\'eorique, 
 % and Center for Astroparticle Physics,
 Universit\'e de Gen\`eve, 
 24 quai Ernest Ansermet, 
 CH-1211 Gen\`eve 4, Switzerland
 }

\vspace*{0.3cm}

$^\rmi{a}$%
{\em
AEC, 
Institute for Theoretical Physics, 
University of Bern, \\ 
Sidlerstrasse 5, CH-3012 Bern, Switzerland \\}

\vspace*{0.3cm}

$^\rmi{b}$%
{\em
KKM BKW AG, M\"uhleberg, Switzerland \\}

\vspace*{0.8cm}

%\mbox{\bf Abstract}

\end{centering}

%\vspace*{0.1cm}
 
\noindent
These lecture notes are meant as a pedagogic guide to cosmological inflation
and the early epochs thereafter. Inflation explains how the seeds for 
density perturbations, which evolved into the largest structures in our
universe, could have formed during a period of exponential
expansion. Apart from density perturbations, also tensor perturbations
are generated, which may be observed as gravitational waves. The
formalism is developed through explicit computations, paying 
attention to general-relativistic gauge invariance, and to
thermalization (the mechanism that converts part of the energy density
driving exponential expansion into the conventional hot big bang). For
the steps best handled numerically or computer-algebraically, simple
{\tt python} scripts are provided.  We aim at an unassuming style,
hopefully accessible to students of theoretical high-energy physics.

\vspace*{1.1cm}

 \centering
 \includegraphics[width=0.55\linewidth]{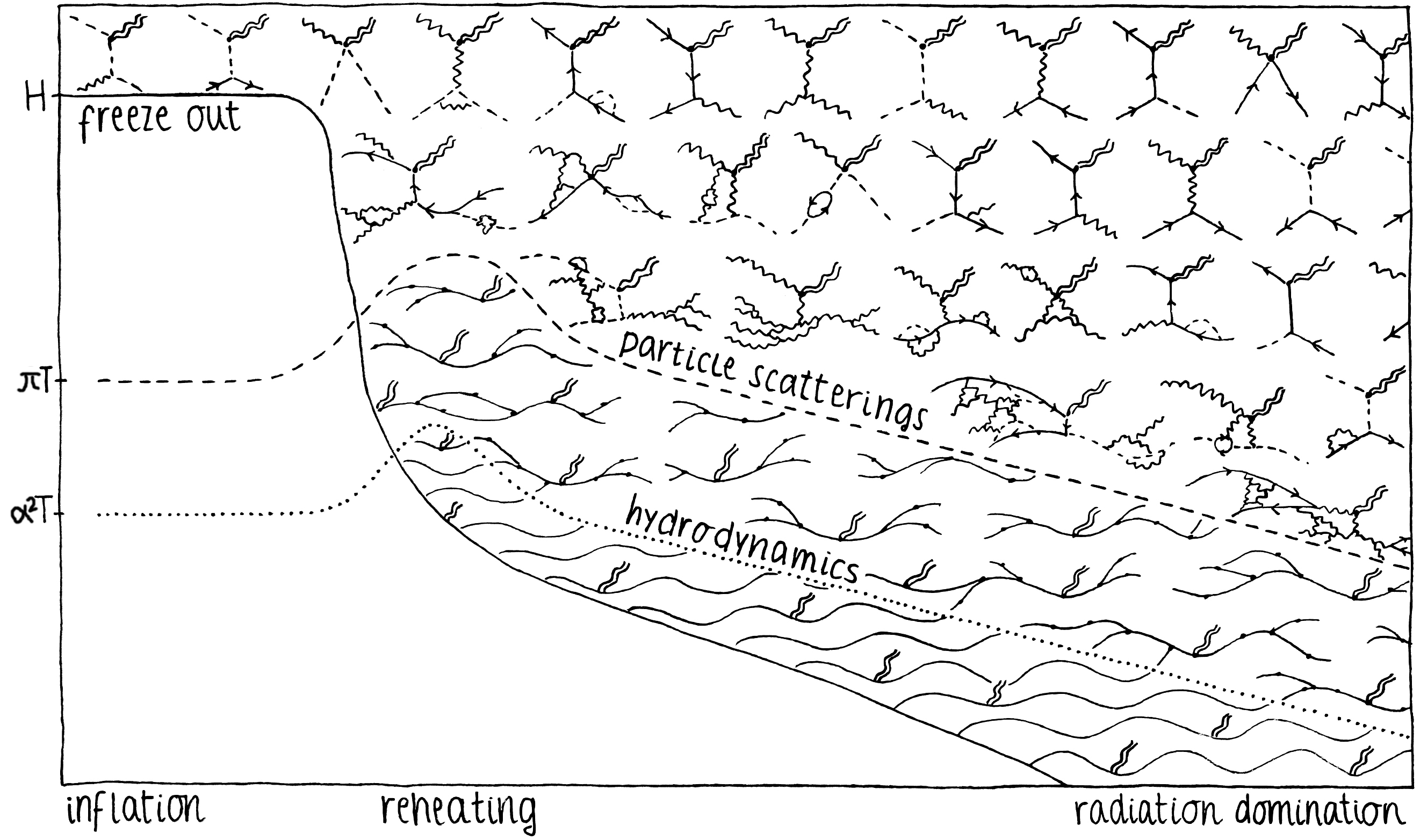} % {fig_scales.pdf}

\vfill

\end{titlepage}

%%%%%%%%%%%%%%%%% EMPTY PAGE AFTER THE COVER %%%%%%%%%%%%%%%%%%%%%% 
% 
% 
% \thispagestyle{empty}
% 
% \mbox{ } 
% 
% \newpage
% 

%%%%%%%%%%%%%%%%%%%% TABLE OF CONTENTS %%%%%%%%%%%%%%%%%%%%%%%%%%%%%%

\pagenumbering{roman}
\setcounter{page}{1}

\tableofcontents

%%%%%%%%%%%%%%%%%%%%%%%%%%%%%%%%%%%%%%%%%%%%%%%%%%%%%%%%%%%%%%%%%%%%%
\newpage

\section*{Preface}
\addcontentsline{toc}{section}{Preface}

These lecture notes 
grew out of the PhD thesis of one of the authors~\cite{Procacci:2023kkj}. 
The latter was, in turn, influenced by unpublished but 
wonderfully detailed lecture notes 
by Hannu Kurki-Suonio~\cite{rev0}. 
Over the years, we have also benefitted 
from many standard books and review articles 
on cosmological perturbations, 
such as refs.~\cite{rev1}--\hspace*{-1.1mm}\cite{rev7}. 
In addition, we have had the chance to present 
parts of this material, notably on primordial gravitational waves, 
at graduate %summer 
schools and small lecture series. 

While preparing lectures, 
we developed the feeling that the basic ingredients
of the inflationary paradigm do not come across in a sufficiently 
transparent manner from existing research monographs, 
despite their otherwise superb quality.
They typically operate, so to say, at a higher level than 
what appears necessary for a first exposure.  
We therefore decided to 
attempt a more elementary treatment, 
specifically for pedagogical purposes,
with the details carefully worked out,  
but with the price that not every modern development is covered. 
That said, we do want to convey a perspective on how  
gravitational-wave observations 
and new astrophysical data may help to probe 
the physics of inflation and reheating in the foreseeable future.

For helpful discussions or collaboration on topics that 
touched upon this work, % (sometimes only remotely), 
we would like to thank (in alphabetical order)
Maria Berti, 
Simone Biondini,
Matthias Blau, 
Dietrich B\"odeker, 
Chiara Caprini, 
Sveva Castello, 
Ruth Durrer,
Miguel Escudero, 
Jacopo Ghiglieri,
Nastassia Grimm,
Armando Hauser, 
Mark Hindmarsh, 
Greg Jackson,  
Keijo Kajantie, 
Philipp Klose, 
Helena Kolesova,
Antonino Midiri,
Germano Nardini, 
Sami Nurmi,
Arttu Rajantie,    
Alica Rogelj,
Stefan Sandner, 
Jan Sch\"utte-Engel, 
Francesco Sorrenti, 
Enrico Speranza,
Anna Tokareva,  
Jorinde van de Vis, 
Sebastian Zell, 
Yan Zhu, 
and many others.  

% \vspace*{5mm}

\noindent
Towards late August 2025, 

\hfill
Mikko Laine (Bern, Switzerland) and  
Simona Procacci (Geneva, Switzerland)

% \vspace*{5mm}

%%%%%%%%%%%%%%%%%%%%%%%%% BIBLIO %%%%%%%%%%%%%%%%%%%%%%%%%%%%%%%%
%

%%%%%%%%%%%%%%%%%%%%%%%%%%%%%%%%%%%%%%%%%%%%%%%%%%%%%%%%%%%%%%%%%%%%%
\newpage 

\section*{Units, notation, and hints on software tools}
\addcontentsline{toc}{section}{Units, notation, and hints on software tools}

We employ units in which the speed of light, $c$, 
the reduced Planck constant, $\hbar$, 
and the Boltzmann constant, $k^{ }_\rmii{B}$, 
equal unity.
The Einstein equations are parametrized
by the Newton constant, $G$, which can 
conveniently be re-expressed via the Planck mass, 
\be
 G \;\equiv\; \frac{1}{\mpl^2}
 \;, \quad 
 \mpl^{ } \;=\; 1.2209 \times 10^{19}_{ }\,\mbox{GeV}
 \;.
 \label{mpl}
 \index{Planck mass}
 \index{$m^{ }_\mathrm{\scriptscriptstyle pl}$ (Planck mass)}
 \index{$G$ (Newton's gravitational constant)}
\ee
In the literature, the so-called
reduced Planck mass, $M^{ }_\rmii{pl} \equiv \mpl^{ }/\sqrt{8\pi}$,
is often employed. \index{reduced Planck mass}
In astrophysics, %
\index{$M^{ }_\mathrm{\scriptscriptstyle pl}$ (reduced Planck mass)} 
distances are expressed in units of a parsec (pc),
\be
 1\,\mbox{pc} \;=\; 3.0857 \times 10^{16}_{ }\,\mbox{m}
 \; \simeq \; 
 3.2616\,\mbox{ly}
 \;, \label{parsec} \index{pc (parsec)}
\ee
where a light year (ly) is the distance that light travels in a year.
Temperatures are measured in energy units,
obtained through a multiplication of Kelvin~(K) 
with the Boltzmann constant,
\be
 k^{ }_\rmii{B} \times (1\hspace*{0.3mm} \mathrm{K} ) 
 \; = \;  
 8.6173 \times 10^{-5}_{ }\hspace*{0.3mm}\mbox{eV}
 \quad \Leftrightarrow \quad
 1\hspace*{0.3mm}\mbox{eV} 
 \; \overset{k^{ }_\rmiii{B}\,=\,1\;\;\;}{=} \;  
 11\,605\,\mbox{K}
 \;. \label{K}
\ee

\vspace*{3mm}

\hrule

\vspace*{3mm}

%%%%%

Indices denoted by small Greek letters describe components 
in the $4$-dimensional space-time, 
$\lbrace \alpha,\beta,\mu,\nu,\rho,\sigma,... \rbrace 
\in \lbrace 0,1,2,3 \rbrace$. 
Small Roman letters are used for spatial indices in three 
dimensions, $\lbrace i, j,k,l,m,n,... \rbrace \in \lbrace 1,2,3\rbrace$.
Derivatives of a generic function $Q$ of an arbitrary variable $x$ and 
coordinates $x^\mu=(x^0,x^i)$ are mostly abbreviated as
\ba
 Q_{,x}^{ }   &\equiv& \partial_x^{ } Q (x, \ldots)	\quad
	\text{derivative with respect to the variable } x 
 \;, \label{der_x} \\[2mm]
%%%%%
 Q_{;\mu}^{ } &\equiv&  D_\mu^{ } Q (x^\mu,\ldots)	\quad
	\text{{\em covariant derivative} with respect to } x^\mu 
 \;. \label{der_mu}
 \index{covariant derivative}
\ea
On a manifold with metric $g^{ }_{\mu\nu}$
and inverse metric $g^{\mu\nu}_{ }$, covariant derivatives read
\ba
 Q^{\alpha\beta\cdots;\nu}_{\rho\sigma\cdots}
 & = & 
 Q^{\alpha\beta\cdots}_{\rho\sigma\cdots;\mu}\, g^{\mu\nu}_{ }
 \;, \label{cov_der_1} \\[2mm]
%%%%%%
 Q^{\alpha\beta\cdots}_{\rho\sigma\cdots;\mu}
 & \equiv & 
 Q^{\alpha\beta\cdots}_{\rho\sigma\cdots,\mu}
 + 
 \Gamma^\alpha_{\mu\nu}\,
 Q^{\nu\beta\cdots}_{\rho\sigma\cdots}
 + 
 \Gamma^\beta_{\mu\nu}\,
 Q^{\alpha\nu\cdots}_{\rho\sigma\cdots}
 -
 \Gamma^\nu_{\mu\rho}\,
 Q^{\alpha\beta\cdots}_{\nu\sigma\cdots}
 -
 \Gamma^\nu_{\mu\sigma}\,
 Q^{\alpha\beta\cdots}_{\rho\nu\cdots}
 + 
 \ldots
 \;, \hspace*{6mm} \label{cov_der_2}
\ea
where $(...)^{ }_{\der\mu} \equiv \partial (...)/ \partial x^\mu_{ }$ is
a partial derivative, 
% (cf.\ \eq\eqref{der_x}), 
and $\Gamma^\alpha_{\mu\nu}$ are the 
{\em Christoffel symbols}. \index{Christoffel symbols}
Repeated indices are summed over, also
when they are purely spatial. 

%%%%%

When the expansion of the universe plays an important role,
it is often useful to work with so-called comoving (or conformal) coordinates, 
and denote them by $\X %\; 
\,\equiv\, %\; 
(\tau, \vec x)$. 
The {\em cosmological scale factor} is denoted by $a$.
\index{$a$ (cosmological scale factor)} 
Physical (or local Minkowskian) coordinates are 
$\R \equiv (t,\vec r)$, 
with $\vec{r} \equiv a\,\vec{x}$.
For spatial momenta, 
$\vec k$ is comoving, 
$\vec p \equiv \vec k/a$ is local Minkowskian. 
The absolute value of the {\em comoving momentum} is $k \equiv |\vec{k}|$.
\index{$k$ (comoving momentum)}
The absolute value of the {\em physical momentum}, $p \equiv |\vec{p}|$, 
\index{$k/a$ (physical momentum)}
unfortunately coincides with the notation used for the 
pressure; given that the dimensions and physical roles 
of these quantities are different, 
it should be clear from the context which is meant. 
More relations between comoving and physical coordinates 
will be introduced in \se\ref{ss:coordinates}.

%%%%%

For Fourier analysis in spatial directions, we use the conventions
\begin{align}
 Q(\,\cdot\,,\vec x) &= \int_\vec{k}\! Q(\,\cdot\,,\vec k)
 \, e^{i\vec{k}\cdot\vec{x}} \ , &
 \int_\vec{k} & \equiv \int\! \frac{\dd^3 \vec k}{(2\pi)^3} \ ,
 \label{fourier_x}  \\[2mm]
%%%%%
 Q(\,\cdot\,,\vec k) &= \int_\vec{x}\! Q(\,\cdot\,,\vec x)
 \, e^{-i\vec{k}\cdot\vec{x}} \ , &  
 \int_\vec{x} &\equiv \int\! \dd^3 \vec x \ .
 \label{fourier_k}
\end{align}
In four-dimensional, conformal and comoving space-time, 
denoting $\K \equiv (\omega,\vec{k})$,  
Fourier analysis is defined as 
\begin{align}
Q(\mathcal{X}) 
	&= \int_\mathcal{K} \! Q(\mathcal{K})
 \, e^{i(-\omega \tau + \vec{k}\cdot\vec{x})} \ , & 
\int_\mathcal{K} 
	&\equiv \int\! \frac{\dd \omega}{2\pi} \int_\vec{k} 
 \,\ , \label{fourier_X} \\[2mm]
%%%%
Q(\mathcal{K}) 
	&= \int_\mathcal{X} \! Q(\mathcal{X})
 \, e^{i(+\omega \tau - \vec{k}\cdot\vec{x})} \ , & 
\int_\mathcal{X} 
	&\equiv \int\! \dd \tau \int_\vec{x} 
 \,\ . \label{fourier_K}
\end{align}
If there is a danger of confusion, we denote Fourier representations
as 
$
 \widetilde Q(\K) \equiv Q(\K)
$.
% With some exceptions, in general we omit hats on quantum operators. 
In local Minkowskian coordinates, the metric signature is taken to 
be $\eta_{00}^{ }=-1$ for the {\em Minkowski metric}, $\eta_{\mu\nu}$.
The scalar product is thus
$
 \K \cdot \X = -\omega\tau + \vec{k}\cdot\vec{x}
$.
\index{$\eta^{ }_{\mu\nu}$ (Minkowski metric)}

\vspace*{5mm}

\hrule

\vspace*{3mm}

In a number of appendices, we provide {\tt python} codes for 
numerical or algebraic computations. 
{\tt Python} is a scripting language, meaning
that the programs are not compiled, but executed line by line. 
The execution can happen via a graphical interface, or as
scripts. 
To give an example, on a {\tt linux} 
operating system, 
the basic kernel and the necessary libraries can be installed as 

\vspace*{-4mm}

{\fontsize{8pt}{10pt}\selectfont
\begin{verbatim}
 sudo apt-get install python3
 sudo apt-get install python3-numpy
 sudo apt-get install python3-scipy
 sudo apt-get install python3-sympy
\end{verbatim}
}

\vspace*{-4mm}

\noindent
Subsequently, a script, if named {\tt *.py},
can be executed from the command line with 
``{\tt 
 python3 *.py}''.
% By issuing just the command {\tt python3}, one gets a prompt {\tt >>>},
% from which commands can be executed one by one. 
Documentation (tutorials and reference manuals) 
can be efficiently found with internet search, with 
{\tt docs.python.org}, 
{\tt numpy.org},
{\tt docs.scipy.org}, and
{\tt docs.sympy.org} being the ``official'' sources.

We have tested our scripts with 
{\tt python} release 3.12.3, 
{\tt numpy 1.26.4}, 
{\tt scipy 1.11.4},
{\tt sympy 1.12.7}. 
New versions can be obtained by running the {\tt install} 
commands again, however the {\tt linux} kernel needs 
to be up-to-date in order for the upgrades to proceed. 

We end by remarking that, at the time of writing, the symbolic 
package {\tt sympy} is not efficient, 
i.e.\ it becomes very
slow with long expressions. However, 
the symbolic routines can be straightforwardly 
transcribed to other languages, such as {\tt Mathematica}. 
All our {\tt python} codes, as well as {\tt Mathematica} transcriptions
of the symbolic routines, are available at 
  \href{https://github.com/laineprocacci/From-inflation-to-hot-big-bang/}%
       {\tt https://github.com/laineprocacci/From-inflation-to-hot-big-bang/}.

%%%%%%%%%%%%%%%%%%%%%%%%%%%%%%%%%%%%%%%%%%%%%%%%%%%%%%%%%%%%%%%%%%%%%
\newpage 

\section*{General outline}
\addcontentsline{toc}{section}{General outline}
\label{literature}

Understanding the origins
of the visible universe is an age-old challenge. 
And yet, we may learn more about this
in the next couple of decades. %. The 
This is thanks to a qualitatively new tool that will become available,
as several gravitational-wave interferometers are foreseen to go 
in operation, opening up new frequency windows for observing
the distant past. 
In addition, 
the properties of the cosmic microwave background, 
and of the large-scale structure in 
galaxies and galaxy clusters, will be
mapped with increasing sensitivity. 

At the same time, on the theoretical side, the predictions 
for early universe observables are not always robust, both because
they are model-dependent, and because 
the simplest computational tools sometimes   
invoke unintended approximations. 
So there is room for progress,    
calling for the attention of the next generations 
of curious and critical minds.

The purpose of these lectures is to offer for a detailed 
hands-on introduction to the earliest
moments of the universe that can presumably be understood {\em without}
having  a theory of quantum gravity at our disposal. The main tool 
is then just general relativity, expanded to linear order in 
perturbations around a cosmological background solution. For certain aspects, 
notably initial conditions, we need to 
recall basic ingredients from quantum field theory as well. 
Statistical physics and elements of 
thermal field theory are employed to study the 
transition from a postulated vacuum-dominated, inflationary, 
universe to the observationally established hot epoch.
We mostly do {\em not} delve into the  
recently popular and interrelated
topics of (quantum) loop corrections and 
(classical) second-order perturbations around cosmological
backgrounds. However, we make an exception
with so-called scalar-induced gravitational waves, which lend us
an opportunity to discuss some of the challenges that arise,
when trying to improve on the accuracy of inflationary predictions
beyond the linear order. 

The overall structure of our presentation is as follows. 
In the first part
\mbox{(chapters~1\bit--\bit4)}, 
we develop the general model-independent 
formalism for 
how small perturbations evolve in a cosmological background, 
and what kind of observable signatures they lead to.
In the second part (chapters 5\bit--\bit10), we turn to the inflationary
paradigm, as a possible explanation for the origin of the perturbations.
The dynamics of inflationary perturbations coupled 
with an emergent thermal plasma and the corresponding observables 
are then studied using the results derived in the first part.

%%%%%%%%%%%%%%%%%%%%%%%%%%%%%%%%%%%%%%%%%%%%%%%%%%%%%%%%%%%%%%%%%%%%%
\newpage %% 

\pagenumbering{arabic}
\setcounter{page}{1}

%%%%%%%%%%%%%%%%%%%%%%%%% CHAPTER / PART I %%%%%%%%%%%%%%%%%%%%%%%%%%%%%%%%%
%

\chapter{Basics of perturbation dynamics in cosmology}

%%%%%%%%%%%%%%%%%%%%%%%%%%% SECTION %%%%%%%%%%%%%%%%%%%%%%%%%%%%%%%%%%
%
\newpage

\section{Einstein equations and the cosmological background solution}
\label{se:bg}

\paragraph{Abstract:}

We review the contents of the Einstein equations of general relativity. 
The ingredients needed for their left-hand side, the Einstein tensor, are 
explained. 
The right-hand side, energy-momentum tensor, is specified for typical
systems appearing in early-universe cosmology (weakly coupled scalar field, 
thermalized plasma, a coupled system). 
The concept of a homogeneous and isotropic
Friedmann-Lema\^itre-Robertson-Walker (FLRW) universe is introduced, 
and prototypical ``background'' solutions of the Einstein equations
are displayed.
We explain why a ``spatially flat'' background is generally adopted
as a sensible initial condition for the universe's evolution, 
even if in the later sections perturbations of
the spatial curvature are introduced and turn out to play a key role. 

\paragraph{Keywords:} 

Cosmic microwave background, blackbody spectrum, 
metric tensor, physical time, conformal time, 
Christoffel symbols, 
Ricci tensor, Ricci scalar, 
Einstein tensor,  
spatial curvature, 
Friedmann equations,  
thermal plasma, 
elementary scalar field, 
Langevin equation,
de Sitter space-time, 
matter domination, 
radiation domination, 
kination.

%\index{tensor perturbations}

%%%%%%%%%%%%%%%%%%%%%%%%%%%%%%%%%%%%%%%%%%%%%%%%%%%%%%%%%%%%%%%%%%%%%%%%%
%
\subsection{Overview of early-universe cosmology}

As in all fields of science, when we attempt to understand 
the early universe, the method is to
use empirical facts as a constraint on the type of mathematical models
that we develop. Arguably, the single most important 
observation is 
the presence of a {\em Cosmic Microwave Background} (CMB) radiation. 
\index{CMB (cosmic microwave background)}
It is an electromagnetic signal that
looks almost the same in every direction in the sky
that we observe. Therefore, 
we assume the universe to be homogeneous and isotropic at %large 
scales as large as our observable patch. 
Furthermore, the CMB displays very accurately 
a {\em blackbody spectrum} (Planck spectrum). From 
\index{blackbody spectrum}
this we deduce that the universe was ``hot'' at early
times, forming a plasma-like state, which emitted the 
CMB photons that we observe. 

We also know that the universe expands. 
The expansion rate is inferred from two types of measurements.
The spectral lines of distant
astrophysical objects show a redshift, 
as if they were moving away from us with a certain velocity. 
Their distance from us 
is estimated from the luminosity of other objects
that are assumed to be well understood, 
such as certain types of supernovea. 
The distance and redshift show an 
approximately linear relationship to each other. 
Extrapolating backwards in time, the objects must have been closer 
to us and to each other at earlier times.

Combining these observations 
has led to the notion of a {\em hot big bang}. 
\index{hot big bang}
There are several other ingredients supporting the paradigm, 
notably successful nucleosynthesis of light elements, a credible
theory for how large-scale structures could form from small initial 
inhomogeneities, and indirect evidence for the presence of a cosmological
neutrino background, having similar thermal characteristics as the CMB. 

Historically, the hot big bang
was envisaged as some kind of an  
{\em initial singularity}, 
\index{initial singularity} 
possibly associated with an infinite temperature. From the
modern perspective, however, we would not like to extrapolate to higher
temperatures than are empirically testable. The reason is that the notion
of a temperature assumes the presence of interactions, so that parts
of the Fock space can be repopulated as the temperature changes. 
Given that interactions require time, 
i.e.\ cannot take place infinitely fast, 
it appears more likely that the initial state of
the universe was some non-thermal (quantum) state, 
which later on equilibrated and reached 
a {\em maximal temperature}. \index{maximal temperature}

Let us add some numbers to this overall picture. 
The hottest epochs 
on which we have indirect evidence are those at which 
{\em big bang nucleosynthesis} (BBN) starts. 
\index{BBN (big bang nucleosynthesis)}
The corresponding temperatures are around
nuclear physics energy scales, 
$T \sim 0.1\,$MeV, and the times, as measured
from the moment at which the temperature would diverge if
we naively extrapolated backwards, are $t > 1$~s. 
At these energies, the behaviour of electrons, positrons, and photons
is described by the well-established theory of 
Quantum Electrodynamics (QED). 
Therefore we are confident 
to find a relativistic plasma
(cf.,\ e.g.,\ refs.~\cite{sw_gc,Schwarz:2003du,ym}), 
where photons interact with ionized matter 
via {\em Compton scattering} \index{Compton scattering}
(sometimes we also refer to its non-relativistic limit,
{\em Thomson scattering}). \index{Thomson scattering}
As the universe cools down to $T
%_\text{dec} 
\sim 3000\,$K, or 0.3~eV, after $%t_\text{dec}
\sim 300\,000$~years, % after $t_\rmii{pl}$, 
free electrons combine with the positive 
protons to form a gas~\cite{Rubakov:2017xzr}, 
in a process that we call \index{recombination} {\em recombination}. %
The universe thus becomes transparent to photons, 
which at this moment {\em decouple}, 
\index{decoupling: photons, CMB}
and start travelling freely across space and time. 
These are the CMB photons that 
we observe 
as background
electromagnetic radiation today. 

Because of the expansion of the universe, resulting in a redshift, 
the decoupled CMB photons are observed to 
have a temperature $T^{ }_\now=2.7255$\hspace*{0.3mm}K 
today~\cite{COBE:1992syq}. 
The subscript $(...)^{ }_\now$ refers to the current value. 
The temperature is almost the same in every direction, with 
fluctuations of the order of $\delta T / T^{ }_\now \sim 10^{-5}_{ }$ 
at small scales~\cite{Planck:2018nkj}. 
Within the visible universe, we observe many structures 
(galaxies, galaxy clusters, etc), but on the average it looks
the same at distances larger than $\sim 200$ Mpc, where the 
astronomical distance unit pc corresponds to about 3 light years, 
or $3.1\times 10^{16}_{ }$\bit{}m
(cf.\ \eq\nr{parsec}).
It is conceivable that the structure seen at distances smaller
than $\sim 200$ Mpc could have formed via gravitational collapse from
early perturbations. Then, 
the early universe would have been almost
homogeneous and isotropic. We return to this issue from 
an {\em a posteriori} perspective at the end of \se\ref{ss:evol_many}.

Given this empirical picture, we describe the early universe
within a {\em perturbative approach}. \index{perturbative approach}
We first determine 
a {\em background solution}, 
which is assumed to be exactly homogeneous and isotropic
(in the language of statistical physics, this could be 
called the {\em mean-field solution}). \index{mean-field solution}  
This is the topic of the present chapter. Subsequently we 
add perturbations, and see how they evolve around the background. 

%%%%%%%%%%%%%%%%%%%%%%%%%%%%%%%%%%%%%%%%%%%%%%%%%%%%%%%%%%%%%%%%%%%%%%%%%
%
\subsection{Choice of coordinates for a homogeneous and isotropic universe}
\label{ss:coordinates}

\index{FLRW metric} %  (Friedmann-Lema{\^{i}}tre-Robertson-Walker)}

We now proceed to a mathematical description of the background solution. 
In a homogeneous and isotropic universe, we expect 
physical quantities $Q$ to only depend on time, and, 
for later purposes, we denote these as 
$Q(\tau, \vec x) \to \bar Q(\tau)$. % (cf.~\eq\eqref{eq_back-pert-dec}).

The basic object of Einstein's general relativity is the space-time metric, 
$g^{ }_{\mu\nu}$, or equivalently, the invariant separation between two
space-time points, 
$
 \dd s^2_{ } = g^{ }_{\mu\nu}\, \dd x^\mu_{ }\, \dd x^\nu_{ }
$, 
where repeated indices are summed over. 
The inverse metric is denoted by $g^{\mu\nu}_{ }$, and it satisfies
\be
 g^{\mu\rho}_{ } g^{ }_{\rho\nu}
 \;=\; \delta^{\mu}_{\nu}
 \;\equiv\; 
 \left\{ 
 \begin{array}{cc}
  1\;, & \mu = \nu \\ 
  0\;, & \mu\neq \nu
 \end{array}
 \right.  
 \;. \label{inverse} 
\ee
An expanding, homogeneous, and isotropic universe is described by the 
Friedmann-Lema{\^{i}}tre-Robertson-Walker (FLRW) metric, 
 \begin{equation}\label{eq_flrw_metric}
 \dd s^2_\rmii{FLRW} 
 \; \equiv \; 
 -\dd t^2 + a^2_{ }(t) \, \dd\vec{x}^2_\kappa                  
 \,, \qquad 
 \dd\vec{x}^2_\kappa    
 \; \equiv \;  
 \frac{\dd r^2}{1-\kappa r^2} + r^2 \, \dd\vec{\Omega}^2
 \,. 
\end{equation}
Here $\kappa$ is a constant representing overall spatial curvature 
(cf.\ \app\ref{app:kappa}), 
and $\dd\vec{\Omega}^2=\dd\theta^2+\sin^2\!\theta \,\dd\phi^2$ 
is the angular distance squared. 
The parameter $a(t)$ is the 
{\em cosmological scale factor}, 
\index{$a$ (cosmological scale factor)}
describing the expansion of the universe, 
and $t$ is the {\em physical time}.  
\index{$t$ (physical time)}
We assume a positive-definite time, $t \in[0,\infty)$, 
even if it is not possible to give a physical definition 
of the moment $t = 0$
(the basic equations are time-translation invariant). 

Another useful coordinate system 
is defined by the {\em conformal time} 
direction, $x^0_{ }=\tau$, such that
\begin{equation}
 \dd\tau  
 \;\equiv\; 
 \frac{\dd t}{a(t)}
 \qquad \Rightarrow \quad 
 \dd s^2_\rmii{FLRW} 
 \; = \;    
  a^2_{ }(\tau) \left[\ -\dd\tau^2_{ } + \dd\vec{x}^2_\kappa \ \right]
 \,. \label{t_tau}
 \index{$\tau$ (conformal time)}
 \index{conformal time -- physical time relation}
\end{equation}
For the conformal time, it is conventional to consider
negative values as well, with the 
initial moment pushed to  $-\infty$.
The name {\em conformal} is motivated by the fact that, 
if the spatial curvature vanishes, $\kappa=0\,$m$^{-2}$, 
the metric in \eq\eqref{t_tau} is related 
to the Minkowski metric by a conformal transformation, 
i.e.\ an overall scaling of dimensionful quantities. 

The relative change of the scale factor $a$ in time measures 
how fast the universe expands, and is called the {\em Hubble rate}.
In the two coordinate systems of \eqs\eqref{eq_flrw_metric} 
and \eqref{t_tau}, it is defined as
\begin{align}
 H   \;\equiv\; \frac{\dot{a}}{a} \,, \quad
 \Hc \;\equiv\; \frac{a'}{a}	  \,, 
	\label{eq_hubble}
 \index{Hubble rate}
 \index{$H$ (Hubble rate in physical time)}
 \index{$\H$ (Hubble rate in conformal time)}
\end{align}
respectively. Here, to distinguish time derivatives 
of an arbitrary physical quantity $Q$, we have introduced the notation
\begin{equation}
 \dot{Q}  \;\equiv\; 	\partial_t^{ }    Q \,,	\quad
 Q\hspace*{0.3mm}' 	 \;\equiv\; 	\partial_\tau^{ } Q	
 \,. \label{Q_deriv}
\end{equation}
With \eq\eqref{t_tau}, we can relate 
the physical Hubble rate and its conformal counterpart,
\begin{align}
\Hc        
 &\;\underset{\rmii{\nr{eq_hubble}}}
  {\overset{\rmii{\nr{t_tau}} \lift }{=}}\;
 aH            \,,   &   \Hc' &\;=\; a^2(H^2 + \dot{H}) \;=\; a\ddot{a}
 \,,\label{eq_HH'}\\[2mm]
%%%%%%%
        Q\hspace*{0.3mm}' 
 &\;\underset{\rmii{\nr{Q_deriv}}}
  {\overset{\rmii{\nr{t_tau}} \lift }{=}}\;
 a\hspace*{0.3mm}\dot{Q} \,,   &
    Q\hspace*{0.3mm}'' &\;=\; a^2(\ddot{Q} + H\dot{Q})     
  \,.\label{eq_HH}
\end{align}

Let us now derive 
the Einstein tensor for a background universe described by the FLRW 
metric with conformal time and generic spatial curvature $\kappa$, 
cf.~\eq\eqref{t_tau}. Using reduced-circumference 
polar coordinates, $(\tau,r,\theta,\phi)$, 
the corresponding background metric is
\begin{align}
 &\bar{g}_{\mu\nu}
 \; 
 \underset{\rmii{\nr{t_tau}}}{
 \overset{\rmii{\nr{eq_flrw_metric}} \lift }{=}}
 \;
 a^2_{ } \, \text{diag}
 \bigl( \, -1 ,\ (1-\kappa r^2_{ })^{-1}_{ } 
 ,\ r^2_{ } ,\ r^2_{ }\sin^2_{ }\theta \,\bigr) \ .
 \label{bg_c}
\end{align}
The {\em Christoffel symbols} 
\index{Christoffel symbols} 
\index{${\Gamma}^\rho_{\mu\nu}$ (Christoffel symbols)}
% corresponding to \eq\nr{bg_c} 
are given by
\begin{equation}\label{eq_ChSy-a}
 \bar{\Gamma}^\rho_{\mu\nu} \; \equiv \; \frac{1}{2} 
 \bar{g}^{\rho\sigma}_{ } ( \bar{g}_{\sigma\mu,\nu}^{ }
  + \bar{g}_{\sigma\nu,\mu}^{ } - \bar{g}_{\mu\nu,\sigma}^{ } )
 \ , \index{Christoffel symbols}
\end{equation}
where $(...)^{ }_{,\mu} \equiv \partial (...) / \partial x^\mu_{ }$.
Noting the symmetry
$
 \bar{\Gamma}^\rho_{\nu\mu} = \bar{\Gamma}^\rho_{\mu\nu}
$, there are $4\times 10$ independent ones. 
Carrying out an explicit computation, 
the non-vanishing among them are
\begin{align}
\bar{\Gamma}^\tau_{\tau\tau} 
 &\;=\; \bar{\Gamma}^r_{\tau r} \;=\; \bar{\Gamma}^\theta_{\tau\theta} \;=\; 
 \bar{\Gamma}^\phi_{\tau\phi} \;=\; \frac{1}{2}a^{-2}_{ }(a^2_{ })'
 \;=\; \Hc 
 \ , \label{eq_chsy-a1}\\[2mm]
%%%
 \bar{\Gamma}^\tau_{rr} &\;=\; \frac{1}{2} 
 \underbrace{(-a^{-2}_{ })}_{\bar g^{\tau\tau}_{ }} 
 \biggl(
 \underbrace{
   -\frac{a^2_{ }}{1-\kappa r^2_{ }}
 }_{-\bar g^{ }_{rr}}
 \biggr)'
 \;=\; \frac{\Hc}{1-\kappa r^2_{ }} 
 \ , \\[2mm]
%%%
 \bar{\Gamma}^\tau_{\theta\theta} &\;=\; \frac{1}{2} 
 (-a^{-2}_{ }) 
 (
 \underbrace{
 -a^2_{ } r^2_{ }
 }_{ -\bar{g}^{ }_{\theta\theta}}
 )' 
 \;=\; r^2_{ }\Hc 
 \ , \\[2mm]
%%%
 \bar{\Gamma}^\tau_{\phi\phi} &\;=\; \frac{1}{2} (-a^{-2}_{ })
 (
 \underbrace{
 -a^2_{ }r^2_{ }\sin^2_{ }\theta
 }_{-\bar g^{ }_{\phi\phi} }
 )' 
 \;=\; r^2_{ }\sin^2_{ }\theta \, \Hc 
 \ , \\[2mm]
%%%%%%%%%%%%%%
 \bar{\Gamma}^r_{rr} &\;=\; \frac{1}{2}
 \underbrace{
 a^{-2}_{ } (1-\kappa r^2_{ })
 }_{\bar g^{rr}_{ } }
 \partial_r^{ } 
 \biggl( 
% \underbrace{
 \frac{a^2_{ }}{1-\kappa r^2_{ }} 
% }_{\bar g^{ }_{rr} }
 \biggr)
 \;=\; \frac{\kappa r}{1-\kappa r^2_{ }} 
 \ , \\[2mm]
%%%
 \bar{\Gamma}^r_{\theta\theta} &\;=\; \frac{1}{2}a^{-2}_{ } (1-\kappa r^2_{ })
 \partial_r^{ } 
 (
% \underbrace{
  -a^2_{ }r^2_{ }
% }_{-\bar g^{ }_{\theta\theta}}
 )
 \;=\; -r\,(1-\kappa r^2_{ }) 
 \ , \\[2mm]
%%%
 \bar{\Gamma}^r_{\phi\phi} &\;=\; \frac{1}{2}a^{-2}_{ } (1-\kappa r^2_{ })
 \partial_r^{ } (-a^2_{ }r^2_{ }\sin^2_{ }\theta)
 \;=\; -r\sin^2_{ }\theta \, (1-\kappa r^2_{ }) 
 \ , \\[2mm]
%%%%%%%%%%%%
 \bar{\Gamma}^\theta_{r\theta} &\;=\; \bar{\Gamma}^\phi_{r\phi} =
  \frac{1}{2}r^{-2}_{ } \partial_r^{ }(r^2_{ }) = r^{-1}_{ } 
 \ ,\\[2mm]
%%%
 \bar{\Gamma}^\theta_{\phi\phi} &\;=\; \frac{1}{2}
 \underbrace{
 a^{-2}_{ }r^{-2}_{ }
 }_{\bar g^{\theta\theta}_{ } }
  \partial_\theta^{ } (
% \underbrace{
 -a^2_{ }r^2_{ }\sin^2_{ }\theta
% }_{-\bar g^{ }_{\phi\phi}}
 )
 \;=\; -\sin\theta \cos\theta
 \ , \\[2mm]
%%%
 \bar{\Gamma}^\phi_{\theta\phi} &\;=\; \frac{1}{2}
 \underbrace{
 a^{-2}_{ } r^{-2}_{ } \sin^{-2}_{ }\theta 
 }_{\bar g^{\phi\phi}_{ }}
 \, \partial_\theta^{ }
 (
% \underbrace{
 a^2_{ }r^2_{ }\sin^2_{ }\theta
% }_{\bar g^{ }_{\phi\phi} }
 )
 \;=\; \frac{\cos\theta}{\sin\theta}  
 \ . \label{eq_chsy-a2}
\end{align}

Given the Christoffel symbols, 
the {\em Ricci tensor} is defined as 
\begin{equation}\label{eq_rt-a}
 \bar{R}_{\mu\nu}^{ } 
 \;\equiv\; 
% \bar{R}^\alpha_{\mu\alpha\nu} 
% \;=\; 
   \bar{\Gamma}^\alpha_{\mu\nu,\alpha}
 - \bar{\Gamma}^\alpha_{\mu\alpha,\nu}
 + \bar{\Gamma}^\beta_{\mu\nu}\bar{\Gamma}^\alpha_{\beta\alpha}
 - \bar{\Gamma}^\beta_{\mu\alpha}\bar{\Gamma}^\alpha_{\nu\beta}
  \ . \index{Ricci tensor}
  \index{${R}_{\mu\nu}^{ }$ (Ricci tensor)}
\end{equation}
Inserting \eqs\eqref{eq_chsy-a1}--\eqref{eq_chsy-a2}, 
the result is diagonal after cancellations, with the entries
\begin{align}
 \bar{R}_{\tau\tau}^{ } &\;=\; 
 \underbrace{
 \Hc' 
 }_{ \bar{\Gamma}^\alpha_{\tau\tau,\alpha} }
 - 
 \underbrace{
 4\Hc' 
 }_{ \bar{\Gamma}^\alpha_{\tau\alpha,\tau} }
 + 
 \underbrace{
 \Hc(4\Hc)
 }_{ \bar{\Gamma}^\beta_{\tau\tau}\bar{\Gamma}^\alpha_{\beta\alpha} }
 - 
 \underbrace{
 4\Hc^2_{ }
 }_{ \bar{\Gamma}^\beta_{\tau\alpha}\bar{\Gamma}^\alpha_{\tau\beta} }
 \;=\;
 -3\Hc' 
 \ , \label{bg_Rdd_tt} \\[2mm]
%%%
 \bar{R}_{rr}^{ } &\;=\; 
 \underbrace{
 \frac{1}{ 1-\kappa r^2_{ }} 
 \biggl( \Hc' + \cancel{\kappa} 
              + \bcancel{\frac{2 \kappa^2_{ }r^2_{ }}
                              {1-\kappa r^2_{ }}}
 \biggr)
 }_{  \bar{\Gamma}^\alpha_{rr,\alpha}  }
 \; - \; 
 \underbrace{
 \biggl[ 
 \frac{1}{ 1-\kappa r^2_{ }} 
 \biggl(         \cancel{\kappa} 
              + \bcancel{\frac{2 \kappa^2_{ }r^2_{ }}
                              {1-\kappa r^2_{ }}}
 \biggr)
  - \cancel{\frac{2}{r^2_{ }}}
 \biggr]
 }_{ \bar{\Gamma}^\alpha_{r\alpha,r} }
 \nonumber\\[2mm]
%%%
 &\quad + 
 \underbrace{
 \frac{1}{ 1-\kappa r^2_{ }} 
 \biggl( 4 \Hc^2_{ } +  2 \kappa 
                     + \bcancel{\frac{\kappa^2_{ }r^2_{ }}
                              {1-\kappa r^2_{ }}}
 \biggr) 
 }_{ \bar{\Gamma}^\beta_{rr}\bar{\Gamma}^\alpha_{\beta\alpha} }
 \; - \;
 \underbrace{
 \biggl[ 
 \frac{1}{ 1-\kappa r^2_{ }} 
 \biggl(      2 \Hc^2_{ } 
              + \bcancel{\frac{\kappa^2_{ }r^2_{ }}
                              {1-\kappa r^2_{ }}}
 \biggr)
  + \cancel{\frac{2}{r^2_{ }}}
 \biggr]
 }_{ \bar{\Gamma}^\beta_{r\alpha}\bar{\Gamma}^\alpha_{r\beta} } 
 \nonumber\\[2mm]
%%%
 &\;=\; \frac{  \Hc' + 2\Hc^2_{ } + 2\kappa  }{1-\kappa r^2_{ }}
 \ , \label{bg_Rdd_rr} \\[2mm]
%%%
 \bar{R}_{\theta\theta}^{ } &\;=\;
 \underbrace{
 r^2_{ }\,(\Hc' + 2 \kappa)
 - ( \cancel{1}-\bcancel{\kappa r^2_{ }} ) 
 }_{  \bar{\Gamma}^\alpha_{\theta\theta,\alpha}  }
 \; + \; 
 \underbrace{
 \cancel{1} + \bcancel{\frac{\cos^2_{ }\theta}{\sin^2_{ }\theta}}
 }_{ -\bar{\Gamma}^\alpha_{\theta\alpha,\theta} }
 \nonumber\\[2mm]
%%%
 &\quad + 
 \underbrace{
 r^2_{ }\,(4 \Hc^2_{ } - \bcancel{\kappa})
 - \cancel{2 ( 1-\kappa r^2_{ } )} 
 }_{ \bar{\Gamma}^\beta_{\theta\theta}\bar{\Gamma}^\alpha_{\beta\alpha} }
 \; - \;
 \underbrace{
 \biggl[ 
 2 r^2_{ }\Hc^2_{ } - \cancel{2 ( 1-\kappa r^2_{ } )}
 + \bcancel{\frac{\cos^2_{ }\theta}{\sin^2_{ }\theta}}
 \biggr]
 }_{ \bar{\Gamma}^\beta_{\theta\alpha}\bar{\Gamma}^\alpha_{\theta\beta} } 
 \nonumber\\[2mm]
%%%
 &\;=\; r^2_{ } (\Hc' +2\Hc^2_{ } + 2\kappa)
 \ , \label{bg_Rdd_thth} \\[3mm] 
%%%
 \bar{R}_{\phi\phi}^{  } &\;=\;
 \underbrace{
 r^2_{ }\sin^2_{ }\theta \,(\Hc' + 2 \kappa)
 - \sin^2_{ }\theta ( \cancel{1} -\bcancel{ \kappa r^2_{ }  } )
 - \cancel{\partial^{ }_\theta [\sin \theta \cos \theta]} 
 }_{  \bar{\Gamma}^\alpha_{\phi\phi,\alpha}  }
 \quad - \; 
 \underbrace{
 0 
 }_{ \bar{\Gamma}^\alpha_{\phi\alpha,\phi} }
 \nonumber\\[2mm]
%%%
 &\quad + 
 \underbrace{
 r^2_{ }\sin^2_{ }\theta \,(4 \Hc^2_{ } - \bcancel{\kappa})
 - \bcancel{2 \sin^2_{ }\theta ( 1-\kappa r^2_{ } )}
 - \cancel{\sin\theta\cos\theta\;\frac{\cos\theta}{\sin\theta}}  
 }_{ \bar{\Gamma}^\beta_{\phi\phi}\bar{\Gamma}^\alpha_{\beta\alpha} }
 \nonumber\\[2mm]
%%%
 &\quad - 
 \underbrace{
 \biggl[ 
   r^2_{ }\sin^2_{ }\theta ( 2 \Hc^2_{ } )
 - \bcancel{2 \sin^2_{ }\theta ( 1-\kappa r^2_{ } )}
 - \cancel{ 2 \sin\theta\cos\theta\;\frac{\cos\theta}{\sin\theta} }
 \biggr]
 }_{ \bar{\Gamma}^\beta_{\phi\alpha}\bar{\Gamma}^\alpha_{\phi\beta} } 
 \nonumber\\[2mm]
%%%
 &\;=\; r^2_{ }\sin^2_{ }\theta (\Hc' + 2\Hc^2_{ } + 2\kappa )
 \ . \label{bg_Rdd_phph}
\end{align}
Raising one index, 
\be
 \mathbin{\bar{R}^\mu_{ }}^{ }_\nu
 \; = \; 
 \bar{g}^{\mu\rho}_{ }\bar{R}^{ }_{\rho\nu} 
 \;, \label{bg_ricciupdown}
\ee
the components of the Ricci tensor simplify to
\begin{align}
 \mathbin{\bar{R}^\tau}_\tau 
 &
 \underset{\rmii{\nr{bg_c}}}
 {\overset{\rmii{\nr{bg_Rdd_tt}} \lift }{=}}
 \;
 3a^{-2}_{ } \Hc' 
 \ , \label{bg_Rud_tt} \\[2mm]
%%%%%
 \mathbin{\bar{R}^r}_r 
 \; 
 \underset{\rmii{\nr{bg_c}}}
 {\overset{\rmii{\nr{bg_Rdd_rr}} \lift }{=}}
 \; 
 \mathbin{\bar{R}^\theta}_\theta
 \; 
 \underset{\rmii{\nr{bg_c}}}
 {\overset{\rmii{\nr{bg_Rdd_thth}} \lift }{=}}
 \; 
 \mathbin{\bar{R}^\phi}_\phi 
 &
 \underset{\rmii{\nr{bg_c}}}
 {\overset{\rmii{\nr{bg_Rdd_phph}} \lift }{=}}
 \;
  a^{-2}_{ }(\Hc' +2\Hc^2_{ } + 2\kappa)
 \ . \label{bg_Rud_rr}
\end{align}
The {\em Ricci scalar} 
(sometimes also referred to as {\em curvature}) 
\index{curvature ($R$): definition}
is therefore 
\begin{equation}
 \bar{R} 
 \;\equiv\; 
 \mathbin{\bar{R}^\mu}_\mu
 \;
 \underset{\rmii{\nr{bg_Rud_rr}}}
 {\overset{\rmii{\nr{bg_Rud_tt}} \lift }{=}}
 \;
 6a^{-2}_{ } ( \Hc' + \Hc^2_{ } + \kappa )
 \ . \label{bg_ricci_scalar}
 \index{Ricci scalar}
 \index{$R$ (Ricci scalar, curvature)}
\end{equation}
This yields the {\em Einstein tensor}, 
for which we consider two versions, 
\be
 \mathbin{\bar{G}^\mu}_\nu
 \; \equiv \; \mathbin{\bar{R}^\mu}_\nu
 - \frac{1}{2} \delta^\mu_\nu \bar{R}
 \;, \qquad
 \bar{G}^{ }_{\mu\nu}
 \; = \; 
 \bar{R}^{ }_{\mu\nu}
 - \frac{1}{2} \bar{g}^{ }_{\mu\nu} \bar{R}
 \;. \label{bg_einsteindowndown}
 \index{$G^{ }_{\mu\nu}$ (Einstein tensor)}
\ee
The former index placement leads to the components 
\begin{align}
 \mathbin{\bar{G}^\tau}_\tau 
 &\;
 \underset{\rmii{\nr{bg_ricci_scalar}}}
 {\overset{\rmii{\nr{bg_Rud_tt}} \lift }{=}}
 \;
 -3a^{-2}_{ } 
 ( \Hc^2_{ } + \kappa ) 
 \ , \label{eq_Gtautaukappa}\\[2mm]
%%%
 \mathbin{\bar{G}^r}_r 
 \;
 = 
% \underset{\rmii{\nr{bg_ricci_scalar}}}{\overset{\rmii{\nr{bg_Rud_rr}}}{=}}
 \;
 \mathbin{\bar{G}^\theta}_\theta 
 \;
 =
% \underset{\rmii{\nr{bg_ricci_scalar}}}{\overset{\rmii{\nr{bg_Rud_rr}}}{=}}
 \;
 \mathbin{\bar{G}^\phi}_\phi 
 &\;
 \underset{\rmii{\nr{bg_ricci_scalar}}}
 {\overset{\rmii{\nr{bg_Rud_rr}} \lift }{=}}
 \;
 -a^{-2}_{ }
 ( 2\Hc' + \Hc^2_{ } + \kappa )
 \ . \label{eq_Grrkappa}
\end{align}
Therefore, evaluating the components of the Einstein equations,
\begin{equation}
 \boxed{
  \quad
  \mathbin{\bar{G}^\mu}_\nu \;=\; 8\pi G\ \mathbin{\bar{T}^\mu}_\nu
  \,, \quad
  \vphantom{\Big|}
  }
 \label{eq_Einstein}
\end{equation}
the off-diagonals vanish. 
We are left with two identities, 
from the $00$ and $ij$-components.

In the literature, the index placement in 
\eq\nr{eq_Einstein} is probably the one most
frequently adopted. Its technical benefit is 
that then ${g^\mu}_{\nu} = \delta^\mu_\nu$ is trivial to all orders, 
simplifying the derivation 
of the Einstein tensor, 
particularly when we go to perturbations. However, 
there is also a drawback, namely that with the mixed index ordering
the Ricci and Einstein 
tensors are in general not symmetric. If we instead consider
\be
 \boxed{
 \quad
    \bar{G}^{ }_{\mu\nu} \;=\; 8\pi G\ \bar{T}^{ }_{\mu\nu}
 \;, \quad
  \vphantom{\Big|}
 }  
 \label{eq_Einstein_symmetric}
\ee
then the tensors are manifestly symmetric. This simplifies 
their decomposition into scalar, vector, and tensor parts, on which
we rely when we study 
perturbations (cf.\ \se\ref{ss:def_perts}). 

%%%%%%%%%%%%%%%%%%%%%%%%%%%%%%%%%%%%%%%%%%%%%%%%%%%%%%%%%%%%%%%%%%%%%%%%%
%
\subsection{Einstein equations for different matter contents}
\label{ss:einstein_background}

In order to solve \eq\nr{eq_Einstein}
or \eq\nr{eq_Einstein_symmetric}, we need to specify the energy-momentum
tensor appearing on the right-hand side. 
To achieve this, physical
input is needed. The simplest system is a thermalized one, because 
in its rest frame its state
is fully characterized by few quantities, a local temperature
and possibly chemical potentials, 
irrespective of the microscopic properties of the particles that 
it is composed of. If the 
system does not equilibrate, specifying its 
energy-momentum tensor requires more information. For instance, we need
to know the spin of the constituents; the simplest example are
spin-0 particles, described by a scalar field. For understanding 
realistic situations, we need to consider multicomponent
systems, with some thermalized degrees of freedom (e.g.,\ electrons and 
photons) and other non-equilibrium ones (e.g.,\ neutrinos and dark matter). 
In the present section, we define the energy-momentum tensor and 
background evolution equations for three prototype systems: 
a thermal plasma,
an elementary scalar field, 
and a two-component system
made of an elementary scalar field 
interacting with a plasma. 

%%%%%%%%%%%%%%%%%%%%%%%%%%%%%%%%%%%%%%%%%%%%%%%%%%%%%%%%%%%%%%%%%%%
%
\subsubsection*{An ideal thermal plasma (or fluid)}

\index{$T^{ }_{\mu\nu}$ (energy-momentum tensor)}
\index{ideal plasma}
\index{perfect fluid}
\index{energy-momentum tensor: perfect fluid}

In the local rest frame of a thermalized plasma, 
the component $- \mathbin{T^0}_0$ of the 
energy-momentum tensor equals the {\em energy density} ($\equiv e$), 
and the components $\mathbin{T^i}_j$, with $i=j$, 
equal the {\em pressure} ($\equiv p$). Denoting by $u^\mu_{ }$ the 
{\em plasma flow velocity},
\index{$u^\mu_{ }$ (plasma flow velocity)} 
and by $\bar{p}$, $\bar{e}$, $\bar{u}^\mu_{ }$ 
the background values of the given quantities,
and assuming homogeneity so that no spatial derivatives appear, 
a covariant form of the energy-momentum tensor reads
\begin{equation}
 \mathbin{\bar{T}^\mu}_\nu
 \;=\;
 (\bar{e}+\bar{p})
 \,\bar{u}^\mu_{ }\bar{u}_\nu^{ }
 + \bar{p}\,\mathbin\delta^\mu_\nu
 \,, \qquad
%%%%%%
 \bar{T}^{ }_{\mu\nu}          
 \;=\; 
 (\bar{e}+\bar{p})
 \,\bar{u}^{ }_\mu\bar{u}^{ }_\nu
  + \bar{p}\,\bar{g}^{ }_{\mu\nu}
 \;. \label{eq_Tmunu_pf}
 \index{$e,\,\bar e$ (energy density)}
 \index{$p,\,\bar p$ (pressure)}
\end{equation}
If $e$ and $p$ are
parametrized by a single quantity, for instance temperature
or chemical potential (but not both at the same time),
their background values 
are related to each other by an {\em equation of state}, 
\index{equation of state}
\begin{equation}
 \bar p  \;=\;    \bar p( \bar e )                        \,, \qquad
 c_s^2 \;\equiv\; \frac{\partial \bar p}{\partial \bar e}
 \,, \label{eq_eos_v1}
 \index{$c_s^2$ (speed of sound squared)}
 \index{equation of state}
\end{equation}
where $c^{ }_s$ is the {\em speed of sound}. Because of isotropy, 
we may assume the background fluid to be at rest, so that $\bar{u}^i=0$. 
Then the background velocity in conformal coordinates is % defined by
\begin{equation}
 \bar{u}^\mu_{ } 
 \; \equiv \;
   \left( \frac{d\tau}{dt},\bm{0} \right) 
 \; \overset{\rmii{\nr{t_tau}}}{=} \; 
 a^{-1}_{ }(1,{\bm 0})                        \,, \qquad\; 
 \bar{u}_\mu^{ }
 \overset{\scriptscriptstyle \bar u^\mu \bar u_\mu\;=\; -1
          \lift }{=} 
 a(-1,\bm{0})                            
 \,. \label{eq_fluid-velocity}
\end{equation}

Inserting \eqs\nr{eq_Tmunu_pf} and \nr{eq_fluid-velocity}
into \eqs\nr{eq_Gtautaukappa} and \nr{eq_Grrkappa}, 
the two identities resulting from the Einstein equations 
of a homogeneous and isotropic universe are 
\begin{align}
 3a^{-2}_{ }(\Hc^2_{ } 		+ \kappa)     	 
  &
  \underset{\rmii{\nr{eq_Tmunu_pf},\nr{eq_fluid-velocity}}}{
  \overset{\rmii{\nr{eq_Gtautaukappa}} \lift }{=}}
  \; 
 8\pi G\, \bar{e}  
           \,, \label{eq_ee-b-00}\\[2mm]
%%%%%%
 -a^{-2}_{ }(2\Hc'+\Hc^2_{ } + \kappa)\delta^{\ibit{}i}_j 
 &
  \underset{\rmii{\nr{eq_Tmunu_pf},\nr{eq_fluid-velocity}}}{
  \overset{\rmii{\nr{eq_Grrkappa}} \lift }{=}}
 \;
 8\pi G\, \bar{p}\ \delta^{\ibit{}i}_j 
 \,. \label{eq_ee-b-ij}
\end{align}
These can be manipulated into
\begin{align}
 \Hc^2_{  } + \kappa 
 &
 \overset{\rmii{\nr{eq_ee-b-00}}}{=} 
 \;
 \frac{8\pi G}{3} \bar{e}a^2_{ }   
          \,, \label{eq_ee-b-1}\\[2mm]
%%%%%
 \Hc'  
 &
 \underset{\rmii{\nr{eq_ee-b-ij}}}
 {\overset{\rmii{\nr{eq_ee-b-00}} \lift }{=}} 
 \; 
 -\frac{4\pi G}{3} (\bar{e}+3\bar{p})\,a^2_{ }
 \,. \label{eq_ee-b-2}
\end{align}

Another piece of information that is frequently helpful is
{\em energy-momentum conservation}. 
\index{energy-momentum conservation}
The Einstein tensor has by construction
a vanishing four-divergence, and correspondingly 
\begin{equation}\label{eq_continuity}
 \mathbin{T^\mu}^{ }_{\nu;\mu}
 \;
 \overset{\rmii{\nr{cov_der_2}}}{=}
 \;
 {T^\mu_{ }}^{ }_{\nu,\mu}
 +
 \Gamma^\mu_{\mu\alpha}{T^\alpha_{ }}^{ }_\nu
 -\Gamma^\alpha_{\mu\nu}{T^\mu_{ }}^{ }_\alpha 
 \;=\; 0
 \;.
\end{equation}
Inserting $\nu = 0$ and making use of 
the Christoffel symbols in \eq\nr{eq_chsy-a1}, 
% \eq\eqref{eq_continuity} 
this amounts to 
\begin{equation}\label{eq_cont-b}
 \mathbin{\bar{T}^\mu}_{0;\mu} 
 \; 
 \underset{\rmii{\nr{eq_chsy-a1}}}
 {\overset{\rmii{\nr{eq_continuity}} \lift }{=}} 
 \;
  \mathbin{\bar{T}^0}'_{0} 
 + \underbrace{4\Hc_{\vphantom{j}}}_{\Gamma^\mu_{\mu\tau}} 
   \mathbin{\bar{T}^0}^{ }_0
  -\underbrace{\Hc_{\vphantom{j}}}_{
       \Gamma^{\tau\vphantom{i}}_{\tau\tau\vphantom{j}}}
   \mathbin{\bar{T}^0}^{ }_0
  -\underbrace{\Hc\delta^i_j}_{\Gamma^i_{j\tau}}
   \,\mathbin{\bar{T}^j}^{ }_i 
 = 0 \quad\Rightarrow\quad 
 \bar{e}\hspace*{0.3mm}' = -3\Hc(\bar{e}+\bar{p})  
 \ .
\end{equation}

The three equations obtained, 
\eqs\nr{eq_ee-b-1}, \nr{eq_ee-b-2} and \nr{eq_cont-b},  
are not independent of each other. 
For instance, \eq\eqref{eq_ee-b-2} follows by differentiating 
\eq\eqref{eq_ee-b-1} with respect to conformal time and using 
the {\em continuity equation} \index{continuity equation}
\eqref{eq_cont-b}. 
To describe the evolution of the background, it is convenient
to choose \eqs\nr{eq_ee-b-1} and \nr{eq_cont-b} as the basic relations.
Converting them to physical time, we thus obtain
\begin{empheq}[box=\fbox]{align}
 \quad   \vphantom{\Bigg|}
 \Hc^2 + \kappa 
 \;
 \overset{\rmii{\nr{eq_ee-b-1}}}{=} 
 \;
 \frac{8\pi G}{3}\bar{e}a^2 
 \hspace*{3mm}
 &
 \hspace*{2mm}
 \underset{\rmii{\nr{eq_HH'} }}{
 \overset{ \tau \leftrightarrow t  \vphantom{ |_q^b } }{\Longleftrightarrow}} 
 \quad
 H^2 + \frac{\kappa}{a^2}  \;=\; \frac{8\pi G}{3}\bar{e} 
 \,, \label{eq_end0-1}\\[2mm] 
%%%%%
 \bar{e}\hspace*{0.3mm}'
 \;
 \overset{\rmii{\nr{eq_cont-b}}}{=}
 \;
 -3\Hc(\bar{e}+\bar{p})     
 \quad 
 &
 \underset{\rmii{\nr{eq_HH'},\nr{eq_HH}}}{
 \overset{ \tau \leftrightarrow t  \vphantom{ |_q^b } }{\Longleftrightarrow}} 
 \quad
 \dot{\bar{e}} + 3 H (\bar{e} + \bar{p} ) \;=\; 0 
 \,. \quad   \vphantom{\Bigg|}
 \label{eq_end0-2}
 \index{Friedmann equations}
\end{empheq}
We refer to this system as the {\em Friedmann equations}. 
\index{Friedmann equations}

%%%%%%%%%%%%%%%%%%%%%%%%%%%%%%%%%%%%%%%%%%%%%%%%%%%%%%%%%%%%%%%%%%%
%
\subsubsection*{An elementary scalar field}

\index{minimally coupled scalar field}

As a second example, we consider a scalar field, $\varphi$, 
\index{$\varphi$ (inflaton, scalar field)}
as constituting the sole matter content of the universe. The scalar
field has a {\em self-interaction potential}, $V(\varphi)$. 
\index{$V$ (self-interaction potential)}
For simplicity
we assume that $\varphi$ is 
{\em ``minimally coupled''} to gravity, implying
that the scalar field action is postulated to contain only the terms
that also appear in Minkowskian space-time. 
Then the corresponding {\em Einstein-Hilbert action} takes 
the form \index{Einstein-Hilbert action}
\begin{align} 
 S 
 \; \supset \;
 \frac{1}{ 16\pi G } \int_\X %  \dd^4x 
 \sqrt{-g} \, R 
 \ - \  
 \int_\X % \dd^4x 
 \sqrt{-g} 
 \left[\ 
 \frac{1}{2}\varphi_{,\mu}\varphi^{,\mu} + V(\varphi)\ 
 \right]
 \ , \label{einstein_hilbert} 
 \index{Einstein-Hilbert action}
\end{align}
where 
$
 g \equiv \det g^{ }_{\mu\nu}
$.  
However, if we include all possible
dimension-4 operators, then the non-minimal term
$\sim \varphi^2_{ }R$ also appears, with its 
associated dimensionless coupling 
(cf.,\ e.g.,\ ref.~\cite{Hertzberg:2010dc} for a review). 
Such terms must generally be included once the theory is 
quantized, given that they can anyways be generated by loop effects. 
The only exception is, if %there is 
a symmetry %which 
forbids them, 
for instance an invariance under a shift $\varphi \to \varphi + c$, 
$c \in \mathbbm{R}$, like in so-called {\em natural inflation}~\cite{ai}.
\index{natural inflation}

With the given action, we can make use of the Hamilton principle, 
leading to the Euler-Lagrange equations of motion. Specifically,
we can  
vary the action with respect to $g^{ }_{\mu\nu}$ and~$\varphi$. 
For the metric variation, we write  
$
 g = \det (\bar g^{ }_{\mu\nu} + \delta g^{ }_{\mu\nu})
 = \det \{(\bar g^{ }_{\mu\nu}) 
 [\mathbbm{1} + (\bar g^{ }_{\mu\nu})^{-1}_{ }( \delta g^{ }_{\mu\nu})]\}
 \approx \bar g \, 
 \{ 1 + \tr[ (\bar g^{ }_{\mu\nu})^{-1}_{ }( \delta g^{ }_{\mu\nu})]\}
 = \bar g \, (1 + \bar g^{\mu\nu}_{ } \delta g^{ }_{\mu\nu} )
$, 
which implies that
$
  %\partial 
  \delta g / 
  %\partial  
  \delta g^{ }_{\mu\nu} = \bar g\, {\bar g}^{\mu\nu}_{ }
$, 
and subsequently 
$
  %\partial 
  \delta \sqrt{-g} / 
  %\partial 
  \delta g^{ }_{\mu\nu} = 
 \frac{1}{2} \sqrt{-\bar g \vphantom{ t } }\, {\bar g}^{\mu\nu}_{ }
$.
Furthermore, expanding the identity in \eq\eqref{inverse} 
at first order in $\delta g_{\mu\nu}$, 
$
 (\bar g^{\mu\rho}_{ } + \delta g^{\mu\rho}_{ })
 (\bar g^{ }_{\rho\nu} + \delta g^{ }_{\rho\nu})
  = 
 \delta^\mu_\nu
$,
leads to 
$
 \delta g^{ }_{\mu\nu} = 
 - \bar g^{ }_{\mu\rho} \bar g^{ }_{\nu\sigma} \delta g^{\rho\sigma}_{ }
$.
Combining these ingredients, we find that
\be
 \frac{ %\partial
 	\delta \sqrt{-g} }{
 		%\partial 
 	\delta g^{\mu\nu}_{} }
 \; = \; 
 -\frac{1}{2} \sqrt{-g}\, g^{ }_{\mu\nu}
 \;. \label{g_variation}
\ee
The variation 
($ 0 = \delta S / \delta g^{\mu\nu}_{ }$)
then yields the Einstein equations 
$
 {G}^{ }_{\mu\nu} = 8\pi G {T}^{ }_{\mu\nu}
$,
where
\begin{equation}
 T^{ }_{\mu\nu}
 \quad
 \equiv
 \quad
 \frac{-2}{\sqrt{-g}} 
 \frac{\delta S|_\varphi}{\delta g^{\mu\nu}_{ }}
 \quad 
 \underset{\rmii{\nr{g_variation}}}{
 \overset{\rmii{\nr{einstein_hilbert}} \lift }{=}} 
 \quad
 \varphi_{,\ibit\mu}\,\varphi_{,\ibit\nu}
 -g^{ }_{\mu\nu} 
 \left( 
 \frac{1}{2}\varphi_{,\alpha}\varphi^{,\alpha} + V 
 \right) 
  \ .
 \label{Tmunu_varphi}
 \index{energy-momentum tensor: scalar field}
\end{equation}
At the same time, 
varying with respect to $\varphi$ ($0 = \delta S / \delta \varphi$)
leads to the scalar field equation
\begin{equation}\label{eq_field-eq}
 (\sqrt{-g}\varphi^{,\mu})_{,\mu}   
 \quad
 \overset{\rmii{\nr{einstein_hilbert}}}{=}
 \quad
   \sqrt{-g} \, V^{ }_{\der\varphi}  
                   \ ,\qquad 
 V^{ }_{\der xy\cdots} \;\equiv\; \partial^{ }_x \partial^{ }_y \cdots V
% V^{ }_{\der\varphi}    \;\equiv\; \frac{ \partial V }{ \partial\varphi }
 \ .
 \index{scalar field equation: vacuum}
\end{equation}

For a background solution, we assume the scalar field 
to be spatially homogeneous, and denote the corresponding value by 
$\varphi(\tau,\bm{x}) \to \bar{\varphi}(\tau)$.
With the metric determinant 
\be
 \sqrt{-\bar{g}(\tau)}
 \; \overset{\rmii{\nr{bg_c}}}{=}\; 
 \frac{ 
 a^4_{ }(\tau)\, r^2_{ } \sin\theta 
 }{ \sqrt{ 1-\kappa r^2_{ } } }
 \quad
 \underset{\rmii{\nr{t_tau}}}{
 \overset{ \tau \leftrightarrow t \vphantom{ |_q^b } }{\Longleftrightarrow}} 
 \quad
 \sqrt{-\bar{g}(t)} \;=\; \frac{ a^3_{ }(t)\, r^2_{ } \sin\theta 
 }{ \sqrt{ 1-\kappa r^2_{ }}  }
 \;, 
\ee 
and noting that 
\be
 \bar\varphi^{,\ibit\mu}_{ }
 \; = \; 
 \bar{g}^{\mu\nu}_{ }
 \bar\varphi^{ }_{,\ibit\nu}
 \; = \; 
 \hspace*{-3mm}
 \overbrace{
 \bar{g}^{\mu\tau\vphantom{t}}_{ } 
  }^{ -a^{-2}_{ } \,\delta^\mu_{\tau\vphantom{t}} }
 \hspace*{-2mm}
 \bar\varphi\hspace*{0.3mm}'
 \; = \; 
 \hspace*{-0mm}
 \overbrace{
 \bar{g}^{\mu t}_{ }
 }^{ - \delta^\mu_t}
 \hspace*{-0mm}
 \dot{\bar\varphi}
 \;, \label{varphi_derivs}
\ee 
where 
$
 \bar{g}^{\mu\nu}_{ }
$ 
is diagonal, only the index choice $\mu=\tau$ or $\mu = t$ gives 
a contribution. We find
\begin{align}
 \bar{g}^{\tau\tau}\sqrt{-\bar{g}}    
 &\;=\; -a^2\frac{r^2 \, \sin\theta}{\sqrt{1-\kappa r^2}} & 
 &\Rightarrow  &  
 (\,\bar{g}^{\tau\tau}\sqrt{-\bar{g}}\,\bar\varphi\hspace*{0.3mm}'\,)' 
 &\;=\; -a^{-2} \sqrt{-\bar{g}} \,
     (\bar\varphi\hspace*{0.3mm}{}'' + 2 \H \bar\varphi\hspace*{0.3mm}{}')
 \ , \label{varphi_kin_tau} \\[2mm]
%%%%%%%
 \bar{g}^{tt}\sqrt{-\bar{g}}          
 &\;=\; -a^3\frac{r^2 \, \sin\theta}{\sqrt{1-\kappa r^2}} & 
 &\Rightarrow & 
 (\,\bar{g}^{tt}\sqrt{-\bar{g}}\,\dot{\bar\varphi}\,)\!\dot{\hphantom{k}} 
 &\;=\; -\sqrt{-\bar{g}} \,
     (\ddot{\bar\varphi} + 3 H\dot{\bar\varphi})        
   \ . \label{varphi_kin_t}
\end{align}
With these, 
the evolution equation % for $\bar \varphi$ 
that follows from \eq\eqref{eq_field-eq} is independent of $\kappa$,
\begin{equation}
 \boxed{ 
 \quad
 \bar{\varphi}'' + 2\Hc\bar{\varphi}'
 + a^2V^{ }_{\der\varphi}( \bar{\varphi} )          
 \;
 \underset{\rmii{\nr{varphi_kin_tau}}}{
 \overset{\rmii{\nr{eq_field-eq}} \lift }{=}} 
 \;
 0
 \qquad 
 \underset{\rmii{\nr{eq_HH}}}{
 \overset{ \tau \leftrightarrow t \vphantom{ |_q^b } }{\Longleftrightarrow}}
 \qquad
 \ddot{ \bar{\varphi} } + 3H\dot{ \bar{\varphi} }
 + V^{ }_{\der\varphi}( \bar{\varphi} ) 
 \;
 \underset{\rmii{\nr{varphi_kin_t}}}{
 \overset{\rmii{\nr{eq_field-eq}} \lift }{=}} 
 \;
 0 \ .
 \quad   \vphantom{\Bigg|}
 }
 \label{eq-F2}
 \index{scalar field equation: background} 
\end{equation}

As far as the energy-momentum tensor from \eq\nr{Tmunu_varphi} goes, 
it is diagonal on the background level, with 
\begin{align}\label{eq_e-p}
  -\mathbin{\bar{T}^0}_0 
&\;
 \underset{\rmii{\nr{varphi_derivs}}}{
 \overset{\rmii{\nr{Tmunu_varphi}} \lift }{=}}
 \;\; 
 \frac{(\bar{\varphi}')^2}{2a^2}+V(\bar{\varphi}) 
 \;\overset{\rmii{\nr{eq_HH'}}}{=}\; 
 \frac{\dot{\bar{\varphi}}^2}{2}+V(\bar{\varphi}) \ , \\[2mm]
%%%%%%%
 \mathbin{\bar{T}^i}_j  
&\;
 \underset{\rmii{\nr{varphi_derivs}}}{
 \overset{\rmii{\nr{Tmunu_varphi}} \lift }{=}}
 \;\; 
 \delta^i_j \biggl[\, 
  \frac{(\bar{\varphi}')^2}{2a^2}-V(\bar{\varphi}) \,\biggr]
 \;\overset{\rmii{\nr{eq_HH'}}}{=}\; 
 \delta^i_j \biggl[\,
 \frac{\dot{\bar{\varphi}}^2}{2}-V(\bar{\varphi}) 
 \,\biggr]
 \ . \label{eq_e-p1}
\end{align}
Using now \eq\eqref{eq_e-p} in connection with \eq\eqref{eq_Gtautaukappa}
yields
\ba 
 \;
 \;
 \Hc^2 + \kappa 
 &
  \underset{\rmii{\nr{eq_Gtautaukappa}}}
  {\overset{\rmii{\nr{eq_e-p}} \lift }{=}}
 &
 \frac{8\pi G}{3} 
 \left[ \frac{(\bar{\varphi}')^2}{2} + a^2V(\bar{\varphi}) \right]
 \label{eq_inflation} \\[2mm]
%%%%%
 \;
  \underset{\rmii{\nr{eq_HH}}}
  {\overset{\rmii{\nr{eq_HH'}}  \vphantom{ |_q^b } }{\Longleftrightarrow}} 
 \; 
 H^2 + \frac{\kappa}{a^2}
 & 
 =
 & 
 \frac{8\pi G}{3}
 \left[ \frac{\dot{\bar{\varphi}}^2}{2} + V(\bar{\varphi}) \right] \ .
 \label{eq_inflation_t}
\ea
From \eqs\nr{eq_e-p1} and \nr{eq_Grrkappa}, we similarly get
\ba
 \;
 \; 
 - ( 2\Hc' + \Hc^2_{ } + \kappa )
 &
  \underset{\rmii{\nr{eq_Grrkappa}}}
  {\overset{\rmii{\nr{eq_e-p1}} \lift }{=}}
 &
 8 \pi G 
 \biggl[\, 
  \frac{(\bar{\varphi}')^2}{2}- a^2_{ }V(\bar{\varphi}) \,\biggr]
 \label{Hp_phi} \\[2mm]
%%%%%
 \;
   \underset{\rmii{\nr{eq_HH}}}
  {\overset{\rmii{\nr{eq_HH'}} \vphantom{ |_q^b }}{\Longleftrightarrow}} 
 \; 
 - 
 \biggl(
 2 \dot{H} + 3 H^2_{ }+ \frac{\kappa}{a^2_{ }} 
 \biggr)
 &
 = 
 & 
 8 \pi G 
 \biggl[\,
 \frac{\dot{\bar{\varphi}}^2}{2}-V(\bar{\varphi}) 
 \,\biggr]
 \;. 
\ea
Eliminating 
$\Hc^2$ with the help of \eq\nr{eq_inflation}, we find 
$\kappa$-independent evolution equations, %(cf.~\eq\eqref{eq_ee-b-2}),
\ba
 \;
 \; 
  \Hc' 
 & 
 \underset{\rmii{\nr{Hp_phi}}}{
 \overset{\rmii{\nr{eq_inflation}} \lift }{=}}
 & 
 -\frac{8\pi G}{3
 } \Bigl[\, (\bar{\varphi}')^2 - a^2 V(\bar{\varphi}) \,\Bigr] 
 \label{eq_Hprime}
 \\[2mm]
%%%%%%%
 \;
  \underset{\rmii{\nr{eq_HH}}}
  {\overset{\rmii{\nr{eq_HH'}}\vphantom{ |_q^b }}{\Longleftrightarrow}}
 \; 
 \dot{H} + H^2 
 & 
  =
 & 
   -\frac{8\pi G}{3}
  \Bigl[\, \dot{\bar{\varphi}}^2 - V(\bar{\varphi}) \,\Bigr]
  \ .
  \label{eq_Hdot}
\ea
Eq.~\nr{eq_Hprime} can also be obtained by taking $\partial^{ }_\tau$
on \eq\nr{eq_inflation} and inserting then \eq\nr{eq-F2}.

%%%%%%%%%%%%%%%%%%%%%%%%%%%%%%%%%%%%%%%%%%%%%%%%%%%%%%%%%%%%%%%%%%%
%
\subsubsection*{A scalar field interacting with a plasma}

\index{scalar field interacting with a plasma}
\index{fluctuation-dissipation relation} 

Finally we consider a system which has simultaneously an elementary
scalar field, which has not equilibrated, and a thermal plasma. If the 
scalar field and the plasma interact with each other, such a system is 
necessarily {\em dissipative} in nature. This means that energy transfer
can take place between the two components: the scalar field feels 
friction, and thereby loses energy to the plasma. At the same 
time, it also experiences
thermal fluctuations, gaining energy from the plasma. 
These processes 
have a relationship to each other, known as the 
{\em fluctuation-dissipation relation}. 
However, they are different kind of processes:
the loss can be associated with a specific differential 
operator, whereas the gain is the result of   
stochastic dynamics, 
void of any ``structure''.

\index{dissipative effective theories}

In the past decades, a framework of 
{\em dissipative effective theories}
has been under development, with the goal of writing down an action 
for a system of the type described, 
generalizing upon \eq\nr{einstein_hilbert}. However, the 
formalism is quite complicated, requiring amongst others
a duplication of degrees of freedom. For our purposes, it is more
straightforward to modify directly the evolution equations.
Let us stress that the evolution equations to be 
introduced should be thought of as 
an {\em effective theory for the slow variables}, 
\index{effective theory for slow variables}
after ``integrating out''
the fast variables. 
As is the case for all effective theories, this implies that the 
description has a limited range of validity.
We return to this in \ch\ref{se:thermal}, 
where a derivation of the effective description is presented.

Let us start from the scalar field evolution equation, previously
in \eq\nr{eq_field-eq}. The presence of a plasma permits for the 
inclusion of two further structures. First of all, the existence 
of a four-velocity, $u^\mu_{ }$, determining the plasma rest frame, 
allows us to define a new
first-order differential operator, $u^\mu_{ }\partial^{ }_\mu$.
Second, the plasma induces a stochastic noise, adding
an inhomogeneous term to the scalar field evolution equation. 
Denoting a covariant derivative by $(...)^{ }_{;\mu}$
(cf.\ \eq\nr{der_mu}), 
the resulting evolution equation, often referred to as the 
{\em Langevin equation}, 
\index{Langevin equation}
thus reads
\be
 {\varphi^{;\mu}_{ }}^{ }_{;\mu} - \Upsilon \, u^{\mu}_{ } \varphi^{ }_{,\mu}
  - V^{ }_{\der\varphi}  
  + \varrho 
 \; = \; 0
 \;. \label{varphi_eq} 
 \index{scalar field equation: thermal}
 \index{Langevin equation}
\ee
Here the dissipative coefficient
$\Upsilon$ is a ``hydrodynamic'' function, similar
to the previous energy density and pressure. Like $V$, 
its form depends on the underlying microscopic
degrees of freedom and 
their interactions with $\varphi$. 
The noise term $\varrho$, on the other hand, describes the 
random kicks that transfer energy from the plasma to $\varphi$.
As it contains no information (or ``no memory''), 
we assume that it correlates as {\em white noise},
\index{white noise} 
\be
 \bigl\langle\,
    \varrho(\X) \varrho(\Y)
 \,\bigr\rangle \; \approx \; 
 \frac{\Omega \, \delta^{(4)}_{ }(\mathcal{X-Y})}{\sqrt{-g}} 
 \;, \quad
 \X \; \equiv \; (x^0_{ },\vec{x})
 \;, \label{varphi_noise}
 \index{noise autocorrelator}
 \index{$\Omega$ (noise autocorrelator)}
\ee
with the {\em noise autocorrelator} $\Omega$ 
depending on the hydrodynamic variables (notably $T$). 
The division in 
\eq\nr{varphi_noise} makes the Dirac-$\delta$ covariant, 
so that 
$
 \int^{ }_\X \sqrt{-g }\, 
 [\, {\delta^{(4)}_{ }(\X)} / {\sqrt{-g}} \,] = 1
$.
The symbol ``$\approx$'' anticipates that 
the Dirac-$\delta$ may need
to be ``regularized'', as discussed after \eq\nr{fl-di} and around the 
end of \app\ref{app:thermal_inflaton}.

As for the energy-momentum tensor of the coupled system, 
its physics is simpler than that of the evolution equation 
in \eq\nr{varphi_eq}, given that a transfer of energy 
from one component  
to another does {\em not affect} the overall energy. Therefore, 
the coefficient $\Upsilon$ and the noise $\varrho$ should
be absent from $T^{ }_{\mu\nu}$. We do remark, however, that 
hydrodynamics is also an effective theory, representing 
an expansion in gradients. %At higher orders in gradients, d
Dissipative coefficients, such as viscosities, as well as 
the associated stochastic noise terms, 
do make an appearance at higher orders in gradients,  
and we return to this in \app\ref{app:viscous}.

Aiming now to combine \eqs\nr{eq_Tmunu_pf}
and \nr{Tmunu_varphi}, 
we write
\be
 T^{ }_{\mu\nu}
   \;
    \underset{\rmii{\nr{Tmunu_varphi}}}{
    \overset{\rmii{\nr{eq_Tmunu_pf}}}{\equiv}} 
   \; 
 \varphi^{ }_{,\ibit\mu}\varphi^{ }_{,\ibit\nu}
 \;-\; \frac{ g^{ }_{\mu\nu}\, 
 \varphi^{ }_{,\alpha} \varphi^{,\alpha}_{ } }{2} 
 \;+\; ( e + p ) \,  u^{ }_{\mu}u^{ }_{\nu}
 \;+\; p \, g^{ }_{\mu\nu}
% + \ldots 
 \;. \label{Tmunu_mixed}
\ee
According to \eq\nr{Tmunu_varphi}, $p \supset - V$, whereas 
$e + p$ contains no $V$. However, thermodynamically, the energy
density is a Legendre transform of the pressure. 
Denoting by $T$ the plasma temperature, we then write 
$e = T s - p \equiv T p^{ }_{,\T} - p $
(cf.\ \se\ref{ss:history}), where $s$ is 
the {\em entropy density}. 
\index{entropy density: definition} 
Therefore, if $V$ is
part of an effective rather than a fundamental theory, and therefore
depends on the temperature 
(representing the {\em free energy density}), then 
\index{free energy density} 
\be
 p \;\equiv\; p^{ }_r - V 
 \;, \quad
 e \;\equiv\; e^{ }_r + V - T V^{ }_{,\T}
 \;, \quad
 e + p \; = \; T s 
 \; = \; T\,(s^{ }_r - V^{ }_{,\T} )
 \;. \label{general_e_p}
 \index{$s$ (entropy density)}
\ee
Here $e^{ }_r$, $p^{ }_r$ and $s^{ }_r$ denote the energy density,
pressure and entropy density
of the ``radiation'' component, respectively, 
defined to be independent
of $\varphi$.
The average values of the combinations in 
\eq\nr{general_e_p}
are denoted by $\bar{e}$ and $\bar{p}$. We remark that, physically, 
these are {\em not} the full energy density and pressure, since the 
contributions from the derivatives 
of $\varphi$ have been kept apart in \eq\nr{Tmunu_mixed}. 

The background field equation, generalizing upon 
\eq\nr{eq-F2}, now becomes
\ba
 \;
 \; 
  \bar\varphi\hspace*{0.3mm}'' 
  +  \bigl( 2 \H  + a \Upsilon \bigr) \bar\varphi\hspace*{0.3mm}'
  + a^2 V^{ }_{\der\varphi} 
  &
  \underset{\rmii{\nr{eq_fluid-velocity},\nr{varphi_eq}}}
  {\overset{\rmii{\nr{eq-F2}} \lift }{=}} 
  &
  0
 \label{bg_varphi_tau} \\[2mm]
%%%
 \;
 \underset{\rmii{\nr{eq_HH}}}{
 \overset{\rmii{\raise1ex\hbox{\nr{eq_HH'}}}}{\Longleftrightarrow}} 
 \;
 \qquad
  \ddot{\bar\varphi} 
  + \,(3 H + \Upsilon)\,\dot{\bar\varphi} + V^{ }_{\der\varphi}
  &
  = 
  &  
  0 
 \;.  
 \label{bg_varphi} 
 \index{scalar field equation: background}
\ea
The noise term, $\varrho$, does not appear on the 
background level, because it is treated as being of the same order
as local perturbations. 
The generalizations of 
\eqs\nr{eq_inflation} and \nr{eq_Hprime} 
are conveniently represented as linear combinations, 
\ba
%%%%
 \; 
 \;
  3 ( \H^2_{ } + \kappa)
  &
  \underset{\rmii{\nr{eq_ee-b-00},\nr{eq_e-p}}}{
  \overset{\rmii{\nr{eq_Gtautaukappa}} \lift }{=}}
  &
  4\pi G
 \bigl[ (\bar\varphi\hspace*{0.3mm}{}')^2_{ }  + 2 a^2 \bar{e}
 \bigr]
 \label{bg_HH_tau} \\[2mm]
%%%%
 \;
 \underset{\rmii{\nr{eq_HH}}}{
 \overset{\rmii{\raise1ex\hbox{\nr{eq_HH'}}}}{\Longleftrightarrow}} 
 \;
 \qquad
 3 \biggl( H^2_{ } + \frac{\kappa}{a^2_{ }} \biggr) 
 &
  = 
 &
  4\pi G \,(\, \dot{\bar\varphi}^2 + 2 \bar{e}\, )
 \;,  \hspace*{6mm}
 \label{bg_HH} 
 \\[3.5mm]
%%%%
 \; 
 \;  
 \H^2 - \H' + \kappa 
 &
  \underset{\rmii{\nr{eq_ee-b-ij},\nr{eq_e-p1}}}{
  \overset{\rmii{\nr{eq_Grrkappa}} \lift }{=}} 
 &
   4 \pi G \bigl[ (\bar\varphi\hspace*{0.3mm}{}')^2_{ }
                                   + a^2(\bar{e} + \bar{p})
                            \bigr]
 \label{bg_ep} \\[2mm] 
%%%
 \;
 \underset{\rmii{\nr{eq_HH}}}{
 \overset{\rmii{\raise1ex\hbox{\nr{eq_HH'}}}}{\Longleftrightarrow}} 
 \;
 \qquad
 \dot{H} - \frac{\kappa}{a^2_{ }}
 &
  = 
 &
 -\,4\pi G \,\bigl(\, 
 \dot{\bar\varphi}^2_{ } + \bar{e} + \bar{p}
 \,\bigr)
 \;. \hspace*{6mm}
 \label{bg_Hprime}
\ea
Finally, the continuity equation from \eq\nr{eq_end0-2} becomes
\ba
%%%%
 \;
 \; 
 -\,
 (\bar\varphi\hspace*{0.3mm}''
 + 2 \H \bar\varphi\hspace*{0.3mm}')\, \bar\varphi\hspace*{0.3mm}' 
 &
 = 
 &
 a^2 \bigl[  \bar{e}\hspace*{0.3mm}' + 3\H (\bar{e} + \bar{p} )\bigr]
 \label{bg_Tmunu_tau}
 \\[2mm]
%%%%%%%%%%%%%%%
 \;
 \underset{\rmii{\nr{eq_HH}}}{
 \overset{\rmii{\raise1ex\hbox{\nr{eq_HH'}}}}{\Longleftrightarrow}} 
 \;
 \qquad
 -\,
 (\ddot{\bar\varphi} + 3 H \dot{\bar\varphi})\, \dot{\bar\varphi}
 &
 = 
 &
 \dot{\bar{e}} + 3 H (\bar{e} + \bar{p} )
 \;.  \label{bg_Tmunu} \hspace*{6mm}
\ea
It is not independent, but can  
be obtained by taking a time derivative of \eq\nr{bg_HH_tau}, 
and employing then
\eq\nr{bg_ep} to eliminate $\H'$
and \eq\nr{bg_HH_tau} to eliminate $3(\H^2_{ }+ \kappa)$.

%%%%%%%%%%%%%%%%%%%%%%%%%%%%%%%%%%%%%%%%%%%%%%%%%%%%%%%%%%%%%%%%%%%%%%%%%
%
\subsection{Examples of analytic background solutions}
\label{ss:bg_rad_mat}

\index{background solution: analytic}

%%%%%%%%%%%%%%%%%%%%%%%%%%%%%%%%%%%%%%%%%%%%%%%%%%%%%%%%%%%%%%%%%%%%%%%%%
%
\subsubsection*{Simplified equations of state}

The evolution equations that we have found in \se\ref{ss:einstein_background}
are coupled first or second-order differential equations. However, 
they are in general non-linear, and not analytically solvable. They also 
contain model-dependent functions, like the equation of state 
$\bar{p} = \bar{p}(\bar{e})$ from \eq\nr{eq_eos_v1}, or the 
self-interaction potential $V(\bar\varphi)$ from 
\eqs\nr{eq-F2}, \nr{eq_inflation} and \nr{eq_Hprime}.

If we make additional assumptions about the model-dependent functions, 
we may obtain simplified equations that can be solved analytically. 
In particular, it is conventional to introduce an
{\it equation of state parameter}, $w$, as 
\index{equation of state parameter}
\be 
  w \;\equiv\; \frac{\bar{p}}{\bar{e}}
  \;,
 \label{def_w}
\ee
and then look for a solution valid for as long as $w$ 
is approximately constant. 

We note from \eqs\nr{bg_HH} and \nr{bg_Hprime} that, if
$\dot{\bar\varphi}^{\ibit 2}_{ }\ll \bar{e}$, then these equations take the 
same form as if the system were an ideal plasma, described by 
\eqs\nr{eq_end0-1} and \nr{eq_end0-2}. 
According to \eq\nr{general_e_p}, $\bar e$ and $\bar p$
include the potential $V$ in this case. 
Specifically, if $\dot{\bar\varphi}^{\ibit 2}_{ }\ll \bar{e}$ and, 
in addition, there is 
no radiation, then \eq\nr{general_e_p} implies that 
$w \stackrel{\rmii{de~Sitter}}{=} -1$. If we employ this assumption
in \eq\nr{eq_end0-2} (i.e.\ we ignore the left-hand side of 
\eq\nr{bg_Tmunu}), then it follows that $\bar{e}$ is approximately constant.
We refer to this situation as {\em de Sitter space-time}.

There are other frequently used equations of state in cosmology.
One is a {\em matter-dominated universe}, 
\index{matter-dominated universe} 
resembling dust, where $\bar{p}=0$
and $w \stackrel{\rmii{matter}}{=} 0$. 
A {\em radiation-dominated universe}
\index{radiation-dominated universe}
is instead approximated with $w \stackrel{\rmii{radiation}}{=} 1/3$, 
which is strictly speaking only correct for non-interacting 
blackbody radiation (cf.\ \eq\nr{p_r}). 
More exotic equations of state also play a role
in some models, for instance {\em kination},
\index{kination}
with $w \stackrel{\rmii{kination}}{=} 1$~\cite{kina1,kina2}.

Once the energy density $\bar{e}$ and the pressure $\bar{p}$
have been related, 
\eqs\eqref{eq_end0-1} and \eqref{eq_end0-2} 
constitute two equations for two unknown variables, $a$ and $\bar{e}$. 
Even before solving these, an important qualitative statement 
can be made. 
From \eq\eqref{eq_HH'}, we learn that if $\Hc'<0$, 
then the expansion of the universe 
is decelerating, $\ddot{a}<0$, whereas $\Hc'>0$
corresponds to an accelerating expansion, $\ddot{a}>0$.
By \eq\eqref{eq_ee-b-2}, after inserting $\bar{p} = w \bar{e}$, 
$w < -1/3$ then leads to an accelerating expansion, 
$w > -1/3$ to a decelerating one. 

%%%%%%%%%%%%%%%%%%%%%%%%%%%%%%%%%%%%%%%%%%%%%%%%%%%%%%%%%%%%%%%%%%%%%%%%%
%
\subsubsection*{De Sitter space-time as an explanation for spatial flatness}

\index{de Sitter expansion} %space-time}

\index{spatial flatness}
\index{geometric curvature}

As a first step towards a solution of the equations, let us 
present an argument for why the parameter~$\kappa$ appearing in them 
could be neglected. 
This is referred to as {\em spatial flatness}: as shown
in \app\ref{app:kappa}, 
the {\em geometric curvature} of a constant-$\tau$ manifold
is proportional to~$\kappa$. 
Spatial flatness 
is considered to be in excellent agreement with empirical 
observations, which indicate that, 
at the current time, % ($\tau^{ }_\now$ or $t^{ }_\now$)
$
 \Omega^{ }_K \equiv 
 | \kappa | % c^2_{ }
   / ( a^2_\now H^2_\now )  
 < 10^{-3}_{ }
$~\cite{Planck:2018nkj}.
Moreover, going to earlier times, 
$1/(a^2_{ } H^2_{ }) \propto \mpl^2 / T^2_{ } $ decreases rapidly
(cf.\ \se\ref{ss:history}), implying that the dimensionless
ratio  
$
 | \kappa | % c^2_{ }
   / ( a^2_{ } H^2_{ } )  
$
is extremely small. 

\index{$\Omega^{ }_K$ (constraint on spatial curvature)}

Let us consider a scenario in which the energy density remains
unchanged at some initial value ($\bar{e}^{ }_i$) for a long time, 
$\bar{e} \equiv \bar{e}^{ }_i \gg \dot{\bar\varphi}^{\ibit 2}_{ }$. 
Physically, we characterize this situation by
saying that the energy density is dominated 
by a {\em cosmological constant}. 
\index{cosmological constant}
Then the Friedmann
equation from \eq\nr{eq_end0-1} takes the form 
\be
 \frac{\dot{a}^2_{ }}{a^2_{ }} + \frac{\kappa}{a^2_{ }}
 \; 
 \overset{\rmii{\nr{eq_end0-1}}}{=} 
 \; 
 \frac{8\pi G \bar{e}^{ }_i}{3} 
 \; \equiv \;
 H_i^2 
 \; \approx \; {\rm constant}
 \;. 
 \label{friedmann_Hi} 
\ee
Here we have introduced the notation $H^{ }_i$ for a quantity which
would agree with the Hubble rate if $\kappa = 0$. Our goal is 
to show that even if $\kappa \neq 0$, the actual Hubble rate soon
approaches the value~$H^{ }_i$, so that effectively $\kappa$
plays no role. 

Mathematically, \eq\nr{friedmann_Hi} is a non-linear first-order
differential equation. It is also symmetric in time reversal, 
$t\to -t$, so the solution should reflect this symmetry: 
if there is an exponentially growing solution, 
there must be an exponentially shrinking solution. 

In order to find a solution, we may first multiply \eq\nr{friedmann_Hi}
with $a^2_{ }$, obtaining
\be
 \dot{a}^2_{ } - H_i^2 a^2_{ } \;=\; -\kappa
 \;. \label{friedmann_Hi2}
\ee
We now take a time 
derivative of \eq\nr{friedmann_Hi2}. Given that the right-hand side
is a constant, this yields
\be
 2 \dot{a} \bigl( \ddot{a} - H_i^2 a ) 
 \; = \; 
 0 \;. 
\ee
Assuming that $\dot{a}\neq 0$, yields a linear differential 
equation of second order, which is immediately solved, 
\be
 \ddot{a} - H_i^2 a \;=\; 0 
 \quad \Rightarrow \quad
 a \;=\; a^{ }_+\, e^{H^{ }_i t}_{ }
   + a^{ }_-\, e^{- H^{ }_i t }_{ } 
 \;, \quad
 \dot{a} \;=\; H^{ }_i \, \Bigl( 
     a^{ }_+\, e^{H^{ }_i t}_{ }
   - a^{ }_-\, e^{- H^{ }_i t }_{ } \Bigr) 
 \;. \label{friedmann_soln1}
\ee
Here $a^{ }_+$ and $a^{ }_-$ are integration constants. 
However, the original \eq\nr{friedmann_Hi2} only requires
one integration constant, so we should substitute 
\eq\nr{friedmann_soln1} back into \eq\nr{friedmann_Hi2}
to obtain the information that we are missing. 
On the left-hand side, most terms cancel, yielding
\be
 -4 H_i^2 a^{ }_+ a^{ }_-
 \; 
   \underset{\rmii{\nr{friedmann_soln1}}}{
   \overset{\rmii{\nr{friedmann_Hi2}} \lift }{=}}
 \; 
  -\kappa
 \quad 
  \Leftrightarrow 
 \quad 
 a^{ }_- \;=\; \frac{\kappa}{4H_i^2 a^{ }_+}
 \;. \label{soln_c2}
\ee
Inserting this into $a$ and $\dot{a}$ from 
\eq\nr{friedmann_soln1} finally gives 
\be
 \frac{\dot{a}}{a}
 \;
   \overset{\rmii{\nr{friedmann_soln1}} \lift }{
   \underset{\rmii{\nr{soln_c2}}}{=}}   
 \; 
 H^{ }_i 
 \, 
 \frac{
    4 a^2_+ H^2_i  -  
    \kappa\, e^{-2H^{ }_i t}_{ }
      }{
    4 a^2_+ H^2_i  + 
    \kappa\, e^{-2H^{ }_i t}_{ }
      }
 \;. \label{friedmann_soln2}
\ee

We see from \eq\nr{friedmann_soln2}
that, if $a^{ }_+ \neq 0$ and 
$|\kappa|/( a_+^2 H_i^2 ) < 4 $, the Hubble rate, 
$\dot{a}/a$,
agrees with $H^{ }_i$ 
up to exponentially small corrections,
as soon as $t \gg H_{i}^{-1}$. 
This is the reason why $\kappa$ can be omitted, if it is assumed that 
the universe underwent a period 
with $\bar{e}^{ }_i \approx\,$constant.
The resulting exponential expansion is 
referred to as {\em inflation}~\cite{Guth:1980zm_copy,ori2_copy,ori3_copy,%
chaotic_copy}. 
\index{inflation: definition}
We will omit $\kappa$ in the rest of this book. However, at early
times, a non-zero value of $\kappa$ can play an important role. 
In particular, 
it is mathematically possible that $\dot{a}/a$ crosses zero, 
transferring from a shrinking to an expanding universe, 
either smoothly (if $\kappa > 0$) 
or through a singularity (if $\kappa < 0$).
We will not elaborate on these possibilities, 
however they have been discussed in the literature 
(cf., e.g., ref.~\cite{Novello:2008ra} for a review). 

%%%%%%%%%%%%%%%%%%%%%%%%%%%%%%%%%%%%%%%%%%%%%%%%%%%%%%%%%%%%%%%%%%%%%%%%%
%
\subsubsection*{A prototype solution for a complete history}

Let us now patch together a complete solution of a possible 
cosmological history. 
At early times, 
assuming $a^{ }_+ \neq 0$
in \eq\nr{friedmann_soln2}, 
setting $\kappa = 0$,  
and solving the equation, we find 
\be
 a(t)
 \;
  \underset{\scriptscriptstyle \kappa\,\to\,0}{
  \overset{\rmii{\nr{friedmann_soln2}} \lift }{=}}
 \;
 a^{ }_i \, e^{ H^{ }_i (t-t^{ }_i) }_{ }
 \;, \label{at_dS}
 \index{$t^{ }_i,\,\tau^{ }_i,\,a^{ }_i,\,H^{ }_i,\,...$ 
 (values at initial moment)}
\ee
where we denote the values of various variables 
at an {\em initial moment} by a subscript $(...)^{ }_i$. 
In principle, $t$ could be defined on the real axis, 
but it is conventional to restrict it to positive values, 
and therefore $t \ge t^{ }_i > 0$. 
As the form of the solution shows, it would be ``natural'' to 
express time in units of $t^{ }_i$, and furthermore to 
choose $t^{ }_i \equiv H^{-1}_i$, because then the variation
of the solution is of $\ord(1)$ 
when the variation of $t/t^{ }_i$ is of $\ord(1)$.

 \index{$t^{ }_e,\,\tau^{ }_e,\,a^{ }_e,\,H^{ }_e,\,...$
 (values at end of inflation)}

We assume the solution in
\eq\nr{at_dS} to hold until a time $t = t^{ }_e$, where the subscript
stands for the {\em end of inflation}. The task then is to determine how
the solution looks like at $t > t^{ }_e$. 
The transition from $t^{ }_e$ to the onset of radiation 
domination 
is called {\em reheating} \index{reheating: definition}
(cf.~\se\ref{sec_T}). 
For a simple sketch, 
we treat the transition 
%in a framework of {\em instantaneous reheating}
{\em instantaneously}, assuming that at $t > t^{ }_e$,
the inflaton field suddenly does not play any role any more
(actually, this assumption can be relaxed, cf.\ 
the discussion following \eq\nr{damped_ho}). 

Assuming that at $t\ge t^{ }_e$, the 
scalar field kinetic energy $\dot{\bar\varphi}^2_{ }$ can be omitted, 
and inserting $\bar{p}=w\bar{e}$ in \eq\nr{eq_end0-2}, 
we consider the background dynamics predicted by 
\be
 0  
 \; 
 \underset{\rmii{\nr{def_w}}}{
 \overset{\rmii{\nr{eq_end0-2}} \lift }{=}} 
 \;
 \dot{\bar{e}} + 3 H \bar{e} (1 + w)
 \; = \; 
 \frac{\partial^{ }_t [\bar{e}\, a^{3(1+w)}_{ }]}{a^{3(1+w)}_{ }} 
 \quad \Rightarrow \quad 
 \bar{e}\, a^{3(1+w)}_{ } \; = \; \mbox{constant}
 \;. \label{bg_Tmunu_appro_t_2}
\ee
It follows from 
\eq\nr{bg_Tmunu_appro_t_2} that 
\be
 \bar{e}
 \;
  \overset{\rmii{\nr{bg_Tmunu_appro_t_2}}}{=}
 \;
 \frac{\bar{e}_e \, a_e^{3(1+w)}}{a_{ }^{3(1+w)}}
 \;, \quad
 t \ge t^{ }_e
 \;. \label{bare_rad}
\ee
Inserting into \eq\nr{eq_end0-1} [with $ \kappa = 0$], 
and making use of the fact that $H$ and $\bar{e}$ are constant 
during the previous de Sitter period, we find
\ba
 H^2_{ }
 \;
  \underset{\rmii{\nr{bare_rad}}}{
  \overset{\rmii{\nr{eq_end0-1}} \lift }{=}} 
 \;
          \overbrace{ 
          \frac{8\pi G \bar{e}_e}{3} }^{\,=\, H_i^2}
          \frac{a^{3(1+w)}_e}{a^{3(1+w)}_{ }}
 & \Leftrightarrow & 
 \dot{a}
 \; = \; 
 H^{ }_i\, a_e^{ \frac{3}{2}(1+w) } a^{\frac{1}{2}(-1-3w) }_{ }
 \nn[2mm]
%%%%
 & \Leftrightarrow & 
 {\rm d}a \, { a^{\frac{1}{2}(1+3w) }_{ }}
 \; = \; 
 H^{ }_i\, a_e^{ \frac{3}{2}(1+w) } {\rm d}t
 \;. \label{HH_t}
\ea
For $w\neq -1$, this can be integrated into 
\ba
 a^{ \frac{3}{2}(1+w) }_{ }
 - 
 a^{ \frac{3}{2}(1+w) }_e
 &
   \underset{\scriptscriptstyle w\,\neq\,-1}{
   \overset{\rmii{\nr{HH_t}} \lift }{=}} 
 &
 a^{ \frac{3}{2}(1+w) }_e
 \frac{
      3(1+w) 
      H^{ }_i (t - t^{ }_e)
      }{2}
 \nn[2mm]
%%%%%%%%%%%%%%%%%%%%
 \Leftrightarrow 
 \quad 
 a 
 &
 =
 & 
 a^{ }_e 
 \biggl[ 1 + 
   \frac{
      3(1+w) 
      H^{ }_i (t - t^{ }_e)
      }{2}
 \biggr]^{\frac{2}{3(1+w)} }_{ }
 \;. 
 \label{soln_bg_t_a}
\ea
For $w=-1$, we go back to \eq\eqref{at_dS}. 

%%%%%%%%%%%%%%%%%%%%%%%%%%%% FIGURE %%%%%%%%%%%%%%%%%%%%%%%%%%%%%%%%%
%
\begin{figure}[t]
    \centering
    \includegraphics[width=0.44\linewidth]{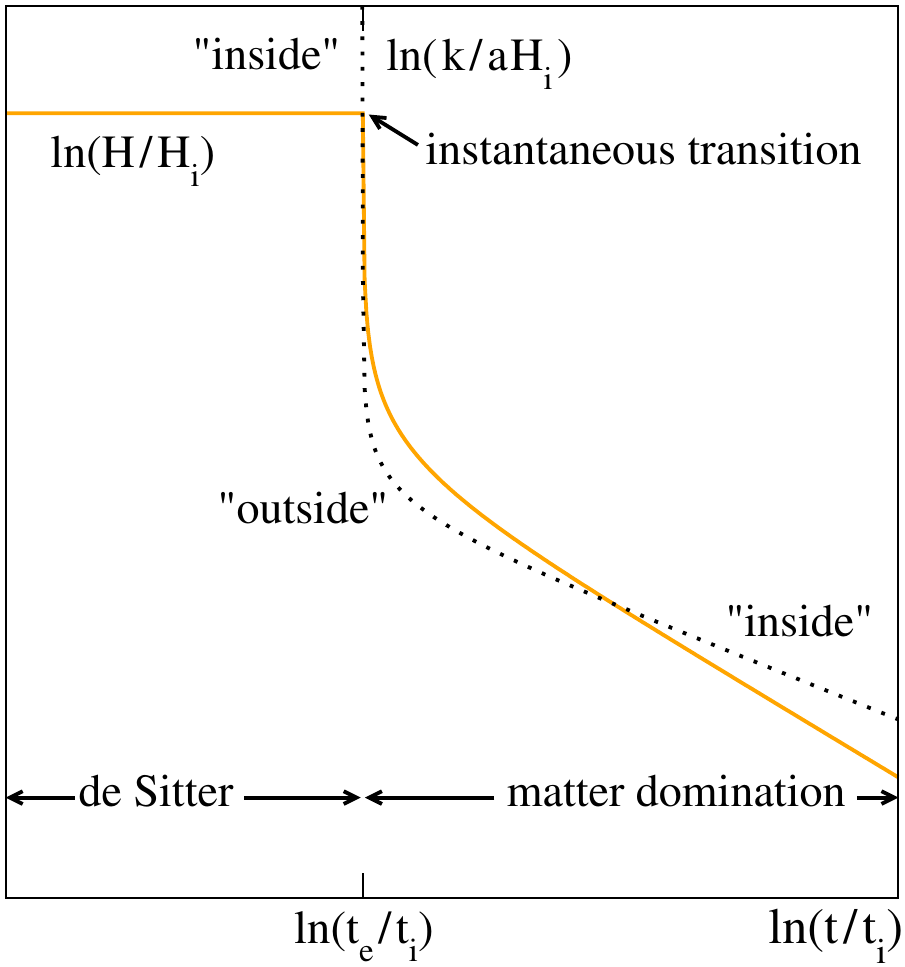}%
    \hspace*{8mm}\vspace*{-2mm}
    \includegraphics[width=0.44\linewidth]{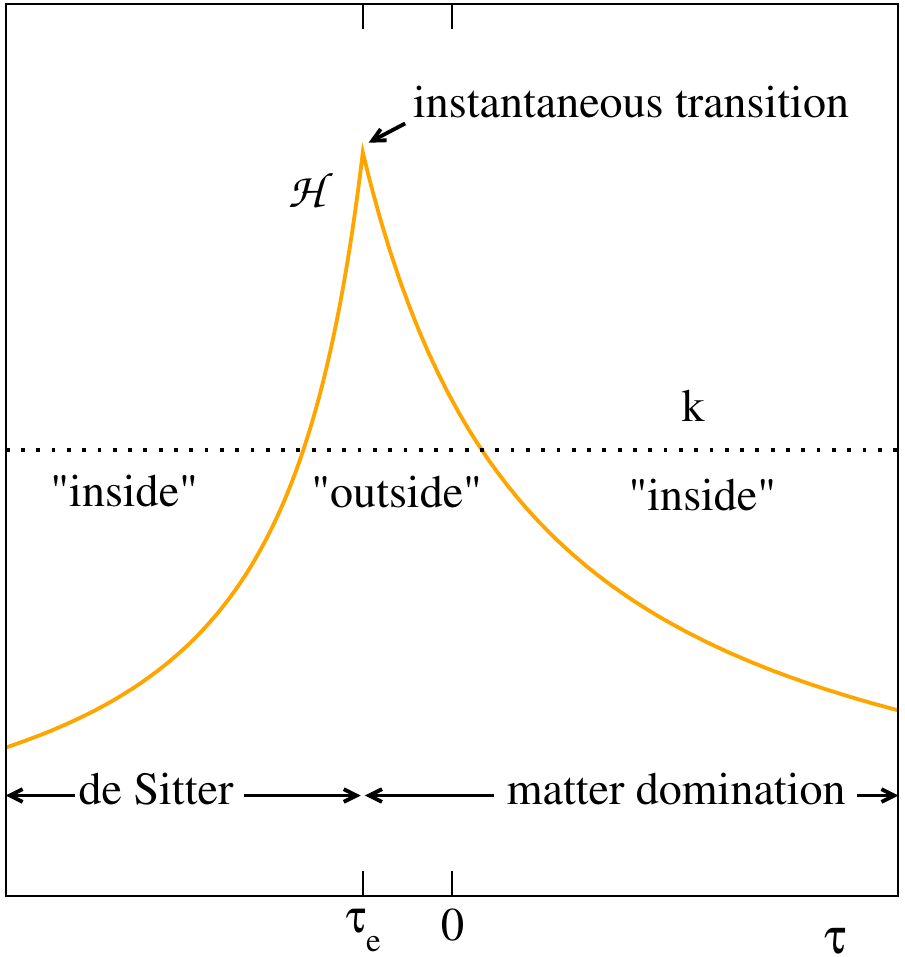}
    \caption{%
     \small
     Sketch of the Hubble rate, $H$, 
     as a function of physical time 
     (left, cf.\ \eq\nr{a_t_mat_1}) and 
     conformal time
     (right, cf.\ \eqs\nr{a_tau_dS_2} and \nr{H_tau_mat_1}).
     By $t^{ }_i$ we denote an initial time, at which the 
     Hubble rate takes the constant value $H^{ }_i \equiv H(t^{ }_i)$, 
     and by $t^{ }_e$ the end of inflation, at which $H$ 
     starts to evolve. 
     Inflation is assumed to end instantaneously, 
     and we have set $w=0$ afterwards (cf.\ \eq\nr{def_w}), 
     which corresponds to matter domination. 
     The dotted line illustrates a comoving 
     momentum, $k/a$ (left) or $k$ (right). 
     The left panel can be compared with a more 
     realistic smoothly evolving 
     numerical solution, shown in 
     \fig\ref{fig:bg_num} on p.~\pageref{fig:bg_num}
     (the axes are logarithmic in both plots).
     }
    \index{cosmic history: sketch (figure)}
    \index{figure: sketch of simple cosmic history}
    \label{fig:history_tau}
\end{figure}
%
%%%%%%%%%%%%%%%%%%%%%%%%%%%%%%%%%%%%%%%%%%%%%%%%%%%%%%%%%%%%%%%%%%%%

The Hubble rate corresponding
to \eq\nr{soln_bg_t_a}, $H = \dot{a}/a$, reads
\be
 H 
 \;
  \overset{\rmii{\nr{soln_bg_t_a}}}{=} 
 \; \frac{2 H^{ }_i}{2 + 3(1+w) H^{ }_i ( t - t^{ }_e ) }
 \;, \quad t \ge t^{ }_e 
 \;. \label{a_t_mat_1} 
\ee    
This illustrates
the prototypical cosmological history in physical time: 
during inflation, the expansion rate, $H$, 
stays constant at $H^{ }_i$ ($w\approx -1$), and afterwards, it 
decreases ($w\ge 0$). 
To understand the solution qualitatively,
and anticipate a key concept of the following chapters,  
we compare $H$ from \eq\nr{a_t_mat_1}
with the evolution of a physical momentum, $p \equiv k/a$
(cf.\ discussion between \eqs\nr{cov_der_2}
and \nr{fourier_x}), with $k$ being constant 
and $a$ as given by \eq\nr{at_dS} or \eqref{soln_bg_t_a}.
During de Sitter expansion, $k/a$ decreases exponentially,
and afterwards as a power law. Whenever a given momentum $k$ 
satisfies $k\gg aH$, we say that it is 
{\em inside the Hubble horizon},
\index{Hubble horizon} meaning that local processes dominate over the 
dilution caused by the expansion. If instead $k\ll aH$, 
the expansion is the dominant effect, and we say that 
the momentum mode is {\em outside the Hubble horizon}.

The behaviour in physical time is sketched in 
\fig\ref{fig:history_tau}(left). The corresponding solution
in conformal time is shown in 
\fig\ref{fig:history_tau}(right), 
with the technical details described
in \app\ref{app:bg_tau}.
In \app\ref{app:num_bg_vac}, 
we also present a numerical solution, which involves 
a smooth rather than an instantaneous transition from the
inflationary period to a matter-dominated epoch. 
We note that in the literature, 
{\em instantaneous reheating} 
\index{reheating: instantaneous}
often denotes 
a sharp transition from inflation to a radiation-dominated era,
however the qualitative shapes of the plots remain the same,
whether $w=1/3$ (radiation domination) or 
$w=0$ (matter domination). 

\index{time-temperature relationship}

Apart from the relationship between time and Hubble rate, it is often
interesting to know the {\em relationship between time and temperature}. 
Let us work this out for the radiation-dominated epoch ($w=1/3$). Then
the energy density scales as $T^4_{ }$ (cf.\ \eq\nr{Hubble}), with
a coefficient conventionally 
parametrized through an 
{\em effective number of light degrees of freedom},
\index{effective number of degrees of freedom} 
$g^{ }_*$ (cf.\ \eq\nr{p_r}). Thereby 
\eq\nr{a_t_mat_1} can be expressed as 
\ba
 H 
 & 
 \underset{\scriptscriptstyle t\;\gg\;t^{ }_e\,,\, H_i^{-1}}{
 \overset{\rmii{\nr{a_t_mat_1}} \lift }{\approx}} 
 & 
 \frac{2}{3(1+w)t}
 \;\hspace*{2.0mm} 
 \overset{w\,=\,\frac{1}{3}}{=}
 \;\hspace*{2.0mm}
 \frac{1}{2t}
 \;
 \underset{\rmii{\nr{p_r}}}{
 \overset{\rmii{\nr{Hubble}}}{\approx}}
 \;
 \sqrt{\frac{8\pi}{3}\frac{g^{ }_*\pi^2_{ }}{30}}
 \, 
 \frac{T^2_{ }}{\mpl^{ }}
 \\[2mm]
%%%%%
 \;
 \Rightarrow 
 \;
 \quad 
 t 
 &
 \approx
 & 
 \frac{3\sqrt{5}}{4\sqrt{g^{ }_* \pi^3_{ }}}
 \, 
 \frac{\mpl^{ }}{T^2_{ }}
 \; 
 \approx
 \; 
 \frac{0.301}{\sqrt{g^{ }_*}}
 \, 
 \frac{\mpl^{ }}{T^2_{ }}
 \;. \label{t_T}
\ea
When we determine the abundance of some cosmic relic, 
it is sometimes practical to integrate over $T$ rather
than $t$, and then the Jacobian following
from \eq\nr{t_T} is needed, 
\be
 \frac{{\rm d}t}{{\rm d}T}
 \; 
 \approx
 \; 
 - \frac{3\sqrt{5}}{2\sqrt{g^{ }_* \pi^3_{ }}}
 \, 
 \frac{\mpl^{ }}{T^3_{ }}
 \;. \label{dt_dT}
\ee
Since we have treated $g^{ }_*$ as 
constant, \eq\nr{dt_dT} is not exact; 
the unapproximated version reads
$
 {\rm d}T/{\rm d}t = - 3 c_s^2(T) T H(T)
$, 
where $c_s^{2}$ 
is the speed of sound squared
(cf.\ \eq\nr{eq_eos_v1}).

\newpage

%%%%%%%%%%%%%%%%%%%%%%%%%%%% start appendices %%%%%%%%%%%%%%%%%%%%%%%%%%%%%%%

%%%%%%%%%%%%%%%%%%%%%%%%%%%%%%%%%%%%%%%%%%%%%%%%%%%%%%%%%%%%%%%%%%%%%%%%%
%
\subsubsection{More on spatial curvature}
\label{app:kappa}

\addcontentsline{toc}{subsection}{\App\ref{app:kappa}: 
More on spatial curvature}

\index{spatial curvature}

We elaborate here on the meaning of the parameter~$\kappa$,
defined in \eq\nr{eq_flrw_metric}, and argued in 
\eq\nr{friedmann_soln2} to be insignificant at late times, 
if the universe underwent de Sitter expansion at early times.

To get started, it is good to think about the nature of different parameters
that characterize our cosmological model. One class of them are Lagrangian 
parameters of elementary particles and fields (masses, coupling constants). 
In an ideal ``reductionist'' philosophy, we should try to determine the
full solution in terms of such parameters, but in practice a more modest
goal needs to be set. As less fundamental parameters we may consider
``derived quantities'', such as an equation of state, in terms of which we
may express the right-hand side of the Einstein equations. Yet another
set of parameters is given by initial conditions. Ideally, we could 
hope to find a solution which is independent of initial conditions
(for example, a fully thermalized state has no memory of where it 
came from). However, 
often some initial conditions are needed --- for instance, 
obtaining a universe of the type that we observe, 
may require choosing a patch of space-time, 
where a sufficiently 
long period of de Sitter expansion took place. 

Coming back to 
the parameter $\kappa$, it is not a Lagrangian parameter and, as long as we 
have no theory of quantum gravity, also not a derived quantity. It rather
characterizes the global topology of the spatial universe
(open/hyperbolic for $\kappa < 0$, 
flat for $\kappa = 0$,  
closed/spherical for $\kappa > 0$). From a general
relativistic point of view, it may be 
viewed as an initial condition. 

To give $\kappa$ a more specific meaning, consider a sub-manifold
$\tau = $ constant of the metric in \eq\nr{eq_flrw_metric}. 
Let us compute the Ricci scalar for such a manifold, 
\be
 R^{ }_\tau \; \equiv \; g^{ij}_{ }R^{ }_{ij}
 \;, \label{ricci_scalar_spatial}
\ee
where it is implicitly 
assumed that only spatial indices are included, also when
deriving the Ricci tensor. The background value of 
$R^{ }_\tau$ is denoted by $\bar{R}^{ }_\tau$. At the background
level, when the metric tensor is diagonal, this simply means
that we eliminate the index $\tau$ and the derivatives 
$(...)^{ }_{,\tau}$ from the derivation in 
\eqs\nr{eq_chsy-a1}--\nr{bg_ricci_scalar}.
This implies that we leave out $\H$ from 
\eq\nr{bg_ricci_scalar}, yielding 
\ba
 \bar{R}^{ }_\tau 
 \; 
 \underset{\scriptscriptstyle \H,\H'\,\to\,0}{
 \overset{\rmii{\nr{bg_ricci_scalar}} \lift }{=}} 
 \; \frac{6 \kappa}{a^2_{ }} 
 \;. \label{bar_R_tau}
\ea
So, indeed, $\kappa$ parametrizes the Ricci scalar of the 
constant-$\tau$ submanifolds, and this version of the 
Ricci scalar is in turn referred to as {\em spatial curvature}. 

\index{spatial curvature}
\index{$R^{ }_\tau$ (spatial curvature)}

\newpage

%%%%%%%%%%%%%%%%%%%%%%%%%%%%%%%%%%%%%%%%%%%%%%%%%%%%%%%%%%%%%%%%%%%%%%%%%
%
\subsubsection{Background solution in conformal time}
\label{app:bg_tau}

\addcontentsline{toc}{subsection}{\App\ref{app:bg_tau}: 
Background solution in conformal time}

\index{$\tau$ (conformal time)}

While the Friedmann equations
\nr{eq_end0-1} and \nr{eq_end0-2} 
are most conveniently solved in 
physical time, conformal time is 
also often made use of. Here we sketch how the background 
equations can be 
solved in conformal time. 

With the same assumptions as before, \eq\nr{bg_Tmunu_appro_t_2}
gets replaced with 
\be
 0  
 \; 
 \underset{\rmii{\nr{def_w}}}{
 \overset{\rmii{\nr{eq_end0-2}} \lift }{=}} 
 \;
 \bar{e}\hspace*{0.3mm}' + 3\H \bar{e} (1 + w)
 \; = \; 
 \frac{\partial^{ }_\tau [\bar{e}\, a^{3(1+w)}_{ }]}{a^{3(1+w)}_{ }} 
 \quad \Rightarrow \quad 
 \bar{e}\, a^{3(1+w)}_{ } \; = \; \mbox{constant}
 \;. \label{bg_Tmunu_appro_2}
\ee
Writing the solution as 
$
 \bar{e} = 
 \bar{e}^{ }_i \, a_i^{3(1+w)} / a_{ }^{3(1+w)}
$
and
inserting into \eq\nr{eq_end0-1} we get
\ba
 \H^2_{ }
 \;\;
   \underset{\rmii{\nr{bg_Tmunu_appro_2}}}{
   \overset{\rmii{\nr{eq_end0-1}}\scriptscriptstyle [\kappa\,=\,0]
            \lift }{=}} 
 \;\;
          \overbrace{ 
          \frac{8\pi G \bar{e}^{ }_i}{3} }^{\,\equiv\, H_i^2}
          \frac{a^{3(1+w)}_i}{a^{1+3w}_{ }}
 & \Leftrightarrow & 
 a' \; = \; H^{ }_i a_i^{ \frac{3}{2}(1+w) } a^{\frac{1}{2}(1-3w) }_{ }
 \nn[2mm]
%%%%
 & \Leftrightarrow & 
 \frac{{\rm d}a}{ a^{\frac{1}{2}(1-3w) }_{ }}
 \; = \; 
 H^{ }_i a_i^{ \frac{3}{2}(1+w) } {\rm d}\tau
 \;, \label{a_simpl_tau}
\ea
which now leads to 
\ba
 && 
 a \; \overset{\rmii{\nr{a_simpl_tau}}}{=} \; 
 a^{ }_i 
 \biggl[ 1 + 
   \frac{
      (1+3w) 
      H^{ }_i a_i^{ } (\tau - \tau^{ }_i)
      }{2}
 \biggr]^{\frac{2}{1+3w} }_{ }
 \;. 
 \label{soln_bg_tau_a}
\ea
This applies both during de Sitter expansion 
and matter or radiation domination, 
with the underlying assumption that $w$ is to a good approximation constant.

Consider first de Sitter expansion, $w=-1$. Then 
\eq\nr{soln_bg_tau_a} yields 
\be
 \mbox{de~Sitter}: \; 
 a 
 \; 
  \underset{\scriptscriptstyle w\;=\;-1}{
  \overset{\rmii{\nr{soln_bg_tau_a}} \lift }{=}} 
 \; 
 \frac{a^{ }_i}{1 + H^{ }_i a^{ }_i (\tau^{ }_i - \tau)}
 \;. \label{a_tau_dS_1} 
\ee
According to this result, $a$ diverges at a time 
$\tau^{ }_\rmi{div} = \tau^{ }_i + 1/(H^{ }_i a^{ }_i)$.
It is conventional to choose 
$
 \tau^{ }_\rmi{div} \equiv 0
$, 
implying
$
 H^{ }_i a^{ }_i \tau^{ }_i = -1
$.
So, \eq\nr{a_tau_dS_1} can be simplified into
\be
% \mbox{de~Sitter}: \; 
 a
 \;
  \overset{\rmii{\nr{a_tau_dS_1}}}{=}
 \;
  - \frac{1}{H^{ }_i \tau }
 \;, \quad 
 \H \;=\; - \frac{1}{\tau}
 \;, \quad \tau \le \tau^{ }_e < 0 
 \;, \label{a_tau_dS_2} 
\ee
where we have denoted by $\tau^{ }_e$ a moment at
which inflation ends and 
the expansion continues 
under matter or radiation domination, with $w\ge 0$.

Let us subsequently work out what happens at $\tau \ge \tau^{ }_e$.
At the initial moment, the scale factor is 
$ a^{ }_e = -1/(H^{ }_i\tau^{ }_e)$ according
to \eq\nr{a_tau_dS_2}.
Therefore, from \eq\nr{soln_bg_tau_a}, 
with $a^{ }_i \to a^{ }_e$ and 
$\tau^{ }_i \to \tau^{ }_e$, 
\ba
% \mbox{matter/radiation}: \; 
 a 
 &
  \overset{\rmii{\nr{soln_bg_tau_a}}}{=} 
 &
 - \frac{1}{H^{ }_i\tau^{ }_e} 
 \biggl[ \frac{3(1+w)}{2} - 
   \frac{
      (1+3w) 
      \tau 
      }{2\tau^{ }_e}
 \biggr]^{\frac{2}{1+3w} }_{ }
 \;, \quad \tau \ge \tau^{ }_e 
 \;, \label{a_tau_mat_1} \\[2mm]
%%%%
 \H 
 & 
  \overset{\rmii{\nr{a_tau_mat_1}}}{=} 
 & \frac{2}{(1+3w)\tau - 3(1+w)\tau^{ }_e}
 \;, \quad \tau \ge \tau^{ }_e 
 \;. \label{H_tau_mat_1} 
\ea
This behaviour is sketched in \fig\ref{fig:history_tau}(right)
on p.~\pageref{fig:history_tau}.

\newpage

%%%%%%%%%%%%%%%%%%%%%%%%%%%%%%%%%%%%%%%%%%%%%%%%%%%%%%%%%%%%%%%%%%%%%%%%%
%
\subsubsection{Numerical background solution}
\label{app:num_bg_vac}

\addcontentsline{toc}{subsection}{\App\ref{app:num_bg_vac}: 
Numerical background solution}

\index{background solution: numerical}

%%%%%%%%%%%%%%%%%%%%%%%%%%%% FIGURE %%%%%%%%%%%%%%%%%%%%%%%%%%%%%%%%%
%
\begin{figure}[t]
    \centering
    \includegraphics[width=0.44\linewidth]{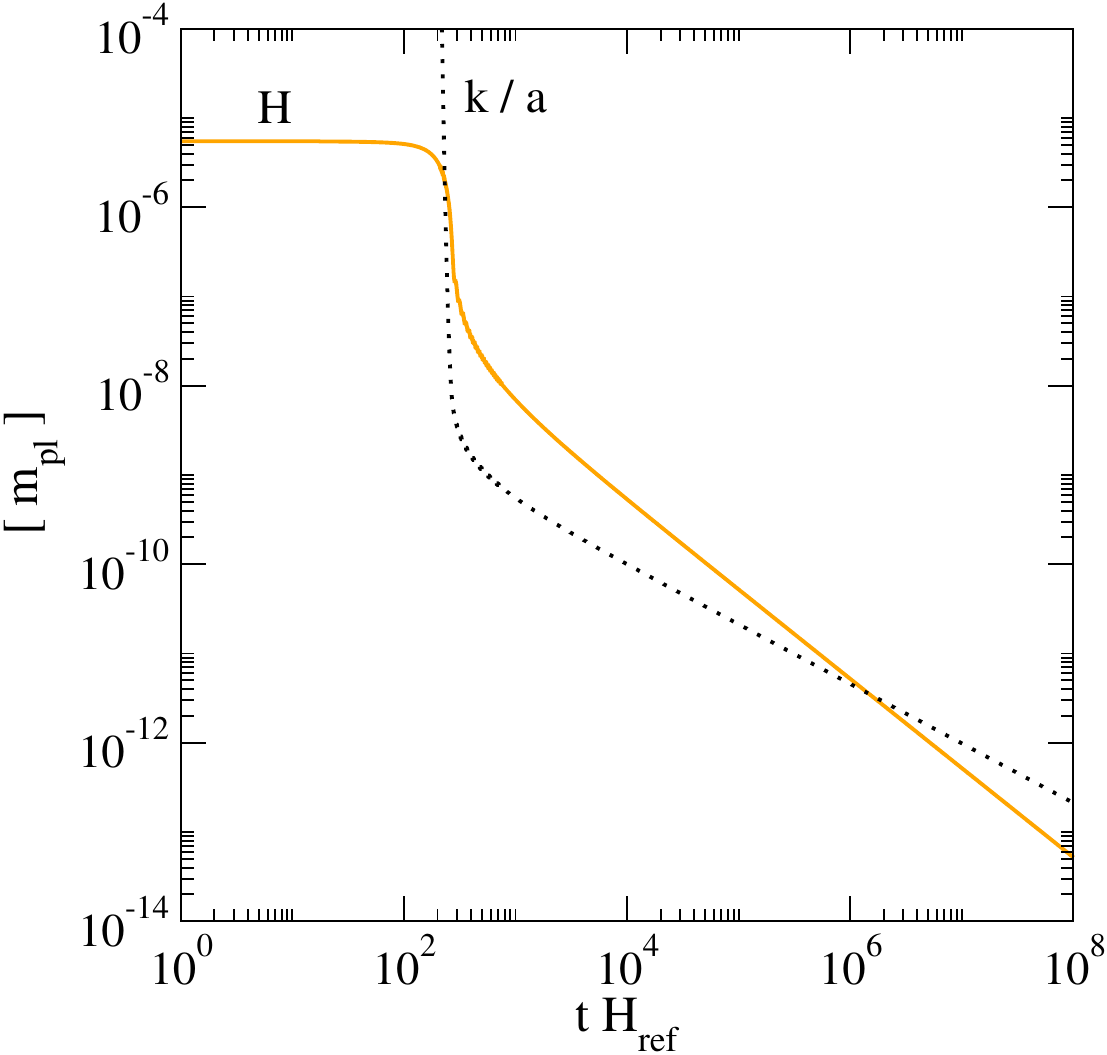}%
    \caption{%
    \small
    Numerical background solution,
    from \eqs\nr{varphi_eq_vac_again}--\nr{Href},  
    in physical time. For illustration, the comoving momentum mode
    has been chosen as $k/a(t^{ }_i) = 7.35 \times 10^{52}_{ }\mpl^{ }$, 
    where the initial time, $t^{ }_i \equiv H_\rmii{ref}^{-1}$,
    is defined according to \eq\nr{Href}.
    At late times, we have gone over from the full equation 
    in \eq\nr{varphi_eq_vac_again} to the simplified 
    version in \eq\nr{dot_e_ave}, but the transition is smooth.  
    This plot can be 
    compared with the analytic expectation from   
    \fig\ref{fig:history_tau}(left) on p.~\pageref{fig:history_tau}.
    A subsequent transition to a radiation-dominated 
    expansion will be illustrated in \fig\ref{fig:bg_thermal}
    on p.~\pageref{fig:bg_thermal}.
    }
    \index{cosmic history: numerics (figure)}
    \index{figure: numerics for cosmic history}
    \label{fig:bg_num}
\end{figure}
%
%%%%%%%%%%%%%%%%%%%%%%%%%%%%%%%%%%%%%%%%%%%%%%%%%%%%%%%%%%%%%%%%%%%%

Between \eqs\nr{at_dS} and \nr{a_t_mat_1},
we presented an analytic background solution
for a prototype system. However, we had to make assumptions
about the equation of state or the magnitude of $\dot{\bar\varphi}$
in order to achieve this. It is not difficult to solve the 
background equations numerically without any such assumptions, 
at least for a finite duration of time (e.g.,\ for interpolating
between periods with different equations of state). Here we
demonstrate how this can be done. 

In order to consider a simple system, we restrict for the moment
to a case in which only a massive scalar field, $\bar\varphi$, is present. 
In this case, the early quasi de Sitter period goes over to 
a matter-dominated expansion at late times
($w=0$), as will be explained below. 

First, let us write down the equations that we are solving. 
The field equation reads 
\be 
 \ddot{\bar\varphi} + 
  3 H \dot{\bar\varphi} +
  V^{ }_{\der\varphi} 
  \; 
  \overset{\rmii{\nr{eq-F2}}}{=} 
  \;
  0
  \;, \label{varphi_eq_vac_again} 
\ee
where for $\kappa = 0$ the Hubble rate is given by 
\be
 H 
 \;
 \underset{\rmii{\nr{mpl}}}{
 \overset{\rmii{\nr{eq_inflation_t}} \lift }{=}}
 \;
 \sqrt{\frac{8\pi}{3}} 
 \frac{\sqrt{ 
 \tfr{1}{2}\! \dot{\bar\varphi}^2 
  + V 
  }}
      { \mpl^{ } }
 \;. \label{num_H}
\ee
As a potential we assume that 
of so-called {\em natural inflation}~\cite{ai}, 
\be
 V
 \;\equiv\;
 m^2 f_a^2\, 
 \biggl[ 1 - \cos\biggl( \frac{\bar\varphi}{f^{ }_a} \biggr) \biggr]
 \;. \label{V} \index{natural inflation}
\ee
Parameter values that yield semi-realistic
observables (cf.~\se\ref{ss:cmb}) are chosen as 
\be
 f^{ }_a \;=\; 1.25 \, \mpl^{ } 
 \;, \quad 
 m \;=\; 1.09 \times 10^{-6}_{ }\, \mpl^{ }  
 \;. \label{params}
\ee

We also need to specify initial conditions. The initial field value
and its time derivative are set to 
\be
 \bar\varphi(t^{ }_{i}) \;=\; 3.5 \,\mpl^{ }
 \;, \quad
 \dot{\bar\varphi}(t^{ }_{i}) \;=\;  
 - \frac{V^{ }_{,\varphi}(\bar\varphi(t^{ }_{i}))}
   {3 H^{ }_\rmii{ref}}
   %{3 H(\bar\varphi(t^{ }_{i}))}
 \;, \label{varphi_i} 
\ee
where $ H^{ }_\rmi{ref} $ is defined in \eq\nr{Href}.
The value $ \bar\varphi(t^{ }_{i}) $ is close enough to the top
of the potential that inflation lasts long, but not eternally. 
The value $ \dot{\bar\varphi}(t^{ }_{i}) $ turns out to play little
role (the solution is an attractor), but it has been chosen so that
the attractor is reached relatively fast. The initial Hubble rate sets 
a ``scale'' in whose units dimensionful quantities can be expressed; 
a particularly convenient choice for this is to 
define a ``reference'' value of the Hubble rate, obtained as if
the potential were quadratic in \eq\nr{num_H}
($V = \tfr{1}{2} \hspace*{-0.5mm} m^2_{ }\bar\varphi^2_{ }$)
and $\dot{\bar{\varphi}}(t^{ }_i) = 0$, 
\be
 H^{ }_\rmi{ref} \; \equiv \; 
  \sqrt{\frac{4\pi}{3}} 
  \frac{m\, \bar\varphi(t^{ }_{i})}{ m^{ }_\rmiii{pl} }
 \;. \label{Href}
\ee
The initial time is chosen as $t^{ }_i \equiv H_\rmi{ref}^{-1}$.
\index{$H_\rmii{ref}^{ }$ (approximate $H$ during slow-roll)}

Apart from solving for $\bar\varphi(t)$, we also solve for the evolution
of the cosmological scale factor. The scale factor is re-parametrized in 
terms of {\em $e$-folds}, \index{$e$-folds}
denoted by $N$, via
\be
 a(t) \;\equiv\; a(t^{ }_i) \, e^{N(t)}_{ }
 \quad \Rightarrow \quad
 \dot{N} \;=\; H
 \;, \quad
 N(t^{ }_i) \;=\; 0
 \;. 
\ee
With a given $a$, we find out how %{\em comoving momenta},
{\em physical momenta},
defined as $k/a(t)$, redshift, 
\be
 \frac{k}{a(t)} \; = \; \frac{k\,  e^{-N(t)}_{ }}{a(t^{ }_i)}
 \;. \label{redshift}
 %\index{comoving momentum}
 \index{$k/a$ (physical momentum)}
\ee

We now explain why the post-inflationary cosmology of this model corresponds
to {\em early matter domination}, 
\index{early matter domination}
in the sense explained in \se\ref{ss:bg_rad_mat}
and sketched in \fig\ref{fig:history_tau}. At late times, $\bar\varphi$
settles around its minimum at $\bar\varphi = 0$. 
Then the potential approximates to 
$ 
 V \approx \tfr{1}{2} \hspace*{-0.5mm} m^2_{ }\bar\varphi^2_{ }
$. 
In this regime, \eq\nr{varphi_eq_vac_again} becomes 
\be
 \ddot{\bar\varphi} + 
 3 H \dot{\bar\varphi} +  
 m^2_{ }\bar\varphi 
  \;\approx\;
  0
 \;. \label{damped_ho}
\ee
By the time when $H \ll m$, this is just a damped 
harmonic oscillator equation of motion, with known trigonometric
solutions, of periodicity $\Delta t \equiv 2\pi / m$. 
If we denote the energy density and pressure appearing in 
\eqs\nr{eq_e-p} and \nr{eq_e-p1}, respectively, as 
\be
 e^{ }_{\bar\varphi}
 \; \overset{\rmii{\nr{eq_e-p}}}{\equiv} \; 
 \frac{\dot{\bar\varphi}^2_{ }}{2} + V
 \;, \quad
 p^{ }_{\bar\varphi}
 \; \overset{\rmii{\nr{eq_e-p1}}}{\equiv} \; 
 \frac{\dot{\bar\varphi}^2_{ }}{2} - V
 \;, \label{e_varphi}
\ee
then an average over the oscillation period yields 
\be
 \biggl\langle\,
 \frac{\dot{\bar\varphi}^2_{ }}{2} 
 \,\biggr\rangle^{ }_{\Delta t}
 \; 
 \overset{H\,\ll\,m}{\approx}
 \;
 \biggl\langle\,
 \frac{m^2_{ }{\bar\varphi}^2_{ }}{2}  
 \,\biggr\rangle^{ }_{\Delta t}
%%%%%
 \quad \Rightarrow \quad 
 \bigl\langle\,
 e^{ }_{\bar\varphi} 
 \,\bigr\rangle^{ }_{\Delta t} 
 \; \approx \; 
 \bigl\langle\,
 \dot{\bar\varphi}^2_{ } 
 \,\bigr\rangle^{ }_{\Delta t} 
 \;, \quad 
%%%%%%
 \bigl\langle\,
 p^{ }_{\bar\varphi} 
 \,\bigr\rangle^{ }_{\Delta t} 
 \; \approx \;
 0 
 \;. \label{varphi_oscs}
\ee
The vanishing of the (averaged) pressure implies that $w\approx 0$.

If we want to solve the evolution equations for a long time, the 
rapid oscillations become costly to handle. However, if we take 
a time derivative of \eq\nr{e_varphi} and make use of 
% $ e^{ }_{\bar\varphi} $ in 
\eq\nr{varphi_eq_vac_again}, we find
\be
 \dot{e}^{ }_{\bar\varphi} 
 \; 
  \overset{\rmii{\nr{e_varphi}} \lift }{
  \underset{\rmii{\nr{varphi_eq_vac_again}}}{=}} 
 \; 
 -3 H \dot{\bar\varphi}^2_{ }
 \; \overset{\rmii{\nr{varphi_oscs}}}{\approx} \; 
 -3 H e^{ }_{\bar\varphi}
 \;, \quad
 H 
 \; 
 \underset{\rmii{\nr{varphi_oscs}}}{
 \overset{\rmii{\nr{num_H}} \lift }{\approx}} 
 \;
 \sqrt{\frac{8\pi}{3}} 
 \frac{\sqrt{e^{ }_{\bar\varphi}\vphantom{h}}}{\mpl^{ }}
 \;. \label{dot_e_ave}
\ee
This system 
contains no oscillations, 
and is simple
to solve for a long period of time.  

We show below a {\tt python} script producing
a benchmark solution of the set of equations 
\eqref{varphi_eq_vac_again}, \eqref{num_H}, 
and \eqref{dot_e_ave}. 
The result is plotted in \fig\ref{fig:bg_num}
on p.~\pageref{fig:bg_num}. We go over to 
\eq\eqref{dot_e_ave} after 50 oscillations
(101 crossings of zero).

\index{code: numerics for cosmic history}

{\fontsize{8pt}{10pt}\selectfont
\begin{verbatim}
# numerical solution for inflation and end of inflation [numerics_bg_cold.py]
#
# import basic tools and integration routines
import numpy as np
from scipy.integrate import solve_ivp

# parameters [mpl]
fa = 1.25          # inflaton decay constant
m = 1.09e-6        # inflaton mass
koa_ref = 7.35e52  # momentum of perturbations at initial time
Pi = np.pi

# potential of inflaton (phi) [mpl^4]
def V(phi): return m*m*fa*fa*( 1 - np.cos(phi/fa) )

# derivative of inflaton potential by phi [mpl^3]
def Vd(phi): return m*m*fa*np.sin(phi/fa)

# total energy density of scalar inflaton field [mpl^4]
def etot(phi,phid): return phid*phid/2 + V(phi)

# Hubble rate [mpl]
def H(energy_density): return np.sqrt(8*Pi*np.abs(energy_density)/3)

# 2nd-order ODE as two 1st-order ODEs: solve for [dot phi, ddot phi, number of efolds]
# derivatives are taken with respect to t = Href*time
def derivatives1(t,y):
    Phi, Phid, Nfolds = y
    dphi_dt = Phid/Href
    ddphi_ddt = ( -3*H(etot(Phi, Phid))*Phid - Vd(Phi) )/Href
    dN_dt = H(etot(Phi, Phid))/Href
    dy_dt = [dphi_dt, ddphi_ddt, dN_dt]
    return dy_dt

# average over fast oscillations at late times => one 1st-order ODE
def derivatives2(t,y):
    e_phi, Nfolds = y
    dy_dt = [-3*H(e_phi)*e_phi/Href,  H(e_phi)/Href]
    return dy_dt

# initial conditions and reference values
phi_0 = 3.5                     # mpl
Href = np.sqrt(4*Pi/3)*m*phi_0  # mpl
phid_0 = - Vd(phi_0)/(3*Href)   # mpl^2

# define the event function counting the zeros of phi
def phi_crossings(t, y):  return y[0]

# integrate up to 50 oscillations (101 crossings of zero)
sol1 = solve_ivp(derivatives1, [1, 1e4], [phi_0, phid_0, 0], events=phi_crossings)
t_match = sol1.t_events[0][100]
time1 = sol1.t[sol1.t <= t_match]
n_match = len(time1)
ephi1 = etot(sol1.y[0], sol1.y[1])
efolds1 = sol1.y[2]

# initial conditions for oscillatory regime
time_0 = time1[n_match-1]
etot_0 = ephi1[n_match-1]
efolds_0 = efolds1[n_match-1]

# integrate in oscillatory regime (matter-dominated era)
sol2 = solve_ivp(derivatives2, [time_0, 1e8], [etot_0, efolds_0], method='DOP853')
ephi2 = sol2.y[0]
efolds2 = sol2.y[1]

# append solutions
time = np.append(time1,sol2.t)
e_phi = np.append(ephi1[:n_match], ephi2)
efolds = np.append(efolds1[:n_match], efolds2)
koa = koa_ref*np.exp(-efolds)

# print to file
np.savetxt('numerics_bg_cold.dat', np.c_[time, e_phi, H(e_phi), efolds, koa],
           fmt='%.6e',newline='\n',
           header='columns: t*H_ref, e_phi/mpl**4, H/mpl, efolds, k_ref/a/mpl')

\end{verbatim}
}

%%%%%%%%%%%%%%%%%%%%%%% end appendices %%%%%%%%%%%%%%%%%%%%%%%%%%%

%%%%%%%%%%%%%%%%%%%%%%%%% BIBLIO %%%%%%%%%%%%%%%%%%%%%%%%%%%%%%%%
%

%%%%%%%%%%%%%%%%%%%%%%%%%%% SECTION %%%%%%%%%%%%%%%%%%%%%%%%%%%%%%%%%
\newpage 

\section{Observable signatures beyond the background solution}
\label{se:obs}

\paragraph{Abstract:}

Our current universe is homogeneous and isotropic on scales that 
are larger than about 1\% of its visible size, 
and spatially flat to a good precision.
Its structure at smaller scales may have formed 
via gravitational collapse of 
early inhomogeneities. 
Reviewing experimental evidences for the universe until 
the emission of the Cosmic Microwave Background (CMB) radiation
and the onset of structure formation, 
we collect the target quantities 
that the theoretical framework 
presented in the book aims to describe. 
The relations between present-day frequencies and
wavelengths and those that play a role 
in the early universe are summarized. 
Finally, we derive basic formulae for 
the statistical description of small perturbations 
on a homogeneous and isotropic background, 
and compare the procedures adopted 
in different contexts. 

\paragraph{Keywords:} 

Scalar power spectrum, 
amplitude $A^{ }_\scalar$, spectral tilt $n^{ }_\scalar$,
tensor power spectrum, 
ratio~$r$ of tensor and scalar spectra, 
adiabatic initial conditions, 
transfer function, 
non-Gaussianity, isocurvature perturbation, 
spectral distortion,
gravitational-wave astronomy, 
interferometer, 
effective number of neutrino degrees of freedom. 

%%%%%%%%%%%%%%%%%%%%%%%%%%%%%%%%%%%%%%%%%%%%%%%%%%%
%
\subsection{Overview}

The quantitative evidence that we have about the early universe comes 
mostly not from 
the homogeneous and isotropic approximation (cf.~\ch\ref{se:bg}), 
but from the fact that, on 
closer inspection, this approximation does not hold exactly. 
This can be inferred by
measuring the local
CMB temperatures in different directions, which show small but highly
correlated fluctuations, 
even at large angular separations. Furthermore, going towards
smaller distances, the CMB fluctuations can be mapped onto 
the structures seen in the roughest distribution of matter. 
The latter can be extracted with several observational techniques.
The 2-point density correlation function extracted from 
spectroscopic galaxy redshift surveys shows a clear peak
at a characteristic distance of $\sim 150$~Mpc, 
associated with the wavelength of 
a single {\em baryon acoustic oscillation} (BAO) that had taken place
by the time that baryons decoupled from photons. A ``forest'' of  
{\em Lyman-$\alpha$ absorption lines}, 
superimposed on the spectra of
distant quasars, reflects the presence of 
neutral hydrogen clouds that lie between
the emitter and the observer. Finally, 
{\em gravitational lensing} (either of CMB photons, or of light emitted by 
galaxies) can tell us about the distribution
not only of matter, but also of {\em dark matter}, \index{dark matter}
which by definition is not directly visible 
in spectroscopic surveys or absorption lines. 

\index{BAO (baryon acoustic oscillations)}
\index{Lyman-$\alpha$ forest}
\index{gravitational lensing}

Going to even smaller scales, astronomical observations reveal
a lot of well-known structure (galaxies, nebulae, stars). 
However, it is more difficult to interpret this %small-scale structure 
in terms of its primordial origin, since complicated
and non-linear physics (gravitational collapse, diffusion, 
magnetohydrodynamics) took place in the intervening period. 

In contrast, 
if we were able to see gravitational waves, they would be agnostic
about the challenges posed by late-time astrophysical
phenomena, given that gravity interacts 
very weakly with matter. Therefore, 
gravitational waves offer for a theoretically 
ideal window to the early universe, 
all the way to the inflationary period, and extending 
over phenomena covering 
an extremely broad range of distance and frequency scales. 
It is speculated that the first signals of 
primordial gravitational waves could have already 
been observed by Pulsar Timing Arrays. 

In this chapter, we summarize the existing measurements of primordial
perturbations in the CMB (technically known as scalar perturbations)
(cf.\ \se\ref{ss:cmb}), 
and discuss the future prospects of observing primordial gravitational
waves (technically known as tensor perturbations)
(cf.\ \se\ref{ss:probes_gw}). 
To make the discussion concrete, 
we also review the basics of cosmological redshift and how it allows
us to probe various epochs of the early universe
(cf.\ \se\ref{ss:history}). 
We end the chapter
with a more foundational elaboration on what so-called power spectra 
are, from the mathematical 
and physical point of view
(cf.\ \se\ref{sec_gaussian}). 

%%%%%%%%%%%%%%%%%%%%%%%%%%%%%%%%%%%%%%%%%%%%%%%%%%%
%
\subsection{The imprints of scalar perturbations on the CMB}
\label{ss:cmb}

\index{CMB (cosmic microwave background)}
\index{temperature: anisotropies}

The observed temperature 
$T^{ }_\now = 2.7255$\hspace*{0.3mm}K of the CMB shows 
small {\em temperature anisotropies}. After subtracting galactic 
foregrounds as well as a 
{\em dipole structure} \index{CMB dipole}
with $\delta T/ T^{ }_\now \sim 10^{-3}$, originating from the motion of
the Earth relative to the CMB reference frame, the remainder is 
of order
\begin{equation}
 \left(  \frac{\delta T}{T^{ }_\now \rule{0pt}{2.2ex} } \right)_\text{obs}  
 \; \sim \;\;  10^{-5}
 \;\; \simeq \;\;   
 \left(  \frac{\delta T}{T^{ }_\now \rule{0pt}{2.2ex} } \right)_\text{intr}
 \! + \; 
 \left(  \frac{\delta T}{T^{ }_\now \rule{0pt}{2.2ex} } \right)_\text{jour} 
 \ . \label{eq_T}
\end{equation}
The left-hand side describes the observational result, while the right-hand
side is its interpretation, 
composed of 
an intrinsic (``intr'') component present at recombination, 
and of a journey (``jour'') component originating
as the photons travel to the present. 
In modern computations, reviewed in \se\ref{ss:evol_many}, 
no such distinction needs to be made, 
and the splitup is also gauge dependent
(cf.\ \ch\ref{se:gauges}),
but it is still helpful for an intuitive picture.

Let us start with the observational side. 
After choosing some coordinate system, $\delta T$ 
can be expanded in terms of 
spherical harmonics, $Y^{ }_{\ell m}$, from which the $\ell =1$ component
is omitted, as it corresponds to a dipole from our relative
movement with respect to CMB, 
\begin{equation}
 \delta T(\theta,\phi) |^{ }_\text{obs}
 \;=\; 
 \sum_{\ell\;\ge\; 2}
 \; 
 \sum_{m\;=\;-\ell}^{\ell}
  a_{\ell m}^{ }Y_{\ell m}^{ }(\theta,\phi)
 \ . \label{eq_deltaT}
\end{equation}
By definition the average value of $\delta T$ vanishes, 
and the physical information is contained in a 2-point function 
of temperature fluctuations, averaged over the sky. 
The coefficients $a_{\ell m}^{ }$ 
of the spherical-harmonics expansion
are independent complex variables. 
Averaged over the azimuthal index $m$, their absolute values squared
are denoted by
\begin{equation}
 C_\ell^{ } \;\equiv\; \frac{1}{2\ell+1}
  \sum_{m\;=\;-\ell}^\ell 
 |a_{\ell m}^{ }|^2_{ } \ .
 \label{C_l}
\end{equation}
We note that a large {\em multipole}, $\ell$, 
\index{CMB multipoles}
corresponds to a small angular scale, $\theta_\ell^{ } \equiv \pi/\ell$. 
In terms of the wavelength, $\lambda \sim 2\pi / k$,
of the underlying density perturbations, 
the angular dependence originates  
by projecting plane 
waves onto spherical harmonics %(see e.g. \cite[p.12]{hannu1}), 
(this is described briefly in the paragraph below \eq\nr{deltaT_fourier}).
Then a large $k$ corresponds to a large~$\ell$.
The total power in the multipole $\ell$ is conventionally defined as
\begin{equation}
 D_\ell^{ } \;\equiv\; \frac{\ell(\ell+1)}{2\pi}\, C_\ell^{ }
 \ . \label{def_D_l}
\end{equation}

\index{Thomson scattering}
\index{$C^{ }_\ell$, $D^{ }_\ell$ (temperature power spectrum)}
\index{CMB polarization}

In addition to the temperature power spectrum, known as $TT$
(so that $D^{ }_\ell \equiv D_\ell^{TT}$), photons
also carry linear polarization, which is a vector. This originates from 
{\em Thomson scatterings} just before decoupling, and is not complete, but
rather on the level $\sim 10\%$ of the total anisotropy~\cite{rees,be}.
As will be discussed in detail later on 
(cf.\ \eq\nr{eq_vec-1o}), any vector can be decomposed into 
a curl-free part, which can in turn be represented with the
help of a scalar potential ($\vec{E} \sim - \nabla A^{ }_0$), 
and a divergence-free part, which can be expressed through
a vector potential ($\vec{B} \sim \nabla \times \vec{A}$). 
Even though these do {\em not} physically correspond
to an electric or magnetic field, but are just parts of 
a single polarization vector, they are 
conventionally referred to as 
an $E$-mode and a $B$-mode, respectively. 
There is information about the cross-correlation $TE$ and 
the power-spectrum $EE$~\cite{Planck:2018nkj_again}. 
Moreover, there is an eager search for a $B$-mode signal
(cf.,\ e.g.,\ refs.~\cite{bicep,spt}), 
as this does not originate directly from scalar perturbations
($\sim A^{ }_0$), and is rather believed to be influenced 
by gravitational waves~\cite{kks,sz}. That said, 
lensing mixes some $E$ modes into $B$ modes~\cite{lensing},
and in addition there is galactic foreground emission
with similar characteristics.

\index{$E$-mode polarization}
\index{$B$-mode polarization}

%%%%%%%%%%%%%%%%%%%%%%%%% FIGURE %%%%%%%%%%%%%%%%%%%%%%%%%%%%%%%%%%%%%%%
%
\begin{figure}
 \centering
 \includegraphics[width=0.98\linewidth]{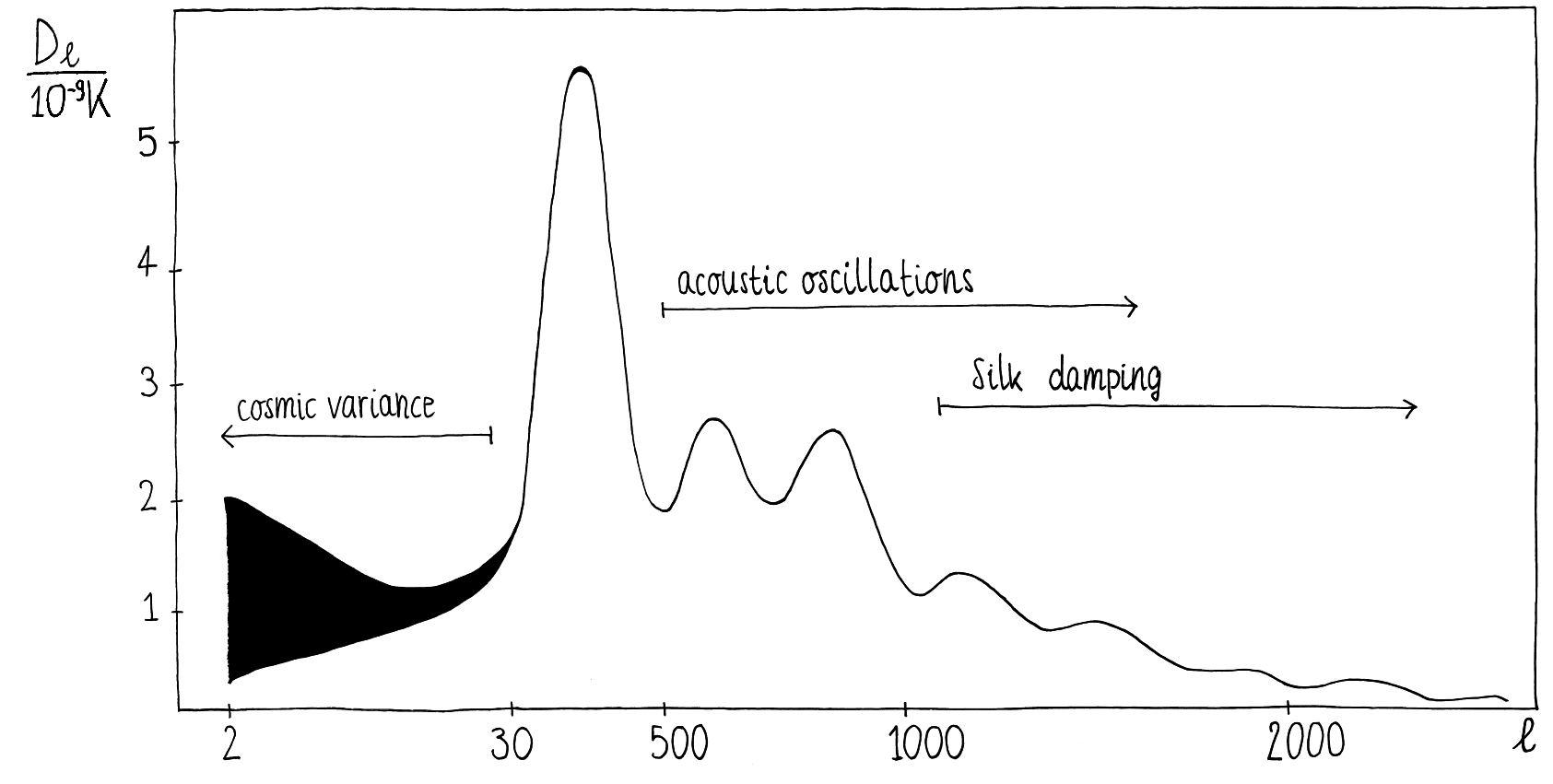}
 \caption{%
 \small 
  A sketch of the angular momentum power spectrum
  of the CMB temperature anisotropies, 
  $D^{ }_\ell$ (cf.~\eq\eqref{def_D_l}), at different multipole moments $\ell$.
 }
 \index{CMB power spectrum (figure)}
 \index{figure: sketch of CMB power spectrum}
 \label{fig:D_l_sketch}
\end{figure} 
%
%%%%%%%%%%%%%%%%%%%%%%%%%%%%%%%%%%%%%%%%%%%%%%%%%%%%%%%%%%%%%

%\subsubsection*{An ansatz for the shape}
As sketched in \fig\ref{fig:D_l_sketch}, 
the measured $D_\ell^{ }$ %(figure \ref{figure_cmb-spectrum}) 
show an oscillatory behaviour, with a global maximum at $\ell \sim 200$. 
The amplitude decreases as $\ell$ increases, starting from $\ell \sim 500$, 
and oscillations are damped after $\ell \sim 1000$. 
This shape can be explained by thinking that, 
just before decoupling, 
perturbations in the energy density propagate 
as {\em acoustic oscillations}, 
\index{acoustic oscillations: definition}
i.e.\ 
as a sequence of compressions and rarefactions of the medium. 
However, waves propagating in an interacting medium get attenuated, 
if their wavelength is small enough. In particular, if the wavelength
is smaller than the photon mean free path, $\lambda_\rmi{free}^{ }$, 
some photons may escape the compressed regions before colliding 
with the free electrons, whereby the wave is smoothened out, 
and the corresponding oscillatory behaviour of $D_\ell^{ }$ is damped. 
In the CMB context this phenomenon is 
known as {\em Silk damping} \cite{Silk:1967kq}. \index{Silk damping}
We note in passing that large-scale structures such as galaxy clusters, 
whose size corresponding to $\ell$ of several thousand, start forming
around inhomogeneities which did not get damped too much. This
is possible if the corresponding particles
feel no pressure forces, as may be the case for dark matter. 
After the decoupling, the apparent frequency and thus the 
temperature of the photons still gets modified, as they propagate 
in an inhomogeneous gravitational potential, through what is known  
as the \textit{Sachs-Wolfe effect} \cite{Sachs:1967er}. 
\index{Sachs-Wolfe effect}

%%%%%%%%%%%%%%%%%%%%%%%%%%

\vspace*{3mm}

Let us now rephrase the qualitative discussion above
in terms of the right-hand side of \eq\nr{eq_T}. 
As might be easy to imagine, %should be clear, 
the physics 
involved is quite complicated. Apart from the hydrodynamic and
collisional phenomena already alluded to, there is also the conceptual
issue that, in the presence of gravitational perturbations, the
definition of a temperature fluctuation is by itself not ``gauge invariant'', 
or physical in the general relativistic sense 
({\em gauge dependence} is defined in \ch\ref{se:gauges}).
\index{gauge dependence in general relativity}
Specifically, the division into intrinsic and journey components in 
\eq\nr{eq_T} depends %separately 
on the gauge chosen, whereas their
sum, the observed anisotropy, is well-defined. 

In order to make the problem tractable, it is conventional to 
{\em factorize} the computation into two parts
(the second captures the intrinsic and journey
components of \eq\nr{eq_T}): 

\bi

\item[(i)] \label{step_i}
As a first step, we consider the early universe until a moment $t^{ }_*(k)$
at which a given momentum mode $k$ {\em crosses outside 
of the Hubble horizon}, \index{horizon crossing} 
i.e.\ $(k/a)(t^{ }_*) \approx H$ 
(see \fig\ref{fig:history_tau} on p.~\pageref{fig:history_tau}).
We then choose a moment 
$t^{ }_\rmi{out} > t^{  }_*$ 
at which the mode is well outside of the Hubble horizon. 
In a time window around~$t^{ }_\rmi{out}$, 
there is a gauge-invariant quantity, called the curvature
perturbation $\R^{ }_\T$ (cf.\  \ch\ref{se:thermal}), 
which is to a very good approximation constant
in time (cf.\ \ch\ref{se:outside}).
Therefore, this quantity sets the target for 
theoretical computations. 

\item[(ii)]
At a certain time after $t^{ }_\rmi{out}$, the mode 
re-enters inside the Hubble horizon
(cf.\ \fig\ref{fig:history_tau} on p.~\pageref{fig:history_tau}). 
Then it starts to undergo oscillations, 
as dictated by viscous hydrodynamics (cf.\ \ch\ref{se:inside}). Viscosities
are the hydrodynamic representation of the mean free path mentioned above,
and lead to the damping of acoustic oscillations. The hydrodynamic 
description is valid as long as the photons stay equilibrated, that
means until their decoupling, at $t = t^{ }_\rmi{dec}$. The decoupling
period itself is usually described by a more microscopic framework than 
hydrodynamics, notably that of Boltzmann equations. 
After their decoupling, the photons ``free stream'' in a curved background
until the present time, when they are observed, at $t = t^{ }_\rmi{obs}$.
In this period, collisions can be omitted from Boltzmann equations, and
photons travel along geodesics. 

\ei

If we stay at linear order in perturbations, 
the result of step~(ii) is frequently called a 
{\em transfer function}, \index{transfer function: scalar perturbations} 
which is a linear mapping from
initial conditions onto physical observables. 
Specifically, for temperature fluctuations, 
we talk about a scalar transfer function, $\mathcal{T}^{ }_{s}$.
It relates the temperature anisotropies
to an initial {\em scalar power spectrum},
\index{scalar power spectrum} 
$\P^{ }_\text{s}$,  as 
\ba
  && 
        D^{ }_\ell(t)
   \; = \; 
        \int_{\vec k}
	\mathcal{T}^{}_\scalar (t,t^{ }_\rmi{out};\ell,k) 
	\, \P^{ }_\scalar (t^{ }_\rmi{out},k) 
 \ , \label{Ps_PRT} \\[2mm]
%%%%%
  && 
	\partial_t \!	\left. 
	\mathcal{T}^{}_\scalar (t,t^{ }_\rmi{out};\ell ,k)
				\right|_{t\;=\;t^{ }_\rmii{out}}
  \; = \;
  0
  \ .  \index{power spectrum: scalar perturbations}
\ea
The oscillating behaviour mentioned in step~(ii) 
above is described by a damped 
cosine function in~$\mathcal{T}^{}_\scalar$, producing ultimately
the features sketched in \fig\ref{fig:D_l_sketch}.

Assuming that the complicated physics of the transfer function
can be taken care of, it is then sufficient to parametrize the 
power spectrum, $\P^{ }_\scalar$, in terms of 
its amplitude and momentum dependence, around some chosen momentum,
$k^{ }_*$, that lies in the domain of the CMB data. Furthermore, 
$\P^{ }_\scalar$ can be identified with a ``universal''
curvature power spectrum,
$\P^{ }_\scalar  = \P^{ }_{\mbit \scriptscriptstyle \R^{ }_\rmiii{\it T}}
 = \P^{ }_{\mbit \scriptscriptstyle \R_\varphi} $
(cf.\ \ch\ref{se:outside}).
In order to 
simplify the notation, 
we normally suppress the time argument $t^{ }_\rmi{out}$,
$
 \P^{ }_\scalar(k)\equiv \P^{ }_\scalar(t^{ }_\rmi{out}, k)
$. 
The parametrization then reads
\begin{equation}
 \P_\scalar^{ }(k)
 =A_\scalar^{ }\, 
 \left( \frac{k}{k_*^{ }} \right)^{n_\iscalar^{ } - 1}
 \ , \label{eq_powerlaw}
 \index{$A^{ }_\scalar$ (scalar amplitude)}
 \index{$n^{ }_\scalar$ (scalar spectral tilt)}
 \index{$k^{ }_*$ (pivot scale)}
\end{equation}
where $A_\scalar^{ }$ is called the {\em scalar amplitude},  
$n_\scalar^{ }$ the {\em spectral tilt}, 
and $k^{ }_*$ the {\em pivot scale}. 
We remark
that the convention for $n^{ }_\scalar$ in \eq\nr{eq_powerlaw}
is unfortunate, as the observed ``almost flat'' spectrum 
corresponds to $n^{ }_\scalar \approx 1$, so that $n^{ }_\scalar - 1$
would be a more natural variable. 
The exactly flat limit, 
$n^{ }_\scalar = 1$, is known as the 
{\em Harrison--Zeldovich spectrum}~\cite{hz_1,hz_2}.
\index{Harrison--Zeldovich spectrum}

The pivot scale appearing
in \eq\nr{eq_powerlaw} is normally chosen as 
$k^{ }_*/a^{ }_\now \equiv (20\,\mbox{Mpc})^{-1}_{ }$, 
which lies in the middle of the 
CMB data domain
(a frequent literature convention is to set $a^{ }_\now = 1$).
The numerical values deduced  
from the CMB experiments 
are~\cite{planck6,Planck:2018jri}
\begin{align}
 A_\scalar^\rmi{obs} &= ( 2.100 \pm 0.030 ) \times 10^{-9}
 \ ,  \label{ex_A_s} \\[2mm]
%%%%%%
 n_\scalar^\rmi{obs} &= 0.9649 \pm 0.0042
 \ ,
 \label{ex_n_s}
\end{align}
with the precision being gradually improved upon
(cf.,\ e.g.,\ ref.~\cite{act}).
The value $ n_\scalar^\rmi{obs} < 1$ implies 
that $\P^{ }_\scalar$ {\em decreases} slowly with increasing $k$. 

%%%%%%%%%%%%%%%%%%%%%%%%%%%%%%%%%%%%%%%%%%%%%%%%%%%%%%%%%%%%%%%%%%%%
%
\subsubsection*{Absence of significant non-Gaussianity}

\index{non-Gaussianity}

Apart from quantities that have been measured, cf.\ 
\eqs\nr{ex_A_s} and \nr{ex_n_s}, there are also 
quantities for which no positive signal has been found yet. This turns
out to 
provide very important empirical information about the early universe. 

A fundamental question concerns the underlying nature of the statistical
ensemble. If %it is Gaussian, 
fluctuations are distributed according 
to {\em Gaussian statistics}, \index{Gaussian statistics}
then this means that all information about
correlation functions of fluctuations is contained already in the 
2-point correlation function (this will be discussed in more detail
in \se\ref{sec_gaussian}). In particular, correlation functions of
an odd number of fluctuations vanish, and correlation functions of
an even number of fluctuations can be related to products of 
2-point correlation functions, 
via {\em Isserlis' or Wick's theorem}.
\index{Isserlis' theorem}
\index{Wick's theorem}

The simplest probe of non-Gaussianity is therefore that we 
consider a 3-point correlator. Suppose that we measure 
temperature anisotropies, 
and analyze them in terms of spherical harmonics, like
in \eq\nr{eq_deltaT}. The 3-point correlator of temperature
fluctuations can be projected onto the overall  
angular momentum mode $\ell$. Recalling the rules of adding
angular momenta, the result would then represent a 
generalization of \eq\nr{C_l}, roughly
\be
  E^{ }_{\ell}
 \; = \; 
  \sum_{\ell_i, m_i}
  c_{\ell}^{\ell_1 m_1; \ell_2 m_2; \ell_3 m_3}
  a^{ }_{\ell_1 m_1} a^{ }_{\ell_2 m_2} a^{ }_{\ell_3 m_3} 
 \;. \label{E_l}
\ee
In particular, taking an angular average over the full sky
corresponds to $\ell(\ell_1,\ell_2,\ell_3)=0$, 
and then the coefficients $c_\rmii{$\ell = 0$}^{\,...}$ are related
to the known Wigner 3-$j$ symbols. 

We may subsequently ask if there
is a statistically significant indication that 
$E^{ }_0 \neq 0$. %and how this could be interpreted. 
If this is the case, the physical interpretation
is not simple, given that non-Gaussian features can be generated both
at early and at late times~\cite{f_NL}. The ultimate goal is, however, 
to measure a parameter, $f^{ }_\rmii{NL}$, that captures the relative
importance of possible non-Gaussianity. 
We give a schematic 
definition around \eq\nr{f_NL}, and for now just note that the experimental 
constraint can be conservatively stated as~\cite{planck9}
\be
 | f^\rmi{obs}_\rmii{NL} | \; < \; 100
 \;. \label{f_NL_obs}
\ee 
Unfortunately, 
this is not considered to be a particularly strong constraint, because 
many viable inflationary models predict 
$ | f^{ }_\rmii{NL} | < 1$
(cf., e.g.,\ ref.~\cite{Lyth:2005fi}).
Since $f^{ }_\rmii{NL}$ is quite sensitive to 
momentum scales larger
than the CMB ones, its constraining power is expected to
grow substantially with the next generation of galaxy surveys,
notably Euclid \cite{Euclid:2024ris}.

%%%%%%%%%%%%%%%%%%%%%%%%%%%%%%%%%%%%%%%%%%%%%%%%%%%%%%%%%%%%%%%%%%%%
%
\subsubsection*{Absence of significant isocurvature perturbations}

\index{isocurvature perturbations}

The picture described around
\eq\eqref{Ps_PRT}, where the result of step~(i)
on p.~\pageref{step_i}
is captured by a single power spectrum, is just the simplest possibility. 
If it is realized, we say that the initial conditions are
{\em adiabatic}
(cf.\ \se\ref{ss:overview_perts}),
meaning that there is no information 
to distinguish between perturbations in different matter
constituents, such as CMB photons and dark matter
(or baryons, or neutrinos). 
In contrast, in some models, additional information is stored in 
so-called {\em isocurvature modes}
(also known as 
entropy or non-adiabatic perturbations). 
At linear order in perturbation theory, a general scalar perturbation 
can then be decomposed into independent adiabatic and isocurvature components.
Denoting the latter by $\I^{ }_i$ for the $i$-th independent component, 
different power spectra like \eq\eqref{eq_powerlaw}
are needed for specifying the full initial conditions,
\begin{equation}
	\P^{ }_\scalar (k) \; \longrightarrow \; 
        \bigl\{\,
	\P^{ }_{\mbit \scriptscriptstyle \R_\rmiii{\it T}^{ }}(k) \,,\, 
	\P^{ }_{\mbit \scriptscriptstyle \I_i}(k) 
        \,\bigr\} \ ,
	%+ \P^{}_\rmii{cross}(k)  
\end{equation}
and in general cross-correlations are needed as well. 
The evolutions at $t > t^{ }_\rmi{out}$, 
i.e.\ the corresponding transfer functions, can also differ, 
though what exactly happens depends on the physical 
nature of the isocurvature mode considered. 

In any case, observational evidence suggests
that there is only 
one underlying power spectrum. Notably, the $TT$, $TE$, and $EE$
power spectra, reflecting the physics of photons, and the BAO data 
on large-scale structure, more sensitive to dark matter, are all
consistent with each other
(isocurvature modes are said to excite out-of-phase
peaks and dips in CMB spectra, which are not observed). 
Through a non-trivial statistical 
analysis, this sets an upper bound on how large the amplitude
of the isocurvature power spectrum can be 
(cf.,\ e.g.,\ ref.~\cite{Enqvist:1999vt}).
In addition, there is theoretical prejudice in favour of adiabaticity, 
given that simple inflationary models make this prediction
(cf.\ \se\ref{ss:outside_isocurv}), and more complicated
multi-field models require tuning to avoid it
(cf.\ \se\ref{ss:outside_other}).
Empirically, the contribution of isocurvature modes to the 
CMB power spectra is constrained to~\cite{Planck:2018jri}
\be 
 \frac{
  \P^{}_{\mbit \scriptscriptstyle \I_\rmii{dm}}(0.04\,k^{ }_*) }{
  \P^{}_{\mbit \scriptscriptstyle \R^{ }_\rmiii{\it T}}(0.04\,k^{ }_*)} \,
  \;<\; 0.03
  \ , 
  \quad \frac{0.04\, k^{ }_*}{a^{ }_\now}
  \; = \; 0.002\, \mbox{Mpc}^{-1}_{ }
  \;, 
 \label{obs_isocurv}
\ee
where ``dm'' refers to cold dark matter. %, and t
The strongest available
constraint comes from the smallest momenta, like for non-Gaussianity, while
the bound is weaker at $k \sim k^{ }_*$,  
and particularly at momenta $ k \gg k^{ }_*$, 
which do not directly influence the CMB. 

%%%%%%%%%%%%%%%%%%%%%%%%%%%%%%%%%%%%%%%%%%%%%%%%%%%%%%%%%%%%%%%%%%%%
%
\subsubsection*{Absence of significant spectral distortions}

\index{CMB spectral distortions}

The shape of the CMB spectrum 
itself may hide interesting information.  
Indeed there are physical processes which 
tend to distort it away from the blackbody form.
Two main types of distortions can be distinguished: 
those due to non-smooth departure 
from thermal equilibrium,
and others that occur later on, when the CMB photons free-stream
through space and time. During the latter epoch, the spectrum
may be influenced by lensing effects or more exotic phenomena, 
like energy insertions from gradually decaying particles. 

In the literature, deviations
from the Planck spectrum are usually parametrized via two modes, called
$\mu$-type and $y$-type distortions. An example for the physics of the
former is that, as the primordial plasma cools down, 
the efficiency of photon number-changing processes 
decreases faster than that of 
{\em Compton scatterings}, \index{Compton scattering}
which keep sustaining 
kinetic equilibrium~\cite{zs}. As a result,  
a small photon chemical potential develops, often denoted as $\mu$, which,
in the language of text-book statistical physics, 
corresponds to the dimensionless
ratio of chemical potential over temperature. 
Assuming a standard radiation-dominated era after inflation, one can  
estimate~(cf.,\ e.g.,\ ref.~\cite{Chluba:2011hw})
\be
 | \mu | \;\sim\; 10^{-8}_{ } \ .
\ee
This target value is much below the 
bound set by the current observational results \cite{Fixsen:1996nj}, 
\be 
 |\mu|^\rmi{obs}_{ } \;<\; 9\times 10^{-5}_{ }
 \ ,
\ee
however, future experiments may turn spectral distortions
into a useful tool. 

%%%%%%%%%%%%%%%%%%%%%%%%%%%%%%%%%%%%%%%%%%%%%%%%%%%%%%%%%%%%%%%%%%%%%%%%%
%
\subsection{Probes of a primordial gravitational-wave background}
\label{ss:probes_gw}

In many theoretical frameworks, notably within the inflationary paradigm, 
the same physical phenomena that produce scalar perturbations,  
also produce tensor perturbations. 
We alternatively refer to the latter as gravitational waves 
(the reason is explained around \eq\nr{einstein_ij_t}, 
and is elaborated upon in \ch\ref{se:gw}). 
Because of the similar dynamics, in order to constrain 
the tensor power spectrum, $\P^{ }_\tensor$, it is 
conventional to adopt a %philosophy 
strategy similar to that for 
$\P^{ }_\scalar$\,. 
Let us assume that the
evolution after Hubble horizon re-entry is 
well understood, and captured by  
a tensor transfer function, $\mathcal{T}^{ }_\tensor$\,. 
Following \eq\nr{eq_powerlaw}, but with a different convention, 
we parametrize the initial conditions with 
a {\em tensor amplitude}, $A^{ }_\tensor$, 
and a {\em tensor tilt}, 
$n^{ }_\tensor$, as  
\be
 \P_\tensor^{ }(k) \; = \; A_\tensor^{ }\, 
 \left( \frac{k}{k_*} \right)^{n_\itensor^{ }}
 \index{$A^{ }_\tensor$ (tensor amplitude)}
 \index{$n^{ }_\tensor$ (tensor tilt)}
 \;. \label{P_tensor}
\ee
Even if the primordial tensor power spectrum has not been observed at
the time of writing, it is being actively searched for. 
Here we mention some of the avenues. 

%%%%%%%%%%%%%%%%%%%%%%%%%%%%%%%%%%%%%%%%%%%%%%%%%%%%%%%%%%%%%%%%%%%%
%
\subsubsection*{Influence on CMB polarization}

\index{CMB polarization}

When photons propagate from their decoupling until observation 
in the presence of tensor perturbations, 
they develop a $B$-mode polarization
(cf.,\ e.g.,\ refs.~\cite{gw4,gw5}).
Such a polarization signal has not been observed yet, setting 
an upper bound on the %gravitational-wave background. 
relative amplitude of the tensor 
power spectrum, or 
{\em tensor-to-scalar ratio}~\cite{Planck:2018jri,act6},
\begin{equation}
 r^\rmi{obs}_\rmii{0.002} 
 \;\equiv\; 
 \frac{\P^{ }_\tensor(0.04\,k^{ }_*)}{\P^{ }_\scalar(0.04\,k^{ }_*)}
% \frac{A_\tensor}{A_\scalar}
 \; < \; 0.038  
 \;,
       \quad \frac{0.04\, k^{ }_*}{a^{ }_\now}
 \; = \; 0.002\, \mbox{Mpc}^{-1}_{ }
 \ , \label{r_obs} \index{$r$ (tensor-to-scalar ratio)}
\end{equation}
where the momentum has been chosen like in \eq\nr{obs_isocurv}.
This turns out to be a powerful constraint, 
because part of the model-dependence of theoretical predictions
cancels in this ratio 
(cf.\ \eqs\nr{r}--\nr{nt}).
% (cf.\ \ch\ref{se:gw}).

%%%%%%%%%%%%%%%%%%%%%%%%%%%%%%%%%%%%%%%%%%%%%%%%%%%%%%%%%%%%%%%%%%%%
%
\subsubsection*{Pulsar timing arrays}

\index{PTA (pulsar timing array)}

As gravitational waves distort time and distance measurements, we can 
obtain constraints on them by measuring the arrival times of signals from
precisely ticking distant emitters. In particular, pulsars have been 
observed since several decades with this 
goal in mind~\cite{nhz1,nhz2,nhz3,nhz4}. However, there are 
plausible astrophysical sources, related  
for instance to supermassive black holes, which may produce
background signals similar to primordial gravitational waves. It remains to be
seen how the significance of the signal evolves, when more data
is gathered in the next decades. 

%%%%%%%%%%%%%%%%%%%%%%%%%%%%%%%%%%%%%%%%%%%%%%%%%%%%%%%%%%%%%%%%%%%%
%
\subsubsection*{Upcoming gravitational-wave detectors}

\index{gravitational-wave interferometers}

A possibility to observe gravitational waves directly is offered
by large laser interferometers
(in the future, atom interferometers may be tested as well, 
cf.,\ e.g.,\ ref.~\cite{atomi}). The hope is that, after subtracting 
a ``foreground'' signal generated by well-understood astrophysical
processes (such as black hole mergers), a stochastic
signal with a distinctive spectral shape will be left over. 
In the 2030's, %a new 
the first orbiting interferometer, 
the Laser Interferometer Space Antenna (LISA), 
\index{LISA (Laser Interferometer Space Antenna)}
should go into operation, with its peak sensitivity 
in the frequency range $f^{ }_\now \sim 10^{-4}_{ }...10^{-1}_{ }$~Hz~% 
\cite{redbook}.

{}From the theoretical perspective, 
what LISA may measure today ($t^{ }_\now$) is the fractional energy 
density carried by gravitational radiation,
referred to as its {\it spectrum}, 
\begin{equation}
 \frac{\dd  \Omega_\rmii{gw,0}}{\dd \ln k}
 \; \equiv \;
 \left. \frac{1}{e^{ }_\rmii{crit}}
 \frac{\dd e_\rmii{gw}}{\dd \ln k}\right|_{t_\inow^{ }} 
 \!\!\overset{\rmii{\eqref{def_T_T}}}{=} 
 \mathcal{T}^{}_\tensor(t^{ }_\inow,t^{ }_\rmi{out},k)
 \, \P^{}_\tensor(k)\ , \qquad 
  e^{ }_\rmi{crit}
 \;\equiv\;
 \bar e(t_\inow^{ }) 
 \;
 \overset{\rmii{\eqref{eq_end0-1}}}{=} 
 \; 
 \frac{3 H_\inow^2}{8 \pi G}
 \ . \label{eq_Ogwdef}
 \index{${\rm d}\Omega^{ }_\rmii{gw}/{\rm d}\ln k$ (gravitational-wave spectrum)} 
 \index{$H^{ }_\inow$ (current Hubble rate)}
\end{equation}
%where t
The {\em current value of the Hubble rate}, $H^{ }_\now$, 
is parametrized in terms 
of the {\em reduced Hubble rate}, $h$, \index{$h$ (reduced Hubble rate)}
as $H_\now^{ }=100\hspace*{0.3mm} h\,$km$\,$s$^{-1}\,$Mpc$^{-1}$. 
Instead of the comoving wave number $k$, the physical variable for 
$\Omega^{ }_{\mathrm{gw}\rmii{,0}}$ is the frequency, $f^{ }_\now$, 
which is related to $k$ by
\begin{equation}
 \frac{k}{a_\inow^{ }} \; = \; 2\pi f_\now^{ } 
 \ . \label{eq_kf0}
\end{equation}
Inserting $h$ and $f^{ }_\now$, 
the projected LISA sensitivity 
is (cf.,\ e.g.,\ ref.~\cite[fig.~10]{Babak:2021mhe})
\begin{equation}
 \frac{ h^2\dd \Omega_\rmii{gw,0}^\rmiii{LISA}}{\dd \ln f^{ }_\inow}
 \;\gsim\; 10^{-13}_{ } \ ,
\end{equation}
but from the cosmological point of view the actual performance 
depends a lot on how astrophysical foregrounds can be subtracted. 

Apart from LISA, there are plans for new generations of 
gravitational-wave interferometers sensitive to higher frequencies, 
such as
TianQin ($f_\now^{ } \sim 10^{-3}_{ }... 10^{0}_{ }$~Hz)~\cite{tianqin},
\index{TianQin}
DECIGO ($f_\now^{ } \sim 10^{-1}_{ }... 10^{1}_{ }$~Hz)~\cite{decigo}, 
\index{DECIGO}
Cosmic Explorer (CE) \index{CE (Cosmic Explorer)}
($f_\now^{ } \sim 10^{1}_{ }... 10^{3}_{ }$~Hz)~\cite{ce}, 
Einstein Telescope (ET) \index{ET (Einstein Telescope)}
($f_\now^{ } \sim 10^{1}_{ }... 10^{3}_{ }$~Hz)~\cite{et}, 
as well as 
Big Bang Observer (BBO), \index{BBO (Big Bang Observer)}
a potential successor of LISA. 
Proposed is also a whole class of so-called 
ultra-high frequency (UHF) detectors
(extending possibly up to $f_\now^{ } \sim 10^{11}_{ }$~Hz
or beyond)~\cite{uhf}. \index{UHF gravitational-wave detectors}
Time will tell what kind of sensitivities
these concepts reach, %however from the physics point of view 
but they would 
definitely offer interesting probes of the earliest moments of
the universe (cf.\ \se\ref{ss:overview_f_0}). 

%%%%%%%%%%%%%%%%%%%%%%%%%%%%%%%%%%%%%%%%%%%%%%%%%%%%%%%%%%%%%%%%%%%%%
%
\subsubsection*{Effective number of neutrino degrees of freedom}

Yet another probe on the energy density that gravitational waves
carry, is offered by the effective number of light degrees of freedom. 
This is strongly constrained, for instance, through its role in 
big bang nucleosynthesis (BBN), 
\index{BBN (big bang nucleosynthesis)}
or the decoupling of CMB photons. 

\label{bbn}

In BBN, fusion processes of neutrons and protons into light 
nuclei start at $T \sim 0.1\,$MeV, when thermal
motion is less energetic than the binding energy~\cite{bbn1,bbn2}. 
The efficiency depends on the relation of the expansion rate and 
of the fusion/fission rates.  
Observed abundances of light elements can therefore impose 
a constraint on viable expansion rates 
via \eq\eqref{eq_end0-1}, which, 
for a radiation-dominated universe ($\bar e \approx e^{ }_r$),
takes the form
\begin{equation}
 H^2_{ } \; = \; \frac{8\pi G}{3}\, e_r \ .
\end{equation}
In this context {\em radiation} means more broadly light, 
{\em relativistic} particles. 
If we modify $e_r$ by adding more relativistic degrees of freedom, 
the expansion of the universe is faster,  
and the abundances change. 
Conventionally, the energy density of additional relativistic components 
is parametrized by an 
{\em effective number of neutrino species}, 
\index{$N^{ }_\rmii{eff}$ (effective number of neutrinos)}
$N_\rmi{eff}$,
\begin{equation}
 e_r^{ } \; = \; e_\gamma^{ } + e_\nu^{ } +   \dots
 \;, \qquad
 \frac{ e^{ }_\nu }{e^{ }_\gamma}
 \; \equiv \; 
 \frac{7}{8}\biggl( \frac{4}{11} \biggr)^{4/3}_{ } N^{ }_\rmi{eff} 
 \;.
 \label{def_Neff}
\end{equation} 
The Standard Model prediction for $N_\rmi{eff}$ 
is (cf., e.g.,~refs.~\cite{Neffm2,Neffm1,Neff0})
\begin{equation}
 N_\rmi{eff}^\rmii{SM} \; = \; 3.044\pm 0.001
 \ . \label{Neff_SM}
\end{equation}
This value is larger than $3$, for instance because the 
neutrinos obtain an extra energy injection 
after their initial decoupling, 
from non-equilibrium
$e^+_{ }e^-_{ } \to \nu\bar\nu$ annihilations.  

As an additional relativistic degree of freedom, 
primordial gravitational waves also influence
the overall energy density. This contribution 
is commonly parametrized as
\begin{equation}\label{eq_Neff}
 \frac{ e^{ }_\rmii{gw} }{e^{ }_\gamma}
 \; \subset \;
 \frac{ \Delta e^{ }_r }{e^{ }_\gamma}
 \; \equiv \; 
 \frac{7}{8}\biggl( \frac{4}{11} \biggr)^{4/3}_{ } \Delta N^{ }_\rmi{eff} 
  \ .
\end{equation}
For a specific gravitational wave source, %model of inflation, 
$e_\rmi{gw}$ can be estimated 
by integrating 
${\rm d} \Omega^{ }_\rmi{gw} / {\rm d} \ln k$ 
in \eq\eqref{eq_Ogwdef} over all~$k$. 
From the joint analysis of CMB measurements and BBN light element abundances 
we have~\cite{planck6,act6}
\begin{equation}
 N_\rmi{eff}^\rmi{obs}
 \; = \; 2.89 \pm 0.11
 \;. \label{Neff_obs}
\end{equation}  
The difference between \eqs\nr{Neff_obs} and \nr{Neff_SM} 
defines the experimental estimate for $\Delta N^{ }_\rmi{eff}$, 
% which is
consistent with zero at the $2\sigma$ level,
and with the upper bound
\begin{equation}
 \Delta N^{ }_\rmi{eff} \;<\; 0.2 \ .
\end{equation} 
The accuracy is expected to improve in the future, 
promoting $\Delta N_\rmi{eff}^{ }$ to a non-trivial constraint on 
the energy density carried by primordial gravitational waves.
We will convert this bound to units of $\Omega^{ }_{\rmi{gw}\rmii{,0}}$
around \eq\nr{Omega_vs_Neff}.

%%%%%%%%%%%%%%%%%%%%%%%%%%%%%%%%%%%%%%%%%%%%%%%%%%%%%
%
\subsection{Cosmological redshift of wavelengths and frequencies}
\label{ss:history}

\index{$f^{ }_\inow$ (current frequency)}

When discussing observations, 
{\em current-day frequencies}, $f^{ }_\now$, 
are regularly expressed in units of Hz, 
and {\em current-day wavelengths}, $\lambda^{ }_\now$, in units of parsec (pc).
In theoretical considerations, we rather employ 
comoving momenta, $k$, or physical momenta, 
$p^{ }_\now = k / a^{ }_\now = 2 \pi / \lambda^{ }_\now$.
We recall here the relations between these quantities, 
and also how they are related to the Hubble rate in the 
early universe. As for the current Hubble rate, we write it as
\begin{align}
 H^{ }_\now 
 &\; = \;  
  h\, \frac{100\,\mbox{km}}{\mbox{s} \,\mbox{Mpc}}
%%%%
 &&\;\;
 \overset{\rmii{\nr{parsec}}\vphantom{ \big| }}{\Rightarrow} 
 \;\; 
 \frac{ H^{ }_\inow }{ h }
 \;
 =
 \;
 3.241 \times 10^{-18}_{ }\,\mbox{Hz}
 \\[2mm]
%%%%
 &\; = \;
 h\, \frac{10^5_{ } c}{2.998\times 10^8_{ }\,\mbox{Mpc}}
%%%%
 &&\;\;
 \overset{c\, =\, 1 \lift }{\Rightarrow} 
 \;\; 
 \frac{ h }{ H_\inow^{ } }
 \;
 =
 \;
  2.998\,\mbox{Gpc}
 \;
 \overset{\rmii{\nr{parsec}}}{\simeq}
 \; 
 9.778 \times 10^9_{ }\,\mbox{ly}
 \;. \label{H0_Mpc} 
 \hspace*{8mm}
\end{align} 
Interpreted as years, 
\eq\nr{H0_Mpc} reflects the age of the universe
(and is close to it when divided by $h\approx 0.7$).

% \index{cosmological scales: overview}

In order to relate present and past scales, 
it is helpful to first determine how temperature evolves in the early
universe (we already discussed this around \eq\nr{t_T}, but now add
the missing ingredients). 
So, let us assume that we can describe the energy-momentum tensor
via an ideal fluid. Then we can take 
\eq\nr{eq_end0-2} as a starting point, 
\be
  \dot{\bar{e}} + 3 H (\bar{e} + \bar{p} ) 
 \;
 \overset{\rmii{\nr{eq_end0-2}}}{=} 
 \; 0 
 \;. 
 \label{friedmann_again}
\ee 

Next, we need to recall some basic thermodynamics. The first law reads
\be
 {\rm d} E \; = \; T {\rm d} S - p\, {\rm d}V + \mu\, {\rm d}N
 \;, \label{1st_law}
\ee
and the overall energy can be written as
\be
 E \; = \; T S - p V + \mu N
 \;. \label{tot_E}
\ee
In cosmology, chemical potentials are usually very small, 
so we set $\mu\to 0$ for simplicity
(that said, \eq\nr{entropy_conservation} also applies with 
$\mu\neq 0$, however in that situation 
it must be supplemented by a conservation
equation for the corresponding comoving particle number, $N$). 
Writing $E = e\, V$ and $S = s\, V$ in \eq\nr{1st_law}, 
where $e$ is the {\em energy density} and 
$s$ is the {\em entropy density}, 
and dividing \eq\nr{tot_E} by volume, we get
\be
 \left\{
 \begin{array}{rcl} 
 \displaystyle
 V {\rm d}e + e\,{\rm d}V
 & % \;
 \underset{\scriptscriptstyle \mu\;\to\;0}{
 \overset{\rmii{\nr{1st_law}} \lift }{=}}
 & % \;
 T V {\rm d}s + T s\, {\rm d}V - p\, {\rm d}V
 \\[3mm]
%%%%%%
 \displaystyle
 e 
 & % \;
 \underset{\scriptscriptstyle \mu\;\to\;0}{
 \overset{\rmii{\nr{tot_E}} \lift }{=}}
 & % \;
 T s - p 
 \end{array}
 \right.
 \quad \Rightarrow \quad 
 {\rm d}e \; = \; T {\rm d}s
 \;. \label{Tds}
 \index{entropy density: definition}
\ee

We may now divide \eq\nr{Tds} by ${\rm d}t$, and apply the result
to the background quantities, as they appear in \eq\nr{friedmann_again}. 
This gives 
\be
 0 
 \; 
 \underset{\rmii{\nr{Tds}}}{
 \overset{\rmii{\nr{friedmann_again}} \lift }{=}} 
 \;  
 T \dot{\bar{s}} + 3 \frac{\dot{a}}{a} T \bar{s} 
 \; = \; 
 \frac{T}{a^3_{ }}\frac{{\rm d} (\bar{s} a^3_{ }) }{{\rm d}t}
 \;, \label{entropy_conservation}
\ee
which is known as the law of {\em entropy conservation}. 
\index{entropy conservation}
The entropy density $\bar{s}$ is in turn proportional to $T^3_{ }$, with 
a slowly evolving coefficient 
(this will be discussed around \eq\nr{p_r}).
Therefore the ratio of scale factors today ($a^{ }_\now$)
and at the end of inflation ($a^{ }_e$), 
which we call the {\em redshift factor}, 
\index{redshift factor: during radiation domination}
can be written as 
\be
 \frac{a^{ }_\inow}{a^{ }_e}
 \;
 \overset{\rmii{\nr{entropy_conservation}}}{=}
 \;
  \frac{T^{ }_e}{T^{ }_\inow} 
   \underbrace{ 
   \biggl( \frac{\bar s^{ }_e / T_e^3}
                {\bar s^{ }_\inow / T_\inow^3 } \biggr)^{1/3}_{ }
   }_{\rm slowly~evolving}
 \;. \label{evolution_a}
\ee

An important point is that
the derivation of \eq\nr{evolution_a} relied on the assumption
of thermal equilibrium. However, the same relation is 
also used when this is {\em not} the case. Notably, {\em neutrinos 
decouple} from the electromagnetic plasma at $T\sim 2$~MeV. Below
this temperature, they still carry energy 
and momentum density, and therefore affect the
overall expansion, but their phase space distribution is non-thermal
(often it is modelled
by saying that neutrinos have a different temperature than 
the electromagnetic plasma; sometimes chemical potentials
are added for further structure). 
It requires a theoretical computation
to determine how the decoupling affects the scale factor. 
Afterwards, we may continue to express the result like in \eq\nr{evolution_a}, 
with~$T$ denoting the temperature of the electromagnetic plasma. In other
words, we {\em redefine the meaning of the entropy density}, so that it
incorporates the contribution from the decoupled neutrinos to the
universe expansion 
(unfortunately, this definition of a non-equilibrium
entropy is not unique). 
At the current temperature, 
identified with that of the CMB background, 
$T^{ }_\now \approx 2.7255$~K, the newly defined 
{\em present-day entropy density}, 
\index{entropy density: current value}
$\bar{s}^{ }_\now$, can be expressed
in terms of the parameter $h^{ }_*$ from \eq\nr{p_r}, 
with a value $h^{ }_{*,0} \approx 3.930$~\cite[table~1]{mea2}. 

\index{neutrino decoupling}
\index{decoupling: neutrinos}

{}From \eq\nr{evolution_a}, we can estimate how many 
{\em $e$-folds} \index{$e$-folds}
took place after inflation. As an example, let us
choose a high temperature after inflation, 
$T^{ }_e = 10^{15}_{ }\,$GeV. 
Assuming that the Standard Model
equation of state~\cite{eos15} 
is the dominant one up to $T^{ }_e$, we obtain
\be
 \ln\biggl( \frac{a^{ }_\inow}{a^{ }_e} \biggr)
 \biggr|^{ }_{T^{ }_e\,=\,10^{15}_{ }\,{\rm GeV}}
 \; 
 \underset{\rmii{\cite{eos15}} \; }{
 \overset{\rmii{\nr{evolution_a}} \lift }{=}} 
 \;  64.7 
 \;. \label{min_e_folds}
\ee
Requiring that the current observable universe was 
causally connected before inflation, it  
is therefore likely 
that inflation lasted at least 65 $e$-folds. In inflationary model
building, it is customary to posit that the pivot scale
(cf.\ \eq\nr{eq_powerlaw}) exited the Hubble horizon 50--60 
$e$-folds before inflation ended, and display the corresponding
variation as a systematic error band related to an unknown reheating
history. 
A way to avoid this uncertainty and 
determine the correct amount of $e$-folds is explained
in the context of \fig\ref{fig:bg_thermal} on p.~\pageref{fig:bg_thermal}.

Next, let us estimate the Hubble rate, from \eq\nr{eq_end0-1}
(setting $\kappa \to 0$). We write it as 
\be
 H
 \;
 \overset{\rmii{\nr{eq_end0-1}}}{=}
 \;
 \sqrt{\frac{8\pi (\bar{e}/T^4_{ })}{3}} \frac{T^2_{ }}{\mpl^{ }}
 \;, \label{Hubble}
 \index{Hubble rate} 
\ee
where the ratio $\bar{e}/T^4_{ }$ is also slowly evolving, 
and (during radiation domination) 
often parametrized through an effective number of light 
degrees of freedom, via \eq\nr{p_r}. 

With these ingredients, we can estimate when a given comoving
momentum mode entered inside the Hubble horizon. We write the 
relevant ratio, evaluated at the end of inflation, as 
\be
 \frac{k}{a^{ }_e H^{ }_e}
 \; \overset{k \,=\, a^{ }_\inow p^{ }_\inow \lift }{=} \; 
 \frac{p^{ }_\inow}{T^{ }_\inow}
 \frac{a^{ }_\inow}{a^{ }_e} \frac{T^{ }_\inow}{H^{ }_e}
 \;\;
 \underset{\rmii{\nr{Hubble}}}
 {\overset{\rmii{\nr{evolution_a}} \lift }{=}}
 \;\;
 \frac{p^{ }_\inow}{T^{ }_\inow}
   \biggl( \frac{\bar s^{ }_e / T_e^3}
                {\bar s^{ }_\inow / T_\inow^3 } \biggr)^{1/3}_{ }
 \sqrt{\frac{3}{8\pi}}
 \frac{\mpl^{ }/T^{ }_e}{\sqrt{\bar{e}/T_e^4}}
 \;. \label{k_aeHe} 
\ee
As examples of numerical values, we obtain
\be
   \biggl( \frac{\bar s^{ }_e / T_e^3}
                {\bar s^{ }_\inow / T_\inow^3 } \biggr)^{1/3}_{ }
 \sqrt{\frac{3}{8\pi}}
 \frac{\mpl^{ }/T^{ }_e}{\sqrt{\bar{e}/T_e^4}}
 \;\; \overset{\rmii{\cite{eos15}~}}{\approx} \;\; 
 \left\{ 
 \begin{array}{ll}
  2.147 \times 10^3_{ }\;,    &  T^{ }_e = 10^{15}_{ }\,\mbox{GeV} \\[2mm] 
  3.145 \times 10^{21}_{ }\;, &  T^{ }_e = 1\,\mbox{MeV}
 \end{array}
 \right.
 \;. \label{redshift_appro}
\ee

Next, we need to express $p^{ }_\now/T^{ }_\now$ in terms of $f^{ }_\now$
or $\lambda^{ }_\now$. In terms of frequency, 
\be
 \frac{p^{ }_\inow}{T^{ }_\inow} 
 \quad
 \underset{\rmi{Hz = 1/s}}{\overset{p^{ }_\inow\,=\,2\pi f^{ }_\inow}{=}}
 \quad
 \frac{2\pi}{\mbox{s}\,T^{ }_\inow} \frac{f^{ }_\inow}{\mbox{Hz}}
 \; \approx \; 
 1.761 \times 10^{-11}_{ }\, \frac{f^{ }_\inow}{\mbox{Hz}}
 \;. \label{p0_f0}
\ee
Combining this with \eq\nr{redshift_appro}, we can find at which moment
today's frequencies entered inside the Hubble horizon. For instance, 
\ba
 \frac{k}{a^{ }_e H^{ }_e} \biggr|^{ }_{T^{ }_e\,=\,10^{15}_{ }\,{\rm GeV}}
 \; = \; 1 
 & \underset{\rmii{\nr{p0_f0}}}
  {\overset{\rmii{\raise1ex\hbox{\nr{k_aeHe},\nr{redshift_appro}}}}
  {\Leftrightarrow}} &
 f^{ }_\now 
 \; = \; 2.645 \times 10^7_{ }\,\mbox{Hz}
 \;, \label{max_gw_vac} \\[2mm]
%%%%%
 \frac{k}{a^{ }_e H^{ }_e} \biggr|^{ }_{T^{ }_e\,=\,1\,{\rm MeV}}
 \; = \; 1 
 & \underset{\rmii{\nr{p0_f0}}}
  {\overset{\rmii{\raise1ex\hbox{\nr{k_aeHe},\nr{redshift_appro}}}}
  {\Leftrightarrow}} &
 f^{ }_\now 
 \; = \; 1.806 \times 10^{-11}_{ }\,\mbox{Hz}
 \;. \label{f0_MeV} 
\ea
% The scaling with $T^{ }_e$ in between these two 
% limits is, to a good approximation, linear.

If we employ wavelengths instead, then 
\be
 \frac{p^{ }_\inow}{T^{ }_\inow}
 \; 
 \overset{p^{ }_\inow \;=\; 2\pi / \lambda^{ }_\inow \lift }{=} 
 \; 
 \frac{2\pi}{{\rm pc}\, T^{ }_\inow} 
 \biggl(\frac{\lambda^{ }_\inow}{\rm pc}\biggr)^{-1}_{ }
 \; \overset{\rmii{\nr{parsec}}}{=}\;
 1.711 \times 10^{-19}_{ } 
 \biggl(\frac{\lambda^{ }_\inow}{\rm pc}\biggr)^{-1}_{ }
 \;. \label{p0_lam0}
\ee
For instance, 
\ba
 \frac{k}{a^{ }_e H^{ }_e} \biggr|^{ }_{T^{ }_e\,=\,10^{15}_{ }\,{\rm GeV}}
 \; = \; 1 
 & \underset{\rmii{\nr{p0_lam0}}}
  {\overset{\rmii{\raise1ex\hbox{\nr{k_aeHe},\nr{redshift_appro}}}}
 {\Leftrightarrow}} &
 \lambda^{ }_\now 
 \; = \; 3.673 \times 10^{-16}_{ }\,\mbox{pc}
 \;, \\[2mm]
%%%%%
 \frac{k}{a^{ }_e H^{ }_e} \biggr|^{ }_{T^{ }_e\,=\,1\,{\rm MeV}}
 \; = \; 1 
 & \underset{\rmii{\nr{p0_lam0}}}
  {\overset{\rmii{\raise1ex\hbox{\nr{k_aeHe},\nr{redshift_appro}}}}
 {\Leftrightarrow}} &
 \lambda^{ }_\now 
 \; = \; 538.1\,\mbox{pc}
 \;.  
\ea
% Again, the scaling in between is almost linear.

Finally, 
we can relate frequencies and wavelengths directly, 
\be
 \frac{f^{ }_\inow}{\mbox{Hz}}
 \; 
 \overset{f^{ }_\inow \;=\; c / \lambda^{ }_\inow \lift }{=}
  \; 
 \frac{c\,{\rm s}}{\rm pc}
 \biggl(\frac{\lambda^{ }_\inow}{\rm pc}\biggr)^{-1}_{ }
 \; \overset{\rmii{\nr{parsec}}}{=} \;
 9.716 \times 10^{-9}_{ }
 \biggl(\frac{\lambda^{ }_\inow}{\rm pc}\biggr)^{-1}_{ }
 \;. \label{f0_lam0}
\ee
The scales probed by the CMB correspond to 
$\lambda^{ }_\now = (10\, ...\, 3\times 10^3_{  })\,$Mpc, which then 
implies $f^{ }_\now = (10^{-15}_{ }\, ...\, 3\times 10^{-18}_{ })\,$Hz. 
It should be clarified that this
is the frequency of a light-like perturbation that leads to 
an anisotropy in temperature measurements in various
directions. It is not the peak frequency of the CMB Planck spectrum, 
which rather lies in the microwave range,  
$f^{ }_\rmii{CMB} \sim 10^{11}_{ }\,$Hz, 
as can be deduced by setting 
$p^{ }_\now \sim T^{ }_\now$ in \eq\nr{p0_f0}.

%%%%%%%%%%%%%%%%%%%%%%%%%%%%%%%%%%%%%%%%%%%%%%%%%%%%%
%
\subsection{What do power spectra mean in different contexts?}
\label{sec_gaussian}

We now turn to somewhat technical issues, needed frequently
in the later chapters. 
Given a set of data for the perturbations 
$\delta Q$ of a physical quantity $Q \in \mathbbm{R}$, 
observed or simulated, we define 
its power spectrum, $\P^{ }_{\mbit\iQ}(k)$, such as $\P^{ }_\scalar$ 
and $\P^{ }_\tensor$ mentioned above, and contrast with 
the determination of power spectra with theoretical methods.

We work in comoving conformal coordinates, 
and denote by $\vec x$ and $\vec k$ the 
positions and momenta, respectively. The time argument 
$\tau$ is often left implicit, 
and unless stated otherwise we assume equal times in the correlations. 
If we place ourselves at $\vec x^{ }_\inow$ and consider 
a light ray travelling in direction $\vec n$, then a plane wave of 
momentum $\vec k$ has the phase 
$e^{i\vec k\cdot (\vec x^{ }_\inow - \Delta \tau\, \vec n) }_{ }$.
The angular momentum that we met above, $\ell$, originates from 
the expansion of 
$e^{ - i \Delta \tau\, \vec k\cdot \vec n }_{ }$
in spherical harmonics
(cf.\ the discussion below \eq\nr{deltaT_fourier}), however here
we work with $\vec x$ and $\vec k$ rather than $\ell$.
To streamline the notation, we define 
\be
 \int_{\tau^{ }_i}
  \; \equiv \; 
	\int_{-\infty}^\tau  \!\dd   \tau_i
  	%\int_{-\infty}^\tau  \!\dd   \tau_2
 \;, \quad
 \int_{\vec{x}_i^{ }}
  \; \equiv \; 
	\int 	\!\dd^3_{ } \vec x_i
	%\int 	\!\dd^3_{ } \vec x_2
 \;, \quad
 \int_{\vec{k}_i^{ }}
 \; \equiv \; 
	\int \!\frac{\dd^3_{ } \vec k_i}{(2\pi)^3}
       % \int \!\frac{\dd^3_{ } \vec k_2}{(2\pi)^3} 
 \;, \quad
 \int_{a_i^{ }, \, b_i^{},\, ...}
 \; \equiv \; 
	\int_{a_i^{ }} \int_{b_i^{ }} \int_{...}   
 \;. 
\ee

%%%%%%%%%%%%%%%%%%%%%%%%%%%%%%%%%%%%%%%%%%%%%%%%%%%%%%%%%%%%%%%

\subsubsection*{Formal definitions and basic properties}

A classical {\em statistical or ensemble average} 
\index{ensemble average}
is defined as the weighted sum over allowed configurations,
\begin{equation}
 \langle \mathcal{O}[Q] \rangle 
 \;\equiv\;
 \sum_i p_i^{ } \, \mathcal{O}[Q^{(i)}_{ }] 
 \ , \label{classical_average}
 \index{statistical average}
\end{equation}
with $p_i^{ }$ being the probability to realize the $i$-th configuration. 
If we assume the system to be {\em translationally invariant}, 
\index{translational invariance}
i.e.\ with no preferred location, then equal-time 
2-point correlations can only depend 
on the relative spatial separation,
\begin{align}
 \langle\, \delta Q(%\tau ,
 \vec x_1^{ })\,\delta Q(%\tau ,
 \vec x_2^{ }) \,\rangle
 =
 \langle\,
 \delta Q(%\tau ,
 \vec x_1^{ }-\vec x_2^{ })\,
 \delta Q(%\tau ,
 \vec 0) \,\rangle
 \ . \label{eq_averageO}
 \index{translational invariance}
\end{align}
Inverse Fourier transforming \eq\nr{eq_averageO}
to comoving momentum yields
\ba
 \langle\, \delta Q(%\tau ,
 \vec k_1^{ })\,\delta Q^*(%\tau ,
 \vec k_2^{ }) \,\rangle
 &
 \underset{\rmii{ }}%\nr{correspondence}}}
 {\overset{\rmii{\nr{fourier_k}}}{=}} 
 & 
 \int_{\vec x_1, \vec x_2} \!\!%\,
 e^{-i(\vec k_1 \cdot\vec x_1-\vec k_2\cdot \vec x_2)}_{ }
 \langle\, \delta Q(%\tau ,
 \vec x_1^{ })\,\delta Q(%\tau ,
 \vec x_2^{ })
 \,\rangle \nonumber\\[2mm]
%%%%%
 & \overset{\rmii{\nr{eq_averageO}}}{=} &
 \int_{\vec x_1, \vec x_2} \!\!%\,
 e^{-i(\vec k_1 \cdot\vec x_1-\vec k_2\cdot \vec x_2)}_{ }
 \langle\, \delta Q(%\tau ,
 \vec x_1^{ }-\vec x_2^{ })\,\delta Q(%\tau ,
 \vec 0)
 \,\rangle \nonumber\\[2mm]
%%%%%
 & \overset{ \scriptscriptstyle
             \vec x_1 \to \vec x_1 + \vec x_2 \vphantom{ \big| } }{=} &
 \! \int_{\vec x_1, \vec x_2} \!\!%\,
 e^{-i(\vec k_1 -\vec k_2) \cdot \vec x_2}_{ }
 e^{-i\vec k_1\cdot \vec x_1}_{ }
 \langle\, \delta Q(%\tau ,
 \vec x_1^{ })\,\delta Q(%\tau ,
 \vec 0)
 \,\rangle \nonumber\\[2mm]
%%%%
 & \overset{\hphantom{\rmiii{  }} }{=} &
 (2\pi)^3_{ } \delta^{\rmi{(3)}}_{ }(\vec{k}_1^{ }-\vec{k}_2^{ })
 \underbrace{ \int_{\vec x_1} %\!\dd^3 \vec{x}_1 \, 
 e^{-i\vec k_1\cdot \vec x_1}
 \langle\, \delta Q(%\tau ,
 \vec x_1^{ })\,\delta Q(%\tau ,
 \vec 0)
 \,\rangle }_{{\rm like}\;\nr{fourier_k}:\;\equiv\; 
 P_{\mbit\rmiii{\it Q}}^{ }(%\tau ,
 \vec k_1^{ })}
 \ . \hspace*{6mm} \label{eq_PQdef}
 \index{$P^{ }_{\mbit\iQ}$ (power spectrum as Fourier transform)}
\ea
The Dirac-$\delta$ can be interpreted as momentum conservation. 
So translational invariance
implies momentum conservation,
in accordance with {\em Noether's theorem}.
 
\index{Noether's theorem}

If we go back to coordinate space, and look at equal positions, 
we obtain 
\ba
 \langle\, \delta Q^2(%\tau ,
 \vec x)%\,\delta Q(%\tau ,\vec x) 
 \,\rangle
 & %\underset{\rmii{\nr{correspondence}}}
 {\overset{\rmii{\nr{fourier_x}}}{=}} &
 \int_{\vec k_1, \vec k_2} \!\!%\,
 e^{i (\vec k_1 - \vec k_2) \cdot \vec x}
 \,\langle\, \delta Q(%\tau ,
 \vec k_1)\,\delta Q^*(%\tau ,
 \vec k_2) \,\rangle
 \label{2pt_fourier_x} \\[2mm]
%%%%%
 & \overset{\rmii{\nr{eq_PQdef}}}{=} &
 \int_{\vec k_1, \vec k_2} \!\!%\,
 e^{i(\vec k_1 - \vec k_2) \cdot \vec x}
 \,(2\pi)^3_{ } \delta^{(3)}_{ }(\vec{k}_1-\vec{k}_2)
 P^{ }_{\mbit\iQ}(%\tau ,
 \vec k_1) 
 \nonumber\\[2mm]
%%%%
 & = &
 \underbrace{
 \int_0^\infty \!\frac{\dd k_1}{k_1}
 }_{ 
  \int_{-\infty}^\infty \dd \ln k^{ }_1
 }
 \, 
 \underbrace{ \frac{k_1^3}{(2\pi)^3}
 \int \! \dd\Omega_{\vec k_1} \,
 %\langle \delta Q(\tau ,\vec k_1) \,\delta Q^*(\tau ,\vec k_1) \rangle 
 	P^{ }_{\mbit\iQ}(%\tau ,
 \vec k_1)
 }_{\equiv\, \P_{\mbit\rmiii{\it Q}}^{ }(%\tau,
	k^{ }_1)}
 \ , \label{eq_angular_average}
 \index{$\P^{ }_{\mbit\iQ}$ (power spectrum after angular average)}
\ea 
where 
$\dd\Omega_{\vec k} \equiv \dd \phi \, \dd \theta \sin\theta$ 
is the {\em infinitesimal solid angle} 
\index{${\rm d}\Omega$ (infinitesimal solid angle)}
in spherical coordinates, and
$\P_{\mbit\iQ}^{ }(k) $ is what we call 
the {\em power spectrum} of $\delta Q$.
\index{power spectrum: general definition} 
We remark that this procedure
is a bit formal, as nothing guarantees that the 
momentum integral in \eq\nr{eq_angular_average} is convergent.
Nevertheless, the {\em integrand} can be defined.  
We also note that the overall outcome is independent of
the position $\vec x$, 
reflecting again translational invariance. 

%%%%%%%%%%%%%%%%%%%%%%%%%%%%%%%%%%%%%%%%%%%%%%%%%%%%%%%%%%%%%%%

\subsubsection*{What if the spatial volume is finite?}

\index{spatial average}

If we carry out a numerical simulation, or inspect the visible
universe, the spatial volume is finite. In this situation
there is a maximal possible wavelength, and a minimal possible momentum. 
In order to maintain the property of translational invariance, it is 
convenient to impose periodic boundary conditions over the finite volume
(this is also natural if we work with angular variables). 
Then momenta get quantized, 
$
 \vec{k} = 2 \pi \vec{m} / L 
$, 
where $\vec{m}\in\mathbbm{Z}^3_{ }$ and the volume has been taken to 
be a box of size $L^3_{ }$. Momentum integrals 
become sums, 
$
 \int_\vec{k} \to \frac{1}{L^3} \sum_\vec{m}
$; 
Dirac-$\delta$'s become Kronecker-$\delta$'s, 
$
 (2\pi)^3_{ }\delta^{(3)}_{ }({\vec k_1 - \vec k_2}) \to 
 \int_\vec{x} e^{ -i\vec{x}\cdot({\vec k_1 - \vec k_2})}
 = 
 L^3_{ }\delta^{ }_{\vec m_1,\vec m_2}
$; 
and the definitions in 
\eqs\nr{eq_PQdef} and \nr{eq_angular_average} 
can be adapted accordingly. 

A large volume offers for an alternative
to the statistical average in \eq\nr{classical_average}, 
namely the {\em spatial average}, as we now illustrate.
Let us consider a generic $2$-point function.
If far-away regions are uncorrelated, it satisfies
\begin{equation}
 \langle \mathcal O(%\tau_1,
 \vec x) \, 
 \mathcal O (%\tau_2,
  \vec y)\rangle  
 \;
 \xrightarrow[]
             { \scriptscriptstyle  
               |\vec x - \vec y| \;\to\; \infty }
 \;
 \langle \mathcal O (%\tau_1,
 \vec x) \rangle 
 \langle \mathcal O (%\tau _2,
 \vec y) \rangle
 \ . \label{discon}
\end{equation} 
In many systems, the approach to this limit is exponentially fast, 
such that we can write
\begin{equation}
 \langle \mathcal O (%\tau _1,
 \vec x) \, 
 \mathcal O (%\tau _2,
 \vec y)\rangle 
 =
 \langle \mathcal O (%\tau _1,
 \vec x) \rangle
 \langle \mathcal O (%\tau _2,
 \vec y) \rangle
 +
 f(%\tau _1,\tau_2,
 |\vec x - \vec y|)\, e^{-|\vec x - \vec y| / \xi} \ ,
 \label{xi_def}
\end{equation}
where $\xi$ is a characteristic 
{\em correlation length}, \index{correlation length}
and the function~$f$ grows at most as a power at large argument. 

Let us now insert $\delta Q^2(\vec x)$
in the role of $\mathcal O(\vec x)$. {}From 
\eq\nr{eq_angular_average} we know that 
$\langle\mathcal O(\vec x)\rangle$ is independent of $\vec{x}$.
From \eq\nr{xi_def} we see that, on average, the values of 
$\delta Q^2(\vec x)$ and $\delta Q^2(\vec y)$ are uncorrelated, 
if $|\vec x - \vec y| \gg \xi$. Therefore, we can view 
$\delta Q^2(\vec x)$ and $\delta Q^2(\vec y)$ as independent 
realizations from the statistical ensemble that defined
\eq\nr{classical_average}. In other words, the statistical
average in \eq\nr{classical_average} can be replaced by a volume
average, provided that the volume is large, $L \gg \xi$,  
\ba
 \langle\, \delta Q^2_{ }(\vec x) \,\rangle 
 &
 \overset{L\; \gg\; \xi }{\approx}
 &
 \frac{1}{L^3} \int_{\vec x} %\!\dd^3 \vec{x} \, 
 \delta Q^2_{ }(%\tau ,
 \vec x)
 \ .
 \label{spat_ave}
\ea
The same procedure also applies to unequal-position correlators, 
like in \eq\nr{eq_averageO}. 

%%%%%

\index{cosmic variance}

A replacement like in \eq\nr{spat_ave} is essential for observational
cosmology. We have only one universe, and cannot take an average over
many statistically independent realizations. 
Instead, we take an angular average over observation directions, 
which physically corresponds to a spatial average over the 
last scattering surface. The result can be compared with theoretical
models based on a statistical ensemble, provided that 
the average contains many uncorrelated domains. 
Instead, when we consider small values of $\ell$, we are 
correlating large angular separations, and our single universe
does not have many independent patches. This
is the issue of {\em cosmic variance}
(cf.\ \fig\ref{fig:D_l_sketch}), and it restricts the
information that is available about the %very 
largest observable structures. 

%%%%%%%%%%%%%%%%%%%%%%%%%%%%%%%%%%%%%%%%%%%%%%%%%%%%%%%%%%%%%%%

\subsubsection*{What if we take a quantum rather than a statistical average?}

\index{quantum average}

The definition in \eq\nr{classical_average} refers to a {\em classical}
statistical average. In contrast, the inflationary paradigm, discussed
in part~II of this book, 
postulates that the fluctuations that we see have 
a quantum-mechanical origin, and the expectation value that we compute
is a (distant-past)
vacuum expectation value in the corresponding Hilbert space. 

In the quantum-mechanical framework, we can still formally
define a correlator like in \eq\nr{2pt_fourier_x}. 
The power spectrum is extracted from the integrand as before
(though the issue of non-convergence of the integral
becomes concrete, because there are vacuum fluctuations at all scales).
Fourier transforms, 
$\delta Q(\vec{k}^{ }_1)$, are replaced by what we call
{\em mode functions}, $\field^{ }_k$, 
which appear with annihilation and creation operators, 
$ w^{ }_\vec{k} $ and $ w^{\dagger}_\vec{q} $, 
so that 
\index{mode function}
\ba
 \bigl\langle\, 
  \field^{\hspace*{0.3mm}2}_{\varphi}(\vec{x})
 \,\bigr\rangle
 & 
 \underset{\rmii{average}}{
 \overset{\rmii{quantum} \lift }{\equiv}}
 & 
 \bigl\langle \bit 0 \bit \big| 
  \field^{ }_{\varphi}(\vec{x})
  \field^{ }_{\varphi}(\vec{x})
 \big| \bit 0 \bit \bigr\rangle
 \nn[2mm]
%%%
 & \overset{\rmii{\nr{mode_expansion}}}{=} &
 \int \! \frac{ {\rm d}^3\vec{k} \, {\rm d}^3_{ }\vec{q} }{(2\pi)^3_{ }} 
 \, 
 \bigl\langle \bit 0 \bit \big| \,
 w^{ }_\vec{k} 
 \, \field^{ }_k
 \,  e^{i\vec{k}\cdot\vec{x}}_{}
 \, w^{\dagger}_\vec{q}
 \, \field^{*}_q
 \, e^{- i\vec{q}\cdot\vec{x}}_{}
 \,\big| \bit 0 \bit \bigr\rangle
 \; 
 \overset{\rmii{\nr{commutators}}}{=} 
 \; 
 \int_\vec{k} | \field^{ }_k |^2_{ } \ . \hspace*{9mm}
 \label{quantum_average}
\ea 
Compared with the derivation leading to \eq\nr{eq_angular_average}, 
the difference is that momentum conservation,  
$\delta^{(3)}_{ }(\vec{k-q})$, does not arise from 
translational invariance (cf.\ \eq\nr{eq_PQdef}), 
but from a commutation relation.  
We return to \eq\nr{quantum_average} around 
\eqs\nr{P_Q_k} and \nr{P_Q_k_again}. 

%%%%%%%%%%%%%%%%%%%%%%%%%%%%%%%%%%%%%%%%%%%%%%%%%%%%%%%%%%%%%%%

\subsubsection*{What if we average over 
the initial rather than the final ensemble?}

Let us now restore time dependence to the problem. 
In theoretical considerations, the observable 
$\delta Q(\tau,\vec x)$ is normally obtained 
from some differential equation,
whose (special) solution at linear order takes the form  
\be
 \mathcal{D}\, \delta Q(\tau ,\vec x) 
 = 
 \rho (\tau ,\vec x) 
 \quad \Rightarrow \quad
 \delta Q(\tau ,\vec x)
 =
 \int_{\tau_1^{ },\,\vec y}
 G^{ }_\iR(\tau ,\tau_1^{ },\vec x , \vec y) \,
 \rho(\tau_1^{ },\vec y) 
 \ . \label{eq_phiGrho} 
 \index{retarded Green's function: general}
 \index{$G^{ }_\iR$ (retarded Green's function)}
\ee
%where 
Since the background is homogeneous and isotropic, 
and the coefficients of $\mathcal{D}$
are determined by the background, we assume that 
the coefficients of $\mathcal{D}$ do not depend on $\vec x$. 
Then $\mathcal{D}$ and thus also the 
{\em retarded Green's function}, 
$G^{ }_\iR \equiv\mathcal{D}^{-1}$, 
are translationally 
and rotationally invariant, and % we can assume 
$G^{ }_\iR(\tau ,\tau_1^{ },\vec x , \vec y)
=G^{ }_\iR(\tau ,\tau_1^{ },|\vec x - \vec y|)$.

It is helpful to transform \eq\nr{eq_phiGrho} 
to comoving momentum space,
\ba
 \delta Q (\tau ,\vec k) 
 & \underset{\rmii{\nr{eq_phiGrho}}}
   {\overset{\rmii{\nr{fourier_k}} \lift }{=}} &
 \int_{\vec x,\,\tau_1^{ },\,\vec y} 
 G^{ }_\iR(\tau ,\tau_1^{ },|\vec x- \vec y|)
 \, \rho (\tau_1^{ },\vec y) \,e^{-i\vec k\cdot\vec x}_{ }
 \nonumber\\[2mm]
%%%%
 & = & 
 \int_{\tau_1^{ }} 
 \underbrace{
 \int_{\vec y} 
 \rho (\tau_1^{ },\vec y) e^{-i\vec k\cdot\vec y}_{ }
 }_{{\rm from}\;\nr{fourier_k}:\;  \rho (\tau_1^{ },\vec k)}
 \; 
 \underbrace{
 \!\int_{\vec x} 
 G^{ }_\iR(\tau ,\tau_1^{ },|\vec x- \vec y|)
 \, e^{-i\vec k\cdot(\vec x -\vec y)}_{ } 
 }_{\equiv\; G^{ }_\iR(\tau ,\tau_1^{ },\ibit k)} 
 \ . \label{G_R_k}
\ea
With the assumptions made, 
the momentum-space Green's function is real,  
\be
 G^*_\iR(\tau ,\tau_1^{ }, k)
 \; \overset{\rmii{\nr{G_R_k}}}{=} \; 
 \int_\vec{x}
 G^{ }_\iR(\tau ,\tau_1^{ },|\vec x|)
 e^{+i\vec k\cdot\vec x }_{ } 
 \;
 \overset{\scriptscriptstyle \vec x\to -\vec x \lift }{=}
 \; 
 G^{ }_\iR(\tau ,\tau_1^{ }, k)
 \;. \label{G_R_real}
\ee

With \eq\nr{G_R_k}, 
we can evaluate the correlator in \eq\nr{2pt_fourier_x}. 
Given that the Green's function~$G^{ }_\iR$ is deterministic, 
the statistical average can be transported  to the initial state. 
Assuming the latter to be translationally invariant, we find
\ba
 && \hspace*{-1.5cm} 
 \langle\, \delta Q^2(\tau ,\vec x) \,\rangle
 \nonumber\\[2mm]
%%%%%
 &
 \underset{ }
 {\overset{\rmii{\nr{2pt_fourier_x}}}{=}}
 &
 \int_{\vec k_1^{ }, \vec k_2^{ }} \!\!%\,
 e^{i(\vec k_1^{ } - \vec k_2^{ })\cdot \vec x }
 \,\langle\, \delta Q(\tau ,
 \vec k_1^{ })\,\delta Q^*(\tau ,
 \vec k_2^{ }) \,\rangle
 \nonumber\\[2mm]
%%%%%
 &
   \underset{\rmii{\nr{G_R_real}}}
   {\overset{\rmii{\nr{G_R_k}} \lift }{=}}
 &
 \int_{\tau^{ }_1,\,\tau^{ }_2,\,\vec{k}_1^{ },\,\vec{k}_2^{ }} \!
	e^{i (\vec k_1^{ }-\vec k_2^{ }) \cdot \vec x}_{ }
        \,
	G^{ }_\iR(\tau ,\tau_1^{ }, k_1^{ })\,
        G^{ }_\iR(\tau ,\tau_2^{ }, k_2^{ })
	\hspace*{-6mm}
          \underbrace{
           \langle\; \rho(\tau_1^{ },\vec k_1^{ })
                   \rho^*(\tau_2^{ },\vec k_2^{ }) \;\rangle
           }_{
           {\rm from}\;\nr{eq_PQdef}:\; 
           (2\pi)^3_{ } 
           \delta^{(3)}_{ }(\vec k_1^{ }-\vec k_2^{ }) P^{ }_{\mbit\rho}
           (\tau_1^{ },\tau_2^{ },\vec k_1^{ })
           }
 \nonumber\\[2mm]
%%%%%
 &
 \underset{\rmii{\nr{eq_wow}}}
 {\overset{\scriptscriptstyle \vec{k}^{ }_\rmiii{1}\to\vec k
  \lift }{=}}
 &
 \boxed{
 \quad   \vphantom{\Bigg|}
	\int_{-\infty}^\infty \!\dd \ln k \,
 \underbrace{
        \int_{\tau^{ }_1,\,\tau^{ }_2}\!%{-\infty}^\tau  \!\dd \tau_1^{ } 
       	%\int_{-\infty}^\tau  \!\dd \tau_2^{ } \,
	G^{ }_\iR(\tau ,\tau_1^{ }, k)\,
        G^{ }_\iR(\tau ,\tau_2^{ }, k)\, 
	\P_{\mbit\rho}(\tau_1^{ },\tau_2^{ },k) }_{
 {\rm from}\;\nr{eq_angular_average}:\; 
    \P_{\mbit\rmiii{\it Q}}^{ }(\tau ,k)}
 \;. \quad
 } 
 \label{noise_average} 
\ea
We have here converted the Fourier transform, $P^{ }_{\mbit\rho}$, 
to the power spectrum, $\P^{ }_{\mbit\rho}$,     
\begin{equation}
 \P^{ }_{\mbit\rho} (\tau^{ }_1,\tau^{ }_2, k) 
 \;\overset{\rmii{\nr{eq_angular_average}}}{\equiv}\;
 \frac{k^3}{(2\pi)^3_{ }} \int \!\dd\Omega_{\vec k} \,
  P^{ }_{\mbit\rho} (\tau_1^{ },\tau_2^{ }, \vec k)
  \ . \label{eq_wow}
\end{equation}
What we learn from \eq\nr{noise_average} is that
at linear order in perturbations, 
the statistical properties of final-state correlators are 
equivalent to those of initial-state correlators, 
except that in general we need 
{\em unequal-time} power spectra on the initial-state side.

We remark that if the source, $\rho$, is stochastic in nature, 
then the average over the source realizations
in \eq\nr{noise_average} could be either 
a statistical (cf.\ \eq\nr{classical_average})
or a quantum one (cf.\ \eq\nr{quantum_average}).
In the statistical case, it can be further simplified
under certain assumptions. 
Notably, if we consider small momenta, and long time differences, 
then the source fluctuations should be uncorrelated in time, 
$\P^{ }_{\mbit\rho}(\tau_1^{ },\tau_2^{ }, k) \approx
\Omega^{ }(\tau^{ }_1,k)\, \delta( \tau_1^{ }-\tau_2^{ })$ 
(cf. \eq\eqref{varphi_noise}), 
meaning that we have {\em white noise}. \index{white noise}
In this case 
\eq\eqref{noise_average} takes a simpler form,
\ba 
 	\P^{ }_{\mbit\iQ}(\tau , k)
 \;
 \underset{\rmii{%\nr{eq_angular_average},
                 \nr{noise_average}}}
 {\overset{k\;\to\;0}{\approx}}
 \; 
        \int_{\tau_1^{ }}
	G^2_\iR(\tau ,\tau_1^{ }, k)\, 
	\Omega(\tau_1^{ }, k) \ .
 \label{eq_spectrum_stochastic} 
\ea
On the other hand, in the quantum case, coherence implies the 
presence of long-time correlations. 
We further elaborate on the notion of a quantum noise
in \se\ref{ss:stochastic_exact}.

%%%%%%%%%%%%%%%%%%%%%%%%%%%%%%%%%%%%%%%%%%%%%%%%%%%%%%%%%%%%%%%

\subsubsection*{What is the role of Gaussian statistics,
and how to include non-Gaussianity?}

\index{non-Gaussianity}

The definitions of the power spectra above do {\em not} require
assumptions about the Gaussian nature of the statistical ensemble;
such assumptions become relevant only when 
we discuss 3-point or 4-point correlators 
and their relations to 2-point correlators. 
On the observational side, there is in general no  
reason for a Gaussian premise. 
On the theoretical side, we did make an additional assumption in 
\eq\nr{eq_phiGrho}, when we restricted to linear order in
perturbations. Then we found that 
each Fourier component, $\delta Q(\tau,{\vec k})$, 
evolves independently, so that the initial and final 
states have the same statistical properties. In particular, 
if the initial statistics is Gaussian, 
the final ensemble inherits this property. 

\label{perturbative}

The linear order in perturbations corresponds to the leading term
of a {\em perturbative approach}, \index{perturbative approach} 
similar to the Feynman diagrams of perturbative quantum field theory. 
In this approach, interactions between the various Fourier modes, which
lead to non-Gaussianities, are treated order-by-order in a Taylor
expansion. This can be applied both to the computation of initial-state 
correlators, as well as to their subsequent evolution until today. 

In order to illustrate the procedure, let us denote the result for 
$\delta Q$, obtained from linear order in perturbation theory, 
by $\delta Q^{\scriptscriptstyle (1)}_{ }$. 
Now, assume that we compute $\delta Q$
to second order, and then take a Fourier transform. 
Assuming translational invariance, we might expect
\ba
 \delta Q^{\scriptscriptstyle (2)}_{ }(\tau,\vec k)
 & \simeq & 
 \delta Q^{\scriptscriptstyle (1)}_{ }(\tau,\vec k)
  \label{f_NL}
  \index{$f^{ }_\rmii{NL}$ (bispectrum)}
 \\[2mm]
%%%%%%%%%%%%%
 & + &  
 \int_{\vec k^{ }_1,\vec k^{ }_2}
 (2\pi)^3_{ } \delta^{(3)}_{ }(\vec k - \vec k^{ }_1 - \vec k^{ }_2)
 \, f^{ }_\rmii{NL}(\tau,\vec k^{ }_1,\vec k^{ }_2)
 \, \delta Q^{\scriptscriptstyle (1)}_{ }(\tau,\vec k^{ }_1)
 \, \delta Q^{\scriptscriptstyle (1)}_{ }(\tau,\vec k^{ }_2)
 \;. \nonumber
\ea
This ansatz is actually simplified, as it assumes
locality in time. In any case, the weight function, 
$
 f^{ }_\rmii{NL}(\tau,\vec k^{ }_1,\vec k^{ }_2)
$,
is called a {\em bispectrum}. \index{bispectrum}
Often it is simplified further, 
for instance by treating it as a constant, rather than as 
a function, which is a drastic assumption. 
But even so, 
if it is non-zero, then a correlation function like
\ba
 && \hspace*{-2.5cm}
 \langle \,\delta Q^{\scriptscriptstyle (2)}_{ }(\tau,\vec{x} )
            \,\delta Q^{\scriptscriptstyle (2)}_{ }(\tau,\vec{x} ) 
            \,\delta Q^{\scriptscriptstyle (2)}_{ }(\tau,\vec{x} ) \,\rangle
 \nn[2mm]
 & = & 
 \int_{\vec{k}^{ }_1,\vec{k}^{ }_2,\vec{k}^{ }_3}
 e^{i( \vec{k}^{ }_1 +\vec{k}^{ }_2 +\vec{k}^{ }_3 )\cdot\vec{x} }_{} 
 \langle    \,\delta Q^{\scriptscriptstyle (2)}_{ }(\tau,\vec{k}^{ }_1 )
            \,\delta Q^{\scriptscriptstyle (2)}_{ }(\tau,\vec{k}^{ }_2 ) 
            \,\delta Q^{\scriptscriptstyle (2)}_{ }(\tau,\vec{k}^{ }_3 )
 \,\rangle
\ea 
is also non-zero, because the terms proportional to 
$f^{ }_\rmii{NL}$ yield 4-point correlators, which can be 
reduced to 2-point correlators with the help of 
Isserlis' or Wick's theorem. 
The 2-point correlators have the properties that we have 
discussed above. Therefore,  
we can obtain a theoretical prediction for a 3-point function, 
to then be compared with \eq\nr{E_l}.  

%%%%%%%%%%%%%%%%%%%%%%%%% BIBLIO %%%%%%%%%%%%%%%%%%%%%%%%%%%%%%%%
%

%%%%%%%%%%%%%%%%%%%%%%%%%%%% SECTION %%%%%%%%%%%%%%%%%%%%%%%%%%%%%%%%%%
\newpage 

\section{Evolution equations for first-order perturbations}
\label{se:pert}

\paragraph{Abstract:}

We describe how the left and right-hand sides of the Einstein
equations can be perturbed to first order around the background
solution. The perturbations are then decomposed into scalar, 
vector, and tensor components under {\em helicity}, 
or two-dimensional rotations. The 
resulting Einstein equations for the different types of 
perturbations are displayed. We demonstrate %explain 
how taking
linear combinations of the Einstein equations, or their
integrals (if an equation relates two total derivatives), 
or their derivatives (like when considering energy-momentum
conservation), can lead to simpler equations. 
We also illustrate 
how anisotropic stress can be 
included within the framework of viscous hydrodynamics, and 
how the steps mentioned can be implemented 
with computer algebra. 

\paragraph{Keywords:}

Linearization, 
scalar perturbations, vector perturbations, tensor perturbations, 
anisotropic stress, bulk and shear viscosities, 
computer algebra.

\index{helicity: definition}

%%%%%%%%%%%%%%%%%%%%%%%%%%%%%%%%%%%%%%%%%%%%%%%%%%%
%
\subsection{Definition of perturbations}
\label{ss:def_perts}

In \ch\ref{se:bg} we have defined a homogeneous 
and isotropic ``background''
solution for an expanding universe. 
However, as explained in \ch\ref{se:obs},
the observed universe is not homogeneous at all scales, 
but displays a definite spectrum
of perturbations. The goal of this chapter is to generalize the equations 
introduced in \ch\ref{se:bg} up to first order in perturbations. 

We assume that any physical quantity $Q(x)$ 
can be decomposed into the sum
\begin{equation}\label{eq_back-pert-dec}
 Q(\tau,\vec x) \; = \; \bar{Q}(\tau) + \delta Q(\tau,\vec x) + \ord(\delta^2)
 \;, 
\end{equation}
of a homogeneous and isotropic background $\bar{Q}(\tau)$, 
and a small (linear) perturbation $\delta Q(\tau,\vec x)$. 
In some cases the background quantity $\bar{Q}(\tau)$ can also vanish.
It is important to realize that 
the validity of the formalism is {\em not} dictated by the zeroth
and first-order terms, which are both accounted for, but rather
by the assumption that the second-order term, which is omitted, 
is small compared with the sum of the zeroth and first-order terms. 
Unfortunately, even if empirically well motivated
(cf.\ \eqs\nr{f_NL_obs} and \nr{f_NL}), 
it is not easy to state the conditions under which this
assumption is mathematically justified; we will return to this issue
in later chapters, notably at the end of \se\ref{ss:soln_dS}, 
and in \se\ref{ss:gw_sigw}.

Within the perturbative method, 
the laws of physics are linearized around the background. 
Subtracting the Einstein equations 
at zeroth order from the ones of the full theory 
we obtain the first-order equations,
\begin{equation}\label{eq_pert-ee}
% \delta {G^\mu}_\nu = 8\pi G\, \delta {T^\mu}_\nu \ ,
 \delta G^{ }_{\mu\nu} \; = \; 8\pi G\, \delta T^{ }_{\mu\nu} \ ,
\end{equation}
where $G$ denotes Newton's gravitational constant. 
This set of equations describes the early evolution 
of primordial perturbations. In a later universe, non-linear
effects become important and are essential for generating
the large-scale structures that we observe today.

The Einstein equations contain different types of fields. We may classify
them as scalars, vectors, and tensors, but attention needs to be paid
to the transformation group that is meant. Roughly speaking,
three different kinds of transformations are relevant: 
3+1-dimensional coordinate transformations; 
3-dimensional spatial rotations; 
and 2-dimensional rotations in a transverse plane. 
In order to make this distinction clear, 
we discuss various types of objects in turn.

We will adopt a very practical 
notation (even if it may require some getting used to), 
namely that when we refer to spatial indices $i,j,k,m$, 
in contrast to space-time indices $\mu,\nu,\rho,\sigma$, 
and they label first-order perturbations 
(e.g.\ $h^{ }_{\mu\nu}$ in \eq\nr{eq_metric-pert}, or 
$\delta T_{\mu\nu}$ in \eq\nr{delta_Tmunu_fluid}), 
or spatial derivatives, 
then they are assumed Euclidean. In other words, 
for a generic tensorial quantity $Q$ we denote
\begin{equation}\label{eq_local-defs}
 \nabla^2_{ } Q    \;\equiv\; \partial^i_{ }\partial_i^{ } Q  
  \,,           \qquad
%%%% 
 Q^i_{ }               \;=\;   \delta^{im}_{ } Q_m^{ } \;=\; Q_i^{ } 
  \,, \qquad
%%%%
 \tr Q_{ij}^{ } \;=\;   \tr Q^{ij}_{ } \;\equiv\; Q^i_i 
 \;.
\end{equation}

%%%%%%%%%%%%%%%%%%%%%%%%%%%%%%%%%%%%%%%%%%%%%%%%%%%%%%%%%%%%%%%%%%%%%%%%%%

\subsubsection*{Scalars}

\index{scalar perturbations}

A quantity is a scalar if it is invariant under all of the 
transformations mentioned above. 
The first-order decomposition of a scalar quantity $A(\tau,\vec x)$ 
according to \eq\eqref{eq_back-pert-dec} gives
\begin{equation}\label{eq_scalar-dec}
 A(\tau,\vec x) \;=\; \bar{A}(\tau) + \delta A(\tau,\vec x) + \ord(\delta^2)
 \ .
\end{equation}

%%%%%%%%%%%%%%%%%%%%%%%%%%%%%%%%%%%%%%%%%%%%%%%%%%%%%%%%%%%%%%%%%%%%%%%%%

\subsubsection*{Vectors}

\index{vector perturbations}

An arbitrary $4$-vector, $B^\mu=(B^0_{ },B^i_{ })$, is made of 
a temporal $3$-scalar component, $B^0_{ }$, 
and a spatial $3$-vector component, $B^i_{ }$. 
Applying the decomposition in \eq\eqref{eq_back-pert-dec}, 
the spatial components of background quantities vanish 
as there is no preferred direction,
\ba
 B^\mu(\tau,\vec x)      
 & =  &
 \bar{B}^\mu(\tau) + \delta B^\mu(\tau,\vec x) + \ord(\delta^2)
 \ , \label{def_vector} \\[2mm]
%%%% 
 \bar{B}^\mu   & = &  (\bar{B}^0,\ {\bm 0}\ ) \ ,\;\ 
 \delta B^\mu  \; = \;  (b^0,\delta B^i) \ ,\;\ 
 b^0       \; \equiv \; \delta B^0
 \ , \label{def_vector_0}
\ea
and $\partial_i\bar{B}^0=0$ for all spatial directions, $i=1,2,3$. 
The spatial perturbation $\delta B^i$ can furthermore be 
decomposed into a divergence-free and a curl-free part,
\begin{equation}\label{eq_vec-1o}
 b^i \;\equiv\; \delta B^i \;=\; b^i_\text{v} + b^i_\text{s} \ , \qquad 
  \partial_i b^i_\text{v} \;=\; 0  \ ,\;\ 
 b^i_\text{s}           
  \;=\; -\partial^i_{ } b  \;=\; -b^{\hspace*{0.3mm},i}_{ } 
 \ .
\end{equation}

The decomposition in \eq\nr{eq_vec-1o} 
can be justified via Helmholtz's theorem, 
but we present 
a constructive proof, this being straightforward. 
Let us introduce a parallel projector,
which in Fourier space acts on a quantity $Q$ as
\be
 \mathbbm{L}^{i}_{m}\, %\phi
 Q(\vec{x})
 \; \overset{\rmii{\nr{fourier_x}}}{\equiv} \; 
 \int_\vec{k} 
 \frac{k^i_{ } k^{ }_m}{\vec{k}^2 } \,
 %\phi
 Q(\vec{k})\,
 e^{i \vec{k}\cdot\vec{x}}_{ }
 \;, 
\ee
provided that $%\phi
Q(\vec{k})$ is not too singular at small $|\vec{k}|$. 
In coordinate space the {\em parallel or longitudinal projector} 
can formally be expressed as 
\be
 \mathbbm{L}^{i}_{m}
 \; = \;
 \frac{\partial^i_{ }\partial^{ }_m}{\nabla^2_{ }}
 \;. \label{def_L}
 \index{projector: parallel to vector} 
 \index{$\mathbbm{L}^{i}_{m}$ (longitudinal projector)}
\ee
A {\em transverse projector}, 
$
 \projP^{i}_{m}
$,  
can then be defined, and has basic properties as 
\be
 \projP^{i}_{m} \; \equiv \; 
 \delta^{\ibit i}_{m} - \mathbbm{L}^{i}_{m}
 \; = \; 
 \delta^{\ibit i}_{m}  -  \frac{\partial^i_{ }\partial^{ }_m}{\nabla^2_{ }}
 \;, \quad
 \partial^{ }_{i} \projP^{i}_{m} \; = \; 0
 \;, \quad
 \projP^{i}_{k}\,
 \projP^{k}_{m}
  \; = \; 
 \projP^{i}_{m}
 \;. \label{def_D}
 \index{projector: transverse vectors} 
 \index{$\projP^{i}_{m}, \projK^{i}_{m}$ (transverse projector)}
\ee
Consider now a general 3-vector, $b^i_{ }$.
It can be written as 
\be
 b^i_{ }
 \; = \; 
 \delta^{\ibit i}_{m} b^m_{ }
 \; \stackrel{\rmii{\nr{def_D}}}{=} \;
 \underbrace{ 
 \projP^{i}_{m} b^m_{ }
 }_{\,\equiv\, b_\rmii{v}^i }
 + 
 \underbrace{ 
 \mathbbm{L}^{i}_{m} b^m_{ }
 }_{\,\equiv\, b_\rmii{s}^i } %- b^{\hspace*{0.3mm},i}_\rmii{ } }
 \;. \label{bi_splitup}
\ee
Here the vector part, $b_\rmi{v}^i$, 
is divergence-free (transverse),  
\be
 \partial^{ }_i b_\rmi{v}^i 
 \;\stackrel{\rmii{\nr{bi_splitup}}}{=}\; 
 \partial^{ }_i 
 \projP^{i}_{m} b^m_{ }
 \;\stackrel{\rmii{\nr{def_D}}}{=}\; 
 0
 \;. \label{vector} 
\ee
The scalar part, $b_\rmi{s}^i$,
can be expressed as a gradient, %longitudinal projection, 
\be
 b_\rmi{s}^i
 \;\equiv\;
 -
 b^{\hspace*{0.3mm},i}_{ }\ , \quad
 b
 \; \underset{\rmii{\nr{def_L}}}
    {\overset{\rmii{\nr{bi_splitup}} \lift }{=}} \; 
 -\frac{\partial^{ }_m}{\nabla^2}\, b^m_{ }
 \;,
\ee
where the overall minus sign
is a convention. 
To summarize, we call $b$ the \textit{scalar} and $b^i_\text{v}$
the \textit{vector} parts of the spatial perturbations 
$\delta B^i$ \cite{Bardeen:1980kt}, whereby these notions now
refer to transformations with respect to the spatial transverse plane.
Altogether $\delta B^\mu_{ }$ splits into 
two scalars, $\lbrace b^0, b\rbrace$, 
and one divergenceless $3$-vector, $b^i_\text{v}$.

%%%%%%%%%%%%%%%%%%%%%%%%%%%%%%%%%%%%%%%%%%%%%%%%%%%%%%%%%%%%%%%%%%%%%%%%%%%%

\subsubsection*{Symmetric tensors}

\index{tensor perturbations}

A central role in Einstein's general relativity is played by
symmetric rank-$2$ tensors, $C^{ }_{\mu\nu}=C^{ }_{\nu\mu}$. 
In four dimensions they have ten independent components. 
We define
\begin{align}
 C^{ }_{\mu\nu} (\tau,\vec x) \;=\; \bar{C}^{ }_{\mu\nu}(\tau)\; + 
 &\ \underbrace{ \delta C^{ }_{\mu\nu}(\tau,\vec x) }\;+\; \ord(\delta^2_{ })
 \label{delta_Cmunu}
 \ ,\\
%%%%%%%%%%
 &\begin{rcases}
 \delta C^{ }_{00}                  \equiv  -2c^{ }_0 \\
 \delta C^{ }_{0i} = \delta C^{ }_{i0}  \equiv  -c^{ }_i  \\
 \delta C^{ }_{ij}                  \equiv 
    -2\delta^{ }_{ij}\bit c^{ }_\rmii{D} + 2\gamma^{ }_{ij} \;\
 \!\! \end{rcases} \, \Rightarrow \,  
 \delta C^{ }_{\mu\nu} = 
 \left(\! \begin{array}{c c}
 -2c^{ }_0 & -c^{ }_j                              
  \\
 %%%%%%%%
 -c^{ }_i  & -2\delta^{ }_{ij}\bit c^{ }_\rmii{D} + 2\gamma^{ }_{ij}
 \\
 \end{array}
 \!\right)
 \ , \label{eq_general-tensor}
\end{align}
where $\gamma^{ }_{ij}=\gamma^{ }_{ji}$ is traceless, $\gamma^i_i=0$. 
The homogeneous and isotropic background satisfies
\begin{equation}
 \bar{C}^{ }_{0i}       \;=\;  \bar{C}^{ }_{i0} \;=\; 0 \ , \qquad 
 \bar{C}^{ }_{ij}       \;=\;  \bar{C}\,\delta^{ }_{ij} \ , \qquad
 \partial_i \bar{C} \;=\; 
   \partial_i \bar{C}^{ }_{00} \;=\; 0 \;\;\  \forall i=1,2,3
 \ . \label{eq_tensor-backgr}
\end{equation}
The $3$-vector $c^{ }_i$ can be decomposed as discussed above,
\begin{equation}\label{eq_tensor-1o-0i}
 c^{ }_i                 
 \; \overset{\rmii{\nr{eq_vec-1o}}}{=} \; 
 c_i^\text{s} + c_i^\text{v} \ , \qquad
 c_i^\text{s}            \;=\;  -\partial_i c \ , \quad 
 \partial^i c_i^\text{v} \;=\;  0 \ ,
\end{equation}
while the traceless $3$-tensor $\gamma^{ }_{ij}$ can be written 
in terms of scalar, transverse vector, and transverse tensor parts
with respect to 2-dimensional transformations,
\begin{align}\label{eq_tensor-ij}
 \gamma^{ }_{ij}                   
 &\;=\; \gamma_{ij}^\text{s} + \gamma_{ij}^\text{v} + \gamma_{ij}^\text{t}
 \ , \\[2mm] 
%%%%
 \gamma^\text{s}_{ij}          
 &\;=\; \left( \partial_i^{ }\partial_j^{ }
 - \delta^{ }_{ij}\nabla^2/3 \right)\gamma 
 \ ,\\[2mm]
%%%%
 \gamma_{ij}^\text{v}          
 &\;=\; -\frac{1}{2}\left( \partial^{ }_i\gamma^\text{v}_j
                     + \partial^{ }_j\gamma^\text{v}_i \right)
 \ , \qquad 
            \partial^i\gamma^\text{v}_i \;=\; 0 \ ,\\[2mm]
 \delta^{ij}_{ }
 \gamma^\text{t}_{ij}       &\;=\; 0 \ ,\;\ \hspace*{3.30cm}
  \partial^i\gamma_{ij}^\text{t} \;=\; 0 
  \ . \label{eq_tensor-ij-x}
\end{align}

Let us verify 
the statements of \eqs\nr{eq_tensor-ij}--\nr{eq_tensor-ij-x}
with the help of the projectors from \eqs\nr{def_L} and \nr{def_D}. 
In general we can write $\delta C^{ }_{ij}$ as 
\ba
 \delta C^{ }_{ij}
 & = &  
 \frac{1}{2}
 \underbrace{
 \bigl(  \delta^{m}_{i}\delta^{n}_{j}
       + \delta^{n}_{i}\delta^{m}_{j}
 \bigr)
 }_{\rm symmetric~in~{\it i\leftrightarrow j}}
  \, \delta C^{ }_{mn}
 + 
 \frac{1}{2}
 \underbrace{
 \bigl(  \delta^{m}_{i}\delta^{n}_{j}
       - \delta^{n}_{i}\delta^{m}_{j}
 \bigr) 
 }_{ \epsilon^{ }_{ijk}\epsilon^{kmn}_{ } }
 \, \delta C^{ }_{mn}
 \;, 
\ea
where $\epsilon^{ }_{ijk}$ is the antisymmetric 
{\em Levi-Civita symbol}.
\index{Levi-Civita symbol: 3-dimensional} 
The contraction 
$
 \epsilon^{kmn}_{ } \delta C^{ }_{mn}
$
vanishes, because we have assumed $\delta C^{ }_{mn}$ symmetric. 
Turning our attention
to the symmetric tensor, we separate its trace part as 
\be
 \delta C^{ }_{ij}
 \; = \; 
 \underbrace{
 \frac{1}{2}
 \biggl(  \delta^{m}_{i}\delta^{n}_{j}
       + \delta^{n}_{i}\delta^{m}_{j}
       - \frac{2}{3}\, \delta^{ }_{ij}\delta^{mn}_{ }
 \biggr)
 }_{ \;\equiv\; \mathbbm{S}^{mn}_{ij} }
  \, \delta C^{ }_{mn}
 + 
  \frac{ \delta^{ }_{ij} }{3}
 \underbrace{
   \delta C^{m}_{m}
 }_{\;\equiv\; -6\, c^{ }_\rmiii{D}}
 \;. \label{def_S}
\ee

In the traceless  symmetric
tensor, $\mathbbm{S}^{mn}_{ij}$, we insert the
projectors from \eq\nr{def_D} in the form 
$
 \delta^{m}_{i} = \projP^{m}_{i} + \mathbbm{L}^{m}_{i}
$.
The quadratic appearances of $\projP$ are pulled apart into
the projector
\be
 \mathbbm{T}^{mn}_{ij}
 \; \equiv \; 
 \frac{1}{2}
 \Bigl(
   \projP^{m}_{i}\projP^{n}_{j}
 + \projP^{n}_{i}\projP^{m}_{j}
 - \projP^{ }_{ij}\projP^{mn}_{ } 
 \Bigr)
 \;, \label{T_ijmn}
 \index{projector: traceless transverse tensor}
 \index{$\mathbbm{T}^{mn}_{ij}$ (traceless transverse projector)}
\ee
which is {\em transverse and traceless}, 
\be
 \partial^{i}_{ } 
 \mathbbm{T}^{mn}_{ij}
 \; = \; 0
 \; = \; 
 \partial^{ }_{m} 
 \mathbbm{T}^{mn}_{ij}
 \;, \quad
 \delta^{ij}_{ } 
 \mathbbm{T}^{mn}_{ij}
 \; = \; 
 0
 \; = \; 
 \delta^{ }_{mn} 
 \mathbbm{T}^{mn}_{ij}
 \;, \quad
 \mathbbm{T}^{mn}_{ij}
 \mathbbm{T}^{kl}_{mn}
 \; = \; 
 \mathbbm{T}^{kl}_{ij}
 \;. \label{props_T}
\ee

As far as quadratic appearances of $\mathbbm{L}$ go, we note 
that 
$
 \mathbbm{L}^{m}_{i}
 \mathbbm{L}^{n}_{j}
 = 
 \mathbbm{L}^{n}_{i}
 \mathbbm{L}^{m}_{j}
 = 
 \mathbbm{L}^{ }_{ij}
 \mathbbm{L}^{mn}_{ }
$.
Making use of this property and the definition in \eq\nr{T_ijmn},
we find
\ba
 \mathbbm{S}^{mn}_{ij}
 &
  \underset{\rmii{\nr{T_ijmn}}}
  {\overset{\rmii{\nr{def_S}} \lift }{=}}
 & 
 \mathbbm{T}^{mn}_{ij}
%%%%%
 \label{orig_t} \\[2mm]
 & + &
 \frac{1}{2}
 \bigl(
   \mathbbm{L}^{m}_{i}\projP^{n}_{j}
 +  
   \mathbbm{L}^{n}_{i}\projP^{m}_{j}
 \bigr)
 + \frac{1}{2}
 \bigl(
   \mathbbm{L}^{m}_{j}\projP^{n}_{i}
 +  
   \mathbbm{L}^{n}_{j}\projP^{m}_{i}
 \bigr)
%%%%
 \label{orig_v} \\[2mm] 
 & + &
  \frac{1}{6}
   \bigl( \projP^{ }_{ij} - 2 \mathbbm{L}^{ }_{ij} \bigr)
   \bigl( \projP^{mn}_{ } - 2 \mathbbm{L}^{mn}_{ } \bigr)
 \;. \label{S_ijmn}
\ea
Once we operate with $\mathbbm{S}^{mn}_{ij}$ on 
$\delta C^{ }_{mn}$, \eq\nr{orig_t}
yields a transverse and traceless tensor, 
\be
 2 \gamma^\rmi{t}_{ij}
 \; = \; 
 \mathbbm{T}^{mn}_{ij} 
 \delta C^{ }_{mn}
 \;. 
\ee
Eq.~\nr{orig_v} yields vectors, 
\ba
 \biggl[ 
 \frac{1}{2}
 \bigl(
   \mathbbm{L}^{m}_{i}\projP^{n}_{j}
 +  
   \mathbbm{L}^{n}_{i}\projP^{m}_{j}
 \bigr)
 + \frac{1}{2}
 \bigl(
   \mathbbm{L}^{m}_{j}\projP^{n}_{i}
 +  
   \mathbbm{L}^{n}_{j}\projP^{m}_{i}
 \bigr)
 \biggr]
 \delta C^{ }_{mn}
 & = & 
 -\bigl( 
  \partial^{ }_i \gamma^\rmi{v}_j
 + 
  \partial^{ }_j \gamma^\rmi{v}_i
  \bigr)
 \;, \hspace*{6mm} \\[2mm]
%%%%%
 \gamma^\rmi{v}_j \; \equiv \;
 -\frac{
  \partial^{m}_{ } \projP^{n}_{j}
 + 
  \partial^{n}_{ } \projP^{m}_{j}
  }{2 \nabla^2_{ }}
 \, 
 \delta C^{ }_{mn}
 \;, \quad
%%%%%
 \partial^{j}_{ } \gamma^\rmi{v}_j 
 & \stackrel{\rmii{\nr{def_D}}}{=} & 0
 \;.
\ea
Eq.~\nr{S_ijmn} yields a scalar, 
\ba
  \frac{1}{6}
   \bigl( \projP^{ }_{ij} - 2 \mathbbm{L}^{ }_{ij} \bigr)
   \bigl( \projP^{mn}_{ } - 2 \mathbbm{L}^{mn}_{ } \bigr)
 \delta C^{ }_{mn}
 \!\! & \overset{\rmii{\nr{def_D}}}{=} & \!\! 
 2 \biggl(
   \partial^{ }_i\partial^{ }_j - \frac{\delta^{ }_{ij}\nabla^2_{ }}{3} 
 \biggr)
 \gamma^\rmi{ }_{ }
 \;, \\[2mm]
%%%%
 \gamma^\rmi{ }_{ }
 \!\!\! & \equiv & \!\!\!
 - \frac{1}{4\nabla^2_{ }} 
 \biggl(
  \delta^{mn}_{ } - \frac{3 \partial^{m}_{ } \partial^{n}_{ }}{\nabla^2_{ }} 
 \biggr)
 \delta C^{ }_{mn}  
 \;.  \hspace*{8mm}
\ea

To summarize, symmetric 4-tensors have four scalar 
degrees of freedom,
$\lbrace c_0^{ }, c_\rmii{D}^{ }, c, \gamma\rbrace$, 
two transverse vectors, 
$\lbrace c_i^\text{v} , \gamma_i^\rmi{v}\rbrace$, 
and one transverse tensor, 
$\gamma_{ij}^\text{t}$, which satisfies four constraints. 
As a crosscheck, this sums up to $4+2\times 2+(6-4)=10$ free variables. 

%%%%%%%%%%%%%%%%%%%%%%%%%%%%%%%%%%%%%%%%%%%%%%%%%%%%%%%%%%%%%%%%%%%%%%%%%
%
\subsection{Einstein tensor} %Perturbed 
\label{ss:delta_Gmunu}

\index{Einstein tensor: general gauge}

Apart from physical quantities, the perturbative expansion 
is also applied to the metric tensor describing 
the geometry of space-time, whose zeroth-order term is given 
by \eq\eqref{t_tau}. 
The indices of proper Lorentz tensors
are raised and lowered using the full, perturbed metric.
However, let us anticipate that % , in perturbation theory,  
we often meet ``non-tensorial'' objects,
for which another convention is chosen 
(cf.\ the discussion around \eq\nr{inv_met}).

The fluctuations of the spatially flat FLRW metric in conformal 
coordinates can be decomposed into their scalar, vector, and tensor
parts with the help of  
\eqs\eqref{delta_Cmunu}--\eqref{eq_tensor-ij-x}. 
We define 
\begin{equation}
 g_{\mu\nu}^{ }
 \; = \;
 \bar{g}_{\mu\nu}^{ } + \delta g_{\mu\nu}^{ } 
 \;
 \overset{\rmii{\nr{t_tau}}}{=}
 \;
 a^2_{ } \left( \eta_{\mu\nu}^{ } + h_{\mu\nu}^{ } \right)
 \ , \label{eq_metric-pert}
 \index{Minkowski metric}
 \index{$\eta^{ }_{\mu\nu}$ (Minkowski metric)}
\end{equation}
where
$\eta^{ }_{\mu\nu} \equiv \mbox{diag}(-$$+$$+$$+)$
is the {\em Minkowski metric}
(we have set $\kappa = 0$ in \eq\nr{t_tau}), 
and $h_{\mu\nu}^{ }$ is its perturbation. 
The scale factor $a$ is normally defined as unperturbed. 

Now, the identification of points in the two different 
space-times, $g_{\mu\nu}^{ }$ and $\bar{g}_{\mu\nu}^{ }$, 
is gauge-dependent, 
as will be discussed in \ch\ref{se:gauges}. 
The quantity 
$
 \delta g_{\mu\nu}^{ } = a^2_{ } h^{ }_{\mu\nu}
$ 
is therefore not a proper tensor. 
When we raise indices in $h^{ }_{\mu\nu}$, this is defined to 
happen with $\eta^{\mu\nu}_{ } \equiv \eta^{ }_{\mu\nu}$.
We note in passing that 
$
 {h^{ }_\alpha}^\beta_{ }
 = 
 h^{ }_{\alpha\mu}\eta^{\mu\beta}_{ }
 = 
 h^{ }_{\mu\alpha}\eta^{\beta\mu}_{ }
 = 
 \eta^{\beta\mu}_{ }h^{ }_{\mu\alpha}
 = 
 {h^\beta_{ }}^{ }_{\alpha}
$, 
so that the mixed index ordering is symmetric. 
We denote it more compactly by $h^\beta_\alpha$.
With this notation, the inverse of the 
metric in \eq\nr{eq_metric-pert} is given by 
\be
 g^{\mu\nu}_{ }
 \;
 =
 \; 
 \bar{g}^{\mu\nu}_{ } + \delta g^{\mu\nu}_{ }
 \;
 = 
 \; 
 a^{-2}_{ }
 \, \bigl(\, 
 \eta^{\mu\nu}_{ } - h^{\mu\nu}_{ } % + h^{\mu\lambda}_{}h^\nu_\lambda 
 \,\bigr)
 +  \ord(\delta^2_{ })
% + \ord(h_{\mu\nu}^2) 
 \;. 
 \label{inv_met}
\ee
This can be verified by an explicit computation,
\ba
 g^{\mu\nu}_{ }g^{ }_{\nu\beta}
 &
 \overset{\rmii{\nr{inv_met}} \lift }{
 \underset{\rmii{\nr{eq_metric-pert}}}{=}}
 & 
 \bigl(\, 
 \eta^{\mu\nu}_{ } - h^{\mu\nu}_{ } % + h^{\mu\lambda}_{}h^\nu_\lambda 
 \,\bigr)
 \bigl(\,
 \eta^{ }_{\nu\beta} + h^{ }_{\nu\beta}
 \,\bigr) 
 +  \ord(\delta^2_{ })
% + \ord(h_{\mu\nu}^2)
 \nn[2mm]
%%%%
 & = & 
 \delta^\mu_\beta
 - \cancel{h^\mu_\beta}
% + \cancel{h^{\mu\lambda}_{ }h^{ }_{\lambda\beta}}
 + \cancel{h^\mu_\beta}
% - \cancel{h^{\mu\nu}_{ }h^{ }_{\nu\beta}}
 +  \ord(\delta^2_{ })
% + \ord(h_{\mu\nu}^2)
 \;. 
\ea

To recover the scalar-vector-tensor decomposition, 
cf.~\eq\eqref{eq_general-tensor},  
we define
\begin{align}
 h^{ }_{00} &\;\equiv\; -2h^{ }_0            
                        \ , \\[2mm]
 h^{ }_{0i} &\;\equiv\; h^{ }_{i0} \;\equiv\; -h^{ }_i    
               \ , \\[2mm]
 h^{ }_{ij} &\;\equiv\; 
   2( -h^{ }_\rmii{D}\,\delta^{ }_{ij} + \vartheta^{ }_{ij} )
 \ , \label{h_notation}
\end{align}
where $\vartheta^{ }_{ij}$ is traceless. 
We note in passing that 
another notation frequently used in the literature 
% see e.g. \cite[p.224]{Weinberg:2008zzc}, 
is $h^{ }_{00}=-2A$, 
$h^{ }_{0i}=h^{ }_{i0}=-B^{ }_i$ and 
$h^{ }_{ij}=2(-D\,\delta^{ }_{ij}+E^{ }_{ij})$, 
or variations thereof, but we reserve capital letters for 
vectors and tensors. 
In matrix form the metric and its inverse are hence given by
\begin{align}
 g^{ }_{\mu\nu} &\;=\; a^2_{ }(\tau) \begin{pmatrix}
 -1-2h^{ }_0 & -h^{ }_j \\[2mm]
 -h^{ }_i    & (1-2h^{ }_\rmii{D})\,\delta^{ }_{ij}+2\vartheta^{ }_{ij}
 \end{pmatrix} 
 \ , \label{g_munu}
 \index{$g^{ }_{\mu\nu}$ (metric)}
 \\[2mm]
%%%%%%%
 g^{\mu\nu}_{ } &\;=\; \frac{1}{a^2_{ }(\tau)} \begin{pmatrix}
 -1+2h^{ }_0 & -h^{ }_j \\[2mm]
 -h^{ }_i    & (1+2h^{ }_\rmii{D})\,\delta^{ }_{ij}-2\vartheta^{ }_{ij}
 \end{pmatrix} 
 + \ord(\delta^2_{ })
 \ . \label{g^munu} 
 \index{metric: perturbed}
\end{align}
The $0i$ and $ij$-components of the metric can be further decomposed, 
cf.~\eqs\eqref{eq_tensor-1o-0i}--\eqref{eq_tensor-ij-x},
\begin{align}
 h^{ }_i                            
 &= -\partial_i^{ }\hspace*{0.3mm} h + h_i^\text{v}\ , & & 
 &\partial^i h_i^\text{v}           &= 0
 \ , \label{eq_metric-pert2}
 \\[2mm]
%%%%%%
 \vartheta^{ }_{ij}                 &= \vartheta_{ij}^\text{s}
 + \vartheta_{ij}^\text{v} + \vartheta_{ij}^\text{t}                 \ ,   
 &\vartheta^\text{s}_{ij}           
 &= \left( \partial^{ }_i\partial^{ }_j-\delta^{ }_{ij}\nabla^2/3 \right)
 \vartheta                               \ , & 
 \label{vartheta_ij_s} \\[2mm]
%%%%%%
 & 
 &\vartheta_{ij}^\text{v}           &=
 -\frac{1}{2}
 \left( \partial^{ }_i\vartheta_j^\rmi{v}
       +\partial^{ }_j\vartheta_i^\rmi{v} \right)           \ , 
 &\partial^i_{ } \vartheta_i^\rmi{v}     &= 0 \ ,
 \label{theta_ij_v} \\[2mm]
%%%%%%
 & 
 &\delta^{ij}_{ } \vartheta^\text{t}_{ij}          &= 0 \ , 
 &\partial^i_{ }\vartheta_{ij}^\text{t} &= 0
 \ . \label{eq_metric-pert3}
\end{align}

\index{$\M_\tau^{ }$ (time slice)}
\index{space-time threads}

In order to give a geometric interpretation 
to the metric perturbations~\cite{Arnowitt:1962hi}, 
let us denote a submanifold of constant $\tau$, 
called a {\em time slice}, by $\M_\tau^{ }$. 
In turn, spatial coordinates separate 
the space-time into $x^i_{ }\,{=}\,$constant {\em threads}. 
At each space-time point $p$, we so find a covector 
$n_\mu^{ }$ normal to the corresponding time slice 
$\M_\tau^{ }$, 
$
 n^{ }_\mu = (n^{ }_0,\vec{0})
$,
and a vector $t^\mu_{ }$ tangent 
to the thread passing through $p$,
$
 t^\mu_{ } = (t^0_{ },\vec{0})
$.
They can be normalized to unit length 
with 
$
 \,g^{\mu\nu}_{ }n_\mu^{ } n_\nu^{ }
 =g_{\mu\nu}^{ }t^\mu_{ } t^\nu_{ }=-1$. 
At first order in perturbations, we find
\begin{align}
 n^\mu_{ } &\;=\; \frac{1}{a}(-1+h^{ }_0,-h^{ }_i) + \ord(\delta^2_{ })
 \ , &n_\mu^{ } &\;=\; a\,(1+h^{ }_0,\bm{0})  + \ord(\delta^2_{ })   
  \ ,\label{eq_n}\\[2mm]
 t^\mu_{ } &\;=\; \frac{1}{a}(1-h^{ }_0,\bm{0})  + \ord(\delta^2_{ })
   \ , &t_\mu^{ } &\;=\; a\,(-1-h^{ }_0,-h^{ }_i)  + \ord(\delta^2_{ })
 \ .\label{eq_t}
\end{align}
We see that because of the non-diagonal terms in the metric, 
$n^\mu_{ }$ and $t^\mu_{ }$ are not parallel, but 
differ by the {\em shift vector}, $h^{ }_i\,$. At the same time, 
$h^{ }_0$ determines the proper time separation 
between time slices, taken along the thread, and
is called a {\em lapse function}.

\index{shift vector}
\index{lapse function}

\vspace*{3mm}

With the metric tensor at hand, the next task is to construct the Einstein
tensor. While in principle straightforward, 
this task is tedious in practice, 
given the many indices and lengthy expressions appearing. 
As a simplification,
one of two strategies is often followed: either we make use of a symbolic 
manipulation package, or we ``choose a gauge'', whereby some of the metric
perturbations are put to zero. The use of a symbolic manipulation 
package is illustrated 
in \app\ref{app:symbolic}, 
whereas a gauge-fixed computation
is shown in \app\ref{app:newton}. In both cases we have
furthermore set $\kappa = 0$, so that Euclidean coordinates can be 
employed in spatial directions. In the remainder of this section, 
we outline the steps and give the results 
of the computation in a general gauge.  

\vspace*{3mm}

As was discussed around \eq\nr{eq_Einstein_symmetric}, 
it has been conventional in the literature 
to determine the Einstein tensor
${G^\mu}_{\nu}$. However, 
it suffers from the problem of not being symmetric
in general. 
We therefore work out $G^{ }_{\mu\nu}$ instead. 

The first step is to determine the Christoffel symbols, 
cf.\ \eq\eqref{eq_ChSy-a}.
%% Christoffel symbols %%%%%%%%%%%%%%%%%%%%%%%%%%%%
%
At first order in perturbations, 
\begin{equation}
 \delta\Gamma^\rho_{\mu\nu} 
 \;
 \overset{\rmii{\nr{eq_ChSy-a}}}{=}
 \; 
 \frac{1}{2} \bar{g}^{\rho\sigma}_{ }( \delta g^{ }_{\sigma\mu,\nu}
 + \delta g^{ }_{\sigma\nu,\mu} - \delta g^{ }_{\mu\nu,\sigma} )
 \;+\;
 \frac{1}{2} \delta g^{\rho\sigma}_{ }( \bar{g}^{ }_{\sigma\mu,\nu}
 + \bar{g}^{ }_{\sigma\nu,\mu} - \bar{g}^{ }_{\mu\nu,\sigma} ) 
 \ . \label{eq_ChSy-p}
\end{equation}
Here the unperturbed Christoffel symbols \index{Christoffel symbols}
with vanishing curvature, 
$\kappa=0$, are 
\begin{align}\label{eq_ChSyu}
 \bar{\Gamma}^0_{0i} \;=\; \bar{\Gamma}^i_{00}
                     \;=\; \bar{\Gamma}^i_{jk} \;=\; 0
 \ , \qquad
 \bar{\Gamma}^0_{00} \;=\; \Hc \ , \qquad
 \bar{\Gamma}^0_{ij} \;=\; \Hc\delta^{ }_{ij} \ , \qquad   
 \bar{\Gamma}^i_{0j} \;=\; \Hc\delta^i_j \ .
\end{align}
%
%% Ricci tensor %%%%%%%%%%%%%%%%%%%%%%%%%%%%%%%%%%%
From the Christoffel symbols we calculate 
the components of the perturbed Ricci tensor, \index{Ricci tensor}
\begin{equation}\label{eq_Ricci-p}
 \delta R_{\mu\nu}^{ } 
% \;=\;
% \delta R^\alpha_{\mu\alpha\nu}
 \;\overset{\rmii{\nr{eq_rt-a}}}{=}\;
 \delta \Gamma^\alpha_{\mu\nu,\alpha}
 - \delta\Gamma^\alpha_{\mu\alpha,\nu}
 + \bar{\Gamma}^\beta_{\mu\nu} \delta\Gamma^\alpha_{\beta\alpha}
 + \bar{\Gamma}^\alpha_{\beta\alpha} \delta\Gamma^\beta_{\mu\nu}
 - \bar{\Gamma}^\beta_{\mu\alpha} \delta\Gamma^\alpha_{\nu\beta}
 - \bar{\Gamma}^\alpha_{\nu\beta} \delta\Gamma^\beta_{\mu\alpha} \ .
\end{equation}
The unperturbed Ricci tensor follows directly from 
\eqs\eqref{eq_rt-a} and \eqref{eq_ChSyu},
\begin{align}\label{eq_Runpert}
 \bar{R}_{00}^{ } \;=\; -3\Hc'\ ,\qquad 
 \bar{R}_{0i}^{ } \;=\; 0\ ,\qquad 
 \bar{R}_{ij}^{ } \;=\; (\Hc'+2\Hc^2)\,\delta^{ }_{ij}\ .
\end{align}
Raising one index of the perturbed Ricci tensor we obtain
\begin{equation}
 \delta {R^\mu}_\nu
 \; 
 \overset{\rmii{\nr{bg_ricciupdown}}}{=} 
 \;
 \bar{g}^{\mu\rho}_{ } \delta R^{ }_{\rho\nu}
 + \delta g^{\mu\rho}_{ } \bar{R}^{ }_{\rho\nu}
 \,. \label{eq_Ricci-p-up-down}
\end{equation}
The perturbed Ricci scalar \index{Ricci scalar}
is obtained by contracting indices,
\begin{align}\label{eq_Ricci-scalar-p}
 \delta R & 
  \;\overset{\rmii{\nr{bg_ricci_scalar}}}{=}\;
 \delta {R^\mu}_\mu
 \;. 
% = \frac{1}{2}(\delta R^0_0+\delta R^i_i)
\end{align} 
The unperturbed trace $\bar{R} = \bar{R}^\mu_{ }{}_\mu$, 
originating from  \eq\eqref{eq_Runpert}, reads 
\begin{equation}
 \bar{R} \;=\; 6a^{-2}(\Hc' + \Hc^2) \ .
\end{equation}
%% Einstein tensor %%%%%%%%%%%%%%%%%%%%%%%%%%%%%%%%
We have now all ingredients to compute 
the perturbed Einstein tensor, 
\begin{equation}
 \delta G^{ }_{\mu\nu} 
 \; \overset{\rmii{\nr{bg_einsteindowndown}}}{=} \; 
 \delta R^{ }_{\mu\nu} - \frac{1}{2} 
 \bigl( \bar{g}^{ }_{\mu\nu} \delta R  + 
       \delta g^{ }_{\mu\nu} \bar{R} \bigr)
 \,. \label{eq_Einstein-pert}
\end{equation}
Proceeding as outlined above and recalling that $\vartheta^{ }_{kk}=0$, 
we find
\ba
 G^{ }_{00}
 & = & 
  3\H^2_{ } 
 + {2}
 \,\bigl[\,
   (\nabla^2_{ } - 3 \H \partial^{ }_\tau ) h^{ }_\rmii{D}
    + \H  h^{ }_{k,k} %% + \H \vartheta'_{kk} 
   \,\bigr]
 + \vartheta^{ }_{kl,kl}
 %% - \vartheta^{ }_{kk,ll}     
 + \ord\bigl(\delta^2_{ }\bigr)
 \;, \label{G_00} 
 \\[3mm]
%%%%%%%%%%%
 G^{ }_{0i}
 & = & 
 2 
 \,\big(\,
 \H h^{ }_{0,i}
 + h'_{\rmii{D},i} 
 \,\bigr)
 + (2 \H'_{ } + \H^2_{ }) h^{ }_i
 + \frac{1}{2}
   \bigl( h^{ }_{i,kk} - h^{ }_{k,ik} \bigr) 
 + \vartheta'_{ki,k} %% - \vartheta'_{kk,i}
 + \ord\bigl(\delta^2_{ }\bigr)
 \;, \label{G_0i} \hspace*{6mm}
 \\[3mm]
%%%%%%%%%%%
 G^{ }_{ij}
 & = & 
 - (2 \H' + \H^2_{ } ) \delta^{ }_{ij} 
 + 2 (2 \H' + \H^2_{ } ) 
 \bigl[\,
  (h^{ }_0 + h^{ }_\rmii{D})\,\delta^{ }_{ij} - \vartheta^{ }_{ij}
 \,\bigr]
 + 2 \H \,h'_{0}\, \delta^{ }_{ij} 
 \nn[2mm]
%%%%%%%%%
 &  & \;+\,
  (\nabla^2_{ }\delta^{ }_{ij} - \partial^{ }_i\partial^{ }_j )
   (h^{ }_0 - h^{ }_\rmii{D} %% + \vartheta^{ }_{kk}
   )
 + 
  (\partial_\tau^2 + 2 \H \partial^{ }_\tau )
  \bigl(\, 
   2 h^{ }_\rmii{D} \delta^{ }_{ij} %%  - \vartheta^{ }_{kk}\delta^{ }_{ij}
   + \vartheta^{ }_{ij} 
  \,\bigr)
  \nn[2mm]
%%%%%%%%%
 &  & \;+\, 
 \fr12 
 (\partial^{ }_\tau + 2 \H )
 \bigl( h^{ }_{i,j} + h^{ }_{j,i} - 2 \delta^{ }_{ij} h^{ }_{k,k} \bigr)
  \nn[2mm]
%%%%%%
 &  & \;+\, \vartheta^{ }_{ik,jk}
 + \vartheta^{ }_{jk,ik}
 - \vartheta^{ }_{ij,kk}
 - \delta^{ }_{ij} \vartheta^{ }_{kl,kl}
 + 
 \ord\bigl(\delta^2_{ }\bigr)
 \;. \label{G_ij} \hspace*{7mm}
\ea
For future reference, we also record the perturbed
Ricci scalar to the same order, which finds use for instance
in the case of a ``non-minimally coupled'' scalar field,
\ba
 \delta R 
 \hspace*{-0.5cm}
 & = & 
 \hspace*{-0.5cm}
% 6\,\frac{a''}{a^3_{ }}
 - \frac{2}{a^2_{ }}
 \biggl[\, 
    3 \H h_0'
  + 6 \bigl( \H' + \H^2_{ } \bigr) h^{ }_0 
  + \nabla^2 ( h^{ }_0 
  - 2 h^{ }_\rmii{D}
  + \vartheta^{ }_{kk}
    )
 \nn[2mm]
%%%%
 &  & \hspace*{1.2cm} +\,
 (\partial^{ }_\tau + 3 \H)
    (
     3 h_\rmii{D}' - h_{k,k}^{ }
  - \vartheta_{kk}'
     )
  - \vartheta^{ }_{kl,kl} 
 \,\biggr]
 \label{ricci_scalar_full} \\[2mm]
%%%%%
 &
   \overset{\rmii{\nr{eq_metric-pert2}--\nr{eq_metric-pert3}}
            \lift }{=}
 & 
 \hspace*{-0.45cm}
 - \frac{2}{a^2_{ }}
 \biggl\{\, 
    3 \H h_0'
  + 6 \bigl( \H' + \H^2_{ } \bigr) h^{ }_0
  + \nabla^2 \biggl[ h^{ }_0 
  - 2 \biggl( h^{ }_\rmii{D} + \frac{\nabla^2\vartheta}{3} \biggr)
%   + \vartheta^{ }_{kk}
    \biggr]
 \nn[2mm]
%%%%
 &  & \hspace*{1.2cm} +\,
    (\partial^{ }_\tau + 3 \H)
    (
     3 h_\rmii{D}' + \nabla^2_{ } h
% - \vartheta_{kk}'
     )
%  - \vartheta^{ }_{kl,kl} 
 \,\biggr\}
 \;. \label{ricci_scalar}
\ea
We see that only scalar perturbations contribute to the 
Ricci scalar at first order. 

Finally, we note that 
if we only include spatial indices in the determination of the Ricci
tensor, in raising and lowering indices, and in taking the trace, 
we get the Ricci scalar
associated with a time slice $\M^{ }_\tau$ 
(cf.\ \eq\nr{ricci_scalar_spatial}). 
The result can be extracted from \eq\nr{ricci_scalar}, 
by eliminating time derivatives ($\H$, $(...)'$) 
and the metric perturbations associated
with $\delta g^{ }_{00}$ ($h^{ }_0$) and $\delta g^{ }_{0i}$ ($h$). 
This yields 
\be
 \delta R^{ }_\tau
 \;
 \underset{\rmii{see~text}}{
 \overset{\rmii{\nr{ricci_scalar}} \lift }{=}}
 \;
 \frac{4}{a^2_{ }} \nabla^2_{ }
 \biggl( h^{ }_\rmii{D} + \frac{\nabla^2\vartheta}{3} \biggr)
 \;. \label{delta_R_tau}
\ee

%%%%%%%%%%%%%%%%%%%%%%%%%%%%%%%%%%%%%%%%%%%%%%%%%%%%%%%%%%%%%%%%%%%%%%%%%
%
\subsection{Energy-momentum tensor} %Perturbed
\label{ss:delta_Tmunu}

For the right-hand side of the Einstein equations, we need to 
specify the energy-momentum tensor. We treat the 
overall energy-momentum tensor as a sum over two different components, 
cf.\ \eq\nr{Tmunu_mixed}. We recall that this does not exclude 
the fluid and the scalar field from 
interacting, through a friction term 
and thermal fluctuations, as dictated
by \eq\nr{varphi_eq}. 

%%%%%%%%%%%%%%%%%%%%%%%%%%%%%%%%%%%%%%%%%%%%%%%%%%%%%%%%%%%%%%%%%%%
%
\subsubsection*{Ideal and non-ideal thermal plasmas}

\index{non-ideal fluid}

Let us start by considering 
an {\em ideal-fluid contribution} to 
the energy-momentum tensor, cf.\ \eq\nr{eq_Tmunu_pf}.  
This means that we stay at leading order 
in a gradient expansion. We write
\begin{equation}
 T^{ }_{\mu\nu} \;=\; \mathbin{\bar T}^{ }_{\mu\nu} 
 + \delta {T}^{ }_{\mu\nu} + \ord(\delta^2_{ }) \ .
\end{equation}
As a symmetric tensor, $T^{ }_{\mu\nu}$ may have 
up to ten degrees of freedom 
in the perturbations, six of which are physical, 
while four are gauge degrees of freedom (cf.\ \ch\ref{se:gauges}). 

For the background tensor,  
we repeat here some results
from \se\ref{ss:einstein_background}, starting with  
the familiar form
\begin{equation}
 \bar{T}^{ }_{\mu\nu}         
 \;
 \overset{\rmii{\nr{eq_Tmunu_pf}}}{=} 
 \;
 (\bar{e}+\bar{p})\,\bar{u}^{ }_\mu\bar{u}^{ }_\nu
 + \bar{p}\,\bar{g}^{ }_{\mu\nu}
 \ , 
 \label{eq_Tmunu_pf_again}
\end{equation}
where $e$ is the energy density, $p$ the pressure, 
and $u^\mu_{ }$ the four-velocity of the fluid. 
If $\bar e$ and $\bar p$
are parametrized by a single quantity, for instance temperature,
the background values of $e$ and $p$ 
are related by the equation of state, 
\begin{equation}\label{eq_eos}
 \bar p 
 \; 
 \overset{\rmii{\nr{eq_eos_v1}}}{=}  
 \; 
  \bar p( \bar e )                        \ , \qquad
  c_s^2 
 \;
 \overset{\rmii{\nr{eq_eos_v1}}}{\equiv}
 \; 
 \frac{\partial \bar p}{\partial \bar e}
 \ ,
\end{equation}
where $c^{ }_s$ is the speed of sound. Perturbations of the 
energy density and pressure are defined as
\be
 p \;=\; \bar{p} + \delta p
 \;, \qquad
 e \;=\; \bar{e} + \delta e
 \;. 
 \label{delta_e_p}
\ee

Because of isotropy, 
we assume the fluid to be at rest at leading order, $\bar{u}^i=0$.
Then the velocity is defined by
\begin{equation}\label{eq_fluid-velocity_again}
 u^\mu       \;=\; \bar{u}^\mu + \delta u^\mu             
     \ , \qquad 
 \bar{u}^\mu \;=\; \left( \frac{d\tau}{dt},\bm{0} \right)            
 \;
 \overset{\rmii{\nr{t_tau}}}{=}
 \; 
 a^{-1}_{ }(1,{\bm 0})                       
  \ , \qquad 
 \bar{u}_\mu 
 \;
 \overset{\rmii{\nr{eq_fluid-velocity}}}{=} 
 \;
 a(-1,\bm{0})                              \ .
\end{equation}
The linear {\em velocity perturbation} is conveniently expressed as  
\be
  v_i^{ } \;\equiv\; v^i_{ } \;\equiv\; % au^i \;=\; 
                                a\,\delta u^i_{ }
  \qquad \Rightarrow \qquad
  \delta u^i_{ } \;=\; a^{-1}_{ } v^i_{ } 
  \;. \label{def_v_i}
  \index{$v^i_{ }$ (velocity perturbation)}
\ee 
For the components of the covector $\delta u_\nu$,  
we find %as functions of $\delta u^\mu$, obtaining
\begin{align}
 \delta u_0^{ } &\;=\; \delta g_{0\mu}^{ }\bar{u}^\mu_{ }
           +\bar{g}^{ }_{0\mu}\delta u^\mu_{ } 
            \;=\; -2ah_0^{ }-a^2_{ }\delta u^0_{ } 
 \ , \label{delta_u_0} \\[2mm]
%%%%%
 \delta u_i^{ } &\;=\; \delta g_{i\mu}^{ }\bar{u}^\mu_{ }
            +\bar{g}^{ }_{i\mu}\delta u^\mu_{ } 
            \;=\; -a\hspace*{0.3mm} h_i^{ }
               +a^2_{ }\delta_{ij}^{ }\delta u^i_{ } 
            \;=\; a\,(-h_i^{ }+v_i^{ })       
   \ . \label{delta_u_i}
\end{align}
The value of $\delta u^0_{ }$ 
is inferred by imposing $u_\mu^{ } u^\mu_{ }=-1$,
\ba
 u_\mu^{ } u^\mu_{ } 
 & = & 
 \overbrace{
 -1
 }^{ 
  \bar u_\mu^{ } \bar u^\mu_{ }
 }
 +
 \overbrace{ 
 a^{-1}_{ }(-2ah_0^{ }-a^2_{ }\delta u^0_{ })
 }^{
  \delta u_\mu^{ } \bar u^\mu_{ }
 }
 -
 \overbrace{ 
 a\,\delta u^0_{ } 
 }^{ 
 \bar u_\mu^{ } \delta u^\mu_{ }
 }
 \hspace*{5mm} \;\Rightarrow\;  
 \delta u^0_{ } \;=\; -a^{-1}h_0^{ }
 \label{u_constraint} \\[2mm]
%%%%%%
 & \Rightarrow & 
 u^\mu_{ }   
 \;
 \underset{\rmii{\nr{u_constraint}}}
 {\overset{\rmii{\nr{def_v_i}} \lift }{=}}
 \; 
 a^{-1}_{ }(1-h_0^{ },v^i_{ }) + \ord(\delta^2_{ }) 
 \ , \label{eq_u^up} \\[2mm]
%%%%%%%
 & \Rightarrow & 
 u_\mu^{ }   
 \;
 \underset{\rmii{}}
 {\overset{\rmii{\nr{delta_u_0}--\nr{u_constraint}} \lift }{=}}
 \; 
 a(-1-h_0^{ },v_i^{ }-h_i^{ }) + \ord(\delta^2_{ }) 
 \ . \label{eq_u_down}
\ea
Recalling the metric from \eq\nr{g_munu}, 
at linear order the perfect fluid (``pf'') part 
of the energy-momentum tensor is thus
\ba
 T^{ }_{\mu\nu}|_\rmi{pf}^{ }
 \hspace*{-4mm}
 & = & 
 \hspace*{-4mm}
 \mathbin{\bar{T}}^{ }_{\mu\nu} + 
 (\delta e + \delta p)
 (a^2_{ }\delta^{ }_{\mu 0} \, \delta^{ }_{\nu 0} ) + 
 (\bar{e} + \bar{p})
 \left( \bar{u}_{\mu} \, \delta u_\nu 
    + \bar{u}^{ }_{\nu} \, \delta u_\mu \right)  
  + \delta p \, \bar{g}_{\mu\nu} 
  + \bar{p}\, \delta g^{ }_{\mu\nu}
 \nn
 \label{eq_emt-perfect-fluid} \\[2mm] 
%%%%%%%
  &
  \underset{\rmii{\nr{eq_u_down}}}
  {\overset{\rmii{\nr{g_munu}},\rmii{\nr{eq_Tmunu_pf_again}}
    \lift }{=}}
  &  
                              a^2_{ }\begin{pmatrix}
                              \bar{e} & 0 \\
                              0        & \bar{p}\, \delta^{ }_{ij}
                              \end{pmatrix} 
  \nn[2mm]
%%%%%%%
  & & \;+\,  
             a^2_{ }
             \begin{pmatrix}
               \delta e  +  2 \bar{e} h_0^{ }    
           & - \bar{e}\, ( v_j^{ } - h_j^{ } ) - \bar{p}\, v_j^{ }
    \\[2mm]
%%%%%%%%%%
             - \bar{e}\, ( v_i^{ } - h_i^{ } ) - \bar{p}\, v_i^{ } 
                &
        ( \delta p -
                  2 \hspace*{0.3mm}\bar{p}
                     \hspace*{0.3mm} h^{ }_\rmii{D} )
                    \, \delta^{ }_{ij}
                + 2 \bar{p}\,\vartheta^{ }_{ij}
             \end{pmatrix} 
 \; + \; \ord(\delta^2_{ })
 \ . \label{delta_Tmunu_fluid}
 \index{energy-momentum tensor: perfect fluid}
\ea

Going beyond ideal fluids, 
i.e.\ including the next order in gradients, further 
terms are introduced in the hydrodynamic description, 
involving the shear and bulk viscosities. 
As dictated by the {\em fluctuation-dissipation theorem},
\index{fluctuation-dissipation relation} 
and exemplified by the scalar field equation in \eq\nr{varphi_eq}, 
the introduction of dissipative coefficients necessitates the 
introduction of the corresponding thermal noise terms. 
We discuss this in more detail 
in \app\ref{app:viscous}. For now we just 
denote an additional contribution to energy-momentum
perturbations by an {\em anisotropic stress tensor}, 
\index{$\Pi^{ }_{ij}$ (anisotropic stress tensor)}
\index{anisotropic stress}
$\Pi^{ }_{ij}$,
\begin{equation}\label{delta_Tmunu_aniso}
 \delta {T}^{ }_{ij}|^{ }_\rmi{tot} 
 \; = \;
 \delta T^{ }_{ij}|_\rmi{pf}^{ } 
 + 
 a^2_{ } \barpPi^{ }_{ij}  \ ,
\end{equation}
noting that, strictly speaking, we have omitted the bulk viscous
part, i.e.\ the term $Z^{ }_{\mu\nu}$ from \eq\nr{Pi_splitup}.
If bulk viscous corrections were to be included, 
the background equations should be modified as well (cf.\ \eq\nr{delta_Z}),
and a full tensor $a^2_{ }\Pi^{ }_{\mu\nu}$ would be needed. However, 
such corrections are expected to be negligible in typical 
cosmological scenarios (cf.\ \eq\nr{bulk_estimates}).
Therefore, we restrict ourselves to the shear viscous part, 
whereby $a^2_{ } \Pi^{ }_{\mu\nu} $ 
can be assumed purely spatial and traceless 
(cf.\ \eqs\nr{delta_T} and \nr{delta_S}).

The vectors and tensors introduced can be decomposed into
their transverse scalar, 
vector, and tensor parts as 
in \eqs\eqref{eq_tensor-1o-0i}--\eqref{eq_tensor-ij-x},
\begin{align}
 v^{ }_i                           &= v_i^\text{s}+v_i^\text{v} \ , 
 &v_i^\text{s}                 &= -\partial^{ }_i v              \ , 
 &\partial^i_{ } v_i^\text{v}       &= 0                         \ , 
 \label{eq_v-v-s}\\[2mm]
%%%%
 \Pi_{ij}^{ }                      &= 
% \frac{\delta^{ }_{ij} \tr\Pi}{3} + 
 \Pi_{ij}^\text{s}+\Pi_{ij}^\text{v}+\Pi_{ij}^\text{t} \ , 
 &\Pi_{ij}^\text{s}            &= 
 (\partial_i^{ }\partial_j^{ }-\delta_{ij}^{ }\nabla^2/3)\Pi       \ , &
 \label{Pi_ij_s} \\[2mm]
%%%%
 &  &\Pi_{ij}^\text{v}         &=
  -\frac{1}{2}(\partial_i^{ }\Pi_j^\rmi{v}+\partial_j^{ }\Pi_i^\rmi{v})  \ , 
 &\partial^i_{ }\Pi_i^\rmi{v}      &= 0                         \ ,
 \label{Pi_ij_v} \\[2mm]
%%%%
 & 
 &\delta^{ij}_{ }\Pi_{ij}^\text{t} &= 0                         \ , 
 &\partial^i_{ }\Pi_{ij}^\text{t}  &= 0                         \ .
 \label{Pi_ij_t}
 \index{$v = \delta v$ (scalar part of velocity perturbation)}
\end{align}

%%%%%%%%%%%%%%%%%%%%%%%%%%%%%%%%%%%%%%%%%%%%%%%%%%%%%%%%%%%%%%%%%%%
%
\subsubsection*{An elementary scalar field (or inflaton)}

Turning to an elementary scalar field, the energy-momentum tensor 
is given in \eq\nr{Tmunu_varphi}. We now write
\be
 \varphi \; = \; \bar\varphi + \delta\varphi
 \;. \label{delta_varphi}
 \index{$\bar\varphi$ (inflaton background)}
 \index{$\delta\varphi$ (inflaton perturbation)}
\ee
At the background level this leads to 
\ba
 \bar{T}^{ }_{\mu\nu}
 & 
 \overset{\rmii{\nr{Tmunu_varphi}}}{=}
 & 
 \bar\varphi_{,\hspace*{0.3mm}\mu}\bar\varphi_{,\hspace*{0.3mm}\nu}
 -\bar{g}^{ }_{\mu\nu} 
 \left( 
 \frac{\bar{g}^{\alpha\beta}_{ }}{2}
 \,\bar\varphi_{,\alpha}\,\bar\varphi_{,\beta} + \bar V 
 \right)  
 \nn[2mm]
%%%%
 &
 \underset{\rmii{\nr{inv_met}}}{
 \overset{\rmii{\nr{eq_metric-pert}} \lift }{=}}
 &  
 \begin{pmatrix}
                              \frac{(\bar{\varphi}')^2_{ } 
                                      \vphantom{ |_q^b} }{2} 
                               + a^2_{ }\bar{V}
                                       & 0 \\
                              0        & \delta^{ }_{ij}\;
                           \Bigl[\,
                              \frac{(\bar{\varphi}')^2_{ } 
                                      \vphantom{ |_q^b} }{2} 
                               - a^2_{ } \bar{V}
                           \,\Bigr]
 \end{pmatrix}
 \;. \label{bg_Tmunu_varphi}
\ea
Linear perturbations originate from 
\ba
 \delta{T}^{ }_{\mu\nu}
 & 
 \overset{\rmii{\nr{Tmunu_varphi}}}{=}
 & 
 \bar\varphi_{,\hspace*{0.3mm}\mu}
 \delta\varphi_{,\hspace*{0.3mm}\nu}
 +\delta\varphi_{,\hspace*{0.3mm}\mu}
 \bar\varphi_{,\hspace*{0.3mm}\nu}
 -\delta{g}^{ }_{\mu\nu} 
 \biggl(  
 - \frac{ (\bar{\varphi}')^2 }{2 a^2}
  + \bar V 
 \biggr) 
 \nn[2mm]
%%%%%%
 & & \;-\, \bar{g}^{ }_{\mu\nu} 
 \left( 
 \frac{  (\bar{\varphi}')^2 }{2}
 \,\delta{g}^{00}_{ } 
 +\bar{g}^{00}_{ }
 \bar\varphi\hspace*{0.3mm}{}'\delta\varphi' 
 + \bar V^{ }_{,\varphi}\delta\varphi 
 \right) 
 \;. \label{delta_T_varphi} 
 \index{energy-momentum tensor: scalar field}
\ea
Inserting the metric perturbations from 
\eqs\nr{g_munu} and \nr{g^munu}, 
we find
\ba
 \delta T^{ }_{00} 
 & \underset{\rmii{\nr{g_munu},\nr{g^munu}}}
    {\overset{\rmii{\nr{delta_T_varphi}} \lift }{=}} &
 \bcancel{2} {\bar\varphi}'\delta\varphi'
 + 2 h^{ }_0  \biggl(  
 - \cancel{ \frac{ (\bar{\varphi}')^2 }{2} }
  + a^2_{ } \bar V 
 \biggr) 
 + 
  2 h^{ }_0 \cancel{ \frac{ (\bar{\varphi}')^2 }{2} }
 -  \bcancel{ {\bar\varphi}'\delta\varphi' } 
 + a^2_{ } \bar{V}^{ }_{,\varphi}\delta\varphi
 \nn[2mm]
%%%%%
 & = & {\bar\varphi}'\delta\varphi'
 + 2 h^{ }_0\,  a^2_{ } \bar V 
 + a^2_{ } \bar{V}^{ }_{,\varphi}\delta\varphi
 \ ,\label{delta_T00_varphi}
 \\[3mm]
%%%%%
  \delta T^{ }_{0i} 
 & \underset{\rmii{\nr{g_munu},\nr{g^munu}}}
    {\overset{\rmii{\nr{delta_T_varphi}} \lift }{=}} &
   {\bar\varphi}' \delta\varphi^{ }_{,i}
 +  h^{ }_i \biggl(  
 - \frac{ (\bar{\varphi}')^2 }{2}
  + a^2_{ } \bar V 
 \biggr) 
  \ , \label{delta_T0i_varphi}
 \\[3mm]
%%%%%
 \delta T^{ }_{ij} 
  & \underset{\rmii{\nr{g_munu},\nr{g^munu}}}
    {\overset{\rmii{\nr{delta_T_varphi}} \lift }{=}} &
  2 ( h^{ }_\rmii{D} \delta^{ }_{ij} - \vartheta^{ }_{ij}) \biggl(  
 - \frac{ (\bar{\varphi}')^2 }{2}
  +  a^2_{ } \bar V 
 \biggr) 
 \nn[2mm]
%%%%
 & & 
 \;-\, \delta^{ }_{ij}
 \biggl(
  2 h^{ }_0 \frac{ (\bar{\varphi}')^2 }{2}
 -  {\bar\varphi}'\delta\varphi' 
 + a^2_{ }  \bar{V}^{ }_{,\varphi}\delta\varphi
 \biggr)
  \ . \label{delta_Tij_varphi}
\ea
The appearances of $\bar{V}$ can be crosschecked against
\eq\nr{delta_Tmunu_fluid}, after recalling 
from \eq\nr{general_e_p} that in the ideal-fluid notation, 
\be
 \bar{e} \;\supset\; \bar V
 \;, \qquad
 \delta e \;\supset\; \bar V^{ }_{,\varphi}\delta\varphi
 \;, \qquad
 \bar{p} \;\supset\; -\bar V
 \;, \qquad 
 \delta p \;\supset\; - \bar V^{ }_{,\varphi}\delta\varphi
 \;. \label{simpl} 
\ee
In the following, to streamline the notation, we write 
$\bar V\to V$.

%%%%%%%%%%%%%%%%%%%%%%%%%%%%%%%%%%%%%%%%%%%%%%%%%%%%%%%%%%%%%%%%%%%%%%%%%
%
\subsection{Einstein equations} %Perturbed 
\label{ssec_Eeqs-p}

Having determined the perturbed Einstein and energy-momentum tensors
in \ses\ref{ss:delta_Gmunu} and \ref{ss:delta_Tmunu}, respectively, 
we can now collect together the perturbed Einstein equations, 
\begin{equation}\label{eq_ee-p}
 \delta G^{ }_{\mu\nu} \;=\; 8\pi G \, \delta T^{ }_{\mu\nu}
 \;. 
\end{equation}
We do this separately 
for scalar (s), vector (v), and tensor (t) perturbations.
For reference, we also recall the background equations once more, 
from \eqs\nr{G_00}, \nr{G_ij}, 
\nr{delta_Tmunu_fluid} and \nr{bg_Tmunu_varphi}, 
which help to simplify some of the Einstein equations:
\be
 3\H^2_{ } 
 \; \underset{\rmii{\nr{delta_Tmunu_fluid},\nr{bg_Tmunu_varphi}}}
    {\overset{\rmii{\nr{G_00}} \lift }{=}} \; 
 8\pi G \biggl[ a^2\bar{e} + \frac{(\bar{\varphi}')^2}{2} \biggr]
 \;, \quad
 -(2\H' + \H^2{ })
 \; \underset{\rmii{\nr{delta_Tmunu_fluid},\nr{bg_Tmunu_varphi}}}
    {\overset{\rmii{\nr{G_ij}} \lift }{=}} \; 
 8\pi G \biggl[ a^2\bar{p} + \frac{(\bar{\varphi}')^2}{2} \biggr]
 \;. \label{bg_again}
\ee
Here we have hidden the appearances of $\bar V$, 
%  and $\bar V^{ }_{,\varphi}$, 
by making use of \eq\nr{simpl}. 

%%%%%%%%%%%%%%%%%%%%%%%%%%%%%%%%%%%%%%%%%%%%%%%%%%%%%%%%%%%%%%%%%%%%%%%%%%%

\subsubsection*{Scalar perturbations}

\index{scalar perturbations}

When we insert the tensor decomposition from 
\eqs\nr{eq_metric-pert2}--\nr{eq_metric-pert3} into 
\eqs\nr{G_00}--\nr{G_ij}, we obtain the scalar parts of
linear perturbations of the Einstein tensor,
\ba
 \delta G^\rmi{s}_{00} & = & 
 2\,\biggl[\,
 - \H \,( 3  h_\rmii{D}' + \nabla^2_{ }h ) 
 + \nabla^2_{ }
   \biggl( h^{ }_\rmii{D} + \frac{\nabla^2_{ }\vartheta}{3} \biggr) 
 \,\biggr]
 \;, \label{delta_G00_s} \\[2mm] 
%%%
 \delta G^\rmi{s}_{0i} & = &  
 \partial^{ }_i\biggl[\, 
 - (2\H' + \H^2_{ }) h
 + 2 \H h^{ }_0 
 + 2 \partial^{ }_\tau 
   \biggl( h^{ }_\rmii{D} + \frac{\nabla^2_{ }\vartheta}{3} \biggr) 
 \,\biggr]
 \;, \label{delta_G0i_s} \\[2mm] 
%%%
 \delta G^\rmi{s}_{ij} & = &
 \delta^{ }_{ij} \, \biggl[\, 
 2 (2\H' + \H^2_{ }) 
 \biggl( 
   h^{ }_0 + h^{ }_\rmii{D} + \frac{\nabla^2_{ }\vartheta}{3}
 \biggr)
  + 2 \H h_0'
  + \nabla^2_{ }
 \biggl( 
   h^{ }_0 - h^{ }_\rmii{D} - \frac{\nabla^2_{ }\vartheta}{3}
 \biggr) 
 \nn[2mm]
%%%
 & & \; + \,  
 2 (\partial^2_\tau + 2 \H \partial^{ }_\tau)
 \biggl( 
   h^{ }_\rmii{D} + \frac{\nabla^2_{ }\vartheta}{3}
 \biggr)
 + (\partial^{ }_\tau + 2 \H) \nabla^2_{ } (h - \vartheta')
  \,\biggr]
 \nn[2mm]
%%%
 & - &
 \partial^{ }_i \partial^{ }_j \biggl[\,
   2 (2\H' + \H^2_{ })\vartheta 
 + h^{ }_0 
 + (\partial^{ }_\tau + 2 \H) (h - \vartheta')
 - h^{ }_\rmii{D} - \frac{\nabla^2_{ }\vartheta}{3}
 \,\biggr] 
 \;. \label{delta_Gij_s}
\ea
Comparing 
$\delta G^\rmi{s}_{00}$ from \eq\nr{delta_G00_s}
with $\delta T^\rmi{s}_{00}$ from \eqs\nr{delta_Tmunu_fluid} 
and \nr{delta_T00_varphi}; 
recalling \eq\nr{simpl};  
and substituting $2 a^2_{ }\bar{e}$
with the help of \eq\nr{bg_again}, we find
\ba
 \boxed{
 \quad 
 - 3 \H^2_{ }h^{ }_0 
 - \H \,( 3  h_\rmii{D}' + \nabla^2_{ }h ) 
 + \nabla^2_{ }
   \biggl( h^{ }_\rmii{D} + \frac{\nabla^2_{ }\vartheta}{3} \biggr) 
 \; = \; 
 4\pi G\, \bigl[\,
   \bar\varphi\hspace*{0.3mm}{}' ( \delta\varphi' 
    % + 2 a^2_{ }\bar{e} h^{ }_0 
    - h^{ }_0 \bar\varphi\hspace*{0.3mm}{}' ) 
    + a^2_{ }\delta e 
    \,\bigr]
% \quad(\mbox{I})
  \;. \quad   \vphantom{\Bigg|}
 } 
\hspace*{-5mm} &&  \nn[2mm] \label{delta_einstein_00} 
 \index{Einstein equations: scalar perturbations}
\ea

As far as 
$\delta G^\rmi{s}_{0i}$ 
from  \eq\nr{delta_G0i_s} and 
$\delta T^\rmi{s}_{0i}$ 
from \eqs\nr{delta_Tmunu_fluid} and 
\nr{delta_T0i_varphi} go, we note that the part
$
 - (2\H' + \H^2_{ }) h^{ }_{,i}
$
can be substituted with the background identity from \eq\nr{bg_again}.
The rest of the equation sets two gradients equal. If we assume that
the perturbations vanish at infinity, so that there is 
no integration constant, we obtain
\be
 \boxed{
 \quad
    \H h^{ }_0 
 +   
   h_\rmii{D}' + \frac{\nabla^2_{ }\vartheta'}{3} 
 \; = \; 
 4\pi G \bigl[\, 
  a^2_{ }( \bar{e} + \bar{p}) (v - h) + {\bar\varphi}' \delta\varphi
  \,\bigr]
% \quad(\mbox{II})
  \;. \quad   \vphantom{\Bigg|}
 }
 \label{delta_einstein_0i}
\ee 

Turning to $\delta G^\rmi{s}_{ij}$ from \eq\nr{delta_Gij_s},
with $\delta T^\rmi{s}_{ij}$ from \eqs\nr{delta_Tmunu_fluid} 
and \nr{delta_Tij_varphi}, let us first consider the ``non-diagonal''
terms, with $i \neq j$. There is a cancellation
between terms containing 
$(2\H' + \H^2_{ })\vartheta$ on the side of $\delta G^\rmi{s}_{ij}$,
and $\vartheta^{ }_{ij}$ in the side of $\delta T^\rmi{s}_{ij}$, 
thanks to the background identity in \eq\nr{bg_again}. The remainder 
can again be integrated, assuming that the integration constants
vanish at spatial infinity. After the cancellation of $\vartheta^{ }_{ij}$, 
there is no source term from \eq\nr{delta_Tmunu_fluid} or 
\nr{delta_Tij_varphi}. The only 
contribution originates from the viscous 
anisotropic stress in \eq\nr{delta_Tmunu_aniso}, which we decompose
according to \eq\nr{Pi_ij_s}, getting 
\be
 \boxed{
 \quad
  h^{ }_0 
 + (\partial^{ }_\tau + 2 \H) (h - \vartheta')
 - h^{ }_\rmii{D} - \frac{\nabla^2_{ }\vartheta}{3}
 \; = \; -\,
  8\pi G a^2_{ }\barpPi 
% \quad(\mbox{III})
 \;. \quad   \vphantom{\Bigg|}
 }
 \label{delta_einstein_didj}
\ee

Finally, in the parts proportional to $\delta^{ }_{ij}$, 
there is a term that cancels thanks to the background identity in 
\eq\nr{bg_again}. In view of the tensor decomposition of 
$\vartheta^{ }_{ij}$ in \eq\nr{vartheta_ij_s}, this concerns
the terms $(2\H' + \H^2_{ })(h^{ }_\rmii{D} + \nabla^2_{ }\vartheta/3)$
on the side of $\delta G^\rmi{s}_{ij}$, 
and 
$
 - [2 a^2_{ }\bar p + (\bar\varphi\hspace*{0.3mm}')^2_{ } ]
 (h^{ }_\rmii{D}\delta^{ }_{ij} - \vartheta^{ }_{ij})
$ on the side of $\delta T^{ }_{ij}$. 
Recalling \eq\nr{simpl} as well, left over is then 
\ba
 & & \hspace*{-1.0cm} 
 \bigl[\, 
   2 (2\H' + \H^2_{ }) 
  + 2 \H \partial^{ }_\tau 
  + \nabla^2_{ } \,\bigr] h^{ }_0
% \nn[2mm]
%%%
% & + & 
 + 
 \bigl[\,
 2 (\partial^2_\tau + 2 \H \partial^{ }_\tau) - \nabla^2_{ }
 \,\bigr]
 \biggl( 
   h^{ }_\rmii{D} + \frac{\nabla^2_{ }\vartheta}{3}
 \biggr)
 \nn[2mm]
 & + &
 (\partial^{ }_\tau + 2 \H) \nabla^2_{ } (h - \vartheta') 
 \; = \;  
 8\pi G  
 \biggl[ a^2_{ } \biggl( \delta p 
   - \frac{\nabla^2_{ }\barpPi}{3} \biggr)
  +  {\bar\varphi}' \bigl( \delta\varphi'
  - h^{ }_0 \bar\varphi\hspace*{0.3mm}{}' \bigr)  
 \biggr]
 \;. \label{delta_einstein_delta_ij}
\ea
We obtain what turns out to be a more helpful relation 
as a linear combination of several equations, defining 
$
 \mbox{\nr{delta_einstein_comb}} 
 \equiv 
 [\mbox{\nr{delta_einstein_delta_ij}}
 - \nabla^2_{ }\mbox{\nr{delta_einstein_didj}}
 - 2 \mbox{\nr{delta_einstein_00}}]/2
$, yielding finally
\ba
 \boxed{ 
 \begin{array}{ccc}
 & & \hspace*{-0.5cm} \displaystyle   \vphantom{\Bigg|}
    2 \bigl( \H' + 2 \H^2_{ } \bigr) 
    h^{ }_0
 + 
  \H \, \bigl(\, h_0' + 3  h_\rmii{D}' + \nabla^2_{ }h \,\bigr) 
 + 
  \bigl(\, \partial^2_\tau + 2 \H \partial^{ }_\tau - \nabla^2_{ } \,\bigr)
 \biggl( 
   h^{ }_\rmii{D} + \frac{\nabla^2_{ }\vartheta}{3}
 \biggr)
 \nn[2mm]
 &  & \displaystyle   \vphantom{\Bigg|}
  =\; 4\pi G  
  a^2_{ } \biggl( \delta p - \delta e 
   + \frac{2}{3} \nabla^2_{ }\barpPi  \biggr)
% \quad(\mbox{IV})
 \;. 
 \end{array}
  }
  \nn[-8mm] \label{delta_einstein_comb}
\ea

Equations~\nr{delta_einstein_00}, 
\nr{delta_einstein_0i}, 
\nr{delta_einstein_didj} and 
\nr{delta_einstein_comb} are independent of each other, 
and constitute the information that 
can be extracted from the Einstein equations. 
It is appropriate to remark, however, that 
in general they do {\em not} suffice to specify the dynamics of the system.
Even if we managed to express 
$\delta p$, $\delta e$ and $\Pi$ in terms of two
independent perturbations, for instance $\delta T$ and $\delta \varphi$,
there are 5 further variables, namely the scalar velocity
$v$ and the metric perturbations $h^{ }_0$, $h^{ }_\rmii{D}$, 
$h$ and $\vartheta$. As will be discussed in \ch\ref{se:gauges}, 
only two linear combinations of the metric perturbations are 
``physical'', i.e.\ cannot be eliminated by gauge (or coordinate)
transformations. This still leaves over 5 physical variables, 
so at least one further equation is needed. 
We return to this in \se\ref{ss:pert_scalar}.

%%%%%%%%%%%%%%%%%%%%%%%%%%%%%%%%%%%%%%%%%%%%%%%%%%%%%%%%%%%%%%%%%%%%%%%%%%%%

\subsubsection*{Vector perturbations}

\index{vector perturbations}

Among \eqs\nr{G_00}--\nr{G_ij}, the last two have vector parts
once we insert \eqs\nr{eq_metric-pert2}--\nr{eq_metric-pert3}, 
\ba
 \delta G^\rmi{v}_{00} & = & 
 0  
 \;, \\[2mm] 
%%%
 \delta G^\rmi{v}_{0i} & = &  
 \bigl(\, 2 \H' + \H^2_{ } \,\bigr) h^\rmi{v}_i
  + \fr12 \nabla^2_{ }\bigl(\, h^\rmi{v}_i - {\vartheta^\rmi{v}_i}' \,\bigr)
 \;, \label{delta_G0i_v} \\[2mm] 
%%%
 \delta G^\rmi{v}_{ij} & = &
 \bigl(\, 2\H' + \H^2_{ } \,\bigr) 
 \bigl(\, \vartheta^\rmi{v}_{i,j} + \vartheta^\rmi{v}_{j,i} \,\bigr)
  + \fr12 \bigl(\, \partial^{ }_\tau + 2 \H \,\bigr)
    \bigl(\,
        h^\rmi{v}_{i,j}
     +  h^\rmi{v}_{j,i}
     - \vartheta^\rmi{v}_{i,j}\hspace*{-1.5mm}' 
     - \vartheta^\rmi{v}_{j,i}\hspace*{-1.5mm}' 
    \,\bigr)  
 \;. \hspace*{6mm} \label{delta_Gij_v}
\ea
Comparing
$\delta G^\rmi{v}_{0i}$ 
from \eq\nr{delta_G0i_v} and 
$\delta T^\rmi{v}_{0i}$ 
from \eqs\nr{delta_Tmunu_fluid} %, \nr{delta_Tmunu_aniso} 
and 
\nr{delta_T0i_varphi}, we note that the part
$
 (2\H' + \H^2_{ }) h^\rmi{v}_{i}
$
can be simplified with the background identity from \eq\nr{bg_again}.
This yields \eq\nr{einstein_0i_v}. 
With 
$\delta G^\rmi{v}_{ij}$ 
from \eq\nr{delta_Gij_v} and 
$\delta T^\rmi{v}_{ij}$ 
from \eqs\nr{delta_Tmunu_fluid}, \nr{delta_Tmunu_aniso} 
and 
\nr{delta_Tij_varphi}, and extracting the vector part
of $\vartheta^{ }_{ij}$ according to \eq\nr{theta_ij_v}, 
the part
$
 (2\H' + \H^2_{ }) ( \vartheta^\rmi{v}_{i,j} + \vartheta^\rmi{v}_{j,i} )
$
cancels. % thanks to the background identity from \eq\nr{bg_again}.
The remainder yields \eq\nr{einstein_ij_v}, 
whereby in total we have
\begin{empheq}[box=\fbox]{align}
   \vphantom{\Bigg|}
 \fr12 \nabla^2_{ }
  \bigl(\, h^\rmi{v}_i - {\vartheta^\rmi{v}_i}' \,\bigr)
 & \;=\;  8\pi G a^2_{ }(\bar{e} + \bar{p})
  \bigl(\, h^\rmi{v}_i - v^\rmi{v}_i \,\bigr) 
 \quad \forall i 
 \;, \label{einstein_0i_v}
 \\[-2mm]
%%%%%%%
 \quad
 (\partial^{ }_\tau + 2 \H)
    \bigl(\,
        h^\rmi{v}_{i,j}
     +  h^\rmi{v}_{j,i}
     - \vartheta^\rmi{v}_{i,j}\hspace*{-1.5mm}' 
     - \vartheta^\rmi{v}_{j,i}\hspace*{-1.5mm}' 
    \;\bigr) 
 & \;=\;  
 - 8\pi G a^2_{ }\, 
 \bigl(\, \barpPi^\rmi{v}_{i,j} + \barpPi^\rmi{v}_{j,i} \,\bigr)
 \quad \forall i,j
 \;. \quad   \vphantom{\Bigg|}
 \label{einstein_ij_v} 
 \index{Einstein equations: vector perturbations}
\end{empheq}
Here $\barpPi^\rmi{v}_i$ is a function of 
$v^\rmi{v}_i - {\vartheta^\rmi{v}_{i}}'$, 
cf.\ \eq\nr{decomposition}.
We note that if we take the divergence $\partial^{ }_j$ from 
\eq\nr{einstein_ij_v} and make use of the transversality of vector
perturbations, we obtain 
\be
 (\partial^{ }_\tau + 2 \H)
 \nabla^2_{ } \bigl(\, h^\rmi{v}_i - {\vartheta^\rmi{v}_i}' \,\bigr)
 \; = \; 
 - 8 \pi G a^2_{ } \nabla^2_{ } \barpPi^\rmi{v}_{i}
 \quad \forall i
 \;. \label{einstein_ij_v_plus}
\ee
With \eqs\nr{einstein_0i_v} and \nr{einstein_ij_v_plus}, 
we have two equations for two transverse vectors, 
$h^\rmi{v}_i - {\vartheta^\rmi{v}_i}'$
and
$h^\rmi{v}_i - v^\rmi{v}_i$.
If we insert \eq\nr{decomposition} 
for $ \barpPi^\rmi{v}_{i} $, then the vector 
$h^\rmi{v}_i - {\vartheta^\rmi{v}_i}'$
can be eliminated, 
and we obtain one diffusive equation 
for the single vector 
$(\bar e + \bar p)(h^\rmi{v}_i - v^\rmi{v}_i)$.

Equation~\nr{einstein_ij_v} is a local first-order time-evolution
equation at each position $\vec{x}$. 
Without anisotropic stress
the equation is easily solved, and leads to a rapid decay, 
\be
   \bigl(\,
        h^\rmi{v}_{i,j}
     +  h^\rmi{v}_{j,i}
     - \vartheta^\rmi{v}_{i,j}\hspace*{-1.5mm}' 
     - \vartheta^\rmi{v}_{j,i}\hspace*{-1.5mm}' 
   \;\bigr)(\tau,\vec{x})
 \; 
 \overset{\Pi^\rmii{v}_{ }\;\to\;0}{\approx}
 \; 
 \frac{c^{ }_{ij}(\vec{x})}{a^2_{ }(\tau)}
 \;, \quad
 \tr c^{ }_{ij} \;=\; 0 
 \;,  \label{soln_v_homog_1}
\ee
where $c^{ }_{ij}$ is an integration constant. 
If $\Pi^\rmi{v}_{ }$ is kept non-zero, vector perturbations
are constantly generated by hydrodynamic fluctuations
(cf.\ \app\ref{app:viscous}),
but subsequently each mode decays, so that their average 
distribution corresponds to that in thermal equilibrium. 
For many considerations, where only non-equilibrium modes
could carry enough energy density to have an observable effect, 
vector perturbations can be neglected.

%%%%%%%%%%%%%%%%%%%%%%%%%%%%%%%%%%%%%%%%%%%%%%%%%%%%%%%%%%%%%%%%%%%%%%%%%%%%

\subsubsection*{Tensor perturbations}

\index{tensor perturbations}

Among the components of $G^{ }_{\mu\nu}$, 
in \eqs\nr{G_00}--\nr{G_ij}, only $G^{ }_{ij}$ has a tensor part, 
\ba
 \delta G^\rmi{t}_{00} & = & 
 0  
 \quad = \quad
 \delta G^\rmi{t}_{0i}  
 \;, \\[2mm] 
%%%
 \delta G^\rmi{t}_{ij} & = &
 -2 (2\H' + \H^2_{ }) \vartheta^\rmi{t}_{ij} 
 + \bigl(\, \partial_\tau^2 + 2 \H \partial_\tau^{ } - \nabla^2_{ } \,\bigr) 
 \vartheta^\rmi{t}_{ij}
 \;. \label{tensor_full}
\ea
Once again, the part proportional to $2\H' + \H^2_{ }$ cancels
against the $\vartheta^{ }_{ij}$-parts from 
\eqs\nr{delta_Tmunu_fluid} and \nr{delta_Tij_varphi},  
on account of \eq\nr{bg_again}.  
Left over is a contribution from the tensor part of 
\eq\nr{delta_Tmunu_aniso}, leading to 
\be
 \boxed{
 \quad 
  \bigl(\, \partial_\tau^2 + 2 \H \partial_\tau^{ } - \nabla^2_{ } \,\bigr)
   \vartheta^\rmi{t}_{ij}
 \; = \; 
 8\pi G a^2_{ } \barpPi^\rmi{t}_{ij} 
 \;. \quad
   \vphantom{\Bigg|}
 } 
 \label{einstein_ij_t}
 \index{Einstein equations: tensor perturbations} 
\ee

\index{d'Alembert operator in FLRW coordinates}

We note that the differential operator
on the left-hand side of \eq\nr{einstein_ij_t} is nothing but 
the {\em d'Alembert operator in FLRW coordinates}. 
Perhaps the easiest
way to see this is to derive the Euler-Lagrange equation for 
a minimally coupled scalar field, cf.\  \eq\nr{eq_field-eq}.
Denoting by $\bar{g}_{\mu\nu}^{ }$ the metric given 
by \eq\eqref{eq_flrw_metric} or \eq\eqref{t_tau}, 
and $\bar{g} \equiv \det \bar{g}_{\mu\nu}$, we obtain in 
physical and conformal time, respectively,
\begin{align}
 \square_\rmii{$H$}^{ } Q &\;\equiv\; 
  \frac{1}{\sqrt{-\bar{g} \vphantom{|} }}\,
  \partial_\mu^{ }
  \bigl(\, \sqrt{-\bar{g} \vphantom{|} }\,
     \bar{g}^{\mu\nu}_{ }\partial_\nu^{ } Q 
  \,\bigr) 
  \nonumber\\[2mm]
%%%%%%
 &\;=\;
 a^{-3}_{ }
 \bigl[ \partial_t^{ }(-a^3_{ } \dot{Q})
  + \partial_i^{ }(a Q^{,i}_{ }) \bigr]
   \;
   =
   \;
     -\ddot{Q} - 3H\dot{Q} + a^{-2}_{ }\nabla^2_{ } Q 
  \;, \label{eq_box-opt} \\[3.5mm]
%%%%%%
  \square_\rmii{$\Hc$}^{ } Q  &\;=\;    a^{-4}_{ }
 \left[ \partial_\tau^{ }(-a^2_{ } Q\hspace*{0.3mm}')
  + \partial_i^{ }(a^2_{ } Q^{,i}_{ }) \right] 
                       \;=\; a^{-2}_{  }
  \left( -Q\hspace*{0.3mm}''
   - 2\Hc Q\hspace*{0.3mm}' + \nabla^2_{ } Q \right) 
 \;. \label{eq_box-opc}
\end{align}
Therefore, the homogeneous version of 
\eq\nr{einstein_ij_t} is a {\em wave equation}, % \index{wave equation} 
$\square_\rmii{$\Hc$}\vartheta_{ij}^\rmi{t} = 0$, 
whose solutions are 
{\em gravitational waves}, propagating 
through the expanding universe.
\index{gravitational waves: definition}

%%%%%%%%%%%%%%%%%%%%%%%%%%%%%%%%%%%%%%%%%%%%%%%%%%%%%%%%%%%%%%%%%%%%%%%%%
%
\subsection{Scalar field equation} %Perturbed 
\label{ss:pert_scalar}

\index{scalar field equation: perturbed}

We have seen in \se\ref{ssec_Eeqs-p} that, 
if both $\delta\varphi$ and $\delta T$ play a role,  
the Einstein equations for scalar
perturbations, \eqs\nr{delta_einstein_00}, 
\nr{delta_einstein_0i}, 
\nr{delta_einstein_didj} and 
\nr{delta_einstein_comb}, are not sufficient for fixing the full
solution. Here we derive the additional equation that
is needed for this purpose. This is given by the scalar field 
equation. At the background level, 
its vacuum form is given in \eq\nr{eq_field-eq}, 
and a form interacting with a fluid in \eq\nr{varphi_eq}. 

The perturbations of the metric are given in 
\eqs\nr{g_munu} and \nr{g^munu}. 
Recalling that $\vartheta^{ }_{ij}$ is traceless
(cf.\ \eq\nr{h_notation}), the 
determinant of $g^{ }_{\mu\nu}$ is 
\ba
 -g 	
 & \overset{\rmii{\nr{g_munu}}}{=} &  
 a^8_{ } ( 1 + 2h_0^{ } )( 1 - 2h_\rmii{D}^{ } )^3_{ }
 + \ord(\delta^2_{ })
 \nn[2mm]
%%%
 \quad \Rightarrow \quad 
 \sqrt{-g} & = &  a^4_{ } ( 1 + h_0^{ } - 3h_\rmii{D}^{ } )
 + \ord(\delta^2_{ })	
 \ . \label{det}
\ea
The derivative from the left-hand side of \eqref{eq_field-eq}, 
taken like in \eq\nr{eq_box-opc} but now to first order in perturbations, 
becomes  
\ba
 &&\hspace*{-2.0cm}
 \bigl(\, \sqrt{-g}\varphi^{,\mu}_{ } \,\bigr)_{,\mu}^{ } 
 \;=\; 
 \bigl[\, 
   \sqrt{-g}\, \bigl(\, g^{00}_{ }\varphi' + g^{0i}_{ }\varphi^{ }_{,i}
   \,\bigr)\,\bigr]'
 + 
 \bigl[\,
   \sqrt{-g}\, \bigl(\, g^{i0}_{ }\varphi' + g^{ij}_{ }\varphi^{ }_{,j}
   \,\bigr)\,\bigr]^{ }_{,i}
 \label{towards_dalemb} \\[2mm]
%%%%%
 & 
 \underset{\rmii{\nr{g^munu}}}{
 \overset{\rmii{\nr{det}} \lift }{=}}
 &
 -\bigl[\, a^2_{ }( 1 - h_0^{ } - 3h_\rmii{D}^{ } )\varphi'
 + a^2_{ } h^{ }_i \hspace*{0.3mm} \varphi^{ }_{,i} \,\bigr]'
 + \bigl[\, a^2_{ }( -h^{ }_i \hspace*{0.3mm} 
   \varphi' + \varphi_{,i}^{ } ) \,\bigr]_{,i}^{ }		
 + \ord(\delta^2_{ }) 
  \nn[2mm]
%%%%
  &=& -a^2_{ }\Bigl[\, ( 1 - h_0^{ } - 3h_\rmii{D}^{ } )
 ( \varphi'' + 2\Hc\varphi' )
 - \nabla^2_{ }\varphi - ( h_0' + 3h_\rmii{D}' - h_{i,i}^{ })\varphi'
  \,\Bigr]
  + \ord(\delta^2_{ })
  \ . \nonumber 
\ea
Here we made use of the fact that 
$\varphi^{ }_{\der i}$ is of $\ord(\delta)$.
Eq.~\nr{towards_dalemb} is divided by $\sqrt{-g}$ from \eq\nr{det}, 
to get ${\varphi^{;\mu}_{ }}^{ }_{;\mu}$.
We also note that according to \eq\nr{eq_metric-pert2}, 
$-h_{i,i}=\nabla^2h$, 
so that only scalar perturbations contribute. 
Inserting $\varphi=\bar{\varphi}+\delta\varphi$,
this gives
\ba
 {\varphi^{;\mu}_{ }}^{ }_{;\mu}
 &
 \underset{\rmii{\nr{det}}}{
 \overset{\rmii{\nr{towards_dalemb}}}{=}} 
 &
 - \frac{1}{a^{2}_{ }}
 \Bigl[\,
 (1-2 h^{ }_0) (\partial_\tau^2 + 2 \H \partial^{ }_\tau) 
 ( \bar{\varphi}+\delta\varphi ) 
 - \nabla^2_{ }\delta\varphi 
 \nn[2mm] 
%%%%%%
 & & 
 \hspace*{8mm}
 \;-\, ( h_0' + 3h_\rmii{D}' +\nabla^2_{ } h)
   \,\bar\varphi\hspace*{0.3mm}{}'
 \,\Bigr]
 \; + \; \ord(\delta^2_{ }) 
 \ .
 \label{dAl_varphi}
\ea

As far as the second term of \eq\nr{varphi_eq} goes, 
we insert the velocity from \eq\nr{eq_u^up}, \nolinebreak to \nolinebreak get
\be
 u^\mu_{ }\varphi^{ }_{,\mu} 
 \; \overset{\rmii{\nr{eq_u^up}}}{=} \; 
 \frac{1}{a^{ }_{ }} 
 \bigl[\, ( 1 - h^{ }_0)\, \varphi'
   + v^{ }_i \bit \varphi^{ }_{,i} \,\bigr] 
  + \ord(\delta^2_{ })
 \;=\;
 \frac{1}{a^{ }_{ }} 
 \bigl[\, ( 1 - h^{ }_0)\, \bar\varphi\hspace*{0.3mm}{}'
  + \delta\varphi' \,\bigr] 
  + \ord(\delta^2_{ })
 \;.  
\ee
The coefficient $\Upsilon$, 
multiplying the second term of \eq\nr{varphi_eq},
also needs to be perturbed, as it can be a function
of the dynamical variables, and the same applies to the potential, 
\be
 \Upsilon \;=\; \bar\Upsilon + \delta\Upsilon + \ord(\delta^2_{ })  
 \;, \qquad
 V^{ }_{,\varphi} \;=\; 
 \bar V^{ }_{,\varphi} + \delta V^{ }_{,\varphi}  + \ord(\delta^2_{ })
 \;. \label{Ups_exp}
\ee
The background equation then yields \eq\nr{bg_varphi}, where we 
had simplified the notation by substituting 
$ \bar\Upsilon \to \Upsilon $ and 
$ \bar V^{ }_{,\varphi} \to V^{ }_{,\varphi} $.
At first order, 
combining \eqs\nr{dAl_varphi}--\nr{Ups_exp} according to 
\eq\nr{varphi_eq}, and 
treating the noise term $\varrho$ as being of the same order
as first-order perturbations
(this is analogous to the discussion below \eq\nr{delta_Tmunu_fluid}), yields
\ba
 &
 \underset{\rmii{\nr{dAl_varphi}--\nr{Ups_exp}}}{
 \overset{\scriptscriptstyle 
 -a^2_{ }\,\times\,\rmii{\nr{varphi_eq}} \lift }{\Rightarrow}} 
 & 
 \delta\varphi'' + 2 \H \delta \varphi'
  - 2 h^{ }_0 
  \hspace*{-9mm}
  \overbrace{ (\bar\varphi\hspace*{0.3mm}{}''
        + 2 \H \bar\varphi\hspace*{0.3mm}{}') }^
            {\nr{bg_varphi_tau}:\;{\rm becomes}\; 
             -a\Upsilon\bar\varphi\hspace*{0.3mm}{}' -a^2 V^{ }_{,\varphi}}
  \hspace*{-9mm}
 - \nabla^2_{ }\delta\varphi 
 - ( h_0' + 3h_\rmii{D}' +\nabla^2_{ } h)\,\bar\varphi\hspace*{0.3mm}{}'
 \nn[2mm]
%%%%%%%
 &&  \hspace*{1.5cm}
 + \, a\, \delta \Upsilon \bar\varphi\hspace*{0.3mm}{}' 
 + a \Upsilon (\,-h^{ }_0\, \bar\varphi\hspace*{0.3mm}{}' + \delta\varphi'\,)
 + a^2_{ } \delta V^{ }_{,\varphi} \;=\; a^2_{ }\varrho  
 \nn[2mm] 
 & \Leftrightarrow & 
 \boxed{
 \quad
 \begin{array}{ccc} \displaystyle   \vphantom{\Bigg|}
 && \hspace*{-1.0cm}
 \delta\varphi'' + ( 2 \H + a \Upsilon )\, \delta \varphi'
 - \nabla^2_{ }\delta\varphi 
 - ( h_0' + 3h_\rmii{D}' +\nabla^2_{ } h)\, \bar\varphi\hspace*{0.3mm}{}'
 \nn[2mm]
 &&  \hspace*{0.5cm} \displaystyle   \vphantom{\Bigg|}
 + \, a ( \delta \Upsilon + h^{ }_0 \Upsilon )\, \bar\varphi\hspace*{0.3mm}{}' 
 + a^2_{ } ( \delta V^{ }_{,\varphi} + 2 h^{ }_0 V^{ }_{,\varphi} )
 \; = \;
 a^2_{ }\varrho   
 \;. 
 \end{array}
 \quad
 }
 \hspace*{1.0cm}
 \nn[-16mm]
 \label{eq_field-eq-pert}
 \\[10mm]
 \nonumber
\ea
With this equation, we have the full 
information needed for determining the time evolution of physical
scalar perturbations. 

%%%%%%%%%%%%%%%%%%%%%%%%%%%%%%%%%%%%%%%%%%%%%%%%%%%%%%%%%%%%%%%%%%%%%%%%%
%
\subsection{Energy-momentum conservation} %Perturbed 

The Einstein tensor has the property 
${G^{ }_{\mu\nu}}^{;\mu}_{ } = 0$, 
consistent with energy-momentum conservation, 
${T^{ }_{\mu\nu}}^{;\mu}_{ } = 0$.
Even though formally the latter relations % (separately for $\nu = 0,i$)
do {\em not} add  
information to the Einstein equations, 
$G^{ }_{\mu\nu} = 8\pi G T^{ }_{\mu\nu}$, 
they may still 
establish a helpful re-organization.
It may be noted that in the limit of 
(non-perturbed) Minkowskian space-time, the Einstein tensor
vanishes, yet the energy-momentum tensor remains non-trivial, 
and energy-momentum conservation yields essential information. 
For example, this is how the equations of hydrodynamics
are obtained, when $T^{ }_{\mu\nu}$
is expressed in terms of fluid variables. 

Inserting the Christoffel symbols, 
energy-momentum conservation reads
\ba
 0 
 & = &  
 \mathbin{T^{ }_{\mu\nu}}^{;\mu}_{ }
 \;
 \overset{\rmii{\nr{cov_der_1}}}{=} 
 \;
 \mathbin{T^{ }_{\mu\nu;\alpha}} g^{\alpha\mu}_{ }
 \;
 \overset{\rmii{\nr{cov_der_2}}}{=} 
 \;
 \bigl( \mathbin{T^{ }_{\mu\nu,\alpha}} 
      - \mathbin{T^{ }_{\mu\beta}}{\Gamma^{\beta}_{\nu\alpha}}
      - \mathbin{T^{ }_{\beta\nu}}{\Gamma^{\beta}_{\mu\alpha}}
 \bigr)
 g^{\mu\alpha}_{ }
 \label{Tmunu;mu}
 \index{energy-momentum conservation}
 \\[3mm]
%%%%%%%%%%%%%%%
 & = & 
 \overbrace{
 \bigl( \mathbin{T_{0\nu}'} 
      - \mathbin{T^{ }_{0\beta}}{\Gamma^{\beta}_{\nu 0}}
      - \mathbin{T^{ }_{\beta\nu}}{\Gamma^{\beta}_{0 0}}
 \bigr)
 g^{0 0}_{ }
 }^{\mu\to 0,\; \alpha\to 0 }
 + 
 \overbrace{
 \bigl( \mathbin{T^{ }_{k \nu,l}} 
      - \mathbin{T^{ }_{k \beta}}{\Gamma^{\beta}_{\nu l}}
      - \mathbin{T^{ }_{\beta\nu}}{\Gamma^{\beta}_{k l}}
 \bigr)
 g^{kl}_{ }
 }^{\mu\to k,\; \alpha\to l }
 \nn[2mm]
 & & \;+\,  
 \underbrace{
  \bigl(
         \mathbin{T^{ }_{0\nu,k}} 
      +  \mathbin{T_{k \nu}'} 
      - \mathbin{T^{ }_{0\beta}}{\Gamma^{\beta}_{\nu k}}
      - \mathbin{T^{ }_{k \beta}}{\Gamma^{\beta}_{\nu 0}}
      - 2 \mathbin{T^{ }_{\beta\nu}}{\Gamma^{\beta}_{0 k}}
 \bigr)
 g^{0 k}_{ }
 }_{
 \mu\to 0,\;\alpha\to k \;{\rm or}\; \mu\to k,\;\alpha\to 0
 }
 \;. \label{dTmn_1}
 \\[-2mm]
 \nonumber 
\ea
At the background level
($g^{00}_{ }\to - 1/a^{2}_{ }$, 
$g^{kl}_{ }\to \delta^{ }_{kl} /a^2_{ }$, 
$g^{0k}_{ }\to 0$), this leads to 
\ba
 0
 &
 \overset{\scriptscriptstyle a^2_{ }\,\times\,\rmii{\nr{dTmn_1}}
  \lift}{=}
 & 
      - \mathbin{\bar T_{0\nu}'} 
      + \mathbin{\bar T^{ }_{0\beta}}{\bar\Gamma^{\beta}_{\nu 0}}
      + \mathbin{\bar T^{ }_{\beta\nu}}{\bar\Gamma^{\beta}_{0 0}}
      + \mathbin{\bar T^{ }_{k \nu,k}} 
      - \mathbin{\bar T^{ }_{k \beta}}{\bar\Gamma^{\beta}_{\nu k}}
      - \mathbin{\bar T^{ }_{\beta\nu}}{\bar\Gamma^{\beta}_{k k}}
 \nn[2mm]
%%%%
 &=&
 \left\{ 
 \begin{array}{ll}
      - \mathbin{\bar T_{00}'} 
      + 2 \mathbin{\bar T^{ }_{00}}{\bar\Gamma^{0}_{00}}
      - \mathbin{\bar T^{ }_{k l}}{\bar\Gamma^{l}_{0 k}}
      - \mathbin{\bar T^{ }_{00}}{\bar\Gamma^{0}_{k k}}
 & 
 (\nu = 0 ) \\[2.5mm]
      + \mathbin{\bar T^{ }_{0 0}}{\bar\Gamma^{0}_{0 i}}
      + \mathbin{\bar T^{ }_{i k}}{\bar\Gamma^{k}_{0 0}}
      + \mathbin{\bar T^{ }_{i k,k}} 
      - \mathbin{\bar T^{ }_{k l}}{\bar\Gamma^{l}_{i k}}
      - \mathbin{\bar T^{ }_{i l}}{\bar\Gamma^{l}_{k k}}
 & 
 (\nu = i )
 \end{array} \right.
 \nn[2mm]
%%%%%%
 & \overset{\rmii{\nr{eq_ChSyu}} \lift }{=} & 
 \left\{ 
 \begin{array}{ll}
      - \mathbin{\bar T_{00}'} 
      - \H ( \mathbin{\bar T^{ }_{00}}
        +  \mathbin{\bar T^{ }_{k k}} )
 & 
 (\nu = 0 ) \\[2mm]
      + \mathbin{\bar T^{ }_{i k,k}} 
 & 
 (\nu = i )
 \end{array} \right.
 \nn[2mm]
%%%%%%
 & \underset{\rmii{\nr{bg_Tmunu_varphi}}}
   {\overset{\rmii{\nr{delta_Tmunu_fluid}} \lift }{=}} & 
 \left\{ 
 \begin{array}{ll}
      - a^2_{ } 
      \bigl[ \bar e\bit' + 3 \H ( \bar e + \bar p ) \big]
      - \bar\varphi\hspace*{0.3mm}{}'\bigl( \bar\varphi\hspace*{0.3mm}{}''
          + 2 \H \bar\varphi\hspace*{0.3mm}{}'\bigr)
 & 
 (\nu = 0 ) \\[2mm]
      0 
 & 
 (\nu = i )
 \end{array} \right.
 \;.
\ea
The $\nu = 0$ part reproduces \eq\nr{bg_Tmunu}.

At first order, \eq\eqref{Tmunu;mu} implies
\ba
 0 
 \; = \; 
 \delta \mathbin{T^{ }_{\mu\nu}}^{;\mu}_{ }
 & \equiv &  
 \bigl( \mathbin{\bar T^{ }_{\mu\nu,\alpha}} 
      - \mathbin{\bar T^{ }_{\mu\beta}}{\bar\Gamma^{\beta}_{\nu\alpha}}
      - \mathbin{\bar T^{ }_{\beta\nu}}{\bar\Gamma^{\beta}_{\mu\alpha}}
 \bigr)
 \delta g^{\alpha\mu}_{ }
 \label{eq_cont-p} \\[3mm]
%%%%%%%
 &  & \;+\,  
 \bigl( \mathbin{\delta T^{ }_{\mu\nu,\alpha}} 
      - \mathbin{\bar T^{ }_{\mu\beta}}\,{\delta \Gamma^{\beta}_{\nu\alpha}}
      - \mathbin{\delta T^{ }_{\mu\beta}}\,{\bar \Gamma^{\beta}_{\nu\alpha}}
      - \mathbin{\bar T^{ }_{\beta\nu}}\,{\delta \Gamma^{\beta}_{\mu\alpha}}
      - \mathbin{\delta T^{ }_{\beta\nu}}\,{\bar \Gamma^{\beta}_{\mu\alpha}}
 \bigr)
 \bar g^{\alpha\mu}_{ }
 \;. \hspace*{8mm} \nonumber
\ea
In order to simplify the expressions, we focus on the 
fluid energy-momentum tensor from 
\eqs\nr{delta_Tmunu_fluid} and \nr{delta_Tmunu_aniso}, 
omitting the scalar
field contribution from \eqs\nr{delta_T00_varphi}--\nr{delta_Tij_varphi}. 
Then, making use of symbolic manipulation, 
as explained in \app\ref{app:symbolic}, 
or carrying out an explicit gauge-fixed computation, 
as shown in \app\ref{app:newton2}, we find
\ba
 0 \;=\; \delta \mathbin{T^{ }_{\mu 0}}^{;\mu}_{ }
 & \underset{\rmii{\nr{delta_Tmunu_aniso}}}
   {\overset{\rmii{\nr{delta_Tmunu_fluid}}}{\supset}} & 
 - \delta e' - 3 \H (\delta e + \delta p)
 + (\bar{e} + \bar{p}) (\, 3 h_\rmii{D}' - v^{ }_{k,k}
 - \vartheta^{\prime}_{kk} \,)
 - \H\, \barpPi^{ }_{kk} 
 % \label{delta_0_Tmunu;mu_full} 
 \nn[1mm]
%%%%%
 & \overset{\rmii{\nr{eq_metric-pert2}--\nr{eq_metric-pert3}}
           \lift }{=} & 
 - \delta e' - 3 \H (\delta e + \delta p)
 + (\bar{e} + \bar{p}) (\, 3 h_\rmii{D}' + \nabla^2_{ } v \,)
 % - \H\, \tr\barpPi 
 \;,  \hspace*{6mm} \label{delta_0_Tmunu;mu} \\[5mm]
%%%%%%%%%%
 0 \;=\; \delta \mathbin{T^{ }_{\mu i}}^{;\mu}_{ }
 & \underset{\rmii{\nr{delta_Tmunu_aniso}}}
   {\overset{\rmii{\nr{delta_Tmunu_fluid}}}{\supset}} & 
 + \delta p^{ }_{,i} + (\bar{e} + \bar{p}) h^{ }_{0,i}
 + (\partial^{ }_\tau + 4\H) [(\bar{e} + \bar{p})(v^{ }_i - h^{ }_i)]
 + \barpPi^{ }_{ik,k}
 \;.  \hspace*{6mm} \nn  \label{delta_i_Tmunu;mu}
\ea
These will play an important role in \ch\ref{se:inside}. 
Eq.~\nr{delta_0_Tmunu;mu} is an evolution equation for 
energy-density perturbations, $\delta e$,
whereas \eq\nr{delta_i_Tmunu;mu} 
is an evolution equation for momentum-density perturbations, 
$(\bar e + \bar p)(v^{ }_i - h^{ }_i)$. 
The full equations, including inflaton perturbations, 
will be needed
in \se\ref{ss:delta_R_thermal}
(cf.\ \eqs\nr{full_delta_0_Tmunu;mu} and 
\nr{full_delta_i_Tmunu;mu}), and are therefore part of 
the computer-algebraic derivation presented 
in \app\ref{app:symbolic}.

We note that \eq\nr{delta_0_Tmunu;mu} 
involves only scalar perturbations~(s). 
In contrast, \eq\nr{delta_i_Tmunu;mu} has both
scalar and vector parts, and can be split into two independent
relations at linear order, after inserting the decompositions from 
\eqs\nr{eq_v-v-s}--\nr{Pi_ij_t}, 
\ba
 \delta p + (\bar{e} + \bar{p}) h^{ }_0
 + (\partial^{ }_\tau + 4\H) [(\bar{e} + \bar{p})(h - v)]
 + \frac{2}{3} \nabla^2_{ }\barpPi 
 & 
 \underset{\rmii{\nr{Pi_ij_s}}}{
 \overset{\rmii{\nr{delta_i_Tmunu;mu}} \lift }{=}} 
 &
 0 
 \;,  
 \label{delta_i_Tmunu;mu_s}
 \\[2mm]
%%%%%%% 
 (\partial^{ }_\tau + 4\H) [(\bar{e} + \bar{p})(v^\rmi{v}_i - h^\rmi{v}_i)]  
 - \frac{1}{2} \nabla^2 \barpPi^\rmi{v}_i 
 & 
 \underset{\rmii{\nr{Pi_ij_v}}}{
 \overset{\rmii{\nr{delta_i_Tmunu;mu}} \lift }{=}} 
 &
 0
 \;. 
 \label{delta_i_Tmunu;mu_v}
\ea
As far as the vector part in \eq\nr{delta_i_Tmunu;mu_v} goes, 
it can also be obtained 
by operating with $\partial^{ }_\tau + 2 \H$ on 
\eq\nr{einstein_0i_v}, and eliminating then the left-hand
side with the help of \eq\nr{einstein_ij_v_plus}.
In analogy with \eq\nr{soln_v_homog_1}, 
\eq\nr{delta_i_Tmunu;mu_v} implies
a rapid decay of $ (\bar{e} + \bar{p})(v^\rmi{v}_i - h^\rmi{v}_i) $
at every spatial position, apart from the equilibrium
fluctuations generated by anisotropic stress. 

With these relations, we have completed the set of equations needed
for determining the evolution of linear perturbations in the early universe.
However, as mentioned below \eq\nr{delta_einstein_comb},
at first sight there are more variables than equations. 
This is related to the possibility to carry out coordinate 
transformations, and we will turn to this in \ch\ref{se:gauges}.

\newpage

%%%%%%%%%%%%%%%%%%%%%%%%%%%% start appendices %%%%%%%%%%%%%%%%%%%%%%%%%%%%%%%

%%%%%%%%%%%%%%%%%%%%%%%%%%%%%%%%%%%%%%%%%%%%%%%%%%%%%%%%%%%%%%%%%
%
\subsubsection{Viscous corrections to the energy-momentum tensor}
\label{app:viscous}

\addcontentsline{toc}{subsection}{\App\ref{app:viscous}: 
Viscous corrections to the energy-momentum tensor}

\index{viscous corrections}

In \eq\nr{delta_Tmunu_aniso}, 
we have added an {\em anisotropic stress} \index{anisotropic stress}
part to perturbations of the energy-momentum tensor. 
However, 
given that it only appears for spatial indices, it is 
not covariant. In this appendix, we explain how anisotropic stress arises
covariantly as a part of viscous corrections to ideal hydrodynamics.
Let us mention in passing that it is also possible to define anisotropic
stress within the framework of Boltzmann equations, 
which is valid beyond thermal equilibrium. 
% but only to leading order in weak interactions. 

\index{constitutive relations for $T^{ }_{\mu\nu}$} 

Hydrodynamics can be viewed as an effective theory, 
originating as we express $T^{ }_{\mu\nu}$ through 
{\em constitutive relations}, 
involving an expansion in gradients. 
The leading term, 
the perfect-fluid expression in \eq\nr{eq_Tmunu_pf_again}, 
is of zeroth order in gradients. Terms of the first order in 
gradients are characterized by two new 
functions, 
the {\em shear viscosity}, \index{shear viscosity}
$\eta$, and 
\index{$\eta$ (shear viscosity)}
the {\em bulk viscosity}, \index{bulk viscosity}
$\zeta$. 
\index{$\zeta$ (bulk viscosity)}
Like in \eq\nr{varphi_eq}, when we
introduce dissipative coefficients, the principle of detailed
balance requires that 
we also introduce 
{\em hydrodynamic fluctuations}~\cite{landau9}. 
\index{hydrodynamic fluctuations}
It is worth 
mentioning that very often the fluctuation part is omitted, 
but this is physically viable only if a macroscopic 
fluid motion sets the system far from equilibrium. 
For near-equilibrium fluids, the fluctuations play an essential role. 
Sometimes they can even be more interesting than dissipation, for instance
when they source something which would otherwise not exist and which
is not easily dissipated away, such as gravitational waves.

\index{Landau-Lifshitz convention}

When we include viscous corrections, so that fluid motions
become diffusive, the notion of a fluid velocity requires a precise
definition. In the so-called {\em Landau-Lifshitz convention}, we require
that $u^\mu_{ }u^\nu_{ }T^{ }_{\mu\nu} = e$, i.e.,\ that viscous 
corrections are orthogonal to $u^{\mu}_{ }$. Then the general
energy-momentum tensor, up to first order in gradients, can be
written as 
\ba
 T^{ }_{\mu\nu} & = & 
 p\, g^{ }_{\mu\nu} + (e + p)\, u^{ }_{\mu}u^{ }_{\nu}
 + a^2_{ }\Pi^{ }_{\mu\nu}
 % \; + \; \ord(\delta^2_{ }) 
 \;, \label{Tmunu_fluid_full} \\[2mm]
%%%%%%%
 a^2_{ }\Pi^{ }_{\mu\nu}
 & \equiv &
 - \eta\, {\projV^{ }_{\mu}}^{\rho}_{ }\,{\projV_{\nu}^{ }}^{\sigma}_{ }
 \Bigl(
  u^{ }_{\rho;\sigma} + u^{ }_{\sigma;\rho}
 - \frac{2 g^{ }_{\rho\sigma}}{3}  u^{\gamma}_{;\gamma}  
 \Bigr) 
% \nn
% & - &
 \; - \;  \zeta\, \projV^{ }_{\mu\nu} u^{\gamma}_{;\gamma}
 \; + \; a^2_{ }S^{ }_{\mu\nu}
% \nn 
% & + &
 \;, \hspace*{6mm} \label{Tmunu_hydro} 
 \index{energy-momentum tensor: viscosities}
 \\[2mm]
%%%%%%
 \projV^{ }_{\mu\nu} 
 & 
 \equiv
 &
 g^{ }_{\mu\nu} + u^{ }_{\mu} u^{ }_{\nu} 
 \;. \label{projV}
 \index{$\projV^{ }_{\mu\nu}$ (projector orthogonal to $u^{\mu}_{ }$)}
 \index{projector: orthogonal to four-velocity}
\ea
Here 
$
 \projV^{ }_{\mu\nu} 
$
is a {\em projector onto directions orthogonal} to $u^{\mu}_{ }$; 
a covariant velocity derivative reads 
$
 u^{ }_{\rho;\sigma} = u^{ }_{\rho,\sigma} 
 - 
 \Gamma^{\alpha}_{\rho\sigma} u^{ }_\alpha
$
(cf.\ \eq\nr{cov_der_2});
and 
$
 u^{\gamma}_{;\gamma} =
 g^{\rho\sigma}_{ } u^{ }_{\rho;\sigma} 
$. 
If we contract
$\Pi^{ }_{\mu\nu}$ with $g^{\mu\nu}_{ }$, the term proportional 
to $\eta$ drops out, 
whereby it represents the traceless part of viscous corrections.
The trace part is represented by the term proportional to $\zeta$.

The term $S^{ }_{\mu\nu}$ in \eq\nr{Tmunu_hydro} 
is {\em not} a function of the four-velocity, 
but rather represents thermal fluctuations, 
similarly to $\varrho$ in \eq\nr{varphi_eq}. 
Whereas viscosities dissipate
energy from macroscopic fluid motion into overall thermal entropy, 
the noise returns energy into random fluid motions. 
In the context of hydrodynamics, we call these
{\em hydrodynamic fluctuations}. We return to them at the 
end of this section. 

We now expand the thermodynamic variables 
in small perturbations like before, 
\be
 e
  \;\overset{\rmii{\nr{delta_e_p}}}{=}\;
 \bar{e} + \delta e  
 \;, \qquad
 p 
   \;\overset{\rmii{\nr{delta_e_p}}}{=}\;
 \bar{p} + \delta p 
 \;, \qquad
 u_\mu^{ } 
  \;\overset{\rmii{\nr{eq_u_down}}}{=}\;
 \bar{u}_\mu^{ } + \delta u_\mu^{ } + \ord(\delta^2_{ })
 \;. \label{e_expansion}
\ee
The viscosities $\eta$ and $\zeta$ could also be expanded, 
but here we replace them by their average values; 
for simplicity we denote these 
by $\eta$ and $\zeta$, rather than 
$\bar\eta$ and $\bar\zeta$. In fact, 
the part of $\Pi^{ }_{\mu\nu}$ proportional to $\eta$
is of $\ord(\delta)$, cf.\ \eq\nr{delta_T}, 
so it is consistent to 
evaluate the coefficient at $\ord(\delta^0_{ })$.

Let us divide the whole 
(first-order-in-gradients) 
anisotropic stress into subparts as 
\be
 \Pi^{ }_{\mu\nu} 
 \quad 
 \overset{\rmii{\nr{Tmunu_hydro}}}{\equiv} 
 \; 
 \underbrace{
 \Sigma^{ }_{\mu\nu}
 }_{{\rm term~with}\,\eta}
 \; + \; 
 \underbrace{
 Z^{ }_{\mu\nu}
 }_{{\rm term~with}\,\zeta}
 \; + \; 
 \underbrace{
 S^{ }_{\mu\nu}
 }_{\rm fluctuations}
 \;. \label{Pi_splitup}
\ee
A somewhat tedious computation shows that 
at $\ord(\delta)$,
only the spatial part
of $\Sigma^{ }_{\mu\nu}$ is non-vanishing, 
\be
 \Sigma^{ }_{ij}
 \; 
 \underset{\rmii{\nr{Pi_splitup}}}{
 \overset{\rmii{\nr{Tmunu_hydro}} \lift }{=}} 
 \; 
 -\frac{\eta}{a}
 \biggl[\, 
   v^{ }_{i,j} + v^{ }_{j,i} + 2 \vartheta_{ij}' 
      - \frac{2}{3}\, \delta^{ }_{ij} 
     \bigl(\, \nabla\cdot\vec{v} + \tr \vartheta' \,\bigr)
 \,\biggr] 
 \;. \label{delta_T}
\ee
Actually, $\tr\vartheta = 0$, 
cf.\ \eqs\nr{vartheta_ij_s}--\nr{eq_metric-pert3}, 
but we have shown its would-be appearance, to make it explicit that 
\eq\nr{delta_T} is traceless. 

Let us carry out a tensor decomposition of \eq\nr{delta_T}. 
We can write 
\be
 \Sigma_{ij}^{ } \;=\; 
 \Sigma_{ij}^\rmi{s} + \Sigma_{ij}^\rmi{v} + \Sigma_{ij}^\rmi{t}
 \;. \label{T_ij_visc}
\ee
Following \eqs\nr{eq_tensor-ij}--\nr{eq_tensor-ij-x},
the scalar, vector, and tensor parts can be expressed as 
\begin{align}
 \Sigma_{ij}^\rmi{s}
  &\; = \; 
 \Bigl( \partial^{ }_i\, \partial^{ }_j - 
 \frac{1}{3} \delta^{ }_{ij} \nabla^2 \Bigr) \Sigma
 \;, 
 & 
 & 
 \\
%%%%%%
 \Sigma_{ij}^\rmi{v} 
 &\; = \;
 - \frac{1}{2} 
 \bigl(\,
   \partial^{ }_i \Sigma^\rmi{v}_j + \partial^{ }_j \Sigma^\rmi{v}_i  
 \,\bigr) 
 \;,
 & \hspace*{-2.0cm} 
 \partial^{i}_{ }\Sigma^\rmi{v}_i 
 &\; = \;
 0 
 \;, \\[2mm] 
%%%%%%
 \delta_{ij}^{ }\Sigma_{ij}^\rmi{t} 
 &\; = \;
 0 
 \;, 
 & \hspace*{-2.0cm}
 \partial^{i}_{ }\Sigma_{ij}^\rmi{t}
 &\; = \;
 0 
 \;, 
\end{align}
similarly to the decomposition of $\vartheta^{ }_{ij}$, 
cf.\ \eqs\nr{vartheta_ij_s}--\nr{eq_metric-pert3}.
If we resolve the velocity to curl-free and divergence-free parts, 
\be
 v^{ }_i 
  \;\overset{\rmii{\nr{eq_v-v-s}}}{=}\; 
 v^\rmi{s}_i + v^\rmi{v}_i 
 \;, \qquad
 v^\rmi{s}_i \;=\; - \partial^{ }_i v  
 \;, \qquad
 \partial^{i}_{ }v^\rmi{v}_i \;=\; 0 
 \;, \label{resolve_v}
\ee
then it follows from \eq\nr{delta_T} that 
\be
 \Sigma \; = \; \frac{2 \eta\, \bigl(\, v - \vartheta' \,\bigr)}{a} 
 \;, \quad
 \Sigma^\rmi{v}_i \; = \; 
 \frac{ 2 \eta\,
  \bigl(\, v^\rmi{v}_{i} - \vartheta^{\rmi{v}\prime}_i \,\bigr) }{a}
 \;, \quad
 \Sigma^\rmi{t}_{ij} \; = \; - 
 \frac{ 2 \eta \,\vartheta^{\rmi{t}\bit\prime}_{ij} }{a}
 \;. \label{decomposition}
\ee

To summarize, a non-vanishing shear viscosity induces anisotropic
stress, $\Sigma^{ }_{ij} \neq 0$. 
However, given that viscous terms represent a derivative 
correction (cf.\ \eq\nr{delta_T}), 
we would expect
this to be a small contribution in a large universe.
In fact, if the corrections were substantial, 
we should worry that the hydrodynamic 
expansion does not converge. We return to this issue 
more concretely in the context of bulk viscous corrections,
in the paragraph below \eq\nr{p_eff}, and then again with 
shear viscous corrections, below \eq\nr{einstein_ij_t_eta}. 

There are important qualitative lessons to 
be learned from \eq\nr{decomposition}. 
If we insert the vector part, 
$\Pi^\rmi{v}_i \supset \Sigma^\rmi{v}_i$, 
into \eq\nr{einstein_ij_v_plus}, 
or the tensor part, 
$\Pi^\rmi{t}_{ij} \supset \Sigma^\rmi{t}_{ij}$, into \eq\nr{einstein_ij_t}, 
we see that $\eta$ induces a {\em friction term},
adding to the ``Hubble friction'' already present. For the tensor
part, this is shown explicitly in \eq\nr{einstein_ij_t_eta}. 
This is similar to the friction~$\Upsilon$ that affects scalar
perturbations in \eq\nr{varphi_eq}.
According to the {\em fluctuation-dissipation relation}, the presence of 
dissipation implies the presence of fluctuations, which can then
also generate vector perturbations and 
gravitational waves. The fluctuation part originates
from $S^{ }_{\mu\nu}$, as discussed around 
\eqs\nr{exp_S} and \nr{noise_hydro}. 

\index{fluctuation-dissipation relation}

\vspace*{3mm}

Next, let us discuss the bulk viscous part from \eq\nr{Pi_splitup}, 
denoted by $Z^{ }_{\mu\nu}$. Despite the simpler appearance in 
\eq\nr{Tmunu_hydro}, the bulk viscous corrections are not simpler
than the shear viscous ones. 
After another tedious computation, we find
\ba
 Z^{ }_{\mu\nu} & = & 
 -\frac{3\zeta}{a}
  \,\left\{\;
 \left( 
   \begin{array}{cc}
       0 & 0 \\[1mm] 
       0 & \H\,\delta^{ }_{ij}
   \end{array}
 \right)
 \right.
 \nn[3mm]
%%%%%
 &  & \;+\,
 \left.
 \left( 
   \begin{array}{cc}
       0 & -\H\, v^{ }_j \\[2mm] 
    -\H\,v^{ }_i & 
      \quad
      2 \H\,\vartheta^{ }_{ij} 
    - \delta^{ }_{ij} 
     \Bigl[
       \H h^{ }_\rmii{0} + h_\rmiii{D}' + 2 \H h^{ }_\rmiii{D}
       - \frac{\nabla\cdot\vec{v}}{3} 
     \Bigr]
   \end{array}
 \right) 
 \;\right\}
 \;. \hspace*{1.0cm} \label{delta_Z}
\ea
The key point, from the first line of \eq\nr{delta_Z}, 
is that a correction arises even at the background level. 
Comparing with \eq\nr{delta_Tmunu_fluid}, and going to 
physical time for a simpler interpretation, 
we see that the background equations now contain an 
{\em effective pressure},
\index{effective pressure}
\be
 \bar{p}^{ }_\rmi{eff} 
 \; \equiv \; 
 \bar{p} - 3\hspace*{0.3mm} \zeta H
 \;. \label{p_eff}
\ee

In order to elaborate on the meaning of \eq\nr{p_eff}, we recall from 
\se\ref{ss:history}, and for $\zeta$ from ref.~\cite{adm} that, 
up to slowly evolving coefficients, 
the three quantities approximate to  
\be
 \bar{p} \;\sim\; T^4_{ } \;, \quad
 H \;\sim\; \frac{T^2_{ }}{\mpl^{ }} \;, \quad
 \zeta 
 \;
 \overset{\rmii{YM}}{\sim}
 \; \frac{ \alpha^2_{ } T^3_{ } }{ \ln(\alpha^{-1}_{ }) }
 \;. 
 \label{bulk_estimates}
\ee
Here we have adopted a bulk viscosity relevant for 
a weakly-coupled Yang-Mills (YM)
plasma, such as the one made of Standard Model particles, 
where $\alpha = g^2_{ }/(4\pi) < 1$ 
is a fine-structure constant related to a gauge group with 
gauge coupling $g$.

We thus see that the corrections from $\zeta$ to 
$
  \bar{p}^{ }_\rmi{eff} 
$
are small as long as  
$
 T < \alpha^{-2}_{ }\ln(\alpha^{-1}_{ }) \mpl^{ }
$. 
This is well satisfied in normal inflationary scenarios; 
in fact, in the regime in which the inequality is violated,  
general relativity
perhaps ceases to be valid as an effective theory. And in any
case, if the correction is as large as the leading term, the 
hydrodynamic expansion breaks down. We thus see that
under reasonable assumptions, bulk viscosity should {\em not}
modify the background expansion significantly
(nevertheless, there is literature
where such a possibility is entertained, 
cf.,\ e.g.,\ ref.~\cite{bulk_lit}). 

\vspace*{3mm}

Let us then turn to the fluctuation part of $\Pi^{ }_{\mu\nu}$, 
denoted by $S^{ }_{\mu\nu}$. Unfortunately, 
because of the tensor structure of the energy-momentum tensor,
it is a non-trivial task to work out the {\em noise autocorrelator}.
\index{noise autocorrelator}
Even if this is a text-book topic~\cite{landau9}, 
for relativistic plasmas the general framework has
been intensely investigated only relatively recently~\cite{kapusta}.  
Here we just state that the result has a close 
analogy with \eq\nr{varphi_noise}, with the value of
the noise autocorrelator
being related to the viscous coefficients through a fluctuation-dissipation
relation similar to that which 
we will work out in \eq\nr{fl-di}, 
\ba
 \bigl\langle 
  S_{\mu\nu}^{ }(\X)
 \bigr\rangle
 & = & 
 0
 \;, \label{exp_S} \\[3mm]
%%%%%%%
 \bigl\langle 
  a^2_{ }S_{\mu\nu}^{ }(\X)\,
  a^2_{ }S_{\rho\sigma}^{ }(\Y)
 \bigr\rangle
 & \approx & 2 T 
 \biggl[
  \eta \, \bigl( \projV_{\mu\rho}^{ } \projV_{\nu\sigma}^{ }
  + \projV_{\mu\sigma}^{ } \projV_{\nu\rho}^{ } \bigr)
 \nn[2mm]
%%%% 
 &  & \;+\,
    \Bigl( \zeta - \frac{2\eta}{3} \Bigr)
    \projV_{\mu\nu}^{ } \projV_{\rho\sigma}^{ }
 \biggr] \frac{\delta^{(4)}_{ }(\mathcal{X-Y})}
 {\sqrt{- \det g^{ }_{\mu\nu}}}
 \;. 
 \label{noise_hydro} 
\ea
Furthermore, up to first order in perturbations, 
\be
 \projV^{ }_{\mu\nu}
 \; 
  \underset{\rmii{\nr{g_munu},\nr{eq_u_down}}}
  {\overset{\rmii{\nr{projV}} \lift }{=}}
 \; 
 a^2_{ }\, 
 \left( 
   \begin{array}{cc} 
     0 & - v^{ }_j \\[2mm] 
    -v^{ }_i & 
    (1-2h^{ }_\rmiii{D})\, \delta^{ }_{ij} + 2 \vartheta^{ }_{ij} 
   \end{array}
 \right)
 + \ord\bigl(\delta^2_{ }\bigr)
 \;. \label{Delta_munu}
\ee
Given that on the average the noise is small, cf.\ \eq\nr{exp_S}, 
the noise is treated as being of similar magnitude as fluctuations, 
$S^{ }_{\mu\nu} \sim \ord(\delta)$. Therefore, in the noise 
autocorrelator, we can employ the leading part of \eq\nr{Delta_munu},
$a^2_{ }\delta^{ }_{ij}$.
This then implies that only the spatial indices carry a non-vanishing
autocorrelator. 
Going simultaneously to momentum space in spatial directions, 
the only non-zero components read
\ba
 \langle\, S_{ij}^{ }(\tau^{ }_1,\vec{k})\, 
           S_{kl}^{ }(\tau^{ }_2,\vec{q}) \,\rangle
 \hspace*{-3mm}
 &
 \underset{\rmii{\nr{fourier_k},\nr{eq_metric-pert}}}{
 \overset{\raise0.5ex\hbox{\rmii{\nr{noise_hydro},\nr{Delta_munu}}}}
 {\approx}}
 &  
 2 T \, \delta(\tau^{ }_1 - \tau^{ }_2)
 \,  (2\pi)^3_{ }\delta^{(3)}_{ }(\vec{k+q}) \,
 \label{delta_S}
 \\[2mm]
%%%%%%%
 \hspace*{-3mm}
 &  &
 \hspace*{-3mm}
 \; \times \,   
 \frac{1}{a^4_{ }}
 \biggl[
 \eta \, \bigl( 
                 \delta^{ }_{ik} \delta^{ }_{jl}
               + \delta^{ }_{il} \delta^{ }_{jk}
         \bigr)
 +       \biggl( 
                 \zeta - \frac{2\eta}{3}
         \biggr) \, 
                 \delta^{ }_{ij} \delta^{ }_{kl} 
 \biggr] 
 + \ord\bigl(\delta^3_{ }\bigr)
 \;. \hspace*{4mm} \nonumber
\ea

An important observation can be made, 
if we contract \eq\nr{delta_S} with 
the traceless transverse projector $\mathbbm{T}^{mn}_{ij}$
from \eq\nr{T_ijmn}. Because of the tracelessness of the projector, 
the part proportional to 
$
 \zeta - {2\eta}/{3}
$
drops out. But the terms proportional to~$\eta$
remain present. Therefore, we can say that the noise has a 
projection onto tensor modes; we may denote this projection by
$
 S^\rmi{t}_{ij} \equiv \mathbbm{T}^{mn}_{ij} S^{ }_{mn}
$.
With this, we may convert the gravitational-wave equation, 
\eq\nr{einstein_ij_t}, 
including the information from \eq\nr{decomposition}, into 
\be
  \boxed{ 
  \quad
  \bigl[\,
   \partial_\tau^2 + 
  \bigl(\, 2 \H
          + 16 \pi G a \eta 
  \,\bigr) \partial_\tau^{ } - \nabla^2_{ }
  \,\bigr]
   \vartheta^\rmi{t}_{ij}
 \; = \; 
 8\pi G a^2_{ } S^\rmi{t}_{ij} 
 \;.  \quad \vphantom{\frac{T'}{q}}
 }
 \label{einstein_ij_t_eta}
 \index{gravitational waves: hydrodynamic regime}
 \index{Einstein equations: tensor perturbations} 
\ee
This shows how 
{\em gravitational waves are dissipated
and produced in the hydrodynamic regime}.
Eq.~\nr{einstein_ij_t_eta} 
represents an analogue of the scalar 
equation \nr{eq_field-eq-pert}, being however 
simpler, because tensor modes do not couple to other perturbations
at linear order.

Let us estimate how important the friction 
in \eq\nr{einstein_ij_t_eta}
could be. Going to physical time, like in \eq\nr{bulk_estimates}, 
the magnitudes of the two frictions
in \eq\nr{einstein_ij_t_eta} are 
\be
 H \;\sim\; \frac{T^2_{ }}{\mpl^{ }}
 \;, \quad
 G\eta 
 \;
 \overset{\rmii{YM}}{\sim}
 \; \frac{T^3_{ }}
 {\mpl^2 \, \alpha^2_{ } \ln(\alpha^{-1}_{ }) }
 \;, 
 \label{shear_estimate}
\ee
where we have adopted the shear viscosity relevant for 
a weakly-coupled Yang-Mills plasma~\cite{amy}. We see that the
correction from $\eta$ is small when
$ 
 T < \alpha^2_{ }\ln(\alpha^{-1}_{ })\mpl^{ }
$.
This is more easily violated than the corresponding
inequality for $\zeta$, nevertheless in normal 
scenarios the viscous corrections should be small
(as for $\alpha$, typical values for Standard Model like theories
at high temperatures are $\alpha\sim 0.01$). 
In contrast, the anisotropic stress induced by decoupled
neutrinos at low temperatures can have a substantial influence
on low-momentum tensor perturbations, as will be discussed
in \se\ref{ss:gw_transfer}.

Apart from the dissipation that $\eta$ induces on the left-hand side, 
the right-hand side of \eq\nr{einstein_ij_t_eta} implies that $\eta$ 
influences the 
production of gravitational waves, from hydrodynamic fluctuations. 
We return to this topic in \se\ref{ss:gw_Tij}
and around \eq\nr{hydro_fluct_5}.

\vspace*{3mm}

We end by remarking that relativistic hydrodynamics
including first-order viscous corrections is said to be
acausal (with the possibility of superluminal modes) and 
unstable (possessing exponentially growing solutions)
(cf.,\ e.g.,\ ref.~\cite{hi-li}). However,  
these problems arise in the domain of large momenta, 
where the gradient expansion breaks down anyways, 
so there is no real reason for concern 
(moreover, hydrodynamic fluctuations 
are omitted from these considerations).
Nevertheless, there are phenomenological recipes to rectify
the issues by incorporating a subset of second-order
corrections beyond ideal hydrodynamics, notably via the so-called 
{\em Mueller-Israel-Stewart theory}~\cite{mue,is-st}. 
\index{Mueller-Israel-Stewart theory}

\newpage

%%%%%%%%%%%%%%%%%%%%%%%%%%%%%%%%%%%%%%%%%%%%%%%%%%%%%%%%%%%%%%%%%
%
\subsubsection{Deriving first-order perturbations with computer algebra}
\label{app:symbolic}

\addcontentsline{toc}{subsection}{\App\ref{app:symbolic}: 
Deriving first-order perturbations with computer algebra}

\index{computer-algebraic methods}
\index{symbolic manipulation}

We show here how the results presented in this chapter
can be obtained with rudimentary symbolic manipulation.
As examples, we evaluate the Ricci scalar from \eq\nr{ricci_scalar_full}, 
and the energy-momentum conservation relations, from 
\eqs\nr{delta_0_Tmunu;mu} and \nr{delta_i_Tmunu;mu}, 
and in more complete form from 
\eqs\nr{full_delta_0_Tmunu;mu} and \nr{full_delta_i_Tmunu;mu}. 
The same methods may be extended to higher orders in perturbations, 
as will be demonstrated 
in \app\ref{app:sigw}. 

%%%%%%%%%%%%%%%%%%%%%%%%%%%%%%%%%%%%%%%%%%%%%%%%%%%%%%%%%%%

The logic of the code is as follows. The metric and its inverse are
inserted from \eqs\nr{g_munu} and \nr{g^munu}. A parameter ``{\tt eps}''
counts the order in $\delta$.
The Christoffel
symbols are computed from \eq\nr{eq_ChSy-a}, the Ricci tensor from 
\eq\nr{eq_rt-a}, and the Ricci scalar from \eq\nr{bg_ricci_scalar}.
The result is expanded to first order in ``{\tt eps}''.
A usual problem with symbolic manipulation is that the output 
does not automatically look simple to the human eye. A key step is 
thus the final one, where we introduce a tentative answer, 
``{\tt testricciscalar}''; insert there ingredients that can 
be identified from the output; re-evaluate the output after
their subtraction, whereby it should now be more
transparent; and keep on
adding terms to the tentative answer, until all structures have been
identified. Of course, this works nicely only if the code is fast, 
which is not the case with the current {\tt sympy} release, however this
will likely improve in the future. Otherwise, the code can 
straightforwardly be transcribed to other 
symbolic languages, such as {\tt Mathematica}. 
All in all, the script looks as follows: 

\index{code: algebra for $R$ (Ricci scalar)}

{\fontsize{8pt}{10pt}\selectfont
\begin{verbatim}
# computer-algebraic determination of the Ricci scalar [symbolic_ricci_scalar.py]
#
# basic coordinates and functions appearing
from sympy import *
i, j, k, l = symbols('i j k l',cls=Idx); x = IndexedBase('x'); eps = symbols('eps')
a = Function('a')(x[0]); H = Function('H')(x[0])
h1 = Function('h1')(x[0],x[1],x[2],x[3]); h2 = Function('h2')(x[0],x[1],x[2],x[3])
h3 = Function('h3')(x[0],x[1],x[2],x[3])
h0 = Function('h0')(x[0],x[1],x[2],x[3]); hD = Function('hD')(x[0],x[1],x[2],x[3])
theta11 = Function('theta11')(x[0],x[1],x[2],x[3]); theta22 = Function('theta22')(x[0],x[1],x[2],x[3])
theta33 = Function('theta33')(x[0],x[1],x[2],x[3]); theta12 = Function('theta12')(x[0],x[1],x[2],x[3])
theta13 = Function('theta13')(x[0],x[1],x[2],x[3]); theta23 = Function('theta23')(x[0],x[1],x[2],x[3])

# partial derivative with respect to x[i]
def d(h,i): return diff(h,x[i],evaluate=True)
def dd(h,i,j): return diff(h,x[i],x[j],evaluate=True)            # doesn't symmetrize as should :(
def dds(h,i,j): return (dd(h,i,j)+dd(h,j,i))/2                   # symmetrized
def ddd(h,i,j,k): return diff(h,x[i],x[j],x[k],evaluate=True)    # doesn't symmetrize as should :(
def ddds(h,i,j,k): return (ddd(h,i,j,k)+ddd(h,i,k,j))/2          # symmetrized in last two arguments

# restrict to order(eps) for possibly more efficient execution
def trunc1(h): return (h.expand()).subs([(eps**5,0),(eps**4,0),(eps**3,0),(eps**2,0)])

# input metric with both indices down
a00 = (-1-2*eps*h0)
a01 = (-eps*h1);  a02 = (-eps*h2); a03 = (-eps*h3)
a10 = a01; a20 = a02; a30 = a03
a11 = (1+2*eps*(theta11-hD)); a22 = (1+2*eps*(theta22-hD)); a33 = (1+2*eps*(theta33-hD)); 
a12 = (2*eps*theta12); a13 = (2*eps*theta13); a23 = (2*eps*theta23); 
a21 = a12; a31 = a13; a32 = a23
gdown=a*a*Matrix([[a00,a01,a02,a03],[a10,a11,a12,a13],[a20,a21,a22,a23],[a30,a31,a32,a33]])

# input metric with both indices up
b00 = (-1+2*eps*h0)
b01 = (-eps*h1); b02 = (-eps*h2); b03 = (-eps*h3) 
b10 = b01; b20 = b02; b30 = b03
b11 = (1+2*eps*(hD-theta11)); b22 = (1+2*eps*(hD-theta22)); b33 = (1+2*eps*(hD-theta33));
b12 = (-2*eps*theta12); b13 = (-2*eps*theta13); b23 = (-2*eps*theta23) 
b21 = b12; b31 = b13; b32 = b23
gup=1/a/a*Matrix([[b00,b01,b02,b03],[b10,b11,b12,b13],[b20,b21,b22,b23],[b30,b31,b32,b33]])

# check inversion
test=(gdown*gup);  print(trunc1(test));  print("inversion checked")

# christoffel symbols
def dgdown(m,n,i): return d(gdown[m,n],i)
def gamma(r,m,n):
    return Rational(1/2)*Sum( gup[r,i]*(dgdown(i,m,n)+dgdown(i,n,m)-dgdown(m,n,i)),(i,0,3)).doit()

# ricci tensor with both indices down
def prericcidown(m,n):
    return ( Sum( d(gamma(k,m,n),k) - d(gamma(k,m,k),n),(k,0,3) ).doit() +
        Sum( gamma(l,m,n)*gamma(k,l,k) - gamma(l,m,k)*gamma(k,n,l),(k,0,3),(l,0,3) ).doit() )
riccidown=trunc1(Matrix(4,4,prericcidown))

# ricci tensor with indices up and down
ricciupdown = trunc1(gup*riccidown)

# ricci scalar
ricciscalar = ricciupdown.trace()

# go over to hubble rate
ricciscalar = ricciscalar.subs([(Derivative(a,x[0]),a*H),(Derivative(a*H,x[0]),a*(d(H,0)+H**2))
                                ,(Derivative(a,x[0],2),a*(d(H,0)+H**2))])

# adjust expression until the difference subtracts to zero 
testricciscalar = a**(-2)*(6*(d(H,0)+H**2) - 2*eps*(
    +3*H*d(h0,0)+6*(d(H,0)+H**2)*h0
    +dd(h0,1,1)+dd(h0,2,2)+dd(h0,3,3)-2*(dd(hD,1,1)+dd(hD,2,2)+dd(hD,3,3))
    +dd(theta11,1,1)+dd(theta22,1,1)+dd(theta33,1,1)
    +dd(theta11,2,2)+dd(theta22,2,2)+dd(theta33,2,2)
    +dd(theta11,3,3)+dd(theta22,3,3)+dd(theta33,3,3)
    +3*dd(hD,0,0)+9*H*d(hD,0) -(dds(h1,0,1)+dds(h2,0,2)+dds(h3,0,3))-3*H*(d(h1,1)+d(h2,2)+d(h3,3))
    -(dd(theta11,0,0)+dd(theta22,0,0)+dd(theta33,0,0))-3*H*(d(theta11,0)+d(theta22,0)+d(theta33,0))
    -(dd(theta11,1,1)+dd(theta22,2,2)+dd(theta33,3,3))
    -2*(dds(theta12,1,2)+dds(theta13,1,3)+dds(theta23,2,3))   )) 
print(expand(a**2*(ricciscalar - testricciscalar)))
print("ricci scalar checked at 1st order")

\end{verbatim}
}

Once all terms from \eq\nr{ricci_scalar_full} are
included in {\tt testricciscalar}, the last line evaluates to zero,  
indicating that
we have identified correctly the zeroth and 
first-order perturbations. Below, we likewise 
illustrate a code that reproduces 
\eqs\nr{delta_0_Tmunu;mu} and \nr{delta_i_Tmunu;mu}, 
starting from an energy-momentum tensor as given 
in \eqs\nr{delta_Tmunu_fluid} and \nr{delta_Tmunu_aniso}, 
or \eqs\nr{full_delta_0_Tmunu;mu} and \nr{full_delta_i_Tmunu;mu}, 
if the scalar field part from 
\eqs\nr{bg_Tmunu_varphi} 
and \nr{delta_T_varphi} is added.

\label{code:enmom}
\index{code: algebra for $ \mathbin{T^{ }_{\mu \nu}}^{;\mu}_{ } = 0$}

{\fontsize{8pt}{10pt}\selectfont
\begin{verbatim}
# computer-algebraic determination of energy-momentum conservation [symbolic_energy_mom_conservation.py]
#
# basic coordinates and functions appearing
from sympy import *
i, j, k, l, m = symbols('i j k l m',cls=Idx); x = IndexedBase('x'); eps = symbols('eps')
a = Function('a')(x[0]); H = Function('H')(x[0])
barphi = Function('barphi')(x[0]); bare = Function('bare')(x[0]); barp = Function('barp')(x[0]) 
h1 = Function('h1')(x[0],x[1],x[2],x[3]); h2 = Function('h2')(x[0],x[1],x[2],x[3])
h3 = Function('h3')(x[0],x[1],x[2],x[3])
h0 = Function('h0')(x[0],x[1],x[2],x[3]); hD = Function('hD')(x[0],x[1],x[2],x[3])
deltaphi = Function('deltaphi')(x[0],x[1],x[2],x[3])
deltae = Function('deltae')(x[0],x[1],x[2],x[3]); deltap = Function('deltap')(x[0],x[1],x[2],x[3]);
v1 = Function('v1')(x[0],x[1],x[2],x[3]); v2 = Function('v2')(x[0],x[1],x[2],x[3])
v3 = Function('v3')(x[0],x[1],x[2],x[3])
theta11 = Function('theta11')(x[0],x[1],x[2],x[3]); theta22 = Function('theta22')(x[0],x[1],x[2],x[3])
theta33 = Function('theta33')(x[0],x[1],x[2],x[3]); theta12 = Function('theta12')(x[0],x[1],x[2],x[3])
theta13 = Function('theta13')(x[0],x[1],x[2],x[3]); theta23 = Function('theta23')(x[0],x[1],x[2],x[3])
pi11 = Function('pi11')(x[0],x[1],x[2],x[3]); pi22 = Function('pi22')(x[0],x[1],x[2],x[3])
pi33 = Function('pi33')(x[0],x[1],x[2],x[3]); pi12 = Function('pi12')(x[0],x[1],x[2],x[3])
pi13 = Function('pi13')(x[0],x[1],x[2],x[3]); pi23 = Function('pi23')(x[0],x[1],x[2],x[3])

# partial derivative with respect to x[i]
def d(h,i): return diff(h,x[i],evaluate=True)
def dd(h,i,j): return diff(h,x[i],x[j],evaluate=True)            # doesn't symmetrize as should :(
def dds(h,i,j): return (dd(h,i,j)+dd(h,j,i))/2                   # symmetrized
def ddd(h,i,j,k): return diff(h,x[i],x[j],x[k],evaluate=True)    # doesn't symmetrize as should :(
def ddds(h,i,j,k): return (ddd(h,i,j,k)+ddd(h,i,k,j))/2          # symmetrized in last two arguments

# restrict to order(eps) for possibly more efficient execution
def trunc1(h): return (h.expand()).subs([(eps**5,0),(eps**4,0),(eps**3,0),(eps**2,0)])

# input metric with both indices down
a00 = (-1-2*eps*h0)
a01 = (-eps*h1);  a02 = (-eps*h2); a03 = (-eps*h3); a10 = a01; a20 = a02; a30 = a03
a11 = (1+2*eps*(theta11-hD)); a22 = (1+2*eps*(theta22-hD)); a33 = (1+2*eps*(theta33-hD)); 
a12 = (2*eps*theta12); a13 = (2*eps*theta13); a23 = (2*eps*theta23); a21 = a12; a31 = a13; a32 = a23
gdown=a*a*Matrix([[a00,a01,a02,a03],[a10,a11,a12,a13],[a20,a21,a22,a23],[a30,a31,a32,a33]])

# input metric with both indices up
b00 = (-1+2*eps*h0)
b01 = (-eps*h1); b02 = (-eps*h2); b03 = (-eps*h3); b10 = b01; b20 = b02; b30 = b03
b11 = (1+2*eps*(hD-theta11)); b22 = (1+2*eps*(hD-theta22)); b33 = (1+2*eps*(hD-theta33));
b12 = (-2*eps*theta12); b13 = (-2*eps*theta13); b23 = (-2*eps*theta23); b21 = b12; b31 = b13; b32 = b23
gup=1/a/a*Matrix([[b00,b01,b02,b03],[b10,b11,b12,b13],[b20,b21,b22,b23],[b30,b31,b32,b33]])

# christoffel symbols
def dgdown(m,n,i): return d(gdown[m,n],i)
def gamma(r,m,n):
    return Rational(1/2)*Sum( gup[r,i]*(dgdown(i,m,n)+dgdown(i,n,m)-dgdown(m,n,i)),(i,0,3)).doit()

# energy-momentum tensor for perfect fluid plus varphi
t00 = d(barphi,0)**2/2 + eps*d(barphi,0)*d(deltaphi,0) + a*a*(bare + eps*(deltae + 2*h0*bare))
t01 = eps*( d(barphi,0)*(d(deltaphi,1) - d(barphi,0)/2*h1) ) + a*a*eps*(bare*(h1-v1)-barp*v1) 
t02 = eps*( d(barphi,0)*(d(deltaphi,2) - d(barphi,0)/2*h2) ) + a*a*eps*(bare*(h2-v2)-barp*v2) 
t03 = eps*( d(barphi,0)*(d(deltaphi,3) - d(barphi,0)/2*h3) ) + a*a*eps*(bare*(h3-v3)-barp*v3) 
t10 = t01; t20 = t02; t30 = t03
t11 = d(barphi,0)*d(barphi,0)/2 + eps*( ((theta11-hD-h0)*d(barphi,0) + d(deltaphi,0))*d(barphi,0)
      ) +  a*a*( barp + eps*(deltap + 2*(theta11-hD)*barp + pi11) )
t22 = d(barphi,0)*d(barphi,0)/2 + eps*( ((theta22-hD-h0)*d(barphi,0) + d(deltaphi,0))*d(barphi,0)
      ) +  a*a*( barp + eps*(deltap + 2*(theta22-hD)*barp + pi22) )
t33 = d(barphi,0)*d(barphi,0)/2 + eps*( ((theta33-hD-h0)*d(barphi,0) + d(deltaphi,0))*d(barphi,0)
      ) +  a*a*( barp + eps*(deltap + 2*(theta33-hD)*barp + pi33) )
t12 = eps*theta12*d(barphi,0)*d(barphi,0) + a*a*eps*(2*theta12*barp + pi12)
t13 = eps*theta13*d(barphi,0)*d(barphi,0) + a*a*eps*(2*theta13*barp + pi13)
t23 = eps*theta23*d(barphi,0)*d(barphi,0) + a*a*eps*(2*theta23*barp + pi23)
t21 = t12; t31 = t13; t32 = t23
tmunu=Matrix([[t00,t01,t02,t03],[t10,t11,t12,t13],[t20,t21,t22,t23],[t30,t31,t32,t33]])

# energy-momentum conservation equation
def preconserve(n,r):                      # second is a dummy variable   
    return ( Sum( gup[m,k]*d(tmunu[m,n],k),(m,0,3),(k,0,3) ).doit() - 
             Sum( gup[m,k]*(tmunu[m,l]*gamma(l,n,k)+
                            tmunu[n,l]*gamma(l,m,k)),(m,0,3),(k,0,3),(l,0,3)).doit() )
conserve=trunc1(Matrix(2,1,preconserve))   # fill only 1st spatial component to save time

# go over to hubble rate
conserve = conserve.subs([(Derivative(a,x[0]),a*H),(Derivative(a*H,x[0]),a*(d(H,0)+H**2))
                          ,(Derivative(a,x[0],2),a*(d(H,0)+H**2))])

# test energy-momentum conservation
# component nu=0
testconserve0 = -(d(bare,0)+3*H*(bare+barp))-a**(-2)*d(barphi,0)*(dd(barphi,0,0)+2*H*d(barphi,0)
    )+eps*( +a**(-2)*(
        + d(barphi,0)*(dd(deltaphi,1,1)+dd(deltaphi,2,2)+dd(deltaphi,3,3)-dd(deltaphi,0,0))
        - (dd(barphi,0,0) + 4*H*d(barphi,0))*d(deltaphi,0)
        + d(barphi,0)**2*(d(h0,0)+3*d(hD,0)-d(h1,1)-d(h2,2)-d(h3,3)
                          -d(theta11,0)-d(theta22,0)-d(theta33,0))
        + 2*d(barphi,0)*(dd(barphi,0,0) + 2*H*d(barphi,0))*h0
        )
    -d(deltae,0)-3*H*(deltae+deltap)
    +(bare+barp)*(3*d(hD,0)-d(v1,1)-d(v2,2)-d(v3,3)
                    -d(theta11,0)-d(theta22,0)-d(theta33,0))
    -H*(pi11+pi22+pi33)
  )
print(expand(conserve[0]-testconserve0))
print("energy-momentum conservation tested: nu=0")

# component nu=1
testconserve1 = eps*(
    -a**(-2)*(dd(barphi,0,0)+2*H*d(barphi,0))*d(deltaphi,1)
    +d(deltap,1)+(bare+barp)*(d(h0,1)+d(v1,0)-d(h1,0)+4*H*(v1-h1))
    +(d(bare,0)+d(barp,0))*(v1-h1)+d(pi11,1)+d(pi12,2)+d(pi13,3)
    #### remnant of non-symmetrized derivatives
    +a**(-2)*d(barphi,0)*(dd(deltaphi,1,0)-dd(deltaphi,0,1))
    )
print(expand(conserve[1]-testconserve1))
print("energy-momentum conservation tested: nu=1")
\end{verbatim}
}

%%%%%%%%%%%%%%%%%%%%%%% end appendices %%%%%%%%%%%%%%%%%%%%%%%%%%%

%%%%%%%%%%%%%%%%%%%%%%%%% BIBLIO %%%%%%%%%%%%%%%%%%%%%%%%%%%%%%%%
%

%%%%%%%%%%%%%%%%%%%%%%%%%%%% SECTION %%%%%%%%%%%%%%%%%%%%%%%%%%%%%%%%%%
\newpage 

\section{Gauge transformations and different gauges}
\label{se:gauges}

\paragraph{Abstract:}

Coordinate covariance is a cornerstone of general relativity. 
At first order in small transformations, it introduces 
degeneracies in the identification of space-time points
in the homogeneous and isotropic description with the
corresponding locations in a perturbed universe. 
These degeneracies are referred to as gauge invariance. 
In the literature, many gauge choices are employed.
We show how the equations can be written in 
a manifestly gauge-invariant form; 
how this permits for powerful 
crosschecks of the computations; and how 
popular gauge choices can be recovered from the general equations. 

\paragraph{Keywords:} 

Gauge transformation, 
Newtonian gauge,
zero-shear gauge,  
longitudinal gauge, 
Poisson gauge, 
comoving gauge, 
spatially flat gauge, 
uniform curvature gauge,
synchronous gauge, 
gauge-invariant observables, 
Bardeen potentials, 
curvature perturbations, 
Einstein tensor and Einstein equations in Newtonian gauge.

%%%%%%%%%%%%%%%%%%%%%%%%%%%%%%%%%%%%%%%%%%%%%%%%%%%
%
\subsection{Definition of gauge transformations}
\label{ss:def_gauge}

\index{gauge transformation}

An intrinsic property of general relativity is the absence 
of a preferred frame. 
Yet, its manifestation within the perturbative approach is non-trivial.
In particular, even if the universe
were exactly homogeneous and isotropic in one coordinate system, we 
could make a small coordinate transformation to a new frame, and in 
the latter our variables would no longer be exactly 
homogeneous (i.e.\ independent of~$\vec{x}$)
and isotropic (i.e.\ independent of an observation direction~$\vec{n}$).
Physical quantities should be identified as 
those having an unambiguous coordinate-independent
meaning. Such quantities are said to be {\em gauge invariant} 
or {\em gauge independent}. 

\index{gauge dependence in general relativity}
\index{active and passive transformations}

We note that when considering transformations, there are two 
alternative points of view. If we talk about 
\textit{active} transformations, the values of physical 
quantities are changed, but we stay at the same coordinate point. 
Instead, in the \textit{passive} picture, which we adopt here, 
physics remains unchanged, but coordinates are transformed, 
leading to a relabelling. 
In the framework of cosmological perturbation theory,
both points of view are valid~\cite{Malik:2008yp}. 

To define the transformations, we envisage that a space-time point $q$, 
or ``event'',
is described in one coordinate system with the coordinates $x^\alpha(q)$.
Then we relabel the same point with the new 
coordinates $\tilde x^\alpha(q)$, such that 
\begin{equation}\label{eq_gaugetr}
 \tilde{x}^\alpha(q) \; = \;   x^\alpha(q) + \xi^\alpha(q) + \ord(\delta^2_{ })
 \ .
\end{equation}
Denoting derivatives with respect to the original system
by $(...)^{ }_{,\alpha} \equiv \partial(...)/\partial x^\alpha_{ }$, 
the transformation
properties of space-time tensors are dictated by 
\begin{equation}
 \Xi^{\tilde{\mu}}_\rho  \;\equiv\; 
   \frac{\partial \tilde{x}^\mu}{\partial x^\rho} 
                       \; = \;   \delta^\mu_\rho + \xi^\mu_{,\rho}
                                + \ord(\delta^2_{ })
  \ ,  \qquad 
 \Xi^\mu_{\tilde{\rho}} \;\equiv\;    
  \frac{\partial x^\mu}{\partial \tilde{x}^\rho} 
                       \; = \;   \delta^\mu_\rho - \xi^\mu_{,\rho}
                                + \ord(\delta^2_{ })
  \ . \label{Xi}
\end{equation}
We may also say that if we consider equivalent values of the 
coordinates, they correspond to different space-time points, 
\begin{equation}\label{eq_coor_transf}
 \tilde{x}^\alpha(\tilde{q})   \;=\;       x^\alpha(q)  
 \qquad \overset{\rmii{\nr{eq_gaugetr}}\vphantom{\big | }}{\Rightarrow} \quad 
 x^\alpha(\tilde{q})           \;=\;    x^\alpha(q) - \xi^\alpha(q)
  + \ord(\delta^2_{ }) \ .
\end{equation}
In the language of the discussion around \eq\nr{eq_t}, 
$\xi^0$ changes the slicing, 
$\xi^i$ the threading of the perturbed universe. 
The non-uniqueness of the coordinate choice leads to 
redundancies in the physical description, 
which can then be interpreted as gauge freedom. 

Let us now consider some general function, 
$Q\equiv Q(\{x^\alpha_{ }\})$. 
We denote its value 
% at the space-time point $\tilde q$ 
in the coordinate system $\{\tilde x^\alpha_{ } \}$
by 
% $\tilde{Q}\equiv\tilde{Q}(\tilde{q})$. 
$\tilde{Q} \equiv \tilde{Q}(\{\tilde x^\alpha_{ }\})$.
The function can be expanded
to first order in perturbations, $Q = \bar Q + \delta Q$ and 
$\tilde Q = \bar{\tilde{Q}} + \delta \tilde Q$. 
If $\tilde q \neq q$, the values and perturbations are not the same. 
However, with the help of \eq\nr{eq_coor_transf}, we can relate
them to each other. The relations depend on $\xi^\alpha_{ }$, and 
since choices of $\xi^\alpha_{ }$ 
represent redundancies in our description, 
we call the relations gauge transformations. 
Below we work out these transformations for 
arbitrary scalar, vector, and tensor quantities. 

\vspace*{3mm}

Starting with scalar quantities
(cf.\ \eq\nr{eq_scalar-dec}), by definition they 
have the same value at the same
point in every coordinate system, 
i.e.\ 
$
 \tilde A(\{\tilde x^\alpha (\tilde q)\}) 
     = 
        A(\{ x^\alpha (\tilde q)\})
$. 
Inserting \eq\nr{eq_coor_transf} on the right-hand side of this
equality, recalling that the background only depends on time, 
and equating the results order by order in perturbations, we find
\begin{equation}
 \tilde{A}                        
 \; = \;
 \bar{\tilde{A}} + \delta\tilde{A} 
 \; 
 \overset{\rmii{\eqref{eq_coor_transf}}}{=}  
 \; 
  \bar{A} - (\partial_\alpha\bar{A})\xi^\alpha
  + \delta A + \ord(\delta^2_{ })
 \quad \Rightarrow \quad  
 \boxed{ 
 \quad
 \delta \tilde{A}      \;\approx\; \delta A - \bar{A}'\xi^0 
 \;.
 \quad   \vphantom{\Big|}
 }
 \label{eq_gauge-scalar}
\end{equation}

Let us pause to stress an important point. Normally, for instance when
we are talking about representations of compact groups, a ``scalar'' 
quantity is one which is invariant under the transformation considered. 
But here, as shown by \eq\nr{eq_gauge-scalar}, scalar quantities do 
transform in a non-trivial way. 
According to \eq\nr{eq_gauge-scalar}, the only quantities which do 
not transform at all, are those whose background value vanishes
or is time-independent. 

Proceeding then to vector perturbations
(cf.\ \eq\nr{def_vector}), 
their transformation 
can be calculated as
\be
 \tilde{B}^\mu 
 \; = \;     
 \Xi^{\tilde{\mu}}_\rho 
 \left[ \bar{B}^\rho - (\partial_\alpha\bar{B}^\rho)\xi^\alpha
 + \delta B^\rho + \ord(\delta^2_{ }) \right] 
 \;
 \overset{\rmii{\nr{Xi}}}{\approx}
 \;
 \bar{B}^\mu - (\partial_\alpha\bar{B}^\mu)\xi^\alpha
 + \bar{B}^\alpha\xi^\mu_{,\alpha} + \delta B^\mu 
\ee
%%%%%
\be
 \Rightarrow \quad 
 \boxed{ 
 \quad
 \delta\tilde{B}^\mu \;\approx\;
  \delta B^\mu - \bar{B}^{\mu\hspace*{0.3mm}\prime}_{ }\xi^0_{ }
               + \bar{B}^0_{ }\xi^{\mu\hspace*{0.3mm}\prime}_{ } 
 \;.
 \quad \vphantom{\Big|}
  }
 \label{eq_vec-gauge}
\ee
For the zeroth component it follows that
\begin{equation}
 \tilde{b}^0 
 \; 
 \overset{\rmii{\nr{eq_vec-gauge}} \lift }
 {\underset{\rmii{\nr{def_vector_0}}}{=}} 
 \; 
 b^0 - \bar{B}^{0\hspace*{0.3mm}\prime}_{ }\xi^0_{ }
     + \bar{B}^0_{ }\xi^{0\hspace*{0.3mm}\prime}_{ }
 \ .
\end{equation}
By decomposing the spatial gauge parameter into its scalar
and vector parts,
\begin{equation}
 \xi^i                    \;=\; -\partial^i\xi + \xi^i_\text{v} \ , \qquad
 \partial_i\xi^i_\text{v} \;=\; 0                               \ ,
 \label{eq_xi-s-v}
\end{equation}
and likewise $\delta B^i = b_\text{s}^i + b_\text{v}^i$ 
(cf.\ \eq\eqref{eq_vec-1o}), 
and recalling $\bar B^{i}_{ } = 0$
(cf.\ \eq\nr{def_vector_0}),
we find the transformations 
\begin{align}
 &\tilde{b}^i_\text{s}
 \;=\; -\partial^i_{ } \left( b + \bar{B}^0_{ }\xi' \right)
    \quad \Rightarrow \quad 
 \tilde{b}             \;=\; b + \bar{B}^0_{ }\xi'           
                    \ ,
 \\[2mm]
%%%%%
 &\tilde{b}^i_\text{v} \;=\; b^i_\text{v}
 + \bar{B}^0_{ }\xi^{i\hspace*{0.3mm}\prime}_\text{v} \ .
\end{align}

The behaviour of tensor perturbations 
(cf.\ \eq\nr{delta_Cmunu})
under small coordinate transformations is 
\ba
 \tilde{C}_{\mu\nu}^{ } 
  &
  = 
  & 
  \Xi^\rho_{\tilde{\mu}}\ \Xi^\sigma_{\tilde{\nu}}\ 
 \left[ 
 \bar{C}_{\rho\sigma}^{ }
 - (\partial_\alpha^{ } \bar{C}_{\rho\sigma}^{ })\xi^\alpha
 + \delta C_{\rho\sigma}^{ } + \ord(\delta^2_{ }) 
 \right] \\[2mm]
%%%%%%%
 &
 \overset{\rmii{\nr{Xi}}}{\approx}
 &  
 \bar{C}_{\mu\nu}^{ }
 - \xi^\rho_{,\mu}\bar{C}_{\rho\nu}^{ }
 - \xi^\rho_{,\nu}\bar{C}_{\mu\rho}^{ }
 - (\partial_\alpha^{ } \bar{C}_{\mu\nu}^{ })
   \,\xi^\alpha_{ } + \delta C_{\mu\nu}^{ }
 \\[2mm]  
%%%%%%%%%%
 & \Rightarrow & 
 \boxed{  
 \quad
 \delta\tilde{C}_{\mu\nu}^{ }
 \;
 \approx
 \;
 \delta C_{\mu\nu}^{ } - \xi^\rho_{,\mu}\bar{C}_{\rho\nu}^{ }
 - \xi^\rho_{,\nu}\bar{C}_{\mu\rho}^{ } 
 - \bar{C}_{\mu\nu}^{\hspace*{0.3mm}\prime}\,\xi^0_{ } 
 \;.
 \quad   \vphantom{\Big|}
  }
 \label{eq_tensor-gauge}
\ea
Using eqs.~\eqref{eq_general-tensor}--\eqref{eq_tensor-ij-x} 
we find how the single components transform. The $00$-component 
of \eq\eqref{eq_tensor-gauge} yields
\begin{equation}
 \tilde{c}_0^{ } 
 \; 
 \underset{\rmii{\nr{eq_general-tensor}}}{
 \overset{\rmii{\nr{eq_tensor-gauge}} \lift }{=}} 
 \;
 c_0^{ } + \bar{C}_{00}^{ } 
 \, \xi^{0\hspace*{0.3mm}\prime}_{ }
 + \frac{1}{2}\bar{C}_{00}^{\hspace*{0.3mm}\prime}\, \xi^0_{ }
 \ . \label{eq_b0-gauge-x}
\end{equation}
For the $0i$-components 
we apply \eqs\eqref{eq_xi-s-v},  
\eqref{eq_tensor-backgr}
and \nr{eq_tensor-1o-0i}, and obtain 
\begin{align}
 \tilde{c}_i^{ }           
 \; 
 \underset{\rmii{\nr{eq_tensor-backgr}}}{
 \overset{\rmii{\nr{eq_tensor-gauge}} \lift }{=}} 
 \;
 c_i^{ } + \bar{C}\,\xi_i'
 + \bar{C}_{00}^{ }\,\xi^0_{,i} 
 \quad 
 \underset{\rmii{\nr{eq_tensor-1o-0i}}}{
 \overset{\raise0.5ex\hbox{\rmii{\nr{eq_xi-s-v}}}}{\Rightarrow }} 
 \quad 
 &\tilde{c}          
 \;
 =
 \;
 c + \bar{C}\,\xi' - \bar{C}_{00}^{ }\,\xi^0_{ }
  \ , \label{eq_c_tilde} \\[2mm]
%%%%%%%%
 &\tilde{c}^\text{v}_i 
 \; 
 =
 \;
 c^\text{v}_i + \bar{C}\,\xi^{\text{v}\hspace*{0.3mm}\prime}_i
 \ . \label{eq_b-gauge-x}
\end{align}
Analogously, 
the transformation behaviour of the $ij$-components is found, 
\begin{align}
 \hspace*{-7mm}
 \underbrace{
 \delta \tilde{C}_{ij}^{ }
 }_{\nr{eq_general-tensor}:\
 -2\delta^{ }_{ij} \tilde{c}^{ }_\rmiii{D} + 2\tilde{\gamma}^{ }_{ij}
 }
 \hspace*{-7mm}
 \;
 \overset{\rmii{\nr{eq_tensor-gauge}}}{=} 
 \;
 \ &\delta C_{ij}^{ } - \bar{C}
 \underbrace{ 
 \biggl(
 \xi_{j,i}^{ } + \xi_{i,j}^{ }
 \overbrace{
 - \frac{2}{3}\,\delta_{ij}^{ }\,\xi^k_{,k} 
 }^{
 \nr{eq_xi-s-v}:\; \frac{2}{3}\,\delta_{ij}^{ } \nabla^2_{ }\xi
 }
 \biggr) 
 }_{\tr\,=\,0} 
 \underbrace{
 - \frac{2}{3}\,\bar{C}\delta_{ij}^{ }\,\xi^k_{,k}
 }_{
 \nr{eq_xi-s-v}:\; \frac{2}{3}\,\bar{C}\delta_{ij}^{ } \nabla^2_{ }\xi
 }
 - \, \bar{C}_{ }^{\hspace*{0.3mm}\prime}\delta_{ij}^{ }\,\xi^0 \\[2mm]
%%%%%
 \underset{\rmii{ }}{
 \overset{\rmii{ }}{\Rightarrow}} 
 \hspace*{3.0mm} 
 &\tilde{c}_\rmii{D}^{ }           
  \;=\; c_\rmii{D}^{ } - \frac{1}{3}\,\bar{C}\,\nabla^2\xi
 + \frac{1}{2}\,\bar{C}_{ }^{\hspace*{0.3mm}\prime}\xi^0
 \ , \label{eq_bD-gauge-x}\\
%%%%%
 \quad 
 &\tilde{\gamma}_{ij}^{ }          \mbit\;=\; \gamma_{ij}^{ }
 - \frac{1}{2}\,\bar{C} 
 \left( 
 \xi_{j,i}^{ } + \xi_{i,j}^{ } 
 \right) 
 - \frac{1}{3}\,\bar{C}\delta_{ij}^{ }\nabla^2_{ }\xi
 \label{xi_tilde_gij} \\[2mm]
%%%%%%%
 &\hspace*{4.0mm} 
 \underset{\rmii{\nr{eq_xi-s-v}}}{
 \overset{\raise0.5ex\hbox{\rmii{\nr{eq_tensor-ij}}}}{\Rightarrow}}
 \; 
 \tilde{\gamma}               \;=\; \gamma + \bar{C}\,\xi             
  \ , \quad 
 \tilde{\gamma}_i^\rmi{v}          
    \;=\; \gamma_i^\rmi{v} + \bar{C}\,\xi^\text{v}_i 
  \ , \quad 
 \tilde{\gamma}_{ij}^\text{t} \;=\; \gamma_{ij}^\text{t}
             \ . \label{eq_beta-gauge-x}
\end{align}

Let us sum up the physical degrees of freedom. 
In addition to two scalar gauge parameters, $\xi^0_{ }$ and $\xi$, 
there are two transverse vector gauge parameters, $\xi^i_\text{v}$. 
So, in total, only two of the four scalar perturbations
introduced in \eq\nr{g_munu} are physical, and likewise, 
only two of the four vector perturbations are physical.

%%%%%%%%%%%%%%%%%%%%%%%%%%%%%%%%%%%%%%%%%%%%%%%%%%%%%%%%%%%%%%%%%%%%%%%%%
%
\subsection{Examples of common gauge choices}
\label{ss:gauge_choices}

\index{gauge choices: examples}

Given the gauge transformations defined in \se\ref{ss:def_gauge}, not all 
the perturbations that we have 
considered in \ch\ref{se:pert} are physical. There are two opposite
philosophies that originate from this realization. One is that we can make
use of the gauge freedom, in order to eliminate some variables. Various
gauges are defined by which variables are eliminated, and in this
section we explain three common choices 
(the list is not exhaustive, and a fourth choice is mentioned
in passing in the caption of table~\ref{table:gauges}). 
The other philosophy is to manipulate the basic equations, so that the
dynamical variables appear as gauge-invariant linear combinations. The latter 
philosophy is explained in \se\ref{ssec_nogauge}, 
and is the one we in general endorse, because it offers for strong
crosschecks of the computations.

To proceed with concrete examples, 
we need to apply \eqs\nr{eq_b0-gauge-x}--\nr{eq_beta-gauge-x} 
to $\delta g^{ }_{\mu\nu}$ from \eq\nr{g_munu}. 
The background values are 
$
 \bar{C}_{00}^{ }=-a^2_{ }
$
and 
$
 \bar{C} = a^2_{ }
$, 
and when we subsequently factor out the overall $a^2_{ }$, 
we effectively need to replace 
$\bar{C}_{00}^{ } \to (-a^2)/a^2=-1$, 
$\bar{C}_{00}'\to -2a'/a = -2 \H$, 
$\bar{C} \to (a^2)/a^2=1$, and 
$\bar{C}'\to 2a'/a = 2\H$ in the transformation laws above.  
These lead to 
\begin{align}
 \tilde{h}_0^{ } 
 &\overset{\rmii{\nr{eq_b0-gauge-x}}}{=} 
 h_0^{ }-\xi^{0\hspace*{0.3mm}\prime}_{ }-\H\xi^0_{ }              
 \ , 
 & 
 &
 & 
 & 
 & \label{eq_h0-gauge} \\[2mm]
%%%%
 \tilde{h}_i^{ }                 
 &\overset{\rmii{\nr{eq_c_tilde}}}{=} 
 h_i^{ }+\xi_i'-\partial_i^{ }\xi^0_{ }  
 & 
 &\overset{\hphantom{\rmii{\nr{eq_beta-gauge-x}}}}{\Rightarrow} 
 &
 \tilde{h}                     
 &\overset{\rmii{\nr{eq_c_tilde}}}{=}
 h+\xi'+\xi^0_{ }
 \ , \label{eq_his-gauge} \\[2mm]
%%%%%
 & 
 & 
 &
 &  
 \tilde{h}_i^\text{v}            
 &\overset{\rmii{\nr{eq_b-gauge-x}}}{=}  
 h_i^\text{v}+\xi_i^{\text{v}\prime} 
 \ , \label{hi_v_xi} \\[2mm]
%%%%%
 \tilde{h}_\rmii{D}^{ }                
 &\overset{\rmii{\nr{eq_bD-gauge-x}}}{=} 
 h_\rmii{D}^{ }-\frac{1}{3}\nabla^2\xi+\H \xi^0_{ } \ , 
 &
 &
 & 
 & \label{eq_hD-gauge} \\[1.5mm]
%%%%%
 \tilde{\vartheta}_{ij}^{ }         
 &\overset{\rmii{\nr{xi_tilde_gij}}}{=}
 \vartheta_{ij}^{ }
 -\frac{1}{2}(\partial_i^{ }\xi_j^{ }+\partial_j^{ }\xi_i^{ })
 -\frac{1}{3}\bit\delta^{ }_{ij} \nabla^2_{ } \xi  
 & 
 &\overset{\rmii{\nr{eq_beta-gauge-x}} \lift }{\Rightarrow} 
 & 
 \tilde{\vartheta}            
 &\overset{\hphantom{\rmii{\nr{eq_beta-gauge-x}}}}{=}  
 \vartheta+\xi             
  \ , \label{eq_hijs-gauge} \\[2mm]
%%%%%%
 &
 &
 &
 & 
 \tilde{\vartheta}_i^\rmi{v}    
 &\overset{\hphantom{\rmii{\nr{eq_beta-gauge-x}}}}{=}
 \vartheta_i^\rmi{v} +\xi_i^\text{v}      
  \ , \label{vartheta_v_xi} \\[2.5mm]
%%%%%%
 &
 &
 &
 &
 \tilde{\vartheta}_{ij}^\text{t} 
 &\overset{\hphantom{\rmii{\nr{eq_beta-gauge-x}}}}{=}  
 \vartheta_{ij}^\text{t}  
  \ .
\end{align}
For vector perturbations, we observe 
from \eqs\nr{hi_v_xi} and \nr{vartheta_v_xi}
the invariance of the combination which played a role 
in \eqs\nr{einstein_0i_v} and \nr{einstein_ij_v}, 
$
 \tilde h^\rmi{v}_i - \tilde\vartheta^{\rmi{v}\hspace*{0.3mm}\prime}_i
 = 
 h^\rmi{v}_i - \vartheta^{\rmi{v}\hspace*{0.3mm}\prime}_i
$.

%%%%%%%%%%%%%%%%%%%%%%%%%%%%%%%%%%%%%%%%%%%%%%%%%%%%%%%%%%%%%%%%%%%%%%%%%%%

\subsubsection*{Newtonian 
(or zero-shear, or longitudinal, or Poisson) gauge}

\index{Newtonian gauge}
\index{zero-shear gauge}
\index{longitudinal gauge}
\index{Poisson gauge}

A much-used and perhaps intuitive gauge choice is the one where 
the perturbed metric in \eq\nr{g_munu}
can be described by an almost diagonal matrix. 
The conformal Newtonian gauge satisfies this 
for scalar perturbations, imposing the two gauge conditions
\begin{equation}
 \boxed{
 \quad 
 h^\rmii{N} \;=\; \vartheta^\rmii{N} \;=\; 0 
 \;. \quad
 \vphantom{\Big|}
 }
 \label{eq_Ngaugedef}
\end{equation}
This gauge is also known as 
the zero-shear, longitudinal, or Poisson gauge. 

We consider the gauge transformation from an arbitrary gauge 
to the Newtonian gauge. From \eqs\eqref{eq_his-gauge} 
and \eqref{eq_hijs-gauge} we get
\begin{equation}
 \xi
 \;   
 \overset{\rmii{\nr{eq_hijs-gauge}}}{=}
 \;
 -\vartheta   \ , \qquad
 \xi^0 
 \;
 \overset{\rmii{\nr{eq_his-gauge}}}{=}
 \;
 \vartheta'-h
 \ . \label{eq_Ngauge}
\end{equation}
The remaining scalar quantities transform to
\begin{empheq}[box=\fbox]{align}
 \quad   \vphantom{\bigg|}
 \phi \;\equiv\; h_0^\rmii{N}     
 &
 \;\underset{\rmii{\nr{eq_Ngauge}}}
   {\overset{\rmii{\nr{eq_h0-gauge}} \lift }{=}}\; 
 h_0^{ } + (\partial^{ }_\tau + \H) (\,h - \vartheta'\,) 
        %% \;=\; h_0+\frac{1}{a}\left[ (h-\vartheta')a \right]'    
        \ ,\label{def_phi}\\[2mm]
%%%%%%%%
 \psi \;\equiv\; h_\rmii{D}^\rmii{N} 
 &
 \;\underset{\rmii{\nr{eq_Ngauge}}}
  {\overset{\rmii{\nr{eq_hD-gauge}} \lift }{=}}\; 
 h_\rmii{D}^{ }+\frac{1}{3}\nabla^2\vartheta+\H (\vartheta'-h)
 \ , \quad   \vphantom{\bigg|}
 \label{def_psi}
 \index{$\phi,\psi$ (Bardeen potentials)}
 \index{Bardeen potentials}
\end{empheq}
which are known as 
the {\em Bardeen potentials}~\cite{Bardeen:1980kt_copy}. 
The Bardeen potentials are gauge invariant: 
this can be proven by using 
the relation $\tilde{\vartheta}'-\tilde{h}=\vartheta'-h-\xi^0$, yielding
\ba
 \tilde{\phi}
 &
  \overset{\rmii{\nr{def_phi}} \lift }
{\underset{\rmii{\nr{eq_h0-gauge}}}{=}} 
 & 
 h_0^{ }-\xi^{0\ibit\prime}-\H\xi^0
 +(\partial^{ }_\tau + \H) (h+\xi^0-\vartheta') 
 \nn[2mm]
%%%%%
 & = & 
                \phi- \cancel{\xi^{0\ibit\prime}}-\bcancel{\H \xi^0}
 + \cancel{\xi^{0\prime}} + \bcancel{\H \xi^0} 
 \;
  =
 \;
 \phi \ , \\[2mm]
%%%%%%
 \tilde{\psi} 
  &
  \overset{\rmii{\nr{def_psi}} \lift }
 {\underset{\rmii{\nr{eq_hD-gauge},\nr{eq_hijs-gauge}}}{=}}   
  &
  h_\rmii{D}^{ }
 -\cancel{ \frac{1}{3}\nabla^2_{ }\xi }
 + \bcancel{ \H \xi^0_{ } }
 +\frac{1}{3}\nabla^2\vartheta
 +\cancel{ \frac{1}{3}\nabla^2_{ }\xi }
 +\H(\vartheta'-h- \bcancel{ \xi^0_{ } }) 
 \;            
 =
 \;
 \psi
 \ .
\ea

The two vector gauge conditions can be imposed, e.g.,~as
\begin{equation}
 \xi^\text{v}_i           \;=\; -\vartheta_i^\rmi{v}              \ ,\quad 
 \partial^i\xi^\text{v}_i \;=\; -\partial^i\vartheta_i^\rmi{v} = 0\  \quad
 \Rightarrow \quad
 \vartheta_i^\rmii{N} |^{ }_\rmi{v}    \;=\; 0                    \ ,\quad 
 h_i^\rmii{N}|_\text{v}^{ }
     \;=\; h_i^\text{v}-\vartheta_i^\rmi{v}{}'
 \ . \label{newton_v}
\end{equation}
Therefore the scalar, 
vector, and tensor parts of the perturbed metric appear as 
\ba
 h_{\mu\nu}^\rmii{N}|_\text{s}^{ }
 &
 \underset{\rmii{\nr{eq_Ngaugedef},\nr{def_phi},\nr{def_psi}}}{
 \overset{\rmii{\nr{eq_metric-pert},\nr{g_munu}}
           \lift }{=}}
 & 
 -2\begin{pmatrix}
 \phi & 0 \\
 0    & \psi\,\delta_{ij}^{ }
 \end{pmatrix} 
 \ , \quad 
 h_{\mu\nu}^\rmii{N}|_\text{v}^{ }
 \;
 \underset{\rmii{\nr{newton_v}}}{
 \overset{\rmii{\nr{eq_metric-pert},\nr{g_munu}}
         \lift }{=}}
 \;
 \begin{pmatrix}
 0                     & -h_j^\rmii{N}|_\text{v}^{ } \\
 -h_i^\rmii{N}|_\text{v}^{ } & 0
 \end{pmatrix} 
 \ , \nn[2mm]
%%%%%%
 h_{\mu\nu}^\rmii{N}|_\text{t}^{ }
 &
 \underset{\rmii{\nr{g_munu}}}{
 \overset{\rmii{\nr{eq_metric-pert}}
           \lift }{=}}
 &
 2\begin{pmatrix}
 0 & 0 \\
 0 & \vartheta_{ij}^\text{t}
 \end{pmatrix} \,. \label{n_gauge_expl}
\ea
This metric is simple enough that the Einstein tensor and 
the Einstein equations can be worked out by hand, 
as we show in 
\apps\ref{app:newton} and \ref{app:newton2}, respectively.

%%%%%%%%%%%%%%%%%%%%%%%%%%%%%%%%%%%%%%%%%%%%%%%%%%%%%%%%%%%%%%%%%%%%%%%%%%%

\subsubsection*{Comoving gauge (with respect to either fluid or scalar field)}

\index{comoving gauge}

While the Newtonian gauge simplifies the left-hand side of 
the Einstein equations, the comoving gauge is geared towards
simplifying the right-hand side, 
particularly if only one fluid (cf.\ \eq\nr{delta_Tmunu_fluid}), 
or only a scalar  
degree of freedom appears (cf.\ \eq\nr{delta_T_varphi}).

Let us start by investigating how the fluid perturbations 
$\delta e$, $\delta p$, $v$ and $\Pi$
transform under gauge transformations. 
Given that $e$ and $p$ are four-scalars, 
they obey \eq\eqref{eq_gauge-scalar},
\begin{equation}
 \delta \tilde{e} \;=\; \delta e-\bar{e}_{ }^{\hspace*{0.3mm}\prime}
 \xi^0_{ } \ , \qquad  
 \delta \tilde{p} \;=\; \delta p-\bar{p}_{ }^{\hspace*{0.3mm}\prime}
 \xi^0_{ }
 \ . \label{eq_e-gt}
\end{equation}
The four-velocity $u^\mu$ behaves as described in \eq\eqref{eq_vec-gauge},
\begin{align}
 &\delta \tilde{u}^0_{}  
 \;
 \overset{\rmii{\nr{eq_vec-gauge}} \lift }{=}
 \; 
 \delta u^0_{ }
 - \bar{u}^{0\hspace*{0.3mm}\prime}_{ }\xi^0_{ }
 + \bar{u}^0_{ }\xi^{0\hspace*{0.3mm}\prime}_{ }
 \;                   
 \overset{\rmii{\nr{eq_fluid-velocity_again}} \lift }{=} 
 \;
         \delta u^0_{ } + \frac{1}{a}(\Hc\xi^0_{ }
      +\xi^{0\hspace*{0.3mm}\prime}_{ }) 
         \ ,\label{eq_u0-gt}\\
%%%%%
 &\delta \tilde{u}^i 
 \;
 \overset{\rmii{\nr{eq_vec-gauge}} \lift }{=}
 \;
 \delta u^i_{ }
 + \bar{u}^0_{ }\xi^{i\hspace*{0.3mm}\prime}_{ }
 \;
 \overset{\rmii{\nr{eq_fluid-velocity_again}} \lift }{=} 
 \;
        \delta u^i_{ } + \frac{1}{a}\xi^{i\hspace*{0.3mm}\prime}_{ }
 \qquad
 \underset{\rmii{\nr{eq_v-v-s},\nr{eq_xi-s-v}}}
 {\overset{\rmii{\nr{def_v_i}} \lift }{\Rightarrow}}\quad 
 \tilde{v}
 \;=\; 
 % \underset{\rmii{ \eqref{eq_v-v-s}}}{\overset{\rmii{\eqref{eq_xi-s-v}}}{=}}
 v + \xi'
 \ .\label{eq_v-gt}
\end{align}
The transformation of $\delta u^0_{ }$ corresponds to that of 
$- a^{-1}_{ }h^{ }_0$, 
cf.\ \eqs\nr{u_constraint} and \nr{eq_h0-gauge}.
Finally, 
the anisotropic stress transforms according to \eq\nr{eq_beta-gauge-x}, 
but since it has no background value ($\bar C = 0$), 
it does {\em not} get transformed. 

A coordinate choice that simplifies the 
energy-momentum tensor corresponds to its rest frame.
The \textit{slicing} introduced below \eq\eqref{eq_metric-pert3} 
is said to be {\em comoving}, if its normal vector aligns with 
the fluid four-velocity,
\begin{equation}
 -\,n^\mu_{ } 
 \;
 \overset{\rmii{\eqref{eq_n}}}{=} 
 \;
  a^{-1}_{ }\bigl(\, 1-h_0^{ },h^i_{ } \,\bigr)
 \;
 \overset{!}{=}
 \;
  u^\mu_{ } 
 \;
 \overset{\rmii{\eqref{eq_u^up}}}{=} 
 \;
 a^{-1}_{ }\bigl(\, 1-h_0^{ },v^i_{ } \,\bigr) \ .
\end{equation}
A comoving slicing does not necessarily exist in general. 
However, for scalar perturbations, 
$v^i_\text{s}=-v_{,i}^{ }$ 
and $h^i_\text{s}=-h_{,i}^{ }$, it always exists, 
as we can impose
\begin{equation}\label{eq_comoving-slicing}
 \text{comoving slicing} \quad \Leftrightarrow \quad  v\;=\;h \ .
\end{equation}
According to \eqs\eqref{eq_his-gauge} and \eqref{eq_v-gt}, 
this is obtained by choosing
\begin{equation}\label{eq_c-gauge1}
 \xi^0_{ } \;=\; v - h \ .
\end{equation}
This does not yet define a (scalar) gauge, 
as $\xi$ is still arbitrary, i.e.\ the threading is left unspecified. 
The \textit{threading} is said to be {\em comoving}, 
if the threads are world lines of comoving observers,
\begin{equation}
 t^\mu_{ }
 \;
 \overset{\rmii{\eqref{eq_t}}}{=} 
 \;
 a^{-1}_{ } \bigl(\, 1-h_0^{ },{\bm 0} \,\bigr)
 \;
 \overset{!}{=}
 \;
  u^\mu_{ }
 \;
 \overset{\rmii{\eqref{eq_u^up}}}{=} 
 a^{-1}_{ }\bigl(\, 1-h_0^{ },v^i_{ } \,\bigr) \ .
\end{equation}
For scalar perturbations this means
\begin{equation}
 \text{comoving threading} \quad \Leftrightarrow \quad  v \;=\; 0 \ .
\end{equation}
According to \eq\eqref{eq_v-gt}, 
we get to comoving threading by the gauge transformation
\begin{equation}\label{eq_c-gauge2}
 \xi' \;=\; -v\;.
\end{equation}
A time-independent part of $\xi$ remains unspecified by this condition. 

The comoving gauge is defined by requiring both comoving 
slicing and comoving threading, such that the threading 
is orthogonal to the slicing,
\begin{equation}
 -\, n^\mu_{ } 
 \;=\; t^\mu_{ } 
 \;=\; u^\mu_{ } \qquad \Leftrightarrow \qquad 
 \boxed{
 \quad
  v^\rmii{c} \;=\; h^\rmii{c} \;=\; 0
 \;. \quad   \vphantom{\Big|}
  }
 \label{comoving_with_fluid}
\end{equation}
We see from \eq\nr{delta_Tmunu_fluid} that in this gauge, 
and in the absence of vector perturbations, 
the components $\delta T^{ }_{0i}$ drop out from 
the fluid energy-momentum tensor. 

A problem with the comoving gauge is that if multiple matter
components are present simultaneously, in our case  
a scalar field and a fluid, 
then $\delta T^{ }_{0i}$ no longer drops out, 
cf.\ \eq\nr{delta_T0i_varphi}. On the other hand, 
if {\em only} a scalar field is present, then, 
according to \eq\nr{eq_gauge-scalar}, 
we could eliminate it by choosing 
$\xi^0 = \delta\varphi/\bar{\varphi}'$. 
Subsequently, $h$ could be eliminated via \eq\nr{eq_his-gauge}.
Then we find 
$ 
 \delta\varphi^\rmii{c} 
% = \delta\varphi-\bar{\varphi}'\frac{\delta\varphi}{\bar{\varphi}'} 
 = 0
$
and 
$\delta T^\rmii{\raise0.5ex\hbox{c}}_{0i} = 0$.
In other words, 
if $\varphi$ evolves in an empty universe, 
constant-time hypersurfaces correspond to constant-$\varphi$ 
hypersurfaces in this gauge, and $\varphi$ is homogeneous. 
This is sometimes employed in the context of inflation. 

%%%%%%%%%%%%%%%%%%%%%%%%%%%%%%%%%%%%%%%%%%%%%%%%%%%%%%%%%%%%%%%%%%%%%%%%%%%

\subsubsection*{Spatially flat 
(or uniform curvature) gauge}

\index{spatially flat gauge}
\index{uniform curvature gauge} 

The spatially flat gauge is defined by vanishing time-slice Ricci scalar
(cf.\ \eq\nr{delta_R_tau}), 
\begin{equation}
 \boxed{
 \quad 
 h_\rmii{D}^\rmii{f} + \frac{1}{3}\nabla^2\vartheta^\rmii{f} \;=\; 0
 \;. \quad   \vphantom{\bigg|}
   }
 \label{eq_flat}
\end{equation}
We can transform to the spatially flat gauge with the scalar gauge 
parameter \linebreak $\xi^0=-\left(h_\rmii{D}+\nabla^2\vartheta/3\right)/\H$, 
cf.~\eqs\eqref{eq_hD-gauge} and \nr{eq_hijs-gauge}. 
The other scalar gauge parameter $\xi$ does not need to be specified
to establish \eq\nr{eq_flat}, however choosing it as $\xi = -\vartheta$, 
we may eliminate $h^{ }_\rmii{D}$ and $\vartheta$ separately, 
cf.\ \eq\nr{eq_hijs-gauge}.
Then the {\em curvature perturbation} associated with the inflaton field, 
denoted by $\R^{ }_\varphi$ in \eq\nr{def_R_varphi}, 
becomes
\begin{equation}\label{eq_R-flat}
 \R^{ }_\varphi 
 \;
 {\overset{\rmii{\eqref{def_R_varphi}}}{=} } 
 \; 
 -\biggl( h_\rmii{D} + \frac{1}{3}\nabla^2\vartheta \biggr)
 - \Hc\frac{\delta\varphi}{\bar{\varphi}'}
 \; 
	\overset{\rmii{\eqref{eq_flat}}}{=} 
 \; 
 -\Hc\frac{\delta\varphi^\rmii{f}}{\bar{\varphi}'}
 \ .
\end{equation}
In the literature, the inflaton perturbation in this gauge is 
often called the {\em Sasaki or Mukhanov variable}, and we denote it by
\begin{equation}\label{def_Q_varphi}
 \Q^{ }_\varphi
 \; \equiv \; 
 - \frac{\bar{\varphi}'}{\Hc} \R^{ }_\varphi
 \; 
 = 
 \; 
 \delta\varphi + \frac{\bar{\varphi}'}{\Hc} 
 \left( h_\rmii{D} + \frac{1}{3}\nabla^2\vartheta \right)
 \; 
  \overset{\rmii{\nr{eq_R-flat}}}{=} 
 \; 
 \delta\varphi^\rmii{f}
 \ .
 \index{Sasaki variable}
 \index{Mukhanov variable}
 \index{$\Q^{ }_\varphi$ (Sasaki or Mukhanov variable)}
\end{equation}
This variable plays an important role in \ch\ref{se:dS}. 

%%%%%%%%%%%%%%%%%%%%%%%%%%%%%%%%%%%%%%%%%%%%%%%%%%%%%%%%%%%%%%%%%%%%%%%%%%%

\subsubsection*{Summary}

In order to organize a bit the various gauge choices and what they 
achieve, let us repeat again the transformation properties 
of scalar quantities,
\ba
 \tilde h^{ }_0 
 & 
 \overset{\rmii{\nr{eq_h0-gauge}}}{=} 
 &
 h^{ }_0 - \xi^{0\hspace*{0.3mm}\prime}_{ } - \H \xi^0_{ }
 \;, \label{gauge1} \\[2.5mm]
%%%
 \tilde{h} 
 &
 \overset{\rmii{\nr{eq_his-gauge}}}{=}
 &
 h + \xi^0_{ } + \xi'
 \;, \label{gauge2} \\[0.5mm] 
%%%
 \tilde h^{ }_\rmii{D}
 &
 \overset{\rmii{\nr{eq_hD-gauge}}}{=}
 &
 h^{ }_\rmii{D} + 
 \H \xi^0_{ }  - \frac{\nabla^2\xi}{3}
 \;,   \label{gauge3} \\[2mm]
%%%
 \tilde\vartheta
 &
 \overset{\rmii{\nr{eq_hijs-gauge}}}{=}
 &
 \vartheta + \xi 
 \;, \label{gauge4} \\[2mm] 
%%%
 \delta\tilde\varphi
 &
 \overset{\rmii{\nr{eq_gauge-scalar}}}{=}
 &
 \delta\varphi - \bar{\varphi}' \xi^0_{ } 
 \;, \label{gauge5} \\[2mm] 
%%%
% \delta\tilde T & = & \delta T - T' \xi^0_{ } 
% \;, \\[2mm]
%%% 
 \tilde v
 &
 \overset{\rmii{\nr{eq_v-gt}}}{=}
 &
 v + \xi'
 \;.  \label{gauge7}
\ea
The influence of the various gauge choices is then as 
tabulated in table~\ref{table:gauges}. We note that in each
case, two scalar perturbations can be put to zero. 

%%%%%%%%%%%%%%%%%%%%%%%% TABLE %%%%%%%%%%%%%%%%%%%%%%%%%%%%%%%%%
%
\begin{table}[t]

\hspace*{-0.1cm}
%%%%%%%%%%%%%%%%%%%%%%%%%%%%%%%%%%%%%%%%%%%%%%%%%%%%%%%%%%%%%%%%
\begin{minipage}[c]{15.2cm}
\small{
\begin{center}
\vspace*{-4mm}
\begin{tabular}{c|c|c|c}
% \hline
  & 
 Newtonian / zero-shear & 
 comoving with fluid & 
 spatially flat / uniform curvature
 \\[2mm]
 \hline
  & & & 
 \\[-4mm] 
%%%%%%%%%%%%%%%%%%%%%%%%%%%%%%%%%%%%%%%%%%%%%%%%
 $\xi^0_{ }$ & 
 $\vartheta'_{ } - h$            & 
 $v - h$            & 
 $-\frac{1}{\H}\bigl( h^{ }_\rmiii{D}
  + \frac{\nabla^2\vartheta \vphantom{ |_q } }{3} \bigr)$
  \\[2mm]
%%%%%%%%%%%%%%%%%%%%%%%%%%%%%%%%%%%%%%%%%%%%%%%%
 $\xi^{ }_{ }$ & 
 $-\vartheta$            & 
 $-\int^\tau\! v$            & 
 $-\vartheta$
  \\[2mm]
%%%%%%%%%%%%%%%%%%%%%%%%%%%%%%%%%%%%%%%%%%%%%%%%
 $\tilde h_0^{ }$ & 
 $h^{ }_0 + (\partial^{ }_\tau + \H)(h-\vartheta'_{ }) = \phi$            & 
 $h^{ }_0 + (\partial^{ }_\tau + \H)(h-v)$            & 
 $h^{ }_0 + (\partial^{ }_\tau + \H)\bigl[
 \frac{1}{\H}\bigl( h^{ }_\rmiii{D}
 + \frac{\nabla^2\vartheta \vphantom{ |_q } }{3} \bigr)
  \bigr] $
  \\[2.5mm]
%%%%%%%%%%%%%%%%%%%%%%%%%%%%%%%%%%%%%%%%%%%%%%%%
 $\tilde h$ & 
 $0$            & 
 $0$            & 
 $h - \vartheta'_{ } 
    - \frac{1}{\H}\bigl( h^{ }_\rmiii{D}
   + \frac{\nabla^2\vartheta \vphantom{ |_q } }{3} \bigr)
 $
  \\[1.5mm]
%%%%%%%%%%%%%%%%%%%%%%%%%%%%%%%%%%%%%%%%%%%%%%%%
 $\tilde h^{ }_\rmii{D}$ & 
 $
  h^{ }_\rmiii{D}
   + \frac{\nabla^2\vartheta \vphantom{ |_q } }{3}
  - 
  \H\,\bigl(h - \vartheta'_{ }\bigr) = \psi
 $            & 
  $
  h^{ }_\rmiii{D} + \frac{\nabla^2\int^\tau_{\vphantom{q}} \! v  }{3}
  - 
  \H\,\bigl(h - v\bigr)
  $            & 
 $ 0 $ 
  \\[2.8mm]
%%%%%%%%%%%%%%%%%%%%%%%%%%%%%%%%%%%%%%%%%%%%%%%%
 $\tilde\vartheta^{ }_{ }$ & 
 $ 0 $             & 
 $\vartheta - \int^\tau\! v $            & 
 $ 0 $ 
  \\[2.5mm]
%%%%%%%%%%%%%%%%%%%%%%%%%%%%%%%%%%%%%%%%%%%%%%%%
 $\delta\tilde\varphi^{ }_{ }$ & 
 $\delta\varphi 
 + \bar\varphi\bit'_{ } \bigl( h - \vartheta'_{ } \bigr)
 $            & 
  $\delta\varphi 
 + \bar\varphi\bit'_{ } \bigl( h - v \bigr)
 $            & 
 $
  \delta\varphi + 
  \frac{\bar\varphi\bit'_{ }}{\H}
  \bigl( h^{ }_\rmiii{D}
  + \frac{\nabla^2\vartheta \vphantom{ |_q } }{3} \bigr)
  = 
  \Q^{ }_\varphi
 $
  \\[2mm]
%%%%%%%%%%%%%%%%%%%%%%%%%%%%%%%%%%%%%%%%%%%%%%%%
 $\tilde v^{ }_{ }$ & 
 $v - \vartheta'_{ }$            & 
 $0$            & 
 $v - \vartheta'_{ }$ 
  \\[2mm]
%%%%%%%%%%%%%%%%%%%%%%%%%%%%%%%%%%%%%%%%%%%%%%%%
% \hline 
\end{tabular} 
\end{center}
}
\end{minipage}

\vspace*{1mm}

\caption[a]{\small
     Examples of gauge choices, in the notation of 
     \eqs\nr{gauge1}--\nr{gauge7}. In each case, two 
     perturbations can be set to zero. 
     By $\phi$ and $\psi$
     we refer to the Bardeen potentials,
     cf.\ \eqs\nr{def_phi} and \nr{def_psi}, 
     and by $\Q^{ }_\varphi$ to 
     a rescaled curvature perturbation, 
     cf.\ \eq\nr{def_Q_varphi}.
     Yet another gauge discussed in the literature
     is the {\em synchronous} one, in which 
     $ \tilde h^{ }_0 = \tilde{h} = 0 $~\cite{Ma:1995ey}. 
     We fix to none of these gauges, but rather keep 
     $\xi^0_{ }$ and $\xi$ general, verifying that
     the results do not depend on the gauge choice. 
     \index{synchronous gauge}
}
\index{gauge choices: summary (table)}
\label{table:gauges}
\end{table}
%
%%%%%%%%%%%%%%%%%%%%%%%%%%%%%%%%%%%%%%%%%%%%%%%%%%%%%%%%%%%%%%%%%

%%%%%%%%%%%%%%%%%%%%%%%%%%%%%%%%%%%%%%%%%%%%%%%%%%%%%%%%%%%%%%%%%%%%%%%%%
%
\subsection{How to eliminate gauge dependence from the equations}
\label{ssec_nogauge}

\index{gauge independence}

Having shown in \se\ref{ss:gauge_choices} how gauges can be chosen, 
we now proceed to discussing the opposite philosophy of keeping the
gauge general, and how this can yield 
strong crosschecks of the computations. 

The basic idea 
is to define combinations of the perturbations which are 
invariant in gauge transformations. Subsequently, we convert
our basic equations into equations for these combinations. 
All gauge non-covariant terms must drop out while we are 
doing this, otherwise we have made an algebraic mistake. 

Let us illustrate the procedure by 
considering the Einstein equation originating
from the traceless (or non-diagonal) scalar part of $G^{ }_{ij}$,
given in \eq\nr{delta_einstein_didj},   
\be
%%%%
 h^{ }_0 
 + \bigl(\, \partial^{ }_\tau + 2 \H \, \bigr) 
   \bigl(\, h - \vartheta' \,\bigr)
 - \biggl( h^{ }_\rmii{D} + \frac{\nabla^2\vartheta}{3} \biggr)
 \;
 \overset{\rmii{\nr{delta_einstein_didj}}}{=}
 \; 
  -\,
 8\pi G a^2_{ }\barpPi
 \;. \label{pert_einstein_ij_traceless} 
%%%%
\ee
The goal is to turn this 
into an equation for the Bardeen potentials, $\phi$ and $\psi$.

First, from \eq\nr{def_phi}, we can solve for 
$
 h^{ }_0 = \phi - ( \partial^{ }_\tau + \H )
 (\, h -\vartheta' \,)
$. 
Second, from \eq\nr{def_psi}, we can solve for 
$
 h^{ }_\rmii{D} + \frac{\nabla^2_{ }\vartheta}{3}
 = 
 \psi + 
 \H \, (\, h - \vartheta' \,)
$. 
Inserting these into \eq\nr{pert_einstein_ij_traceless}, 
all appearances of $ h - \vartheta' $ cancel, and 
we simply get
\be
 \phi -\psi
 \;
 \underset{\rmii{\nr{def_phi},\nr{def_psi}}}{
 \overset{\rmii{\nr{pert_einstein_ij_traceless}} \lift }{=}}
 \; 
   -\,
 8\pi G a^2_{ }\barpPi
 \;. \label{demo_1}
\ee
As discussed below \eq\nr{eq_v-gt}, 
the anisotropic stress is gauge invariant. 
Therefore, this 
is a gauge-invariant relation between gauge-invariant
quantities. 

Actually, we can be a bit more precise concerning
the anisotropic stress. 
The explicit expression for its shear
viscous part, $\Pi \supset \Sigma$ where $\Sigma$
is from \eq\nr{decomposition}, 
shows that it contains 
$
 v - \vartheta' = - (\psi + \R^{ }_v)/\H
$, 
where $\R^{ }_v$ is from \eq\nr{def_R_v}. Therefore
\eq\nr{demo_1} represents a relationship between the 
gauge-invariant quantities
$\phi$, $\psi$, and $\R^{ }_v$. 

Let us contrast \eq\nr{demo_1} with a direct derivation of the 
corresponding Einstein equation in the Newtonian gauge, 
cf.\ \eq\nr{eq_ee-p-c2}. While the result is the same,  
if we had only derived \eq\nr{eq_ee-p-c2}, we would have no 
internal crosscheck on the correctness of the computation.
Instead, when we derived the same relation
from \eq\nr{pert_einstein_ij_traceless}, 
a strong crosscheck was offered by the exact 
cancellation of $  h - \vartheta'  $. 

Another similar consideration concerns the 
perturbed Ricci scalar, $\delta {R}$.
A general expression is given in \eq\nr{ricci_scalar}, 
and a Newtonian result is derived in \eq\nr{delta_R_scalar}. If we set 
$h, \vartheta \to 0$ in \eq\nr{ricci_scalar}, 
\eq\nr{delta_R_scalar} is reproduced. 
However, \eq\nr{ricci_scalar} contains more information.
If we substitute 
$
 h^{ }_0 = \phi - ( \partial^{ }_\tau + \H )
 (\, h -\vartheta' \,)
$
and 
$
 h^{ }_\rmii{D} + \frac{\nabla^2_{ }\vartheta}{3}
 = 
 \psi + 
 \H \, (\, h - \vartheta' \,)
$, 
then almost all appearances of 
$
 (\, h - \vartheta' \,)
$
cancel, however we are left over with 
\be
 \delta {R} |^{ }_\rmii{\nr{ricci_scalar}}
 \;
 = 
 \;
 \delta {R}^\rmii{N}_{ } |^{ }_\rmii{\nr{delta_R_scalar}}
 \underbrace{ \;-\; 
 \frac{6}{a^2_{ }}
 \bigl(\, \H'' - 2 \H^3_{ } \,\bigr)}_{-\,6(a''/a^3)' 
 \,\underset{\rmii{\nr{bg_ricci_scalar}}}{=}\,
 -\bar R\ibit'}
 \bigl(\, h - \vartheta' \,\bigr)
 \;. \label{demo_R} 
\ee
If we insert the gauge transformations from 
\eqs\nr{gauge2} and \nr{gauge4}, and make use of the 
gauge independence of $\phi$ and $\psi$, we obtain
\be
 \delta \tilde{R} |^{ }_\rmii{\nr{ricci_scalar}}
 \; = \; 
 \delta {R}^\rmii{N}_{ } |^{ }_\rmii{\nr{delta_R_scalar}}
 -\bar R\bit'
 \bigl(\, \tilde h - \tilde \vartheta' \,\bigr)
 \;
 \underset{\rmii{\nr{demo_R}}}
 {\overset{\rmii{\nr{gauge2},\nr{gauge4}} \lift }{=}}
 \; 
 \delta {R} |^{ }_\rmii{\nr{ricci_scalar}}
 -\bar R\bit' \xi^0_{ }
 \;. 
\ee
So again we find a strong crosscheck, namely that the Ricci scalar
transforms in gauge transformations as a scalar should, 
according to \eq\nr{eq_gauge-scalar}. 

The procedure that led to \eq\nr{demo_1} can be repeated
for the three other Einstein equations, as well as the scalar
field equation and energy-momentum conservation equations. 
However, we then need to introduce further gauge-invariant variables. 
A convenient
set is offered by so-called 
{\em curvature perturbations}, 
\index{$\R^{ }_\iQ$ (curvature perturbation of flavour $Q$)}
$\R^{ }_\iQ$, \index{curvature perturbations: definition}
defined from the Ricci scalar of a constant-$\tau$ slice as 
\be
 \delta R^{ }_\tau 
 \; 
 \overset{\rmii{\nr{delta_R_tau}}}{=} 
 \;
 \frac{4}{a^2_{ }} \nabla^2_{ }
 \biggl(\;
   h^{ }_\rmii{D} + \frac{\nabla^2\vartheta}{3}
 \;\biggr)
 \; 
 \equiv 
 \; 
 -\, \frac{4}{a^2_{ }} \nabla^2_{ }
 \biggl(\;
  \R^{ }_\iQ + \H\, \frac{\delta Q}{\bar Q\hspace*{0.3mm}{}'}
 \;\biggr)
 \;, \label{def_R_x}
\ee
where $Q$ is one of the scalar quantities 
entering the Einstein equations
via the energy-momentum tensor,
and $\R^{ }_\iQ$ is gauge invariant.
If we refer to $R^{ }_\tau$ as spatial curvature,  
the nomenclature of referring to $\R^{ }_\iQ$
as a curvature perturbation is a bit imprecise, 
however it has established itself. 
% In any case, 
Examples of various 
``flavours'' of curvature perturbations are 

\vspace*{-5mm}

\begin{empheq}[box=\fbox]{align}
 \quad   \vphantom{\Bigg|}
 \R^{ }_\varphi
 & \equiv \;
 - \biggl( 
  h^{ }_\rmii{D} + \frac{\nabla^2\vartheta}{3} 
 \biggr)
  - \H\,\frac{\delta\varphi}{\bar\varphi\hspace*{0.3mm}{}'}
 \vphantom{\frac{T'}{q}}
 \label{def_R_varphi}
 \;, \\[2mm] 
%%%
 \R^{ }_v
 & \equiv \;
 - \biggl( 
  h^{ }_\rmii{D} + \frac{\nabla^2\vartheta}{3} 
   \biggr)
 + \H \, (h - v)
 \;, \quad \label{def_R_v} \\[2mm] 
%%%
 \R^{ }_\T
 & \equiv \;
 - \biggl( 
  h^{ }_\rmii{D} + \frac{\nabla^2\vartheta}{3} 
   \biggr)
 - \H\, \frac{\delta T}{\bar T\bit' \lift}
 \;, \label{def_R_T} \\[2mm]
%%%
 \R^{ }_e 
 & \equiv \; 
 - \biggl( 
  h^{ }_\rmii{D} + \frac{\nabla^2\vartheta}{3} 
   \biggr)
 - \H\, \frac{\delta e}{\bar e\bit'}
 \;.   \vphantom{\Bigg|}
 \label{def_R_e} 
\end{empheq}

\vspace*{-1mm}

\noindent
As follows from \eqs\nr{gauge1}--\nr{gauge7}, these are gauge invariant, 
and for $\R^{ }_\varphi$ we also demonstrate 
this explicitly in \eq\nr{eq_R}. 
We note that if
$\delta e$ can be expressed as a function of the other 
dynamical variables, like $\delta\varphi$ and $\delta T$, then 
$\R^{ }_e$ can be represented as a linear combination of 
$\R^{ }_\varphi$ and $\R^{ }_\T$.
To streamline the notation, we set $\bar T \to T$.

A further class of gauge-invariant quantities 
is obtained by taking differences of \eqs\nr{def_R_varphi}--\nr{def_R_e}.
In particular, a 
{\em relative energy density perturbation} 
\index{gauge-invariant energy perturbations}
is defined as 
\be
 \Delta \; \equiv \; 
 \frac{\delta e}{\bar e}  + 
 \frac{\bar{e}^{\bit\prime}_{ }}
      {\bar{e}}(h-v)
 \; = \; 
 \frac{\bar e^{\hspace*{0.3mm}\prime}_{ }}
      {\bar e \bit\H} \bigl(\, \R^{ }_v - \R^{ }_e \,\bigr)
 \;. \label{Delta_v}
\ee
We also note that in the classic literature, 
following ref.~\cite{zeta}, 
the {\em curvature perturbation related to energy density},
$\R^{ }_e$, is often denoted by 
\be
 \zeta \; \equiv \; 
 \R^{ }_e 
 \; 
  \underset{\rmii{\nr{def_psi}}}
 {\overset{\rmii{\nr{eq_Ngaugedef}} \lift }{=}} 
 \; 
 -\psi 
 - \H\, \frac{\delta e^\rmiii{N}_{ }}{\bar e^{\bit\prime}_{ } }
 \;, \label{zeta}
 \index{$\zeta$ (curvature perturbation)}
\ee
where we also indicated the frequently appearing 
Newtonian gauge representation. 

For us, the task now
is to re-organize the Einstein equations so that 
the dynamical variables are $\phi$, $\psi$, $\R^{ }_\varphi$, 
$\R^{ }_v$, and $\R^{ }_\T$. Given that there are five
variables, we need five independent equations; 
one was already given in \eq\nr{demo_1}.
We will return to the remaining four 
equations in the later chapters
(cf.\ \eqs\nr{dot_R_varphi}, \nr{dot_E_v}, 
\nr{dot_E_T}, \nr{dot_psi}). 

Let us end this section by demonstrating explicitly the gauge
independence of $\R^{ }_\varphi$ from \eq\nr{def_R_varphi}
and $\Delta$ from \eq\nr{Delta_v}. For the former, 
the gauge 
transformation properties in \eqs\nr{gauge3}--\nr{gauge5} yield
\begin{equation}
 \tilde{\R}^{ }_\varphi
 \; 
 \underset{\rmii{\nr{gauge3}--\nr{gauge5}}}{
 \overset{\rmii{\nr{def_R_varphi}} \lift }{=}} 
 \;   
 \R^{ }_\varphi
 + 
   \overbrace{
   \bcancel{\frac{1}{3}\nabla^2\xi}
 - \cancel{\Hc\xi^0_{ }}
   }^{{\rm from}\; \tilde h^{ }_\rmiii{D}}
 - \overbrace{
   \bcancel{\frac{1}{3}\nabla^2\xi}
   }^{{\rm from}\;\tilde\vartheta}
 - \overbrace{
   \cancel{\Hc \bigl(\, -\xi^0_{ } \,\bigr)} 
   }^{{\rm from}\;\delta\tilde\varphi }
 \;=\;
  \R^{ }_\varphi 
 \;\overset{\rmii{\nr{eq_R-flat}}}{=} \;
 - \frac{\H}{\bar\varphi\hspace*{0.3mm}{}'} \, \delta\varphi^\rmii{f}_{ }
 \;.  \label{eq_R}
\end{equation}
For the latter, 
\eqs\nr{eq_e-gt}, \nr{gauge2} and \nr{gauge7} imply
\be
 \tilde{\Delta} 
 \;
 \underset{\rmii{\nr{gauge2},\nr{gauge7}}}{
 \overset{\rmii{\nr{Delta_v},\nr{eq_e-gt}} \lift }{=}} 
 \;  
 \underbrace{ \frac{\delta e}{\bar{e}} 
  - \frac{\bar{e}\bit'}{\bar{e}}\,\cancel{\xi^0_{ }}
 }_{{\rm from}\;\delta \tilde e}
  + \frac{\bar{e}\bit'}{\bar{e}} 
 \underbrace{ \bigl(\, h-v+\cancel{\xi^0_{ }} \,\bigr)
              }_{{\rm from}\;\tilde h - \tilde v}     
 \;=\;
 \Delta
 \;
 \overset{\rmii{\nr{comoving_with_fluid}}}{=}
 \;
 \frac{\delta e^\rmii{c}}{\bar{e}}   
  \ . \label{eq_Delta}
\ee

\newpage

%%%%%%%%%%%%%%%%%%%%%%%%%%%% start appendices %%%%%%%%%%%%%%%%%%%%%%%%%%%%%%%

%%%%%%%%%%%%%%%%%%%%%%%%%%%%%%%%%%%%%%%%%%%%%%%%%%%%%%%%%%%%%%%%%%%%%%%%%
\subsubsection{Deriving the Einstein tensor in the Newtonian gauge}
\label{app:newton}

\addcontentsline{toc}{subsection}{\App\ref{app:newton}: 
Deriving the Einstein tensor in the Newtonian gauge}

\index{Einstein tensor: Newtonian gauge}

If we choose the conformal Newtonian gauge from 
\eq\nr{n_gauge_expl}, implying
\ba
 g^\rmii{N}_{\mu\nu} 
 & 
 \underset{\rmii{\nr{eq_metric-pert},\nr{n_gauge_expl}}}{
 \overset{\rmii{\nr{g_munu}} \lift }{=}} 
 & 
 a^2_{ }(\tau) \begin{pmatrix}
 -1-2\phi & -h^\rmi{v}_j \\[2mm]
 -h^\rmi{v}_i    & (1-2\psi)\,\delta^{ }_{ij}+2\vartheta^\rmi{t}_{ij}
 \end{pmatrix} 
 \ , \label{g_munu_N}\\[2mm]
%%%%%%%
 g^{\mu\nu\hspace*{0.3mm}}_{\rmii{N}} 
 &
 \overset{\rmii{\nr{g^munu}}\lift }{=}
 &
 \frac{1}{a^2_{ }(\tau)} \begin{pmatrix}
 -1+2\phi & -h^\rmi{v}_j \\[2mm]
 -h^\rmi{v}_i    & (1+2\psi)\,\delta^{ }_{ij}-2\vartheta^\rmi{t}_{ij}
 \end{pmatrix} 
 \;+\; \ord(\delta^2_{ })
 \ , \label{g^munu^N}
\ea
then the Einstein tensor
can be derived by hand. Here we show how this goes. 
To streamline the notation, the superscript
$(...)^\rmii{N}_{ }$ is suppressed until \eq\nr{delta_R_scalar}. 

Starting with 
the perturbed Christoffel symbols from \eq\eqref{eq_ChSy-p}, {\it viz.} 
\be
 \delta\Gamma^\rho_{\mu\nu} 
 \;
 \overset{\rmii{\nr{eq_ChSy-p}}}{=}
 \; 
 \frac{1}{2} \bar{g}^{\rho\sigma}_{ }( \delta g^{ }_{\sigma\mu,\nu}
 + \delta g^{ }_{\sigma\nu,\mu} - \delta g^{ }_{\mu\nu,\sigma} )
 \;+\;
 \frac{1}{2} \delta g^{\rho\sigma}_{ }( \bar{g}^{ }_{\sigma\mu,\nu}
 + \bar{g}^{ }_{\sigma\nu,\mu} - \bar{g}^{ }_{\mu\nu,\sigma} ) 
 \ , \label{def_gamma_again}
\ee
they become
\begin{align}
 \delta \Gamma^0_{00} 
 &= 
 \frac{a^{-2}}{2}
 \bigl\lbrace\;
 - \overbrace{ 
  (-2a^2\phi)'
  }^{\delta g^{ }_{00,0} }
 + 2\phi\,
 \overbrace{
 [2(-a^2)' - (-a^2)']
 }^{\bar{g}^{ }_{00,0} }
 \;\bigr\rbrace
 \;=\; \cancel{2\Hc\phi} + \phi' - \cancel{2\Hc\phi} 
 \;=\; \phi' 
 \ , \label{d_gamma_000} \\[2mm]
%%%%%
 \delta \Gamma^0_{0i} 
 &= 
 \frac{a^{-2}}{2}
 \bigl[\;
 - \overbrace{ 
   (-2a^2_{ }\phi)_{,i}
   }^{\delta g^{ }_{00,i} }
 - h_j^{ }
   \overbrace{ 
   (a^2_{ }\delta^{ }_{ij})'
   }^{\bar{g}^{ }_{ij,0} }
 \;\bigr]
 \;=\; \phi_{,i}^{ } - \Hc h_i^{ }
 \ , \label{d_gamma_00i} \\[2mm]
%%%%%
 \delta \Gamma^0_{ij} 
 &= 
 \frac{a^{-2}}{2}
 \bigl\lbrace\;
 -[\;
   \overbrace{
   (-a^2_{ }h_i^{ })_{,j}^{ }
   }^{\delta g^{ }_{0i,j} }
 + \overbrace{
   (-a^2_{ }h_j^{ })_{,i}^{ }
   }^{\delta g^{ }_{0j,i} }
 - \overbrace{
   (-2a^2_{ }\psi\,\delta_{ij}^{ } + 2a^2\vartheta_{ij}^{ })'
   }^{\delta g^{ }_{ij,0} }
  \;]
 + 2\phi\, 
   \overbrace{ 
   [-(a^2_{ } \delta^{ }_{ij})'\,]
   }^{ -\bar{g}^{ }_{ij,0} }
 \;\bigr\rbrace
 \nonumber\\[2mm]
 &= 
 \frac{1}{2}(h_{i,j}^{ }+h_{j,i}^{ })
 - \delta_{ij}^{ }[\psi' + 2\Hc(\phi+\psi)]
 + \vartheta_{ij}' + 2\Hc\vartheta_{ij}^{ }
 \ , \label{d_gamma_0ij} \\[2mm]
%%%%%
 \delta \Gamma^i_{00} 
 &= 
 \frac{a^{-2}}{2} 
 \bigl\lbrace\;
  [\;
   2 \overbrace{
  (-a^2_{ }h_i^{ })'
  }^{\delta g^{ }_{i0,0} }
 - \overbrace{ 
   (-2a^2_{ }\phi)_{,i}^{ }
  }^{\delta g^{ }_{00,i} }
  \;]
 - h_i^{ } 
  \overbrace{ 
  (-a^2_{ })'
 }^{\bar{g}^{ }_{00,0} }
 \;\bigr\rbrace
 \;=\;
%  -2\Hc h_i-h_i' + \phi_{,i} + \Hc h_i \nonumber\\[2mm] &= 
 \phi_{,i}^{ } - \Hc h_i^{ } - h_i'
 \ , \label{d_gamma_i00}
\end{align}
%%%%%%
\begin{align}
 \delta \Gamma^i_{0j} 
 &= 
 \frac{a^{-2}}{2} 
 \bigl\lbrace\; 
 [\;
   \overbrace{
   (-a^2_{ } h_i^{ })_{,j}^{ }
   }^{\delta g^{ }_{i0,j} }
 + \overbrace{
   (-2a^2_{ }\psi\,\delta_{ij}^{ } + 2a^2_{ }\vartheta_{ij}^{ })'
   }^{\delta g^{ }_{ij,0} }
 - \overbrace{ 
   (-a^2_{ } h_j^{ })_{,i}^{ }
   }^{\delta g^{ }_{0j,i} }
 \;]
 + 2(\psi\,\delta_{ik}^{ } - \vartheta_{ik}^{ })
 \overbrace{
 (a^2_{ }\delta_{kj}^{ })'
 }^{\bar{g}^{ }_{kj,0} }
 \;\bigr\rbrace
 \nonumber\\[2mm]
 &= 
 \frac{1}{2}(h_{j,i}^{ }-h_{i,j}^{ }) + 
 \delta_{ij}^{ }(-\psi' - \cancel{2\Hc\psi} + \cancel{2\Hc\psi} ) + 
 \vartheta_{ij}' + \cancel{2\Hc\vartheta_{ij}^{ }}
 - \cancel{2\Hc\vartheta_{ij}^{ }}
 \nonumber\\[2mm]
 &= 
 \frac{1}{2}(h_{j,i}^{ }-h_{i,j}^{ })
 - \delta_{ij}^{ }\,\psi' + \vartheta_{ij}'
 \ , \label{d_gamma_i0j} \\[2mm]
%%%%%%
 \delta \Gamma^i_{jk} 
 &= 
 \frac{1}{2}
 \bigl[\;
   \overbrace{
   2(\vartheta_{ij}^{ } - \psi\,\delta_{ij}^{ } )_{,k}^{ }
   }^{ a^{-2}_{ }\delta g^{ }_{ij,k}}
 + \overbrace{
   2(\vartheta_{ik}^{ } - \psi\,\delta_{ik}^{ } )_{,j}^{ }
   }^{ a^{-2}_{ }\delta g^{ }_{ik,j}}
 - \overbrace{
   2(\vartheta_{jk}^{ } - \psi\,\delta_{jk}^{ } )_{,i}^{ }
   }^{ a^{-2}_{ }\delta g^{ }_{jk,i}}
 \;\bigr]
 - \frac{a^{-2}}{2}
  h_i^{ }\, [\, - \overbrace{
                  (a^2_{ }\delta_{jk}^{ })'
                  }^{\bar{g}^{ }_{jk,0} }
             \,] 
 \nonumber\\[2mm]
 &= 
 -\, \delta_{ij}^{ }\,\psi_{,k}^{ }
 - \delta_{ik}^{ }\,\psi_{,j}^{ }
 + \delta_{jk}^{ }\,\psi_{,i}^{ }
 + \vartheta_{ij,k}^{ }
 + \vartheta_{ik,j}^{ }
 - \vartheta_{jk,i}^{ }
 + \Hc h_i^{ }\,\delta_{jk}^{ }
 \ .
\end{align}
The unperturbed Christoffel symbols can be found in \eq\eqref{eq_ChSyu}.
%% Ricci tensor %%%%%%%%%%%%%%%%%%%%%%%%%%%%%%%%%%%
\allowdisplaybreaks
The resulting perturbed Ricci tensor,
from \eq\eqref{eq_Ricci-p}, is 
\ba
 \delta R_{00}^{ }
 &
 \overset{\rmii{\nr{eq_Ricci-p}}}{=}
 &
    \delta \Gamma^\alpha_{00,\alpha}
  - \delta \Gamma^\alpha_{0\alpha,0}
  + \bar{\Gamma}^\beta_{00} \delta\Gamma^\alpha_{\beta\alpha}
  + \bar{\Gamma}^\alpha_{\beta\alpha} \delta\Gamma^\beta_{00}
  - \bar{\Gamma}^\beta_{0\alpha} \delta\Gamma^\alpha_{0\beta}
  - \bar{\Gamma}^\alpha_{0\beta} \delta\Gamma^\beta_{0\alpha}
 \nonumber\\[3mm]
%%%
 &=&
   \overbrace{
   \cancel{\phi''}
   }^{\delta \Gamma^0_{00,0}}
   \; 
   \overbrace{
 +\; \nabla^2_{ }\phi
   }^{\delta \Gamma^i_{00,i} }
   \; 
   \overbrace{
 - \; \cancel{\phi''}
   }^{- \delta \Gamma^0_{00,0} } 
   \;  
   \overbrace{
 + \; 3\psi''
   }^{- \delta \Gamma^i_{0i,0} }
 +
   \overbrace{
   \Hc\delta\Gamma^\alpha_{0\alpha}
 + 4\Hc\delta\Gamma^0_{00}
 - 2\Hc\delta\Gamma^0_{00}
 - 2\Hc\delta\Gamma^i_{0i}
   }^{{\rm from}\;\nr{eq_ChSyu}}
 \nonumber\\[3mm]
%%%
 &=&
   3\psi'' + \nabla^2_{ }\phi + 3\Hc(\phi'+\psi')
 \ , \label{d_R_00} \\[3mm]
%%%%%%%%%%%%%%%%%
 \delta R_{0i}^{ }
 &
 \overset{\rmii{\nr{eq_Ricci-p}}}{=}
 & 
    \delta \Gamma^\alpha_{0i,\alpha}
  - \delta\Gamma^\alpha_{0\alpha,i}
  + \bar{\Gamma}^\beta_{0i}\delta\Gamma^\alpha_{\beta\alpha}
  + \bar{\Gamma}^\alpha_{\beta\alpha}\delta\Gamma^\beta_{0i}
  - \bar{\Gamma}^\beta_{0\alpha}\delta\Gamma^\alpha_{i \beta}
  - \bar{\Gamma}^\alpha_{i \beta}\delta\Gamma^\beta_{0\alpha}
 \nonumber\\[3mm]
%%%
 &=&
    \overbrace{
    \cancel{\phi'_{,i}} -\Hc'h_i^{ } - \Hc h_i' 
    }^{\delta\Gamma^0_{0i,0} }
    \; 
    \overbrace{
  +\; \frac{1}{2}\nabla^2_{ } h_i^{ } - \psi'_{,i}
    }^{\delta\Gamma^j_{0i,j} }
    \; 
    \overbrace{
  -\; \cancel{\phi'_{,i}}
    }^{-\delta\Gamma^0_{00,i} }
    \; 
    \overbrace{
  +\; 3\psi'_{,i}
  }^{-\delta\Gamma^j_{0j,i} }
 \nonumber\\[2mm]
%%%
 &&
  + 
  \overbrace{
   \Hc\delta\Gamma^\alpha_{i\alpha}
  + 4\Hc\delta\Gamma^0_{0i}
 - \Hc\delta\Gamma^0_{i0}
 - \Hc\delta\Gamma^j_{ij}
 - \Hc\delta_{ij}\delta\Gamma^j_{00}
 - \Hc\delta_{ij}\delta\Gamma^0_{0j}
   }^{{\rm from}\;\nr{eq_ChSyu}}
 \nonumber\\[3mm]
%%%
 &=&
  - \Hc'h_i^{ }
  - \cancel{\Hc h_i'}
  + \frac{1}{2}\nabla^2_{ } h_i^{ }
  + 2\psi'_{,i}
  + 3\Hc(\phi_{,i}^{ } - \Hc h_i^{ })
  - \Hc(\phi_{,i}^{ }-\Hc h_i^{ }-\cancel{h_i'})
 \nonumber\\[3mm]
%%%
 &=&
   2(\psi'+\Hc\phi)_{,i}^{ }
   + \frac{1}{2}\nabla^2_{ } h_i^{ }
   - (\Hc'+2\Hc^2)h_i^{ }
 \ , \label{d_R_0i} \\[3mm]
%%%%%%%%%%%%%%%
 \delta R_{ij}^{ }
 &
 \overset{\rmii{\nr{eq_Ricci-p}}}{=}
 & 
   \delta \Gamma^\alpha_{ij,\alpha}
 - \delta\Gamma^\alpha_{i\alpha,j}
 + \bar{\Gamma}^\beta_{ij}\delta\Gamma^\alpha_{\beta\alpha}
 + \bar{\Gamma}^\alpha_{\beta\alpha}\delta\Gamma^\beta_{ij}
 - \bar{\Gamma}^\beta_{i\alpha}\delta\Gamma^\alpha_{j \beta}
 - \bar{\Gamma}^\alpha_{j \beta}\delta\Gamma^\beta_{i\alpha}
 \nonumber\\[3mm]
%%%
 &=&
   \overbrace{
    \frac{1}{2}(h_{i,j}^{ }+h_{j,i}^{ })'
  - \delta_{ij}^{ }\psi''
 + \vartheta_{ij}''
 - 2\Hc'[(\phi+\psi)\delta_{ij}^{ } - \vartheta_{ij}^{ }]
 - 2\Hc[(\phi+\psi)\delta_{ij}^{ } - \vartheta_{ij}^{ }]'
 }^{\delta \Gamma^0_{ij,0}}
 \nonumber\\[2mm]
%%%
 &&
   \overbrace{
 - 2\psi_{,ij}^{ }
 + \delta_{ij}^{ }\nabla^2_{ }\psi
 - \nabla^2_{ }\vartheta_{ij}
   }^{\delta \Gamma^k_{ij,k} }
   \;
   \overbrace{
 - \; \phi_{,ij}^{ }
 + \cancel{\Hc h_{i,j}^{ }}
   }^{-\delta\Gamma^0_{i0,j} }
   \; 
   \overbrace{
 + \; 3 \psi_{,ij}^{ }
 - \cancel{\Hc h_{i,j}^{ }}
   }^{-\delta\Gamma^k_{ik,j} }
 \nonumber\\[2mm]
%%%
 &&
 +
   \overbrace{
   \Hc\delta_{ij}^{ } \delta\Gamma^\alpha_{0\alpha}
 + 4\Hc\delta\Gamma^0_{ij}
 - \Hc(2\delta\Gamma^0_{ij}
 + \delta\Gamma^i_{0j}
 + \delta\Gamma^j_{0i})
   }^{{\rm from}\;\nr{eq_ChSyu}}
 \nonumber\\[3mm]
%%%
 &=&
  \frac{1}{2}(h_{i,j}^{ }+h_{j,i}^{ })'
 - \delta_{ij}^{ } \psi''
 + \vartheta_{ij}''
 - 2\Hc'[(\phi+\psi)\delta_{ij}^{ } - \vartheta_{ij}^{ }]
 - 2\Hc[(\phi+\psi)\delta_{ij}^{ } - \vartheta_{ij}^{ }]'
 \nonumber\\[2mm]
%%%
 &&
  +\psi_{,ij}^{ }
  + \delta_{ij}^{ }\nabla^2_{ }\psi
  - \nabla^2_{ }\vartheta_{ij}^{ }
  - \phi_{,ij}^{ }
  + \Hc\delta_{ij}^{ }(\phi-3\psi)'
  + \Hc(h_{i,j}^{ }+h_{j,i}^{ })
 \nonumber\\[2mm]
%%%
 &&
  -\cancel{2\Hc(\delta_{ij}^{ }\psi-\vartheta_{ij}^{ })'}
  - 4\Hc^2[(\phi+\psi)\delta_{ij}^{ } - \vartheta_{ij}^{ }]
% - \cancel{\frac{\Hc}{2}(h_{j,i}^{ }-h_{i,j}^{ })}
% \nonumber\\[2mm]
% &\quad\
 +\cancel{2\Hc(\delta_{ij}^{ }\psi-\vartheta_{ij}^{ })'}
% + \cancel{\frac{\Hc}{2}(h_{j,i}^{ }-h_{i,j}^{ })}
 \nonumber\\[3mm]
%%%
 &=&
  -\delta_{ij}^{ } \left[ \psi'' + \Hc(\phi'+5\psi')
 + 2(\Hc'+2\Hc^2)(\phi+\psi)
  - \nabla^2_{ }\psi \right]
 + (\psi-\phi)_{,ij}^{ }
 \nonumber\\[2mm]
%%%
 &&
 + \frac{1}{2}(h_{i,j}'+h_{j,i}')
 + \Hc(h_{i,j}^{ }+h_{j,i}^{ })
 + \vartheta_{ij}''
 + 2\Hc\vartheta_{ij}'
 - \nabla^2_{ }\vartheta_{ij}^{ }
 + 2(\Hc'+2\Hc^2)\vartheta_{ij}^{ }
 \ . \nn \label{d_R_ij}
\ea
The unperturbed version 
$\bar{R}^{ }_{\mu\nu}$ 
is given in \eq\eqref{eq_Runpert}.
Let us now raise one index of the perturbed Ricci tensor 
as in \eq\eqref{eq_Ricci-p-up-down},
\be
 \delta {R^\mu}_\nu
 \; 
 \overset{\rmii{\nr{eq_Ricci-p-up-down}}}{=} 
 \;
   \bar{g}^{\mu\rho}_{ } \delta R^{ }_{\rho\nu}
 + \delta g^{\mu\rho}_{ } \bar{R}^{ }_{\rho\nu}
 \,. \label{delta_ricciupdown_again}
\ee
We obtain
\begin{align}
 \delta {R^0}_0 &= -a^{-2}_{ }
 \bigl[\;
    \overbrace{
     3\psi''
  + \nabla^2_{ }\phi
  + 3\Hc(\phi'+\psi')
    }^{{\rm from}\; \bar{g}^{00}_{ }\, \delta R^{ }_{00} }
  \; +
    \hspace*{-3mm}
    \overbrace{
    6\Hc'\phi
    }^{{\rm from}\; \delta g^{00}_{ }\, \bar{R}^{ }_{00}  }
    \hspace*{-3mm}
  \;\bigr]
 \ , \label{eq_rt0}\\[2mm]
%%%%%
 \delta {R^0}_i &= -a^{-2}_{ }
 \biggl[\;
   \overbrace{ 
   2\psi_{,i}'
 + 2\Hc\phi_{,i}^{ }
 + \frac{1}{2}\nabla^2_{ } h_i^{ }
 - \cancel{(\Hc'+2\Hc^2)h_i^{ }}
    }^{{\rm from}\; \bar{g}^{00}_{ }\, \delta R^{ }_{0i} }
 \; + \;
  \overbrace{
   \cancel{(\Hc'+2\Hc^2)h_i^{ }}
    }^{{\rm from}\; \delta g^{0j}_{ }\, \bar{R}^{ }_{ji}  }  
 \;\biggr]
 \nonumber\\[2mm]
%%%
 &= -a^{-2}_{ }
  \left[
     2(\psi'+\Hc\phi)_{,i}^{ }
  + \frac{1}{2}\nabla^2_{ } h_i^{ }
  \right]
 \ ,\\[2mm]
%%%%%
 \delta {R^i}_0 &= a^{-2}_{ }
 \biggl[\;
   \overbrace{
     2\psi_{,i}'
 + 2\Hc\phi_{,i}
 + \frac{1}{2}\nabla^2_{ } h_i^{ }
 - (\Hc'+2\Hc^2) h_i^{ }
   }^{ {\rm from}\; \bar{g}^{ij}_{ }\, \delta R^{ }_{j0} }
 \; + \; 
 \hspace*{-3mm}
 \overbrace{ 
 3\Hc'h_i^{ }
    }^{{\rm from}\; \delta g^{i0}_{ }\, \bar{R}^{ }_{00}  }  
 \;\biggr]
  \nonumber\\[2mm]
%%%%
 &= a^{-2}_{  }
 \left[
   2(\psi'+\Hc\phi)_{,i}^{ }
 + \frac{1}{2}\nabla^2_{ } h_i^{ }
  + 2(\Hc'-\Hc^2)h_i^{ }
 \right]
 \ ,\\[2mm]
%%%
 \delta {R^i}_j &= a^{-2}_{ }
 \bigl[\;
  \hspace*{-3mm}
   \overbrace{
   \delta R_{ij}^{ }
   }^{ {\rm from}\; \bar{g}^{ik}_{ }\, \delta R^{ }_{kj} }
  \hspace*{-3mm}
 \; + \; 
   \overbrace{
   2(\Hc'+2\Hc^2)(\psi\delta_{ij}^{ } - \vartheta_{ij}^{ })
    }^{{\rm from}\; \delta g^{ik}_{ }\, \bar{R}^{ }_{kj}  }  
 \;\bigr]
 \nonumber\\[3mm]
%%%%
 &= -\,a^{-2}_{ }
 \left[
   \psi''
  + \Hc(\phi'+5\psi')
 + 2(\Hc'+2\Hc^2)\phi
 - \nabla^2_{ } \psi
 \right]\delta_{ij}^{ }
 \nonumber\\[2mm]
%%%%
 &\quad +\, a^{-2}_{ }
 \bigl[\,
    (\psi-\phi)_{,ij}^{ }
  + \frac{1}{2}(h_{i,j}'+h_{j,i}')
  + \Hc(h_{i,j}^{ }+h_{j,i}^{ })
  + \vartheta''_{ij}
  + 2\Hc\vartheta_{ij}'
  - \nabla^2_{ }\vartheta_{ij}^{ }
 \,\bigr]
 \ . \label{eq_rtij}
\end{align}
The perturbed Ricci scalar follows from \eq\eqref{eq_Ricci-scalar-p},
\begin{align}
 \frac{1}{2}\delta R^\rmii{N}_{ }
 &
 \,
 \overset{\rmii{\nr{eq_Ricci-scalar-p}}}{=}
 \,
 \frac{1}{2}\delta {R^\mu}_\mu
 \;
  =
 \;
 \frac{1}{2}\bigl(\, \delta {R^0}_0+\delta {R^i}_i \,\bigr)
  \nonumber\\[2mm]
%%%%
 &
 \underset{\rmii{\nr{eq_rtij}}}{
 \overset{\rmii{\nr{eq_rt0}}}{=}}
 \; 
 \frac{a^{-2}_{ }}{2}
 \Bigl[\;
  \overbrace{
  -3\psi''
  - \nabla^2_{ }\phi
  - 3\Hc(\phi'+\psi')
  - 6\Hc'\phi
 }^{{\rm from}\; \delta {R^0_{ }}^{ }_0 }
 \nonumber\\[2mm]
%%%%
 &\qquad\ \qquad\
 \underbrace{
 - 3\psi''
 - 3\Hc(\phi'+5\psi')
 - 6(\Hc'+2\Hc^2)\phi
 + 3\nabla^2_{ }\psi
 + \nabla^2_{ }(\psi-\phi)
 }_{{\rm from}\; \delta {R^i_{ }}^{ }_i }
  \;\Bigr]
 \nonumber\\[3mm]
%%%%
 &=-a^{-2}_{ }
 \bigl[\;
  3\psi''
 + \nabla^2_{ }(\phi-2\psi)
 + 3\Hc(\phi'+3\psi')
 + 6(\Hc'+\Hc^2)\,\phi
 \; \bigr]
 \,. \label{delta_R_scalar}
\end{align}

%% Einstein tensor %%%%%%%%%%%%%%%%%%%%%%%%%%%%%%%%

\pagebreak

We now have all the ingredients to evaluate the components 
of \eq\eqref{eq_Einstein-pert}, 
\be
 \delta G^\rmii{N}_{\mu\nu} 
 \; \overset{\rmii{\nr{eq_Einstein-pert}}}{=} \; 
 \delta R^\rmii{N}_{\mu\nu} - \frac{1}{2} 
 \bigl(\,
       \bar{g}^{ }_{\mu\nu}\, \delta R^\rmii{N}_{ }
  + 
       \delta g^\rmii{N}_{\mu\nu}\, \bar{R} 
 \,\bigr)
 \,, \label{delta_einstein_again} 
\ee
and distinguish the parts 
involving scalar (s), vector (v), or tensor (t) metric perturbations.
{}From \eqs\nr{bg_ricci_scalar}, \nr{g_munu},  
\nr{d_R_00}, and
\nr{delta_R_scalar}, 
\ba
 \delta G^\rmii{N}_{00}  
 & \overset{\rmii{\nr{delta_einstein_again}}}{=} &
 \overbrace{
  \cancel{3\psi''} + \bcancel{\nabla^2_{ }\phi}
 + 3\Hc(\cancel{\phi'}+\psi') 
 }^{{\rm from}\; \delta R^\rmiii{N}_{00}\;{\rm in}\;\nr{d_R_00} }
 \; + \;
 \hspace*{-5mm}
 \overbrace{
 \bcancel{6 (\H' + \H^2_{ })}
 }^{{\rm from}\; \bar R\;{\rm in}\;\nr{bg_ricci_scalar} \qquad }
 \hspace*{-16mm}
 \overbrace{
 \phi
  }^{\quad\qquad{\rm from}\;\delta g^\rmiii{N}_{00}\;{\rm in}\;\nr{g_munu_N}}
 \nn[2mm]
%%%%%
 &  & \;-\, \underbrace{
 \bigl[  
  \cancel{3\psi''} + \nabla^2_{ }( \bcancel{\phi} - 2\psi )
 + 3\Hc(\cancel{\phi'}+3\psi') + \bcancel{6(\Hc'+\Hc^2)\phi} 
 \bigr]}_{{\rm from}\; \delta R^\rmiii{N}_{ }\;{\rm in}\;\nr{delta_R_scalar}}
 \nn[2mm]
%%%%
 & = & 
 \underbrace{ 2( \nabla^2_{ } - 3 \H \partial^{ }_\tau )\psi }_\text{s}
 \ , \label{delta_G_00_N} 
\ea
which agrees with the Newtonian gauge limit of \eq\nr{G_00}. 
Inserting instead $\delta R^{ }_{0i}$ from \eq\nr{d_R_0i}, we find
\ba
%%%%%%%%%%%%%
 \delta G^\rmii{N}_{0i} 
 & \overset{\rmii{\nr{delta_einstein_again}}}{=} & 
 \overbrace{
 2(\psi'+\Hc\phi)_{,i}^{ }
 + \frac{1}{2}\nabla^2_{ }h_i^\rmii{N} - (\Hc'+2\Hc^2) h_i^\rmii{N}
 }^{ {\rm from}\; \delta R^\rmiii{N}_{0i}\;{\rm in}\;\nr{d_R_0i}  }
 \; + \;  
 \hspace*{-3mm}
 \overbrace{
 3 (\H' + \H^2_{ }) 
 }^{ {\rm from}\; \bar R\;{\rm in}\;\nr{bg_ricci_scalar}\qquad }
 \hspace*{-16mm}
 \overbrace{
  h^\rmii{N}_i
  }^{\quad\qquad{\rm from}\;\delta g^\rmiii{N}_{0i}\;{\rm in}\;\nr{g_munu_N}}
 \nn[2mm]
%%%%%%
 & = & 
 \underbrace{ 2(\psi'+\Hc\phi)_{,i} }_\text{s} + 
 \underbrace{ \frac{1}{2} \nabla^2_{ } h^\rmii{N}_i
  + (2\Hc'+\Hc^2)h^\rmii{N}_i }_\text{v} 
 \ , \label{delta_G_0i_N}
\ea
which agrees with the Newtonian gauge limit of \eq\nr{G_0i}. 
Finally, inserting $\delta R^{ }_{ij}$ from 
\eq\nr{d_R_ij}, we obtain
\ba
 \delta G^\rmii{N}_{ij} 
 & \overset{\rmii{\nr{delta_einstein_again}}}{=} &
 \overbrace{
 -\, \delta_{ij}^{ }\, \bigl[ \psi'' + \Hc(\phi'+5\psi')
 + 2(\Hc'+2\Hc^2)(\phi+\psi) - \nabla^2_{ }\psi \bigr]
 + (\psi-\phi)^{ }_{,ij}
 }^{ {\rm from}\; \delta R^\rmiii{N}_{ij}\;{\rm in}\;\nr{d_R_ij} }
 \nn[2mm]
%%%%
 & &
 +\,
 \overbrace{
 \frac{1}{2}(\partial^{ }_\tau + 2 \H)(h_{i,j}^\rmii{N}+h_{j,i}^\rmii{N})
 + \bigl[
    \partial_\tau^2 + 2 \H \partial^{ }_\tau - \nabla^2_{ }
    + 2(\Hc'+2\Hc^2)
    \bigr]\,\vartheta_{ij}^\rmii{N} 
 }^{ {\rm from}\; \delta R^\rmiii{N}_{ij}\;{\rm in}\;\nr{d_R_ij} }
 \nn[2mm]
%%%
 & & +\, \delta^{ }_{ij}\,
 \underbrace{ 
 \bigl[ 
  3\psi'' + \nabla^2_{ }(\phi-2\psi)
 + 3\Hc(\phi'+3\psi') + 6(\Hc'+\Hc^2)\phi
  \bigr]
 }_{ {\rm from}\; \delta R^\rmiii{N}_{ }\;{\rm in}\;\nr{delta_R_scalar} }
 \nn[2mm]
%%%%
 & & 
 + 
 \underbrace{
 6 (\H' + \H^2_{ }) 
 }_{ {\rm from}\; \bar R\;{\rm in}\;\nr{bg_ricci_scalar}\quad }
 \hspace*{-6mm} 
 \underbrace{
 (\psi\,\delta^{ }_{ij} - \vartheta^\rmii{N}_{ij})
 }_{\quad{\rm from}\;\delta g^\rmiii{N}_{ij}\;{\rm in}\;\nr{g_munu_N}}
 \hspace*{6mm}
 \nn[2mm]
%%%%%
 & = &
 \overbrace{
 \delta_{ij}^{ }\, \bigl[ 2 \psi'' + 2 \Hc(\phi'+2\psi')
 + 2(2 \Hc'+\Hc^2)(\phi+\psi) + \nabla^2_{ }(\phi - \psi) \bigr]
 + (\psi-\phi)^{ }_{,ij}
 }^\text{s}
 \nn[2mm]
%%%%%
 & &
 +\,
 \underbrace{
 \frac{1}{2}(\partial^{ }_\tau + 2 \H)(h_{i,j}^\rmii{N}+h_{j,i}^\rmii{N})
 }_\text{v}
 + 
 \underbrace{
 \bigl[
    \partial_\tau^2 + 2 \H \partial^{ }_\tau - \nabla^2_{ }
    - 2(2\Hc'+\Hc^2)
    \bigr]\,\vartheta_{ij}^\rmii{N}
 }_\text{t} 
 \ . \label{delta_G_ij_N}
\ea
This agrees with the Newtonian gauge limit of \eq\nr{G_ij}. 

\newpage

%%%%%%%%%%%%%%%%%%%%%%%%%%%%%%%%%%%%%%%%%%%%%%%%%%%%%%%%%%%%%%%%%%%
%
\subsubsection{Einstein equations for a fluid in the Newtonian gauge}
\label{app:newton2}

\addcontentsline{toc}{subsection}{\App\ref{app:newton2}: 
Einstein equations for a fluid in the Newtonian gauge}

\index{Einstein equations: Newtonian gauge}

In this appendix
we complement $\delta G^\rmii{N}_{\mu\nu}$ from 
\app\ref{app:newton} by equating
it with $8\pi G \delta T^\rmii{N}_{\mu\nu}$, 
and thereby setting up the Einstein
equations. To simplify the task, we consider a fluid, from 
\eqs\nr{delta_Tmunu_fluid} and \nr{delta_Tmunu_aniso}, 
omitting the scalar field contribution from 
\eqs\nr{delta_T00_varphi}--\nr{delta_Tij_varphi}. 
In addition, we restrict ourselves to scalar perturbations.

If we equate $\delta G^\rmii{N}_{00}$ from 
\eq\nr{delta_G_00_N} with 
$
 8\pi G \delta T^\rmii{N}_{00} = 
 8 \pi G a^2_{ }(\delta e^\rmii{N}_{ } + 2 \bar{e} \phi)
$
from \eq\nr{delta_Tmunu_fluid}, and make use of the background
identity 
$
 8\pi G a^2_{ }\bar{e}  = 3\H^2_{ }
$
from \eq\nr{bg_again}, we get 
\be
 \nabla^2\psi-3\Hc(\psi'+\Hc\phi)\;=\; 4\pi G a^2_{ }\delta e^\rmii{N}_{ }
 \ , \label{delta_einstein_00_N}
\ee
which is a gauge-fixed version of \eq\nr{delta_einstein_00}. 
The scalar part of $\delta G^\rmii{N}_{0i}$ from 
\eq\nr{delta_G_0i_N} and the gauge-fixed value
$
 \delta T^\rmii{N}_{0i} = a^2_{ }(\bar{e} + \bar{p}) v^\rmii{N}_{\der i}
$
from \eq\nr{delta_Tmunu_fluid}, where the scalar part of the 
velocity was obtained from \eq\nr{eq_v-v-s}, can be integrated into 
\be
 \boxed{ 
 \quad
 \psi'+\Hc\phi \;=\; 4\pi Ga^2_{ }(\bar{e}+\bar{p})v^\rmii{N}_{ } 
 \;, \quad \vphantom{\Big|}
 }
 \label{delta_einstein_0i_N}
\ee
which is a gauge-fixed version of \eq\nr{delta_einstein_0i}.
If we insert \eq\nr{delta_einstein_0i_N} into \eq\nr{delta_einstein_00_N},
and leave only the Laplacian on the left-hand side, we find
the constraint equation
\begin{equation}
 \nabla^2\psi \;=\; 4\pi G a^2_{ } 
 \underbrace{ [\; \delta e^\rmii{N}_{ } 
           +3\Hc(\bar{e}+\bar{p})v^\rmii{N}_{ } \;] 
  }_{{\rm via}\;\nr{eq_cont-b}:\; \bar{e}\,\Delta\;{\rm from}\;\nr{Delta_v} }
 \qquad\Leftrightarrow\qquad
 \boxed{ 
 \quad
 \nabla^2\psi \;=\; 4\pi Ga^2_{ }\bar{e}\,\Delta 
 \;. \quad \vphantom{\Big|}
 }
 \label{eq_newton}
\end{equation}
Eq.~\eqref{eq_newton} has the appearance of Newtonian gravity,
with $\psi$ playing the role of the gravitational potential. 

The remaining Einstein equations come from the $ij$-components. 
The traceless scalar parts 
of \eqs\nr{delta_G_ij_N} and \nr{delta_Tmunu_aniso}
(via \eq\nr{Pi_ij_s}) yield 
\be
 \left( \partial_i^{ }\partial_j^{ }
 -\frac{1}{3}\delta_{ij}^{ }\nabla^2_{ } \right)
 (\psi-\phi) \; = \; 
8\pi G a^2_{ }\left( \partial_i^{ }\partial_j^{ }
 -\frac{1}{3}\delta_{ij}^{ }\nabla^2_{ } \right)\barpPi
 \ . \label{eq_ee-s-ij-1}
\ee
Integrating with vanishing boundary conditions, 
we obtain a second constraint equation,
\be
 \boxed{ 
 \quad
 \psi-\phi                
 \; = \; 8\pi G a^2\barpPi
 \;.
 \quad \vphantom{\Big|}
  }
 \label{eq_ee-p-c2} 
\ee 
This corresponds to \eq\nr{delta_einstein_didj}.  
Finally, after making
use of the background identity 
$
 -8\pi G a^2_{ } \bar{p} = 2 \H' + \H^2_{ }
$ 
from \eq\nr{bg_again}, the spatial trace
parts of \eqs\nr{delta_G_ij_N} and \nr{delta_Tmunu_fluid}, 
$
 [\,\delta T^\rmii{N}_{ij}\,]^{ }_\rmi{pf}
 \supset
 a^2_{ } \,
 ( \delta p^\rmii{N}_{ } -
                              2 \hspace*{0.3mm}\bar{p}
                                \hspace*{0.3mm}\psi )
                                \, \delta^{ }_{ij}
$, 
give the evolution equation  
\be
 \boxed{
 \quad 
 \psi''+\Hc(\phi' + 2\psi')
 +(2 \H' + \H^2_{ })\,\phi
 +\frac{1}{3}\nabla^2(\phi-\psi)
     \;=\;  4\pi Ga^2_{ } \delta p^\rmii{N}_{ }
 \;. 
 \quad \vphantom{\Bigg|}
 }
 \label{eq_ee-p-ev2}
\ee
This contains equivalent time-evolution information as 
\eq\nr{delta_einstein_delta_ij}.

For completeness, we also write down the Newtonian gauge versions
of the energy-momentum conservation laws from 
\eqs\nr{delta_0_Tmunu;mu} and 
\nr{delta_i_Tmunu;mu_s}, obtaining 
\begin{empheq}[box=\fbox]{align}
 \quad \vphantom{\Bigg|}
 &\nu = 0: \,   
 & \delta e^{\rmii{N}\prime}_{ }
 + 3\Hc \bigl(\, \delta e^\rmii{N}_{ }+\delta p^\rmii{N}_{ } \,\bigr)
 - (\bar{e}+\bar{p})
   \bigl(\, 3\psi' + \nabla^2_{ } v^\rmii{N}_{ } \,\bigr) 
 &\;=\; 0
 \,, 
 \label{eq_cont-p1}\\[0mm]
%%%%%%%%%%%%%%%%
 &\nu = i: \,   
 & -\delta p^\rmii{N}_{ } + (\bar{e}+\bar{p})
  \bigl(\, 4\Hc v^\rmii{N}_{ }-\phi \,\bigr)
 + \bigl[\, (\bar{e}+\bar{p})v^{\rmii{N}}_{ } \,\bigr]'
  - \frac{2}{3}\nabla^2\barpPi 
 &\;=\; 0
 \,. \quad \vphantom{\Bigg|}
 \label{eq_cont-p2}
\end{empheq}

%%%%%%%%%%%%%%%%%%%%%%% end appendices %%%%%%%%%%%%%%%%%%%%%%%%%%%

%%%%%%%%%%%%%%%%%%%%%%%%% BIBLIO %%%%%%%%%%%%%%%%%%%%%%%%%%%%%%%%
%

%%%%%%%%%%%%%%%%%%%%%%%%%%%% CHAPTER / PART II %%%%%%%%%%%%%%%%%%%%%%%%
%
\newpage

\chapter{The inflationary paradigm --- 
generation of cosmological perturbations}
\label{ch_inflation}

\setcounter{section}{4}

%%%%%%%%%%%%%%%%%%%%%%%%%%%% SECTION %%%%%%%%%%%%%%%%%%%%%%%%%%%%%%%%%%
\newpage 

\section{Idealized initial conditions and solution for
curvature perturbations}
\label{se:dS}

\paragraph{Abstract:}

Combining Einstein equations and 
a scalar field evolution equation, 
obtained in chapter~\ref{se:pert}, 
we derive 
a Mukhanov-Sasaki equation for 
a gauge-invariant curvature perturbation, 
defined in chapter~\ref{se:gauges}.
Proceeding to its solution, we review the case in which 
the inflationary period corresponds to an almost de Sitter space-time. 
We show, analytically, how the amplitude and the phase of the 
perturbations can be fixed in the distant-past vacuum, and how
the perturbations evolve as a function of conformal time, until they cross 
outside of the Hubble horizon. We demonstrate that this produces
an almost scale-invariant spectrum of curvature perturbations, 
which can subsequently be compared with observational 
data, as they were reviewed in chapter~\ref{se:obs}. 

\paragraph{Keywords:}  Mukhanov-Sasaki equation, Bunch-Davies vacuum, 
ultra-slow-roll regime, slow-roll regime,
quasi de Sitter space-time, 
exiting or crossing outside of the Hubble horizon, 
scale-invariant spectrum of curvature perturbations.

%%%%%%%%%%%%%%%%%%%%%%%%%%%%%%%%%%%%%%%%%%%%%%%%%%%
%
\subsection{Deriving the Mukhanov-Sasaki equation}
\label{ss:sasaki}

In \ch\ref{se:pert}, we have derived a general 
set of evolution equations for
first-order cosmological perturbations.
The goal of the present chapter is to solve these in a special
limit. 
Until \ch\ref{se:thermal}, 
we assume the absence of a fluid, so that the energy-momentum 
tensor is dominated by a scalar field, cf.\ 
\eqs\nr{delta_T00_varphi}--\nr{delta_Tij_varphi}. The first step
is to rephrase the equations in a gauge-invariant form, 
as outlined in \se\ref{ssec_nogauge}. This leads to a single
equation, cf.\ \eq\nr{final_vacuum},
for a gauge-invariant variable, $ \field^{ }_\varphi $, 
which according to \eqs\nr{aQ_vs_Q} and \nr{Q_vs_R}
is proportional to
the curvature perturbation, $\R^{ }_\varphi$, 
that we have introduced in \eq\nr{def_R_varphi}. 

Our starting point is the scalar field evolution equation in vacuum, 
originally given in \eq\nr{eq_field-eq}. In terms of \eq\nr{varphi_eq}, 
we drop the terms induced by interactions with a fluid, namely 
the friction $\Upsilon$ and the noise $\varrho$
(the computation is repeated in their presence
in \se\ref{ss:delta_R_thermal}), 
considering then just
\be
 {\varphi^{;\mu}_{ }}^{ }_{;\mu} 
  - V^{ }_{\der\varphi}  
 \; \underset{\scriptscriptstyle \Upsilon,\,\varrho\,\to\, 0}
   {\overset{\rmii{\nr{varphi_eq}} \lift }{=}} \; 
 0
 \;. \label{varphi_eq_0} 
\ee
Writing the scalar field as 
$
 \varphi = \bar\varphi + \delta\varphi
$, 
where $\bar\varphi$ only depends on time, 
the background (i.e.\ zeroth order) field equation, 
governing the behaviour of $\bar\varphi$, was derived 
in \eq\nr{eq-F2}, 
\ba
 && \bar\varphi\hspace*{0.3mm}{}''
 + 2 \H \bar\varphi\hspace*{0.3mm}{}'
 + a^2_{ } V^{ }_{\der\varphi}
 \; \overset{\rmii{\nr{eq-F2}}}{=} \; 
 0 
 \;, \label{bg_scalar} 
\ea
where 
$
 (...)' \equiv \partial^{ }_\tau(...)
$
and 
$
 \H \equiv a'/a
$.
Higher derivatives of $\bar\varphi$ can be obtained 
by taking time derivatives of \eq\nr{bg_scalar}. 
In particular, taking
one time derivative, and then re-substituting 
$\bar\varphi\hspace*{0.3mm}{}''$ via \eq\nr{bg_scalar}, we find
a relation that is needed later in this section, 
\ba
 \mbox{\nr{bg_scalar}}' 
 & \Rightarrow & 
 \bar\varphi\hspace*{0.3mm}{}'''
 + 2 \bigl(\, \H'\bar\varphi\hspace*{0.3mm}{}'
 + \H \bar\varphi\hspace*{0.3mm}{}'' \,\bigr)
 + a^2 \bigl(\, 2 \H V^{ }_{\der\varphi} 
  + V^{ }_{\der\varphi\varphi}\, \bar\varphi\hspace*{0.3mm}{}' \,\bigr)
 \; = \; 0 
 \nn[2mm]
%%%%
 & \overset{\rmii{\nr{bg_scalar}}\vphantom{\big | }}{\Rightarrow} & 
 \bar\varphi\hspace*{0.3mm}{}''' + 
 \bigl(\, 2 \H' - 4 \H^2 
   + a^2  V^{ }_{\der\varphi\varphi}\, \bigr)\,
  \bar\varphi\hspace*{0.3mm}{}' 
 \; = \; 
 0 
 \;. \label{bg_scalar_2}
\ea

Going to first order in perturbations, 
the equation satisfied by 
$\delta\varphi$ is derived in \se\ref{ss:pert_scalar}.
Dropping again $\Upsilon$ and $\varrho$, and writing 
$\delta V^{ }_{,\varphi} = V^{ }_{,\varphi\varphi} \delta\varphi$, 
\eq\nr{eq_field-eq-pert} 
reduces to
\ba
 \delta\varphi'' + 2 \H \delta\varphi' - \nabla^2 \delta\varphi 
 - \bigl(\, h_0' + 3 h_\rmiii{D}' + \nabla^2 h \,\bigr)
  \bar\varphi\hspace*{0.3mm}{}'
 + a^2 V^{ }_{\der\varphi\varphi} \, \delta\varphi 
 + 2 h^{ }_0\, a^2 V^{ }_{\der\varphi} 
 \;
 \underset{\scriptscriptstyle \Upsilon,\,\varrho\,\to\, 0}
 {\overset{\rmii{\nr{eq_field-eq-pert}} \lift }{=}}
 \;
 0 
 \;. \label{pert_scalar}
\ea
It will be convenient to rewrite this for the rescaled variable
$
 \delta\widehat\varphi \, \equiv \, a \delta\varphi
$. 
Noting % from $\delta\varphi = \delta\widehat\varphi / a$ 
that
\ba
 \delta\varphi\hspace*{0.3mm}' & = & \frac{1}{a} 
  \Bigl(\,  \delta\widehat\varphi\hspace*{0.5mm}'
 - \H\, \delta\widehat\varphi \,\Bigr)
 \;, \label{rsc_1}
 \\[2mm]
%%%%%%
 \delta\varphi'' + 2 \H \delta\varphi\hspace*{0.3mm}' & = &
 \frac{1}{a} 
  \Bigl(\,  \delta\widehat\varphi\hspace*{0.5mm}'' 
 - \frac{a''}{a} \delta\widehat\varphi \,\Bigr)
 \;, \label{rsc_2}
\ea 
and moving the terms involving metric perturbations
to the right-hand side, we get
\ba
 \boxed{ 
 \quad
 \delta\widehat\varphi{\bit}'' - \nabla^2 \delta\widehat\varphi 
 + \biggl( a^2 V^{ }_{\der\varphi\varphi}  - \frac{a''}{a}\biggr) 
  \, \delta\widehat\varphi 
 \;\underset{\rmii{\nr{rsc_2}}}
   {\overset{\rmii{\nr{pert_scalar}} \lift }{=}}\; 
 \bigl(\, h_0' + 3 h_\rmiii{D}' + \nabla^2_{ } h \,\bigr)
  a \bar\varphi\hspace*{0.3mm}{}' 
 -2 h^{ }_0\, a^3_{ } V^{ }_{\der\varphi} 
 \;. 
 \quad   \vphantom{\Bigg|}
 }
 \label{rescaled_pert_scalar}
\ea

{}From \eq\nr{rescaled_pert_scalar},
we see that the metric 
perturbations 
$ 
 h_0' + 3 h_\rmii{D}' + \nabla^2 h
$
and 
$h^{ }_0$ 
need to be eliminated, 
if we want to obtain a closed equation for 
$\delta\widehat\varphi$. 
For this we need to make use of perturbed Einstein equations. 

At the background level, the Einstein equations read, 
from \eqs\nr{eq_inflation} and \nr{eq_Hprime}, 
\ba
 && 
 \H^2_{ } 
 \; 
 \underset{\scriptscriptstyle \kappa\,=\, 0}{
 \overset{\rmii{\nr{eq_inflation}}\vphantom{ \big | }}{=}}
 \; 
 \frac{4\pi G}{3}
 \bigl[ (\bar\varphi\hspace*{0.3mm}{}')^2_{ }  + 2 a^2 V
 \bigr]
 \;, \quad 
 \H' 
 \;\overset{\rmii{\nr{eq_Hprime}}\vphantom{ \big | }}{=}\; 
 \frac{8\pi G}{3}
 \bigl[ -  (\bar\varphi\hspace*{0.3mm}{}')^2_{ } + a^2 V
 \bigr]
 \label{bg_einstein} 
 \;.
\ea
Here we have set $\kappa = 0$, as also done
when deriving the perturbed Einstein equations
(cf.\ the discussion around \eq\nr{friedmann_soln2}, 
and \app\ref{app:kappa}).
Taking a linear combination and then a time derivative, yields the 
further useful relations 
\ba
 \mbox{\nr{bg_einstein}}
 & \Rightarrow & 
 \H^2 - \H' \; = \; 4 \pi G (\bar\varphi\hspace*{0.3mm}{}')^2_{ }
 \quad  \Rightarrow \quad
 2\H\H' - \H'' \; = \; 8\pi G \bar\varphi\hspace*{0.3mm}{}'
                              \bar\varphi\hspace*{0.3mm}{}'' 
 \;. \label{bg_einstein_2}
\ea

At the first order in perturbations, the variable $h^{ }_0$ is conveniently
eliminated via the $0i$ components of the Einstein equations, 
cf.\ \eq\nr{delta_einstein_0i}. 
Dropping the medium part, this leads to 
\be
  \H h^{ }_0
 + \biggl( h^{ }_\rmii{D} + \frac{\nabla^2\vartheta}{3} \biggr)'
 \; \overset{\rmii{\nr{delta_einstein_0i}}\vphantom{ \big | }}
  {\underset{\scriptscriptstyle T\to 0}{=}} \; 
  \frac{ 4 \pi G \bar\varphi\hspace*{0.3mm}{}' }{a}\, \delta\widehat\varphi
 \;. \label{pert_einstein_0i} 
\ee
For eliminating 
$ 
 h_0' + 3 h_\rmii{D}' + \nabla^2 h
$, 
we make use of the linear combination of the Einstein equations
given in \eq\nr{delta_einstein_comb}. Recalling the 
$e\supset V$ and $p\supset -V$ (cf.\ \eq\nr{simpl}), this yields 
\ba
 && \hspace*{-1.5cm}
 2 \bigl( \H' + 2 \H^2 \bigr) h^{ }_0
 + \H \bigl(\, h_0' + 3 h_\rmii{D}' + \nabla^2_{ } h \,\bigr)
 \nn[2mm] 
 & + &
  \bigl(\, \partial^{2}_\tau 
    + 2  \H \partial^{ }_\tau
  - \nabla^2_{ } \,\bigr)
    \biggl( h^{ }_\rmii{D} + \frac{\nabla^2\vartheta}{3} \biggr)
 \; \overset{\rmii{\nr{delta_einstein_comb}}\vphantom{ \big | }}
  {\underset{\scriptscriptstyle T\to 0}{=}} \;  
 - 8 \pi G  
   a V^{ }_{\der\varphi}
   \, \delta\widehat\varphi
 \;. \label{from_einstein_sum}
\ea

Therefore, 
we can solve for $h^{ }_0$ from \eq\nr{pert_einstein_0i},
and for 
$ 
 h_0' + 3 h_\rmii{D}' + \nabla^2 h
$
from \eq\nr{from_einstein_sum}.
It is helpful to introduce the shorthand notation 
\be
 \boxed{
 \quad
 X \; \equiv \; h - \vartheta'
 \;, \quad
 Y \; \equiv \;  h_\rmii{D}^{ }+\frac{\nabla^2_{ }\vartheta}{3}
 \;, \quad   \vphantom{\Bigg|}
 }
 \label{shorthand}
 \index{$X$, $Y$ (gauge-variant metric perturbations)}
\ee
where $X$ has already been 
employed to derive \eqs\nr{demo_1} and \nr{demo_R}. 
We remark in passing that $X$ and $Y$ transform only with $\xi^0_{ }$ under
gauge transformations (cf.\ \eqs\nr{gauge2}--\nr{gauge4}), 
$\tilde X = X + \xi^0_{ }$, 
$\tilde Y = Y + \H \xi^0_{ }$.
In any case, the results for $h^{ }_0$ and 
$ 
 h_0' + 3 h_\rmii{D}' + \nabla^2 h
$
become
\ba
  h^{ }_0
 & \overset{\rmii{\nr{pert_einstein_0i}}}{=} & 
 - \frac{  Y' }{\H}
 + \frac{ 4 \pi G \bar\varphi\hspace*{0.3mm}{}' }{a \H}
   \, \delta\widehat\varphi
 \;, 
 \label{h0_repl}
 \\[2mm]
%%%%%
  h_0' + 3 h_\rmii{D}' + \nabla^2 h 
 & \overset{\rmii{\nr{from_einstein_sum}}}{=} &  
 - 2 \biggl( 2 \H + \frac{\H'}{\H} \biggr) h^{ }_0
 - \frac{ \bigl( \partial^{2}_\tau 
    + 2  \H \partial^{ }_\tau
  - \nabla^2 \bigr) Y }{\H}
 - \frac{ 8 \pi G  
   a V^{ }_{\der\varphi}}{\H}
   \, \delta\widehat\varphi
 \;. 
 \nn \label{hD_repl}
\ea

We now express the inflaton perturbation as 
\be
 \delta\widehat\varphi
 \; 
  =
 \;
 \field^{ }_\varphi
 - 
 \frac{a \bar\varphi\hspace*{0.3mm}{}'}{\H}
  \, Y
 \;, \label{delta_widehat_varphi}
\ee
where we have defined 
(cf.\ \eqs\nr{eq_R-flat} and \nr{def_Q_varphi})
\ba
 \field^{ }_\varphi
 & \equiv & 
 a \Q^{ }_\varphi
 \;, \label{aQ_vs_Q} \\[3mm]
 \index{$\field^{ }_\varphi$ (conformal version of $\Q^{ }_\varphi$)}
%%%%%
 \Q^{ }_\varphi 
 & \overset{\rmii{\nr{def_Q_varphi}}}{\equiv} & 
 \delta\varphi + 
 \frac{\bar\varphi\hspace*{0.3mm}{}'}{\H}
 \biggl( 
 \underbrace{
  h^{ }_\rmii{D} + \frac{\nabla^2\vartheta}{3}
 }_{\nr{shorthand}:\; Y}
 \biggr)
 \; \overset{\rmii{\nr{eq_R-flat}}}{=} \; 
 -\frac{ \bar\varphi\bit' }{ \H }
 \R^{ }_\varphi
 \;. \label{Q_vs_R} 
\ea
Here $ \R^{ }_\varphi $ 
is the gauge-invariant curvature perturbation, 
defined in \eq\nr{def_R_varphi}.

When we insert \eqs\nr{h0_repl}--\nr{delta_widehat_varphi} 
into \eq\nr{rescaled_pert_scalar}, there are many appearances of $Y$. 
On the left-hand side (L), 
\ba
 \mbox{\nr{rescaled_pert_scalar}}^{ }_\rmii{L}
 & \supset & 
 - 
 \biggl(\frac{a \bar\varphi\hspace*{0.3mm}{}'}{\H}\biggr)''
  \, Y
 - 2
 \biggl(\frac{a \bar\varphi\hspace*{0.3mm}{}'}{\H}\biggr)'
  \, Y'
 - 
 \frac{a \bar\varphi\hspace*{0.3mm}{}'}{\H}
  \,
 \biggl( \partial_\tau^2 - \nabla^2_{ }
 + a^2 V^{ }_{\der\varphi\varphi}  - \frac{a''}{a}
 \biggr)\,Y
 \;. \hspace*{8mm} \label{ms_Y_lhs}
\ea
On the right-hand side (R), 
\ba
 \mbox{\nr{rescaled_pert_scalar}}^{ }_\iR
 & \overset{\rmii{\nr{h0_repl}--\nr{delta_widehat_varphi}} \lift }
   {\underset{\rmii{ }}{\supset}} & 
 \overbrace{
 2 \biggl[ 
 \frac{a \bar\varphi\hspace*{0.3mm}{}'}{\H}
 \biggl( 2 \H + \frac{\H'}{\H} \biggr)
 + \frac{a^3_{ }V^{ }_{\der\varphi}}{\H}
 \biggr]
 }^{{\rm coefficient~of}\;(-h^{ }_0)H}
%%%
 \overbrace{
 \biggl[ Y' 
 + \frac{4\pi G (\bar\varphi\hspace*{0.3mm}{}')^2_{ }}{\H} Y\biggr]
 }^{{\rm from}\;(-h^{ }_0)H\;{\rm via}\;\nr{h0_repl}}
 \nn[2mm]
%%%%%
 &  & \; -\, 
 \frac{a \bar\varphi\hspace*{0.3mm}{}'}{\H}
 \bigl(\, \partial^{2}_\tau 
    + 2  \H \partial^{ }_\tau
  - \nabla^2_{ } \,\bigr) Y
 +  \biggl(\frac{a \bar\varphi\hspace*{0.3mm}{}'}{\H}\biggr)^2_{ }
    \, 8 \pi G a V^{ }_{\der\varphi} \, Y
 \;. \label{ms_Y_rhs} 
\ea
Subtracting the terms and inserting 
$
 a''/a = \H' + \H^2_{ }
$, 
$(\partial_\tau^2 - \nabla^2_{ })Y$ cancels, 
and we find
\ba
 \mbox{\nr{rescaled_pert_scalar}}^{ }_\iR
 - 
 \mbox{\nr{rescaled_pert_scalar}}^{ }_\rmii{L}
 & \underset{\rmii{\nr{ms_Y_lhs}}}
   {\overset{\rmii{\nr{ms_Y_rhs}}}{\supset}} &
 \biggl[ 
  2 \biggl(\frac{a \bar\varphi\hspace*{0.3mm}{}'}{\H}\biggr)'
 + 
  a \bar\varphi\hspace*{0.3mm}{}' \biggl( 
  4 - 2 + \frac{2\H'}{\H^2_{ }}
  \biggr)
 + \frac{2 a^3_{ }V^{ }_{\der\varphi}}{\H}
 \biggr]\, Y'
 \nn[2mm]
%%%%%
 & + & 
 \biggl\{ 
   \biggl(\frac{a \bar\varphi\hspace*{0.3mm}{}'}{\H}\biggr)''
 + 
   \frac{a \bar\varphi\hspace*{0.3mm}{}'}{\H}
 \biggl[
   a^2 V^{ }_{\der\varphi\varphi}  - \H' - \H^2_{ }
 \nn[2mm]
%%%%%
 &  & \;+\,  
  \biggl( 2 \H + \frac{\H'}{\H} \biggr)
  \frac{8\pi G (\bar\varphi\hspace*{0.3mm}{}')^2_{ }}{\H}
 + \frac{16\pi G a^2_{ }
 V^{ }_{\der\varphi}\bar\varphi\hspace*{0.3mm}{}'}{\H}
 \biggr]
 \biggr\}\, Y
 \;. \hspace*{8mm} \label{subtr_Y}  
\ea
We now use the background identity in \eq\nr{bg_scalar} to write
\be
   \biggl(\frac{a \bar\varphi\hspace*{0.3mm}{}'}{\H}\biggr)'
 \; =  \; 
  \frac{a \bar\varphi\hspace*{0.3mm}{}''}{\H}  + 
   a \bar\varphi\hspace*{0.3mm}{}' \biggl( 1 - \frac{\H'}{\H^2} \biggr)
 \; \overset{\rmii{\nr{bg_scalar}}}{=} \;
   a \bar\varphi\hspace*{0.3mm}{}' \biggl( - 1 - \frac{\H'}{\H^2} \biggr)
   - \frac{a^3_{ }V^{ }_{\der\varphi}}{\H}
 \;, \label{mg_1}
\ee
and see that the coefficient of $Y'$ cancels in \eq\nr{subtr_Y}.
So we are on the right track! 

To simplify the coefficient of $Y$, 
we take the second derivative of \eq\nr{mg_1}, obtaining
\ba
   \biggl(\frac{a \bar\varphi\hspace*{0.3mm}{}'}{\H}\biggr)''
 & \overset{\rmii{\nr{mg_1}}}{=}  & 
     a \bar\varphi\hspace*{0.3mm}{}'' \biggl( - 1 - \frac{\H'}{\H^2} \biggr)
  +  a \bar\varphi\hspace*{0.3mm}{}' \biggl[-\H - \frac{\H'}{\H}
       - \frac{\H''}{\H^2} + \frac{2(\H')^2_{ }}{\H^3}
       - \frac{a^2_{ }V^{ }_{\der\varphi\varphi}}{\H} \biggr]
 \nn[2mm]
 &  & \;-\,
 a^3_{ }V^{ }_{\der\varphi} \biggl( 3 - \frac{\H'}{\H^2_{ }} \biggr)
 \label{dd_hatm_1} \\[2mm]
%%%%%
 &
  \underset{\rmii{\nr{bg_einstein_2}}}
  {\overset{\rmii{\nr{bg_scalar}} \lift }{=}} 
 &
  \frac{ a \bar\varphi\hspace*{0.3mm}{}' }{\H}
  \biggl[\,
         \biggl( \H + \frac{\H'}{\H} \biggr) 
  \;
  \overbrace{
       \biggl( 2 \H 
  +\frac{a^2_{ }V^{ }_{\der\varphi}}{\bar\varphi\hspace*{0.3mm}{}'} \biggr)
   }^{~\rm from~\bar\varphi\hspace*{0.3mm}{}'' }
  - \H^2_{ } - \H'
  + 
  \overbrace{ 
  \frac{8\pi G \bar\varphi\hspace*{0.3mm}{}'
  \bar\varphi\hspace*{0.3mm}{}'' - 2 \H \H'}{\H}
  }^{\rm from~-\H''}
  \nn[2mm]
%%%%%
   &  & \;+\,
    \frac{2(\H')^2_{ }}{\H^2} - a^2_{ }V^{ }_{\der\varphi\varphi}
     - \frac{a^2_{ }V^{ }_{\der\varphi}}{\bar\varphi\hspace*{0.3mm}{}'}
      \biggl( 3\H - \frac{\H'}{\H} \biggr)
   \biggr]
 \\[2mm]
%%%%%
 &
  \underset{\rmii{\nr{bg_einstein_2}}}
  {\overset{\rmii{\nr{bg_scalar}} \lift }{=}} 
 &
  \frac{ a \bar\varphi\hspace*{0.3mm}{}' }{\H}
  \biggl[
  \;
   \H^2 - \H' + \frac{2(\H')^2_{ }}{\H^2}
  + 
       \overbrace{ \biggl( \H + \frac{\H'}{\H}
                         - 3\H + \frac{\H'}{\H}  \biggr) }^
                   {{\rm becomes}\;-8\pi G
          (\bar\varphi\hspace*{0.3mm}{}')^2_{ }/\H}
       \frac{a^2_{ }V^{ }_{\der\varphi}}{\bar\varphi\hspace*{0.3mm}{}'} 
  \nn[2mm]
%%%%%
   &  & \;-\,  a^2_{ }V^{ }_{\der\varphi\varphi}
     - \frac{8\pi G (\bar\varphi\hspace*{0.3mm}{}')^2_{ }}{\H}
  \overbrace{
       \biggl( 2 \H 
  +\frac{a^2_{ }V^{ }_{\der\varphi}}{\bar\varphi\hspace*{0.3mm}{}'} \biggr)
   }^{~\rm from~\bar\varphi\hspace*{0.3mm}{}'' }
   \biggr]
 \\[2mm]
%%%%%
 &
  \overset{\rmii{\nr{bg_einstein_2}}}{=} 
 &
  \frac{ a \bar\varphi\hspace*{0.3mm}{}' }{\H}
  \biggl[
  \;
   \H^2 + \H' +
   \hspace*{-5mm}
   \overbrace{ 
   \frac{2\H' (\H' - \H^2_{ })}{\H^2} }^
   {{\rm becomes}\; -\H' 8\pi G (\bar\varphi\hspace*{0.3mm}{}')^2_{ } / \H^2}
   \hspace*{-5mm}
  - 16 \pi G (\bar\varphi\hspace*{0.3mm}{}')^2_{ }
  - \frac{16 \pi G a^2_{ }
          V^{ }_{\der\varphi}\bar\varphi\hspace*{0.3mm}{}'}{\H}
  -  a^2_{ }V^{ }_{\der\varphi\varphi}
   \biggr]
 \;. 
 \nn[2mm]
 \label{mgc_2}
\ea
From here, we see that 
the coefficient of $Y$ also cancels in \eq\nr{subtr_Y}.

The vanishing of \eq\nr{subtr_Y} constitutes 
a check of gauge invariance
in the sense of \se\ref{ssec_nogauge}.
To summarize, once we replace 
$ 
 \delta\widehat\varphi 
$
by the gauge-invariant
$
 \field^{ }_\varphi
$, 
and make use of background identities, 
all appearances of 
$
  Y = h^{ }_\rmii{D} + \frac{\nabla^2\vartheta}{3}  
$
cancel. 
Of course, had we chosen the spatially flat case, 
cf.\ \eq\nr{eq_flat}, 
then appearances of  
$
  h^{ }_\rmii{D} + \frac{\nabla^2\vartheta}{3}  
$
would have cancelled all along. Thus the end result would have
been the same, and the computation would have been simpler, 
but we would have missed a powerful crosscheck. Also, gauge-fixed
computations run the danger that we might inadvertently consider
gauge-variant quantities; if this happens, and we combine results
from different gauges, the results are no longer correct.  

\vspace*{3mm}

Let us finally work out the physical (gauge-invariant) 
evolution equation. 
We again insert \eqs\nr{h0_repl}--\nr{delta_widehat_varphi} 
into \eq\nr{rescaled_pert_scalar}, but now we can omit
all appearances of $Y$, since they have been verified to cancel. 
Then 
\ba
 \mbox{\nr{rescaled_pert_scalar}}^{ }_\rmii{L}
 & \overset{\rmii{\nr{delta_widehat_varphi}}\vphantom{\big | }}
 {\underset{\scriptscriptstyle Y\,\to\, 0}{=}} & 
 \field_\varphi^{\hspace*{0.3mm}\prime\prime}
 - \nabla^2_{ } \field^{ }_\varphi 
 + \biggl( a^2_{ } V^{ }_{\der\varphi\varphi}  - \frac{a''}{a}\biggr) 
  \, \field^{ }_\varphi
 \;,
 \label{ms_1} 
 \\[2mm]
%%% 
 \mbox{\nr{rescaled_pert_scalar}}^{ }_\iR
 & \overset{\rmii{\nr{h0_repl}--\nr{delta_widehat_varphi}}\vphantom{\big | }}
            {\underset{\scriptscriptstyle Y\,\to\, 0}{=}} & 
 \overbrace{
 -2 \biggl(
 2 \H + \frac{\H'}{\H} 
 + \frac{a^2_{ }V^{ }_{\der\varphi}}{\bar \varphi'}
 \biggr)
 }^{
 {\rm coefficient~of}\; h^{ }_0\, a \bar\varphi\hspace*{0.3mm}{}'
 }
 \hspace*{-3mm}
 \overbrace{
 \frac{4\pi G (\bar\varphi\hspace*{0.3mm}{}')^2_{ }}{\H} \field^{ }_\varphi
 }^{
 {\rm from}\; h^{ }_0 \, a \bar\varphi\hspace*{0.3mm}{}'
 \;{\rm via}\;\nr{h0_repl}
 }
 \;
 -
 \;
 \frac{8\pi G a^2_{ }V^{ }_{\der\varphi}\bar\varphi\hspace*{0.3mm}{}'}{\H}
   \field^{ }_\varphi
 \;. \hspace*{8mm}
 \label{ms_2}
\ea
Moving all the terms on the left-hand side, this can be expressed as 
\be
 \mbox{\nr{rescaled_pert_scalar}}^{ }_\rmii{L}
 - 
 \mbox{\nr{rescaled_pert_scalar}}^{ }_\iR
 \; 
 = 
 \; 
 \boxed{ 
 \quad
 \bigl[\,
 \partial_\tau^2 - \nabla^2 
 + \widehat{m}^2_{ }(\tau) 
 \,\bigr]
 \field^{ }_\varphi
 \; = \; 0
  \;.
 \quad \vphantom{\frac{T'}{q}}
 }
 \label{final_vacuum}
 \index{curvature perturbations: equation}
 \index{Mukhanov-Sasaki equation: in $\tau$}
\ee
This is known as the {\em Mukhanov-Sasaki equation}~\cite{sasaki,mukhanov}.
Furthermore, 
the mass parameter can be simplified by making use of the 
same information that was used for showing gauge 
invariance in \eq\nr{mgc_2}
(inserting again
$
 a''/a = \H' + \H^2_{ }
$),
\ba
 \widehat{m}^2_{ }(\tau) 
 & 
  \underset{\rmii{\nr{ms_2}}}
  {\overset{\rmii{\nr{ms_1}} \lift }{\equiv}} 
 &
 a^2 V^{ }_{\der\varphi\varphi}  - \H' - \H^2_{ }
 + 
 2 \biggl(
 2 \H + \frac{\H'}{\H} 
 + \frac{a^2_{ }V^{ }_{\der\varphi}}{\bar \varphi'}
 \biggr)
 \frac{4\pi G (\bar\varphi\hspace*{0.3mm}{}')^2_{ }}{\H} 
 + \frac{8\pi G a^2_{ }V^{ }_{\der\varphi}\bar\varphi\hspace*{0.3mm}{}'}{\H}
 \nn[2mm]
%%%%%
 & = & 
  a^2 V^{ }_{\der\varphi\varphi}  - \H' - \H^2_{ }
 + \frac{\H'\, 8\pi G (\bar\varphi\hspace*{0.3mm}{}')^2_{ }}{\H^2_{ }} 
 + 16 \pi G (\bar\varphi\hspace*{0.3mm}{}')^2_{ }
 + \frac{16\pi G a^2_{ }V^{ }_{\der\varphi}\bar\varphi\hspace*{0.3mm}{}'}{\H}
 \nn[2mm]
%%%%%
 & \overset{\rmii{\nr{mgc_2}}\vphantom{ \big | }}{\Rightarrow} & 
 \boxed{
 \quad
 \widehat{m}^2_{ }(\tau) 
 \; = \;
 - \frac{\H}{a\bar\varphi\bit'}
   \biggl( \frac{a\bar\varphi\bit'}{\H} \biggr)''
 \;. 
 \quad
 }
 \label{final_vacuum_2}
\ea
\index{$\widehat{m}^2$ (effective mass parameter)}

We conclude this section by remarking 
that even though the inflaton potential $V(\varphi)$ 
does not appear explicitly 
in \eq\nr{final_vacuum_2}, the effective 
mass squared, $\widehat{m}^2_{ }$, still depends on it via the 
background solution. For instance, 
suppose that we go to Minkowskian space-time by 
setting $a = a^{ }_0 + a^{ }_1 \tau$
and considering the limit $a_1^{ }\to 0$. 
Then $\H \to a^{ }_1 / a^{ }_0$, and 
$\widehat{m}^2_{ } 
\to -\bar\varphi\hspace*{0.3mm}{}'''/\bar\varphi\hspace*{0.3mm}{}'$
from \eq\nr{final_vacuum_2}. The third derivative
can be obtained from \eq\nr{bg_scalar_2}, where
$\lim_{a_1\to 0}( \H' - 2 \H^2 ) = 0$.
Therefore 
$\lim_{a_1\to 0} \widehat{m}^2_{ } = a^2_{ } V^{ }_{,\varphi\varphi}$, 
reproducing the usual Minkowskian definition of mass squared as the curvature
of a potential.  

%%%%%%%%%%%%%%%%%%%%%%%%%%%%%%%%%%%%%%%%%%%%%%%%%%%%%%%%%%%%%%%%%%%%%%%%%
%
\subsection{Initial conditions from the Bunch-Davies vacuum}
\label{ss:initial}

\index{initial conditions from Bunch-Davies vacuum}

\index{Bunch-Davies vacuum}

Equation~\nr{final_vacuum} is a second-order partial differential 
equation. We can Fourier transform it in spatial
directions, and then we are faced with a second-order ordinary differential
equation in time. In order to specify a solution, two initial conditions
are needed, the value of~$\field^{ }_\varphi$ and its
time derivative, $\field^{\hspace*{0.3mm}\prime}_\varphi$. 
The idea of the inflationary paradigm is that  
when a given mode is deep inside the Hubble horizon, its 
perturbations are localized quantum fluctuations, not 
affected by the expansion of the universe. In terms of 
\eq\nr{final_vacuum_3}, this means that 
$\lim_{\tau\to -\infty} \widehat{m}^2_{ }(\tau) = 0$
(cf.\ \eq\nr{estimates}),  
so that we have a time-independent Minkowskian-looking
equation. Thus we require that $\lim_{\tau\to-\infty}\field^{ }_\varphi$
corresponds to a canonically normalized 
{\em quantum field}. \index{quantum field} 
The corresponding 
{\em ground state} \index{ground state}
at $\tau\to-\infty$ is referred to as a Bunch-Davies
vacuum~\cite{bd}. 
Effectively, what we do is to replace
the classical $\field^{ }_\varphi$ by a quantum-mechanical operator, 
and re-interpret the time-dependent 
problem as a determination of the corresponding
{\em mode function}. \index{mode function}

To proceed, let us simplify the notation, dropping
the subscript from $\field^{ }_\varphi$, so that the equation 
reads
\be
 \field\bit{}'' 
 + 
   \underbrace{ 
   \bigl[\, 
       -\nabla^2 
       +\widehat{m}^2_{ }(\tau)
    \,\bigr]}_
   {\;\equiv\; \hat\epsilon^{2}_{\vec{x}}(\tau) } 
   \field 
 \; 
 \overset{\rmii{\nr{final_vacuum}}}{=} 
 \; 
 0
 \;. \label{final_vacuum_3} 
\ee
In general, a solution of \eq\nr{final_vacuum_3}
can be found with a {\em mode expansion}, 
\be
 \field(\tau,\vec{x})
 \;=\; 
 \int \! \frac{{\rm d}^3\vec{k}}{\sqrt{(2\pi)^{3}_{ }}} \, 
 \Bigl[ \,
    w^{ }_\rmii{\vec{k}}
    \, \field^{ }_k(\tau) 
    \, e^{ i \vec{k}\cdot\vec{x}}_{ }   
  + 
    w^{\dagger}_\rmii{\vec{k}}
    \, \field^{\,*}_k(\tau) 
    \, e^{ - i \vec{k}\cdot\vec{x}}_{ }   
 \, \Bigr]
 \;, \label{mode_expansion}
 \index{mode expansion}
\ee
where $k \equiv |\vec{k}|$ and the mode function, 
$\field^{ }_k$, 
satisfies 
\be
 \field\bit{}_k'' 
 + 
 \underbrace{ 
 \bigl[\, 
   k^2 
       +\widehat{m}^2_{ }(\tau)
%       - \frac{\H}{a\bar\varphi\bit'}
%   \biggl( \frac{a\bar\varphi\bit'}{\H} \biggr)''
 \,\bigr] 
 }_{\;\equiv\; \hat\epsilon^{2}_k(\tau) }
 \field^{ }_k 
 \; 
 \underset{\rmii{\nr{mode_expansion}}}
 {\overset{\rmii{\nr{final_vacuum_3}}}{=}} 
 \; 
 0
 \;. \label{mode_functions} 
\ee
\index{$\hat{\epsilon}_k^2$ (energy-like variable)}
The coefficients $w^{ }_\vec{k}$ and $w^\dagger_\vec{k}$
are quantum-mechanical operators, normalized as 
\be
 [\, w^{ }_\vec{k} \;,\; w^{ }_\vec{q} \,]
 \; = \; 
 [\, w^{ \dagger }_\vec{k} \;,\; w^{ \dagger }_\vec{q} \,]
 \; = \; 
 0
 \;, \qquad
 [\, w^{ }_\vec{k} \;,\; w^{ \dagger }_\vec{q} \,]
 \; = \; 
 \delta^{(3)}_{ }(\vec{k-q})
 \;. \label{commutators}
\ee
To specify the full solution, 
the task is to fix the normalization
and first derivative of the mode function
at some initial time. 

Let us define a {\em Wronskian} \index{Wronskian}
between two mode functions as 
\be
 \W\bigl[\,
    \field^{\scriptscriptstyle (1)}_k,
    \field^{\scriptscriptstyle (2)}_k
   \,\bigr]
 \; \equiv \; 
 \field^{\scriptscriptstyle (1)}_k 
 \, (\field^{\scriptscriptstyle (2)}_k)'
 - 
 (\field^{\scriptscriptstyle (1)}_k)' 
 \,  \field^{\scriptscriptstyle (2)}_k
 \;. \label{wronskian_def}
 \index{$\W$ (Wronskian)}
\ee
It follows from \eq\nr{mode_functions} that the Wronskian
is independent of time, 
\be
 \W_{ }^{\hspace*{0.3mm}\prime} 
 \; \overset{\rmii{\nr{wronskian_def}}}{=} \;
 \field^{\scriptscriptstyle (1)}_k 
 \, (\field^{\scriptscriptstyle (2)}_k)''
 - 
 (\field^{\scriptscriptstyle (1)}_k)''
 \,  \field^{\scriptscriptstyle (2)}_k
 \; \overset{\rmii{\nr{mode_functions}}}{=} \; 
 - \hat\epsilon^{2}_k(\tau) \, 
 \bigl[\,
 \field^{\scriptscriptstyle (1)}_k
 \, \field^{\scriptscriptstyle (2)}_k
 - 
 \field^{\scriptscriptstyle (1)}_k 
 \,  \field^{\scriptscriptstyle (2)}_k
 \,\bigr]
 \; = \; 
 0
 \;. \label{W_indep_tau}
\ee
Its correct value can be deduced by inspecting 
the {\em canonical commutation relation} 
\index{canonical commutation relations}
following from \eq\nr{mode_expansion}, 
\ba
 & & \hspace*{-2.5cm}
 \bigl[\,
   \field(\tau,\vec{x}) \;,\; \partial^{ }_\tau \field(\tau,\vec{y}) 
 \,\bigr]
 \nn[2mm]
%%%%
 & \overset{\rmii{\nr{mode_expansion}}}{=} & 
 \int \! \frac{{\rm d}^3_{ }\vec{k}\, {\rm d}^3_{ }\vec{q}}
              {(2\pi)^3_{ }} \,
 \Bigl\{\, 
 \bigl[\, w^{ }_\vec{k}\,,\,w^\dagger_\vec{q}\,\bigr]
 \field^{ }_k(\tau)\,
 \field^{*\prime}_q(\tau)\,
 e^{i\vec{k}\cdot\vec{x} - i \vec{q}\cdot\vec{y}}_{ } 
 \nn[2mm]
%%%%
 & & \hspace*{1.8cm}
 + \, 
 \bigl[\, w^{\dagger}_\vec{k}\,,\,w^{ }_\vec{q}\,\bigr]
 \field^{*}_k(\tau)\,
 \field^{\hspace*{0.3mm}\prime}_q(\tau)\,
 e^{i\vec{q}\cdot\vec{y} - i\vec{k}\cdot\vec{x} }_{ } 
 \,\Bigr\} 
 \nn[2mm]
%%%%
 & \overset{\rmii{\nr{commutators}}}{=} & 
 \int \! \frac{{\rm d}^3_{ }\vec{k}}
              {(2\pi)^3_{ }} \,
 \Bigl\{\, 
 \field^{ }_k(\tau)\,
 \field^{*\prime}_k(\tau)\,
 e^{i\vec{k}(\cdot\vec{x} - \vec{y})}_{ } 
 - 
 \field^{\hspace*{0.3mm}\prime}_k(\tau)\,
 \field^{*}_k(\tau)\,
  e^{i\vec{k}\cdot(\vec{y} - \vec{x})}_{ } 
 \,\Bigr\}
 \nn[2mm]
%%%%
 & \underset{\rmii{\nr{mode_functions}}}
            {\overset{\scriptscriptstyle \vec{k} \,\to\, -\vec{k}}{=}}& 
 \int \! \frac{{\rm d}^3_{ }\vec{k}}
              {(2\pi)^3_{ }} \,
 \W[\field^{ }_k(\tau), 
 \field^{\,*}_k(\tau)]
 \,  
  e^{i\vec{k}\cdot(\vec{x} - \vec{y})}_{ } 
 \;.
\ea
For the last equality, we observe from \eq\nr{mode_functions} 
that the mode functions are even in $\vec{k}\to -\vec{k}$.
It then follows that we obtain the canonical value, 
$i \delta^{(3)}_{ }(\vec{x-y})$, 
if we set 
\be
 \W\bigl[\, 
    \field^{ }_k(\tau), 
    \field^{\,*}_k(\tau)
    \,\bigr]
 \;=\; i \quad \forall k
 \;.  \label{wronskian_value}
\ee
This fixes one of the integration constants. 

The other initial condition is that the
mode function $\field^{ }_k$, associated with $w^{ }_\vec{k}$, 
should correspond to the ``{\em forward-propagating}'' 
or ``{\em positive-energy}'' mode at early times.
\index{forward-propagating mode}
\index{positive-energy mode} 
The logic is that $w^{ }_\vec{k}$ corresponds to 
an annihilation operator, and a vacuum state is
defined by 
$
 w^{ }_\vec{k} | 0 \rangle = 0 
$ 
for every~$\vec{k}$. 
Since $\field^{ }_k$ multiplies $w^{ }_\vec{k}$ in 
\eq\nr{mode_expansion}, putting the positive-energy 
modes in $\field^{ }_k$ guarantees that there are 
no forward-propagating particles
in the initial state~$| 0 \rangle$. 
In concrete terms, recalling the Schr\"odinger equation, 
this is implemented as 
a ``root'' of \eq\nr{mode_functions},  
\be
 \lim_{\tau\to-\infty}
 i \partial^{ }_\tau \field^{ }_k 
 \;=\; 
 +  
 \lim_{\tau\to-\infty}
       \hspace*{-2mm}
       \underbrace{
           %\sqrt{ 
           \hat\epsilon^{ }_k(\tau)
           %}
        }_
       {{\rm from}\;\nr{mode_functions}}
       \hspace*{-3mm}
       \, \field^{ }_k
 \;. \label{derivative_value}
\ee 
Thereby the value and the first derivative of the mode
function are related to each other. With this additional
information, the solution is fully specified. 
We note that $ \field^{ }_k \in \mathbbm{C}$.

%%%%%%%%%%%%%%%%%%%%%%%%%%%%%%%%%%%%%%%%%%%%%%%%%%%%%%%%%%%%%%%%%%%%%%%%%
%
\subsection{Solution during slow-roll expansion}
\label{ss:soln_dS}

\index{quasi de Sitter space-time}

In \se\ref{ss:initial} we have determined the  
initial conditions satisfied by
the mode functions, $\field^{ }_k(\tau)$.
We can then solve for their time dependence, from 
\eq\nr{mode_functions}. 
In this section we do this by assuming approximately 
de Sitter expansion, $H \approx $ constant, which in physical 
time is expressed as 
\be
 \biggl| \frac{\Delta H}{H} \biggr|
 \; \approx \; 
 \biggl| \frac{\Delta t \, \dot{H}}{H_{ }} \biggr|
 \; \overset{\scriptscriptstyle ~~\Delta t\,\equiv\, H^{-1}_{ }
             \lift }{=} \; 
 \biggl| \frac{\dot{H}}{H^2_{ }} \biggr|
 \quad \ll \quad 1
 \;. \label{quasi_de_Sitter}
\ee 
We remark that among
the earliest examples of a quantum-mechanical computation 
in such a context, 
though not explicitly for inflation, 
can be found in ref.~\cite{mc}, whereas some of the first
inflationary computations were presented in 
refs.~\cite{inf1,inf2,inf3,zeta_copy}.

For an exactly constant $H$,  the background 
solution was determined around \eq\nr{a_tau_dS_2},
\be
 a 
 \; \overset{\rmii{\nr{a_tau_dS_2}}}{=} \; 
 -\frac{1}{H \tau}
 \;, \quad \tau \in (-\infty,\tau^{ }_e) 
 \;. \label{a_eta_2}
\ee
Here we adopt this functional form also for the case that $H$
may change adiabatically with time. 
We refer to this scenario as a {\em quasi de Sitter space-time}. 
Let us first elaborate on the conceptual foundation of this 
approximation in our context.  

In general, 
the background solution is determined by \eq\nr{bg_scalar}, 
\begin{equation}
 \bar{\varphi}'' + 2\Hc\bar{\varphi}'
 + a^2V^{ }_{\der\varphi}( \bar{\varphi} )          
 \;\overset{\rmii{\nr{bg_scalar}}}{=}\;
 0 
 \qquad 
 \underset{\rmii{\nr{eq-F2}}}{
 \overset{ \tau \leftrightarrow t \vphantom{\big | } }{\Longleftrightarrow}} 
 \qquad
 \ddot{ \bar{\varphi} } + 3H\dot{ \bar{\varphi} }
 + V^{ }_{\der\varphi}( \bar{\varphi} )
 \; = \; 0 \ . \label{eq-F2_again}
\end{equation}
Given that $H$ is approximately constant but $\H$ is not, it is 
easier to inspect the version on the right. 
The Hubble rate $H$ is itself a function
of $\bar\varphi$ and $\dot{\bar\varphi}$, as dictated
by \eq\nr{eq_inflation_t}, so the equation is non-linear.

Solutions of non-linear differential equations are often
classified according to their asymptotic behaviour at late times, 
known as {\em fixed points}. It may be observed that 
\eq\nr{eq-F2_again} does possess a fixed point corresponding to
exact de Sitter expansion. Namely, 
a constant $H$ could be obtained if we set 
$ V^{ }_{,\varphi} =0 $, i.e.\ the potential is just a cosmological
constant. The equation 
$
 \ddot{ \bar{\varphi} } + 3H\dot{ \bar{\varphi} } = 0
$
then has two solutions, 
$
 \dot{\bar\varphi}(t) = 0
$
and 
$
 \dot{\bar\varphi}(t) = \dot{\bar\varphi}(t^{ }_i)
 e^{-3 H(t- t _i)}
$.
The second solution approaches the first one exponentially fast. 
Afterwards, we find a static background field, $\dot{\bar\varphi} = 0$, 
implying eternal de Sitter expansion. 
However, strictly speaking exact de Sitter expansion 
is {\em not} a cosmologically
relevant trajectory, because observational evidence points to 
a power-like rather than exponential expansion,  
during the period when CMB was formed.

We note in passing that if a solution satisfying 
$
 \ddot{ \bar{\varphi} } + 3H\dot{ \bar{\varphi} } \approx 0
$
is present for a finite period of time, this dynamics
is known as {\em ultra-slow-roll expansion}. 
\index{ultra-slow-roll regime}
Cosmological models having this property have been built
(cf.,\ e.g.,\ refs.~\cite{usr1,usr2}), 
however mostly not for the inflationary dynamics affecting CMB modes, 
but rather for some later period, influencing shorter wavelengths. 
 
Another simple solution can be found if we look at a situation in which
$\dot{\bar\varphi}$ and 
$V^{ }_{,\varphi}$ do not vanish but are small. Adopting the language
from the harmonic oscillator, we envisage finding ourselves in 
an {\em overdamped regime}, \index{overdamped regime}
with $ |\ddot{\bar\varphi}| \ll |3 H \dot{\bar\varphi}| $. 
Then \eq\nr{eq-F2_again} reduces to a first-order differential equation, 
\be
 3H\dot{ \bar{\varphi} }
 + V^{ }_{\der\varphi}( \bar{\varphi} ) 
 \; \approx \; 0 
 \;. \label{slow_roll}
\ee
{}From here it follows that 
\be
 \frac{\bar\varphi\hspace*{0.3mm}{}'}{\H} 
 \;
 \overset{\rmii{\nr{t_tau}}}{=}
 \; 
 \frac{\dot{\bar\varphi}}{H} 
 \; \overset{\rmii{\nr{slow_roll}}}{\approx} \;
 - \frac{V^{ }_{\der\varphi}}{3 H^2}
 \;. 
 \label{const}
\ee 
If the potential and the initial value of $\bar\varphi$
satisfy certain conditions, known as 
{\em slow-roll} constraints 
(cf.\ \se\ref{ss:slow-roll}), 
then the condition 
$ |\ddot{\bar\varphi}| \ll |3 H \dot{\bar\varphi}| $ 
indeed remains valid for a relatively long period of time. 

Furthermore, it turns out that the ratio in \eq\nr{const} only evolves
slowly, and in a first approximation 
can be treated as a constant (cf.\ \se\ref{ss:slow-roll}).
If $\bar\varphi\hspace*{0.3mm}{}'/\H$ is approximately constant, and 
$a$ evolves according to \eq\nr{a_eta_2}, 
then it follows that 
the effective mass parameter from \eq\nr{final_vacuum_2} becomes 
\be
 \widehat{m}^2_{ }(\tau) 
 \;
 \overset{\rmii{\nr{final_vacuum_2}}}{=}
 \; 
 - \frac{\H}{a\bar\varphi\hspace*{0.3mm}{}'} 
 \biggl( \frac{a\bar\varphi\hspace*{0.3mm}{}'}{\H} \biggr)''
 \hspace*{-1mm}
 \;
 \overset{\scriptscriptstyle
 \;\;\;\bar\varphi\hspace*{0.3mm}{}'/\H\;\approx\;\rmii{constant}}
 {\approx}
 \hspace*{-1mm}
 \;
 - \frac{a''}{a} 
 \;
 \overset{\rmii{\nr{a_eta_2}}}{=}
 \; 
 - \frac{2}{\tau^2}
 \;. \label{estimates}
\ee
Even though this is an approximation for curvature perturbations, 
the same effective mass squared,  
$-a''/a$,
is obtained in an exact form for some 
other excitations in de Sitter space-time, such as gravitons, or 
massless minimally coupled ``spectator'' scalar fields, 
which have no background value.
With a slight abuse of notation, we therefore write
the equation for mode functions, \eq\nr{mode_functions}, 
as an equality, 
\be
 \biggl( \partial^{2}_\tau
  + 
 \underbrace{
 k^2_{ } - \frac{2}{\tau^2} 
 }_{\,\approx\, \hat\epsilon^{2}_k(\tau) }
 \biggr) \, 
 \field^{ }_{k} 
 \;
 {=} \; 0 
 \;. \label{eq_hatQ_vac}
\ee
An explicit solution can be found, 
\be
  \field^{ }_k \;=\; 
  \frac{e^{-i k \tau}}{\sqrt{2 k}}
  \biggl( 1 - \frac{i}{k\tau} \biggr)
 \;, 
 \label{Q_k}
\ee
with another solution given by the complex 
conjugate, $\field^{*}_k$. 

Let us check that the integration constants 
have been fixed correctly in \eq\nr{Q_k}. 
Taking time derivatives, we find 
\be
  \field^{\hspace*{0.3mm}\prime}_k \;=\; 
  \frac{e^{-i k \tau}}{\sqrt{2 k}}
  \biggl( -i k - \frac{1}{\tau} + \frac{i}{k\tau^2_{ }} \biggr)
  \;, \quad
  \field^{*\prime}_k \;=\; 
  \frac{e^{i k \tau}}{\sqrt{2 k}}
  \biggl( i k - \frac{1}{\tau} - \frac{i}{k\tau^2_{ }} \biggr)
  \;. \label{d_Q_k}
\ee
The Wronskian from \eq\nr{wronskian_def} then becomes
\ba
 \W\bigl[\,
   \field^{ }_k \, ,  
   \field^{*}_k  
 \,\bigr] 
 & \overset{\rmii{\nr{wronskian_def}}}{=} & \frac{1}{2k}
 \biggl[\,
  \overbrace{
  \biggl( 1 - \frac{i}{k\tau} \biggr)
  }^{{\rm from}\;\nr{Q_k}}
  \overbrace{
  \biggl( i k - \frac{1}{\tau} - \frac{i}{k\tau^2_{ }} \biggr)
  }^{{\rm from}\;\nr{d_Q_k} }
    - 
  \biggl( 1 + \frac{i}{k\tau} \biggr)
  \biggl( - i k - \frac{1}{\tau} + \frac{i}{k\tau^2_{ }} \biggr)
 \,\biggr]
 \nn[2mm]
%%%%
 & = & 
 \frac{i}{k}\im
 \biggl[\,
  \biggl( 1 - \frac{i}{k\tau} \biggr)
  \biggl( i k - \frac{1}{\tau} - \frac{i}{k\tau^2_{ }} \biggr)
 \,\biggr]
 \nn[2mm]
%%%%
 & = & 
 \frac{i}{k}\im
 \biggl[\,
   i k - \cancel{\frac{1}{\tau}} - \cancel{\frac{i}{k\tau^2_{ }}} 
  + \cancel{\frac{1}{\tau}} + \cancel{\frac{i}{k\tau^2_{ }}}
 - \frac{1}{k^2_{ }\tau^3_{ }}
 \,\biggr]
 \; = \; i
 \;, 
\ea
indicating that the normalization condition 
from \eq\nr{wronskian_value} is respected. 

Furthermore, if $k|\tau|\gg 1$, \eq\nr{eq_hatQ_vac} implies
\be
 - i\, \hat\epsilon^{ }_k(\tau)
 \;
   \overset{\rmii{\nr{eq_hatQ_vac}}}{=}
 \;
 - i \sqrt{  k^2_{ } - \frac{2}{\tau^2} }
 \; 
  \overset{\scriptscriptstyle k\;\gg\;1/|\tau|
           \lift }{\approx} 
 \; 
 - i k \, 
 \biggl(\, 
   1 - \frac{1}{k^2_{ }\tau^2_{ }} + ...
 \,\biggr)
 \; 
 = 
 \;
 -i k + \frac{i}{ k\tau^2_{ }} 
 + ... 
 \;. \label{O_k_exp}
\ee
This leads to 
\ba
 -i\, \hat\epsilon^{ }_k(\tau) \, \field^{ }_k
 & 
 = 
% \underset{\rmii{\nr{O_k_exp}}}{\overset{\rmii{\nr{Q_k}}}{=}} 
 & 
  \frac{e^{-i k \tau}}{\sqrt{2 k}}
  \overbrace{
  \biggl(\,  -i k + \frac{i}{ k\tau^2_{ }} 
 + ... 
 \,\biggr)}^{{\rm from}\;\nr{O_k_exp}}
  \overbrace{
  \biggl(\, 1 - \frac{i}{k\tau} \,\biggr)
  }^{{\rm from}\;\nr{Q_k}}
 \nn[2mm]
%%%%
 & = &
 \underbrace{
  \frac{e^{-i k \tau}}{\sqrt{2 k}}
  \biggl(\,
  -i k
  - \frac{1}{\tau}
  + \frac{i}{ k\tau^2_{ }}
 }_{\field'_k\;{\rm from}\;\nr{d_Q_k} }
  + \frac{1}{k^2_{ }\tau^3_{ }}
 + ...  
  \,\biggr)
 \;.
\ea
Therefore, \eq\nr{derivative_value} is satisfied
in the domain $k|\tau|\gg 1$,
and we conclude that \eq\nr{Q_k} 
represents a correctly normalized forward-propagating
mode function.

The next task is to compute the variance of the perturbations, 
defined like in \eq\nr{eq_angular_average}. The expectation value is 
taken in the Bunch-Davies vacuum, making use of the creation and
annihilation operators from \eq\nr{commutators}. We obtain
\ba
 \bigl\langle\, 
  \field^{\bit2}_{\varphi}(\tau,\vec{x})
 \,\bigr\rangle
 & \overset{\rmii{\nr{quantum_average}}}{\equiv} & 
 \bigl\langle\, 0 \bit \big| 
  \field^{ }_{\varphi}(\tau,\vec{x})
  \field^{ }_{\varphi}(\tau,\vec{x})
 \big| \bit 0 \,\bigr\rangle
 \label{2pt_field} \\[2mm]
%%%
 & \overset{\rmii{\nr{mode_expansion}}}{=} &
 \int \! \frac{ {\rm d}^3\vec{k} \, {\rm d}^3_{ }\vec{q} }{(2\pi)^3_{ }} 
 \, 
 \bigl\langle\, 0 \bit \big| \,
 w^{ }_\vec{k} 
 \, \field^{ }_k(\tau)
 \,  e^{i\vec{k}\cdot\vec{x}}_{}
 \, w^{\dagger}_\vec{q}
 \, \field^{*}_q(\tau)
 \, e^{- i\vec{q}\cdot\vec{x}}_{}
 \,\big| \bit 0 \,\bigr\rangle
 \nn[2mm]
%%%
 & \overset{\rmii{\nr{commutators}}}{=} & 
 \int_\vec{k} | \field^{ }_k(\tau) |^2_{ }
 \; \overset{\rmii{\nr{Q_k}}}{=} \; 
 \int_{-\infty}^{+\infty} \! {\rm d}\ln k \,
 \underbrace{   
 \frac{k^3_{ }}{2\pi^2}
 \frac{1}{2k} \biggl( 1 + \frac{1}{k^2_{ }\tau^2_{ }}\biggr)
 }_{{\rm from}\;\nr{eq_angular_average}:
  \;\P_{\mbit \scriptscriptstyle \widehat{\Q}^{ }_\varphi} (\tau,k) }
 \;. \label{P_Q_k}
\ea 
Subsequently, 
the {\em power spectrum for curvature
perturbations} from \eq\nr{Q_vs_R},
noting that
$
 \R^{ }_\varphi = - H \field^{ }_\varphi / (a\dot{\bar\varphi})
$, 
is obtained as 
\be
 \P^{ }_{\mbit \scriptscriptstyle \R^{ }_\varphi} (\tau,k) 
 \; \underset{\rmii{\nr{aQ_vs_Q},\nr{Q_vs_R}}}
    {\overset{\rmii{\nr{P_Q_k}} \lift }{ = }} \; 
 \biggl( \frac{H}{2\pi\dot{\bar\varphi}} \biggr)^2_{ }
 \biggl( \frac{k^2_{ }}{a^2_{ }} + \frac{1}{a^2_{ }\tau^2_{ }}\biggr)
 \; \overset{\rmii{\nr{a_eta_2}}}{=} \;
 \biggl( \frac{H}{2\pi\dot{\bar\varphi}} \biggr)^2_{ }
 \biggl( \frac{k^2_{ }}{a^2_{ }} + H^2_{ }\biggr)
 \;. \label{P_R_prefinal}
\ee

Now, as illustrated in \fig\ref{fig:history_tau}
on p.~\pageref{fig:history_tau}, the physical momentum
$k/a$ equals the Hubble rate $H$ at a certain
moment, which is denoted by $\tau^{ }_*$, and
decreases very fast below it afterwards, $k/a \ll H$. 
In terms of \eq\nr{eq_hatQ_vac}, this means that the 
time-dependent term $-2/\tau^2_{ }$ becomes more 
important than $k^2_{ }$, and vacuum-like 
oscillations cease. 
As will be 
illustrated numerically in 
\fig\ref{fig:P_R_varphi} on p.~\pageref{fig:P_R_varphi},
and shown in 
\ch\ref{se:outside}, 
the curvature power spectrum
{\em freezes out} \index{freeze out: curvature perturbations}
at this moment, becoming a constant. So, 
if we denote $H^{ }_* \equiv H(\tau^{ }_*)$, 
$a^{ }_* \equiv a(\tau^{ }_*)$,  
and $\dot{\bar\varphi}^{ }_* \equiv \dot{\bar\varphi}(\tau^{ }_*)$, 
then the prediction for the curvature power spectrum can be expressed as 
\be
 \P^{ }_{\mbit \scriptscriptstyle \R^{ }_\varphi}(k)
 \; \overset{\rmii{\nr{P_R_prefinal}}}{\approx} \;    
 \biggl( \frac{H_*^{\hspace*{0.3mm}2}}
              {2\pi\dot{\bar{\varphi}}^{ }_* } \biggr)^2_{ }
 \bigg|^{ }_{H^{ }_* \,=\, k/a^{ }_*}
 \;. \label{P_R_final}
 \index{power spectrum: curvature perturbations} 
 \index{curvature power spectrum: basic result}
\ee 

The prediction for the curvature power spectrum can now be compared
with the experimentally measured power spectrum.
In particular, at the pivot scale, $k^{ }_*$,
we can extract the {\em scalar amplitude} as (cf.\ \eq\nr{eq_powerlaw})
\ba
 A^{ }_\scalar 
 & 
 \overset{\rmii{\nr{eq_powerlaw}}}{=} 
 & 
 \P^{ }_{\mbit \scriptscriptstyle \R^{ }_\varphi}(k^{ }_*)  
 \;. \label{As} 
 \index{$A^{ }_\scalar$ (scalar amplitude)}
\ea
This can be compared with \eq\nr{ex_A_s}.
We remark in passing that the asterisk in $k^{ }_*$
(signifying the pivot scale)
has no relation to that in $\tau^{ }_*$
(signifying the moment of horizon crossing for any $k$).
For the {\em spectral tilt}, we write
\ba 
%%%
 n^{ }_\scalar 
 & 
  \overset{\rmii{\nr{eq_powerlaw}}}{=} 
 & 
 1 + \frac{{\rm d}\ln 
 \P^{ }_{\mbit \scriptscriptstyle \R_\varphi}(k) }
        {{\rm d}\ln k}
 \biggr|^{ }_{k\, =\, k^{ }_*}
 \; = \; 
 1 + \frac{{\rm d}\ln 
 \P^{ }_{\mbit \scriptscriptstyle \R_\varphi}(k) }
 {{\rm d} t}
 \frac{{\rm d}t}{{\rm d}\ln k}
 \biggr|^{ }_{k\, =\, k^{ }_*}
 \;. 
 \index{$n^{ }_\scalar$ (scalar spectral tilt)}
 \label{n_s_x1}
\ea
For the relation between the crossing time 
and the momentum mode $k$,
the definition $k(t^{ }_*) = a^{ }_* H^{ }_*$ in \eq\nr{P_R_final} yields 
\be
 \frac{{\rm d}\ln k}{{\rm d}t}
 \; \overset{\rmii{\nr{P_R_final}} \lift }{=}  \;
 \frac{\dot{a}^{ }_* H^{ }_* + a^{ }_* \dot{H}^{ }_*}{k}
 \; \overset{\scriptscriptstyle k \,=\, a^{ }_* H^{ }_*
             \lift }{=} \; 
 \frac{H_*^{\ibit 2} + \dot{H}^{ }_*}{H^{ }_*}
 \;. \label{dlnk_dt}
\ee
This then results in
\ba
 \boxed{
 \quad
 n^{ }_\scalar 
 \; 
 \underset{\rmii{\nr{n_s_x1},\nr{dlnk_dt}}}{
 \overset{\rmii{\nr{P_R_final}}}{\approx}} 
 \; 
 1 + \frac{2H^{ }_*}{H^{\ibit 2}_{*} + \dot{H}^{ }_*}
 \biggl(
  \frac{2\dot{H}^{ }_*}{H^{ }_*} 
 - 
  \frac{\ddot{\bar{\varphi}}^{ }_* }{\dot{\bar{\varphi}}^{ }_* }
 \biggr)
  \;,
 \quad   \vphantom{\Bigg|}
 }
  \label{ns} 
\ea
which can be compared with \eq\nr{ex_n_s}. 
Given that we have assumed 
$
 |\dot{H}^{ }_*| \ll H^{\ibit 2}_*
$
(cf.\ \eq\nr{quasi_de_Sitter})
and 
$
 |\ddot{\bar{\varphi}}^{ }_* | \ll |H\, \dot{\bar{\varphi}}^{ }_* |
$
(cf.\ \eq\nr{slow_roll}), 
we find an almost flat spectrum, 
$ n^{ }_\scalar \approx 1 $, 
corresponding to a
{\em Harrison--Zeldovich spectrum} \index{Harrison--Zeldovich spectrum}
as anticipated below \eq\nr{eq_powerlaw}.

\vspace*{3mm}

We end this chapter with three remarks:

\bi
 
\item
Tensor perturbations generated
by vacuum fluctuations undergo a similar dynamics as the curvature
perturbations that have been discussed above. We return to this topic
in \se\ref{ss:gw_vac}, obtaining then a prediction
also for the tensor-to-scalar ratio $r$ from \eq\nr{r_obs}.
Thereby we can contrast the predictions originating from 
a given inflationary potential $V$ with all stringent
observational bounds. 

\item
Vector perturbations obey a diffusive rather than 
a (second-order) wave equation, 
cf.\ the discussion around \eqs\nr{einstein_0i_v} and \nr{einstein_ij_v}. 
Therefore their dynamics differs qualitatively from that of
the curvature and tensor perturbations. 

\item
If the inflationary potential $V$ is not quadratic, we may 
treat interactions perturbatively, and proceed to compute also higher-point
correlation functions of the type in \eq\nr{2pt_field}~\cite{jm,sw_qm}
(see also ref.~\cite{uzan} for the simple case
of a ``spectator'' scalar field). This has evolved
into a large field of research in the meanwhile. Such corrections
represent a primordial source for the {\em non-Gaussianities} and 
{\em bispectrum} that were discussed around \eq\nr{f_NL}.

\index{non-Gaussianity}
\index{bispectrum}

\ei

\newpage

%%%%%%%%%%%%%%%%%%%%%%%%% BIBLIO %%%%%%%%%%%%%%%%%%%%%%%%%%%%%%%%
%

%%%%%%%%%%%%%%%%%%%%%%%%%%%% SECTION %%%%%%%%%%%%%%%%%%%%%%%%%%%%%%%%%%
\newpage 

\section{Scalar power spectrum in a general cold inflation scenario}
\label{se:cold}

\paragraph{Abstract:}

Moving on from the idealized quasi de Sitter space-time to a more
general situation in which the Hubble rate is 
a function of time, we show how 
the equation for the curvature  
perturbation can be set up in a manageable form. 
The concept of an overdamped ``slow-roll'' regime is elaborated upon, 
and we show how observable quantities can be derived as 
a power series in small parameters. 
We demonstrate how the basic equations can be rephrased through 
the so-called stochastic formalism,
introducing the notion of quantum noise. 
We also explain how the stochastic formalism can be simplified under
a number of further assumptions, leading to a framework that has
been used in the literature
for simulating non-linear aspects of inflationary dynamics. 

\paragraph{Keywords:} 
cold inflation, 
Mukhanov-Sasaki equation in physical time, 
slow-roll parameters, 
stochastic formalism, 
quantum noise, 
numerical solution, 
non-linear simulations.

\index{single-field inflation}

%%%%%%%%%%%%%%%%%%%%%%%%%%%%%%%%%%%%%%%%%%%%%%%%%%%
%
\subsection{Setup in physical time}
\label{ss:setup_num}

Let us start by recapitulating the equations that we are trying
to solve. The unknown variable of the Mukhanov-Sasaki equation is 
the dimensionless $\field^{ }_\varphi$ (cf.\ \eq\nr{final_vacuum}), 
obtained by rescaling the curvature perturbation 
$\R^{ }_\varphi$ (cf.\ \eq\nr{def_R_varphi}), 
\be
 \R^{ }_\varphi 
 \;
 \underset{\rmii{\nr{Q_vs_R}}}
 {\overset{\rmii{\nr{aQ_vs_Q}} \lift }{=}} 
 \;
 -\frac{\H}{a \bar\varphi\hspace*{0.3mm}'}
 \field^{ }_\varphi
 \;. \label{rescaling}
\ee 
Representing $\field^{ }_\varphi$ via the 
corresponding mode function, $\field^{ }_k$ 
(cf.\ \eq\nr{mode_expansion}), the power
spectrum of $\field^{ }_\varphi$ is proportional to 
$| \field^{ }_k |^2_{ }$ (cf.\ \eq\nr{P_Q_k}).
The mode function satisfies 
\be
 \boxed{
 \quad
 \field\bit{}_k'' 
 + 
 \bigl[\, 
   k^2 +\widehat{m}^2_{ }(\tau)
 \,\bigr] 
 \field^{ }_k 
 \; 
 \overset{\rmii{\nr{mode_functions}}}{=} 
 \; 
 0
 \;, \quad
 \widehat{m}^2_{ }(\tau) 
 \; \overset{\rmii{\nr{final_vacuum_2}}}{\equiv} \; 
       - \frac{\H}{a\bar\varphi\bit'}
   \biggl( \frac{a\bar\varphi\bit'}{\H} \biggr)''
 \;. 
 \quad
 }
 \label{mode_functions_again} 
 \index{$\widehat{m}^2$ (effective mass parameter)}
\ee
One benefit of the rescaling in \eq\nr{rescaling} is that no 
first-order time derivatives appear in \eq\nr{mode_functions_again}. 
Therefore one of the initial conditions, 
originating from the interpretation of 
$\field^{ }_\varphi$ as a quantum-mechanical vacuum fluctuation, 
can be expressed as 
a normalization condition that is satisfied at all times, 
\be
 \W\bigl[\, 
  \field^{ }_k(\tau), 
  \field^{\,*}_k(\tau)
 \,\bigr]
 \overset{\rmii{\nr{wronskian_value}}}{=} 
 i \quad \forall k,\tau
 \;,  \label{wronskian_value_again}
\ee
where $\W$ denotes the Wronskian (cf.\ \eq\nr{wronskian_def}).
The other initial condition 
can be viewed as a ``root'' of \eq\nr{mode_functions_again}, 
 \be
 \field^{\hspace*{0.3mm}\prime}_k 
 \;
 \underset{\scriptscriptstyle \tau \;\to\; -\infty}
 {\overset{\rmii{\nr{derivative_value}} \lift }{=}} 
 \; 
 - i 
 \underbrace{
 \sqrt{    k^2 +\widehat{m}^2_{ }(\tau) }
 }_{ \,\equiv\, \hat\epsilon^{ }_k(\tau) }
 \, \field^{ }_k
 \;, \label{derivative_value_again}
 \index{$\hat{\epsilon}_k^2$ (energy-like variable)}
\ee 
where we furthermore expect 
$
 \lim_{\tau\to-\infty}
 \widehat{m}^2_{ }(\tau)
 = 
 0
$ 
(cf.\ \eq\nr{estimates}).
We note that according to \eqs\nr{wronskian_value_again}
and \nr{derivative_value_again}, $\field^{ }_k \in \mathbbm{C}$.

Now, even if the equations above are not too complicated, 
corresponding just to a harmonic oscillator with 
a time-dependent mass, their solution requires in general 
numerical methods. The reason is that the parameter
$\widehat{m}^2_{ }(\tau)$ is obtained from a solution of 
{\em non-linear} differential equations, and 
therefore cannot be represented in closed form. 
A practical procedure is therefore to solve for 
$\widehat{m}^2_{ }(\tau)$ and 
$\field^{ }_k(\tau)$
simultaneously, starting from given initial values. 
However, in order to do this efficiently, it turns out to 
be helpful to change variables, as we describe in the following. 

In a first step, we go back to the curvature
perturbation $\R^{ }_\varphi$ from \eq\nr{rescaling}, 
but now with the mode functions, getting
\ba
  \field^{ }_k
  & \overset{\rmii{\nr{rescaling}}}{=} & 
 -\frac{a \bar\varphi\hspace*{0.3mm}'}{\H}
  \R^{ }_k
  \;, \label{Qk_vs_Rk} \\[2mm]
%%%%%%%%%%% 
  \field^{\bit\prime}_k
  & = & 
 -\biggl( \frac{a \bar\varphi\hspace*{0.3mm}'}{\H} \biggr)'
  \R^{ }_k
 -\frac{a \bar\varphi\hspace*{0.3mm}'}{\H}
  \R^{\prime}_k
  \;, \label{Q_k_d} \\[2mm]
%%%%%%%%%%% 
  \field^{\bit\prime\prime}_k
  & = & 
 -\biggl( \frac{a \bar\varphi\hspace*{0.3mm}'}{\H} \biggr)''
  \R^{ }_k
 -2\, \biggl( \frac{a \bar\varphi\hspace*{0.3mm}'}{\H} \biggr)'
  \R^{\prime}_k
 -\frac{a \bar\varphi\hspace*{0.3mm}'}{\H}
  \R^{\prime\prime}_k
  \;. \label{Q_k_dd}
\ea
Substituting these in \eq\nr{mode_functions_again}, 
and multiplying the whole with 
$
 -{\H}/({a \bar\varphi\hspace*{0.3mm}'})
$, 
yields
\be
 \R^{\prime\prime}_k
 + 2\, 
 \frac{\H}{a \bar\varphi\hspace*{0.3mm}'}
 \biggl( \frac{a \bar\varphi\hspace*{0.3mm}'}{\H} \biggr)'
 \R^{\prime}_k
 + \biggl[\,
   \cancel{
      \frac{\H}{a \bar\varphi\hspace*{0.3mm}'}
  \biggl( \frac{a \bar\varphi\hspace*{0.3mm}'}{\H} \biggr)'' 
   } 
  + k^2_{ }
 - \cancel{
      \frac{\H}{a \bar\varphi\hspace*{0.3mm}'}
  \biggl( \frac{a \bar\varphi\hspace*{0.3mm}'}{\H} \biggr)'' 
   }
 \,\biggr] \, \R^{ }_k
 \; = \; 0
 \;. \label{R_k_1} 
\ee

In a second step, like for 
the numerical solution of the background equations
in \app\ref{app:num_bg_vac}, 
it is beneficial to transform to physical time. 
{}From \eq\nr{t_tau}, we recall that 
${\rm d}\tau = a^{-1}_{ }(t)\ibit {\rm d}t$, 
and physical time derivatives are denoted by
\be
 (...)^{\textstyle .}_{ } \; \equiv \; 
 \partial^{ }_t (...) \; = \; 
 \frac{1}{a}\,(...)^{\prime}_{ }
 \quad \Leftrightarrow \quad 
 (...)^{\prime}_{ } 
 \;
 \overset{\rmii{\nr{eq_HH}}}{=}
 \; 
 a\, (...)^{\textstyle .}_{ }
 \;. \label{time_derivs}
\ee
For the derivatives of $\R^{ }_k$ we thus obtain
$
 \R^{\prime}_k
 = 
 a \dot{\R}^{ }_k
$, 
and 
\ba
%%%
 \R^{\prime\prime}_k
 &
 \overset{\rmii{\nr{time_derivs}}}{=}
 & 
 a\,\bigl(\, a \dot{\R}^{ }_k \,\bigr)^{\textstyle .}_{ }
 \; = \; 
 a^2_{ } \,
 \biggl(
  \ddot{\R}^{ }_k
  + 
  \hspace*{-2mm}
  \underbrace{
  \frac{\dot{a}}{a} 
  }_{ H }
  \hspace*{-1mm}  
  \dot{\R}^{ }_k  
 \biggr)
 \;. \label{R_k_dd}
\ea
Substituting these in \eq\nr{R_k_1}, 
and multiplying the whole equation
with $1/a^2_{ }$, yields
\be
 \boxed{
 \vphantom{\Biggr|}
 \quad
 \ddot{\R}^{ }_k 
 + 
 \biggl[\, 
 \underbrace{
   \frac{\dot{a}}{a} + 
  2\,\frac{H}{a\dot{\bar\varphi}}
  \biggl( \frac{a\dot{\bar\varphi}}{H} \biggr)^{\textstyle .}_{ }
 }_{\equiv\; 2 \F \;+\; 3 H}
 \,\biggr]
 \,  \dot{\R}^{ }_k 
 + \frac{k^2_{ }}{a^2_{ }} \, \R^{ }_k 
 \; = \; 0
 \;, 
 \quad
 }
 \label{eq_R_k}
 \index{curvature perturbations: equation}
\ee
where we have defined 
\be
 \F(t) \; \equiv \; 
 \frac{H}{\dot{\bar\varphi}}
 \biggl( \frac{\dot{\bar\varphi}}{H} \biggr)^{\textstyle .}_{ }   
 \; = \;  \frac{\ddot{\bar\varphi}}{\dot{\bar\varphi}} 
 - \frac{\dot H}{H}  
 \;. \label{cal_F} 
\ee
Here $\ddot{\bar\varphi}$ is given by \eq\nr{eq-F2}, 
$H$ by \eq\nr{eq_inflation_t}, 
and $\dot{H}$ by \eq\nr{eq_Hdot}. For a physical 
interpretation, it is useful to note that \eq\nr{cal_F}
represents the relative rate of change of the variable
$\dot{\bar\varphi}/H$
from \eq\nr{const}, which is approximately constant in 
the slow-roll regime. Equation~\nr{eq_R_k} is 
the {\em Mukhanov-Sasaki equation in physical time}. 
\index{Mukhanov-Sasaki equation: in $t$}

Let us remark that both \eq\nr{mode_functions_again} and \eq\nr{eq_R_k}
have important special properties which underline the value of using
$\field^{ }_k$ as a variable in conformal time and 
$\R^{ }_k$ in physical time. Specifically, \eq\nr{eq_R_k} includes
a first-order time derivative, implying that in physical time, 
the Wronskian of $\R^{ }_k$ is not constant 
(cf.\ \eq\nr{wronskian_def}). On the other hand, 
the mass term is absent from \eq\nr{eq_R_k}, which turns out to 
have a very helpful implication for the late-time evolution,
as will be discussed
in \ses\ref{ss:slow-roll} and \ref{ss:outside_curv}.

We still need to transcribe the initial conditions 
from \eqs\nr{wronskian_value_again} 
and \nr{derivative_value_again} into the new variables.
Given that \eq\nr{wronskian_value_again} concerns overall
normalization, let us recall from \eq\nr{P_Q_k} 
that the power spectrum corresponding to 
$\R^{ }_\varphi$ is obtained from the 
mode function, $\R^{ }_k$, via
\be
 \boxed{ 
 \quad
 \P^{ }_{\mbit \scriptscriptstyle \R^{ }_\varphi}
 \; \overset{\rmii{\nr{P_Q_k}}}{=} \; 
 \frac{k^3_{ }}{2\pi^2_{ }}
 \, | \R^{ }_k |^2_{ }
 \; \equiv \; 
 \bigl|\, 
  \bigl[\, \R^{ }_k \,\bigr]^{ }_\rmi{rescaled}
 \,\bigr|^2_{ }
 \;.
 \quad \vphantom{\Bigg|}
 } 
 \label{rescale}
\ee
In the second step, we noted that  
it is practical to ``rescale'' the mode function by 
$
 k^{3/2}_{ } / (\sqrt{2}\pi)
$, 
so that its absolute value squared yields directly 
the power spectrum. 

Let $t^{ }_i$ denote a moment at which initial
conditions are set, and $f^{ }_i \equiv f(t^{ }_i)$ the
values of various functions at that time. 
Given that the variable $k/a$ 
in \eq\nr{eq_R_k} is exponentially large at early times
(cf.\ \fig\ref{fig:history_tau}(right) on 
p.~\pageref{fig:history_tau}), it is not possible
to take $t^{ }_i$ arbitrarily small in practice. Rather, it is 
sufficient to take $t^{ }_i$ from a domain 
in which $k/a^{ }_i \gg H^{ }_i$,
so that $k/a^{ }_i$ represents the dominant 
scale in the evolution, and
\eq\nr{derivative_value_again} becomes
\be
 \partial^{ }_t {\field\hspace*{0.6mm}}^{ }_{\hspace*{-0.6mm}k}(t^{ }_i) 
 \;
 \underset{\rmii{\nr{time_derivs}}}
 {\overset{\rmii{\nr{derivative_value_again}}}{\approx}}
 \;
  -i \frac{k}{a_i^{ }}
      \field^{ }_k(t^{ }_i)
 \;. \label{field_ini_1}
\ee
We rewrite \eq\nr{field_ini_1} in terms of $\R^{ }_k$
by making use of \eq\nr{Qk_vs_Rk}. Furthermore,  
we note that we 
can approximate $a \dot{\bar\varphi}/H$ as constant, 
because 
$\dot{a}^{ }_i / a^{ }_i = H^{ }_i \ll k/a^{ }_i$, 
and similarly for the relative 
time dependence of $\dot{\bar\varphi}/H$, 
given by $\cal{F}$ from \eq\nr{cal_F}. 
Then \eq\nr{field_ini_1} implies 
\be
 \boxed{ 
 \quad
 \bigl[\, \partial^{ }_t {\R}^{ }_k \,\bigr]^{ }_\rmi{rescaled} (t^{ }_i) 
 \; 
   \underset{\rmii{\nr{field_ini_1}}}{
   \overset{\scriptscriptstyle k/a^{ }_i \;\gg\; H^{ }_i
            \lift }{\approx}}
 \; 
  -i \frac{k}{a_i^{ }}
 \bigl[\, \R^{ }_k \,\bigr]^{ }_\rmi{rescaled} (t^{ }_i) 
 \;.
 \quad \vphantom{\Bigg|}
 }
 \label{R_ini_1}
\ee 
Under the same approximation, the Wronskian from 
\eq\nr{wronskian_value_again} can be written as 
\ba
 i
 \; = \; 
 \W \bigl[\, \field^{ }_k,\field^{*}_k \,\bigr]
 & \approx &
 \biggl( \frac{a^{ }_i\dot{\bar\varphi}^{ }_i}{H^{ }_i} \biggr)^2_{ }
 \times 
 2 i a^{ }_i
 \im \bigl( \R^{ }_k \dot{\R}^{*}_k \bigr)
 \nn[2mm]
%%%%
 &
 \underset{\rmii{\nr{rescale}}}
 {\overset{\rmii{\nr{R_ini_1}} \lift }{=}} 
 & 
 \biggl( \frac{a^{ }_i\dot{\bar\varphi}^{ }_i}{H^{ }_i} \biggr)^2_{ }
 \times 
 2 i k 
 \frac{2\pi^2_{ }}{k^3_{ }}
 \,\bigl|\, 
 \bigl[\, \R^{ }_k \,\bigr]^{ }_\rmi{rescaled} (t^{ }_i)
 \,\bigr|^2_{ } 
 \;. \label{wron_again}
\ea
Choosing the overall phase so that the $R^{ }_k(t^{ }_i) > 0$
(assuming $\dot{\bar\varphi}^{ }_i < 0$), we obtain
\be
 \boxed{
 \quad 
 \bigl[\, \R^{ }_k \,\bigr]^{ }_\rmi{rescaled}
 (t^{ }_i)
 \; 
   \underset{\rmii{\nr{wron_again}}}{
   \overset{\scriptscriptstyle k/a^{ }_i \;\gg\; H^{ }_i
            \lift }{\approx}}
 \; 
 - \frac{H^{ }_i}{\dot{\bar\varphi}^{ }_i} 
 \,\frac{ 1 }{2\pi}\, \frac{k}{a^{ }_i}
 \;.
 \quad \vphantom{\Bigg|}
 }  
 \label{R_ini_2} 
\ee

An algorithm for solving \eq\nr{eq_R_k} numerically, with initial
values fixed according to \eqs\nr{R_ini_1} and \nr{R_ini_2},
is given in \app\ref{app:num_R_varphi}.

%%%%%%%%%%%%%%%%%%%%%%%%%%%%%%%%%%%%%%%%%%%%%%%%%%%
%
\subsection{Simplifications in the slow-roll regime}
\label{ss:slow-roll}

\index{slow-roll regime}

Apart from numerical solutions, 
\eq\nr{eq_R_k} can also be used as a starting point for 
analytic approximations. We already discussed them 
in \se\ref{ss:soln_dS}, where a background
solution was obtained from \eq\nr{slow_roll}, and subsequently
the power spectrum was determined, leading to 
\eq\nr{P_R_final}. Here we want to identify small parameters,
which will be called {\em slow-roll parameters},  
and formalize the computation 
as a Taylor expansion in them~\cite{liddle}.

In the present section, 
we remain at the zeroth or first order in 
the slow-roll expansion, showing how predictions
such as \eq\nr{P_R_final} can be justified. 
However, we remark in passing that 
it is possible to go up to high orders
in the slow-roll expansion, even though formulating
this systematically requires modified definitions of 
the expansion parameters~\cite{n3lo}. 

While the curvature power spectrum originates by solving 
for scalar perturbations, 
% from \eq\nr{eq_R_k}, the key to 
the slow-roll approximation operates 
mostly on the side of the (non-linear) background equations.
The conventional logic for determining the slow-roll predictions
is to first assume ourselves to be in a specific regime,  
and then work out consistency criteria for
when this assumption is self-consistent. 
So, we start by {\em assuming} that 
$ \ddot{ \bar{\varphi} } \approx 0$
in \eq\nr{eq-F2_again}, implying 
\be
 \dot{\bar\varphi} 
 \;
 \overset{\rmii{\nr{slow_roll}}}{\approx}
 \; 
 - \frac{V^{ }_{,\varphi}}{3 H}
 \;. \label{assume}
\ee
The slow-roll parameters are defined as 
\be
 \epsilon^{ }_\rmii{$V$}
 \; \equiv \; 
 \frac{1}{16\pi G}
 \biggl( \frac{V^{ }_{,\varphi}}{V} \biggr)^2_{ }
 \;, \quad
 \eta^{ }_\rmii{$V$}
 \; \equiv \; 
 \frac{1}{8\pi G} \frac{V^{ }_{,\varphi\varphi}}{V}
 \;, \label{slowroll_params}
 \index{slow-roll parameters}
 \index{$\epsilon^{ }_\rmii{$V$}$, $\eta^{ }_\rmii{$V$}$
 (slow-roll parameters)}
\ee
where the subscript $(...)^{ }_\rmii{$V$}$ 
refers to the potential~$V$ (there is another
convention in which the slow-roll parameters are determined in terms
of the Hubble rate). The time at which the parameters are
evaluated is left implicit for now, and they are
treated as being \mbox{approximately} constant around that moment. 

We note that, since $G = 1/\mpl^2$ (cf.\ \eq\nr{mpl}), 
$ \epsilon^{ }_\rmii{$V$} $
and
$ \eta^{ }_\rmii{$V$} $
in \eq\nr{slowroll_params}
can be small only if variations in the field $\varphi$
are at the Planck scale,  
$\mpl^{ }\partial^{ }_\varphi < 1$. Naively this 
requires {\em trans-Planckian field 
values}, \index{trans-Planckian field values}
$\varphi > \mpl^{ }$, 
which is considered a foundational challenge for 
inflationary model building. 

For the Hubble rate, 
it now follows from \eq\nr{eq_inflation_t} that 
\ba
 H^2_{ }
 & \overset{\rmii{\nr{eq_inflation_t}} \lift }
  {\underset{\scriptscriptstyle \kappa\; =\; 0}{=}} & 
 \frac{8\pi G V}{3}
 \biggl( 1 + \frac{\dot{\bar\varphi}^2_{ }}{2 V} \biggr) 
 \;
 \overset{\rmii{\nr{assume}}}{\approx} 
 \;
 \frac{8\pi G V}{3}
 \biggl( 1 + \frac{V^2_{,\varphi}}{18 V H^2} \biggr) 
 \nn[2mm]
%%%%%%%
 &
 \overset{\rmii{\nr{slowroll_params}}}{=} 
 &
 \frac{8\pi G V}{3}
 \biggl( 1 + \frac{8\pi G V }{3 H^2}
 \frac{ \epsilon^{ }_\rmii{$V$} }{3} \biggr)
 \; \overset{\epsilon^{ }_\rmiii{\it V} \;\ll\; 1
             \lift }{\approx} \; 
 \frac{8\pi G V}{3}
 \biggl( 1 + 
 \frac{ \epsilon^{ }_\rmii{$V$} }{3} \biggr)
 \;. \label{slowroll_HH}
\ea
The last equality shows an approximate solution 
of the quadratic equation for $H^2_{ }$.

To verify the consistency of our approximation,  we need to 
determine the time derivatives of the key quantities. 
{}From \eqs\nr{eq_Hdot} and \nr{eq_inflation_t}, we obtain
\be
 \dot{H}
 \;
  \underset{\rmii{\nr{eq_inflation_t}}}{
  \overset{\rmii{\nr{eq_Hdot}} \lift }{=}} 
 \; 
 -4\pi G \dot{\bar\varphi}^2_{ } 
 \;
 \overset{\rmii{\nr{assume}}}{\approx}
 \;
 -\frac{4\pi G V^2_{,\varphi}}{9H^2_{ }}  
 \;       
 \overset{\rmii{\nr{slowroll_params}}}{=} 
 \;
 -\biggl( \frac{8 \pi G V }{3 H} \biggr)^2_{ }        
  \epsilon^{ }_\rmii{$V$} 
 \;
 \overset{\rmii{\nr{slowroll_HH}}}{\approx} 
 \; 
 - H^2_{ }  
  \epsilon^{ }_\rmii{$V$} 
 \;. \label{slowroll_Hdot}
\ee
Taking a time derivative of \eq\nr{assume}, yields
\be 
 \ddot{\bar\varphi}
 \;
 \overset{\rmii{\nr{assume}}}{\approx}
 \;
 -\frac{V^{ }_{,\varphi\varphi} \dot{\bar\varphi}}{3 H}
 + \frac{V^{ }_{,\varphi} \dot{H}}{3 H^2_{ }}
 \;
 \underset{\rmii{\nr{slowroll_Hdot}}}
 {\overset{\rmii{\nr{slowroll_params}}
    \lift  }{\approx}}
 \;
 -\frac{8\pi G V \dot{\bar\varphi} }{3 H} \,\eta^{ }_\rmii{$V$}
 - \frac{V^{ }_{,\varphi} }{3} \,\epsilon^{ }_\rmii{$V$}
 \;
 \underset{\rmii{\nr{slowroll_HH}}}
 {\overset{\rmii{\nr{assume}}
    \lift  }{\approx}}
 \; 
 ( \epsilon^{ }_\rmii{$V$} - \eta^{ }_\rmii{$V$} ) H \dot{\bar\varphi}
 \;. \label{slowroll_ddot}
\ee
Now, comparing $\ddot{\bar\varphi}$ with $3 H\dot{\bar\varphi}$, 
\eq\nr{slowroll_ddot} shows that 
$\ddot{\bar\varphi}$ is indeed subdominant in the regime
$
 \epsilon^{ }_\rmii{$V$}, |\eta^{ }_\rmii{$V$}| \ll 1
$.
Therefore, the assumption in \eq\nr{assume} is self-consistent. 
For $\mathcal{F}$ from \eq\nr{cal_F}, \eqs\nr{slowroll_Hdot} 
and \nr{slowroll_ddot} imply that
$
 \mathcal{F} \approx 
 ( 2 \epsilon^{ }_\rmii{$V$} - \eta^{ }_\rmii{$V$}) H \ll H
$.
So, \eq\nr{eq_R_k} becomes
\be
 \ddot{\R}^{ }_k 
 + 
 \bigl[\, 
 3 + 2 ( 2 \epsilon^{ }_\rmii{$V$} - \eta^{ }_\rmii{$V$})
 \,\bigr]
 \, H 
 \,  \dot{\R}^{ }_k 
 + \frac{k^2_{ }}{a^2_{ }} \, \R^{ }_k 
 \;
 \underset{\rmii{\nr{slowroll_Hdot},\nr{slowroll_ddot}}}{
 \overset{\rmii{\nr{eq_R_k}}  \lift  }{\approx}}
 \; 0
 \;. \label{eq_R_k_appro}
\ee
Furthermore, \eq\nr{slowroll_Hdot} shows that the change of 
$H$ in a {\em Hubble time}, \index{Hubble time}
$\Delta t \equiv H^{-1}_{ }$, is 
$
 \Delta H = H^{-1}_{ } \dot{H} \approx - H \epsilon^{ }_\rmii{$V$}
 \ll H
$, so that $H$ is approximately constant. 
In other words, we are in
quasi de Sitter space-time, 
in accordance with \eq\nr{quasi_de_Sitter}.

We can then ask what is the curvature power spectrum in the 
regime governed by \eqs\nr{eq_R_k_appro},
\nr{R_ini_1}, and \nr{R_ini_2}. It is non-trivial to do this
up to first order in the slow-roll parameters, because 
then the first-order time evolution of $H$ needs to be accounted for. 
Let us instead do this up to {\em zeroth order} in the slow-roll 
parameters. 

The key insight is that 
according to \fig\ref{fig:history_tau}(right)
on p.~\pageref{fig:history_tau}, $k/a$ decreases very
rapidly below the Hubble rate at a certain moment, which is 
denoted by $t^{ }_*$ and called {\em horizon exit} \index{horizon exit}
(by convention, $k/a^{ }_* = H^{ }_*$ at $t^{ }_*$). 
If the horizon exit happens in the slow-roll regime,
and we evaluate $H$ at zeroth order, 
then at $t> t^{ }_*$, \eq\nr{eq_R_k_appro} reads
$
 \ddot{\R}^{ }_k + 3 H^{ }_* \dot{\R}^{ }_k \approx 0
$.
The general solution is 
\be
 \R^{ }_k \;\approx\; c^{ }_1 + c^{ }_2\, e^{-3 H^{ }_* t}_{ } 
 \;, \label{soln_R_k_appro} 
\ee
implying that $\R^{ }_k$ settles to a constant, $c^{ }_1$. The value of 
$c^{ }_1$ can be deduced from \eq\nr{R_ini_2}, by pushing
the initial time, $t^{ }_i$, close to $t^{ }_*$.  
The absolute value squared
of \eq\nr{R_ini_2} at this moment
reproduces the power spectrum in
\eq\nr{P_R_final}. Even if these considerations are a bit sloppy, 
it is comforting that the correct result is recovered. 

The predictions for the CMB observables, from 
\eqs\nr{As} and \nr{ns}, can now be expressed 
in terms of the slow-roll parameters. 
Straightforward but somewhat tedious substitutions lead to 
expressions widely used in the literature, 
\ba
 A^{ }_\scalar 
 &
 \underset{\rmii{\nr{As}}}{
 \overset{\rmii{\nr{P_R_final}}}{\approx}}
 & 
 \frac{H^4_*}{4\pi^2_{ }\dot{\bar\varphi}^2_*}
 \; 
 \overset{\rmii{\nr{assume}}}{\approx}
 \; 
 \frac{9 H^6_*}{4\pi^2_{ }V^2_{,\varphi}}
 \; 
 \overset{\rmii{\nr{slowroll_params}}}{\approx}
 \; 
 \frac{9 H^6_*}{64\pi^3_{ }G V^2_{ } \epsilon^{ }_\rmii{$V$}}
% \;
% \frac{128\pi G^3}{3} \frac{V^3}{V_\varphi^2} 
 \; 
 \overset{\rmii{\nr{slowroll_HH}}}{\approx}
 \;  
 \frac{8 G^2 V}{3\epsilon^{ }_\rmii{$V$}}
 \;, \label{As_slowroll}
 \index{$A^{ }_\scalar$ (scalar amplitude)}
 \\[2mm] 
%%%
 n^{ }_\scalar & \underset{\rmii{\nr{slowroll_Hdot}}}
           {\overset{\rmii{\nr{ns}}}{\approx}} & 
 1 + \frac{2}{H^{ }_{*}}
 \biggl(
  \frac{2\dot{H}^{ }_*}{H^{ }_*} 
 - 
  \frac{\ddot{\bar{\varphi}}^{ }_* }{\dot{\bar{\varphi}}^{ }_* }
 \biggr)
 \; 
 \underset{\rmii{\nr{slowroll_ddot}}}
           {\overset{\rmii{\nr{slowroll_Hdot}}}{\approx}}
 \; 
% 1 + 2
% \bigl[
%  -2 \epsilon^{ }_\rmii{$V$} 
%  - (  \epsilon^{ }_\rmii{$V$} - \eta^{ }_\rmii{$V$} )
% \bigr]
% \; = \;
 1 - 6 \epsilon^{ }_\rmii{$V$} + 2 \eta^{ }_\rmii{$V$}  
 \;. \label{ns_slowroll} 
 \index{$n^{ }_\scalar$ (scalar spectral tilt)}
\ea
For the tensor-to-scalar ratio $r$, mentioned in 
\eq\nr{r_obs} and defined
in \eq\nr{r}, the corresponding prediction is 
worked out in \eq\nr{r_slowroll}.

\index{ultra-slow-roll regime}

As a final remark, 
the quasi de Sitter regime breaks down when 
$\epsilon^{ }_\rmii{$V$} \gsim 1$, cf.\ \eq\nr{slowroll_Hdot}. From 
\eq\nr{slowroll_HH}, this happens when 
$\dot{\bar\varphi}^{\ibit 2}_{ }/2 \gsim V$. The slow-roll
approximation breaks down also 
if $|\eta^{ }_\rmii{$V$}|$ becomes large. 
In particular, in the 
{\em ultra-slow-roll regime}, 
discussed in a paragraph 
preceding \eq\nr{slow_roll}, we still 
have $\epsilon^{ }_\rmii{$V$} \ll 1$
according to \eq\nr{slowroll_params}, 
given that~$V^{ }_{\der\varphi}$ had been set to zero. 
But otherwise the estimates
above cannot be used, 
as the assumption introduced in \eq\nr{assume}
is not respected
(the background equation reads
$\ddot{\bar\varphi}\approx -3 H \dot{\bar\varphi}$). 
 
%%%%%%%%%%%%%%%%%%%%%%%%%%%%%%%%%%%%%%%%%%%%%%%%%%%%%%%%%%%%%%%%%%%%%%%%%
%
\subsection{Stochastic derivation with quantum noise}
\label{ss:stochastic_exact}

\index{stochastic formalism: exact}

We now proceed to re-formulating the initial-value problem
in \se\ref{ss:setup_num} in a way that the information contained
in the initial conditions at $t^{ }_i$ 
(cf.\ \eqs\nr{R_ini_1} and \nr{R_ini_2}),   
can instead be imposed at a later time, 
for instance at $t^{ }_*$. 
Even if this may feel a bit
like a ``trick'', the result has a nice physical interpretation, and has also
paved the way for an approximation scheme that is used 
in non-linear numerical simulations (cf.\ \se\ref{ss:stochastic_appro}).
For the present section, 
we return to conformal time, 
and denote rescaled quantities with a wide hat, ``$\widehat{...}$''.
The equation to solve is 
the original $\vec{x}$-space version 
of \eq\nr{mode_functions_again}, 
given in \eq\nr{final_vacuum_3}.

The starting point is to 
introduce a {\em window function}, \index{window function}
or rather a distribution, 
$W^{ }_k(\tau)$, which splits
the comoving momentum space into ultraviolet (UV) and
infrared (IR) parts. The window is
such that in the distant past, $\tau\to -\infty$, 
all modes % are UV ones, and 
are selected by $W^{ }_k$, i.e.\  
\be
 W^{ }_k(-\infty) \; = \; 1
 \;, \quad
 W^{\prime}_k(-\infty) \; = \; 0 \quad \forall k 
 \;. \label{window_1}
\ee
Later on, 
physical momenta redshift towards the IR. 
We consider a final moment, 
$\tau^{ }_{\sfin}$, at which  
the modes of interest are in the IR and
fall out of the window, 
so that
 \be
 W^{ }_k(\tau^{ }_{\sfin}) \; \approx \; 0
 \;. \label{window_2}
\ee
Often the UV part is identified as ``{\em inside the Hubble horizon}''
and the IR part as ``{\em outside the Hubble horizon}'', which requires 
that $ \tau^{ }_{\sfin} $ is chosen inside a specific time window
(cf.\ figure~\ref{fig:history_tau}(right) on p.~\pageref{fig:history_tau}).
\index{Hubble horizon}

Following ref.~\cite{ref_6_snk}, 
we now split the mode expansion in \eq\nr{mode_expansion} 
into UV and IR parts,
\be
  \field(\tau,\vec{x}) 
  \; =  \;
 \field^{ }_{<}(\tau,\vec{x})
 + \field^{ }_{>}(\tau,\vec{x})
 \;, \label{field_splitup}
\ee
with the short-distance (UV) part given by
\ba
 \field^{ }_{<}(\tau,\vec{x}) 
 & \equiv & 
 \int \! \frac{{\rm d}^3\vec{k}}{\sqrt{(2\pi)^{3}_{ }}} \, 
 W^{ }_k(\tau) \, 
 \Bigl[ \,
    w^{ }_\rmii{\vec{k}}
    \, \field^{ }_k(\tau) 
    \, e^{ i \vec{k}\cdot\vec{x}}_{}   
  + 
    w^{\dagger}_\rmii{\vec{k}}
    \, \field^{\,*}_k(\tau) 
    \, e^{ - i \vec{k}\cdot\vec{x}}_{}   
 \, \Bigr]
 \;. \label{delta_varphi_splitup_4}
\ea
Inserting \eq\nr{field_splitup} into \eq\nr{final_vacuum_3},  
a {\em quantum noise} is defined as \index{quantum noise}
\be
 \bigl[\, \partial_\tau^2 + \hat\epsilon^{2}_\vec{x}(\tau) \,\bigr]\, 
 \field^{ }_{>}(\tau,\vec{x})
  \; = \; 
 \widehat\varrho_\iQ^{ }(\tau,\vec{x})
 \;, \quad
 \widehat\varrho_\iQ^{ }(\tau,\vec{x}) 
 \; \equiv \; 
 - 
 \bigl[\, \partial_\tau^2 + \hat\epsilon^{2}_\vec{x}(\tau) \,\bigr]\, 
 \field^{ }_{<}(\tau,\vec{x})
 \;. \label{delta_varphi_eq_splitup_2}
\ee
The idea is that the long-distance modes are sourced by the 
quantum noise rather than by initial conditions, 
which is why the formalism is called ``stochastic''. 

Operating on \eq\nr{delta_varphi_splitup_4}, and noting that several
terms drop out, thanks to the equation satisfied by the mode functions, 
the noise can be expressed as  
\ba
 \widehat\varrho_\iQ^{ }(\tau,\vec{x}) & = & 
 - \int \! \frac{{\rm d}^3\vec{k}}{\sqrt{(2\pi)^{3}_{ }}} \, 
 \Bigl[ \,
    w^{ }_\rmii{\vec{k}}
    \, \widehat{f}^{ }_k(\tau) 
    \, e^{ i \vec{k}\cdot\vec{x}}_{}   
  + 
    w^{\dagger}_\rmii{\vec{k}}
    \, \widehat{f}^{*}_k(\tau) 
    \, e^{ - i \vec{k}\cdot\vec{x}}_{}   
 \, \Bigr]
 \;, \label{xi_3} \\[2mm]
%%%%
  \widehat{f}^{ }_k(\tau) & \equiv &
   W''_{k}(\tau) 
  \field^{ }_k(\tau) 
 + 2 W'_{k}(\tau) \field\bit{}'_k(\tau)
 \;. \label{f_k_new_2} 
\ea
The derivatives $W''_k$ and $W'_k$ are to be understood in the sense of 
distributions, as we ultimately integrate over $\tau$ (see below).

Defining a {\em retarded Green's function}, as 
\ba
 \bigl[\, \partial_\tau^2 + \hat\epsilon^{2}_{\vec{x}}(\tau) \,\bigr]
 G^{ }_{|\vec{x-z}|}(\tau;\tau^{ }_i)
 & = & 
 \delta(\tau - \tau^{ }_i)\,\delta^{(3)}_{ }(\vec{x-z})
 \;, \label{greens_x} 
 \index{retarded Green's function: general}
 \\[2mm]
%%%%%%
 G^{ }_{|\vec{x-z}|}(\tau;\tau^{ }_i)
 & \equiv & 0
 \quad \mbox{for} \quad \tau < \tau^{ }_i
 \;, \label{boundary_x} 
\ea
our long-distance IR solution from \eq\nr{delta_varphi_eq_splitup_2} is  
\be
 \field^{ }_{>}(\tau,\vec{x})
 \;=\; \int_{-\infty}^{\tau} \! {\rm d}\tau^{ }_i \, 
   \int_\vec{z} \,
 G^{ }_{|\vec{x-z}|}(\tau;\tau^{ }_i) \,
 \widehat\varrho_\iQ^{ }(\tau^{ }_i,\vec{z})
 \;. \label{soln_with_greens}
\ee
In order to transform the solution to momentum space, we define
\be
 G^{ }_k(\tau;\tau^{ }_i)
 \;
 \overset{\rmii{\nr{fourier_k}}}{\equiv}
 \;
 \int_\vec{x} e^{-i\vec{k}\cdot(\vec{x-z})}_{ }
   \,
   G^{ }_{|\vec{x-z}|}(\tau;\tau^{ }_i) 
 \;, \label{def_G_k}
\ee
which fulfils
\ba
 \bigl[\, \partial_\tau^2 + \hat\epsilon^{2}_{k}(\tau) \,\bigr]
 G^{ }_{k}(\tau;\tau^{ }_i)
 & \underset{\rmii{\nr{def_G_k}}}{\overset{\rmii{\nr{greens_x}}}{=}} &
 \delta(\tau - \tau^{ }_i)
 \label{greens_k_2} 
 \\[2mm]
 G^{ }_{k}(\tau;\tau^{ }_i)
 & \equiv & 0
 \quad \mbox{for} \quad \tau < \tau^{ }_i
 \;. \label{boundary_k} 
\ea
Then we get 
\ba
 \field^{ }_{>}(\tau,\vec{k}) 
 & \overset{\rmii{\nr{fourier_k}}}{\equiv} &
 \int_{\vec{x}} e^{-i\vec{k}\cdot\vec{x}}_{ }
 \,\field^{ }_{>}(\tau,\vec{x}) 
%%%%%
 \nn 
 & \stackrel{\rmii{\nr{soln_with_greens}}}{=} &
 \int_{-\infty}^{\tau} \! {\rm d}\tau^{ }_i \, 
   \int_{\vec{x},\vec{z}} \,
 e^{ i\vec{k}\cdot ( {\vec{z} - \vec{x}}) }_{ } \,
 G^{ }_{|\vec{x-z}|}(\tau;\tau^{ }_i) \,
 e^{ -i\vec{k}\cdot\vec{z} }_{ }\,
 \widehat\varrho_\iQ^{ }(\tau^{ }_i,\vec{z})
%%%%%
 \nn 
 & \underset{\rmii{\nr{def_G_k}}}
   {\overset{\rmii{\nr{xi_3}} \lift }{=}} &
 - \sqrt{(2\pi)^3_{ }}
 \int_{-\infty}^\tau \! {\rm d}\tau^{ }_i \, 
 G^{ }_k(\tau;\tau^{ }_i) \,
 \Bigl[ \,
    w^{ }_\rmii{\vec{k}}
    \, \widehat{f}^{ }_k(\tau^{ }_i) 
  + 
    w^{\dagger}_\rmii{$-$\vec{k}}
    \, \widehat{f}^{*}_k(\tau^{ }_i) 
 \, \Bigr]
 \;. \label{greens_4}
\ea

Now, like in the quantum-mechanical problem with a Dirac-$\delta$
potential, the initial conditions for $G^{ }_k$
at $\tau = \tau_i^+$
can be obtained by imposing
continuity at $\tau_i^{ }$, % with \eq\nr{boundary_k}, 
and by figuring out the correct
value of the first derivative by integrating over \eq\nr{greens_k_2}. 
This yields
\be
 \lim_{\tau \to \tau^{+}_i}
  G^{ }_{k}(\tau;\tau^{ }_i) \; = \; 0
 \;, \quad
 \lim_{\tau \to \tau^{+}_i}
 \partial^{ }_\tau G^{ }_{k}(\tau;\tau^{ }_i) \; = \; 1
 \;. \label{G_R_k_again}
\ee
Remarkably, the Green's function
satisfying these initial conditions can be expressed in terms of 
the mode functions from \eq\nr{mode_functions_again}, recalling that 
the Wronskian of a mode function and its complex conjugate
obeys \eq\nr{wronskian_value_again}. The solution reads 
\be
 G^{ }_{k}(\tau;\tau^{ }_i) \;=\; 
 2 \im \bigl[\,  
        \field\bit{}^*_k(\tau)
        \field\bit{}^{ }_k(\tau^{ }_i)              
     \,\bigr]
 \;, \qquad \tau \;>\; \tau^{ }_i
 \;, \label{greens_2}
\ee
because
\ba
 2\, \partial^{ }_\tau
   \im \bigl[\,  
        \field\bit{}^*_k(\tau)
        \field\bit{}^{ }_k(\tau^{ }_i)              
     \,\bigr]
 & = & 
 \frac{1}{i} 
 \bigl[\,
        \field\bit{}^{*\prime}_k(\tau)
        \field\bit{}^{ }_k(\tau^{ }_i)              
       - 
        \field\bit{}^{\prime}_k(\tau)
        \field\bit{}^{*}_k(\tau^{ }_i)              
 \,\bigr]
 \nn[2mm]
 &
  \overset{\rmii{\nr{wronskian_def}} \lift }{=}
 & 
 \frac{1}{i} \, \W\bigl[\, \field\bit{}^{ }_k(\tau^{ }_i) \,,\,
                    \field\bit{}^{*}_k(\tau) \,\bigr]
 \;
 \xrightarrow[ \rmii{\nr{wronskian_value_again}} ]
             { \scriptscriptstyle  
               \tau\;\to\; \tau^{ }_i\; \lift }
 \; 1
 \;. \label{W_vs_Im}
\ea

Let us then inspect the integral over $\tau^{ }_i$ in \eq\nr{greens_4}. 
Inserting \eq\nr{f_k_new_2}, and carrying out a partial integration, 
we find
\ba
 && \hspace*{-2.0cm}
 \int_{-\infty}^\tau \! {\rm d}\tau^{ }_i \, 
 G^{ }_k(\tau;\tau^{ }_i) \,
 \widehat{f}^{ }_k(\tau^{ }_i) 
 \;
 \overset{\rmii{\nr{f_k_new_2}} \lift }{=}
 \;   
 \int_{-\infty}^\tau \! {\rm d}\tau^{ }_i \, 
   G^{ }_k(\tau;\tau^{ }_i) \,
 \,\Bigl[\,
   W''_k(\tau^{ }_i) \,
   \field\bit{}^{ }_k(\tau^{ }_i)
 + 
 2 W'_k(\tau^{ }_i)
   \field\bit{}'_k(\tau^{ }_i)
 \,\Bigr] 
%%%%%%%%%%%%
 \nn[2mm] 
 &
 \overset{\rmii{\nr{window_1}} \lift }{
 \underset{\rmii{\nr{G_R_k_again}}}{=}}
 &   
 \int_{-\infty}^\tau \! {\rm d}\tau^{ }_i \, 
 W'_k(\tau^{ }_i)
 \,\Bigl[\,
 - \partial^{ }_{\tau_i}
   G^{ }_k(\tau;\tau^{ }_i) \,
   \field\bit{}^{ }_k(\tau^{ }_i)
 + 
   G^{ }_k(\tau;\tau^{ }_i) \,
   \field\bit{}'_k(\tau^{ }_i)
 \,\Bigr] 
%%%%%%%%%%%%
 \nn[2mm] 
 & \underset{\rmii{\nr{W_vs_Im}}}
   {\overset{\rmii{\nr{greens_2}} \lift }{=}} & 
 \int_{-\infty}^\tau \! {\rm d}\tau^{ }_i \, 
 \frac{ W'_k(\tau^{ }_i) }{i} 
 \biggl\{ 
   \bigl[\,
     - \cancel{ 
     \field\bit{}^{*}_k(\tau)
     \field\bit{}_k'(\tau^{ }_i)
              }
     + 
     \field\bit{}^{ }_k(\tau)
     \field\bit{}^{*}_k{}'(\tau^{ }_i)
   \,\bigr]
   \, 
   \field\bit{}^{ }_k(\tau^{ }_i)
%%%%%%%%%%%%
 \nn[2mm] 
 & & \hspace*{2.6cm}
 + \, 
   \bigl[\, \bcancel{ 
     \field\bit{}^{*}_k(\tau)
     \field\bit{}^{ }_k(\tau^{ }_i)
                    }
     - 
     \field\bit{}^{ }_k(\tau)
     \field\bit{}^{*}_k(\tau^{ }_i)
   \,\bigr]
   \, 
   \field\bit{}_k'(\tau^{ }_i)
  \biggr\} 
%%%%%%%
 \nn[2mm] 
 & = & 
 \int_{-\infty}^\tau \! {\rm d}\tau^{ }_i \, 
 \frac{ W'_k(\tau^{ }_i) }{i} \,
 \W\bigl[\, 
     \field\bit{}^{ }_k(\tau^{ }_i)
      \,,\,
     \field\bit{}^{*}_k(\tau^{ }_i)
 \,\bigr] \,
 \field\bit{}^{ }_k(\tau)
%%%%%%%
 \nn[2mm] 
 & \stackrel{\rmii{\nr{wronskian_value_again}} \lift }{=} & 
 \bigl[\, 
   W^{ }_k(\tau) - W^{ }_k(-\infty)
 \,\bigr] \,
 \field\bit{}^{ }_k(\tau)
 \;  
 \overset{\rmii{\nr{window_1}}}{=}
 \; 
 - 
 \bigl[\, 
   1 -  W^{ }_k(\tau) 
 \,\bigr] \,
 \field\bit{}^{ }_k(\tau)
 \;. \label{Wprime} 
\ea
If we choose a late time moment, 
according to \eq\nr{window_2}, 
% and recall \eq\nr{window_1}, 
\eq\nr{greens_4} becomes
\be
 \field^{ }_{>}(\tau^{ }_{\sfin},\vec{k}) 
 \; 
 \underset{\rmii{\nr{Wprime},\nr{window_2}}}{
 \overset{\rmii{\nr{greens_4}}}{=}}
 \;  
 \sqrt{(2\pi)^3_{ }} \,
 \Bigl[ \,
    w^{ }_\rmii{\vec{k}}
    \, \field^{ }_k(\tau^{ }_{\sfin}) 
  + 
    w^{\dagger}_\rmii{$-$\vec{k}}
    \, \field^{*}_k(\tau^{ }_{\sfin}) 
 \, \Bigr]
 \;. \label{hatphi>}
\ee

Actually, \eq\nr{hatphi>} could have been written 
down immediately, without a derivation, 
as it is nothing but a Fourier 
transform of the full field
operator from \eq\nr{mode_expansion}. 
This should not be surprising, as we have made no approximation
along the way. However, the derivation suggests a non-trivial 
interpretation. If we inspect the form in \eq\nr{xi_3}, 
and assume that $W^{ }_k(\tau^{ }_i)$ is a 
(perhaps smoothened) step function, 
so that $W_k'(\tau^{ }_i)$ is sharply localized, then 
$
 \widehat\varrho_\iQ^{ }(\tau^{ }_1,\vec{x}^{ }_1)
$
and
$
 \widehat\varrho_\iQ^{ }(\tau^{ }_2,\vec{x}^{ }_2)
$
with $\tau^{ }_1 \neq \tau^{ }_2$
are generated from different momentum shells, 
$k^{ }_1 \neq k^{ }_2$. As the corresponding 
$w^{ }_{\vec{k}_1}$ and $w^{\dagger}_{\vec{k}_2}$ commute, 
the noise can be viewed as ``classical''. 
This is said to imply that initial quantum fluctuations
have become classical ones, when they have redshifted deep into
the IR domain. However, this is not strictly true: 
the full quantum information (both amplitude and phase)
about the initial conditions is still stored in the quantum noise
(technically, this is taken care of by the appearence 
of both $W''_k$ and $W'_k$ in \eq\nr{f_k_new_2}). 

Conventionally, the dividing line between UV and IR modes
is adopted as the  
{\em crossing of the Hubble horizon}, \index{horizon crossing} 
i.e.\ at $k \sim \H$. 
So, according to \eq\nr{a_tau_dS_2}, 
\be
 W^{ }_k(\tau) 
 \; 
 \underset{\rmii{\nr{a_tau_dS_2}}}{
 \overset{\rmii{de~Sitter} \lift }{\simeq}}
 \; 
 \theta\biggl( k + \frac{\epsilon}{\tau} \biggr)
 \;, \label{W_theta}
 \index{window function}
\ee
where $\epsilon \sim 1$ is a parameter which should 
drop out from physical results. For completeness
we remark that with such
a distribution, 
$W'_k$ is Dirac-$\delta$ and 
$W''_k$ is singular, but this is not a 
problem, given that all derivatives of $W^{ }_k$ are 
eliminated by the integrations in \eq\nr{Wprime}.

%%%%%%%%%%%%%%%%%%%%%%%%%%%%%%%%%%%%%%%%%%%%%%%%%%%%%%%%%%%%%%%%%%%%%%%%%
%
\subsection{A pragmatic version of the stochastic formalism}
\label{ss:stochastic_appro}

\index{stochastic formalism: pragmatic}

The stochastic formalism described in \se\ref{ss:stochastic_exact}
is exactly equivalent to the full quantum-mechanical treatment of 
the problem, but does not offer any practical advantages for 
determining the solution numerically. 
There exists, however, a simplified version
of the stochastic formalism, elaborated upon in ref.~\cite{stochastic}, 
whose status is a bit different. In some sense, it is conceived to 
be an effective theory~\cite{sdseft1}. 
This means that when we derive the 
formalism, approximations or restrictions of generality
are made, notably that:
\bi

\item 
We assume to find ourselves in the slow-roll regime
(cf.\ \se\ref{ss:slow-roll}).

\item
The window function is taken from \eq\nr{W_theta}, 
and written in physical time as  
\be
 W^{ }_k(t) \; \equiv \; 
 \theta\bigl(\, k - \epsilon\, a(t) H \,\bigr)
 \;, \label{W_theta_2}
\ee
where $H$ is assumed to be slowly varying (in the following, constant).

\item
We choose $\epsilon \ll 1$ in \eq\nr{W_theta_2}, 
so that modes fall out of the 
window only once they are far outside the Hubble horizon.
Philosophically, 
this choice makes the construction similar in spirit
to the so-called {\em separate universes} approach to 
cosmological perturbations~\cite{separate,uk}.

\item
The simple mode function from \eq\nr{Q_k} is used for describing
the UV modes. 

\item
A gauge is fixed, frequently by going over to the
so-called $\delta N$ formalism~\cite{deltaN}. 

\ei
In the literature,
there are attempts to relax some of these assumptions, 
nevertheless certain compromises appear necessary in
order to obtain the desired simplifications. 

\index{separate universes picture}

On the other hand, the effective description
is conjectured to have merits as well, notably that it could be 
valid beyond the linear order in perturbations 
for treating the IR modes. 
This is apparent already in the choice of variables: 
rather than splitting the inflaton field into three parts, 
$\bar\varphi$, $\delta\varphi^{ }_{>}$ and $\delta\varphi^{ }_{<}$, 
as would correspond to the approach of \se\ref{ss:stochastic_exact}, 
the IR part $\bar\varphi + \delta\varphi^{ }_{>}$ is treated as
a single variable. 
Simulations of this framework are being used for probing
the likelihood of rare (non-Gaussian)
large fluctuations, which might  
collapse into {\em primordial black holes}. 
It is fair to say, however, 
that the question of whether the formalism
faithfully accounts for all non-linearities has
an unclear answer at present. 

\index{primordial black holes}

In the following, we illustrate some basic ingredients of the 
simplified stochastic formalism. We remain, however, at linear order
in perturbations, and use  
the gauge-invariant variable corresponding to $\delta\varphi$,
$\Q^{ }_\varphi$ from \eq\nr{Q_vs_R}. In \se\ref{ss:sasaki},
we saw that the ideal variable in conformal time is its 
rescaled version, $\field^{ }_\varphi$, cf.\ \eq\nr{aQ_vs_Q}, 
whereas in \se\ref{ss:setup_num}, for 
considerations in physical time, the best variable turned out to be 
the curvature perturbation $\R^{ }_\varphi$, from \eq\nr{rescaling}. 
In the present section, it is advantageous to 
mix physical and conformal times (see below), 
and work with $\Q^{ }_\varphi$, 
rather than the previous $\field^{ }_\varphi$ or $\R^{ }_\varphi$. 

In order to obtain an evolution equation for $\Q^{ }_\varphi$, 
we start with the Mukhanov-Sasaki equation from 
\eqs\nr{final_vacuum} and \nr{final_vacuum_2}, 
\be
  \bigl[\,
 \partial_\tau^2 - \nabla^2 
 + \widehat{m}^2_{ }(\tau) 
 \,\bigr]
 \field^{ }_\varphi
 \;
 \overset{\rmii{\nr{final_vacuum}}}{=}
 \;
 0
 \;, \quad
  \widehat{m}^2_{ }(\tau) 
 \;
 \overset{\rmii{\nr{final_vacuum_2}}}{=}
 \;
 - \frac{\H}{a\bar\varphi\bit'}
   \biggl( \frac{a\bar\varphi\bit'}{\H} \biggr)''
 \;. \label{mu-sa_1}
\ee
Similarly to \eqs\nr{Qk_vs_Rk}--\nr{Q_k_dd}, 
we substitute 
\ba
 \field^{ }_\varphi & \overset{\rmii{\nr{aQ_vs_Q}}}{=} & a \Q^{ }_\varphi
 \;, \\[2mm]
%%%
 \field^{\bit\prime}_\varphi & = & 
 a' \Q^{ }_\varphi + a \Q^{\bit\prime}_\varphi
 \;, \\[2mm]
%%% 
 \field^{\bit\prime\prime}_\varphi & = & 
 a'' \Q^{ }_\varphi + 2 a' \Q^{\bit\prime}_\varphi
 + a \Q^{\bit\prime\prime}_\varphi
 \;, 
\ea
and divide then the whole equation by $a$, leading to 
\be
 \Q^{\bit\prime\prime}_\varphi
 + 2 \H \Q^{\bit\prime}_\varphi
 + \biggl(
    \frac{a''}{a} - \nabla^2_{ } + \widehat{m}^2_{ } 
   \biggr) \Q^{ }_\varphi
 \; = \; 0
 \;. \label{mu-sa_2}
\ee
Then we go to physical time, like in 
\eq\nr{R_k_dd}, 
\be
 \Q^{\bit\prime}_\varphi
 \; \overset{\rmii{\nr{time_derivs}}}{=} \;  
 a \dot{\Q}^{ }_\varphi
 \;,\qquad
%%%
 \Q^{\bit\prime\prime}_\varphi
 \; \overset{\rmii{\nr{R_k_dd}}}{=} \; 
 a^2_{ } \,
 \biggl(
  \ddot{\Q}^{ }_\varphi
  + \frac{\dot{a}}{a} 
  \dot{\Q}^{ }_\varphi  
 \biggr)
 \;. 
\ee
Dividing by $a^2_{ }$, \eq\nr{mu-sa_2} gets thereby converted into
\be
 \ddot{\Q}^{ }_\varphi
 + 3 H \dot{\Q}^{ }_\varphi
 + \biggl(\,
 \frac{\ddot{a}}{a} + \frac{\dot{a}^2_{ }}{a^2_{ }}
 - \frac{\nabla^2_{ }}{a^2_{ }}
 + \frac{\widehat{m}^2_{ }}{a^2_{ }}
 \,\biggr)\,\Q^{ }_\varphi \;=\; 0 
 \;. \label{mu-sa_3}
\ee
For the mass parameter appearing here, we write
\ba
 \frac{\widehat{m}^2_{ }}{a^2_{ }} 
 &
  \overset{\rmii{\nr{mu-sa_1}}}{=} 
 &
 - \frac{H}{a^3_{ }\dot{\bar\varphi}}
 \times a \partial^{ }_t
 \biggl[\, 
 \hspace*{-5mm}
 \overbrace{
 a \partial^{ }_t \biggl(\, 
  \frac{a\dot{\bar\varphi}}{H}
 \,\biggr) 
 }^{ a\dot{a}\ibit 
    ( {\dot{\bar\varphi}} / {H} ) 
 \ibit + \ibit a^2_{ } 
    \partial^{ }_t
    ( {\dot{\bar\varphi}} / {H} ) % ^{.}_{ }
 }
 \hspace*{-5mm}
 \,\biggr]
 \nn[2mm]
%%%%%
 & = & 
 - \frac{H}{a^2_{ }\dot{\bar\varphi}}
 \biggl[\,
 \dot{a}^2_{ } \frac{\dot{\bar\varphi}}{H} 
 + a\ddot{a} \frac{\dot{\bar\varphi}}{H} 
 + 3 a\dot{a} \biggl( \frac{\dot{\bar\varphi}}{H} \biggr)^{\textstyle .}_{ }
 + a^2_{ } \biggl( \frac{\dot{\bar\varphi}}{H} \biggr)^{\textstyle ..}_{ }
 \,\biggr]
 \nn[2mm]
%%%%%
 & = & 
 - \frac{\dot{a}^2_{ }}{a^2_{ }}
 - \frac{\ddot{a}}{a} 
 - 3 H  \times \frac{H}{\dot{\bar\varphi}}
 \biggl( \frac{\dot{\bar\varphi}}{H} \biggr)^{\textstyle .}_{ }   
 -     \frac{H}{\dot{\bar\varphi}}
 \biggl( \frac{\dot{\bar\varphi}}{H} \biggr)^{\textstyle ..}_{ }   
 \;. \label{mass_t}
\ea
Recalling \eq\nr{cal_F}, we find 
\ba
 \F  
 & \overset{\rmii{\nr{cal_F}}}{=} & 
 \frac{H}{\dot{\bar\varphi}}
 \biggl( \frac{\dot{\bar\varphi}}{H} \biggr)^{\textstyle .}_{ }   
 \; = \;  \frac{\ddot{\bar\varphi}}{\dot{\bar\varphi}} 
 - \frac{\dot H}{H}  
 \\[2mm]
%%%%%%
 \Rightarrow\quad
 \dot{\F}
 & = &
 \underbrace{
 \frac{\dot{H}}{\dot{\bar\varphi}}
 \biggl( \frac{\dot{\bar\varphi}}{H} \biggr)^{\textstyle .}_{ }   
 -
 \frac{H \ddot{\bar\varphi}}{(\dot{\bar\varphi})^2_{ }}
 \biggl( \frac{\dot{\bar\varphi}}{H} \biggr)^{\textstyle .}_{ }   
 }_{-\F^2_{ }}
 +\,
 \frac{H}{\dot{\bar\varphi}}
 \biggl( \frac{\dot{\bar\varphi}}{H} \biggr)^{\textstyle ..}_{ }   
 \;. \label{cal_F_dot}
\ea 
Inserting \eqs\nr{mass_t} and \nr{cal_F_dot}, 
and going also to momentum space, 
\eq\nr{mu-sa_3} becomes
\be
 \boxed{
 \quad 
 \ddot{\Q}^{ }_k 
 + 3 H \dot{\Q}^{ }_k 
 + \biggl(\,
   \frac{k^2_{ }}{a^2_{ }}
 - 3 \F H 
 - \dot{\F}
 - \F^2_{ }
   \,\biggr) \Q^{ }_k
 \; = \; 0
 \;.
 \quad \vphantom{\Bigg|}
 }
 \label{mu-sa_4}
\ee

We now recall from \eqs\nr{slowroll_Hdot} 
and \nr{slowroll_ddot} and the discussion below them that
in the slow-roll regime, 
$
 \mathcal{F} \approx 
 ( 2 \epsilon^{ }_\rmii{$V$} - \eta^{ }_\rmii{$V$}) H \ll H
$.
Therefore,  
$k^2_{ }/a^2_{ }$
dominates the third term in \eq\nr{mu-sa_4}
as long as $k/a \ge H$. 
Under this assumption, the previously determined 
conformal-time mode function from \eq\nr{Q_k}
indeed solves \eq\nr{mu-sa_4}. 
To show this, we consider 
\be
 \Q^{ }_k(\tau)
 \; 
 \overset{\rmii{\nr{aQ_vs_Q}}}{=} 
 \; 
 \frac{\field^{ }_k(\tau)}{a}
 \; 
 \overset{\rmii{\nr{Q_k}}}{=}
 \;
   \frac{e^{-i k \tau}_{ }}{\sqrt{2 k}}
  \biggl( \tau - \frac{i}{k} \biggr)
 \frac{1}{a\tau}
 \; 
  \overset{\rmii{\nr{a_eta_2}}}{\approx}
 \; 
 - 
 \frac{H}{\sqrt{2 k}}
  \biggl( \tau - \frac{i}{k} \biggr)
 e^{-i k \tau}_{ }
 \;,
 \label{Q_k_t}
\ee
where we inserted 
$H \approx -1/(a\tau)$ from \eq\nr{a_eta_2}. Being in 
quasi de Sitter space-time, $H$ is treated 
as approximately constant. Then 
\ba
 \partial^{ }_t  \Q^{ }_k(\tau)
 &
 \overset{\rmii{\nr{time_derivs}}}{=}
 & 
 \frac{\Q^{\bit\prime}_k(\tau)}{a}
 \; \overset{\rmii{\nr{Q_k_t}}}{\approx} \; 
 - 
 \frac{H}{a \sqrt{2 k}}
 \biggl[ \bcancel{1} 
  - ik \biggl( \tau - \bcancel{\frac{i}{k}} \biggr)
 \biggr] 
 e^{-i k \tau}_{ }
 \nn[2mm]
%%%%
 &
 \underset{\scriptscriptstyle H \;\approx\; -1/(a\tau)}{
 \overset{\rmii{\nr{a_eta_2}} \lift }{\approx}}
 & 
 \frac{H^2_{ }}{\sqrt{2 k}}
 \bigl( 
  - ik \tau^2_{ }
 \bigr) 
 e^{-i k \tau}_{ }
 \;, \label{Q_k_t_dot}
 \\[3mm]
%%%%%
 \partial^{2}_t  \Q^{ }_k(\tau)
 &
 \underset{\rmii{\nr{Q_k_t_dot}}}{
 \overset{\rmii{\nr{time_derivs}} \lift }{=}}
 & 
 \frac{H^2_{ }}{a \sqrt{2 k}}
 \bigl( 
  - 2 ik \tau
  - k^2_{ } \tau^2_{ }
 \bigr) 
 e^{-i k \tau}_{ }
 \nn[2mm]
%%%%
 &
 \underset{\scriptscriptstyle H \;\approx\; -1/(a\tau)}{
 \overset{\rmii{\nr{a_eta_2}} \lift }{\approx}}
 &
 \frac{H^3_{ }}{\sqrt{2 k}}
 \bigl( 
   2 ik \tau^2_{ }
  + k^2_{ } \tau^3_{ }
 \bigr) 
 e^{-i k \tau}_{ }
 \;. \label{Q_k_t_ddot}
\ea
It follows that 
\ba
 && \hspace*{-2.5cm}
 \biggl( \partial^2_t + 3 H \partial^{ }_t + \frac{k^2_{ }}{a^2_{ }}\biggr)
 \Q^{ }_k(\tau) 
 \nn[2mm]
%%%%%
 & \approx &
 \frac{H^3_{ }}{\sqrt{2 k}}
 \biggl[ 
  \overbrace{
   \cancel{ 2 ik \tau^2_{ } }
  + \bcancel{k^2_{ } \tau^3_{ }}
  }^{{\rm from}\;\nr{Q_k_t_ddot} }
  -
  \hspace*{-3mm}
  \overbrace{ 
  \cancel{ 3 ik \tau^2_{ } } 
  }^{{\rm from}\;\nr{Q_k_t_dot} }
  \hspace*{-3mm}
  - 
  \overbrace{
  \underbrace{
    \frac{k^2_{ }}{a^2_{ }H^2_{ }}
  }_{\approx\; k^2_{ }\tau^2_{ } }
    \biggl( \bcancel{\tau} - \cancel{ \frac{i}{k} } \biggr)
  }^{{\rm from}\;\nr{Q_k_t} }
 \biggr] 
 e^{-i k \tau}_{ }
 \; \approx \; 0
 \;. \label{Q_k_t_eom} 
\ea
{}From \eq\nr{Q_k_t_dot},
we also obtain the information
that $\partial^{ }_t \Q^{ }_k(\tau) \approx 0$ when $\tau \to 0^-_{ }$, 
i.e.\ that $\Q^{ }_k$ is constant when the modes are far outside of 
the Hubble horizon. 

Equipped with this knowledge, we take inspiration from 
\se\ref{ss:stochastic_exact}, and define IR perturbations
in analogy with the mode expansion in \eq\nr{delta_varphi_splitup_4}, i.e.\ 
by subtracting the UV part from the full mode expansion, 
\be
 \Q^{ }_{>}(t,\vec{x}) 
 \; \equiv \; 
 \int \! \frac{{\rm d}^3\vec{k}}{\sqrt{(2\pi)^{3}_{ }}} \, 
 \bigl[ 1 -  W^{ }_k(t) \bigr]\, 
 \Bigl[ \,
    w^{ }_\rmii{\vec{k}}
    \, \Q^{ }_k(\tau) 
    \, e^{ i \vec{k}\cdot\vec{x}}_{}   
  + 
    w^{\dagger}_\rmii{\vec{k}}
    \, \Q^{\,*}_k(\tau) 
    \, e^{ - i \vec{k}\cdot\vec{x}}_{}   
 \, \Bigr]
 \;, \label{Q_larger} 
\ee
where the window function is from \eq\nr{W_theta_2}.
Let us consider a time derivative of this operator. Because of the 
weight $1-W^{ }_k$, each momentum $k$ contributes only once the mode
is far outside of the Hubble horizon, but 
as follows from \eq\nr{Q_k_t_dot} and was discussed below
\eq\nr{Q_k_t_eom}, $\Q^{ }_k(\tau)$ is approximately
constant in this domain. 
Therefore the time derivative acts only on the window
function, leading to 
\ba
 \partial^{ }_t \bigl[ 1 -  W^{ }_k(t) \bigr]
 & 
  \underset{\scriptscriptstyle H\;\approx\;{\rm const}}
 {\overset{\rmii{\nr{W_theta_2}} \lift }{\approx}}
 & 
 \epsilon \dot{a}(t) H \delta \bigl( k - \epsilon a(t) H \bigr)
 \nn[2mm]
%%%%
 & = & 
 \epsilon \dot{a}(t) H \delta 
 \bigl\{ k - 
 \epsilon [\,a(t^{ }_*) + \dot{a}(t^{ }_*)(t-t^{ }_*)\,] H \bigr\}
 \nn[2mm]
%%%%
 & = & 
 \frac{\epsilon \dot{a}(t) H}{ | \epsilon \dot{a}(t^{ }_*) H |}
 \,\delta\bigl(\, t - t^{ }_*(k)\, \bigr)
 \; \overset{\scriptscriptstyle \dot{a} \, > \, 0 \lift }{=} \; 
 \delta\bigl(\, t - t^{ }_*(k)\, \bigr)
 \;, 
 \label{dirac_delta}
\ea 
where $t^{ }_*(k)$ is the moment at which 
a given momentum exits the Hubble horizon, 
i.e.\ 
$
 k \equiv \epsilon\hspace*{0.3mm} a(t^{ }_*) H
$.
All in all, the time derivative of the IR mode function reads
\ba
 \dot{\Q}^{ }_{>}(t,\vec{x}) 
 & 
 \approx
 & 
 \varrho^{ }_\iQ(t,\vec{x})
 \;, \label{dot_Q_larger_eom} \\[2mm]
%%%%%%%%
 \varrho^{ }_\iQ(t,\vec{x})
 & 
  \underset{\rmii{\nr{dirac_delta}}}{\overset{\rmii{\nr{Q_larger}}}
  {\equiv}} 
 & 
 \int \! \frac{{\rm d}^3\vec{k}}{\sqrt{(2\pi)^{3}_{ }}} \, 
 \delta \bigl(\, t - t^{ }_*(k)\, \bigr)
 \Bigl[ \,
    w^{ }_\rmii{\vec{k}}
    \, \Q^{ }_k(\tau) 
    \, e^{ i \vec{k}\cdot\vec{x}}_{ }   
  + 
    w^{\dagger}_\rmii{\vec{k}}
    \, \Q^{\,*}_k(\tau) 
    \, e^{ - i \vec{k}\cdot\vec{x}}_{ }   
 \, \Bigr]
 \;. \hspace*{5mm} \label{dot_Q_larger} 
\ea
Here, in analogy with \eq\nr{delta_varphi_eq_splitup_2}, 
we have introduced the notion of a {\em quantum noise}. \index{quantum noise}

In the literature, the conventional
next step is to determine the autocorrelator 
of the quantum noise in coordinate space.
We obtain~\cite{stochastic}
\ba
 && \hspace*{-1.5cm}
 \langle 0 | 
 \varrho^{ }_\iQ(t^{ }_1,\vec{x}^{ }_1)
 \varrho^{ }_\iQ(t^{ }_2,\vec{x}^{ }_2)
 | 0 \rangle
 \nn[2mm]
%%%%%
 & \overset{\rmii{\nr{dot_Q_larger}}}{=} & 
 \int \! 
 \frac{{\rm d}^3\vec{k}\, {\rm d}^3\vec{q} }{(2\pi)^{3}_{ }}
 \, 
 \delta \bigl(\, t^{ }_1 - t^{ }_*(k)\, \bigr)
 \delta \bigl(\, t^{ }_2 - t^{ }_*(q)\, \bigr)
% \nn[2mm]
%%%%%
% & \times & 
   \langle 0|
    w^{ }_\rmii{\vec{k}}
    w^{\dagger}_\rmii{\vec{q}}
   | 0 \rangle
    \, \Q^{ }_k(\tau^{ }_1) 
       \Q^{\,*}_q(\tau^{ }_2) 
    \, e^{ i ( \vec{k}\cdot\vec{x}_1 - \vec{q}\cdot\vec{x}_2)}_{ }   
 \nn[3mm]
%%%%%
 & \overset{\rmii{\nr{commutators}}}{=} & 
 \int \! 
 \frac{{\rm d}^3\vec{k}}{(2\pi)^{3}_{ }}
 \, 
 \delta \bigl(\, t^{ }_1 - t^{ }_*(k)\, \bigr)
 \delta \bigl(\, t^{ }_2 - t^{ }_1 \, \bigr)
% \nn[2mm]
%%%%%
% & \times & 
    \, | \Q^{ }_k(\tau^{ }_1) |^2_{ } 
    \, e^{ i \vec{k}\cdot( \vec{x}_1 - \vec{x}_2)}_{ }   
 \nn[-2mm]
%%%%%
 &
  \underset{\rmii{\nr{dirac_delta}}}
  {\overset{\rmii{\nr{Q_k_t}}}{\approx}}
 & 
 \delta (\, t^{ }_1 - t^{ }_2 \,)
 \int_0^\infty \! \frac{{\rm d}k\, k^2_{ }}{4\pi^2_{ }}
 \overbrace{ 
 \underbrace{
 \epsilon {a}(t^{ }_1) H^2_{ } }_{kH}
 \delta \bigl( k - \epsilon a(t^{ }_1) H \bigr)
 }^{\delta (\, t^{ }_1 - t^{ }_*(k)\, )\; {\rm via}\;\nr{dirac_delta}  }
 \;
 \overbrace{
  |\Q^{ }_k(\tau_1^{ })|^2_{ } 
  }^{
 \nr{Q_k_t}:\; 
 \frac{H^2_{ }}{2k^3_{ }  \lift  } 
  }
 \;
 \overbrace{
 \int_{-1}^{+1} \! {\rm d} z \,
 e^{i k |\vec{x}_1 - \vec{x}^{ }_2| z }_{ }
 }^{ 
 \frac{ 
 2 \sin(k |\vec{x}_1 - \vec{x}^{ }_2|) \lift }
     { k |\vec{x}_1 - \vec{x}^{ }_2|  \lift }
 }
 \nn[3mm]
%%%%%
 & = & 
 \delta (\, t^{ }_1 - t^{ }_2 \,)
 \, 
 \frac{H^3_{ }}{4\pi^2_{ }} 
 \,
 \frac{ 
 \sin[\, \epsilon a(t^{ }_1) H |\vec{x}_1 - \vec{x}^{ }_2| \,] }
     {   \epsilon a(t^{ }_1) H |\vec{x}_1 - \vec{x}^{ }_2|     }
 \;. \label{auto_stoch_x}
\ea
In the penultimate step, we made use of the asymptotics of 
\eq\nr{Q_k_t} outside of the Hubble horizon. 
We thus find {\em white noise} \index{white noise}
in the time direction, 
similarly to \eq\nr{varphi_noise}. 
However, 
the oscillatory behaviour of \eq\nr{auto_stoch_x}
in the spatial directions, and the strong dependence on 
the arbitrary parameter~$\epsilon$, look artificial. 
Nevertheless, the $\epsilon\to 0$ limit 
(corresponding to $k\to 0$
according to \eq\nr{W_theta_2}) exists, 
and is often used as a quantum noise for the evolution
of the background field, $\bar\varphi$. 

The meaning of the noise becomes 
more transparent if we go to momentum space. 
Fourier transforming \eq\nr{dot_Q_larger}, we find
\ba
 \varrho^{ }_\iQ(t,\vec{k})
 & \overset{\rmii{\nr{fourier_k}}}{=} & 
 \int \! {\rm d}^3_{ }\vec{x} \,  \varrho^{ }_\iQ(t,\vec{x})\, 
 e^{-i\vec{k}\cdot\vec{x} }_{ }
 \nn[2mm]
%%%%%
 &
 \overset{\rmii{\nr{dot_Q_larger}}}{=}
 & 
 \sqrt{ (2\pi)^{3}_{ } }  
 \,\delta \bigl(\, t - t^{ }_*(k)\, \bigr)
 \Bigl[ \,
    w^{ }_\rmii{\vec{k}}
    \, \Q^{ }_k(\tau) 
  + 
    w^{\dagger}_\rmii{$-$\vec{k}}
    \, \Q^{\,*}_k(\tau) 
 \, \Bigr]
 \;. \label{dot_Q_larger_k} 
\ea
Then the autocorrelator reads
\ba
 && \hspace*{-2.5cm}
 \bigl\langle \bit 0 \bit \big| 
 \varrho^{ }_\iQ(t^{ }_1,\vec{k})
 \varrho^{ }_\iQ(t^{ }_2,\vec{q})
 \big| \bit 0 \bit  \bigr\rangle
 \nn[2mm]
%%%%%
 & \overset{\rmii{\nr{dot_Q_larger_k}}}{=} & 
 (2\pi)^3_{ }
 \delta \bigl(\, t^{ }_1 - t^{ }_*(k)\, \bigr)
 \delta \bigl(\, t^{ }_2 - t^{ }_*(q)\, \bigr)
 \,
   \langle 0|
    w^{ }_\rmii{\vec{k}}
    w^{\dagger}_\rmii{$-$\vec{q}}
   | 0 \rangle
    \, \Q^{ }_k(\tau^{ }_1) 
    \, \Q^{\,*}_q(\tau^{ }_2) 
 \nn[2mm]
%%%%%
 & \overset{\rmii{\nr{commutators}}}{=} & 
 (2\pi)^3_{ }\delta^{(3)}_{ }(\vec{k+q}) \,
 \delta \bigl(\, t^{ }_1 - t^{ }_*(k)\, \bigr)
 \delta \bigl(\, t^{ }_2 - t^{ }_*(q)\, \bigr)
 \, 
 \underbrace{
  |\Q^{ }_k(\tau_1^{ })|^2_{ } 
  }_{
 \nr{Q_k_t}:\; \frac{H^2_{ }}{2k^3_{ }  \lift } 
  }
 \;. \label{auto_stoch_k}
\ea
Here we again made use of the small-$\tau$ asymptotics of \eq\nr{Q_k_t}. 
What \eq\nr{auto_stoch_k} signifies is that the noise autocorrelator
is zero most of the time. It is non-vanishing only at the moment 
when a given mode exits the Hubble horizon. 

More concretely, we may interpret 
\eq\nr{dot_Q_larger_eom} as an evolution equation, 
but viewing as the underlying variable  
not the IR field itself, $\Q^{ }_{>}(t,\vec{x})$, 
but rather the corresponding
mode function, which we denote by $\Q^{ }_{k>}(t)$.
Then \eq\nr{auto_stoch_k} implies that 
\be
 \dot{\Q}^{ }_{k>}(t)
 \; 
   \underset{\rmii{\nr{auto_stoch_k}}}
  {\overset{\rmii{\nr{dot_Q_larger_eom}}}{\simeq}} 
 \; 
 \delta \bigl(\, t^{ } - t^{ }_*(k)\, \bigr)
 \, \frac{H}{\sqrt{2 k^3_{ }}}
 \;. \label{eom_stock_lin}
\ee
Integrating this equation in time 
with a vanishing initial condition, 
${\Q}^{ }_{k>}$ jumps to its final value as the corresponding mode
crosses into the IR domain. 
The power spectrum following from \eq\nr{eom_stock_lin} reads 
\be
 \P^{ }_{\scriptscriptstyle \Q_\varphi} (t^{ }_\rmi{out},k)
 \; 
 \overset{t^{ }_\rmii{out}\,>\, t^{ }_* \vphantom{\big | } }{=} 
 \; 
 \frac{k^3_{ }}{2\pi^2_{ }} \, 
 \frac{H^2_{ }}{2 k^3_{ }}
 \; = \; 
 \frac{H^2_{ }}{(2\pi)^2_{ }}
 \;, \label{P_stoch}
\ee
which agrees with the second term of \eq\nr{P_R_prefinal}.

To summarize, we have shown that, staying at linear order 
and restricting ourselves to the slow-roll regime, 
the simplified version of the 
stochastic formalism actually corresponds to a 
{\em non-stochastic} evolution equation, \eq\nr{eom_stock_lin}.
The strength of the stochastic formalism 
lies in its presumed generalization beyond the linear order, 
however this property is not easy to prove rigorously.
In the non-linear case, different momenta are not 
independent, and the evolution equation is typically 
solved in coordinate rather
than momentum space, with a noise autocorrelator
reminiscent of \eq\nr{auto_stoch_x}. 
In practice, 
the equation becomes non-linear
if the Hubble rate appearing on the right-hand
side is made a function of $\Q^{ }_\varphi$, 
as would be natural in the separate universes
picture~\cite{separate,uk}.

\newpage

%%%%%%%%%%%%%%%%%%%%%%%%%%%% start appendices %%%%%%%%%%%%%%%%%%%%%%%%%%%%%%%

%%%%%%%%%%%%%%%%%%%%%%%%%%%%%%%%%%%%%%%%%%%%%%%%%%%%%%%%%%%%%%%%%%%%%%%%%
%
\subsubsection{Numerical solution for curvature power spectrum}
\label{app:num_R_varphi}

\addcontentsline{toc}{subsection}{\App\ref{app:num_R_varphi}: 
Numerical solution for curvature power spectrum}

\begin{figure}[t]
    \hspace*{-0.2cm}
    \includegraphics[width=0.47\linewidth]{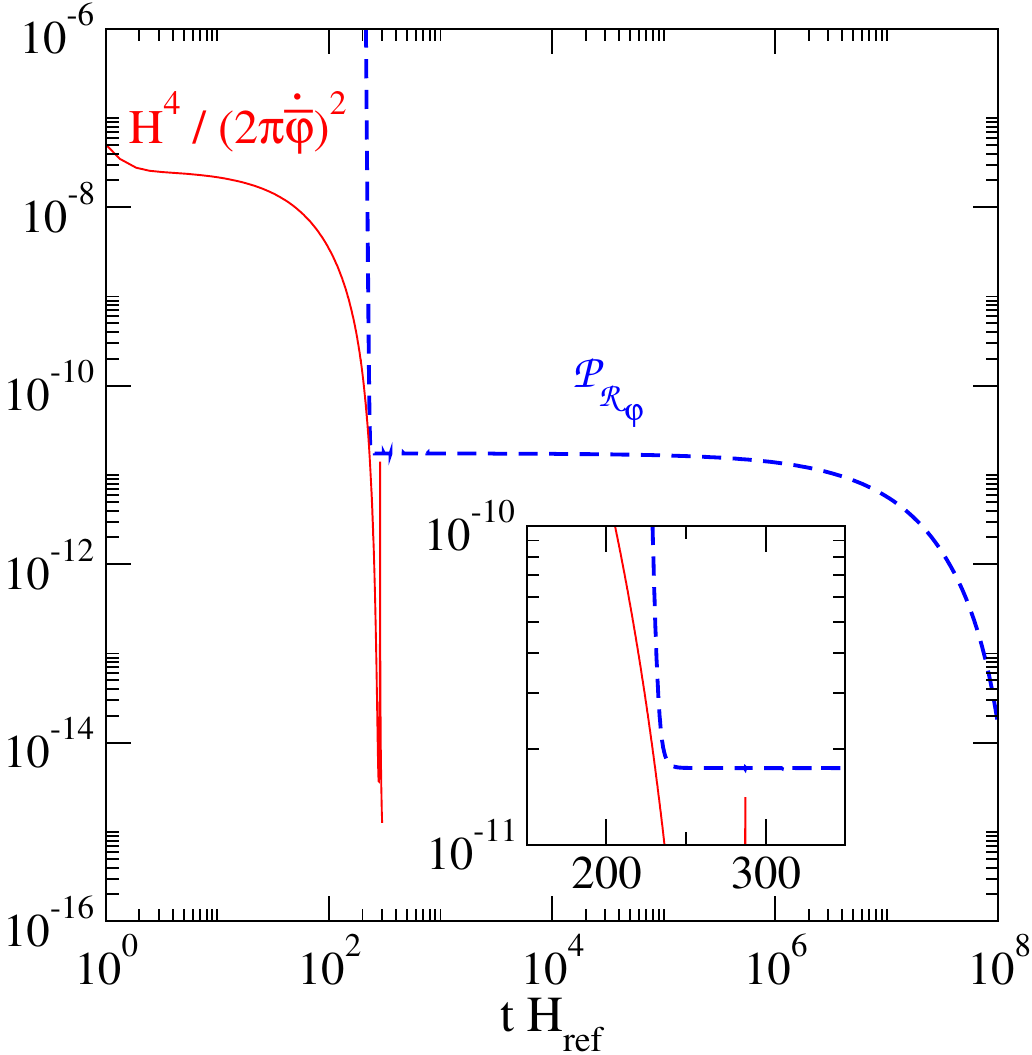}%

    \vspace*{-7.25cm}

    \hspace*{7.8cm}
    \includegraphics[width=0.47\linewidth]{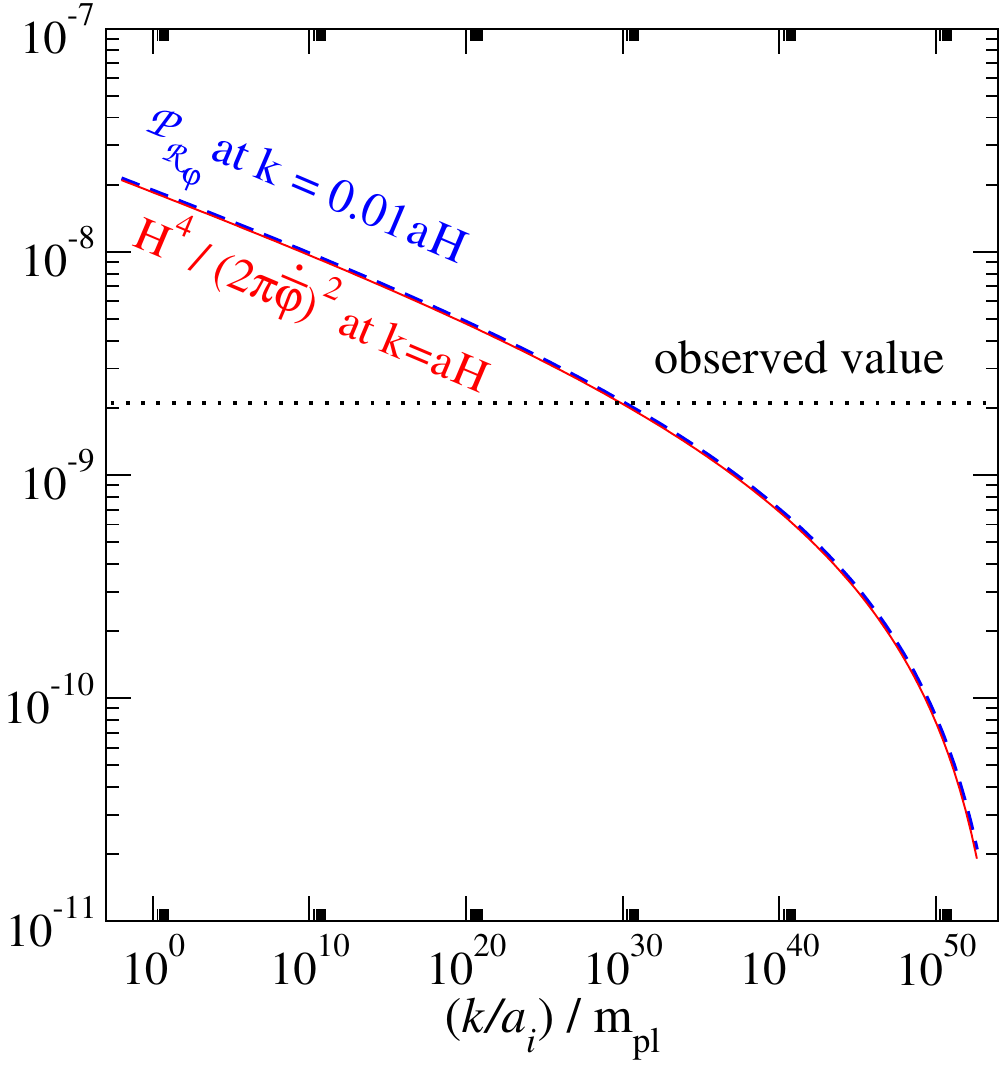}

    \caption{%
    \small
    Left: The dashed blue line shows 
    a numerical solution of 
    \eq\nr{eq_R_k}, 
    with initial conditions fixed through
    \eqs\nr{R_ini_1} and \nr{R_ini_2}, 
    and the result normalized according to 
    \eq\nr{rescale}. The parameters 
    are like in \fig\ref{fig:bg_num}
    on p.~\pageref{fig:bg_num}.
    The solid red curve
    shows the result from \eq\nr{P_R_final} but {\em without}
    fixing to the moment $H^{ }_*$ of horizon crossing.  
    We see how $\P^{ }_{\mbit \scriptscriptstyle \R^{ }_\varphi}$ is 
    normally much larger than $H^4_{ }/(2\pi\dot{\bar\varphi})^2_{ }$,
    due to the term $k^2/a^2$ in \eq\nr{P_R_prefinal}. However, 
    when $k/a$ becomes smaller than~$H$, 
    $\P^{ }_{\mbit \scriptscriptstyle \R^{ }_\varphi}$ freezes out to the 
    value that it had at that moment. Afterwards,
    $H^4_{ }/(2\pi\dot{\bar\varphi})^2_{ }$ continues
    evolving, becoming oscillatory
    when $\dot{\bar{\varphi}}$ starts crossing zeros.
    In contrast, $\P^{ }_{\mbit \scriptscriptstyle \R^{ }_\varphi}$ stays
    fixed, until the mode crosses back inside the 
    Hubble horizon. 
    The initial decrease of 
    $H^4_{ }/(2\pi\dot{\bar\varphi})^2_{ }$ is due to the fact that 
    the initial condition for $\dot{\bar\varphi}$ from \eq\nr{varphi_i}
    is approximate, because $H^{ }_\rmii{ref}$ is not the correct
    Hubble rate, and it takes a few Hubble times to adjust to the
    attractor trajectory. 
    Right: 
    The dependence of $\P^{ }_{\mbit \scriptscriptstyle \R^{ }_\varphi}$,
    evaluated at 
    $k = 0.01\,a H$, and $H^4_{ }/(2\pi\dot{\bar\varphi})^2_{ }$, 
    evaluated at $k = a H$, on the initial momentum, 
    fixed at $t = t^{ }_i = H_\rmii{ref}^{-1}$,
    cf.\ \eq\nr{Href}. 
    We find that 
    $\P^{ }_{\mbit \scriptscriptstyle \R^{ }_\varphi}$ 
    at $ k = 0.01\, aH$ is $(2-5)\%$
    larger than 
    $H^4_{ }/(2\pi\dot{\bar\varphi})^2_{ }$ at $k = a H$.
    The results are compared with the observed
    $\P^{ }_{\mbit \scriptscriptstyle \R^{ }_\varphi}(k^{ }_*)$,
    given in \eq\nr{ex_A_s}. 
    Whether the $k$ at which the observed value is reached indeed
    corresponds to $k^{ }_*$,  
    depends on the reheating history after inflation, 
    and we return to this in  
    \fig\ref{fig:bg_thermal} on p.~\pageref{fig:bg_thermal}.
    }
    \index{curvature perturbations: numerics (figure)} 
    \index{figure: numerics for curvature perturbations}
    \label{fig:P_R_varphi}
\end{figure}

In this appendix we show how the evolution equation~\nr{eq_R_k}, 
with initial conditions fixed according to 
\eqs\nr{R_ini_1} and \nr{R_ini_2}, 
can be solved numerically, in order to obtain the curvature
power spectrum from \eq\nr{rescale}. 
The numerical result is plotted in \fig\ref{fig:P_R_varphi}. 

%%%%%%%%%%%%%%%%%%%%%%%%%%%% FIGURE %%%%%%%%%%%%%%%%%%%%%%%%%%%%%%%%%
%
%
%%%%%%%%%%%%%%%%%%%%%%%%%%%%%%%%%%%%%%%%%%%%%%%%%%%%%%%%%%%%%%%%%%%%

It is worth stressing that the numerical solution
of \eq\nr{eq_R_k} poses a number of challenges, if we want to 
integrate over a long period of time
(a discussion of dedicated softwares
can be found in ref.~\cite{numerics}). 
At early times, $\R^{ }_k$ undergoes
rapid phase oscillations. These are expensive to treat, and therefore
we can start the integration of $\R^{ }_k$ only when $k/(a^{ } H^{ })$ 
is moderate, say $k/(a^{ } H^{ }) \sim 10^3_{ }$. 
Once the modes exit the Hubble
horizon, the rapid phase oscillations cease, and $\R^{ }_k$ freezes out. 
Since its frozen value is the main variable of our interest, this dynamics
has to be determined precisely. 
After a while, the background field, $\bar\varphi$,
may start to oscillate. Then the coefficient $\F$ in \nr{eq_R_k} can become
singular, cf.\ \eq\nr{cal_F}. The singular value needs to be regularized,
but care is
required, as we want to render the equation solvable without 
changing the nature of the solution. 
A procedure for this, making use of complexified variables,  
has been discussed in ref.~\cite{alica}.
Finally, at late times, it is costly to treat every single 
background oscillation individually. Then we can go over to an averaged 
solution, as explained in \app\ref{app:num_bg_vac}. 
Below we show a {\tt python} script which
implements these ingredients (the evolution has been split into
4~periods, as explained in the script, so that the accuracy
of each can be tuned separately if necessary).  

\index{code: numerics for curvature perturbations}

{\fontsize{8pt}{10pt}\selectfont
\begin{verbatim}
# numerical solution for curvature perturbations in the cold regime [numerics_curvature.py]
#
# import basic tools and integration routines
import numpy as np
from scipy.integrate import solve_ivp 

# parameters [mpl]
fa = 1.25          # inflaton decay constant
m = 1.09e-6        # inflaton mass
koa_ref = 7.35e52  # momentum of perturbations at initial time
delta = 1.e-15     # regulator for circumventing poles [mpl^2]
Pi = np.pi

# potential of inflaton (phi) [mpl^4]
def V(phi): return m*m*fa*fa*( 1 - np.cos(phi/fa) )

# derivative of inflaton potential by phi [mpl^3]
def Vd(phi): return m*m*fa*np.sin(phi/fa)

# total energy density of inflaton field [mpl^4]
def etot(phi,phid): return phid*phid/2 + V(phi)

# Hubble rate [mpl]
def H(energy_density): return np.sqrt(8*Pi*np.abs(energy_density)/3)

# time derivative of Hubble rate [mpl^2]
def dotH(phid): return - 4*Pi*phid*phid

# solve for [dot phi, ddot phi, number of efolds] for the background evolution
# derivatives are taken with respect to t = Href*time
def derivatives1(t,y):
    Phi, Phid, Nfolds = y
    Hubble = H(etot(Phi, Phid));        dN_dt = Hubble/Href
    dphi_dt = Phid/Href;                ddphi_ddt = ( -3*Hubble*Phid - Vd(Phi) )/Href
    dy_dt = [dphi_dt, ddphi_ddt, dN_dt]
    return dy_dt

# solve for [dot phi, ddot phi, number of efolds, dot Rk, ddot Rk] including curvature perturbations
def derivatives2(t,y):
    Phi, Phid, efolds, reRk, imRk, reRkd, imRkd = y
    Hubble = H(etot(Phi, Phid));        dHubble = dotH(Phid);             dN_dt = Hubble/Href
    dphi_dt = Phid/Href;                ddphi_ddt = ( -3*Hubble*Phid - Vd(Phi) )/Href
    recalF = ddphi_ddt*Href*np.real(1/(Phid + 1j* delta)) - dHubble/Hubble
    imcalF = ddphi_ddt*Href*np.imag(1/(Phid + 1j* delta))
    dreRk_dt = reRkd/Href;              dimRk_dt = imRkd/Href
    ddreRk_ddt = - ( 2*(recalF*reRkd-imcalF*imRkd) + 3*Hubble*reRkd 
                     + koa_ref*koa_ref*np.exp(-2*efolds)*reRk )/Href
    ddimRk_ddt = - ( 2*(recalF*imRkd+imcalF*reRkd) + 3*Hubble*imRkd 
                     + koa_ref*koa_ref*np.exp(-2*efolds)*imRk )/Href
    dy_dt = [dphi_dt, ddphi_ddt, dN_dt, dreRk_dt, dimRk_dt, ddreRk_ddt, ddimRk_ddt]
    return dy_dt

# solve for [e_phi, number of efolds, dot Rk, ddot Rk] after averaging over oscillations
def derivatives3(t,y):
    e_phi, efolds, reRk, imRk, reRkd, imRkd = y
    Hubble = H(e_phi);                  dHubble = -4*Pi*e_phi;            dN_dt = Hubble/Href
    recalF =  - dHubble/Hubble;         imcalF = 0
    dreRk_dt = reRkd/Href;              dimRk_dt = imRkd/Href
    ddreRk_ddt = - ( 2*(recalF*reRkd-imcalF*imRkd) + 3*Hubble*reRkd 
                     + koa_ref*koa_ref*np.exp(-2*efolds)*reRk )/Href
    ddimRk_ddt = - ( 2*(recalF*imRkd+imcalF*reRkd) + 3*Hubble*imRkd 
                     + koa_ref*koa_ref*np.exp(-2*efolds)*imRk )/Href
    dy_dt = [-3*Hubble*e_phi/Href, dN_dt, dreRk_dt, dimRk_dt, ddreRk_ddt, ddimRk_ddt]
    return dy_dt

# initial conditions and reference values
phi_0 = 3.5                             # mpl
Href = np.sqrt(4*Pi/3)*m*phi_0          # mpl
phid_0 = - Vd(phi_0)/(3*Href)           # approximate slow-roll value for initial derivative [mpl^2]
N_0 = 0                                 # initial e-folds
koaH_start = 1e3                        # when to start solving for perturbations
time_end = 250                          # choose large enough that desired koaH is reached

# event function monitoring k/aH (here _not_ the correct H, rather the initial approximation)
def koaH(t, y):  return koa_ref*np.exp(-y[2])/Href - koaH_start

# integrate without curvature perturbations until k/aH obtains prescribed value koaH_start
sol1 = solve_ivp(derivatives1, [1, time_end], [phi_0, phid_0, N_0], events=koaH, rtol=1e-10)
t_match1 = sol1.t_events[0][0]          # point at which desired value was reached
time1 = sol1.t[sol1.t <= t_match1];     n_match1 = len(time1)
phi1 = sol1.y[0][:n_match1];            phid1 = sol1.y[1][:n_match1]
ephi1 = etot(phi1, phid1);              efolds1 = sol1.y[2][:n_match1]
reRk1 = np.ones(n_match1);              imRk1 = np.zeros(n_match1) 
slowroll1 = H(ephi1)**4/(2*Pi*phid1)**2

# define initial conditions for curvature perturbations at moment t_match1
time_0 = t_match1
phi_0 = sol1.y_events[0][0][0]; phid_0 = sol1.y_events[0][0][1]; N_0 = sol1.y_events[0][0][2]
reRk_0 = -H(etot(phi_0,phid_0))/(2*Pi*phid_0)*koa_ref*np.exp(-N_0);  imRk_0 = 0.
dreRk_0 = 0.;                           dimRk_0 = - reRk_0* koa_ref*np.exp(-N_0)
print("start Rk-evolution at t/tref = ",time_0,", with k/aH approx ",koaH_start)
koaH_start = 10                         # re-adjust target k/aH
time_end = 250                          # choose large enough that desired koaH is reached

# integrate with curvature perturbations until k/aH obtains prescribed value koaH_start
sol2 = solve_ivp(derivatives2, [time_0,time_end],
                 [phi_0, phid_0, N_0, reRk_0, imRk_0, dreRk_0, dimRk_0],
                 events=koaH,t_eval=np.linspace(time_0,time_end,1000),
                 atol=1e-12,rtol=1e-13,method='DOP853')  
t_match2 = sol2.t_events[0][0]          # point at which desired value was reached
time2 = sol2.t[sol2.t <= t_match2];     n_match2 = len(time2)
phi2 = sol2.y[0][:n_match2];            phid2 = sol2.y[1][:n_match2]
ephi2 = etot(phi2,phid2);               efolds2 = sol2.y[2][:n_match2]
reRk2 = sol2.y[3][:n_match2];           imRk2 = sol2.y[4][:n_match2]
slowroll2 = H(ephi2)**4/(2*Pi*phid2)**2

# define initial conditions for next range (crossing outside of Hubble horizon)
time_0 = t_match2
phi_0 = sol2.y_events[0][0][0] ;  phid_0 = sol2.y_events[0][0][1];  N_0 = sol2.y_events[0][0][2]
reRk_0 = sol2.y_events[0][0][3];        imRk_0 = sol2.y_events[0][0][4]
dreRk_0 = sol2.y_events[0][0][5];       dimRk_0 = sol2.y_events[0][0][6]
print("start horizon crossing at t/tref = ",time_0,", with k/aH approx ",koaH_start)
time_end = 1e3                          # choose large enough that 10 full oscillations take place

# define the event function counting the zeros of phi
def phi_crossings(t, y):  return y[0]

# integrate up to 10 oscillations (21 crossings of zero)
sol3 = solve_ivp(derivatives2, [time_0,time_end],
                 [phi_0, phid_0, N_0, reRk_0, imRk_0, dreRk_0, dimRk_0],
                 events=phi_crossings,t_eval=np.linspace(time_0,time_end,1000),
                 atol=1e-12,rtol=1e-13,method='DOP853')  
t_match3 = sol3.t_events[0][20]         # point at which desired value was reached
time3 = sol3.t[sol3.t <= t_match3];     n_match3 = len(time3)
phi3 = sol3.y[0][:n_match3];            phid3 = sol3.y[1][:n_match3]
ephi3 = etot(phi3,phid3);               efolds3 = sol3.y[2][:n_match3]
reRk3 = sol3.y[3][:n_match3];           imRk3 = sol3.y[4][:n_match3]
slowroll3 = H(ephi3)**4/(2*Pi*phid3)**2

# define initial conditions for averaged regime
time_0 = t_match3;                      time_end = 1e8
phi_0 = sol3.y_events[0][20][0];        phid_0 = sol3.y_events[0][20][1]
etot_0 = etot(phi_0,phid_0);            N_0 = sol3.y_events[0][20][2]
reRk_0 = sol3.y_events[0][20][3];       imRk_0 = sol3.y_events[0][20][4] 
dreRk_0 = sol3.y_events[0][20][5];      dimRk_0 = sol3.y_events[0][20][6]
print("start averaged regime at t/tref = ",time_0)

# integrate in averaged regime (matter-dominated era)
sol4 = solve_ivp(derivatives3, [time_0, time_end],
                 [etot_0, N_0, reRk_0, imRk_0, dreRk_0, dimRk_0],
                 atol=1e-12,rtol=1e-13,method='DOP853')
time4 = sol4.t;                         n_match4 = len(time4)
ephi4 = sol4.y[0];                      efolds4 = sol4.y[1]
reRk4 = sol4.y[2];                      imRk4 = sol4.y[3];            slowroll4 = np.zeros(n_match4)

# assemble together the complete solution
time = np.concatenate((time1,time2,time3,time4)); e_phi = np.concatenate((ephi1,ephi2,ephi3,ephi4))
efolds = np.concatenate((efolds1,efolds2,efolds3,efolds4)); koa = koa_ref*np.exp(-efolds)
reRk = np.concatenate((reRk1,reRk2,reRk3,reRk4)); imRk = np.concatenate((imRk1,imRk2,imRk3,imRk4))
slowroll = np.concatenate((slowroll1,slowroll2,slowroll3,slowroll4))

# print to file
np.savetxt('numerics_curvature.dat',
           np.c_[time, e_phi, H(e_phi), efolds, koa, reRk**2+imRk**2, slowroll],fmt='%.6e',newline='\n',
           header='columns: t*H_ref, e_phi/mpl**4, H/mpl, efolds, k_ref/a/mpl, P_R, slowroll')
\end{verbatim}
}

%%%%%%%%%%%%%%%%%%%%%%% end appendices %%%%%%%%%%%%%%%%%%%%%%%%%%%

%%%%%%%%%%%%%%%%%%%%%%%%% BIBLIO %%%%%%%%%%%%%%%%%%%%%%%%%%%%%%%%
%

%%%%%%%%%%%%%%%%%%%%%%%%%%%% SECTION %%%%%%%%%%%%%%%%%%%%%%%%%%%%%%%%%%
\newpage 

\section{Evolution equations in the presence of a thermalizing plasma}
\label{se:thermal}

\paragraph{Abstract:}

As time goes by, other matter 
components than the inflaton field 
play an increasingly important role. 
The expectation is that some of them should be Standard Model 
particles, interacting fairly strongly with each other, 
and eventually thermalizing, setting up the required 
environment for big-bang nucleosynthesis. 
The equilibrated system is called a primordial plasma, 
while the equilibration process, culminating
in a radiation-dominated universe, is known as ``reheating''.
Introducing generic couplings between the 
inflaton and the plasma, 
we write down the corresponding background and 
perturbed equations. 
We show how a ``seed'' temperature may emerge
as a fixed point of the background solution already during
the slow-roll stage of inflation. 
We demonstrate how reheating  
influences inflationary predictions, through the
overall redshift between horizon crossing 
and the late universe. 
We indicate how
interactions damp initial quantum fluctuations, 
but also generate new thermal fluctuations,
via thermal noise. 

\paragraph{Keywords:}

temperature, 
Langevin equation, 
equipartition, 
smooth reheating,
redshift, 
inflaton and plasma equilibration rates, 
thermal noise, 
fluctuation-dissipation relation,  
Rayleigh-Jeans divergence, 
quantum statistical physics, 
coupled curvature perturbations.

%%%%%%%%%%%%%%%%%%%%%%%%%%%%%%%%%%%%%%%%%%%%%%%%%%%
%
\subsection{What is temperature?}
\label{sec_T}

\index{temperature: definition}

When we discuss thermal effects, which ultimately lead 
to the notions of a {\em hot big bang} \index{hot big bang}
and the generation of a {\em primordial plasma}, \index{primordial plasma}
the question of how
temperature is defined is a central one. 
In text-book statistical physics, a starting point
is offered by the {\em micro\-canonical ensemble}. 
\index{microcanonical ensemble}
Let us inspect a closed system
of a total energy $E$. 
The {\em microcanonical partition function}, $\Omega(E)$, 
\index{$\Omega(E)$ (microcanonical partition function)}
counts the
number of states in an energy interval $(E-\Delta E,E]$. The Boltzmann
entropy is defined as $S(E) \equiv \ln \Omega(E)$, and the temperature as
\be
 \frac{1}{T} 
 \;
 \equiv
 \;
 \frac{ \partial S(E) }{ \partial E }
 \;. \label{def_T}
 \index{$T$ (temperature)}
\ee 
Multiplying by $T{\rm d}E$, this  
yields the first law of thermodynamics, ${\rm d} E = T {\rm d}S + ...\,$.

There is, however, an implicit assumption in this definition
of the temperature. The relations above make no reference to interactions, 
only to the counting of states in the vicinity
of a given energy (this could be done even
in a non-interacting system). The implicit assumption is that of 
{\em equipartition}: \index{equipartition}
all those states should be filled with equal 
probability, $1/\Omega(E)$. 
This implies that the thermal system contains minimal
information: 
if we carry out 
physical computations, which require reactions between states, 
all possibilities allowed by energy conservation are  
treated on equal footing, and averaged over.  
 
In the real world, establishing equipartition from a given 
initial state takes time. For instance,
let us think of a system made of 
unstable heavy particles (such as inflatons)
and their decay products 
(perhaps the much lighter Standard Model particles).
In an initial decay, the large energy released is  
carried mostly by the momenta of the decay products. But this is only
one of the possible states of the same energy. Another would be that 
the energy is redistributed evenly between, say, a hundred light particles.
Combinatorially, counting the directions of the momenta, there are many
more states of the latter type. In thermal equilibrium, 
{\em all possible final states}
should be filled with equal probability. 

\index{$\Gamma$ (plasma equilibration rate)}

It should be clear from the picture described that 
in any realistic setting,
it is impossible to carry out an exact computation of how the initial 
energy released in the decay gets redistributed
to all possible final states. Rather, we try
to capture the efficiency of this process by defining an 
{\em equilibration rate}, $\Gamma$. We could compute $\Gamma$ by 
first estimating a would-be $T$ from the thermodynamic definition. 
Then, we can ask how efficiently the processes proceed within a 
{\em Hubble time}, \index{Hubble time} 
$\Delta t \equiv H^{-1}_{ }$. If $\Gamma \Delta t \gg 1$, we can 
be confident that equipartition is established, and make use of
the usual tools of equilibrium statistical physics. 
If $\Gamma \Delta t \ll 1$, there is no time for equipartition, and 
we say that the multiparticle system is ``out of equilibrium''. 
The latter type of systems are in general hard to handle.   

\begin{figure}[t]

    \centering
    \includegraphics[width=0.85\linewidth]{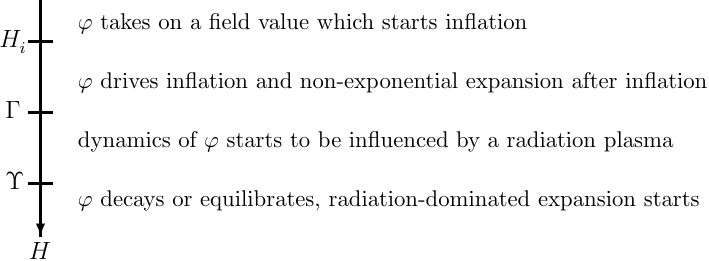}

 \caption[a]{\small
  A sketch of a typical timeline of how inflation ends, 
  if $\Upsilon \ll \Gamma$. Here $\Upsilon$ is the equilibration 
  rate of the low-momentum modes of the inflaton, and $\Gamma$ is
  the equilibration rate of the plasma particles, with physical 
  momenta $p \sim T$.
  As the Hubble rate
  decreases, any interacting particles that are present 
  tend to equilibrate when $\Gamma > H$, and form a plasma.
  In the end, when $\Upsilon > H$, the inflaton background decays,  
  and the inflaton particles may equilibrate with the plasma.
   }
 \index{reheating: sketch (figure)}
 \index{figure: sketch of reheating history}
 \label{fig:timeline}
\end{figure}

Even if out-of-equilibrium systems are somewhat intractable,
there is one generic idea that helps to classify them, 
and that is that we may view them as a collection of 
separate {\em subsystems}. In the example above, 
inflatons constitute one subsystem, 
the decay products another one.
In full equilibrium, every reaction, including the initial decay, 
can also go in the opposite
direction: many low-energy Standard Model particles could merge
together, to form one inflaton particle. 
The rate of the inflaton
equilibration processes, which we denote by $\Upsilon$, 
is often much smaller than $\Gamma$.
One reason is that, 
for simple model building, 
it is beneficial
if inflaton-plasma interactions do not induce large radiative
corrections to the inflaton potential. This can be achieved if  
the inflaton couples weakly to the Standard Model, 
for instance only via a higher-dimensional operator, 
suppressed by powers of $\mpl^{ }$. 
Such models possess the hierarchy $\Upsilon \ll \Gamma$. 
As illustrated in \fig\ref{fig:timeline}, this implies the presence
of different epochs in the history of the universe, 
as the Hubble rate $H$ decreases.

%%%%%%%%%%%%%%%%%%%%%%%%%%%%%%%%% FIGURE %%%%%%%%%%%%%%%%%%%%%%%%%%%%%%%%%
%%%%%%%%%%%%%%%%%%%%%%%%%%%%%%%%%%%%%%%%%%%%%%%%%%%%%%%%%%%%%%%%%%%%%%%%%%%

The remainder of this chapter is organized as follows.  
In \se\ref{sec_eom}, we show how $\Upsilon$ can be computed from 
a certain Green's function, 
given a coupling of $\varphi$ to a radiation plasma. 
In \se\ref{sec_varrho}, we discuss what can be said about
the other plasma-induced ingredient in \eq\nr{varphi_eq}, 
the thermal noise $\varrho$. The background evolution equations satisfied
by the radiation plasma are elaborated upon in 
\se\ref{sec_e-eom}, whereas in 
\se\ref{ss:delta_R_thermal}, we derive the evolution equations 
for curvature perturbations in 
the presence of $\Upsilon$ and $\varrho$. 
\App\ref{app:thermal_inflaton} 
illustrates
what happens when $\Upsilon \gg H$, so that the inflaton itself
may equilibrate, and \app\ref{app:num_bg_thermal} 
shows a numerical solution in the presence
of a radiation plasma, from the inflationary epoch 
through reheating
until a late universe. 
We aim to keep the discussion model independent, 
which means that it is to be understood in an effective theory
sense. 

The details of how the universe heated up
after inflation~\cite{reheat1,reheat2} are, however, 
model dependent. Let us briefly mention a few notions 
that appear frequently in the literature:
\bi

\item
An example of a radiation plasma equilibration rate,
$\Gamma$, is that of a non-Abelian gauge theory. It has been studied
extensively, motivated by 
heavy-ion collision experiments
(but also cosmology, 
cf.,\ e.g.,\ ref.~\cite{equil_cosmo}). 
In terms of microscopic processes, 
a necessary ingredient is to estimate 
the ensemble average of a cross section for 
``large-angle'' scattering, so that momenta are efficiently 
redistributed, leading to 
{\em kinetic equilibration}. \index{equilibration: kinetic}
In the limit of a weak self-coupling of the plasma, 
$\alpha \equiv g^2_{ }/(4\pi) \ll 1$ (implying that 
its temperature is well above the confinement scale),
and considering physical momenta around the 
{\em thermal energy scale}, $p \sim T$,
\index{thermal energy scale}  
the result is 
$\Gamma \sim \alpha^2_{ } T/\ln(\alpha^{-1}_{ })$~\cite{equil}.
The rate of {\em helicity equilibration} has been studied
in less detail, but is believed to be 
of order $\alpha^3_{ } T$ at weak coupling
(cf., e.g., ref.~\cite{hook}). 
\index{equilibration: helicity}

\item \index{reheating: general considerations}
The process of heating up is often called {\em reheating}. 
The name originates from the fact that early models
of inflation \cite{Guth:1980zm,ori2,ori3} involved 
a phase transition taking place in the hot early universe
at the corresponding critical temperature. 
After inflation, the universe must {\em heat up again},
to produce the hot environment observed in the CMB. 
However, as an alternative, 
an exponential expansion 
can also be driven by the energy density of a slowly evolving 
scalar field (cf.,\ e.g.,\ ref.~\cite{chaotic}), 
or other effects (cf.,\ e.g.,\ ref.~\cite{aas}), 
and no pre-existing thermal state is needed.  

\item
The original stages of how the inflaton loses energy to other
degrees of freedom, before a proper thermal state is established, 
are often referred to as 
{\em preheating}~\cite{Kofman:1997yn,bbr}. \index{preheating}
Nowadays preheating is typically studied by solving numerically
a coupled set of non-linear classical field equations, for the
inflaton field and for some ``spectators'', which model the 
constituents of a radiation plasma. However, classical dynamics  
is a good approximation only for large occupation numbers, 
relevant for low-energy bosons, thanks to their Bose enhancement. 
Quantum mechanics was discovered by Planck 
to explain the spectrum of blackbody radiation, and correspondingly, 
quantum mechanics is  
essential for the main stage of the heating-up process.  

\item
The normal tool for studying equilibration, once the occupation numbers 
are already close to unity, is 
{\em kinetic theory} \index{kinetic theory} 
({\em Boltzmann equations}). \index{Boltzmann equations} 
However, this framework 
is not equivalent to the full quantum field theory, either. 
In fact, even for weakly coupled systems, it must often be combined
with various ``resummations'', in order to account for the emergence
of ``quasiparticle'' states, whereby the effective masses carried by
different excitations are modified by interactions 
with the thermal ensemble
(see, e.g.,\ ref.~\cite{equil}). These effects can have a large 
qualitative effect, because modified masses may permit to open or close
scattering channels, changing the ways in which
equipartition can be established. 

\index{Gibbons-Hawking temperature}

\item
A famous concept from quantum field theory in a curved 
background, notably in de Sitter space-time, is that of a 
{\em Gibbons-Hawking temperature}~\cite{gh}. 
A Gibbons-Hawking temperature is obtained for any observer,
independent of their velocity, and of the interactions
that are present. 
In contrast, as discussed at the beginning of this section, 
for the physics that we are interested in, 
a temperature is a meaningful notion when interactions have time 
to re-distribute energy insertions. 
This leads to the corresponding 
equilibration rates, $\Gamma$ and $\Upsilon$. 
Once interactions are present, they
also establish a {\em preferred frame}, 
the plasma rest frame: any moving
excitation loses energy until it comes to rest with respect to
this frame. The absence of 
these properties indicates that the Gibbons-Hawking temperature, 
though it can be defined, 
does {\em not} 
capture the plasma physics that is of relevance to us here.

\index{gravitational particle production}
\index{Bogolyubov transformation}

\item
A notion somewhat related to the Gibbons-Hawking temperature
is that of cosmological {\em gravitational particle production}
(for reviews see, e.g., refs.~\cite{parker,ford,kolo}).
When the background metric is time-dependent, ``instantaneous'' 
vacuum states can be defined at different times, 
and they are not equivalent to each other. 
What looked to be a vacuum state at an initial time seems 
to contain many particles
if considered from the viewpoint of another vacuum. 
The semiclassical transition from one to another vacuum 
is called a {\em Bogolyubov transformation}. The particles produced
when mode functions initialized in 
a Bunch-Davies vacuum are projected onto a post-reheating vacuum
state might be interpreted as dark matter, for instance.  
In the particle-physics language, the same physics can be
captured by Feynman diagrams
of $n\to 2$ processes, where $n\ge 2$. 
Here $n$ refers to inflaton particles, 
annihilating into an off-shell ``$s$-channel'' 
gravitational perturbation
($
 > \hspace*{-2.1mm} - \hspace*{-2.1mm} < \bit
$), 
which in turn decays into dark matter particles
(and in general Standard Model particles as well).
The particles produced have a non-thermal spectrum, 
so that this mechanism does {\em not} explain reheating, 
though it could produce an initial state from which it can start. 

\ei

We end by anticipating that
the idea of many subsystems, mentioned above, applies not only
to stages of the heating-up process, but also to a much later
universe. For instance, a precise understanding of large-scale structure
formation requires treating dark matter, neutrinos, baryons, 
and photons, as decoupled degrees of freedom.

%%%%%%%%%%%%%%%%%%%%%%%%%%%%%%%%%%%%%%%%%%%%%%%%%%%
%
\subsection{How to estimate the inflaton equilibration rate, $\Upsilon$?}
\label{sec_eom}

\index{$\Upsilon$ (inflaton equilibration rate)}
\index{equilibration: inflaton}

The dynamics of a {\em scalar field interacting with a plasma} can
under certain circumstances be
described by a {\em Langevin equation}, given in 
\eqs\nr{varphi_eq} and \nr{varphi_noise}, {\it viz.}
\be
 {\varphi^{;\mu}_{ }}^{ }_{;\mu} - \Upsilon \, u^{\mu}_{ } \varphi^{ }_{,\mu}
  - V^{ }_{\der\varphi}  
  + \varrho 
 \; \overset{\rmii{\nr{varphi_eq}}}{=} \; 0
 \;, \quad
 \bigl\langle\,
    \varrho(\X) \varrho(\Y)
 \,\bigr\rangle 
 \; \overset{\rmii{\nr{varphi_noise}}}{\approx} \; 
 \frac{\Omega \, \delta^{(4)}_{ }(\mathcal{X-Y})}{\sqrt{-g}} 
 \;.
 \label{langevin}
 \index{Langevin equation}
 \index{scalar field interacting with a plasma}
 \index{scalar field equation: thermal}
\ee
Here the matching coefficients $\Upsilon$ and $\Omega$ 
describe the interactions of $\varphi$ with 
the plasma, in particular $\Upsilon$ can be referred
to as its {\em equilibration rate} (to be precise, it is the 
equilibration rate of the momentum modes $k/a \ll T$; that
for $k/a \ge T$ is in general different, cf.,\ e.g.,\ ref.~\cite{jacopo}). 
Eq.~\nr{langevin} should be viewed as a (dissipative)
{\em low-energy
effective theory}, in which the details of the plasma decomposition
do not matter. However, we may backtrack to a more
fundamental description, 
notably an action such as in \eq\nr{einstein_hilbert}, 
and ask how $\Upsilon$ and $\varrho$ originate in that context?

Specifically, we consider a Lagrangian of the form 
\begin{equation}
 \mathcal{L} 
 \;=\;
 -\frac{1}{2}\ibit g^{\mu\nu}_{ }
   \partial_\mu^{ } \varphi \, \partial_\nu^{ } \varphi
 - V_0^{ }(\varphi)
 - \varphi J
 + \mathcal{L}_\rmi{bath}^{ }
  \ , \label{L_bath}
\end{equation}
where $V^{ }_0$ is the inflaton self-interaction potential, 
$J$ is an operator made of plasma fields, 
and $\mathcal{L}_\rmii{bath}$ is a Lagrangian describing their kinetic
terms and self-interactions. As discussed above, 
we assume that the coupling between $\varphi$
and the plasma is weak; for instance, $J$ could be suppressed
by the Planck mass, $\mpl^{ }$.
This then leads to a hierarchy $\Upsilon \ll \Gamma$.

If we have simultaneously 
$\Upsilon \ll \Gamma$, $k/a \ll \Gamma$, and $\widehat{m}/a \ll \Gamma$, 
where $\widehat{m}$ is from \eq\nr{mode_functions_again}, 
then the evolution of~$\varphi$, as described by \eq\nr{langevin},
is slow compared with the equilibration dynamics of the plasma.
The background field, $\bar\varphi$, then follows the evolution equation
\begin{equation}
 \ddot{\bar \varphi} + 3H\dot{ \bar \varphi}
 + V^{ }_{,\varphi}(\bar{\varphi}) 
 \; 
 \underset{\rmii{\nr{bg_varphi}}}{
 \overset{\rmii{\nr{langevin}} \lift }{=}} 
 \; 
 -\Upsilon\dot{ \bar \varphi}
% + \ord(\dddot{\bar \varphi}, \dot{\bar \varphi}^2)
  \ . \label{eq_phi_eom1}
\end{equation}
We now want to determine the values of $V$ and $\Upsilon$ by 
{\em matching},  \index{matching of effective couplings} 
i.e.\ by looking at the equation of motion
following from \eq\nr{L_bath}, and setting it in the same
form as \eq\nr{eq_phi_eom1}.
In order to simplify the matching, we work in local Minkowskian 
coordinates, in the plasma rest frame. By the equivalence principle, 
the resulting evolution equation can be written in a covariant 
form, and finally evaluated in an expanding FLRW universe. 

In a Minkowskian frame, the Euler-Lagrange equation 
for $\bar{\varphi}$ from \eq\eqref{L_bath} reads
\begin{equation}\label{eqEoM}
 \ddot{ \bar \varphi } + V^{ }_{0,\varphi} ( \bar \varphi )
 \; = \;
 - \langle J(t)\rangle^{ }_{\bar\varphi}
 \ .
\end{equation}
The right-hand side is non-trivial, because 
the average value of $J$ depends on the slow
variables of the problem, notably $\bar{\varphi}$ and $T$. 
To find this dependence, we can invoke a {\em linear response}
argument. Let us assume that a heat bath is present from 
a time $t=0$ onwards. 
The heat bath dynamics is considered in the canonical
formalism, so that the Hamiltonian reads 
\begin{equation}
 \hat{H}
 \; = \;
 \hat{H}_\rmi{bath} + \bar{\varphi}\,\hat{J}
 \ . \label{eq_Hamilton}
\end{equation}
With the help of the 
{\em density matrix} of the heat bath, $\hat{\rho}(t)$, 
\index{$\hat{\rho}$ (density matrix)}
the average value of $\hat{J}$ can be written as
\begin{equation}\label{eq_averageJ}
 \langle \hat{J}(t) \rangle^{ }_{\bar\varphi} 
 \;\equiv\; 
 \tr[\, \hat{\rho}(t) \hat{J}(t) \,] \ .
\end{equation}
The density matrix satisfies the 
{\em Liouville-von Neumann equation}, 
\begin{equation}
 i\partial_t^{ } \hat{\rho}(t)
 \;=\;
 [\, \hat{H}(t),\hat{\rho}(t) \,]
 \ . \label{eq_schroed_densitym}
 \index{Liouville-von Neumann equation}
 \index{$\hat{H}$ (Hamiltonian)}
\end{equation} 
It is helpful to go over to an {\em interaction picture}, 
in which the time evolution of the operators is determined
by an unperturbed Hamiltonian 
($\hat{H}^{ }_\rmii{bath}$ in \eq\nr{eq_Hamilton}), 
whereas that of states originates from a perturbation
($\bar{\varphi}\,\hat{J}$ in \eq\nr{eq_Hamilton}).
The corresponding density matrix, 
$
 \hat{\rho}^{ }_\iI
$,
and operator,
$
 \hat{J}^{ }_\iI
$,
are obtained by substituting
\be
 \hat{\rho}(t)
 \; \equiv \; 
 e^{-i \hat{H}^{ }_\rmii{bath} t}_{ } 
 \,\hat{\rho}^{ }_\iI(t)\,
 e^{i \hat{H}^{ }_\rmii{bath} t}_{ }
 \;, \quad
 \hat{J}(t)
 \; \equiv \; 
 e^{-i \hat{H}^{ }_\rmii{bath} t}_{ } 
 \,\hat{J}^{ }_\iI(t)\,
 e^{i \hat{H}^{ }_\rmii{bath} t}_{ }
 \label{I_pict}
 \index{interaction picture}
\ee
in \eqs\nr{eq_Hamilton} and \nr{eq_schroed_densitym}.
Then the Liouville-von Neumann equation becomes
\be
 i\partial_t^{ } \hat{\rho}^{ }_\iI(t)
 \; 
 \underset{\rmii{\nr{I_pict}}}{
 \overset{\rmii{\nr{eq_schroed_densitym}}}{=}}
 \;
 \bar\varphi(t)\,
 [\, \hat{J}^{ }_\iI(t),\hat{\rho}^{ }_\iI(t) \,] 
 \;, \label{LN_I}
\ee
which can be solved for $\hat{\rho}^{ }_\iI$ 
iteratively in the perturbation, 
\ba
 \hat{\rho}^{ }_\iI(t) 
 &
 \overset{\rmii{\nr{LN_I}}}{=}
 & 
 \hat{\rho}^{ }_\iI(0)
 -i \int_0^t \dd t' \, \bar{\varphi}(t') \, 
 [\, \hat{J}^{ }_\iI(t'),\hat{\rho}^{ }_\iI(0) \,]
 +\ord(\bar\varphi^2_{ }J^{2}_\iI)
 \;. \label{rho_soln}
\ea 
% where the initial time has been chosen as $t = 0$.
Inserting the result in \eq\eqref{eq_averageJ}, where
$
  \tr[\hat{\rho}(t) \hat{J}(t)] 
 = 
  \tr[\hat{\rho}^{ }_\iI(t) \hat{J}^{ }_\iI(t)] 
$
by the cyclic property of the trace,
we find
\ba
 \langle \hat{J}(t) \rangle^{ }_{\bar\varphi} 
 & = & 
 \tr[\, \hat{\rho}^{ }_\iI(0) \hat{J}^{ }_\iI(t) \,] 
 -i \int_0^t \dd t' \, \bar{\varphi}(t') \, 
 \tr\bigl\{\,  
 [\, \hat{J}^{ }_\iI(t'),\hat{\rho}^{ }_\iI(0) \,] \,
 \hat{J}^{ }_\iI(t) \,\bigr\}
 +\ord(\bar\varphi^2_{ }J^{3}_\iI)
 \nn[2mm]
%%%%
 & = & 
 \tr[\, \hat{\rho}^{ }_\iI(0) \hat{J}^{ }_\iI(t) \,] 
 -i \int_0^t \dd t' \, \bar{\varphi}(t') \, 
 \tr\bigl\{\, \hat{\rho}^{ }_\iI(0)\,
 [\, \hat{J}^{ }_\iI(t) , \hat{J}^{ }_\iI(t') \,] 
 \,\bigr\}
 +\ord(\bar\varphi^2_{ }J^{3}_\iI)
 \;. \hspace*{6mm} \label{J_expansion}
\ea
According to \eq\nr{rho_soln}, at time $t=0$, 
the density matrix $ \hat{\rho}^{ }_\iI $ does not depend
on~$\bar\varphi$. Given that the equilibration dynamics of the
heat bath has been assumed faster than the evolution of 
$\varphi$, we may assume its initial density matrix to be a thermal one, 
\be
 \hat{\rho}^{ }_\iI(0)  
 \;
 \overset{\rmii{\nr{I_pict}}}{=} 
 \;
 \hat{\rho}(0)
 \;\equiv\;
 \frac{ e^{ - { \hat{H}^{ }_\rmii{bath} } / { T } }_{ } }{\Z}
 \;, \quad
 \Z \; \equiv \; \tr \bigl(\,
  e^{ - { \hat{H}^{ }_\rmii{bath} } / { T } }_{ }
  \,\bigr)
 \;, \label{rho_thermal}
\ee
where $\Z$ is the 
{\em canonical partition function}.
\index{$\Z$ (canonical partition function)}
We define
\ba
 \langle \hat{J}(t) \rangle^{ }_0 
 &
 \overset{\rmii{\nr{J_expansion}}}{\equiv}
 & 
 \tr \bigl[ 
 \hat{\rho}^{ }_\iI(0)
  \hat{J}^{ }_\iI(t)
 \bigr]
 \;
 \underset{\rmii{\nr{rho_thermal}}}{
 \overset{\rmii{\nr{I_pict}} \lift }{=}}
 \; 
 \tr \bigl[ 
 \hat{\rho}(0)
  \hat{J}(t)
 \bigr] 
 \;, \label{def_J0} \\[2mm]
%%%% 
 G_\iR^{ }(t-t')
 &
 \overset{\rmii{\nr{J_expansion}}}{\equiv}
 & 
 \theta(t-t') 
 \tr \bigl\{\,
 \hat{\rho}^{ }_\iI(0)\,
 i\, [\, \hat{J}^{ }_\iI(t),\hat{J}^{ }_\iI(t') \,]
 \,\bigr\}
 \nn[1mm]
%%%
 &
 \underset{\rmii{\nr{rho_thermal}}}{
 \overset{\rmii{\nr{I_pict}} \lift }{=}}
 &
 \theta(t-t') 
 \tr \bigl\{\,
 \hat{\rho}(0)\,
 i\, [\, \hat{J}(t),\hat{J}(t') \,]
 \,\bigr\}
 \;, 
 \label{def_GR_again}
 \index{retarded Green's function: quantum}
\ea
where $G^{ }_\iR$ is a {\em retarded Green's function}, of the 
type that we have met before but now in the context of 
quantum statistical physics. 
Then \eq\nr{J_expansion} becomes
\be
 \langle \hat{J}(t) \rangle^{ }_{\bar\varphi} 
 \;
 \underset{\rmii{\nr{def_J0},\nr{def_GR_again}}}{
 \overset{\rmii{\nr{J_expansion}} \lift }{=}}
 \; 
 \langle \hat{J}(t) \rangle^{ }_0 
 -\int_0^{t} \dd t'\bar{\varphi}(t')\, G_\iR^{ }(t-t')
 \;
 +
 \;
 \ord(\bar\varphi^2_{ }J^{3}_\iI)
 \;. \label{J_expansion_2} 
\ee
This is called a {\em linear response} \index{linear response}
relation, given that
$\bar\varphi$ is expanded to linear order. 

Returning now to \eq\nr{eqEoM}, let us assume that we can
define a global symmetry, $\bar\varphi \to - \bar\varphi$, 
which guarantees that in a stable vacuum, 
we have $\langle \bar\varphi \rangle = 0$. In order
to keep the full Lagrangian invariant under this symmetry, so that the
symmetry is respected by quantum corrections, $J$ needs to be odd.
This then implies that  
$ \langle \hat{J}(t) \rangle^{ }_0  = 0 $.
Thereby \eq\nr{eqEoM} combined with 
\eq\nr{J_expansion_2} 
can be turned into an effective equation of motion, 
\be
  \ddot{ \bar \varphi } + V^{ }_{0,\varphi} ( \bar \varphi )
 -\int_0^{t} \dd t'\bar{\varphi}(t')\, G_\iR^{ }(t-t')
 \; \underset{\rmii{\nr{J_expansion_2}}}
    {\overset{\rmii{\nr{eqEoM}}}{\approx}} \; 
 0 
 \;.  \label{eff_eom}
\ee

In order to match \eqs\nr{eq_phi_eom1} and \nr{eff_eom},
let us transform them to frequency space, recalling that
the time evolution starts at $t=0$,  
\begin{align}
 \bar \varphi(\omega )
 &\;\equiv\; 
 \int_0^\infty \dd t \, e^{i\omega t}\, \bar \varphi (t)
 \ , 
 & 
 \bar\varphi(t) \theta (t)
 &\;=\;
 \int_{-\infty}^\infty \frac{\dd \omega}{2\pi}
 e^{-i\omega t}\, \bar \varphi(\omega)
 \ ,\\
%%%%%%%%%%%%%%%%%
 G_\iR^{ }(\omega)
 &\;=\;
 \int_0^{\infty} \dd t \, e^{i\omega t}\, G_\iR^{ }(t)
 \ ,
 & 
 G_\iR^{ }(t) 
 &\;=\; 
  \int_{-\infty}^\infty \frac{\dd \omega}{2\pi}
 e^{-i\omega t}\,
 G_\iR^{ }(\omega) 
 \ . \label{eqFourierCR}
\end{align}
The one-sided Fourier transforms of $\dot{\bar \varphi}$ 
and $\ddot{\bar \varphi}$ yield
\ba
 \int_0^{\infty} \dd t \, \dot{\bar{\varphi}}(t)\, e^{i\omega t}
 & = & 
 -  {\bar \varphi}(0) 
 -  i\omega \bar \varphi (\omega) 
 \;, \\[2mm]
 \int_0^{\infty} \dd t \, \ddot{\bar{\varphi}}(t)\, e^{i\omega t}
 & = & 
 -\dot{\bar \varphi}(0) + i\omega\bar{\varphi}(0)
 - \omega^2 \bar \varphi (\omega)
 \ ,
\ea
and for a convolution we find
\ba
 \int_0^\infty \! {\rm d}t\, e^{i\omega t}_{ }
 \int_0^t \! {\rm d}t'
 \, \bar\varphi(t')\, G_\iR^{ }(t - t')
 & = & 
 \overbrace{
 \int_0^\infty \! {\rm d}t\, \int_0^t \! {\rm d}t'\,
 }^{ 
 \int_0^\infty \! {\rm d}t'\, \int_{t'}^\infty \! {\rm d}t\,
 }
 \, 
 e^{i\omega t'}_{ } \bar\varphi(t')
 \, 
 e^{i\omega (t - t')}_{ } G_\iR^{ }(t - t') 
 \nn[2mm]
%%%%%
 &
 \overset{\scriptscriptstyle t\;=\;t' + \tilde t \lift }{=}
 & 
 \underbrace{
 \int_0^\infty \! {\rm d}t'
 \, 
 e^{i\omega t'}_{ } \bar\varphi(t')
 }_{ \bar\varphi(\omega) }
 \, 
 \underbrace{
 \int_{0}^\infty \! {\rm d}\tilde t
 \,
 e^{i\omega \tilde t}_{ }\, G_\iR^{ }(\tilde t) 
 }_{ G_\iR^{ }(\omega) }
 \;. 
\ea
Furthermore, we denote the Fourier transform of the
self-interaction ``force'' by 
\be
 \mathcal{F}^{ }_{\omega}[V^{ }_{0,\varphi}]
 \; \equiv \; 
 \int_0^{\infty} \dd t \, V^{ }_{0,\varphi} \, e^{i\omega t}
 \;. 
\ee
Then we find (in a Minkowskian frame, 
so that the Hubble rate vanishes) \pagebreak
\begin{align}
 \mbox{\nr{eq_phi_eom1}}:& 
 & 
  -\dot{\bar \varphi}(0) + i\omega\bar{\varphi}(0)
  -\omega^2 \bar \varphi (\omega)
  -  \Upsilon \,[\,
    {\bar \varphi}(0) 
  +  i\omega \bar \varphi (\omega)  
  \,] 
 + 
 \mathcal{F}^{ }_\omega [ V^{ }_{,\varphi}(\bar{\varphi}) ] 
 &\; \approx \; 0
 \;, \hspace*{6mm} \label{w_match_1} \\[2mm]
%%%
 \mbox{\nr{eff_eom}}:& 
 & 
  -\dot{\bar \varphi}(0) + i\omega\bar{\varphi}(0)
  -\omega^2 \bar \varphi (\omega)
  - {\bar \varphi}(\omega)\, G^{ }_\iR(\omega) 
  + 
 \mathcal{F}^{ }_\omega [ V^{ }_{0,\varphi}(\bar{\varphi}) ] 
 &\; \approx \; 0
 \;. \label{w_match_2}
\end{align}
The terms related to initial conditions should
play no role for the late-time dynamics. 
If we write 
$
 G^{ }_\iR(\omega) 
 = 
 \re G^{ }_\iR(\omega) + 
 i \im G^{ }_\iR(\omega)
$, 
a comparison of \eqs\nr{w_match_1} and \nr{w_match_2} yields
\ba
 \mathcal{F}^{ }_\omega [ V^{ }_{,\varphi}(\bar{\varphi}) ] 
 &
 \underset{\rmii{\nr{w_match_2}}}{
 \overset{\rmii{\nr{w_match_1}}}{\approx}}
 & 
 \mathcal{F}^{ }_\omega [ V^{ }_{0,\varphi}(\bar{\varphi}) ] 
 - 
 {\bar \varphi}(\omega) \re G^{ }_\iR(\omega)
 \;, \label{re_GR} \\[2mm]
%%%% 
 \Upsilon
 &
 \underset{\rmii{\nr{w_match_2}}}{
 \overset{\rmii{\nr{w_match_1}}}{\approx}}
 & 
 \frac{\im G^{ }_\iR(\omega)}{\omega}
 \;. \label{im_GR}
\ea
To be precise, \eq\nr{re_GR} is only valid at linear
order in $\bar\varphi$, as otherwise the $\omega$-dependence
of $\mathcal{F}^{ }_\omega$ could contain both a real and 
an imaginary part. In principle, effects non-linear 
in~$\bar\varphi$ can also be incorporated~\cite{db}.

The friction coefficient in \eq\nr{im_GR} has an important property, 
namely that 
\be
 \Upsilon \;\ge\; 0
 \;. \label{Ups_ineq}
\ee
On a general level, this can be related to the ``optical theorem'', 
% \index{optical theorem}
according to which imaginary parts of scattering amplitudes 
are related to their absolute values squared. More concretely, 
we could evaluate the Green's function from 
\eq\nr{def_GR_again} in the eigenbasis of $\hat H^{ }_\rmii{bath}$,
$\hat{H}^{ }_\rmii{bath} |n\rangle = E^{ }_n |n\rangle$, 
resulting in a weighted sum over transition matrix 
elements like $|\langle m | \hat{J} | n \rangle|^2_{ }$.
Even if this is a nice exercise, 
the details play no role in practical computations, 
and thus we do not show them here. 

The interpretation of \eqs\nr{re_GR} and \nr{im_GR} is not entirely 
trivial. Eq.~\nr{re_GR} suggests that 
$\re G^{ }_\iR \equiv -\delta m_\T^2$ plays
the role of a thermal mass correction, and 
\eq\nr{im_GR} shows how the friction coefficient 
$\Upsilon$ arises. However, 
these relations are functions of $\omega$, and it must be
asked how $\omega$ is chosen. If the inflaton oscillates around the
minimum of its potential, so that 
$
 V^{ }_{0,\varphi} \approx m^2_{ }\, \bar\varphi
$,  
then it is natural to identify 
the frequency with the one of the oscillations, 
$\omega\to m$, 
and the thermally corrected mass reads 
$m_\T^2 = m^2_{ } + \delta m_\T^2(m)$. 
On the other hand, before the oscillation period, 
if the inflaton dynamics is slow compared with plasma equilibration rate,
as indicated above \eq\nr{eq_phi_eom1}, then we can approximate the 
plasma Green's function by the leading term of its expansion 
around $\omega = 0$. Generically, both 
$\lim_{\omega \to 0} \re G^{ }_\iR$ and 
$\lim_{\omega \to 0} \im G^{ }_\iR/\omega$
are non-zero. 

Equations~\nr{re_GR} and \nr{im_GR} underline an important
challenge for coupling the inflaton field to other
degrees of freedom. If we introduce an operator
like in \eq\nr{L_bath}, we see that the retarded correlator 
associated with $J$ yields simultaneously a mass correction
and a friction coefficient. If a plasma is present already during
the inflationary period, this may be regarded as a problem~\cite{linde}, 
given that large mass corrections may spoil the desired  
inflationary predictions, based on $V^{ }_0$.
At the same time, we would like to have a non-zero 
$\Upsilon$, in order to efficiently heat up the universe
after inflation. These conflicting requirements pose 
(sometimes ignored) constraints on viable inflationary models. 

%%%%%%%%%%%%%%%%%%%%%%%%%%%%%%%%%%%%%%%%%%%%%%%%%%%
%
\subsection{Which role is played by the thermal noise, $\varrho$?}
\label{sec_varrho}

\index{$\varrho$ (thermal noise)}

We now turn to the other part of the effective Langevin description
(cf.\ \eq\nr{langevin})
that is related to the interactions between the inflaton and the
radiation plasma, 
namely the {\em thermal noise}, $\varrho$. \index{thermal noise}
Before describing the influence
of the noise within the Langevin equation, let us specify
the ``target'' for what the noise should achieve. 

We place ourselves in a Minkowskian frame like in \se\ref{sec_eom}, 
and furthermore consider 
a thought experiment leading to $\Upsilon \Delta t \gg 1$, 
so that the inflaton has equilibrated. It can still experience
fluctuations around its global minimum, $\bar\varphi = 0$. We assume
that the curvature of the potential 
is $m^2_{ } \equiv V^{ }_{,\varphi\varphi}$
(we have redefined $m_\T^2\to m^2_{ }$ for notational simplicity). 
Then the inflaton fluctuations
have a 2-point correlator related to that determined in 
\eq\nr{2pt_field}, but now modified by 
thermal corrections~(cf.,\ e.g.,\ ref.~\cite[eq.~(3.39)]{basics}), 
\be
 \langle
    \delta \widehat\varphi(t,\vec{x})\,
    \delta \widehat\varphi(t,\vec{y})
 \rangle
 \; \underset{\rmii{ }}{\overset{\rmii{Gaussianity}}{=}} \; 
 \int \! \frac{{\rm d}^3_{ }\vec{k}}{(2\pi)^3_{ }} \,
 \frac{
         e^{i\vec{k}\cdot(\vec{x-y}) }_{ } 
      }{2 \hat{\epsilon}^{ }_k } 
 \, \biggl[\,
   \underbrace{ 
   1
   }_{ 
   \begin{array}{c}
   \scriptscriptstyle
   \,\Leftrightarrow\,\rmii{\nr{P_Q_k}} \\[-2mm]
   \scriptscriptstyle \,{\rm at}\, |k\tau|\,\gg\,1
   \end{array}
   }
   + 
   \; 
   \underbrace{
    \frac{2}{e^{ \hat{\epsilon}^{ }_k / (a T) }_{ } - 1}
   }_{  
   \;\equiv\; 
    2 n^{ }_\rmiii{B}
      ( \hat{\epsilon}^{ }_k / a)
   }  
 \,\biggr]
 \;, \label{2pt_quantum} 
\ee 
where the Gaussian approximation refers to the perturbative approach
(cf.\ the discussion on p.~\pageref{perturbative}).
In \eq\nr{2pt_quantum},  
$
 \hat{\epsilon}^{ }_k \equiv \sqrt{k^2_{ } + a^2_{ }m^2_{ } }
$,
and 
$\nB^{ }$ is the {\em Bose distribution}. 
\index{$\nB^{ }$ (Bose distribution)}

In order to simplify
the notation further, we go over to non-conformal (physical) coordinates
in the following, denoted by
\be
 \vec{r} \;\equiv\; a\hspace*{0.3mm} \vec{x} \;, \quad
 \vec{s} \;\equiv\; a\hspace*{0.3mm} \vec{y} \;, \quad
 \vec{p} \;\equiv\; \frac{\vec{k}}{a} \;, \quad
 \epsilon^{ }_p \;\equiv\; \sqrt{p^2_{ } + m^2_{ }} 
 \;.  \label{mink_not}
\ee
Recalling also $\delta\widehat\varphi = a\ibit \delta\varphi$, 
\eq\nr{2pt_quantum} becomes
\be
 \langle
    \delta \varphi(t,\vec{r})\,
    \delta \varphi(t,\vec{s})
 \rangle
 \;
 \underset{\rmii{\nr{mink_not}}}
 {\overset{\rmii{\nr{2pt_quantum}} \lift }{=}}
 \; 
 \int \! \frac{{\rm d}^3_{ }\vec{p}}{(2\pi)^3_{ }} \,
 \frac{
         e^{i\vec{p}\cdot(\vec{r-s}) }_{ } 
      }{2 {\epsilon}^{ }_p } 
 \, \bigl[\,
   1 + 
    2 \nB^{ }( \epsilon^{ }_p)
 \,\bigr]
 \;. \label{spectrum_full}
\ee 

Next, we would like to compute the same correlator
as in \eq\nr{spectrum_full} from the
Langevin equation, \eq\nr{langevin}. In a local Minkowskian 
coordinate system, around the global minimum, and boosting to 
the plasma rest frame, the perturbations satisfy
\be
 \bigl(\, \partial_t^2 - \nabla^2_\vec{r} 
 + \Upsilon \partial^{ }_t + m^2_{ } \,\bigr) \delta\varphi (t,\vec{r}) 
 \;
 \underset{\varphi \,=\, \bar\varphi \,+\, \delta\varphi}{
 \overset{\rmii{\nr{langevin}} \lift }{=}}  
 \; 
 \varrho (t,\vec{r})
 \;. 
\ee 
As this is a linear partial differential equation, its general solution
is a sum of the general solution of the homogeneous equation and a special
solution of the inhomogeneous one. We assume the solution of the 
homogeneous equation to be represented in momentum space, and its
integration constants to be fixed like for 
a quantum-mechanical mode function, 
from \eqs\nr{wronskian_value} and \nr{derivative_value},
though it now gradually decays, 
due to the dissipative coefficient~$\Upsilon$. Let us denote
this solution by 
$
 \delta\varphi^{ }_\vac
$. 
The special solution originates from the noise, $\varrho$, 
and we call it the classical one, 
$
 \delta\varphi^{ }_\rmi{cl}
$.
The general solution is then 
\be
 \delta \varphi \;=\; \delta \varphi^{ }_\vac + \delta \varphi^{ }_\rmi{cl}
 \;. 
 \label{full_soln}
\ee

Let us stress that 
we treat both parts of $\delta\varphi$ as complex numbers
(not operators), 
in accordance with the classical nature of the Langevin equation.
However, the initial value of 
$ \delta \varphi^{ }_\vac $ is normalized like 
a quantum-mechanical mode function, from
\eqs\nr{wronskian_value} and \nr{derivative_value}. 
In quantum mechanics, the power spectrum is 
obtained from the absolute value squared of the mode function, 
cf.\ \eq\nr{P_Q_k}, whereas in the classical description, there are
no creation and annihilation operators, and the power spectrum
originates like in \eq\nr{eq_PQdef}. 
Either way, in the absence of $\Upsilon$, 
$ \delta \varphi^{ }_\vac $ 
accounts for the vacuum part of \eq\nr{spectrum_full}, 
$1/(2\epsilon^{ }_p)$.

On the other hand, the special solution, $\delta \varphi^{ }_\rmi{cl}$, 
can be obtained with a Green's function. To streamline the notation, 
we denote
\be
 \R \; \equiv \; (t,\vec{r})
 \;, \quad 
 \int_\R
 \;
 \underset{\rmii{\nr{mink_not}}}{
 \overset{\rmii{\nr{fourier_K}}}{\equiv}}
 \;
 \int \! {\rm d}^4_{ }\R
 \;, \quad
 \P \; \equiv \; (\omega,\vec{p})
 \;, \quad
 \int_\P 
 \; 
 \underset{\rmii{\nr{mink_not}}}{
 \overset{\rmii{\nr{fourier_X}}}{\equiv}} 
 \; \int_{-\infty}^{\infty}
 \! \frac{{\rm d} \omega}{2\pi} 
 \int \! 
 \frac{{\rm d}^3_{ }\vec{p}}{(2\pi)^3_{ }}
 \;. 
\ee
With different letters we denote coordinate 
($\mathcal{R,S,U,V}$) or momentum space ($\mathcal{P,Q}$). 

With this notation, the special solution is given by
\be
 \delta\varphi^{ }_\rmi{cl}(\R)
 \; = \; \int_{\mathcal{U}}
 G^{ }_\iR(\mathcal{R-U}) \varrho(\mathcal{U})
 \;, \label{varphi_cl}
\ee
where the 
{\em retarded Green's function}, 
\ba
 &&
  \bigl(\, \partial_t^2 - \nabla^2_\vec{r} 
 + \Upsilon \partial^{ }_t + m^2_{ } \,\bigr)
 G^{ }_\iR(\R) 
 \;=\;  
 \delta^{(4)}_{ }(\R)
 \;, \label{G_R_R}
\ea
can be represented in momentum space,
\ba
 &&
 G^{ }_\iR(\R) 
 \; \overset{\rmii{\nr{fourier_X}}}{=} \; 
 \int_\P e^{-i\omega t + i\vec{p}\cdot\vec{r}}_{ }\, \widetilde G^{ }_\iR(\P)
 \;, \quad
 \widetilde G^{ }_\iR(\P) 
 \; \overset{\rmii{\nr{G_R_R}}}{=} \; 
 \frac{1}
 {-\omega^2_{ } - i \omega \Upsilon + \epsilon_p^2 }
 \;. \label{greens} 
 \index{retarded Green's function: thermal}
 \index{$G^{ }_\iR$ (retarded Green's function)}
\ea
With the Green's function, the contribution of the classical 
fluctuations to the 2-point correlator becomes 
\ba
 \langle\,  \delta\varphi^{ }_\rmi{cl}(\R)
 \, \delta\varphi^{ }_\rmi{cl}(\mathcal{S}) \,\rangle
 & \overset{\rmii{\nr{varphi_cl}}}{=} & 
 \int_{\mathcal{U,V}}
 G^{ }_\iR(\mathcal{R-U})\,G^{ }_\iR(\mathcal{S-V})\,
 \langle\,
  \varrho(\mathcal{U}) \varrho(\mathcal{V}) 
 \,\rangle
 \nn[2mm]
%%%
 & \overset{\rmii{\nr{langevin}}}{=} & 
 \Omega 
 \int_{\mathcal{U}}
 G^{ }_\iR(\mathcal{R-U})\,G^{ }_\iR(\mathcal{S-U})\,
 \nn[2mm]
%%%
 & \overset{\rmii{\nr{greens}}}{=} & 
 \Omega 
 \int_{\mathcal{P,Q}}
 \underbrace{ 
 \int_{\mathcal{U}}
 e^{i\P\cdot(\mathcal{R-U})+i\Q\cdot(\mathcal{S-U})}_{ }
 }_{
 (2\pi)^4_{ }\delta^{(4)}_{ }(\P + \Q)\, e^{i\P\cdot(\mathcal{R-S})}_{ } 
 }
 \, \widetilde G^{ }_\iR(\P) \widetilde G^{ }_\iR(\Q)
 \nn[2mm]
%%%
 & \underset{\rmii{time}}{\overset{\rmii{equal}\lift }
  {\longrightarrow}} & 
 \Omega  
 \int_{-\infty}^{\infty} \! \frac{{\rm d}\omega}{2\pi}
 \int_\vec{p} 
 e^{i\vec{p}\cdot(\vec{r-s})}_{ }
 \,
 \widetilde G^{ }_\iR(\P) \widetilde G^{ }_\iR(-\P) 
 \;. \label{2pt_1}
\ea
We can search for the poles of the Green's functions, 
\ba
 [-\widetilde G^{ }_\iR(\P)]^{-1}_{ }
 & = & 
 \omega^2_{ } + i \omega\Upsilon - \epsilon_p^2
 \; = \; 
 \prod_{\sigma = \pm}
 \Bigl( 
  \omega + \tfr{i\Upsilon}{2}
 +\sigma \sqrt{\epsilon_p^2 - \tfr{\Upsilon^2}{4}}
 \Bigr) 
% \Bigl( 
%  \omega + \tfr{i\Upsilon}{2}
% - \sqrt{\epsilon_p^2 - \tfr{\Upsilon^2}{4}}
% \Bigr) 
 \hspace*{6mm} \nn[2mm]
%%%%%%%
 & \Rightarrow & 
 \mbox{poles in the lower half-plane if $\epsilon_p^2 > 0$} 
 \;, \label{poles_lower} \\[2mm]
%%%%%%%
 [-\widetilde G^{ }_\iR(-\P)]^{-1}_{ }
 & = & 
 \omega^2_{ } - i \omega\Upsilon - \epsilon_p^2
 \; = \; 
 \prod_{\sigma = \pm}
 \Bigl( 
  \omega - \tfr{i\Upsilon}{2}
  +\sigma \sqrt{\epsilon_p^2 - \tfr{\Upsilon^2}{4}}
 \Bigr) 
% \Bigl( 
%  \omega - \tfr{i\Upsilon}{2}
% - \sqrt{\epsilon_p^2 - \tfr{\Upsilon^2}{4}}
% \Bigr) 
 \hspace*{6mm} \nn[2mm]
%%%%%%
 & \Rightarrow & 
 \mbox{poles in the upper half-plane if $\epsilon_p^2 > 0$} 
 \;. \label{poles_upper}
\ea
We integrate over $\omega$ in \eq\nr{2pt_1} by closing the contour
in the upper half-plane, obtaining
\ba
 && \hspace*{-2.0cm}
 \langle\,  \delta\varphi^{ }_\rmi{cl}(\R)
 \, \delta\varphi^{ }_\rmi{cl}(\mathcal{S}) \,\rangle
 \nn[2mm]
%%%%
 & \underset{\rmii{time}}{\overset{\rmii{equal}}{\longrightarrow}} & 
 \Omega  
 \int_\vec{p} 
 e^{i\vec{p}\cdot(\vec{r-s})}_{ }
 \frac{2\pi i}{2\pi}
 \biggl[\,
 \underbrace{ 
  \frac{1}{i\Upsilon(i\Upsilon + 2\sqrt{\cdots})2\sqrt{\cdots}}
 }_{\rm from~pole~at~\omega \;=\; \frac{i\Upsilon}{2} + \sqrt{\cdots} }
 \; + \;  
 \underbrace{ 
  \frac{1}{(i\Upsilon - 2\sqrt{\cdots})i\Upsilon (-2\sqrt{\cdots})}
 }_{\rm from~pole~at~\omega \;=\; \frac{i\Upsilon}{2} - \sqrt{\cdots} }
 \,\biggr] 
 \nn[2mm]
%%%%% 
 & = & 
 \frac{\Omega}{\Upsilon}
 \int_\vec{p} 
 e^{i\vec{p}\cdot(\vec{r-s})}_{ }
 \biggl[\, 
  \frac{\bcancel{i\Upsilon} - 2 \sqrt{\cdots}}
   {-\cancel{\Upsilon^2_{ }}
   - 4 (\epsilon_p^2 - \cancel{\frac{\Upsilon^2}{4}} )}
 \; - \; 
  \frac{\bcancel{i\Upsilon} + 2 \sqrt{\cdots}}
   {-\cancel{\Upsilon^2_{ }}
   - 4 (\epsilon_p^2 - \cancel{\frac{\Upsilon^2}{4}} )}
 \,\biggr]\frac{1}{2\sqrt{\cdots}}
 \nn[2mm]
%%%%% 
 & = & 
 \frac{\Omega}{\Upsilon}
 \int_\vec{p} 
 e^{i\vec{p}\cdot(\vec{r-s})}_{ }
 \,\frac{1}{2\epsilon_p^2}
 \;. \label{2pt_2}
\ea

Let us compare \eq\nr{2pt_2} with the target expression
in \eq\nr{spectrum_full}. We recall that $\delta\varphi^{ }_\vac$
is responsible for $1/(2\epsilon^{ }_p)$. Obviously, \eq\nr{2pt_2}
does {\em not} agree with the other part, containing 
the Bose distribution. However, the low-energy
expansion of the Bose distribution reads 
\be 
 \nB^{ }(\epsilon^{ }_p)
 \; = \; 
 \frac{1}{e^{\epsilon^{ }_p/T } - 1}
 \; \overset{\epsilon^{ }_p \;\ll\; T}{\approx} \;
 \frac{1}{
    \frac{\epsilon_p^{ } \lift }{T  \lift }
  + \frac{\epsilon_p^2 \lift }{2T^2_{ } \lift }
  + ...}
 \; \overset{\epsilon^{ }_p \;\ll\; T}{\approx} \;
 \frac{T}{\epsilon^{ }_p} -\frac{1}{2} + ...
 \;. \label{bose_taylor} 
\ee
It follows that 
\be
  \langle
    \delta \varphi(t,\vec{r})\,
    \delta \varphi(t,\vec{s})
 \rangle
 \; \underset{\rmii{\nr{bose_taylor}}}{
     \overset{\rmii{\nr{spectrum_full}} \ilift }{\supset}} \; 
 \int_\vec{p} \,
 e^{i\vec{p}\cdot(\vec{r-s}) }_{ } 
 \biggl( \frac{T} {\epsilon_p^2} + ... \biggr) 
 \;. \label{spectrum_appro}
\ee

To summarize, 
we see that \eqs\nr{2pt_2} and \nr{spectrum_appro} match, 
if the {\em noise autocorrelator} is 
\be
 \Omega \;=\; 2 T \Upsilon
 \;. \label{fl-di}
 \index{fluctuation-dissipation relation}
 \index{noise autocorrelator}
\ee
This relation between $\Omega$ and $\Gamma$
is known as the {\em fluctuation-dissipation theorem}. 
Making use of \eq\nr{fl-di} guarantees that the noise $\varrho$
drives $\varphi^{ }_\rmi{cl}$ towards the thermal state. 
At the same time, the approximate nature of the matching
underlines that the Langevin description is ``only'' an 
effective theory, valid for inflaton 
excitations with energies $\epsilon^{ }_p \ll T$. 
We elaborate on this point around the end 
of \app\ref{app:thermal_inflaton}, 
on p.~\pageref{rayleigh-jeans}.
The possibility to keep the full Bose distribution in the 
noise autocorrelator, in order to extend the validity regime
of the Langevin description, has been discussed in ref.~\cite{alica2}.  

%%%%%%%%%%%%%%%%%%%%%%%%%%%%%%%%%%%%%%%%%%%%%%%%%%%%%%%%%%%%%%%%%%%%%%%%%
%
\subsection{Temperature evolution and its influence on CMB observables}
\label{sec_e-eom} 

\index{temperature: evolution equation}

Having discussed what temperature means, in \se\ref{sec_T}, and how
thermal effects can be incorporated via the friction coefficient,
$\Upsilon$, in \se\ref{sec_eom}, and the noise, 
$\varrho$, in \se\ref{sec_varrho}, 
we now ask which value
the temperature can take, assuming that it is a meaningful 
notion ($\Gamma > H$), and which role it plays
for CMB observables 
(the last point is discussed at the end of
this section). 
To proceed, we need an 
evolution equation for the radiation plasma. 

The key input for determining the temperature evolution 
comes from overall energy conservation. 
At the background level, the conservation implies
(cf.\ \eq\nr{bg_Tmunu}) 
\be
 \dot{\bar{e}} + 3 H (\bar{e} + \bar{p} )
 \;
 \overset{\rmii{\nr{bg_Tmunu}}}{=}
 \; 
  -\,
 (\ddot{\bar\varphi} + 3 H \dot{\bar\varphi})\, \dot{\bar\varphi}
 \;, \label{bg_Tmunu_again} 
\ee
where, according to \eq\nr{general_e_p}, 
\ba
 \bar{e}
 &
 \overset{\rmii{\nr{general_e_p}}}{\equiv}
 &
 \bar e^{ }_r + V  - T V^{ }_{,\T}
 \;, \quad
 \bar{p}
 \;
 \overset{\rmii{\nr{general_e_p}}}{\equiv}
 \;
 \bar p^{ }_r - V 
 \;, \label{e_p_again}
 \\[2mm]
%%%%%%%
 \dot{\bar e}
 &
 \overset{\rmii{\nr{e_p_again}}}{=}
 &
       \dot{\bar e}^{ }_r 
     + V^{ }_{,\varphi}\,\dot{\bar\varphi}
     + \cancel{ V^{ }_{,\T} \dot{T} }
     - \cancel{ \dot{T} V^{ }_{,\T} } 
     - T \dot{V}^{ }_{,\T}
 \;. \label{dot_bare} 
\ea

We can simplify \eq\nr{bg_Tmunu_again} by combining it with 
the evolution equation for $\bar\varphi$. Concretely, 
let us multiply \eq\nr{eq_phi_eom1}, {\it viz.}
\be
 \boxed{
 \quad
  \ddot{\bar\varphi} 
 \, + \,(3 H + \Upsilon)\,\dot{\bar\varphi} \, + V^{ }_{\der\varphi}
 \;
 \overset{\rmii{\nr{eq_phi_eom1}}}{=}
 \; 0 
 \;,
 \quad \vphantom{\Bigg|} 
 }
 \label{bg_varphi_again}
\ee 
by $\dot{\bar\varphi}$, and 
add it to \eq\nr{bg_Tmunu_again}. 
Then, with the help of \eq\nr{dot_bare}, we get 
\be
 \boxed{ 
 \quad
       \dot{\bar e}^{ }_r 
     + \cancel{ V^{ }_{,\varphi}\,\dot{\bar\varphi} }
     - T \dot{V}^{ }_{,\T}
 + 3 H (\bar e^{ }_r + \bar p^{ }_r - T V^{ }_{,\T} )
 \;
 \underset{\rmii{ }}{
 \overset{\rmii{\nr{bg_Tmunu_again}--\nr{bg_varphi_again}}
          \lift }{=}}
 \; 
 \cancel{ V^{ }_{,\varphi}\,\dot{\bar\varphi} }
 + \Upsilon \dot{\bar\varphi}^2_{ } 
 \;.
 \quad \vphantom{\Bigg|} 
 }
 \index{evolution equations: inflaton and fluid}
 \label{eom_e_r}
\ee
Here, given that the radiation energy density is a function of $T$ only, 
we could furthermore write 
$
 \dot{\bar e}^{ }_r = \dot{T} \bar c^{ }_r
$, 
where 
$
 \bar c^{ }_r \equiv \bar e^{ }_{r,\T}
$
is the {\em heat capacity}. \index{heat capacity} 
Therefore, \eq\nr{eom_e_r} determines the time evolution
of the temperature. To have the complete set of 
equations in one place, let us also repeat
the Hubble rate (cf.\ \eq\nr{bg_HH}) once again, 
\begin{empheq}[box=\fbox]{align}
%%%%%
 \quad
  \frac{8\pi}{3\mpl^2} 
  \,\biggl(\, \frac{\dot{\bar\varphi}^2_{ }}{2} + \bar{e}\, \biggr)
 & 
 \;
 \underset{\scriptscriptstyle \kappa\,=\,0}{
 \overset{\rmii{\nr{bg_HH},\nr{mpl}} \lift }{=}}
 \;
  H^2_{ }  
 \;.
  \quad \vphantom{\Bigg|} 
 \label{bg_HH_again}
\end{empheq}

We recall from \eq\nr{Tds} that
if the system has no 
chemical potentials related to conserved charges, then 
\be
 \bar e^{ }_r  + \bar p^{ }_r
 \;
  \overset{\rmii{\nr{Tds}}}{=}
 \;
 T \bar s^{ }_r
 \;, \label{entropy}
 \index{entropy density: definition}
\ee
where $s^{ }_r = \bar{p}^{ }_{r,\T}$ 
is the {\em entropy density}. In \eq\nr{eom_e_r}, 
the term $-V^{ }_{,\T}$ represents 
a contribution to the entropy density. The negative sign 
means that storing free energy density in $\bar \varphi$, 
whose value carries definite information, decreases 
the entropy density. However, as mentioned in the last
paragraph of \se\ref{sec_eom}, inflationary models
typically have $V^{ }_{,\T} \approx 0$. 

\vspace*{3mm}

Let us now ask whether \eq\nr{eom_e_r} could have a stationary
solution, $\dot{\bar e}^{ }_r \approx 0$? We also set 
$V^{ }_{,\T} \approx 0$ for simplicity. Furthermore, 
could this happen already during the slow-roll stage of 
inflation, when $\ddot{\bar\varphi} \approx 0$? 
Solving for $\dot{\bar\varphi}$ from \eq\nr{bg_varphi_again}, 
substituting the answer in \eq\nr{eom_e_r}, and inserting 
\eq\nr{entropy}, we find
\be
 T \bar{s}^{ }_r 
 \; 
 \underset{\rmii{\nr{bg_varphi_again}--\nr{entropy}}}{
 \overset{\rmii{stationary} \lift }{\approx}}  
 \; 
 \frac{\Upsilon}{3 H}
 \biggl( \frac{V^{ }_{,\varphi}}{3 H + \Upsilon }\biggr)^2_{ }
 \;. \label{stationary} 
\ee
The left-hand side of this equation is a rapidly varying function
of $T$, while the right-hand side is normally slowly varying. 
Therefore, a solution can be 
found~(cf.,\ e.g.,\ refs.~\cite{fixed_pt,warm_sm1,warm_sm2}), 
and it turns out to be a stable 
{\em fixed point}, \index{thermal fixed point}  
in the sense that
if the initial temperature is above or below the stationary value, 
it adjusts to it. An example of a numerical solution that 
displays these features is shown 
in \app\ref{app:num_bg_thermal}. 

However, whenever making use of $T$,  
we have to keep in mind the discussion from \se\ref{sec_T}: 
temperature is a self-consistent notion only if 
we can also show that the plasma self-interaction rate satisfies
$\Gamma \gg H$ during the period considered.
Here $\Gamma$ should be evaluated
at the thermal energy scale, $p \sim T$, 
because such momenta contribute most 
to the thermal energy density and pressure
(here we have in mind ultrarelativistic particles, 
with masses $m \ll T$). 
Given that the parameters entering $\Gamma$ are different from those
entering $\Upsilon$ and $H$, 
it is a model-dependent question when 
$\Gamma \gg H$ is satisfied. 

An example of 
a simple estimate of $\Gamma$ and $H$ can be obtained by considering
a radiation-dominated universe, in which the Hubble rate is parametrically
of magnitude 
\be
 H 
 \;
 \overset{\rmii{\nr{bg_HH_again}}}{\sim}
 \;
 \frac{\sqrt{\bar e}}{\mpl^{ }}
 \;
 \overset{\rmii{\nr{p_r}}}{\sim}
 \;
 \frac{T^2_{ }}{\mpl^{ }}
 \;. \label{T_estimate}
\ee
At the same time, for kinetic equilibration,
$\Gamma \sim \alpha^2_{ } T/\ln(\alpha^{-1}_{ })$ 
according to the discussion
in \se\ref{sec_T}, 
with $\alpha < 1$ referring to a fine-structure
constant of non-Abelian gauge interactions~\cite{equil}. 
So, we see that $\Gamma \gg H$ is satisfied if 
$T \ll \alpha^2_{ } \mpl^{ }/\ln(\alpha^{-1}_{ })$. 
For a numerical value, we could insert $\alpha\sim 0.01$ for QCD, 
obtaining $T \ll 10^{15}_{ }\,$GeV.

\index{maximal temperature}

The consideration below \eq\nr{T_estimate} is, however, 
not a strict criterion 
for when we can talk about a temperature. In principle,  
the notion of a temperature may be meaningful already before
the universe entered a radiation-dominated epoch, as the
inflaton loses energy to the other particles, 
which attain a would-be temperature
according to \eq\nr{def_T}. 
We may then ask what the 
{\em maximal temperature} of the universe could be. 
For this, we may look for 
a solution of \eq\nr{eom_e_r}, with $\dot{T} = 0$
at $T = T^{ }_\rmii{max}$ (together with $\ddot{T} < 0$). 
However, now we {\em cannot} use
the slow-roll approximation on the right-hand side, 
given that $T^{ }_\rmii{max}$ could be reached
after the end of inflation. 
Therefore, \eq\nr{stationary} should be rephrased as 
\be
 \bigl(\, T \bar{s}^{ }_r \,\bigr)^{ }_\rmi{max} 
 \; \overset{\rmii{ }}{\simeq} \; 
 \biggl(
   \Upsilon\,
   \frac{ \dot{\bar\varphi}^{\ibit 2}_{ } }{3 H}
 \biggr)^{ }_\rmi{max}
 \;. \label{maximal} 
\ee
The solution is model-dependent, through 
the properties of the radiation plasma (via~$s^{ }_r$), 
through its interactions with the inflaton 
(via~$\Upsilon$), and through inflaton dynamics
(via~$\dot{\bar\varphi}^{\ibit 2}_{ }$). 
In some models, the solution coincides with 
the stationary value from \eq\nr{stationary}.
If the plasma is strongly self-interacting
(``confining''), $T^{ }_\rmii{max}$ 
could be particularly high,
as its entropy density is otherwise
exponentially suppressed~\cite{Tmax}. 

After having reached $T^{ }_\rmii{max}$, the temperature starts to 
decrease. For a while, $\dot{\bar\varphi}^{\ibit 2}_{ }$ and~$V$ can 
dominate the total energy density. If $V$ can be approximated as 
quadratic around its minimum, 
this leads to a matter-dominated epoch, 
as discussed in \app\ref{app:num_bg_vac}. 
The matter-dominated epoch stops when
$\Upsilon  \Delta t \approx \Upsilon / H \gg 1$. Then we enter
the radiation-dominated era, where \eq\nr{T_estimate} applies.
This whole process may be called
{\em smooth reheating}. \index{reheating: smooth} 
With this concept 
we underline the distinction
to {\em instantaneous reheating}, which is often employed as a simplified
model (cf.\ \fig\ref{fig:history_tau} on p.~\pageref{fig:history_tau}). 

\vspace*{3mm}

Three further remarks are in order. First, 
if $\Upsilon = 0$, like in cold inflation, 
there is no source term for the radiation plasma, 
and any possible initial temperature just redshifts away. 
However, if $\Upsilon$ is proportional to a positive 
power of $T$, the solution $T=0$ 
corresponds to an unstable fixed point. 
Mathematically, just a small perturbation drives
the system to the solution of \eq\nr{stationary}. 
 
Second, we have not yet specified the radiation 
{\em thermodynamic functions}. 
In general, they can be parameterized as 
\begin{equation} 
    e^{ }_r    \;=\;  \frac{  g^{ }_* \pi^2_{ } T^4_{ } }{ 30 }  \ , \qquad
    s^{ }_r    \;=\;  \frac{2 h^{ }_* \pi^2_{ } T^3_{ } }{ 45 }  \ , \qquad
    c^{ }_r    \;=\;  \frac{2 i^{ }_* \pi^2_{ } T^3_{ } }{ 15 }  
  \ , 
  \label{p_r}
  \index{thermodynamic functions}
\end{equation}
with the pressure obtained as 
$
 p^{ }_r = T s^{ }_r - e^{ }_r
$
(for small chemical potentials).
\index{effective number of degrees of freedom}
The functions $g^{ }_*$, $h^{ }_*$ and $i^{ }_*$ are referred to 
as {\em effective numbers of massless degrees of freedom}. If the 
plasma were non-interacting, and its constituents were massless bosons, 
these functions would be constant integers, 
with $g^{ }_* = h^{ }_* = i^{ }_*$.
In realistic systems, the functions vary slowly and are close
to each other, unless the system becomes strongly interacting, 
like QCD at temperatures (0.1...1.0)~GeV, in which case they evolve
fast. The determination of these functions for 
the Standard Model is a topic
of its own (cf.,\ e.g.,\ ref.~\cite{eos15_copy}). 

Third, reheating dynamics plays an important role 
for the inflationary curvature power spectrum, 
because it influences the {\em redshift} 
\index{redshift factor: during reheating}
between early-universe and
present-day momenta. We illustrate this with a concrete example 
in \fig\ref{fig:bg_thermal} on p.~\pageref{fig:bg_thermal}.
We note that in the literature, if reheating dynamics is not addressed, 
it is conventional to vary the time at which the pivot scale
exited the Hubble horizon between 50--60 
$e$-folds before inflation ends, 
and display this variation as an error band.
Here the {\em end of inflation} is \index{inflation: ending}
defined as the moment at which the slow-roll parameter 
$\epsilon^{ }_\rmii{$V$}$ from \eq\nr{slowroll_params}
grows to be of order unity or, 
in physical terms, when the change of the Hubble rate
during a Hubble time is as large as the Hubble rate itself, 
$
 |\Delta H| \equiv
 |\Delta t \, \dot{H}| \equiv
 | \dot{H} / H | 
 \sim H
$
(cf.\ \eq\nr{slowroll_Hdot}).

%%%%%%%%%%%%%%%%%%%%%%%%%%%%%%%%%%%%%%%%%%%%%%%%%%%%%%%%%%%%%%%%%%%%%
%
\subsection{Evolution equations for curvature perturbations}
\label{ss:delta_R_thermal}

\index{evolution equations: coupled perturbations}
\index{curvature perturbations: equation}

Having determined the background equations in the presence
of a plasma and discussed their solution
(cf.\ \se\ref{sec_e-eom}), the next challenge is to work out the evolution
equations for the perturbations. Restricting to scalar 
perturbations, this amounts to generalizing
the derivation in \se\ref{ss:sasaki} to include the effects
from $\Upsilon$, $\varrho$, $\delta e$, 
$\delta p$, and $v \equiv \delta v$. 
As a starting point, 
we take 
\eqs\nr{delta_einstein_00}, 
\nr{delta_einstein_0i}, 
\nr{delta_einstein_didj}, 
\nr{delta_einstein_comb} and 
\nr{eq_field-eq-pert}. 

To be concrete, a closed set for inflaton and metric perturbations 
is constituted by
\ba
 a^2_{ }\varrho   
 & \overset{\rmii{\nr{eq_field-eq-pert}}}{=} &
 \delta\varphi'' + ( 2 \H + a \Upsilon )\, \delta \varphi'
 - \nabla^2_{ }\delta\varphi 
 - ( h_0' + 3h_\rmii{D}' +\nabla^2_{ } h)
   \, \bar\varphi\hspace*{0.3mm}{}'
 \nn[2mm]
%%%%
 & & \; + \,   
   \, a ( \delta \Upsilon + h^{ }_0 \Upsilon )
   \bar\varphi\hspace*{0.3mm}{}' 
 + a^2_{ } ( \delta V^{ }_{,\varphi} + 2 h^{ }_0 V^{ }_{,\varphi} )
 \;, \label{delta_varphi_full} \\[3mm]
%%%%%%%%
 - \,( h_0' + 3  h_\rmii{D}' + \nabla^2_{ }h )
 & \overset{\rmii{\nr{delta_einstein_comb}}}{=} &
    2 \biggl( 2 \H + \frac{ \H' }{\H}  \biggr) 
    h^{ }_0
 + \frac{1}{\H}
  (\partial^2_\tau + 2 \H \partial^{ }_\tau - \nabla^2_{ } ) Y
 \nn[2mm]
%%%
 &  & \; - \, 
  \frac{ 4\pi G a^2_{ } }{\H}
  \biggl( \delta p - \delta e 
   + \frac{2}{3} \nabla^2_{ }\barpPi  \biggr)
 \;, \label{h0_hD_h_full} \\[3mm]
%%%%%%%%
 h^{ }_0 
 & \overset{\rmii{\nr{delta_einstein_0i}}}{=} &
 -\frac{ Y' }{\H} 
 + 
 \frac{ 4\pi G }{\H} \bigl[\, 
  a^2_{ }( \bar{e} + \bar{p}) (v - h) + {\bar\varphi}' \delta\varphi
  \,\bigr]
 \;.  \label{h0_full}
\ea
Here, gauge dependence, which will be demonstrated to cancel, 
is put into one of the quantities introduced in \eq\nr{shorthand},
related to spatial curvature (cf.\ \eq\nr{delta_R_tau}),
\be
 Y \; \overset{\rmii{\nr{shorthand}}}{\equiv} \;
 h^{ }_\rmii{D} + \frac{\nabla^2\vartheta}{3}
 \;. \label{def_Y} 
\ee

As a basic variable suitable for a treatment in conformal time, 
we adopt the gauge-invariant field perturbation from 
\eqs\nr{aQ_vs_Q} and \nr{Q_vs_R}, 
\be
  \field^{ }_\varphi
 \; \underset{\rmii{\nr{Q_vs_R}}}{\overset{\rmii{\nr{aQ_vs_Q}}}{\equiv}} \; 
 a\,
 \biggl(\,
 \delta\varphi + 
 \frac{\bar\varphi\hspace*{0.3mm}{}'}{\H}\, Y 
 \,\biggr)
 \; \overset{\rmii{\nr{def_R_varphi}}}{=} \; 
  -\frac{ a \bar\varphi\bit' }{ \H }
 \R^{ }_\varphi
 \;. \label{def_field_varphi}
\ee
The velocity and temperature perturbations are written in terms
of the gauge-invariant curvature perturbations, 
$\R^{ }_v$ and $\R^{ }_\T$, defined in 
\eqs\nr{def_R_v} and \nr{def_R_T}, respectively. However, it will turn
out to be helpful to express the curvature part in terms of 
$\R^{ }_\varphi$, and represent the velocity and temperature 
in terms of {\em isocurvature perturbations}, i.e.\ differences
of curvature perturbations in which 
the spatial curvature term, $Y$, drops out,  
\begin{empheq}[box=\fbox]{align}
 \quad \vphantom{\Bigg|}
 \E^{ }_v &
 \underset{\rmii{\nr{def_R_v}}}{
 \overset{\rmii{\nr{def_R_varphi}}}{\equiv}}
 \;
 (\bar{e} + \bar{p}) (\mathcal{R}^{ }_v - \mathcal{R}^{ }_\varphi)
 \vphantom{\frac{T'}{q}}
 \;, \label{def_E_v} 
 \\[2mm]
%%%
 \E^{ }_\T &
 \underset{\rmii{\nr{def_R_T}}}{
 \overset{\rmii{\nr{def_R_varphi}}}{\equiv}}
 \;
 \bar{e}^{ }_{\der\T}\, \dot{T}\,
 (\mathcal{R}^{ }_\T - \mathcal{R}^{ }_\varphi)
 \; \underset{\rmii{\nr{eq_HH}}}{\overset{\rmii{\nr{eq_HH'}}}{=}} \; 
 \frac{ \bar{e}^{ }_{\der\T}\, {T}\ibit' }{a} \, 
 (\mathcal{R}^{ }_\T - \mathcal{R}^{ }_\varphi)
 \;. \quad  \vphantom{\Bigg|}
 \label{def_E_T}
 \index{isocurvature perturbations}
 \index{$\E^{ }_v$, $\E^{ }_{\scriptscriptstyle T}$ (isocurvature perturbations)}
\end{empheq}

We now express the right-hand sides of
\eqs\nr{delta_varphi_full}--\nr{h0_full} in terms of the variables
in \eqs\nr{def_field_varphi}--\nr{def_E_T}. For $\delta\varphi$ and 
its derivatives, \eq\nr{def_field_varphi} yields
\ba
 \delta\varphi
 &
 \overset{\rmii{\nr{def_field_varphi}}}{=}
 & 
 \frac{1}{a}\, \field^{ }_\varphi  - 
 \frac{\bar\varphi\hspace*{0.3mm}'}{\H} \, Y
 \;, \label{d_phi_p0} \\[2mm]
%%%%
 \delta\varphi' & = & 
 \frac{1}{a}\,
 \bigl(\, \field^{\hspace*{0.3mm}\prime}_\varphi
  - \H \field^{ }_\varphi \,\bigr)
  - 
 \biggl( \frac{\bar\varphi\hspace*{0.3mm}'}{\H} \biggr)' \, Y
  - 
 \biggl( \frac{\bar\varphi\hspace*{0.3mm}'}{\H} \biggr) \, Y'
 \;, \label{d_phi_p1} \\[2mm]
%%%%
 \delta\varphi'' & = & 
 \frac{1}{a}\,
 \Bigl[\, 
    \field^{\hspace*{0.3mm}\prime\prime}_\varphi
  - 2\H \field^{\hspace*{0.3mm}\prime}_\varphi 
  + (\H^2_{ } - \H') \field^{ }_\varphi
  \,\Bigr]
 \nn[2mm]
 & &  
 \; - \, 
 \biggl( \frac{\bar\varphi\hspace*{0.3mm}'}{\H} \biggr)'' \, Y
  - 
 2 \biggl( \frac{\bar\varphi\hspace*{0.3mm}'}{\H} \biggr)' \, Y'
  - 
 \biggl( \frac{\bar\varphi\hspace*{0.3mm}'}{\H} \biggr) \, Y''
 \;. \label{d_phi_p2}
\ea
For $\delta\Upsilon$, which may be a function of $\varphi$ and $T$, 
we find
\ba
 \delta\Upsilon & = & 
 \Upsilon^{ }_{\der\varphi}\, \delta\varphi 
 +  
 \Upsilon^{ }_{\der\T}\, \delta T
 \nn[2mm]
 %%%%
 & \underset{\rmii{\nr{def_R_T}}}
   {\overset{\rmii{\nr{d_phi_p0}} \lift }{=}} & 
 \Upsilon^{ }_{\der\varphi}\, 
 \biggl(
   \frac{1}{a}\, \field^{ }_\varphi  - 
 \frac{\bar\varphi\hspace*{0.3mm}'}{\H} \, Y 
 \biggr)
 -  
 \Upsilon^{ }_{\der\T}\, \frac{T'}{\H}
 \bigl( \R^{ }_\T - \R^{ }_\varphi + \R^{ }_\varphi + Y \bigr)
 \nn[2mm]
 %%%%
 & \underset{\rmii{\nr{def_E_T}}}
  {\overset{\rmii{\nr{def_field_varphi}} \lift }{=}} & 
 \frac{\Upsilon^{ }_{\der\varphi}}{a}\, 
 \biggl(
    \field^{ }_\varphi  - 
 \frac{a \bar\varphi\hspace*{0.3mm}'}{\H} \, Y 
 \biggr)
 -  
 \frac{ a \Upsilon^{ }_{\der\T} }{\H\bar{e}^{ }_{\der\T}}
 \,\E^{ }_\T
 + 
 \frac{ \Upsilon^{ }_{\der\T} T'}{a \bar\varphi\hspace*{0.3mm}'}
 \biggl(
    \field^{ }_\varphi  - 
 \frac{a \bar\varphi\hspace*{0.3mm}'}{\H} \, Y 
 \biggr)
 \nn[2mm]
 %%%%
 & \overset{}{=} & 
  \frac{\Upsilon'}{a \bar\varphi\hspace*{0.3mm}'}\, 
 \biggl(
    \field^{ }_\varphi  - 
 \frac{a \bar\varphi\hspace*{0.3mm}'}{\H} \, Y 
 \biggr)
 -  
 \frac{ a \Upsilon^{ }_{\der\T} }{\H\bar{e}^{ }_{\der\T}}
 \,\E^{ }_\T
 \;, \label{d_Ups}
\ea
where in the last step we made use of
$
 \Upsilon' = \Upsilon^{ }_{\der\varphi}\bar\varphi\hspace*{0.3mm}{}'
      + \Upsilon^{ }_{\der\T} T'
$.
In complete analogy, 
\ba
 \delta V^{ }_{\der\varphi} & = & 
  \frac{V_{\der\varphi}'}{a \bar\varphi\hspace*{0.3mm}'}\, 
 \biggl(
    \field^{ }_\varphi  - 
 \frac{a \bar\varphi\hspace*{0.3mm}'}{\H} \, Y 
 \biggr)
 -  
 \frac{ a V^{ }_{\der\varphi\T} }{\H\bar{e}^{ }_{\der\T}}
 \,\E^{ }_\T
 \;, \label{d_V_T} \\[2mm]
%%%%
 \delta p - \delta e & = & 
 \frac{\bar{p}\hspace*{0.3mm}' - \bar{e}\hspace*{0.3mm}'}
      {a \bar\varphi\hspace*{0.3mm}'}\, 
 \biggl(
    \field^{ }_\varphi  - 
 \frac{a \bar\varphi\hspace*{0.3mm}'}{\H} \, Y 
 \biggr)
 +   
 \frac{ a }{\H }
 \biggl( 1 - \frac{\bar{p}^{ }_{\der\T} }{ \bar{e}^{ }_{\der\T} } \biggr)
 \,\E^{ }_\T
 \;. \label{de_m_dp}
\ea
Finally, for the velocity perturbation, 
\ba
 (\bar{e} + \bar{p})(v-h) 
 & \underset{\rmii{\nr{def_Y}}}
   {\overset{\rmii{\nr{def_R_v}} \lift }{=}} & 
 - \frac{\bar e + \bar p}{\H}
 \bigl(
   \R^{ }_v - \R^{ }_\varphi + \R^{ }_\varphi + Y 
 \bigr)
 \nn[2mm] 
%%%%
 & \overset{\rmii{\nr{def_field_varphi}} \lift }
  {\underset{\rmii{\nr{def_E_v}}}{=}} &
 \frac{\bar{e} + \bar{p} }
      {a \bar\varphi\hspace*{0.3mm}'}\, 
 \biggl(
    \field^{ }_\varphi  - 
 \frac{a \bar\varphi\hspace*{0.3mm}'}{\H} \, Y 
 \biggr)
 - \frac{1}{\H} \, \E^{ }_v
 \;. \label{d_vmh}
\ea

Inserting the results in \eqs\nr{d_phi_p0}--\nr{d_vmh} 
into \eqs\nr{delta_varphi_full}--\nr{h0_full}, 
the expressions become lengthy. However, they can be simplified 
by making use of the background identities 
\nr{bg_varphi_tau}--\nr{bg_Tmunu}. 
As an example, $h^{ }_0$ from \eq\nr{h0_full} becomes
\be
 h^{ }_0 
 \; 
 \underset{\rmii{\nr{d_phi_p0},\nr{d_vmh}}}
 {\overset{\rmii{\nr{h0_full}} \lift }{=}}
 \;
   - \frac{Y'}{\H}
   + 
     \overbrace{
     \frac{4\pi G [a^2_{ }(\bar e + \bar p) + 
     (\bar\varphi\hspace*{0.3mm}')^2_{ } ]}
    { \H a \bar\varphi\hspace*{0.3mm}'}
    }^{{\rm from}~\nr{bg_ep}:\;
       (\H - \frac{\H'}{\H})/( a \bar\varphi\hspace*{0.3mm}') }
 \biggl(
    \field^{ }_\varphi  - 
 \frac{a \bar\varphi\hspace*{0.3mm}'}{\H} \, Y 
 \biggr) 
   -\frac{4\pi G a^2_{ }}{\H^2_{ }} \, \E^{ }_v
 \;. \label{h0_full_rev}
\ee
Another frequent simplification is obtained 
by collecting the terms multiplying $h^{ }_0$ from 
\eqs\nr{delta_varphi_full} and \nr{h0_hD_h_full}, whereby 
they can be expressed as 
\ba
  a^2_{ }\varrho   
 & 
  \overset{\rmii{\nr{delta_varphi_full}}}
  {\underset{\rmii{\nr{h0_hD_h_full}}}{\supset}}
 & 
 \biggl[\, 
    2 \bar\varphi\hspace*{0.3mm}' \biggl( 2 \H + \frac{ \H' }{\H}  \biggr) 
    + a \bar\varphi\hspace*{0.3mm}' \Upsilon
    + 2 a^2_{ } V^{ }_{\der\varphi}  
 \,\biggr] h^{ }_0
 \nn[2mm]
 %%%%
 & 
 \overset{\rmii{\nr{simple_der}}}{=} 
 & 
 - \biggl[\,
    2 \H \biggl( \frac{\bar\varphi\hspace*{0.3mm}'}{\H} \biggr)'
    + a \bar\varphi\hspace*{0.3mm}' \Upsilon
   \,\biggr] h^{ }_0
 \;. \label{h0_terms}
\ea
Here we made use of 
\ba
 - \biggl( \frac{\bar\varphi\hspace*{0.3mm}'}{\H} \biggr)'
 & = &
 - \frac{\bar\varphi\hspace*{0.3mm}''}{\H}
 + \frac{\H' \bar\varphi\hspace*{0.3mm}'}{\H^2_{ }}
 \; 
   \underset{\rmii{ }}
   {\overset{\rmii{\nr{bg_varphi_tau}}}{=}} 
 \;
 \biggl(a \Upsilon + 2 \H + \frac{\H'}{\H} \biggr)
 \frac{\bar\varphi\hspace*{0.3mm}'}{\H}
 + 
 \frac{a^2 V^{ }_{\der\varphi}}{\H}
 \;, \label{simple_der}
\ea
which is a generalization of \eq\nr{mg_1} to $\Upsilon > 0$.

Given the complicated expressions, gauge invariance once again
offers for a valuable crosscheck. Almost immediately, it can 
be seen that $Y''$ from \eqs\nr{h0_hD_h_full} and 
\nr{d_phi_p2} cancel against each other, and the same happens
for $\nabla^2_{ }Y$. A less trivial crosscheck originates by
inspecting the coefficient of $Y'$, 
\be
  a^2_{ }\varrho   
  \; 
   \overset{\rmii{\nr{delta_varphi_full}}}{\supset}
  \; 
  \biggl[\, 
  \overbrace{
   - \cancel{ 2\, \biggl( \frac{\bar\varphi\hspace*{0.3mm}'}{\H} \biggr)' }
  }^{\delta\varphi''\;{\rm from}\;\nr{d_phi_p2}}
  \overbrace{
   -\, (\bcancel{2\H} + \cancel{a\Upsilon}  )
   \biggl( \frac{\bar\varphi\hspace*{0.3mm}'}{\H} \biggr) 
  }^{\delta\varphi'\;{\rm from}\;\nr{d_phi_p1}}
  \;
  \overbrace{
   + \,\bcancel{ 2\, \bar\varphi\hspace*{0.3mm}' } 
  }^{{\rm from}\;\nr{h0_hD_h_full}}
  \;
  \overbrace{
   +\, \cancel{ 2\, \biggl( \frac{\bar\varphi\hspace*{0.3mm}'}{\H} \biggr)' }
   + \cancel{ \frac{ a \bar\varphi\hspace*{0.3mm}'}{\H} \Upsilon }
  }^{{\rm from}\;\nr{h0_full_rev},\,\nr{h0_terms}}
  \,\biggr] \, Y'
  \;. 
\ee
The most non-trivial part is the cancellation of the coefficient
of $Y$. To verify it, we generalize \eq\nr{mgc_2}
to include $\Upsilon$. For this, we take a derivative of 
\eq\nr{simple_der}, 
\ba
 - \biggl( \frac{\bar\varphi\hspace*{0.3mm}'}{\H} \biggr)''
 & 
   \overset{\rmii{\nr{simple_der}}}{=} 
 &
 \biggl(a \Upsilon + 2 \H + \frac{\H'}{\H} \biggr)
 \biggl( \frac{\bar\varphi\hspace*{0.3mm}'}{\H} \biggr)'
 \nn[2mm]
%%%%
 & & \; + \, 
 \biggl(a \H \Upsilon + a \Upsilon' 
      + 2 \H' + \frac{\H''}{\H}
      - \frac{\H^{\prime\hspace*{0.3mm}2}_{ }}{\H^2_{ }} \biggr)
 \frac{\bar\varphi\hspace*{0.3mm}'}{\H}
 \nn[2mm]
%%%%
 & & \; + \, 
 \biggl( 
 2 -  
 \frac{\H'}{\H^2_{ }}
 \biggr)
 a^2 V^{ }_{\der\varphi}
 + 
 \frac{a^2 V^{\prime}_{\der\varphi}}{\H}
 \;. \label{double_der}
\ea
In addition we need a new background identity,  
relating $\H\hspace*{0.3mm}''$ to the difference
$
 \bar{p}\hspace*{0.3mm}' - \bar{e}\hspace*{0.3mm}'
$ appearing in \eq\nr{de_m_dp}.
Subtracting
\eqs\nr{bg_HH_tau} and \nr{bg_ep}; omitting once again $\kappa$;
and taking a time derivative,  we find
\ba
 2 \H^2_{ } + \H ' 
 \;
 \underset{\rmii{\nr{bg_ep}}}
 {\overset{\rmii{\nr{bg_HH_tau}} \lift }{=}} 
 \;
 4\pi G a^2_{ }(\bar{e} - \bar{p})
 & \overset{\rmii{\raise1.5ex\hbox{$\partial^{ }_\tau$}}}{\Rightarrow} & 
 4 \H \H' + \H '' = 2 \H \overbrace{ 
    4 \pi G a^2_{ } (\bar{e} - \bar{p})
  }^{ 2 \H^2_{ } + \H ' } 
 + \, 4\pi G a^2_{ }(\bar{e}\hspace*{0.3mm}' - \bar{p}\hspace*{0.3mm}')
 \nn[3mm]
%%%%
 & \Rightarrow &
 \H'' + 2 \H \H' - 4 \H^3_{ }
 \; = \; 
 4\pi G a^2_{ }(\bar{e}\hspace*{0.3mm}' - \bar{p}\hspace*{0.3mm}')
 \;.  \label{bg_vT_2}
\ea
Through a repeated use of \eqs\nr{simple_der}, 
\nr{double_der} and \nr{bg_vT_2}, the 
complete cancellation of the coefficient of~$Y$ 
can indeed be verified. 

It remains to work out the physical terms. 
Given that 
these steps offer no new insight beyond what was met
in the context of deriving 
the vacuum Mukhanov-Sasaki equation, 
leading to \eq\nr{final_vacuum}, 
we refrain from showing them explicitly, 
but provide a brief outline.  
The isocurvature
$\E^{ }_\T$ originates from the last terms 
of \eqs\nr{d_Ups}--\nr{de_m_dp},
and $\E^{ }_v$ from 
\eqs\nr{h0_full_rev} and \nr{h0_terms}.
The anisotropic stress, $\nabla^2_{ }\Pi$, 
only appears in \eq\nr{h0_hD_h_full}, 
so it does not cancel. 
The operator acting on~$\field^{ }_\varphi$ is fairly complicated. 
It has 
similarities with the operator acting on~$Y$, except that there
is no cancellation. However, \eqs\nr{simple_der}, \nr{double_der}
and \nr{bg_vT_2} can again be employed 
in order to combine terms. 
Appearances of $\H$ and $\H'$  
can be hidden by letting derivatives act on 
$(a\bar\varphi\hspace*{0.3mm}'/\H)$ rather than 
$(\bar\varphi\hspace*{0.3mm}'/\H)$, 
like in \eq\nr{final_vacuum_2}.

Proceeding along these lines, and putting the stochastic noise
on the right-hand side as is conventionally done,  
the final gauge-invariant relation in conformal time reads
\ba
  \overbrace{
   \biggl\{\, 
      \partial_\tau^2
    - \nabla^2_{ }
    - \frac{\H}{a\bar\varphi\hspace*{0.3mm}'}
      \biggl( \frac{a\bar\varphi\hspace*{0.3mm}'}{\H} \biggr)''_{ } 
    +  a \Upsilon \biggl[\, 
    \partial^{ }_\tau 
    - \frac{\H}{a\bar\varphi\hspace*{0.3mm}'}
      \biggl( \frac{a\bar\varphi\hspace*{0.3mm}'}{\H} \biggr)'_{ } 
   \,\biggr]
   \,\biggr\}
   \, \widehat{\Q}^{ }_\varphi
  }^{\;\equiv\; L^{ }_\varphi }
 & &  
  \nn[2mm] 
%%%%% 
 - \, 
 \biggl\{
  \frac{4\pi G a}{\H} \biggl( 1 - \frac{\bar{p}^{ }_{\der\T}}
                                       {\bar{e}^{ }_{\der\T}} \biggr) 
 + 
 \frac{\Upsilon^{ }_{\der\T}\, \bar\varphi\hspace*{0.3mm}'
               + a V^{ }_{\der\varphi\T}}
      {\bar{e}^{ }_{\der\T}\, \bar\varphi\hspace*{0.3mm}'}
 \biggr\}
 \, \frac{a^3_{ } \bar\varphi\hspace*{0.3mm}' }{\H} 
 \, \E^{ }_\T
 & & 
  \nn[2mm] 
%%%%% 
 + \, 
 \biggl\{\, 
 \Upsilon  
 +  
 \frac{2\H}{a\bar\varphi\hspace*{0.3mm}'}
      \biggl( \frac{\bar\varphi\hspace*{0.3mm}'}{\H} \biggr)'_{ }
 \,\biggr\}
 \, \frac{4\pi G a^4_{ } \bar\varphi\hspace*{0.3mm}' }{\H^2_{ }}
 \, \E^{ }_v
 & & 
  \nn[2mm] 
%%%%% 
 \, - \, 
 \frac{8\pi G a^3_{ } \bar\varphi\hspace*{0.3mm}'}{3\H}
 \, \nabla^2_{ }\Pi
%%%%%
  & = & 
   a^3_{ }\varrho
   \;. \label{eq_hatQ}
\ea
For later convenience, 
the first line has been given a special name.

Next, we transcribe the equation to physical time, and 
simultaneously from $\field^{ }_\varphi$
to the curvature perturbation $\R^{ }_\varphi$, 
according to \eq\nr{def_field_varphi}. First, substituting
$
 \field^{ }_\varphi 
 = 
   -({ a \bar\varphi\hspace*{0.3mm}' }/{ \H })
 \R^{ }_\varphi
$
and following \eqs\nr{Qk_vs_Rk}--\nr{Q_k_dd}, 
\ba
   \biggl\{\, 
      \partial_\tau^2
    - \frac{\H}{a\bar\varphi\hspace*{0.3mm}'}
      \biggl( \frac{a\bar\varphi\hspace*{0.3mm}'}{\H} \biggr)''_{ } 
   \,\biggr\}
   \,\field^{ }_\varphi
 &
 \underset{\rmii{\nr{Q_k_dd}}}{
 \overset{\rmii{\nr{Qk_vs_Rk}} \lift }{=}}
 & 
 - \frac{a\bar\varphi\hspace*{0.3mm}'}{\H} 
  \biggl\{\,
  \R^{\prime\prime}_\varphi
  + 
  \frac{2 \H}{a\bar\varphi\hspace*{0.3mm}'}
      \biggl( \frac{a\bar\varphi\hspace*{0.3mm}'}{\H} \biggr)'_{ }
  \,\R^{\prime}_\varphi
  \,\biggr\}
 \;,
 \label{Q_R_rel_1} \\[2mm]
%%%% 
   \biggl\{\, 
      \partial_\tau^{ }
    - \frac{\H}{a\bar\varphi\hspace*{0.3mm}'}
      \biggl( \frac{a\bar\varphi\hspace*{0.3mm}'}{\H} \biggr)'_{ } 
   \,\biggr\}
   \,\field^{ }_\varphi
 & 
 \underset{\rmii{\nr{Q_k_d}}}{
 \overset{\rmii{\nr{Qk_vs_Rk}} \lift }{=}}
 & 
 - \frac{a\bar\varphi\hspace*{0.3mm}'}{\H} 
  \bigl\{\,
  \R^{\prime}_\varphi
  \,\bigr\}
  \;. \label{Q_R_rel_2}
\ea
This implies that 
\be
 L^{ }_\varphi
 \;
  \underset{\rmii{\nr{Q_R_rel_1},\nr{Q_R_rel_2}}}
 {\overset{\rmii{\nr{eq_hatQ}} \lift }{=}}
 \; 
 - \frac{a\bar\varphi\hspace*{0.3mm}'}{\H} 
  \biggl\{\,
  \partial_\tau^2
  - \nabla^2_{ } 
  + \biggl[\,
     a \Upsilon + 
  \frac{2 \H}{a\bar\varphi\hspace*{0.3mm}'}
      \biggl( \frac{a\bar\varphi\hspace*{0.3mm}'}{\H} \biggr)'_{ }
  \,\biggr]\,\partial^{ }_\tau
  \,\biggr\}
  \,  \,\R^{ }_\varphi  
 \;. \label{L_varphi_1}
\ee
Second, going to physical time by 
following \eqs\nr{time_derivs} and \nr{R_k_dd}, 
\be
 L^{ }_\varphi
 \;
  \underset{\rmii{\nr{time_derivs},\nr{R_k_dd}}}
 {\overset{\rmii{\nr{L_varphi_1}} \lift }{=}}
 \; 
 - \frac{a^3_{ }\dot{\bar\varphi}}{H} 
  \biggl\{\,
  \partial_t^2
  - \frac{\nabla^2_{ }}{a^2_{ }} 
  + \biggl[\,
     \frac{\dot{a}}{a}
     + \Upsilon 
  + 
  \frac{2 H}{a\dot{\bar\varphi}}
      \biggl( \frac{a\dot{\bar\varphi}}{H} \biggr)^{\textstyle .}_{ }
  \,\biggr]\,\partial^{ }_t
  \,\biggr\}
  \,  \,\R^{ }_\varphi  
 \;. \label{L_varphi}
\ee
Subsequently, multiplying the whole with
$
 -H / (a^3_{ }\dot{\bar\varphi})
$, 
and recalling the definition 
$
 \mathcal{F}
 = 
 {\ddot{\bar\varphi}} / {\dot{\bar\varphi}} 
 - {\dot H} / {H}
$ 
from \eq\nr{cal_F}, 
the equation becomes 
\begin{empheq}[box=\fbox]{align}
 \vphantom{\Bigg|} 
 \biggl\{\,
 \partial_t^2
 + 
    \bigl(\, \Upsilon+ 2 \mathcal{F}  + 3 H \,\bigr) 
    \partial^{ }_t 
 -  \frac{\nabla^2_{ }}{a^2_{ }} 
 \,\biggr\}\, \R^{ }_\varphi 
 &  
 \nn[2mm]
%%%%
 \quad
 \,+\, 
   \biggl\{\,
   \frac{4\pi G}{H}
 \biggl( 1 - \frac{\bar{p}^{ }_{\der\T}}{\bar{e}^{ }_{\der\T}}\biggr)
   + \frac{
     \Upsilon^{ }_{\der\T} \, \dot{\bar\varphi} 
   + 
     V^{ }_{\der\varphi\T} 
     }
     { \bar{e}^{ }_{\der\T} \, \dot{\bar\varphi} }
   \,\biggr\} 
   \, \E^{ }_\T 
 &  
 \nn[2mm]
%%%%
 \,-\,   
   \biggl\{\,
       \frac{4\pi G ( \Upsilon  +  2 \mathcal{F} )}{H}
   \,\biggr\}
   \,\E^{ }_v 
 \,+\, 
 \frac{8\pi G}{3}\nabla^2_{ } \Pi
 & 
 \;
 \underset{\rmii{\nr{L_varphi}}}{
 \overset{\rmii{\nr{eq_hatQ}} \lift }{=}}
 \;  
 \,-\, \frac{\varrho H}{\dot{\bar\varphi}}
 \;. 
 \hspace*{5mm}  \vphantom{\Bigg|} 
 \label{dot_R_varphi}
\end{empheq}
This generalizes \eq\nr{eq_R_k} to an environment
in which a thermal plasma is present, and the evolution
of curvature perturbations 
is influenced by $\Upsilon$ and $\varrho$. 

As for the anisotropic stress
in \eq\nr{dot_R_varphi}, $\Pi$, we recall from 
\eq\nr{decomposition} that it contains 
\ba
 \Pi 
 & 
  \overset{\rmii{\nr{decomposition}} \lift }{\supset}
 & 
 \frac{2 \eta ( v - \vartheta' )}{a}
 \;
   \underset{\rmii{\nr{def_R_v}}}
   {\overset{\rmii{\nr{def_psi}} \lift }{=}} 
 \; 
 - \frac{2\eta}{a \H} \bigl(\, \psi + \R^{ }_v \,\bigr)
 \; 
   \overset{\rmii{\nr{eq_HH'}} \lift }{=} 
 \; 
 - \frac{2\eta}{a^2_{ } H}
 \bigl(\, \psi + \R^{ }_\varphi + \R^{ }_v - \R^{ }_\varphi \,\bigr)
 \nn[2mm]
%%%%
 & 
  \overset{\rmii{\nr{def_E_v}}}{=} 
 & 
 - \frac{2\eta}{a^2_{ } H}
 \biggl(\, 
    \psi + \R^{ }_\varphi
  + \frac{\E^{ }_v}{\bar e + \bar p}
 \,\biggr)
 \;, \label{Pi_gi}
\ea
where $\eta$ is the shear viscosity. 
In addition, $\Pi$ contains a hydrodynamic noise term 
(cf.\ \eqs\nr{Pi_splitup} and \nr{noise_hydro}), which could
be moved to the right-hand side of the equation, where it plays
a role similar to $\varrho$. 
All in all, \eq\nr{dot_R_varphi} contains four dynamical 
variables ($\R^{ }_\varphi$, $\E^{ }_\T$, $\E^{ }_v$ and $\psi$), 
and further equations are needed for specifying the solution. 

%%%%%%%%%%%%%%%%%%%%%%%%%%%%%%%%%%%%%%%%%%%%%%%%%%%%%%%%%%%

\vspace*{3mm}

To obtain equations for the isocurvature perturbations, 
$\E^{ }_v$ and $\E^{ }_\T$, we start from generalizations of 
the energy-momentum conservation in 
\eqs\nr{delta_0_Tmunu;mu} and \nr{delta_i_Tmunu;mu}, 
including now both inflaton and fluid perturbations, 
\ba
 0 \;=\; \delta \mathbin{T^{ }_{\mu 0}}^{;\mu}_{ }
 & 
 \underset{\rmii{p.~\pageref{code:enmom}}}{
 \overset{\rmii{code~on} \lift }{=}} 
 & 
 \frac{1}{a^2_{ }}
 \Bigl\{\,
 \bar\varphi\hspace*{0.3mm}'
 \bigl(\, \nabla^2_{ } - \partial_\tau^2 \,\bigr) \delta\varphi
 - 
 (\, \bar\varphi\hspace*{0.3mm}''
         + 4 \H \bar\varphi\hspace*{0.3mm}' \, ) \, 
                \delta\varphi^{\prime}_{ }
 \nn[2mm]
%%%%%
 &  & \; + \,
 ( \bar\varphi\hspace*{0.3mm}' )^2_{ }
 (\, h_{0}' + 3 h_\rmii{D}' + \nabla^2_{ } h\,)
 + 2  \bar\varphi\hspace*{0.3mm}' 
 (\, \bar\varphi\hspace*{0.3mm}''
         + 2 \H \bar\varphi\hspace*{0.3mm}' \, )
 \, h^{ }_0
 \,\Bigr\}
 \nn[2mm]
%%%%%
 &  & \; - \,
 \delta e' - 3 \H (\delta e + \delta p)
 + (\bar{e} + \bar{p}) (\, 3 h_\rmii{D}' + \nabla^2_{ } v \,)
 \;,  \hspace*{6mm} \label{full_delta_0_Tmunu;mu} \\[3mm]
%%%%%%%%%%
 0 \;=\; \delta \mathbin{T^{ }_{\mu i}}^{;\mu}_{ }
 & 
 \underset{\rmii{p.~\pageref{code:enmom}}}{
 \overset{\rmii{code~on} \lift }{=}} 
 & 
 -\frac{1}{a^{2}_{ }}(\, \bar\varphi\hspace*{0.3mm}''
         + 2 \H \bar\varphi\hspace*{0.3mm}' \, ) \, 
                \delta\varphi^{ }_{,i}
 \nn[2mm]
%%%%
 &  & \; + \,
 \delta p^{ }_{,i} + (\bar{e} + \bar{p}) h^{ }_{0,i}
 + (\partial^{ }_\tau + 4\H) [(\bar{e} + \bar{p})(v^{ }_i - h^{ }_i)]
 + \barpPi^{ }_{ik,k}
 \;.  \hspace*{6mm} \label{full_delta_i_Tmunu;mu}
\ea
We can eliminate second time derivatives from \eq\nr{full_delta_0_Tmunu;mu}
by inserting $\delta\varphi\hspace*{0.3mm}''$ from \eq\nr{delta_varphi_full}. 
As for \eq\nr{full_delta_i_Tmunu;mu}, it contains both a scalar and 
a vector part. The vector part retains the form 
in \eq\nr{delta_i_Tmunu;mu_v} in the presence of $\delta\varphi$, 
whereas the scalar part from \eq\nr{delta_i_Tmunu;mu_s} gets 
modified. Collecting together, the equations 
reduce to 
\ba
 0 
 &
 \underset{\rmii{\nr{delta_varphi_full}}}{
 \overset{\rmii{\nr{full_delta_0_Tmunu;mu}} \lift }
  {=}} 
 & 
 \frac{1}{a^2_{ }}
 \Bigl\{\,
 \overbrace{
 (\, - \bar\varphi\hspace*{0.3mm}''
         - 2 \H \bar\varphi\hspace*{0.3mm}'
     + a \Upsilon \bar\varphi\hspace*{0.3mm}' \, ) 
 }^{ 
  {\rm from~\nr{bg_varphi_x}}:\;
               2 a \Upsilon \bar\varphi\hspace*{0.3mm}'
                           + a^2_{ } V^{ }_{,\varphi}
  } \, 
                \delta\varphi^{\prime}_{ }
 \; + \; 
 a (\bar\varphi\hspace*{0.3mm}')^2_{ }
 \hspace*{-3mm}
 \overbrace{ \delta\Upsilon  }^{
  {\rm insert~\nr{d_Ups}}
  }
 \hspace*{-3mm}
 + a^2_{ } \bar\varphi\hspace*{0.3mm}' (\, 
 \hspace*{-3mm}
 \overbrace{ \delta V^{ }_{,\varphi} }^{
  {\rm insert~\nr{d_V_T}}
  }
 \hspace*{-3mm}
 \;-\; \varrho \,)
 \nn[2mm]
%%%%%
 &  & \;+\,
  \bar\varphi\hspace*{0.3mm}' 
 \overbrace{
 [\, 2 \bar\varphi\hspace*{0.3mm}''
         + (4 \H + a \Upsilon )\bar\varphi\hspace*{0.3mm}'
         + 2 a^2_{ } V^{ }_{,\varphi}
  \,]}^{ 
  {\rm from~\nr{bg_varphi_x}:}\;
  - a \Upsilon \bar\varphi\hspace*{0.3mm}'
 }
 \hspace*{-3mm}
 \overbrace{ h^{ }_0 }^{
  {\rm insert~\nr{h0_full_rev}}
  }
 \hspace*{-3mm}
 \,\Bigr\}
 \nn[2mm]
%%%%%
 &  & \;-\,
 \delta e' - 3 \H 
 \overbrace{
 (\delta e + \delta p)
 }^{ {\rm insert~\nr{de_m_dp}}}
 + (\bar{e} + \bar{p}) 
 \bigl[\, 
  \overbrace{ h_0' + 3 h_\rmii{D}' + \nabla^2_{ } h }^{
  {\rm insert~\nr{h0_hD_h_full}}
  }
  + \overbrace{ \nabla^2_{ }(v-h) }^{
  {\rm insert~\nr{d_vmh}}
  }
   - h_0' \,\bigr]
 \;, \label{full_0} \\[3mm]
%%%%%%%%%%%%%%%%%%%%%%%%%%%%
 0
 & 
  \underset{\rmii{\nr{Pi_ij_s}}}{
  \overset{\rmii{\nr{full_delta_i_Tmunu;mu}} \lift }{=}} 
 & 
 -\frac{1}{a^{2}_{ }}(\, \bar\varphi\hspace*{0.3mm}''
         + 2 \H \bar\varphi\hspace*{0.3mm}' \, ) \, 
 \hspace*{-3mm}
               \overbrace{ \delta\varphi^{ }_{ } }^{
  {\rm insert~\nr{d_phi_p0}}
  }
 \nn[2mm]
%%%%
 & & \;+\,
 \hspace*{-3mm}
 \overbrace{ \delta p }^{
  {\rm insert~\nr{de_m_dp}}
  }
 \hspace*{-3mm}
 + [\, \overbrace{ (\bar{e} + \bar{p})(h - v) }^{
  {\rm insert~\nr{d_vmh}}
  }
    \,]'
 + (\bar{e} + \bar{p}) 
   [\, 
 \hspace*{-4mm}
 \overbrace{ h^{ }_0 }^{
  {\rm insert~\nr{h0_full_rev}}
  }
 \hspace*{-4mm}
  +\;  4\H 
 \hspace*{-2mm} \overbrace{ (h - v) }^{
  {\rm insert~\nr{d_vmh}}
  } \,]
 + \frac{2}{3} \nabla^2_{ }\barpPi  
 \;. \hspace*{7mm}
 \label{full_i}
\ea
Here at several points we employed the background identity 
from \eq\nr{bg_varphi_tau}, 
\be
   \bar\varphi\hspace*{0.3mm}'' 
  +  \bigl( 2 \H  + a \Upsilon \bigr) \bar\varphi\hspace*{0.3mm}'
  + a^2 V^{ }_{\der\varphi} 
  \;
 \overset{\rmii{\nr{bg_varphi_tau}}}{=}
 \; 0
 \;. \hspace*{6mm}
 \label{bg_varphi_x}
\ee

Let us first tackle \eq\nr{full_i}, which leads to a simpler 
analysis than \eq\nr{full_0}. We start with the coefficient of~$Y$
(cf.\ \eq\nr{def_Y}), which should cancel. We find
\ba
 \mbox{\nr{full_i}}^{ }_\rmii{R}
 & \supset & 
 \biggl[\,
   \overbrace{ 
   \frac{(\, \bar\varphi\hspace*{0.3mm}''
         + 2 \H \bar\varphi\hspace*{0.3mm}' \, ) \,
   \bar\varphi\hspace*{0.3mm}' }{a^{2}_{ } \H} 
   }^{
   {\rm from}\;\delta\varphi
   }
  \overbrace{
  - \frac{\bcancel{\bar{p}}\hspace*{0.3mm}'}{\H}
  }^{
  {\rm from}\;\delta p 
  }
 \overbrace{
 + \frac{\bar e\hspace*{0.3mm}' + \bcancel{\bar p}\hspace*{0.3mm}'}{\H} 
 - \bcancel{\frac{(\bar e + \bar p)\H'}{\H^2_{ }}} 
 + \cancel{\frac{\bar e + \bar p}{\H} \partial^{ }_\tau}     
 + 4 (\bar e + \bar p)
 }^{
 {\rm from}\;(\bar e + \bar p)(h-v)
 }
 \nn[2mm]
%%%
 & & 
 \overbrace{
 - \cancel{\frac{\bar e + \bar p}{\H} \partial^{ }_\tau}
 - \biggl( 1 - \bcancel{\frac{\H'}{\H^2_{ }}} \biggr)(\bar e + \bar p)
 }^{
 {\rm from}\; h^{ }_0
 }
 \,\biggr]\, Y
 \nn[3mm]
%%%%% 
 & = & 
 \biggl[\,
   \frac{(\, \bar\varphi\hspace*{0.3mm}''
         + 2 \H \bar\varphi\hspace*{0.3mm}' \, ) \,
   \bar\varphi\hspace*{0.3mm}' }{a^{2}_{ } \H} 
 + 
   \frac{\bar e\hspace*{0.3mm}' + 3\H(\bar e + \bar p)}{\H}  
 \,\biggr]\, Y
 \; 
   \underset{\rmii{}}{\overset{\rmii{\nr{bg_Tmunu_tau}}}{=}} 
 \;
 0 
 \;. 
\ea 
Encouraged by the cancellation of gauge dependence, 
we can proceed to the simplification of 
the physical terms. They can be combined into
\ba
 0 
 & 
   \underset{\scriptscriptstyle Y\;\to\;0}{
   \overset{\rmii{\nr{full_i}} \lift }{=}}
 & 
 \biggl[\,
 \overbrace{
  - \frac{a \bar p^{ }_{\der\T}}{\H\bar e^{ }_{\der\T}}
 }^{ 
  {\rm from}\;\delta p
 }
 \,\biggr] \E^{ }_\T
 \; + \; \frac{1}{\H}\,
 \biggl[\,
  \overbrace{
     \partial^{ }_\tau - \frac{\H'}{\H} + 4\H 
  }^{
   {\rm from}\; (\bar e + \bar p)(h-v)
  }
 \;
 \overbrace{
 -\, \frac{4\pi G a^2_{ }(\bar e + \bar p)}{\H}
 }^{
  {\rm from}\; h^{ }_0 \,;\;
  {\rm insert~\nr{bg_ep}}
 }
 \,\biggr] \E^{ }_v
 \nn[2mm]
%%%%%
 &  & \;+\,
 \biggl[\,
   \overbrace{
   -\,
   \frac{ \bar\varphi\hspace*{0.3mm}''
         + 2 \H \bar\varphi\hspace*{0.3mm}' }{a^{3}_{ }} 
   }^{
   {\rm from}\;\delta\varphi
   }
   \; 
   \overbrace{
   + \;
   \frac{\bcancel{\bar p}\hspace*{0.3mm}'}
        { a\bar\varphi\hspace*{0.3mm}'}
   }^{ 
   {\rm from}\;\delta p
   }
   \; 
   \overbrace{
   - \; 
   \frac{\bar e + \bar p}
        { a\bar\varphi\hspace*{0.3mm}'}\,\partial^{ }_\tau 
   - 
   \frac{\bar e\hspace*{0.3mm}' + \bcancel{\bar p}\hspace*{0.3mm}'}
        { a\bar\varphi\hspace*{0.3mm}'}
   + 
   \frac{(\bar e + \bar p)\bar\varphi\hspace*{0.3mm}''}
        { a(\bar\varphi\hspace*{0.3mm}')^2_{ }}
   - 3 \H\, 
   \frac{\bar e + \bar p}{ a\bar\varphi\hspace*{0.3mm}'}
   }^{ 
   {\rm from}\; (\bar e + \bar p)(h-v)   
   }
 \nn[2mm]
%%%%%
 & & \qquad 
 \overbrace{
 +\;
 \biggl(\H - \frac{\H'}{\H} \biggr)
   \frac{\bar e + \bar p}{ a\bar\varphi\hspace*{0.3mm}'}
 }^{
  {\rm from}\; (\bar e + \bar p)\, h^{ }_0
 }
 \,\biggr]\,\field^{ }_\varphi
 \; + \; 
 \frac{2}{3} \nabla^2_{ }\Pi
 \nn[3mm]
%%%%%
 & 
     \underset{\rmii{}}{\overset{\rmii{\nr{bg_Tmunu_tau}}}{=}} 
 & 
 \biggl[\,
  - \frac{a \bar p^{ }_{\der\T}}{\H\bar e^{ }_{\der\T}}
 \,\biggr]\, \E^{ }_\T
 \; + \; 
 \frac{1}{\H}\,
 \biggl[\,
     \partial^{ }_\tau + 3\H
   + \frac{4\pi G ( \bar\varphi\hspace*{0.3mm}' )^2_{ }}{\H}
 \,\biggr]\, \E^{ }_v
 \nn[2mm]
%%%%%
 & & \;+\,
 \frac{\bar e + \bar p}{ a\bar\varphi\hspace*{0.3mm}'}
 \biggl[\,
 -\, \partial^{ }_\tau 
  \underbrace{ 
  +\; \frac{ \bar\varphi\hspace*{0.3mm}'' }
           { \bar\varphi\hspace*{0.3mm}' }
  + \H 
  - \frac{\H'}{\H}
  }_{
  \textstyle
  \frac{\H }
       {a \bar\varphi\hspace*{0.3mm}' }
  \bigl( 
  \frac{a \bar\varphi\hspace*{0.3mm}' }
       {\H }
  \bigr)'
%
%  \frac{\H \lift }
%       {a \bar\varphi\hspace*{0.3mm}'  \lift }
%  \bigl( 
%  \frac{a \bar\varphi\hspace*{0.3mm}'  \lift }
%       {\H  \lift }
%  \bigr)'
%
  }
 \,\biggr] \,\field^{ }_\varphi
 \; + \; 
 \frac{2}{3} \nabla^2_{ }\Pi
 \;. 
 \label{dot_E_v_tau}
\ea 
As a final step, we go over to the variable $\R^{ }_\varphi$ and 
to physical time, by inserting \eqs\nr{Q_R_rel_2} and \nr{time_derivs}, 
whereby we find
\begin{empheq}[box=\fbox]{align}
 \vphantom{\Bigg|} 
  \quad 
   \biggl\{\,
      \partial^{ }_t + 3 H +
       \frac{4\pi G \dot{\bar\varphi}^2_{ }}{H}
   \,\biggr\}
   \,\E^{ }_v 
 \,-\, 
   \biggl\{\,
   \frac{\bar{p}^{ }_{\der\T}}{\bar{e}^{ }_{\der\T}}
   \,\biggr\} 
   \, \E^{ }_\T 
 &  
 \nn[2mm]
%%%%
 \,+\,   
 \Bigl\{\,
    \bigl(\, \bar e + \bar p \,\bigr) \partial^{ }_t 
 \,\Bigr\}\, \R^{ }_\varphi 
 \,+\, 
 \frac{2 H}{3}\nabla^2_{ } \Pi
 & 
 \;
 \underset{\rmii{\nr{Q_R_rel_2},\nr{time_derivs}}}{
% \overset{\rmii{$H\times$\nr{dot_E_v_tau}}}{=}}
 \overset{\scriptscriptstyle H\times \nr{dot_E_v_tau} \lift }{=}}
 \;  
 0 
 \;. \hspace*{5mm}  \vphantom{\Bigg|} 
 \label{dot_E_v}
\end{empheq}
This is an equation for the same variables as \eq\nr{dot_R_varphi}, 
however the two equations are linearly independent. Given that 
$\dot{\E}^{ }_v$ appears, \eq\nr{dot_E_v} can be said to determine
the time evolution of $\E^{ }_v$. 
 
\vspace*{3mm}

Next, we attack the most complicated relation, from \eq\nr{full_0}.
Two new ingredients are needed for it, 
$\delta e'$ and $h_0'$. From \eq\nr{de_m_dp}, 
\ba
 \delta e & \overset{\rmii{\nr{de_m_dp}}}{=} & 
 \frac{\bar{e}\hspace*{0.3mm}'}
      {a \bar\varphi\hspace*{0.3mm}'}\, 
 \biggl(
    \field^{ }_\varphi  - 
 \frac{a \bar\varphi\hspace*{0.3mm}'}{\H} \, Y 
 \biggr)
 -   
 \frac{ a }{\H }
 \,\E^{ }_\T
 \label{delta_e}
 \\[3mm]
%%%%%
 \Rightarrow \; 
 \delta e'
 & = & 
 \biggl[\,
  \frac{\bar{e}\hspace*{0.3mm}''}
       {a \bar\varphi\hspace*{0.3mm}'}
  + 
  \frac{\bar{e}\hspace*{0.3mm}'}
      {a \bar\varphi\hspace*{0.3mm}'}
  \biggl(\,
    \partial^{ }_\tau - \H - \frac{ \bar\varphi\hspace*{0.3mm}'' }
                               { \bar\varphi\hspace*{0.3mm}'  }
  \,\biggr)
 \,\biggr]\,\field^{ }_\varphi
 \nn[2mm]
%%%%%
 & & \; - \,
 \biggl[\,
  \frac{\bar{e}\hspace*{0.3mm}''}
       {\H}
  + 
  \frac{\bar{e}\hspace*{0.3mm}'}
      {\H}
  \biggl(\,
    \partial^{ }_\tau - \frac{ \H' }
                          { \H  }
  \,\biggr)
 \,\biggr]\, Y
 \; - \; 
 \frac{a}{\H}
 \biggl(\,
 \partial^{ }_\tau + \H - \frac{ \H' }
                            { \H  }
 \,\biggr)\,\E^{ }_\T
 \;. 
 \label{delta_e_prime}
\ea
{}From \eq\nr{h0_full_rev}, 
\ba
 h^{ }_0 
 & 
   \overset{\rmii{\nr{h0_full_rev}}}{=}
 & 
\biggl( \H - \frac{ \H' }
                   { \H  }
 \biggr)
 \frac{\field^{ }_\varphi}{a \bar\varphi\hspace*{0.3mm}'} 
 \; - \;
 \biggl(
   \frac{1}{\H}\, \partial^{ }_\tau + 1 - \frac{\H'}{\H^2_{ }}
 \biggr)\,Y
  \; - \; 
  \frac{4\pi G a^2_{ }}{\H^2_{ }}\, \E^{ }_v
  \label{h0_full_rev2} \\[3mm]
%%%%%%
 \Rightarrow \; 
 h_0' 
 &  
   =
 & 
 \frac{1}{a \bar\varphi\hspace*{0.3mm}'}
 \biggl[\,
 \biggl( \H - \frac{ \H' }
                   { \H  }
 \biggr)
 \,\partial^{ }_\tau 
 \; 
 + 
 \; 
   2 \H'
   - \H^2_{ }
   - \frac{ \H'' } { \H  }
   + \frac{ \H^{\prime\bit 2}_{ } } { \H^2_{ }  }
 \; 
  -
 \; 
  \biggl( \H - \frac{ \H' }
                   { \H  }
  \biggr)
  \frac{ \bar\varphi\hspace*{0.3mm}''}{\bar\varphi\hspace*{0.3mm}' }
 \,\biggr]\, \field^{ }_\varphi
 \nn[2mm]
%%%%%
 & & \; - \,
 \biggl[\, 
 \frac{1}{\H} \,
 \partial^2_\tau 
 + 
 \biggl( 
   1 - \frac{2 \H'}{\H^2_{ }} 
 \biggr) \, \partial^{ }_\tau
 + \frac{2 \H^{\prime\bit2}_{ }}{\H^3_{ }}
 - \frac{\H''}{\H^2_{ }}
 \,\biggr]\, Y
 \nn[2mm]
%%%%%
 & & \; - \,
 \frac{4\pi G a^2_{ }}{\H^2_{ }}\,
 \biggl[\, 
  \partial^{ }_\tau + 2 \biggl( \H  - \frac{\H'}{\H} \biggr)
 \,\biggr]\,
 \E^{ }_v
%%%%
 \label{h0_full_prime}
 \;. 
\ea
In order to eliminate $\bar e\hspace*{0.3mm}''$ 
from \eq\nr{delta_e_prime},
we combine \eqs\nr{bg_Tmunu_tau} 
and \nr{bg_varphi_x} into
\ba
 \bar{e}\hspace*{0.3mm}'
 &
   \underset{\rmii{}}{\overset{\rmii{\nr{bg_Tmunu_tau}}}{=}} 
 &
 - 3 \H (\bar e + \bar p) 
 - \frac{ \bar\varphi\hspace*{0.3mm}' }{a^{2}_{ } }
   \hspace*{-3mm}
   \overbrace{ 
   \bigl(\, 
         \bar\varphi\hspace*{0.3mm}''
         + 2 \H \bar\varphi\hspace*{0.3mm}' 
   \,\bigr) 
   }^{ 
   {\rm from~\nr{bg_varphi_x}:}\;
               - a \Upsilon \bar\varphi\hspace*{0.3mm}'
                           - a^2_{ } V^{ }_{,\varphi}
   }
  \nn[2mm]
%%%%%
 & = & 
 - 3 \H (\bar e + \bar p) 
 + \frac{\Upsilon (\bar\varphi\hspace*{0.3mm}' )^2_{ } }{a}
 + \bar\varphi\hspace*{0.3mm}'\, V^{ }_{,\varphi}
 \label{bare_prime} \\[3mm]
%%%%%%%
 \Rightarrow \; 
 \bar{e}\hspace*{0.3mm}''
 &
 = 
 & 
 - 3 \H' (\bar e + \bar p) 
 - 3 \H (\bar e\hspace*{0.3mm}' + \bar p\hspace*{0.3mm}' ) 
 + \frac{(\Upsilon' - \H \Upsilon )(\bar\varphi\hspace*{0.3mm}' )^2_{ } }{a}
 + \frac{2 \Upsilon \bar\varphi\hspace*{0.3mm}'
                    \bar\varphi\hspace*{0.3mm}'' }{a}
 \nn[2mm]
%%%%%
 & & \; + \,
   \bar\varphi\hspace*{0.3mm}''\, V^{ }_{,\varphi}
 + \bar\varphi\hspace*{0.3mm}'\, V^{\prime}_{,\varphi}
 \;. \label{bare_prime2}
\ea
{}From \eq\nr{h0_full_prime}, it is advantageous to eliminate 
$\H''$, 
by making use of \eq\nr{bg_vT_2}, 
and 
$\E_v'$, 
by making use of \eq\nr{dot_E_v_tau}.
In addition, \eqs\nr{bg_HH_tau} 
and \nr{bg_ep} are needed. 

It is inevitable that when we insert \eqs\nr{delta_e_prime}
and \nr{h0_full_prime} into \eq\nr{full_0}, the expressions 
become lengthy. It is then all the more important that we have
an efficient crosscheck available, from the cancellation of~$Y$.
Most simply, 
$(\partial_\tau^2 - \nabla^2_{ })Y$
cancels between its appearances in 
$ h_0' + 3  h_\rmii{D}' + \nabla^2_{ }h $
(cf.\ \eq\nr{h0_hD_h_full}), 
$ \nabla^2_{ }(v-h)$
(cf.\ \eq\nr{d_vmh}), 
and 
$- h_0'$
(cf.\ \eq\nr{h0_full_prime}).
The cancellation of $Y'$  
requires a bit more effort, 
making use of \eq\nr{bare_prime}, 
whereas the coefficient of~$Y$ contains 
$\bar{e}\hspace*{0.3mm}''$ and $\H''$, 
so that \eqs\nr{bare_prime2} and \nr{bg_vT_2} need to be employed.

There are also many cancellations between the physical terms. 
Notably, after having eliminated $\E_v'$, 
by making use of \eq\nr{dot_E_v_tau}, the anisotropic stress, 
$\nabla^2_{ }\Pi$, drops out. 
By setting $\field_\varphi^{\hspace*{0.3mm}\prime}$ into the combination
$
 \field_\varphi^{\hspace*{0.3mm}\prime} - 
   \frac{\H}{a \bar\varphi\hspace*{0.3mm}'}
  \bigl( 
  \frac{a \bar\varphi\hspace*{0.3mm}'}{\H}
  \bigr)'
 \field^{ }_\varphi
$, 
all other non-derivative appearances of 
$
 \field^{ }_\varphi
$
cancel. By a repeated use of background identities, 
the coefficients of 
$\field_\varphi^{\hspace*{0.3mm}\prime}$ 
and
$\E^{ }_v$
are seen to be proportional to the same combination, 
$
 \Upsilon ( \bar\varphi\hspace*{0.3mm}' )^2_{ }/a^2_{ }
 + 
 8\pi G a (\bar e + \bar p)\bar e/\H
$.

Finally, we go once again to physical time, 
by inserting \eqs\nr{Q_R_rel_2} and \nr{time_derivs}, 
as well as \eq\nr{eq_HH'}, 
in the form 
$
 \H' - \H^2_{ } = a^2_{ }\dot{H}
$.
This then leads to our third basic equation, 
\ba
 \boxed{ 
 \begin{array}{rc} 
 \displaystyle   \vphantom{\Bigg|^b_q} 
%%%
   \biggl\{\,
   \partial^{ }_t 
  + 3 H 
 \biggl( 1 + \frac{\bar{p}^{ }_{\der\T}}{\bar{e}^{ }_{\der\T}}\biggr)
  + \frac{4\pi G(\bar e + \bar p) - \dot{H}}{H}
  - \frac{ \dot{\bar\varphi} \, (
     \Upsilon^{ }_{\der\T} \, \dot{\bar\varphi} 
   + 
     V^{ }_{\der\varphi\T} ) 
     }
     { \bar{e}^{ }_{\der\T} }
   \,\biggr\} 
   \, \E^{ }_\T 
%%%
 &  
 \nn[4mm]
%%%%
 \displaystyle   \vphantom{\Bigg|^b_q} 
%%%
 - \biggl\{\,
 \Upsilon \dot{\bar\varphi}^2_{ }
 + \frac{8\pi G(\bar e + \bar p)\,\bar e}{H}
 \,\biggr\} 
 \biggl\{\, 
   \dot{\R}^{ }_\varphi - \frac{4\pi G}{H}\, \E^{ }_v
 \,\biggr\}
 - \frac{\nabla^2_{ }}{a^2_{ }} 
 \Bigl\{\, 
   (\bar e + \bar p) \R^{ }_\varphi + \E^{ }_v
 \,\Bigr\}
%%%
 & \;=\;  
 \varrho\, \dot{\bar\varphi}\bit H 
 \;. 
 \end{array}
 }
 \nn[2mm] 
 \label{dot_E_T}
\ea
This is linearly independent of \eqs\nr{dot_R_varphi} and \nr{dot_E_v},
and given the appearance of $\dot{\E}^{ }_\T$, 
can be said to determine the time evolution of $\E^{ }_\T$. 

%%%%%%%%%%%%%%%%%%%%%%%%%%%%%%%%%%%%%%%%%%%%%%%%%%%%%%%%%%%%%%%%%

\vspace*{3mm}

Let us summarize the situation so far. 
In \eqs\nr{dot_R_varphi}, \nr{dot_E_v} and \nr{dot_E_T},  
we have three independent equations for three variables,  
$\R^{ }_\varphi$, $\E^{ }_v$ and $\E^{ }_\T$.
If the anisotropic stress, $\Pi$, can be omitted, 
i.e.\ if shear viscous corrections are small
(cf.\ the discussion below \eq\nr{shear_estimate}), 
then this is sufficient for specifying the time
evolution of the three physical perturbations. 
However, if the anisotropic stress plays a role, 
then, as shown in \eq\nr{Pi_gi},
one of the Bardeen potentials, $\psi$ (cf.\ \eq\nr{def_psi}), 
also makes an appearance. 
In \eq\nr{demo_1}, 
we already obtained one equation for the Bardeen potentials, 
\be
 \boxed{
 \quad
 \phi -\psi 
 +  8\pi G a^2_{ }\barpPi
 \; 
 \overset{\rmii{\nr{demo_1}}}{=}
 \; 
 0
  \;.
  \quad \vphantom{\Bigg|} 
 }
 \label{dot_phi}
\ee
However, since this contains 
the other Bardeen potential, $\phi$
(cf.\ \eq\nr{def_phi}), we also need a fifth equation, 
in order to fully specify the time evolution of the 
perturbations of a non-perfect fluid coupled
to an inflaton field. 

A possible starting point for deriving the remaining equation
is given by \eq\nr{h0_full_rev2}. Even if we have already employed
this equation as a part of other computations, we can still 
extract novel information from it. Notably, we express the 
left-hand side of \eq\nr{h0_full_rev2} in terms of the Bardeen potential 
from \eq\nr{def_phi}, 
\be
 h^{ }_0 \; \overset{\rmii{\nr{def_phi}}}{=} \; 
 \phi - (\, \partial^{ }_\tau + \H \,)X
 \;, \quad
 X
 \; \overset{\rmii{\nr{shorthand}}}{\equiv} \; 
         h-\vartheta' 
 \;. \label{h0_subst}
\ee
On the right-hand side of 
\eq\nr{h0_full_rev2}, we express the gauge-variant combination~$Y$
from \eq\nr{def_Y} in terms of $\psi$ via \eq\nr{def_psi}, 
\be
 Y \; \underset{\rmii{\nr{shorthand}}}
      {\overset{\rmii{\nr{def_psi}} \lift }{=}} \; 
 \psi + \H\, X % (\, h-\vartheta' \,)
 \;. \label{Y_subst}
\ee 
Substituting \eqs\nr{h0_subst} and \nr{Y_subst}
into \eq\nr{h0_full_rev2}, where we also insert
$
 \field^{ }_\varphi 
 = 
   -({ a \bar\varphi\hspace*{0.3mm}' }/{ \H })
 \R^{ }_\varphi
$
according to \eq\nr{def_field_varphi}, 
it can be verified that all appearances of $X$ % $h-\vartheta'$ 
and $X'$ % $\partial^{ }_\tau (h-\vartheta')$ 
drop out. The remaining terms give
\ba
 \phi
 & 
   \underset{\rmii{\nr{h0_subst},\nr{Y_subst}}}{
   \overset{\rmii{\nr{h0_full_rev2},\nr{def_field_varphi}}
            \lift }{=}}
 & 
\biggl(       \frac{ \H' }
                   { \H^2_{ }  } - 1 
 \biggr)
 \R^{ }_\varphi 
 \; - \;
 \biggl(
   \frac{1}{\H}\, \partial^{ }_\tau + 1 - \frac{\H'}{\H^2_{ }}
 \biggr)\,\psi
  \; - \; 
  \frac{4\pi G a^2_{ }}{\H^2_{ }}\, \E^{ }_v
  \;. 
\ea
In this equation, all the quantities appearing 
are gauge-invariant. 
Moving the terms to the left-hand side, and going
over to physical time, 
with 
$
 \H'/\H^2_{ } = 1 + \dot{H}/H^2_{ }
$
from \eq\nr{eq_HH'}, we obtain our fifth and final relation, 
\be
 \boxed{ 
 \quad  \vphantom{\Bigg|} 
 \biggl( \partial^{ }_t - \frac{\dot{H}}{H} \biggr) \psi
 +  H\phi  
 - \frac{\dot{H}}{H}\, \R^{ }_\varphi 
 + \frac{4\pi G}{H}\, \E^{ }_v \; = \; 0
  \;. \quad
 }
 \label{dot_psi}
\ee
Given that 
$\dot{\psi}$ appears, \eq\nr{dot_psi} can be said to determine
the time evolution of $\psi$. 

We now have the full set of equations needed for determining 
the time evolution of the coupled system at linear order. 
The solutions following from 
\eqs\nr{dot_R_varphi}, \nr{dot_E_v}, \nr{dot_E_T}, 
\nr{dot_phi} and \nr{dot_psi} will be discussed
in \ch\ref{se:outside} (for Hubble horizon exit)
and \ch\ref{se:inside} (for times after Hubble horizon re-entry).

%%%%%%%%%%%%%%%%%%%%%%%%%%%%%%%%%%%%%%%%%%%%%%%%%%%

\newpage

%%%%%%%%%%%%%%%%%%%%%%%%%%%% start appendices %%%%%%%%%%%%%%%%%%%%%%%%%%%%%%%

%%%%%%%%%%%%%%%%%%%%%%%%%%%%%%%%%%%%%%%%%%%%%%%%%%%%%%%%%%%%%%%%%%%%%%%%%
%
\subsubsection{Perturbative thermodynamics for a thermalized inflaton}
\label{app:thermal_inflaton}

\addcontentsline{toc}{subsection}{\App\ref{app:thermal_inflaton}: 
Perturbative thermodynamics for a thermalized inflaton}

\index{inflaton thermodynamics}

If $\Upsilon \gg H$, the inflaton field
thermalizes (cf.\ \fig\ref{fig:timeline} on p.~\pageref{fig:timeline}).
In this regime (if it gets realized for all~$k$), it is 
possible to adopt a ``complementary'' viewpoint, in which the inflaton
is actually part of the radiation plasma. In the present appendix, 
we elaborate on this from a few different perspectives. We start by 
showing two methods to determine the contributions of a thermalized
inflaton to the energy density and pressure. Afterwards,  
we recall why our effective description, 
based on the Langevin equation, ultimately fails in this regime, 
so that more advanced tools are needed for obtaining accurate results. 

The first method goes through the effective potential. 
In \eq\nr{L_bath}, we met $V^{ }_0$, the ``bare'' self-interaction 
potential appearing in a Lagrangian. In the effective-theory description
of \eq\nr{eq_phi_eom1}, this had been replaced by $V$. Through a matching
computation, we saw that~$V$ obtains a thermal mass correction from 
interactions with the medium, cf.\ \eq\nr{re_GR}. However, through
the formalism of thermal field theory, we can also determine the 
influence of the medium on $V$ to all orders in $\bar\varphi$, 
not only to linear order, as we had done to arrive at \eq\nr{re_GR}. 
In fact, important effects arise already at {\em zeroth order}
in $\bar\varphi$, through an overall modification of the
energy density and pressure. 
To be specific, let us consider a time at which the inflaton 
is settled around its global minimum, $\bar\varphi = 0$, but still
experiences thermal fluctuations. In statistical physics, the
effect of fluctuations is obtained by 
computing the {\em canonical partition function}, $\Z$. From $\Z$, 
we can extract the {\em free energy density}, 
and this is exactly what is called the {\em effective potential}, 
$V^{ }_\rmi{eff}$.

\index{$\Z$ (canonical partition function)}
\index{free energy density}
\index{effective potential}

The leading non-trivial contribution to the effective potential
is referred to as a ``1-loop'' contribution. To compute it in the
perturbative approach, we treat the inflaton as a weakly coupled 
massive scalar field, $V_0^{ }\approx m^2 \varphi^2/2$. 
Then the 1-loop correction 
reads~(cf.,\ e.g.,\ ref.~\cite[eq.~(9.26)]{basics}) 
\begin{equation} 
 V_\rmi{eff}^\rmii{(1)}
 \; = \; 
 -\lim_{L \to\infty} \frac{T}{L^3} \, \ln \Z^\rmii{(1)}_{ }
 \; = \;
 \int \! \frac{\dd^3 \vec{p}}{(2\pi)^3_{ }} 
 \biggl[\, 
    \frac{\epsilon_{p}^{ }}{2} 
  + T \ln \left( 
        1 - e^{-\epsilon_{p}^{ }/T}_{ }
    \right)
 \,\biggr]^{ }_{\epsilon_{p}^{ } = \sqrt{p^2 + m^2}}
 \ . \label{eq_Veff1}
\end{equation}
The first, temperature-independent term, requires regularization
and renormalization; let us assume that it modifies the parameters
appearing in $V_0^{ }$. The second term 
in \eq\nr{eq_Veff1} gives the thermal effects
that we are interested in. 

Changing integration variables as
\be
 \dd^3 \vec{p} \;=\; 4\pi \dd p \, p^2_{ }
 \;=\; 4\pi \dd \epsilon_{p}^{ } 
 \, \epsilon_{p}^{ } \sqrt{\epsilon_{p}^2 - m^2}
 \;, \label{measure}
\ee
the contribution to the pressure 
according to \eq\nr{general_e_p} is then 
\begin{align}
 p^{ }_\varphi 
 &
 \;
 \underset{\rmii{\nr{eq_Veff1},\nr{measure}}}{
 \overset{\rmii{\nr{general_e_p}}}{\supset}}
 \; 
 - 
 \frac{T}{2\pi^2}
 \int_m^\infty \! \dd \epsilon_{p}^{ }\,
  \epsilon_{p}^{ } \sqrt{\epsilon_{p}^2 - m^2}\,
 \ln\left( 1 - e^{-\epsilon_{p}^{ } / T}_{} \right)
 \;. \label{p_varphi} 
\end{align}
For the energy density from \eq\nr{general_e_p}, 
we need a Legendre transform. Noting that 
\ba
 \bigl(1 - T\partial^{ }_{\T} \bigr)
 \bigl[\, T \ln \bigl( 1 - e^{ -\epsilon^{ }_p/T}_{ } \bigr)\, \bigr]
 & = & 
 -T^2_{ } \partial^{ }_{\T} \ln \bigl( 1 - e^{ -\epsilon^{ }_p/T}_{ } \bigr)
 \; = \; 
 - \cancel{ T^2_{ }}
  \frac{\; - \,\epsilon^{ }_p\, e^{ -\epsilon^{ }_p/T}_{ } }
  { 1 - e^{ -\epsilon^{ }_p/T}_{ } }
   \cancel{ \frac{1}{T^2_{ }} }
 \nn[2mm] 
%%%%
 & = & 
 \frac{\epsilon^{ }_p }{e^{\epsilon^{ }_p/T}_{ } - 1}
 \; \equiv \; \epsilon^{ }_p\, \nB^{ }(\epsilon^{ }_p)
 \;, \hspace*{6mm} \label{legendre}
\ea 
we find 
\begin{align}
 e^{ }_\varphi 
 &
 \; 
 \underset{\rmii{\nr{legendre}}}{
 \overset{\rmii{\nr{p_varphi}}}{\supset}} 
 \; 
 \frac{1}{2\pi^2}
 \int_m^\infty \! \dd \epsilon_{p}^{ }
 \, \epsilon_{p}^2 \sqrt{\epsilon_{p}^2 - m^2}\,
 \nB^{ }(\epsilon_{p}^{ })
 \;. \label{eq_e-varphi}
\end{align}

Even if the integral representations of $p^{ }_\varphi$ 
and $e^{ }_\varphi$ in \eqs\nr{p_varphi} and \nr{eq_e-varphi}
are rapidly convergent, neither 
can be evaluated analytically. However, by expanding the integrands in 
$e^{-\epsilon^{ }_p/T}_{ }$, sum representations in terms
of modified Bessel functions can be found. Analytic results become
available by sending $m/T\to 0$, and then 
\eqs\nr{p_varphi} nor \nr{eq_e-varphi} amount just to adding
one light degree of freedom to the radiation plasma, i.e.\ changing
$g^{ }_* \to g^{ }_* + 1$ and $h^{ }_* \to h^{ }_* + 1$ 
in \eq\nr{p_r}.

\vspace*{3mm}

Let us now inspect the same results from a different perspective. 
In a flat Minkowskian frame, the energy-momentum tensor of the scalar
field from \eq\nr{Tmunu_varphi}, after the replacement 
$V \to V^{ }_0 \to \tfr{1}{2} m^2_{ }\varphi^2_{ }$, has the components
\ba
 T^{ }_{00}
 &
 \overset{\rmii{\nr{Tmunu_varphi}}}{=}
 &
 \frac{1}{2}
 \bigl( \dot{\varphi}^2_{ }
   + {\textstyle \sum_l} \ibit
   \varphi^{ }_{,\hspace*{0.3mm}l}\ibit
   \varphi^{ }_{,\hspace*{0.3mm}l}
       + m^2_{ }\varphi^2_{ } \bigr)
 \;, \label{T_00_2nd} \\[2mm]
%%%%%
 T^{ }_{ij} 
 &
 \overset{\rmii{\nr{Tmunu_varphi}}}{=}
 &
 \varphi^{ }_{,i} \varphi^{ }_{,j}
 + \delta^{ }_{ij} \, 
 \frac{1}{2}
 \bigl( \dot{\varphi}^2_{ } - 
 {\textstyle \sum_l} \ibit
 \varphi^{ }_{,\hspace*{0.3mm}l} \ibit
 \varphi^{ }_{,\hspace*{0.3mm}l}
       - m^2_{ }\varphi^2_{ } \bigr)
 \;. \label{T_ij_2nd}
\ea

Placing ourselves in the same situation as before, with 
$\bar\varphi \to 0$, the $\varphi$ here is $\delta\varphi$.
We can average over 
the thermal noise influencing $\delta\varphi$, 
denoting the results 
by~$
 \langle T^{ }_\rmii{$00$} \rangle 
$
and~$
 \langle T^{ }_{ij} \rangle
$. 
Employing the ``target'' 
expression from \eq\nr{spectrum_full}
for the 2-point correlator of $\delta\varphi$; inserting a plane-wave
time dependence; and only including the convergent thermal 
parts, we obtain
\ba
 \langle T^{ }_{00} \rangle
 & \underset{\rmii{\nr{spectrum_full}}}
  {\overset{\rmii{\nr{T_00_2nd}}}{\supset}} & 
 \int \! \frac{{\rm d}^3_{ }\vec{p}}{(2\pi)^3_{ }} \,
 \frac{\nB^{ }( \epsilon^{ }_p)}{{\epsilon}^{ }_p}
 \frac{1}{2}
 \bigl( \epsilon_p^2 + p^2_{ } + m^2_{ } \bigr) 
 \; = \; 
 \int \! \frac{{\rm d}^3_{ }\vec{p}}{(2\pi)^3_{ }} \,
 {\epsilon}^{ }_p\, \nB^{ }( \epsilon^{ }_p) 
 \;. \label{T00_exp}
%%%% 
\ea 
This agrees with \eq\nr{eq_e-varphi} when we make use of \eq\nr{measure}. 

We have to work a bit harder
for the spatial components, $\langle T^{ }_{ij} \rangle$.
Inserting the thermal part of \eq\nr{spectrum_full}, 
and making use of isotropy, we get 
\ba
 \langle T^{ }_{ij} \rangle
 & \underset{\rmii{\nr{spectrum_full}}}
  {\overset{\rmii{\nr{T_ij_2nd}}}{\supset}} & 
 \int \! \frac{{\rm d}^3_{ }\vec{p}}{(2\pi)^3_{ }} \,
 \frac{\nB^{ }( \epsilon^{ }_p)}{{\epsilon}^{ }_p}
 \biggl[\,
 p^{ }_i\, p^{ }_j + \delta^{ }_{ij} \ibit
 \frac{1}{2}
 \bigl(\, 
    \cancel{\epsilon_p^2} \; \bcancel{ - p^2_{ } - m^2_{ } }
 \,\bigr) 
 \,\biggr]
 \nn[2mm]
%%%%
 & \overset{\rmii{isotropy} \lift }{=} &  
 \int \! \frac{{\rm d}^3_{ }\vec{p}}{(2\pi)^3_{ }} \,
 \frac{p^2_{ } \delta^{ }_{ij}}{3\epsilon^{ }_p} \, \nB^{ }( \epsilon^{ }_p) 
 \;. \label{Tij_exp}
%%%% 
\ea 
Similarly to \eq\nr{legendre}, we now note that 
\ba
 \partial^{ }_{p}
 \bigl[\, T \ln \bigl( 1 - e^{ -\epsilon^{ }_p/T}_{ } \bigr)\, \bigr]
 & = & 
  \cancel{ T }
  \frac{ - e^{ -\epsilon^{ }_p/T}_{ } }
  { 1 - e^{ -\epsilon^{ }_p/T}_{ } }
   \frac{(- \partial^{ }_p \epsilon^{ }_p ) }{ \cancel{T} } 
 \; = \; 
 \frac{p}{ \epsilon^{ }_p }\, \nB^{ }(\epsilon^{ }_p)
 \;. \hspace*{6mm} \label{ibp_Tij}
\ea 
Therefore 
\ba
 \langle T^{ }_{ij} \rangle
 & \underset{\rmii{\nr{ibp_Tij}}}
   {\overset{\rmii{\nr{Tij_exp}}}{\supset}} & 
 \frac{\delta^{ }_{ij}}{2\pi^2_{ }}
 \int_0^\infty \! \dd p \, \frac{p^3_{ }}{3}
 \partial^{ }_{p}
 \bigl[\, T \ln \bigl( 1 - e^{ -\epsilon^{ }_p/T}_{ } \bigr)\, \bigr] 
 \nn[2mm]
%%%%
 & = &  - 
 \frac{\delta^{ }_{ij}}{2\pi^2_{ }}
 \int_0^\infty \! \dd p \, p^2_{ }
 \, T \ln \bigl( 1 - e^{ -\epsilon^{ }_p/T}_{ } \bigr)
 \;, 
\ea
where we carried out a partial integration. 
This agrees with \eq\nr{p_varphi}. 

To summarize, 
if we go to {\em second order in perturbations},
and assume that the inflaton field has equilibrated, 
its fluctuations contribute to the energy density and pressure
just like a massive scalar field that is part of the 
radiation plasma. 

\index{second-order perturbations}
\index{Rayleigh-Jeans divergence}

Now, however, we come to an issue. Our actual effective-theory
framework is not based on \eq\nr{spectrum_full}, but rather on 
the Langevin equation, which yields \eq\nr{spectrum_appro}. 
This corresponds to replacing 
$\nB^{ }(\epsilon^{ }_p) \to T/\epsilon^{ }_p$
in \eqs\nr{T00_exp} and \nr{Tij_exp}. 
Then the integrals over the momenta 
are UV-divergent. This is a reflection of the famous 
{\em Rayleigh-Jeans divergence} of classical field theory. 

The analysis in this book stays mostly at linear order
in perturbations. Then, if we restrict ourselves to small
enough comoving momenta, the Rayleigh-Jeans divergence 
is of no concern. However, it is important to keep in mind
that equilibration implies a transfer of energy, from the
initial quantum state towards thermal fluctuations,
which carry a typical momentum $p\sim T$
for a massless particle ($m \ll T$), 
or $p\sim \sqrt{2 m T}$ 
for a massive one ($m \gg T$). 
If we encounter 
a phenomenon which is sensitive 
to the thermal energy scales, $\epsilon^{ }_p \sim T$, 
we have to employ quantum-statistical tools, in order to 
obtain physically meaningful results. Notably, this can be
important if we consider thermally induced 
{\em non-Gaussianities} in power spectra (cf.,\ e.g.,\ ref.~\cite{mehrdad}), 
or {\em second-order effects} producing gravitational waves
(cf.\ the end of \se\ref{ss:gw_sigw}
on p.~\pageref{double_count}).
In principle, the issue is also relevant for 
the autocorrelator of the thermal noise (cf.\ \eq\nr{fl-di}), 
because at early times, any comoving momentum lies in the domain
$p = k/a \gg T$. 
At linear order, the last issue can be rectified by
modifying the noise autocorrelator, 
as mentioned below \eq\nr{fl-di}. 

\label{rayleigh-jeans}

\index{non-Gaussianity}
\index{second-order perturbations}

%%%%%%%%%%%%%%%%%%%%%%%%%%%%%%%%%%%%%%%%%%%%%%%%%%%

\newpage 

\subsubsection{Numerical solution for smooth reheating} 
\label{app:num_bg_thermal}

\addcontentsline{toc}{subsection}{\App\ref{app:num_bg_thermal}: 
Numerical solution for smooth reheating}

%%%%%%%%%%%%%%%%%%%%%%%%%%%% FIGURE %%%%%%%%%%%%%%%%%%%%%%%%%%%%%%%%%
%
\begin{figure}[t]
    \hspace*{-2mm}
    \includegraphics[width=0.49\linewidth]{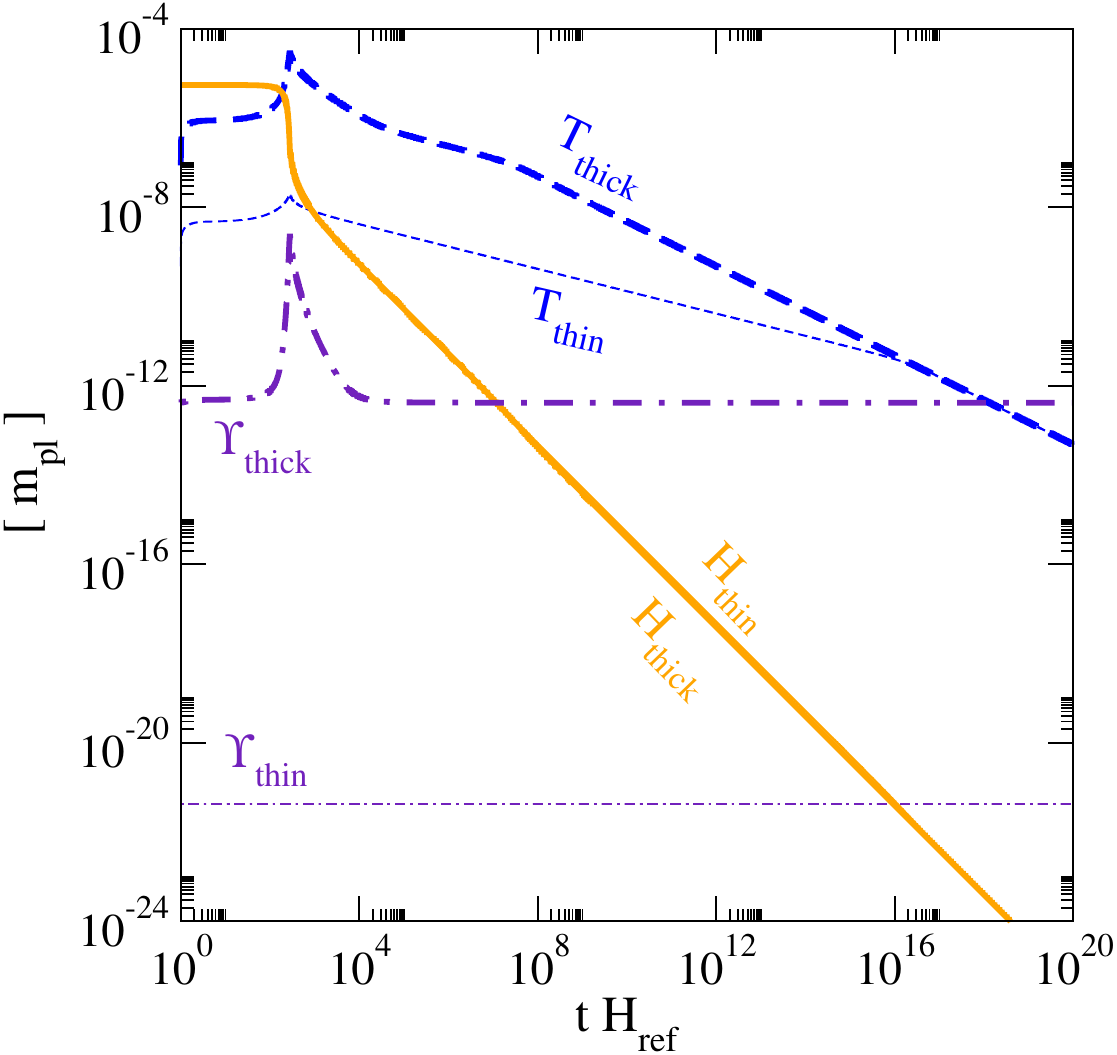}%

    \vspace*{-7.0cm}

    \hspace*{7.6cm}
    \includegraphics[width=0.49\linewidth]{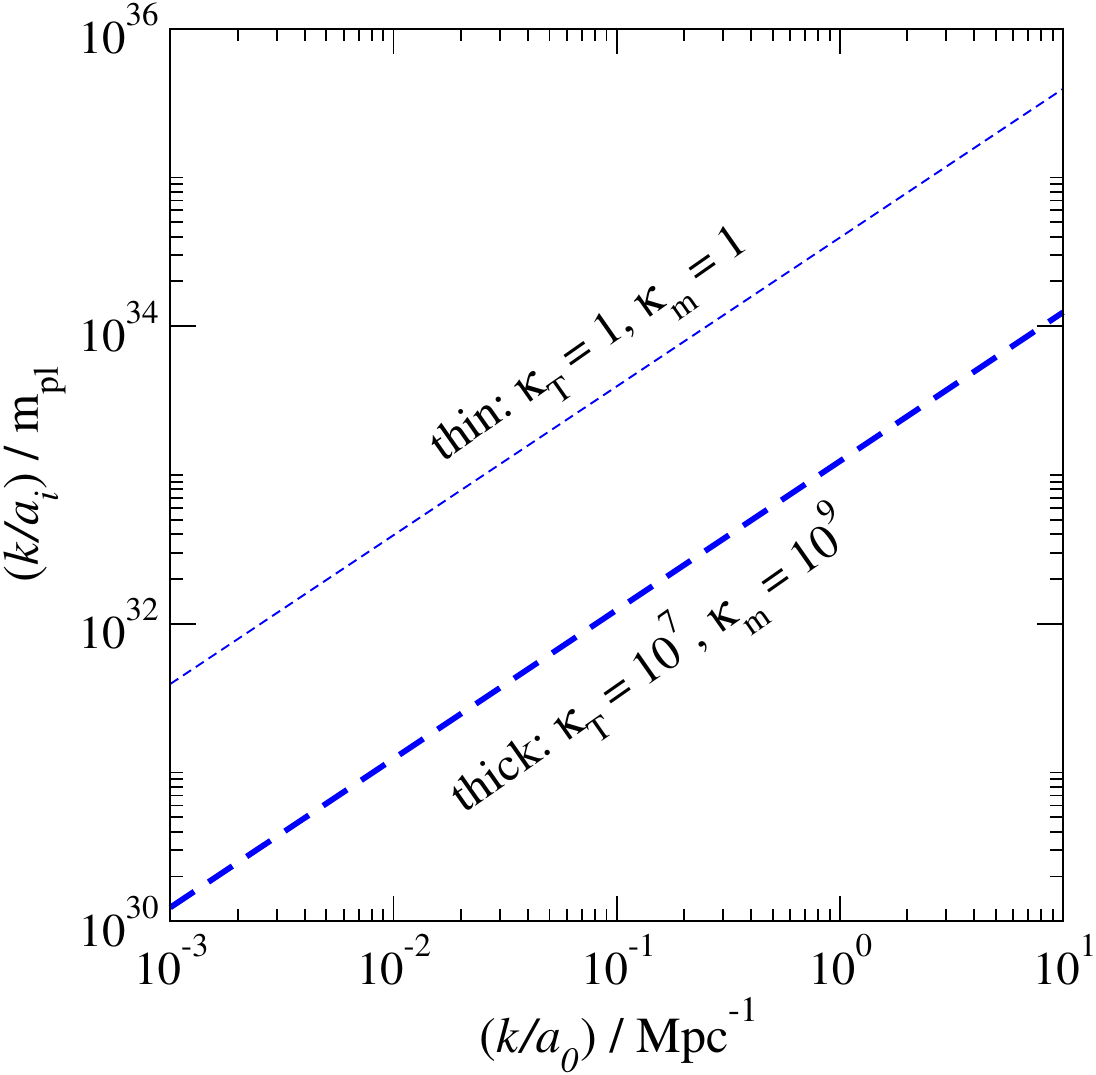}

    \caption{%
    \small
    Left: A numerical solution for the reheating dynamics described by 
    \eqs\nr{bg_varphi_again}--\nr{bg_HH_again}.
    The vacuum potential is like in 
    \eqs\nr{V} and \nr{params}, and $H^{ }_\rmii{ref}$
    is from \eq\nr{Href},  
    whereas the equation of state of the plasma is parametrized according
    to \eq\nr{ex_plasma}, and 
    the friction according to \eq\nr{ex_friction}. 
    If the friction is small
    (thin lines),
    the universe is matter-dominated 
    until $ t\hspace*{0.3mm} H^{ }_\rmii{ref} \approx 10^{16}_{ }$ 
    (cf.\ discussion around \eq\nr{damped_ho}),   
    when $\Upsilon \approx H$ and $T$ starts to decrease faster. 
    If the friction is larger (thick lines), a higher maximal temperature
    is reached, and radiation domination starts 
    already at $ t\hspace*{0.3mm} H^{ }_\rmii{ref} \approx 10^{7}_{ }$.  
    Right:
    mapping from
    current momenta in the CMB range, 
    $(k/a^{ }_0)/\mbox{Mpc}^{-1}_{ }$,   
    onto initial momenta at $t = H_\rmii{ref}^{-1}$, 
    $(k/a^{ }_i)/\mpl^{ }$, 
    as determined from \eq\nr{micro_macro}. 
    With a larger friction (thick lines), 
    the matter-domination period is shorter, 
    and less $e$-folds separate inflation
    and present day. Consequently, we should consider
    smaller initial momenta. Given that the curvature power spectrum
    depends on the momentum mode considered
    (cf.\ \fig\ref{fig:P_R_varphi} on p.~\pageref{fig:P_R_varphi}), 
    this influences inflationary predictions. 
    In the literature, reheating dynamics is often not addressed
    quantitatively, and the results rather contain 
    an error band, obtained by requiring that the pivot scale
    (cf.\ \eq\nr{eq_powerlaw}) exited the Hubble horizon 50--60 
    $e$-folds before inflation ended. 
    }
    \index{reheating: numerics (figure)}
    \index{figure: numerics for reheating history}
    \label{fig:bg_thermal}
\end{figure}
%
%%%%%%%%%%%%%%%%%%%%%%%%%%%%%%%%%%%%%%%%%%%%%%%%%%%%%%%%%%%%%%%%%%%%

We show here how the background equations 
\nr{bg_varphi_again}--\nr{bg_HH_again} can be solved numerically. 
The functions $e^{ }_r$ and $p^{ }_r$ are parametrized
according to \eq\nr{p_r}, with 
\be
  g^{ }_* \;=\; h^{ }_* \;=\; i^{ }_* \;=\; 106.75 
  \;. \label{ex_plasma}
\ee
The number 106.75 corresponds to the Standard Model in the limit
that all interactions are weak enough to be insignificant, and the 
temperature is high enough for masses to be negligible
(the fractional number originates from fermionic contributions).
For the friction we take the ansatz 
\be
 \Upsilon 
 \; \equiv \; 
 \frac{
    \kappa^{ }_\T\, (\pi T)^3_{ } + 
    \kappa^{ }_m\, m^3_{ }
 }{(4\pi)^3_{ } f_a^2}
 \;.  \label{ex_friction}
\ee
Otherwise the setup is identical to that described
in \app\ref{app:num_bg_vac}.
A {\tt python} script is displayed below, and it produces the
data as shown in \fig\ref{fig:bg_thermal}.

Having available a background solution from inflation until reheating, 
we can determine how momenta {\em redshift 
across different epochs}.
Let us denote by $a^{ }_\switch$ a moment when the reheating dynamics
has ended, and we can switch to an extrapolation of Standard Model 
thermodynamics, to describe the dominant plasma component
(we choose $T^{ }_\switch = 10^{-12}_{ }\mpl^{ }$ for the switching point). 
Then we can write a current momentum mode (in astrophysical units) as
\be
 \frac{( k / a^{ }_\inow )}{\mbox{\small Mpc}^{-1}_{ }}
 \;\; 
 = 
 \; 
 \underbrace{
 \frac{( k / a^{ }_i )}{\mpl^{ }}
 }_{{\rm like}\;{\rm fig.}\;\ref{fig:P_R_varphi}}
 \,
 \underbrace{
 \biggl( \frac{a^{ }_i}{a^{ }_\switch} \biggr)
 }_{e^{-N^{ }_\switch }_{ } }
 \,
 \underbrace{
 \biggl( \frac{a^{ }_\switch}{a^{ }_\inow} \biggr)
 }_{{\rm like}\;\nr{min_e_folds}}
 \, 
 \underbrace{
 \mpl^{ }\, \mbox{Mpc}
 }_{ 1.9092 \times 10^{57}_{ } }
 \;. \label{micro_macro}
 \index{redshift factor: during reheating}
\ee 
For the chosen $T^{ }_\switch  = 10^{-12}_{ }\mpl^{ }$, 
a computation like in \eq\nr{min_e_folds} yields 
$a^{ }_\switch/a^{ }_0 \approx e^{- 46.5 }_{ }$.
The factor $a^{ }_i/a^{ }_\switch 
\equiv e^{- N^{ }_\switch}_{ }$ needs to be determined
by integrating the $H(t)$ produced by 
a numerical computation, like that in \fig\ref{fig:bg_thermal}(left).
Putting everything together, we obtain \fig\ref{fig:bg_thermal}(right), 
which shows the initial momentum that corresponds to a given 
physical value today.
The corresponding data file is generated at the end of the 
{\tt python} script. %_e

\index{code: numerics for reheating history}

{\fontsize{8pt}{10pt}\selectfont
\begin{verbatim}
# numerical solution for reheating and/or warm inflation [numerics_bg_warm.py]
#
# import basic tools and integration routines
import numpy as np
from scipy.integrate import solve_ivp

# parameters [mpl]
fa = 1.25;       m = 1.09e-6;    Pi = np.pi     # inflaton decay constant and mass, shorthand for pi
gstar = 106.75;  hstar = gstar;  istar = gstar  # energy density, entropy density, heat capacity

# parameters for the cases considered
time_end = 1e20        # time sufficiently large that the system has reheated
case = 'thin' # 'thick'
if case == 'thin': 
    kappaT = 1e0;      kappam = 1e0;     T_0 = 1.e-9        # initial seed temperature [mpl]
else:
    kappaT = 1e7;      kappam = 1e9;     T_0 = 1.e-8        

# potential of inflaton (phi) [mpl^4]
def V(phi): return m*m*fa*fa*( 1 - np.cos(phi/fa) ) 

# derivative of inflaton potential by phi [mpl^3]
def Vd(phi): return m*m*fa*np.sin(phi/fa)

# friction coefficient [mpl]
def Ups(T): return (kappaT*Pi*Pi*Pi*T*T*T + kappam*m*m*m)/(4*4*4*Pi*Pi*Pi)/(fa*fa)

# radiation energy density [mpl^4]
def er(T): return gstar*Pi*Pi*T*T*T*T/30

# radiation entropy density [mpl^3]
def sr(T): return 2*hstar*Pi*Pi*T*T*T/45

# radiation heat capacity [mpl^3]
def cr(T): return 2*istar*Pi*Pi*T*T*T/15

# energy density of inflaton field [mpl^4]
def efield(phi,phid): return phid*phid/2 + V(phi)

# Hubble rate [mpl]
def H(energy_density): return np.sqrt(8*Pi*np.abs(energy_density)/3)

# evolution at early times: solve for [phi, dot phi, efolds, T]
# derivatives are taken with respect to t=Href*time
def derivatives1(t,y):
    Phi, Phid, Nfolds, T = y
    Hubble = H(efield(Phi, Phid)+er(T))
    dphi_dt = Phid/Href;      ddphi_ddt = (-(3*Hubble+Ups(T))*Phid - Vd(Phi))/Href
    dN_dt = Hubble/Href;      dT_dt = (Ups(T)*Phid*Phid-3*Hubble*T*sr(T))/cr(T)/Href
    dy_dt = [dphi_dt, ddphi_ddt, dN_dt, dT_dt]; return dy_dt

# evolution in averaged regime: solve for [e_phi, efolds, T]
def derivatives2(t,y):
    e_phi, Nfolds, T = y
    Hubble = H(e_phi+er(T))
    dephi_dt = (-(3*Hubble+Ups(T))*e_phi)/Href
    dN_dt = Hubble/Href;      dT_dt = (Ups(T)*e_phi-3*Hubble*T*sr(T))/cr(T)/Href
    dy_dt = [dephi_dt, dN_dt, dT_dt];  return dy_dt   

# evolution after e_phi has become insignificant: solve for [efolds, T]
def derivatives3(t,y):
    Nfolds, T = y
    dN_dt = H(er(T))/Href;    dT_dt = (-3*H(er(T))*T*sr(T))/cr(T)/Href
    dy_dt = [dN_dt, dT_dt];  return dy_dt   

# initial conditions and reference values [mpl]
phi_0 = 3.5;   Href = np.sqrt(4*Pi/3)*m*phi_0;   phid_0 = - Vd(phi_0)/(3*Href);   N_0 = 0

# define the event function counting the zeros of phi
def phi_crossings(t, y): return y[0]  

# integrate up to 10 oscillations (21 crossings of zero)
sol1 = solve_ivp(derivatives1, [1, 1e4], [phi_0, phid_0, N_0, T_0],
                 events=phi_crossings,method='DOP853',atol=1e-12,rtol=1e-13) 
t_match = sol1.t_events[0][20]         # time at which condition was met
time1 = sol1.t[sol1.t <= t_match];     n_match1 = len(time1)  
ephi1 = efield(sol1.y[0][:n_match1], sol1.y[1][:n_match1]);
efolds1 = sol1.y[2][:n_match1];        T1 = sol1.y[3][:n_match1]

# initial conditions for averaged regime
time_0 = t_match;   ephi_0 = efield(sol1.y_events[0][20][0], sol1.y_events[0][20][1])
efolds_0 = sol1.y_events[0][20][2];   T_0 = sol1.y_events[0][20][3] 
print("go over to averaged evolution at t/t_ref = ",time_0)

# define the event function locating when e_field = e_r/500
def phi_decays(t, y): return y[0]-er(y[2])/500
setattr(phi_decays,'terminal',True)    # for stopping immediately when condition is met

# integrate in averaged regime until phi decays
if(ephi_0 > er(T_0)/500):
    sol2 = solve_ivp(derivatives2, [time_0, time_end], [ephi_0, efolds_0, T_0],
                     events=phi_decays,method='DOP853',atol=1e-12,rtol=1e-13)  
    t_match = sol2.t_events[0][0]      # time at which condition was met
    time2 = sol2.t[sol2.t <= t_match]; n_match2 = len(time2)  
    ephi2 = sol2.y[0][:n_match2];      efolds2 = sol2.y[1][:n_match2];    T2 = sol2.y[2][:n_match2]
    time_0 = t_match;                  efolds_0 = sol2.y_events[0][0][1]; T_0 = sol2.y_events[0][0][2] 
else:
    t_match = time_0;                  time2 = np.array([])
    ephi2 = np.array([]);              efolds2 = np.array([]);            T2 = np.array([])
print("go over to evolution without e_phi at t/t_ref = ",time_0)

# parameters for matching to Standard Model
mplMpc = 1.9092e57                     # conversion between microscopic and macroscopic scales 
Tswitchmpl = 1.e-12                    # temperature at which we switch to Standard Model [mpl]
efolds_after_switch = 46.5             # efolds from Tswitch until today

# define the event function locating the switching temperature
def Tswitch(t, y): return y[1]-Tswitchmpl  

# integrate until temperature is reached
sol3 = solve_ivp(derivatives3, [time_0, time_end], [efolds_0, T_0],
                 events=Tswitch,method='DOP853',atol=1e-12,rtol=1e-13)  
t_match = sol3.t_events[0][0]          # time at which condition was met
time3 = sol3.t[sol3.t <= t_match];     n_match3 = len(time3)  
ephi3 = np.zeros(n_match3);            efolds3 = sol3.y[0][:n_match3];    T3 = sol3.y[1][:n_match3]

# collect information about switching point 
efolds_till_switch = sol3.y_events[0][0][0]
print("Tswitch reached at t/tref = ",t_match," after ",efolds_till_switch," e-folds")
coeff = np.exp(-efolds_till_switch)*np.exp(-efolds_after_switch)*mplMpc

# initial conditions for the last domain
time_0 = t_match;                      efolds_0 = sol3.y_events[0][0][0]; T_0 = sol3.y_events[0][0][1] 

# integrate till the very end
sol4 = solve_ivp(derivatives3, [time_0, time_end], [efolds_0, T_0],
                 method='DOP853',atol=1e-12,rtol=1e-13)  
time4 = sol4.t;                        n_match4 = len(time4)  
t_match = time4[n_match4-1]            # time reached at the end
ephi4 = np.zeros(n_match4);            efolds4 = sol4.y[0];               T4 = sol4.y[1]
print("integration terminated at t/tref = ",t_match," after ",efolds4[n_match4-1]," e-folds")

# assemble together the complete solution
time = np.concatenate((time1,time2,time3,time4)); e_phi = np.concatenate((ephi1,ephi2,ephi3,ephi4))
efolds = np.concatenate((efolds1,efolds2,efolds3,efolds4)); T = np.concatenate((T1,T2,T3,T4))
e_r = er(T)

# print out background solution
np.savetxt('numerics_bg_warm.dat',
           np.c_[time, e_phi, H(e_phi+e_r), T, Ups(T), efolds],fmt='%.6e',newline='\n',
           header='columns: t*H_ref, e_phi/mpl**4, H/mpl, T/mpl, Ups/mpl, efolds')

# print out relation of microscopic and macroscopic momentum scales
koampl = np.geomspace(1e-10,1e60,100); koaMpc = coeff*koampl
np.savetxt('numerics_redshift.dat',
           np.c_[koaMpc,koampl],fmt='%.6e',newline='\n',
           header='columns: k/a_0*Mpc, k/a_i/mpl')
\end{verbatim}
}

%%%%%%%%%%%%%%%%%%%%%%% end appendices %%%%%%%%%%%%%%%%%%%%%%%%%%%

%%%%%%%%%%%%%%%%%%%%%%%%% BIBLIO %%%%%%%%%%%%%%%%%%%%%%%%%%%%%%%%
%

%%%%%%%%%%%%%%%%%%%%%%%%%%%% SECTION %%%%%%%%%%%%%%%%%%%%%%%%%%%%%%%%%%
\newpage 

\section{What happens when scalar modes exit the Hubble horizon?}
\label{se:outside}

\paragraph{Abstract:}

The equations
for scalar perturbations from \ch\ref{se:thermal}
have a fairly complicated structure. 
We explain how their solution gets simplified
once the wavelength redshifts enough that the mode 
``exits the Hubble horizon'', whereby it effectively decouples
from local dynamics. Subsequently, 
the nature of the background solution may change, 
from vacuum-energy dominated expansion, 
via a possible intermediate period, ultimately 
into a radiation-dominated universe. 
A key feature of the inflationary paradigm is that 
the cosmological perturbations visible in the CMB 
are outside of the Hubble horizon throughout these 
processes, and are almost unaffected by them, or ``frozen''.
We also explain how the frozen solution may have a richer
structure in multi-field inflationary models, leading to the 
appearance of so-called isocurvature perturbations. 

\paragraph{Keywords:} 
horizon exit, 
freezing out of curvature or adiabatic perturbations, 
isocurvature perturbations, 
non-adiabatic modes, 
entropy perturbations, 
multi-field inflation, 
spectator fields, 
dark matter isocurvature.

%%%%%%%%%%%%%%%%%%%%%%%%%%%%%%%%%%%%%%%%%%%%%%%%%%%
%
\subsection{Overview}
\label{ss:overview_perts}

It is an important property of the inflationary paradigm
that curvature perturbations ``freeze out''
when they exit the Hubble horizon. 
For a system characterized by 
a single curvature perturbation, we have already
seen this in \ch\ref{se:cold}  
(cf.\ the discussion between \eqs\nr{slowroll_ddot} and
\nr{As_slowroll}). 
If several 
``flavours'' of curvature perturbations are present
simultaneously
(like in \eqs\nr{def_R_varphi}--\nr{def_R_e}),  
they freeze out to the same value. 
As a consequence, differences between various flavours of 
curvature perturbations, which are sometimes called
{\em isocurvature perturbations}, 
vanish outside of the Hubble horizon. 
The purpose of the present section is 
to enlarge the setup considered in \ch\ref{se:cold}
by the addition of a radiation plasma, and  
establish these properties. % more generally. 

\index{adiabatic perturbations}

At this point, it is good to clarify the terminology that is 
used in cosmological literature. 
Focussing on scalar perturbations, 
there are several of them 
as discussed in \ch\ref{se:thermal}
(in the minimal case, metric perturbations, 
$\delta\varphi$, $v \equiv \delta v$, 
and $\delta T$), but they are coupled to each
other, via the Einstein equations and energy-momentum conservation. 
A special role is played by gauge-invariant linear
combinations of scalar perturbations that are  called curvature
perturbations, or {\em adiabatic perturbations}. The latter term may
sound confusing, since at the time of their generation, 
the system is normally assumed cold (non-thermal), 
so that entropy should play no role. 
Indeed the nomenclature comes more from the side of the later 
universe, when the same perturbations re-enter the Hubble horizon
(cf.\ \ch\ref{se:inside}), at a time when the universe is 
radiation-dominated. Then adiabatic perturbations are
the ones which have a ``common origin'', coming from a single 
curvature perturbation, which at late times 
splits into many separate components, 
no longer in equilibrium with each other  
(dark matter, neutrinos, baryonic matter, and photons).

{\em Non-adiabatic perturbations}, in contrast, 
are defined as differences of various types of 
curvature perturbations, and are also referred to as
isocurvature perturbations, or simply 
{\em entropy perturbations}. 
As we discuss in this chapter, 
when the modes are outside of the
Hubble horizon, isocurvature perturbations normally vanish. 
But before they exit, or 
after they re-enter, they are non-zero. 
We thus have to watch out for the period that is considered. 
If it is said that 
``there is no isocurvature'' or that ``perturbations are
adiabatic'', this normally refers to the time at which 
the perturbations are outside of the Hubble horizon. 
In other words, even if the primordial 
scalar perturbations were adiabatic, 
entropy perturbations still take a non-zero value at late
times, when the modes re-enter inside the Hubble horizon. 
``Adiabaticity'' then refers to their {\em initial conditions}. 

\index{adiabatic initial conditions}

Before proceeding to the detailed discussion, 
we recall an intuitive argument
about the nature of a background solution 
and the corresponding 
perturbations, and contrast it with 
the much stronger freeze-out statement that
applies to curvature perturbations. 
Let $Q$ be a field or an ensemble of fields, 
which satisfies the non-linear evolution equation 
\be
 (\partial_t^2 - \nabla^2_\vec{r}) Q 
 + \mathcal{A}(Q) \partial^{ }_t Q 
 + \mathcal{B}(Q) \;=\; 0 
 \;. 
\ee
We write $Q = \bar{Q} + \delta Q$, where $\nabla^{ }_\vec{r}\,\bar Q = 0$.
The background and first-order equations read 
\ba
 \partial_t^2 \bar Q 
 + \mathcal{A}(\bar Q) \partial^{ }_t \bar Q 
 + \mathcal{B}(\bar Q) & = & 0 
 \;, \label{ex_0} \\[2mm]
%%%%
 \bigl[\, \partial_t^2 - \nabla^2_\vec{r} 
 + \mathcal{A}(\bar Q) \partial^{ }_t  
 + \mathcal{A}'(\bar Q) \partial^{ }_t \bar Q  
 + \mathcal{B}'(\bar Q) \,\bigr]\, \delta Q & = & 0 
 \;, \label{ex_1} 
\ea
respectively. On the other hand, if we take a time derivative
of \eq\nr{ex_0}, we find 
\ba
  \bigl[\, \partial_t^2 
 + \mathcal{A}(\bar Q) \partial^{ }_t  
 + \mathcal{A}'(\bar Q) \partial^{ }_t \bar Q  
 + \mathcal{B}'(\bar Q) \,\bigr]\, \partial^{ }_t \bar Q & = & 0 
 \;. \label{ex_2} 
\ea
Comparing \eqs\nr{ex_1} and \nr{ex_2}, we see that the time
derivative of the background 
solution, $\partial^{ }_t \bar Q$, already determines one of 
the possible perturbations, $\delta Q$, specifically a mode  
satisfying $\nabla^2_\vec{r}\, \delta Q = 0$. In 
the cosmological context, 
this corresponds to comoving momenta that have 
exited the Hubble horizon, $k/a \ll H$. 

Now, in the slow-roll regime, 
$\dot{\bar\varphi}$ is almost constant
(cf.\ \se\ref{ss:slow-roll}), 
in the sense
that its relative change within a Hubble time 
($ \Delta t \, \equiv \, H^{-1}_{ } $) is small, 
\be
 \frac{\Delta \dot{\bar\varphi}}{\dot{\bar\varphi}}
 \; \approx \; 
 \frac{\Delta t\, \ddot{\bar\varphi}}{\dot{\bar\varphi}}
 \; \underset{\rmii{\nr{slowroll_ddot}}}
    {\overset{\;\;\Delta t \, \equiv \, H^{-1}_{ }}{=}} \; 
 \; \approx \; 
 \epsilon^{ }_\rmii{$V$} - \eta^{ }_\rmii{$V$}
 \; \ll \; 1
 \;. 
\ee
Then, if we insert $\varphi$ in the role of $Q$ in \eq\nr{ex_1}, 
and consider modes with $k/a \ll H$ in the slow-roll regime, 
we see that $\delta\varphi$ is almost constant. 

However, what actually happens for 
curvature perturbations is much more remarkable: they stay
constant even when $\bar\varphi$ starts to oscillate fast, 
as long as $k/a \ll H$. To see this, we have to inspect
in detail their evolution equation, which in fact does {\em not}  
have exactly the same form as the equation 
satisfied by $\dot{\bar\varphi}$.

The rest of this chapter is organized as follows. 
In \se\ref{ss:outside_curv}, we show that in single-field
inflationary models, 
the curvature perturbation $\R^{ }_\varphi$ freezes out
while it is outside
of the Hubble horizon. 
In \se\ref{ss:outside_isocurv}, we elaborate on how this implies
that isocurvature perturbations vanish outside of the Hubble horizon. 
Despite the prediction of single-field models that isocurvature 
perturbations should vanish, they are broadly discussed in the 
literature. 
Therefore, in \se\ref{ss:outside_other}, we show how 
isocurvature perturbations can be non-vanishing 
in inflationary setups involving multiple scalar fields. 

%%%%%%%%%%%%%%%%%%%%%%%%%%%%%%%%%%%%%%%%%%%%%%%%%%%%%%%%%%%%%%%%%%%%%%%%%
%
\subsection{Freezing out of curvature perturbations}
\label{ss:outside_curv}

\index{freeze out: curvature perturbations}

We start this section on a general note. Given the 
very simple freeze-out result,  
it may be wondered if there is a simple intuitive explanation
for this behaviour.
Indeed physical arguments are presented in the literature,  
but unfortunately they depend on the gauge chosen. 
Therefore standard proofs rely on
a mathematical determination of all independent 
solutions of the underlying 
differential equations~(cf.,\ e.g.,\ refs.~\cite{weinberg,lyth}). 
This is then also the strategy that we adopt in the following.

Let us consider the evolution equation for the
gauge-invariant curvature perturbation~$\R^{ }_\varphi$ 
from \eq\nr{dot_R_varphi}. 
We convert it into an equation for the corresponding
{\em mode function}, $\R^{ }_k$. 
The mode functions are related, but not exactly equivalent, 
to Fourier transforms as defined in \eq\nr{fourier_k}. 
An example of a mode expansion is 
given in \eq\nr{mode_expansion}, with the mode function
denoted by $\field^{ }_k(\tau)$. 
If we take a Fourier transform
of this expression, according to \eq\nr{fourier_k},
we get 
\be
 \field^{ }_{ }(\tau^{ }_{ },\vec{k}) 
 \;
 \underset{\rmii{\nr{fourier_k}}}{
 \overset{\rmii{\nr{mode_expansion}} \lift }{=}}
 \;  
 \sqrt{(2\pi)^3_{ }}
 \Bigl[ \,
    w^{ }_\rmii{\vec{k}}
    \, \field^{ }_k(\tau^{ }_{ }) 
  + 
    w^{\dagger}_\rmii{$-$\vec{k}}
    \, \field^{*}_k(\tau^{ }_{ }) 
 \, \Bigr]
 \;. \label{field_veck}
\ee
The vacuum expectation value, similar to \eq\nr{2pt_field}
but now in momentum space, gives
\be
  \bigl\langle\, 0 \big|\, 
  \field^{ }_{\varphi}(\tau,\vec{k})
  \field^{ }_{\varphi}(\tau,\vec{q})
 \,\big| 0 \,\bigr\rangle
 \; 
 \overset{\rmii{\nr{field_veck}} \lift }{
 \underset{\rmii{\nr{commutators}}}{=}} 
 \; 
 (2\pi)^3_{ }\delta^{(3)}_{ }
 (\vec k + \vec q) \, 
 | \field^{ }_k(\tau^{ }_{ }) |^2_{ }
 \;. \label{2pt_field_kq}
\ee 
This can be compared with the corresponding 
statistical average, from \eq\nr{eq_PQdef}. We see that,
if we use mode functions, the momentum conservation constraint
is factored out, so that the mode functions give 
directly the power spectra, 
\be
 P^{ }_{\mbit \scriptscriptstyle\field}(\vec{k})
 \;
 \underset{\rmii{\nr{2pt_field_kq}}}
 {\overset{\rmii{\nr{eq_PQdef}} \lift }{=}}
 \; 
 | \field^{ }_k(\tau^{ }_{ }) |^2_{ }
 \;, \quad
%%%%%%
 \P^{ }_{\mbit \scriptscriptstyle\field}(k)
 \;
 \overset{\rmii{\nr{eq_angular_average}}}{=}
 \; 
 \frac{k^3_{ }}{2\pi^2_{ }}
 | \field^{ }_k(\tau^{ }_{ }) |^2_{ }
 \;. \label{P_Q_k_again}
\ee
We remark that  
the latter relation appears also in \eq\nr{P_Q_k}, 
and subsequently in \eq\nr{rescale}, 
where we make use of it to define a rescaled mode function, 
$[\, \R^{ }_k \,]^{ }_\rmi{rescaled}$.

Returning now to \eq\nr{dot_R_varphi}, on the left-hand side
it makes no difference whether we carry out a Fourier transform
or go to mode functions, as in both cases we just need to replace
$\nabla^2_{ } \to - k^2_{ }$. However, we have to be careful with 
the inhomogeneous terms on the right-hand side
(originating from $\varrho$ and the noise part of $\Pi$). 
Fourier transforming
the noise correlator from \eq\nr{varphi_noise}, we find
\ba
 \bigl\langle\,
 \varrho(t_1^{ },\vec{k}) \,
 \varrho(t_2^{ },\vec{q}) \,
 \,\bigr\rangle 
 &
 \underset{\rmii{\nr{fourier_k}}}{
 \overset{\rmii{\nr{varphi_noise}} \lift }{=}}
 & 
 \int\! {\rm d}^3_{ }\vec{x} 
 \int\! {\rm d}^3_{ }\vec{y}
 \,  
 \overbrace{
 \bigl\langle\,
 \varrho(t_1^{ },\vec{x}) \,
 \varrho(t_2^{ },\vec{y}) \,
 \,\bigr\rangle
 }^{
 \Omega\,\delta(t^{ }_1 - t^{ }_2) \delta^{(3)}_{ }(\vec{x-y})
 / a^3_{ }(t^{ }_1) 
 } 
 \, 
 e^{-i(\vec k \cdot \vec x + \vec q \cdot \vec y) }_{ }
 \nn[2mm] 
%%%%%
 & = &
 \frac{ \Omega\,\delta(t^{ }_1 - t^{ }_2) }
      { a^3_{ }(t^{ }_1) }
 \, 
 \int\! {\rm d}^3_{ }\vec{x} \,
 e^{-i(\vec k + \vec q) \cdot \vec x }_{ }
 \nn[2mm]
%%%%%
 & = & 
 (2\pi)^3_{ }\delta^{(3)}_{ }
 (\vec k + \vec q) \,
 \underbrace{
 \frac{ \Omega\,\delta(t^{ }_1 - t^{ }_2) }
      { a^3_{ }(t^{ }_1) }
 }_{
 \equiv\, 
 \langle\, 
  \varrho^{ }_k(t^{ }_1)\, \varrho^{ }_k(t^{ }_2)
 \,\rangle 
 }
 \;, \label{noise_mode}
\ea
where we inserted
$
 \sqrt{-\bar{g}^{\hphantom{|}} } = a^3_{ }
$
for the determinant of the background metric in physical time. 
The definition of the {\em noise mode function}, $\varrho^{ }_k$, 
in \eq\nr{noise_mode}, draws on an analogy with \eq\nr{2pt_field_kq}, 
i.e.\ on factoring out the momentum conservation constraint. 

With these ingredients, we convert \eq\nr{dot_R_varphi} 
into an equation for the curvature and noise mode functions,
$\R^{ }_k$ and $\varrho^{ }_k$. In order to keep the discussion
simple, we do this by first {\em assuming} that the 
isocurvature perturbations vanish, 
$\E^{ }_v, \E^{ }_\T \to 0$. 
This assumption will be justified {\em a posteriori}, 
in \se\ref{ss:outside_isocurv}. In the said limit, we get
\ba
 \biggl\{\,
 \partial_t^2
 + 
    \bigl(\, \Upsilon+ 2 \mathcal{F}  + 3 H \,\bigr) 
    \partial^{ }_t 
 +  \frac{k^2_{ }}{a^2_{ }} 
 \,\biggr\}\, \R^{ }_k 
 \,-\, 
 \frac{8\pi G}{3}k^2_{ } \Pi^{ }
 &
   \underset{\scriptscriptstyle \E^{ }_{v},\,\E^{ }_\rmiii{\it T}\,\to\,0}
  {\overset{\rmii{\nr{dot_R_varphi}}}{\approx}} 
 &  
 \,-\, \frac{\varrho^{ }_k H}{\dot{\bar\varphi}}
 \;, \hspace*{5mm} 
 \label{dot_R_varphi_again}
 \index{mode function}
\ea
where $\mathcal{F}$ is defined in \eq\nr{cal_F}, 
and $\Pi$ is the mode function corresponding
to the anisotropic stress in \eq\nr{Pi_gi}. 
We put no subscript in $\Pi$ as it is always preceded by $k^2_{ }$, 
so that the context should be clear. 
Our goal is to determine  
the solution of \eq\nr{dot_R_varphi_again}
after $k/a$ goes outside of the 
Hubble horizon, $k/a \ll H$ (cf.\ \fig\ref{fig:bg_num}
on p.~\pageref{fig:bg_num} for an illustration).
We recall from \eq\nr{decomposition}, and the discussion
a few lines below \eq\nr{demo_1}
(or from \eq\nr{Pi_gi}), that 
\be
 \Pi
 \;
 \overset{\rmii{\nr{Pi_splitup}}}{\supset}
 \; 
 \Sigma
 \;
 \overset{\rmii{\nr{decomposition}}}{=}
 \;
 \frac{2 \eta ( v - \vartheta' )}{a} 
 \; 
 \overset{\rmii{\nr{def_psi}} \lift }{
 \underset{\rmii{\nr{def_R_v}}}{=}}
 \; 
 - \frac{2 \eta ( \psi + \R^{ }_v )}{a^2_{ }H}
 \;, \label{Pi_again}
\ee
so that the momentum appears as $k^2_{ }/a^2_{ }$ also in front of
viscous corrections. 

\vspace*{3mm}

Given that \eq\nr{dot_R_varphi_again} is a linear differential
equation, 
its solution can be represented as a sum of the general 
homogeneous solution and a special solution of the 
inhomogeneous equation. Adopting the notation from \eq\nr{full_soln}, 
we denote this splitup as  
\be
 \R^{ }_k \;=\; 
 \R^\vac_k
 + 
 \R^\rmi{cl}_k
 \;, \label{full_soln_R}
\ee
where the ``vac'' part refers to the homogeneous equation. 
Its integration constants are fixed through
the initial conditions from 
\eqs\nr{R_ini_1} and \nr{R_ini_2}. 

If we set $k/a\to 0$, 
the homogeneous equation takes the form 
\be
 \Bigl\{\,
 \partial_t^{ }
 + 
    \bigl(\, \Upsilon + 2 \mathcal{F} + 3 H \,\bigr) 
 \,\Bigr\} 
 \, \dot{\R}^\vac_k
 \; \underset{\rmii{\nr{dot_R_varphi_again},\nr{Pi_again}}}
   {\overset{\scriptscriptstyle k\;\ll\;aH
    \lift }{ \approx }} \; 
 0
 \;. \label{dot_R_qm}
\ee
There is a trivial solution,  
\be
 \dot{\R}^\vac_k(t) \; = \; 0
 \;, \label{R_qm_1}
\ee
and a non-trivial one, 
\be 
 \dot{\R}^\vac_k(t) 
 \; \approx \; 
 \dot{\R}^\vac_k(t^{ }_*)
 \, \exp\Bigl\{\,
 - \int_{t^{ }_*}^t \! {\rm d} t'\,  
  \bigl(\, \Upsilon + 2 \mathcal{F} + 3 H \,\bigr) 
 \,\Bigr\}
 \;, \label{R_qm_2}
\ee
where $t^{ }_* = t^{ }_*(k)$ 
is the moment at which a mode exits the horizon, 
i.e.\ 
$
 k \equiv a(t^{ }_*) H(t^{ }_*)
$. 
To analyze \eq\nr{R_qm_2}, we specialize
to the slow-roll regime, even if a full 
numerical solution of \eq\nr{dot_R_varphi_again} shows that
the result applies more generally. 
In the slow-roll regime, as is shown below \eq\nr{slowroll_ddot}, 
$
 \mathcal{F} \approx 
 ( 2 \epsilon^{ }_\rmii{$V$} - \eta^{ }_\rmii{$V$}) H \ll H
$. 
Therefore 
\be
 \Upsilon + 2 \mathcal{F} + 3 H  
 \;\approx\; 
 \Upsilon + 3 H 
 \;\overset{\rmii{\nr{Ups_ineq}}}{\ge}\; 3 H 
 \;>\; 0
 \;. 
\ee
This implies that \eq\nr{R_qm_2} decreases exponentially, 
if $H$ is approximately constant
and $(t-t^{ }_*)\, H \gg 1$. Therefore, the {\em only} solution
outside of the Hubble horizon reads
\be
 \R^\vac_k
 \; \underset{\rmii{\nr{R_qm_1},\nr{R_qm_2}}}
   {\overset{\scriptscriptstyle k\;\ll\;aH
            \lift }{\approx}} \; 
 \mbox{constant}
 \;. \label{soln_R_const}
\ee

Let us then turn to the special solution of the inhomogeneous equation, 
i.e.\ the function $\R^\rmi{cl}_k$ from \eq\nr{full_soln_R}. 
As the average of the noise 
vanishes, the average of $\R^\rmi{cl}_k$ also vanishes. Therefore, the
noise term cannot change the freeze-out dynamics in a qualitative way. 

We can say a bit more 
by estimating the size of the mean-squared
noise kicks. After the rescaling in \eq\nr{rescale},
which leads to the initial conditions in 
\eqs\nr{R_ini_1} and \nr{R_ini_2}, 
the noise appearing in \eq\nr{dot_R_varphi_again} 
has the {\em autocorrelator} 
\be
 \frac{H(t^{ }_1)H(t^{ }_2)}
 {\dot{\bar\varphi}(t^{ }_1) \dot{\bar\varphi}(t^{ }_2)}
 \bigl\langle\, 
  \bigl[\, \varrho^{ }_k(t^{ }_1) \,
        \, \varrho^{ }_k(t^{ }_2) \,\bigr]^{ }_\rmi{rescaled} 
 \,\bigr\rangle 
 \;
    \underset{\rmii{\nr{fl-di},\nr{noise_mode}}}
   {\overset{\rmii{\nr{rescale}}  \lift  }{=}}
 \; 
 \delta(t^{ }_1 - t^{ }_2) \,
 \biggl[
 2 T \Upsilon \,
 \frac{H^2_{ }}{\dot{\bar\varphi}^2_{ }} \,
 \frac{ (k/a)^3_{ } }
      { 2\pi^2_{ } }
 \, \biggr]_{t\;=\;t^{ }_1}
 \;, 
 \label{noise_R}
 \index{noise autocorrelator}
\ee
where we inserted
$\Omega = 2 T \Upsilon$ from 
\eq\nr{fl-di}. 
The coefficient 
$ {H}/ {\dot{\bar\varphi}}$ 
is slowly varying, as long as we are in the slow-roll regime. 
If we are in the oscillatory regime, 
so that $\dot{\bar\varphi}$ has zeros, 
the autocorrelator in \eq\nr{noise_R} looks singular, 
and needs to be regularized; however, after doing this with the help of 
complexified variables, the average value of the noise autocorrelator 
is small~\cite[eq.~(4.41)]{alica_copy}.
As discussed around \eq\nr{stationary}, the plasma temperature, $T$,
could also be slowly varying, to the extent that we can define it. 
Instead, the momentum dependence redshifts 
away rapidly in \eq\nr{noise_R}, as $k^3_{ }/a^3_{ }$.
We therefore conclude that the thermal noise 
has a fast decreasing amplitude outside of the Hubble horizon, 
so that it adds only very 
small ripples on top of $\R^\rmi{vac}_k$ 
(cf.\ \eq\nr{soln_R_const}). 

To summarize, \eq\nr{soln_R_const} represents actually
the general solution outside of the Hubble horizon, 
showing that curvature perturbations indeed freeze out.

%%%%%%%%%%%%%%%%%%%%%%%%%%%%%%%%%%%%%%%%%%%%%%%%%%%%%%%%%%%
%
\subsection{Vanishing of isocurvature perturbations}
\label{ss:outside_isocurv}

\index{isocurvature perturbations}

For the argument in \se\ref{ss:outside_curv}, 
we assumed the isocurvature
perturbations, $\E^{ }_v$ and $\E^{ }_\T$, to vanish. 
This assumption is justified rather succinctly
in the literature, again from 
the solutions of gauge-fixed evolution equations, rather than from 
an intuitively clear physical phenomenon. In the following, 
even if we do not succeed in offerring the latter, 
we provide a gauge-invariant demonstration for why the
isocurvature perturbations vanish at the same time as 
$\R^{ }_\varphi$ freezes out. The implicit assumption 
we make is that a non-vanishing background temperature exists, 
so that $\E^{ }_v$ and $\E^{ }_\T$, defined in 
\eqs\nr{def_E_v} and \nr{def_E_T}, are in principle non-trivial
variables. However, the result is more general, 
and in \se\ref{ss:outside_other} we demonstrate it 
for a system involving two scalar fields 
but no temperature at all.

Let us return back to \eq\nr{dot_R_varphi}, but keep now 
the isocurvature perturbations in place. 
We again go to mode functions, 
like in \eq\nr{dot_R_varphi_again}, 
and outside of the Hubble horizon, setting $k/a \ll H$. 
In order to avoid double subscripts, we add no index $k$ 
to $\E^{ }_v$ and $\E^{ }_\T$.
Furthermore, 
we complement \eq\nr{dot_R_varphi} 
with the two other evolution equations, 
from \eqs\nr{dot_E_v} and \nr{dot_E_T}, inspected under 
the same assumptions. We note that the anisotropic stress contribution
is proportional to $k^2_{ }/a^2_{ }$, 
cf.\ \eqs\nr{dot_R_varphi_again} and \nr{Pi_again}, 
so it drops out. Consequently, 
the Bardeen potentials, $\phi$ and $\psi$,
decouple from the dynamics, and we do not need to include them.

Now, we have demonstrated in \se\ref{ss:outside_curv} that, 
outside the Hubble horizon, when $k/a \ll H$,
$\R^{ }_k$ is constant. Then 
suppose that we search for
a {\em stationary solution} also for $\E^{ }_\T$ and $\E^{ }_v$,
setting all time derivatives to zero. 
Then the coupled set of our three equations becomes
\ba
% \nn[2mm]
%%%%
% \,+\, 
   \biggl\{\,
   \frac{4\pi G}{H}
 \biggl( 1 - \frac{\bar{p}^{ }_{\der\T}}{\bar{e}^{ }_{\der\T}}\biggr)
   + \frac{
     \Upsilon^{ }_{\der\T} \, \dot{\bar\varphi} 
   + 
     V^{ }_{\der\varphi\T} 
     }
     { \bar{e}^{ }_{\der\T} \, \dot{\bar\varphi} }
   \,\biggr\} 
   \, \E^{ }_{\T} 
% & & 
% \nn[2mm]
%%%%
 \,-\,   
   \biggl\{\,
       \frac{4\pi G ( \Upsilon  +  2 \mathcal{F} )}{H}
   \,\biggr\}
   \,\E^{ }_{v} 
% \,+\, 
% \frac{8\pi G}{3}\nabla^2_{ } \Pi
 & 
  \underset{\partial^{ }_t,\frac{k}{a}\,\to\, 0}
  {\overset{\rmii{\nr{dot_R_varphi}} \lift }{=}} 
 &   
 \!-\, \frac{\varrho^{ }_k H}{\dot{\bar\varphi}}
 \;, \hspace*{5mm} \label{dot_R_varphi_stat}
 \\[3mm] 
%%%%%%%%%%%%%%%%%%%%%%%%%%%%%%%%%%%%%%%%%%%%%%%%%%%
% \Bigl\{\,
%    \bigl(\, \bar e + \bar p \,\bigr) \partial^{ }_t 
% \,\Bigr\} \R^{ }_\varphi 
% \,+\,   
   \biggl\{\,
    %  \partial^{ }_t +
   3 H +
       \frac{4\pi G \dot{\bar\varphi}^2_{ }}{H}
   \,\biggr\}
   \,\E^{ }_{v} 
% & &  
% \nn[2mm]
%%%%
 \,-\, 
   \biggl\{\,
   \frac{\bar{p}^{ }_{\der\T}}{\bar{e}^{ }_{\der\T}}
   \,\biggr\} 
   \, \E^{ }_{\T} 
% \,+\, 
% \frac{2 H}{3}\nabla^2_{ } \Pi
 & 
  \underset{\partial^{ }_t,\frac{k}{a}\,\to\, 0}
  {\overset{\rmii{\nr{dot_E_v}} \lift }{=}}  
 &  
 0 
 \;, \hspace*{5mm} \label{dot_E_v_stat}
 \\[3mm]
%%%%%%%%%%%%%%%%%%%%%%%%%%%%%%%%%%%%%%%%%%%%%%%%%%%%%%%
   \biggl\{\,
  % \partial^{ }_t 
  % +
  3 H 
 \biggl( 1 + \frac{\bar{p}^{ }_{\der\T}}{\bar{e}^{ }_{\der\T}}\biggr)
  + \frac{4\pi G(\bar e + \bar p) - \dot{H}}{H}
  - \frac{ \dot{\bar\varphi} \, (
     \Upsilon^{ }_{\der\T} \, \dot{\bar\varphi} 
   + 
     V^{ }_{\der\varphi\T} ) 
     }
     { \bar{e}^{ }_{\der\T} }
   \,\biggr\} 
   \, \E^{ }_{\T} 
 & &  
 \nn[3mm]
%%%%
 \;+\; \frac{4\pi G}{H}\,
 \biggl\{\,
 \Upsilon \dot{\bar\varphi}^2_{ }
 + \frac{8\pi G(\bar e + \bar p)\,\bar e}{H}
 \,\biggr\} 
% \biggl\{\, 
   % dot{\R}^{ }_\varphi
 \,\E^{ }_{v}
% \,\biggr\}
% +  \frac{\nabla^2_{ }}{a^2_{ }} 
% \Bigl\{\, 
%   (\bar e + \bar p) \R^{ }_\varphi + \E^{ }_v
% \,\Bigr\}
 &
  \underset{\partial^{ }_t,\frac{k}{a}\,\to\, 0}
  {\overset{\rmii{\nr{dot_E_T}} \lift }{=}} 
 &  
 \varrho^{ }_k \bit \dot{\bar\varphi} \bit H 
 \;. 
 \hspace*{12mm}
 \label{dot_E_T_stat}
\ea
This is a system of three linear equations for two unknowns, 
$\E^{ }_{v}$ and $\E^{ }_{\T}$. The coefficients are, 
in general, non-degenerate. 
The average of the noise is zero, and its mean-squared
amplitude is small outside of the Hubble horizon, as 
discussed below \eq\nr{noise_R}. Omitting the right-hand 
side, the only solution is zero, 
$\E^{ }_{v} = \E^{ }_{\T} = 0$, because the system is 
overconstrained. This completes our argument that 
isocurvature perturbations vanish outside of the Hubble horizon. 
The vanishing of 
$\E^{ }_v$ and $\E^{ }_\T$ 
in turn implies that $\R^{ }_\varphi$, $\R^{ }_v$ and $\R^{ }_\T$
freeze out to the same value
(cf.\ \eqs\nr{def_E_v} and \nr{def_E_T}).

%%%%%%%%%%%%%%%%%%%%%%%%%%%%%%%%%%%%%%%%%%%%%%%%%%%%%%%%%%%%%%%%%%%%%%%%%
%
\subsection{Multi-field inflation, spectator fields, and
dark matter isocurvature}
\label{ss:outside_other}

\index{single-field inflation}
\index{multi-field inflation}
\index{spectator fields}
\index{dark matter isocurvature}
\index{isocurvature perturbations}
\index{curvaton scenario}

The arguments presented in \ses\ref{ss:outside_curv}
and \ref{ss:outside_isocurv} are specific to 
{\em single-field inflation}. The case that multiple scalar
fields participate in inflationary dynamics was 
considered early on (cf.,\ e.g.,\ refs.~\cite{kl,sm,tw}), 
and the situation was noticed to change. 
This type of models 
gained prominence particularly in the context of the
{\em curvaton scenario}~\cite{curv1,curv2,curv3}, 
in which the late-time
adiabatic perturbations are generated not by the quantum
fluctuations of the inflaton, 
but by the fluctuations of another scalar field,
which is subdominant at the time of inflation. The latter
field is conjectured to have a very small equilibration rate, 
$
 \Upsilon^{ }_\rmi{curvaton} \ll \Upsilon^{ }_\rmi{inflaton}
$. 
Therefore it comes to dominate the energy density at a late
time, when
$
 \Upsilon^{ }_\rmi{curvaton} \ll H \ll \Upsilon^{ }_\rmi{inflaton}
$
(cf.\ \fig\ref{fig:timeline} on p.~\pageref{fig:timeline}),
and can generate cosmological perturbations then. 
In this setup, one of the scalar fields always acts as 
a {\em spectator}, but the roles change as time goes by.

From the point of view of model building, it may be particularly
interesting if the curvaton field represents, or decays into, 
{\em dark matter}. In such a situation, it might also be natural
to consider simultaneously multiple
fields {\em and} fluids (cf.,\ e.g.,\ ref.~\cite{noh}). 
In the present section, we restrict ourselves to the 
simplest non-standard case, 
illustrating how the evolution of
curvature and isocurvature perturbations changes if we introduce
a second scalar field, denoted by $\chi$.

We thus repeat the 
analysis from \se\ref{ss:sasaki}, 
but now looking for evolution equations
for two (minimally coupled) scalar fields. 
The equations of motion, from \eq\nr{varphi_eq_0}, become
\ba
 {\varphi^{;\mu}_{ }}^{ }_{;\mu} 
  - V^{ }_{\der\varphi}  
 \;
 \overset{\rmii{\nr{varphi_eq_0}}}{=}
 \; 0
 \;, 
 && 
 {\chi^{;\mu}_{ }}^{ }_{;\mu} 
  - V^{ }_{\der\chi}  
 \; = \; 0
 \;, \label{chi_eq_0} 
\ea
where $V = V(\varphi,\chi)$ yields an inflationary
solution for $\varphi$ or $\chi$.
Both fields are expanded around their background values, 
$
 \varphi = \bar\varphi + \delta\varphi
$
and
$
 \chi = \bar\chi + \delta\chi
$.
The background equations then take the form in \eq\nr{bg_scalar}, 
\ba
   \bar\varphi\bit''
 + 2 \H \bar\varphi\bit' + a^2 V^{ }_{\der\varphi}
 \;
 \overset{\rmii{\nr{bg_scalar}}}{=}
 \;
 0
 \;,  
 && \bar\chi\bit''
 + 2 \H \bar\chi\bit' + a^2 V^{ }_{\der\chi}
 \;
 =
 \;
 0 
 \;. \label{bg_chi} 
\ea
Rescaling the fields like before, 
$
 \delta\widehat\varphi \, \equiv \, a\ibit \delta\varphi
$
and
$
 \delta\widehat\chi \, \equiv \, a\ibit \delta\chi
$, 
the first-order perturbations satisfy a generalization of 
\eq\nr{rescaled_pert_scalar},
\ba
 && \hspace*{-1.5cm}
 \delta\widehat\varphi{\bit}''
 - \nabla^2 \delta\widehat\varphi 
 + \biggl( a^2_{ } V^{ }_{\der\varphi\varphi}
 - \frac{a''}{a}\biggr) 
  \, \delta\widehat\varphi
 + a^2_{ } V^{ }_{\der\varphi\chi}
  \, \delta\widehat\chi
 \nn[2mm]
%%%%
 && \hspace*{1.5cm}
 \;
 \overset{\rmii{\nr{rescaled_pert_scalar}}}{=}
 \;   
 \bigl( h_0' + 3 h_\rmiii{D}' + \nabla^2 h \bigr)
 a \bar\varphi\hspace*{0.3mm}{}' 
 -2 h^{ }_0\, a^3 V^{ }_{\der\varphi} 
 \;. 
 \label{rescaled_pert_chi}
\ea
The other equation is obtained with the exchange
$\varphi\leftrightarrow \chi$.

On the side of the Einstein equations, the source terms for the 
metric and its perturbations change, because both scalar fields 
appear in the energy-momentum tensor. At the background level, 
\eq\nr{bg_einstein} gets replaced with 
\be
 \H^2_{ }
 \;
  \overset{\rmii{\nr{bg_einstein}}}{=}
 \;
 \frac{4\pi G}{3}
 \bigl[ (\bar\varphi\hspace*{0.3mm}')^2_{ }
      + (\bar\chi\hspace*{0.3mm}')^2_{ }  + 2 a^2_{ } V
 \bigr]
 \;, \quad 
 \H'
 \;
  \overset{\rmii{\nr{bg_einstein}}}{=}
 \;
 \frac{8\pi G}{3}
 \bigl[ - (\bar\varphi\hspace*{0.3mm}')^2_{ }
        - (\bar\chi\hspace*{0.3mm}')^2_{ }
   + a^2_{ } V
 \bigr]
 \label{bg_einstein_chi} 
 \;.
\ee
We also need linear combinations
of these relations, % in \eq\nr{bg_einstein_chi}, 
as well as $\H''$ (cf.\ \eq\nr{bg_einstein_2}), finding
\ba
 \H^2 - \H'
 & \overset{\rmii{\nr{bg_einstein_chi}}}{=} & 
 4 \pi G 
 \bigl[ (\bar\varphi\hspace*{0.3mm}')^2_{ }
      + (\bar\chi\hspace*{0.3mm}')^2_{ }  
 \bigr] 
 \;, \quad
 2 \H^2 + \H' 
 \; \overset{\rmii{\nr{bg_einstein_chi}}}{=} \; 
 8 \pi G a^2_{ } V
 \hspace*{8mm}
 \label{bg_einstein_chi_3}
 \\[2mm]
%%%%%%
 \Rightarrow \quad 
 2\H\H' - \H'' 
 & \overset{\rmii{\nr{bg_einstein_chi_3}}}{=} & 
 8\pi G 
 \bigl(\, 
   \bar\varphi\hspace*{0.3mm}' \bar\varphi\hspace*{0.3mm}'' 
 + 
   \bar\chi\hspace*{0.3mm}' \bar\chi\hspace*{0.3mm}'' 
 \,\bigr)
 \;. \label{bg_einstein_chi_2}
\ea

At the first order in perturbations, we employ  
the shorthand notation 
$
 Y \equiv 
 h^{ }_\rmii{D} + {\nabla^2\vartheta} / {3}
$
from \eq\nr{shorthand}.
Then \eqs\nr{h0_repl} and \nr{hD_repl} 
get replaced with 
\ba
  h^{ }_0
 & \overset{\rmii{\nr{h0_repl}}}{=} & 
 - \frac{  Y' }{\H}
 + \frac{ 4 \pi G  }{a \H}\, 
 \bigl(\, 
  \bar\varphi\hspace*{0.3mm}' \delta\widehat\varphi
 + 
  \bar\chi\hspace*{0.3mm}' \delta\widehat\chi
 \,\bigr)
 \;, 
 \label{h0_repl_chi}
 \\[2mm]
%%%%%
  h_0' + 3 h_\rmii{D}' + \nabla^2 h 
 & \overset{\rmii{\nr{hD_repl}}}{=} &  
 - 2 \biggl( 2 \H + \frac{\H'}{\H} \biggr) h^{ }_0
 - \frac{ \bigl( \partial^{2}_\tau 
    + 2  \H \partial^{ }_\tau
  - \nabla^2 \bigr) Y }{\H}
 \nn[2mm]
 & & \; 
 - \, \frac{ 8 \pi G  
   a }{\H}
  \bigl(\,
   V^{ }_{\der\varphi}
   \, \delta\widehat\varphi
  + 
   V^{ }_{\der\chi}
   \, \delta\widehat\chi
  \,\bigr)
 \;. 
 \label{hD_repl_chi}
\ea
The field perturbations are reparametrized 
like in \eq\nr{delta_widehat_varphi},
\be
 \delta\widehat\varphi 
 \;
 \overset{\rmii{\nr{delta_widehat_varphi}}}{=}
 \; \field^{ }_\varphi
 - 
 \frac{a \bar\varphi\hspace*{0.3mm}'}{\H}
  \, Y
 \;, \quad
  \delta\widehat\chi \;=\; \field^{ }_\chi
 - 
 \frac{a \bar\chi\hspace*{0.3mm}'}{\H}
  \, Y
 \;. \label{delta_widehat_chi}
\ee

Inserting \eqs\nr{h0_repl_chi}--\nr{delta_widehat_chi} 
into \eq\nr{rescaled_pert_chi}, the first task is once again 
to verify the cancellation of $Y$. This goes like in 
\eqs\nr{ms_Y_lhs}--\nr{subtr_Y}.
The coefficient of $Y'$ can be seen to cancel with the help
of the background identity from \eq\nr{mg_1}, which remains
unchanged, 
\be
   \biggl(\frac{a \bar\varphi\hspace*{0.3mm}{}'}{\H}\biggr)'
 \; 
 \underset{\rmii{\nr{mg_1}}}
 {\overset{\rmii{\nr{bg_chi}} \lift }{=}} 
 \;
   a \bar\varphi\hspace*{0.3mm}{}'
   \biggl( - 1 - \frac{\H'}{\H^2_{ }} \biggr)
   - \frac{a^3_{ }V^{ }_{\der\varphi}}{\H}
 \;. \label{mg_1_chi}
\ee
To verify the cancellation of the coefficient of $Y$, 
we need to take a second derivative from \eq\nr{mg_1_chi}.
Proceeding like in \eqs\nr{dd_hatm_1}--\nr{mgc_2},
we find
\ba
   \biggl(\frac{a \bar\varphi\hspace*{0.3mm}{}'}{\H}\biggr)''
 & \underset{\rmii{\nr{mgc_2}}}
   {\overset{\rmii{\nr{mg_1_chi}}  \lift }{=}}
  & 
  \frac{ a \bar\varphi\hspace*{0.3mm}' }{\H}
  \biggl\{\,
  \;
   \H^2 + \H' -
   \biggl( 
    2 + \frac{\H'}{\H^2_{ }} 
   \biggr)
    8 \pi G 
    \bigl[\,
    (\bar\varphi\hspace*{0.3mm}')^2_{ }
    + 
    (\bar\chi\hspace*{0.3mm}')^2_{ }
    \,\bigr]
 \nn[2mm]
%%%%%%
 && \hspace*{1.0cm}
 \; - \,
 \frac{16 \pi G a^2_{ }
   V^{ }_{\der\varphi}\bar\varphi\hspace*{0.3mm}'}{\H}
  -  a^2_{ }V^{ }_{\der\varphi\varphi}
   \,\biggr\}
 \nn[2mm]
%%%%%
 &  & \; - \, 
  \frac{ a \bar\chi\hspace*{0.3mm}' }{\H}
  \biggl\{\,
    \frac{8 \pi G a^2_{ }
    (\,
     V^{ }_{\der\varphi}\bar\chi\hspace*{0.3mm}'
    + 
     V^{ }_{\der\chi}\bar\varphi\hspace*{0.3mm}'
    \,)
    }{\H}
  +  a^2_{ }V^{ }_{\der\varphi\chi}
   \,\biggr\}
 \;. 
 \label{mgc_2_chi}
\ea
When this is inserted, the coefficient of $Y$ cancels exactly.

\vspace*{3mm}

Subsequently, we insert the physical (i.e.\ non-$Y$) terms
from \eqs\nr{h0_repl_chi}--\nr{delta_widehat_chi}
into \eq\nr{rescaled_pert_chi}, generalizing
on \eqs\nr{ms_1}--\nr{final_vacuum}. The expressions can be 
simplified significantly by making use of \eq\nr{mgc_2_chi}, 
and the corresponding relation with 
$\varphi\leftrightarrow\chi$. In particular, we are led to 
define effective masses like in \eq\nr{final_vacuum_2},
\be
 \widehat{m}^2_{\varphi}(\tau) 
 \;
 \overset{\rmii{\nr{final_vacuum_2}}}{\equiv}
 \;
 - \frac{\H}{a\bar\varphi\bit'}
   \biggl( \frac{a\bar\varphi\bit'}{\H} \biggr)''
 \;, \quad
 \widehat{m}^2_{\chi}(\tau) 
 \; \equiv \;
 - \frac{\H}{a\bar\chi\bit'}
   \biggl( \frac{a\bar\chi\bit'}{\H} \biggr)''
 \;. 
 \label{final_vacuum_2_chi}
\ee 
Then \eq\nr{final_vacuum} gets generalized into 
\ba
 0 & = & 
 \Bigl[\,
 \partial_\tau^2 - \nabla^2 
 + \widehat{m}^2_{\varphi}(\tau) 
 \,\Bigr]
 \field^{ }_\varphi
 \label{final_vacuum_chi} \\[2mm]
%%%%%
 &  & \;+\,
 \biggl[\, 
     a^2_{ }V^{ }_{\der\varphi\chi}
  + 
    \hspace*{-6mm}
    \underbrace{
    \biggl( 
    2 + \frac{\H'}{\H^2_{ }} 
   \biggr)}_{
    {\rm from}\;\nr{bg_einstein_chi_3}:\,
    {8\pi G a^2_{ }V } / { \H^2_{ }}
    }
    \hspace*{-6mm}
    8 \pi G 
     \bar\varphi\hspace*{0.3mm}'
     \bar\chi\hspace*{0.3mm}'
  +  
    \frac{8 \pi G a^2_{ }
    (\,
     V^{ }_{\der\varphi}\bar\chi\hspace*{0.3mm}'
    + 
     V^{ }_{\der\chi}\bar\varphi\hspace*{0.3mm}'
    \,)
    }{\H}
 \,\biggr]
 \biggl(\,
  \field^{ }_\chi - 
  \frac{\bar\chi\bit'}{\bar\varphi\bit'}\, 
  \field^{ }_\varphi 
 \,\biggr) 
 \;. \nonumber
\ea
There is a corresponding equation for $\field^{ }_\chi$, 
obtained with $\varphi\leftrightarrow\chi$.

The final step is to go over to curvature perturbations,
like in \eqs\nr{aQ_vs_Q} and \nr{Q_vs_R},  
\be
  \field^{ }_\varphi
  \;
  \underset{\rmii{\nr{Q_vs_R}}}{
  \overset{\rmii{\nr{aQ_vs_Q}} \lift }{=}}
  \;
 -\frac{a \bar\varphi\hspace*{0.3mm}'}{\H}
  \R^{ }_\varphi
  \;, \quad
  \field^{ }_\chi
  \; = \;
 -\frac{a \bar\chi\hspace*{0.3mm}'}{\H}
  \R^{ }_\chi
  \;. 
  \label{Qk_vs_Rk_chi} 
\ee
Time derivatives transform in this substitution 
as indicated in \eq\nr{Q_k_dd}.
Subsequently, we multiply \eq\nr{final_vacuum_chi} with 
$
 -{\H}/({a \bar\varphi\hspace*{0.3mm}'})
$, yielding 
a generalization of \eq\nr{R_k_1}. 
Finally we transform to physical time, according to 
\eqs\nr{time_derivs} and \nr{R_k_dd},
and multiply the equation with $1/a^2_{ }$. 
Defining
\be
 \F^{ }_\varphi 
 \;
   \overset{\rmii{\nr{cal_F}}}{\equiv} 
 \; 
  \frac{\ddot{\bar\varphi}}{\dot{\bar\varphi}} 
 - \frac{\dot H}{H}  
 \;, \quad
 \F^{ }_\chi 
 \;\;
   {\equiv} 
 \;\; 
  \frac{\ddot{\bar\chi}}{\dot{\bar\chi}} 
 - \frac{\dot H}{H}  
 \;, \label{cal_F_chi}
\ee
and introducing a mixed combination of the potential
and its derivatives, 
\be
 \mathcal{V}^{ }_{\varphi\chi}
 \; \equiv \; 
     V^{ }_{\der\varphi\chi}
  +  
    \frac{8 \pi G 
    }{H}
    \bigl(\,
     V^{ }_{\der\varphi} \ibit \dot{\bar\chi}
    + 
     V^{ }_{\der\chi} \ibit \dot{\bar\varphi}
    \,\bigr)
  + 
    \biggl( \frac{8\pi G}{H} \biggr)^2_{ }
    V 
     \dot{\bar\varphi} \ibit
     \dot{\bar\chi}
 \;, \label{V_mixed}
\ee
yields the pair of equations
\begin{empheq}[box=\fbox]{align}
 \quad  \vphantom{\Bigg|^b_q} 
 \biggl[\,
 \partial_t^2
 + 
 \bigl(\, 2 \F^{ }_{\varphi} + 3 H  \,\bigr) \,\partial^{ }_t
 - \frac{\nabla^2_{ }}{a^2_{ }} 
 \,\biggr]
 \, \R^{ }_\varphi
 + 
 \mathcal{V}^{ }_{\varphi\chi}
 \, 
 \frac{\dot{\bar\chi}}{\dot{\bar\varphi}}
 \, 
 \bigl(\, 
 \R^{ }_\chi - \R^{ }_\varphi
 \,\bigr)
 & \; = \; 0 
 \;,
 \label{dot_R_chi_1} \\[3mm]
%%%%%%%%%%%%%%%%%%%%%%%%%%%%%%%%%%%%%%%%%%%%%%%%%%
 \biggl[\,
 \partial_t^2
 + 
 \bigl(\, 2 \F^{ }_{\chi} + 3 H  \,\bigr) \,\partial^{ }_t
 - \frac{\nabla^2_{ }}{a^2_{ }} 
 \,\biggr]
 \, \R^{ }_\chi
 + 
 \mathcal{V}^{ }_{\varphi\chi}
 \, 
 \frac{\dot{\bar\varphi}}{\dot{\bar\chi}}
 \, 
 \bigl(\, 
 \R^{ }_\varphi - \R^{ }_\chi 
 \,\bigr)
 & \; = \; 0 
 \;. 
 \quad  \vphantom{\Bigg|^b_q} 
 \label{dot_R_chi_2}
 \index{evolution equations: two scalar fields}
\end{empheq}

\vspace*{3mm}

Let us elaborate on the meaning 
of \eqs\nr{dot_R_chi_1} and \nr{dot_R_chi_2}.
First of all, if we look at a stationary solution, with 
$
 \ddot{\R}^{ }_\varphi = \dot{\R}^{ }_\varphi = 0 
 = \ddot{\R}^{ }_\chi = \dot{\R}^{ }_\chi
$, 
and consider modes outside of the Hubble horizon, 
with $k/a \ll H$, then 
the first terms of \eqs\nr{dot_R_chi_1} and \nr{dot_R_chi_2} vanish.
Then, if both scalar fields are displaced from their minima
and are evolving, 
so that 
$
 {\dot{\bar\varphi}} \neq 0 
$
and
$ 
 {\dot{\bar\chi}} \neq 0
$, 
we find $\mathcal{V}^{ }_{\varphi\chi} \neq 0$. In this situation, 
the last terms of \eqs\nr{dot_R_chi_1} and \nr{dot_R_chi_2}
guarantee that $\R^{ }_\varphi = \R^{ }_\chi$. So, we recover 
the result of \se\ref{ss:outside_isocurv}, namely that 
isocurvature perturbations vanish outside of the Hubble horizon. 

On the other hand, suppose that $\chi$ is at its minimum 
or oscillates around it, 
so that~$\dot{\bar\chi}^{\ibit 2}_{ }$ carries little energy density.
Then the last term in \eq\nr{dot_R_chi_1}, 
weighted by $\dot{\bar\chi}$, should have little
influence, and we expect that $\R^{ }_\varphi$ evolves
independently of $\R^{ }_\chi$.
Conversely, if $\dot{\bar\varphi}^{\ibit 2}_{ }$ does not carry
much energy density, 
\eq\nr{dot_R_chi_2} indicates that $\R^{ }_\chi$ evolves
independently of~$\R^{ }_\varphi$.
This is how a curvaton field
could play a role after the inflaton field
has ceased to be active 
(i.e. $H \ll \Upsilon^{ }_\rmi{inflaton}$).

In model building, the coefficients of the last terms
in \eqs\nr{dot_R_chi_1} and \nr{dot_R_chi_2} 
might be non-zero during some period of time, but so small that
they do not efficiently drive $\R^{ }_\varphi - \R^{ }_\chi$
to zero. Then some amount
of an isocurvature perturbation is generated.
Consequently, it requires
a quantitative study to verify whether this amount remains within 
the observationally allowed window 
(cf.\ the discussion around \eq\nr{obs_isocurv}).

%%%%%%%%%%%%%%%%%%%%%%%%% BIBLIO %%%%%%%%%%%%%%%%%%%%%%%%%%%%%%%%
%

%%%%%%%%%%%%%%%%%%%%%%%%%%%% SECTION %%%%%%%%%%%%%%%%%%%%%%%%%%%%%%%%%%
\newpage 

\section{What happens when scalar modes re-enter the Hubble horizon?}
\label{se:inside}

\paragraph{Abstract:}

Eventually, all the modes that have observational significance today, 
need to cross the horizon again, ``re-entering'' the domain governed by
causal physics. This happens earlier for larger momenta, so that modes
above the CMB scale (corresponding to shorter distances)
re-enter before recombination. 
We derive the evolution equations satisfied by temperature
and velocity perturbations after re-entry, and show how their solution
leads to the emergence of acoustic oscillations. We discuss how the 
original curvature perturbation, $\R^{ }_\varphi$, decouples from the
evolution, thereby exiting the stage. 
We indicate how non-relativistic perturbations can undergo
gravitational collapse via Jeans instability, initiating the formation
of the first large-scale structures in the universe. 
Finally, we provide references to frequently used softwares
for a realistic study of CMB physics. 

\paragraph{Keywords:} 
acoustic oscillations, 
gravitational collapse, 
Jeans instability, 
transition from radiation to matter domination, 
dark matter, 
scalar transfer function, 
Einstein-Boltzmann solver, 
multicomponent universe, 
CMB spectrum, 
$N$-body simulations.

%%%%%%%%%%%%%%%%%%%%%%%%%%%%%%%%%%%%%%%%%%%%%%%%%%%
%
\subsection{Overview}

The key result of \ch\ref{se:outside} is that, when modes exit 
the Hubble horizon, they experience kind of a memory loss, at least 
in single-field inflationary models: many types of initial curvature
perturbations merge into a single time-independent function of the
comoving momentum, $k$. 
Furthermore, this function can be computed as
the solution of a single differential equation
(cf.\ \eq\nr{dot_R_varphi_again}), 
or a set of equations
(cf.\ \eqs\nr{dot_R_varphi}, \nr{dot_E_v}, \nr{dot_E_T}). 
When the modes re-enter the Hubble horizon later on, the opposite
dynamics takes place: the original curvature mode proliferates
into many coupled perturbations, which undergo fast oscillations, 
with evolving phases and gradual damping.
At the end, these oscillations are observed as the 
consecutive peaks in the CMB spectrum~(cf., e.g., ref.~\cite{peebles}). 

In the present chapter, we illustrate this late dynamics for 
a system in which three curvature perturbations are present,
namely $\R^{ }_\varphi$, $\R^{ }_v$ and 
$\R^{ }_\T$, as defined in \eqs\nr{def_R_varphi}--\nr{def_R_T}.
In \se\ref{ss:evol_v_T},
we derive a coupled set of evolution equations for 
$\R^{ }_v$ and $\R^{ }_\T$, under the assumption
that $\bar\varphi$ gives a subdominant contribution 
to the overall energy density. 
In \se\ref{ss:osc}, we show how the 
coupled set leads to quasi-periodic solutions, known as acoustic
oscillations. In \se\ref{ss:evol_varphi}, we return to  
$\R^{ }_\varphi$, and show that while still present,
it decouples from the other evolutions. In \se\ref{ss:jeans}, 
we enrich the setup by introducing a second fluid, and show
how it can experience Jeans instability
(cf., e.g., refs.~\cite{lss_1,lss_2,efst}). 
Finally, in \se\ref{ss:evol_many}, we briefly describe more
complicated frameworks, normally based on Boltzmann equations, 
that are being used for describing the post-inflationary evolution
of scalar perturbations in the real world, to be matched onto
observed CMB spectra.  

%%%%%%%%%%%%%%%%%%%%%%%%%%%%%%%%%%%%%%%%%%%%%%%%%%%%%%%%%%%%%%%%%%%%%%%%%
%
\subsection{Evolution equations for temperature and velocity perturbations}
\label{ss:evol_v_T}

\index{evolution equations: after re-entry}

Let us focus on a situation in which the modes considered re-enter inside
the Hubble horizon once the inflaton field no longer dominates the energy
density. 
Physically, this means that we exclude the very largest momenta 
(shortest distances) from the consideration, as they may re-enter
already during an earlier epoch, for instance when inflaton
oscillations dominate the energy density
(cf.\ \app\ref{app:num_bg_vac}). 
Under this assumption, the scalar perturbations $v \equiv \delta v$ 
and $\delta T$ are more important than $\delta\varphi$, and 
we can therefore consider a simplified set of evolution equations.
% The background evolution is also simplified, and corresponds to 
% the standard cosmological scenario, as discussed in \se\ref{ss:history}.

For the simplified setup, the starting point is given by 
\eqs\nr{delta_0_Tmunu;mu} and \nr{delta_i_Tmunu;mu_s}, 
\ba
  - \delta e' - 3 \H (\delta e + \delta p)
 + (\bar{e} + \bar{p}) (\, 3 h_\rmii{D}' + \nabla^2_{ } v \,)
 % - \H\, \tr\barpPi 
 & 
 \overset{\rmii{\nr{delta_0_Tmunu;mu}}}{=}
 &
 0 
 \;, \label{eom_vT_1} \\[2mm]
%%%%%
  \delta p + (\bar{e} + \bar{p}) h^{ }_0
 + (\partial^{ }_\tau + 4\H) [(\bar{e} + \bar{p})(h - v)]
 + \frac{2}{3} \nabla^2_{ }\barpPi
 & 
 \overset{\rmii{\nr{delta_i_Tmunu;mu_s}}}{=}
 &
 0 
 \;. \label{eom_vT_2}
\ea
The anisotropic stress $\Pi$ could be expressed in terms of 
dissipative coefficients as in  
\eqs\nr{decomposition} and \nr{delta_S}, but it is easier to 
display the expressions with $\Pi$, keeping in mind
that it is a gauge-invariant variable.

Let us first consider \eq\nr{eom_vT_2}. We insert here $h^{ }_0$ from 
\eq\nr{delta_einstein_0i}, making use of the notation
$
 Y \equiv h_\rmii{D}^{ } + {\nabla^2_{ }\vartheta}/{3}
$
from \eq\nr{shorthand}, and omit the background 
$ {\bar\varphi}' $, whereby $h^{ }_0$ reads 
\be
 h^{ }_0 
 \; \underset{\rmii{\nr{shorthand}}}
    {\overset{\rmii{\nr{delta_einstein_0i}}
     \lift }{ \approx }} \; 
 - \frac{Y'}{\H}
 + \frac{ 4\pi G 
  a^2_{ }( \bar{e} + \bar{p}) (v - h) }{\H} 
 \;. \label{vT_h0}
\ee
From \eqs\nr{def_R_v} and \nr{def_R_T}, the other perturbations
can be expressed as 
\ba
 h - v 
 &
 \overset{\rmii{\nr{def_R_v}}}{=}
 &
 \frac{\R^{ }_v + Y }{\H} 
 \;, \label{vT_hv} \\[2mm]
%%%%%
 \delta T
 &
 \overset{\rmii{\nr{def_R_T}}}{=}
 &
 - \frac{T' ( \R^{ }_\T + Y )}{\H} 
 \;. \label{vT_deltaT}
\ea
We express $\delta p$ in terms of $\delta T$
via  $\delta p \approx \bar{p}^{ }_{ ,\T}\, \delta T$, 
and collect together the terms acting on 
$ Y $ on the left-hand side (L) of \eq\nr{eom_vT_2}, 
\ba
 \mbox{\nr{eom_vT_2}}^{ }_\rmii{L} & \supset & 
 \biggl\{
  - \overbrace{
     \cancel{ \frac{\bar{p}^{ }_{\der\T} T' }{\H} }
     }^{{\rm from}\;\delta p}
  -
    \overbrace{
    \bcancel{ \frac{(\bar{e} + \bar{p})\partial^{ }_\tau}{\H} }
  - \frac{4\pi G a^2_{ }(\bar{e}+\bar{p})^2_{ }}{\H^2_{ }}
    }^{{\rm from}\; h^{ }_0\;{\rm via}\;\nr{vT_h0}}
  +
    \overbrace{
    \frac{\bar{e}\hspace*{0.3mm}'+\cancel{ \bar{p}' }}{\H}
  - \frac{(\bar{e} + \bar{p})\H'}{\H^2_{ }} 
  + \bcancel{ \frac{(\bar{e} + \bar{p})\partial^{ }_\tau}{\H} }
    }^{{\rm from}\; h-v \;{\rm via}\;\nr{vT_hv}}
 \nn[3mm] 
%%%%
 &  & \hphantom{\biggl\{}
 + 4 (\bar{e} + \bar{p})
 \biggr\} 
 \, Y
 \;. \label{vT_canc_1}
\ea
As background identities, we can make use of 
\eq\nr{bg_ep}, with $\kappa\to 0$ and 
$ (\bar\varphi\hspace*{0.3mm}{}')^2_{ } \ll  a^2(\bar{e} + \bar{p})$, and of 
\eq\nr{bg_Tmunu_tau}, 
\be
 4 \pi G  a^2_{ }(\bar{e} + \bar{p})
  \; 
 \overset{\rmii{\nr{bg_ep}}}{\approx}
 \; 
 \H^2 - \H'
 \;, \quad
 \bar{e}\hspace*{0.3mm}' + 3\H (\bar{e} + \bar{p} )
 \;
 \overset{\rmii{\nr{bg_Tmunu_tau}}}{\approx}
 \;
 0 
 \;. \label{bg_vT}
\ee
Inserting these in \eq\nr{vT_canc_1}, all terms
cancel. This demonstrates that the evolution equation is 
gauge invariant, and  
offers for a valuable crosscheck of the computation. 

For the remaining terms, multiplying the whole equation with $\H$, we get 
\ba
 \H \times \mbox{\nr{eom_vT_2}}^{ }_\rmii{L}
 &
  \overset{Y\;\to\;0}{=}
 & 
 -  
  \overbrace{ 
  \bar{p}\hspace*{0.3mm}'\, \R^{ }_\T
  }^{{\rm from}\;\delta p }
 - (\bar{e} + \bar{p})
  \hspace*{-2mm}
  \overbrace{
  \biggl(\H - \bcancel{\frac{\H'}{\H}} \biggr)
  \R^{ }_v}^{{\rm from}\;h^{ }_0\;{\rm via}\;\nr{vT_h0},\,\nr{bg_vT}}
  \hspace*{-2mm}
 + 
  \overbrace{
   (\bar{e}\hspace*{0.3mm}' + \bar{p}\hspace*{0.3mm}') \R^{ }_v
 + (\bar{e} + \bar{p}) \R^{\prime}_v
  }^{{\rm from}\; h-v \;{\rm via}\;\nr{vT_hv}}
 \nn[3mm]
%%%%%
 &  & \;+\,
 \overbrace{
 (\bar{e} + \bar{p})\biggl( 4\H - \bcancel{\frac{\H'}{\H}} \biggr)
 \R^{ }_v
  }^{{\rm from}\; h-v \;{\rm via}\;\nr{vT_hv}}
 + \frac{2}{3} \H \,\nabla^2_{ }\barpPi
 \nn[3mm]
%%%%%%
 & \overset{\rmii{\nr{bg_vT}}}{=} & 
 \bar{p}\hspace*{0.3mm}'\, \bigl(\, \R^{ }_v - \R^{ }_\T \,\bigr)  
 + (\bar{e} + \bar{p}) \R^{\prime}_v
 + \frac{2}{3} \H \,\nabla^2_{ }\barpPi
 \; \overset{\rmii{\nr{eom_vT_2}}}{=} \; 0
 \;. \label{dtau_R_v_rad}
\ea
Going over to physical time 
($ \bar{p}\hspace*{0.3mm}' = a \dot{\bar p}$, 
$\H = a H$)
and to momentum space
($\nabla^2_{ }\Pi\to - k^2_{ } \Pi$), 
\eq\nr{dtau_R_v_rad} turns into 
the evolution equation that is shown in \eq\nr{dot_R_v_rad}.

Turning then to \eq\nr{eom_vT_1}, we rewrite 
the last term in a longer but more convenient form, 
\be
 3 h_\rmii{D}' + \nabla^2_{ } v
 \;=\; 
 \nabla^2_{ } ( v - h) 
 +\, 
 \underbrace{
 \nabla^2_{ } h
 + 3 h_\rmii{D}' + h_0' }_{{\rm from}\;\nr{delta_einstein_comb}}
 \,- 
 \hspace*{-3mm}
 \underbrace{
 h_0'}_{{\rm from}\;\nr{vT_h0}}
 \;.  \label{vT_comb}
\ee
Therefore, we need to evaluate
\ba
 \nabla^2_{ } h
 + 3 h_\rmii{D}' + h_0'
 &
 \overset{ \rmii{\nr{delta_einstein_comb}} }{\approx}
 & 
  -  2 \biggl( 2 \H + \frac{\H'}{\H}  \biggr) h^{ }_0
  - \frac{1}{\H}
  (\partial^2_\tau + 2 \H \partial^{ }_\tau - \nabla^2_{ } )
 \, Y
 \nn[2mm]
%%%%
 &  & \;+\,
  \frac{ 4\pi G a^2_{ }}{\H}  
 \biggl[  \biggl( \bar{p}^{ }_{\der\T}
                - \bar{e}^{ }_{\der\T} \biggr) \delta T
   + \frac{2}{3} \nabla^2_{ }\barpPi  
 \biggr]
 \;, \label{vT_comb_prime} \\[3mm]
%%%%%
 h_0'
 &
 \underset{\rmii{\nr{bg_vT}}}
 {\overset{\rmii{\nr{vT_h0}} \lift }{\approx}}
 & 
 \frac{\H' - \H \partial^{ }_\tau }{\H^2_{ }}\, Y'
 \; + \; 
 \biggl( 1  -\frac{\H'}{\H^2_{ }}\biggr)[\H (v - h)]'
 \nn[2mm]
%%%%%%
 &  & \;+\,
   \biggl( \frac{2\H^{\prime\hspace*{0.3mm}2}_{ }}{\H^3_{ }}
 - \frac{\H''}{\H^2_{ }}\biggr)
 [\H (v - h)]
 \;. \label{vT_h0_prime} \hspace*{7mm}
\ea
Given that $\H''$ appears, we also need an additional
background identity, given in \eq\nr{bg_vT_2}. 

When we insert \eqs\nr{vT_comb_prime} and \nr{vT_h0_prime} into
\eq\nr{vT_comb}, and then the whole into \eq\nr{eom_vT_1}, the 
expression becomes quite complicated. In this situation, it is all 
the more important to have the powerful principle of gauge invariance
as a crosscheck that we are not making mistakes along the way. 
Without displaying the details, we note that it is remarkable how 
all different structures
in the differential operator acting on 
$
 Y 
$
cancel, after we make use of the background identities 
in \eqs\nr{bg_vT} and \nr{bg_vT_2}. 

Let us then proceed to the physical terms. 
Inserting \eqs\nr{vT_h0}--\nr{vT_deltaT} and 
\nr{vT_comb_prime}, \nr{vT_h0_prime} into \eq\nr{eom_vT_1},
we find
\ba
 \mbox{\nr{eom_vT_1}}^{ }_\rmii{L}
 \hspace*{-2mm}
  &
  \overset{Y\;\to\;0}{=}
  & 
 \overbrace{
 \biggl(\frac{\bar{e}\hspace*{0.3mm}'\R^{ }_\T}{\H}\biggr)'
 }^{{\rm from}\;\delta e\hspace*{0.3mm}' }
 + 
 \overbrace{
   3 (\bar{e}\hspace*{0.3mm}' + \bar{p}\hspace*{0.3mm}') \R^{ }_\T 
 }^{{\rm from}\; -3 \H (\delta e + \delta p) }
 -
 \hspace*{-2mm}
 \overbrace{
 \frac{(\bar{e}+\bar{p}) \nabla^2_{ } \R^{ }_v}{\H}
 }^{{\rm from}\; \nabla^2_{ }(v-h) \;{\rm via}\;\nr{vT_hv} }
 \hspace*{-2mm}
 \nn[3mm]
%%%%%
 &  & \;+\,
   (\bar{e}+\bar{p})
   \biggl\{\,
   \overbrace{
   2 \biggl( 2\H + \frac{\H'}{\H} \biggr)
     \biggl( 1 - \frac{\H'}{\H^2_{ }} \biggr)
   \R^{ }_v
   }^{{\rm from\,}h^{ }_0\,{\rm via~\nr{vT_comb_prime},\,\nr{vT_h0}} }
   + 
   \overbrace{
   \frac{ 4\pi G a^2_{ }(\bar{e}\hspace*{0.3mm}' - \bar{p}\hspace*{0.3mm}') }
        {\H^2_{ }} \R^{ }_\T
   }^{{\rm from\,}\delta T\,{\rm via~\nr{vT_comb_prime}} }
 \nn[3mm]
%%%%%
 &  & \;+\,
   \overbrace{
   \biggl( 
   1 - \frac{\H'}{\H^2_{ }} 
   \biggr)\R^{\prime}_v
   + 
   \biggl( \frac{2\H^{\prime\hspace*{0.3mm}2}_{ }}{\H^3_{ }}
   -\frac{\H''}{\H^2_{ }}\biggr)
   \R^{ }_v
   }^{{\rm from\,}h_0'\;{\rm via}\;\nr{vT_h0_prime},\,\nr{vT_hv}}
   \,\biggr\}
 + \frac{2}{3} \frac{4\pi G a^2_{ }(\bar{e} + \bar{p})}{\H} \nabla^2_{ }\Pi
 %% - \H\,\tr\Pi
 \nn[3mm]
%%%%%%%
 & \overset{\rmii{\nr{bg_vT}}}{=} & 
 - 3 [(\bar{e} + \bar{p}) \R^{ }_\T]'
 + 3 (\bar{e}\hspace*{0.3mm}' + \bar{p}\hspace*{0.3mm}') \R^{ }_\T 
 + (\bar{e} + \bar{p}) \biggl\{\, 
 - \frac{\nabla^2_{ }\R^{ }_v}{\H}
 +    \biggl( 
   1 - \frac{\H'}{\H^2_{ }} 
   \biggr)\R^{\prime}_v
 \nn[3mm]
%%%%%%%%
 &  & \;+\,
 \biggl( 4\H  - \frac{2\H'}{\H}
 - \cancel{\frac{2\H^{\prime\hspace*{0.3mm}2}_{ }}{\H^3_{ }}}
 + \cancel{\frac{2\H^{\prime\hspace*{0.3mm}2}_{ }}{\H^3_{ }}}
  -\frac{\H''}{\H^2_{ }}
  \biggr) \R^{ }_v
   + 
   \frac{ 4\pi G a^2_{ }(\bar{e}\hspace*{0.3mm}' - \bar{p}\hspace*{0.3mm}') }
        {\H^2_{ }} \R^{ }_\T
  \biggr\}
 \nn[3mm]
%%%%%%%
 &  & \;+\,
  \frac{8\pi G a^2_{ }(\bar{e} + \bar{p})}{3\H} \nabla^2_{ }\Pi
 %%- \H\,\tr\Pi
 \nn[3mm]
%%%%%%%
 & \overset{\rmii{\nr{bg_vT}} \lift }
 {\underset{\rmii{\nr{bg_vT_2}}}{=}} &
 (\bar{e} + \bar{p}) \biggl\{ 
 - 3 \R^{\prime}_\T - \frac{\nabla^2_{ }\R^{ }_v}{\H}
 + \frac{4\pi G a^2_{ }(\bar{e} + \bar{p})}{\H^2_{ }} \R^{\prime}_v
 + \frac{4\pi G a^2_{ }(\bar{e}\hspace*{0.3mm}'
                      - \bar{p}\hspace*{0.3mm}')}{\H^2_{ }}
   \bigl( \R^{ }_\T - \R^{ }_v \bigr)
 \biggr\}  
 \nn[3mm]
%%%%%%%
 &  & \;+\, 
  \frac{8\pi G a^2_{ }(\bar{e} + \bar{p})}{3\H} \nabla^2_{ }\Pi
 %%- \H\,\tr\Pi
 \nn[3mm]
%%%%%%%
 & \overset{\rmii{\nr{dtau_R_v_rad}}}{=} & 
 (\bar{e} + \bar{p}) \biggl\{ 
 - 3 \R^{\prime}_\T - \frac{\nabla^2_{ }\R^{ }_v}{\H}
 + \frac{4\pi G a^2_{ } \bar{e}\hspace*{0.3mm}'}{\H^2_{ }}
   \bigl( \R^{ }_\T - \R^{ }_v \bigr)
 \biggr\}  
 %%- \H\,\tr\Pi
 \; \overset{\rmii{\nr{eom_vT_1}}}{=} \; 0
 \;. \label{ac_der_2}
\ea
Going subsequently to physical time
($\tau\to t$) and to comoving momentum 
space
($\vec x\to \vec k$), 
and making use of the background identity in \eq\nr{bg_vT}
once more, we obtain \eq\nr{dot_R_T_rad}.

To summarize, the evolution equations for curvature
perturbations in a plasma-dominated universe read
\begin{empheq}[box=\fbox]{align}
 \quad  \vphantom{\Bigg|} 
 \dot{\R}^{ }_v
 &
 \underset{\rmii{\nr{dot_E_v_decoupling}}}{
 \overset{\rmii{\nr{dtau_R_v_rad}} \ilift }{\approx}}
 \;
 \underbrace{ 
 \frac{\dot{\bar{p}}}{\bar{e}+\bar{p}}
 }_{ 
  \;\approx\;\dot{T} / T
 }
 \,\bigl(\, \R^{ }_\T - \R^{ }_v \,\bigr)
 + 
 \frac{2 H\, k^2_{ }\barpPi }{3 (\bar{e} + \bar{p})}
 \;, \label{dot_R_v_rad} 
 \\[3mm]
%%%%%%%%%%%%%%%%%%%%
 \dot{\R}^{ }_\T
 &
 \underset{\rmii{\nr{dot_E_T_decoupling}}}{
 \overset{\rmii{\nr{ac_der_2}} \ilift }{\approx}}
 \; 
 \frac{\R^{ }_v}{3 H} \frac{k^2_{ }}{a^2_{ }} 
 \;
 \underbrace{ 
 - \;
 \frac{4\pi G(\bar{e} + \bar{p})   
 }{H}
 }_{ 
  \;\approx\; + \, \dot{H} / H
 }
 \,\bigl(\, \R^{ }_\T - \R^{ }_v \,\bigr)
 \;. \label{dot_R_T_rad} 
 \quad  \vphantom{\Bigg|} 
\end{empheq}
Here we also anticipated that an alternative derivation
can be found in \eqs\nr{dot_E_v_decoupling} and \nr{dot_E_T_decoupling}.
For the terms proportional to $ \R^{ }_\T - \R^{ }_v $, 
we indicated simpler interpretations for the coefficients, 
by making use of thermodynamic identities from \eq\nr{entropy}
as well as a background identity from \eq\nr{bg_Hprime}, 
once again simplified by taking $\kappa\to 0$ and 
$\dot{\bar\varphi}^{\ibit 2}_{ } \ll \bar{e} + \bar{p}$.

Equations \nr{dot_R_v_rad} and \nr{dot_R_T_rad} manifest the property
discussed in \se\ref{ss:outside_isocurv}. 
Namely, 
if we look at momenta outside the Hubble horizon, 
$k/a \ll H$, then viscous corrections  (cf.\ \eq\nr{Pi_again})
as well as the first term on the right-hand side 
of \eq\nr{dot_R_T_rad} drop out. 
Then the system has a stationary solution, 
i.e.\ $\dot{\R}^{ }_v \approx 0 \approx \dot{\R}^{ }_\T$, 
with vanishing isocurvature perturbations, 
i.e.\ $\R^{ }_v \approx \R^{ }_\T$. 
According
to \ch\ref{se:outside}, the initial value for 
$\R^{ }_v$ and $\R^{ }_\T$
can be equated with $\R^{ }_\varphi$, 
which in turn can be determined from the inflationary dynamics. 

\vspace*{3mm}

We end this section by noting that the anisotropic stress, $\Pi$, 
could become important in \eq\nr{dot_R_v_rad},
once the momenta re-enter the Hubble horizon, i.e.\ 
$k/a \gg H$. As argued around
\eq\nr{shear_estimate}, the most important part of the 
anisotropic stress is probably a term proportional to 
the shear viscosity, given in \eq\nr{Pi_again}.
That term contains the Bardeen potential $\psi$, 
so we need an equation for $\psi$. 

One of the equations for the Bardeen potentials is given in 
\eq\nr{demo_1} (see also~\eq\nr{dot_phi_again}), but it contains
both $\phi$ and $\psi$, and does not suffice on its own. To derive
another equation, 
we can take \eq\nr{vT_h0} as a starting point
(we could also proceed from \eq\nr{dot_psi}, by considering
the limit $\dot{\bar\varphi}^{\ibit 2}_{ } \ll \bar{e} + \bar{p}$).
In order to simplify the equations,
we make use of the notation in \eq\nr{shorthand}. 
On the left-hand side of \eq\nr{vT_h0}, we can then insert 
\be
 h^{ }_0 
 \; 
  \underset{\rmii{\nr{shorthand}}}
  {\overset{\rmii{\nr{def_phi}} \lift }{=}}
 \; 
 \phi - X' - \H X 
 \;, \label{h0_again}
\ee
and on the right-hand side of \eq\nr{vT_h0}, 
we use the two representations
\ba
 Y 
 &
  \underset{\rmii{\nr{shorthand}}}
  {\overset{\rmii{\nr{def_psi}} \lift }{=}} 
 &
 \psi + \H X
 \;, \label{Y_aagain} \\[2mm]
%%%%
 v - h 
 & 
   \underset{\rmii{}}{\overset{\rmii{\nr{vT_hv}}}{=}}
 & 
 -\frac{1}{\H}
 \bigl(\,
   Y + \R^{ }_v 
 \,\bigr)
 \;. \label{vmh_aagain}
\ea
In addition we need the background identity from \eq\nr{bg_vT}.
All in all,  \eq\nr{vT_h0} multiplied by $\H$ yields
\ba
 \overbrace{
 \H\phi - \cancel{\H X'} - \bcancel{\H^2_{ } X}
 }^{
 {\rm from}\; \H h^{ }_0\;{\rm via}\;\nr{h0_again} 
 }
 & 
    \overset{\scriptscriptstyle \H\,\times\,\rmii{\nr{vT_h0}}
             \vphantom{\big | } }{=}
 & 
 \overbrace{
 -\psi' - \bcancel{\H' X} - \cancel{\H X'}
 }^{{\rm from}\; -Y' \;{\rm via}\;\nr{Y_aagain}}
% \nn[2mm]
%%%%
% & & 
 \;
 - 
 \,
 \overbrace{
 (\H^2_{ } - \H')
 }^{4\pi G a^2_{ }(\bar e + \bar p)}
 \hspace*{2mm}
 \times
 \hspace*{-3mm}
 \overbrace{
 \frac{ \psi + \bcancel{\H X} + \R^{ }_v}{\H}
 }^{{\rm from}\; v-h \;{\rm via}\;\nr{Y_aagain},\,\nr{vmh_aagain}}
 \hspace*{-3mm}
 \;. \nn 
 \label{psi_derivation}
\ea
Going over to physical time, 
with $\psi' = a \dot{\psi}$, 
$\H = a H$, 
and $\H' - \H^2_{}  = a^2_{ }\dot{H}$, we get
\begin{empheq}[box=\fbox]{align}
 \vphantom{\Bigg|} 
 \phi -\psi 
 +  8\pi G a^2_{ }\barpPi
 &\; \overset{\rmii{\nr{demo_1}} }{=}
 \; 
 0
 \;,
 \label{dot_phi_again}
 \\[0mm]
 %%%%%%
 \quad
 \dot\psi +  H\phi
 - \frac{\dot{H}}{H}\, \bigl(\, \psi + \R^{ }_v \,\bigr)
 &\; \overset{\rmii{\nr{psi_derivation}}  }{=}
  \;
  0
  \;.
 \quad  \vphantom{\Bigg|} 
 \label{dot_psi_again}
\end{empheq}
Combining \eqs\nr{dot_phi_again} and \nr{dot_psi_again} 
with \eqs\nr{dot_R_v_rad} and \nr{dot_R_T_rad},
we now have a complete set of equations also for 
the situation that anisotropic stress plays a role. 
It is interesting that
in the constraint relation in \eq\nr{dot_phi_again},  
$\Pi$ does {\em not} come with a coefficient $k^2_{ }$,
unlike in the time evolution equations, 
and $a^2_{ }$ rather cancels against
the $1/a^2_{ }$ in \eq\nr{Pi_again}.

%%%%%%%%%%%%%%%%%%%%%%%%%%%%%%%%%%%%%%%%%%%%%%%%%%%%%%%%%%%

%%%%%%%%%%%%%%%%%%%%%%%%%%%%%%%%%%%%%%%%%%%%%%%%%%%%%%%%%%%%%%%%%%%%%%%%%
%
\subsection{The origin of acoustic oscillations}
\label{ss:osc}

\index{acoustic oscillations: evolution equations}

The next task is to solve \eqs\nr{dot_R_v_rad} and \nr{dot_R_T_rad}.
To simplify our life a bit, we omit the viscous corrections
appearing in the anisotropic stress, 
so that \eqs\nr{dot_phi_again} and \nr{dot_psi_again}
do not need to be included. 
The physics that anisotropic stress describes 
is nevertheless interesting: it leads to the  
damping of the acoustic oscillations, with the damping rate
growing as $k^2_{ }$ with the momentum. This is related to 
the {\em Silk damping} discussed in \se\ref{ss:cmb}.

\index{Silk damping}

We now focus on times 
$t \ge t^{ }_\rmi{out} \gg t^{ }_e$, where $t^{ }_e$ denotes
the end of inflation, and $t^{ }_\rmi{out}$ a time at which
the mode considered is well outside of the Hubble horizon. 
Following \eqs\nr{soln_bg_t_a}
and  \nr{a_t_mat_1}, we adopt a simplified equation of state to 
describe the background evolution. 
Moreover, it is convenient to define the parameter
\be
 \alpha \; \equiv \; \frac{2}{3(1+w)}
 \; = \; 
 \left\{ 
 \begin{array}{lll}
    \frac{1}{2} \;, & w = \frac{1}{3} & \mbox{(radiation domination)} \\[2mm]
    \frac{2}{3} \;, & w = 0           & \mbox{(matter domination)}
 \end{array}
 \right.
 \;. \label{def_alpha}
\ee
The scale factor and the Hubble rate then behave as 
\be
 a 
 \;\overset{\rmii{\nr{soln_bg_t_a}}}{\approx}\;
  a^{ }_\rmi{out}  
 \biggl(\frac{t}{t^{ }_\rmii{out}}\biggr)^\alpha_{ }
 \;, \qquad
 H
 \;\overset{\rmii{\nr{a_t_mat_1}}}{\approx}\;
 \frac{\alpha}{t}
 \;, \qquad
 t \;\ge\; t^{ }_\rmi{out} \;\gg\; t^{ }_e
 \;. \label{osc_appro_1}
\ee
Furthermore, from \eq\nr{evolution_a}, we know that $T \propto 1 / a$.
Therefore the ratios appearing in \eqs\nr{dot_R_v_rad} and \nr{dot_R_T_rad}
scale as 
\be
 \frac{\dot{T}}{T}
 \; 
 \underset{\rmii{\nr{osc_appro_1}}}{
 \overset{\rmii{\nr{evolution_a}} \lift }{\approx}}
 \;
 -\frac{\alpha}{t}
 \;, \qquad
 \frac{\dot{H}}{H}
 \; 
 \overset{\rmii{\nr{osc_appro_1}}}{\approx}
 \; 
 -\frac{1}{t}
 \;. \label{osc_appro_2}
\ee

For the momentum appearing in \eq\nr{dot_R_T_rad}, we write
\be
 \frac{k^2_{ }}{a^2_{ } H} \; = \;
 \biggl( \frac{k}{a H} \biggr)^2_\rmii{out}
 \biggl( \frac{a^{ }_\rmii{out}}{a} \biggr)^2_{ }
 \frac{H^{2}_\rmii{out}}{H}
 \;
 \overset{\rmii{\nr{osc_appro_1}}}{\approx}
 \; 
 \biggl( \frac{k}{a H} \biggr)^2_\rmii{out}
 \biggl( \frac{t^{ }_\rmii{out}}{t} \biggr)^{2\alpha}_{ }
 \frac{\alpha\, t}{t^2_\rmii{out}}
 \;. \label{koaH_out}
\ee
It is also helpful to take 
\be
 x \; \equiv \; \ln\biggl( \frac{t}{t^{ }_\rmii{out}}\biggr)
 \; \ge \; 0
 \;, \qquad \; 
 {\dd }x \; = \; \frac{\dd t}{t}
 \;, \qquad
 \partial^{ }_t \; = \; \frac{1}{t} \bit \partial^{ }_x
 \;, \label{def_x}
\ee
as an integration variable. 
Then \eqs\nr{dot_R_v_rad} and \nr{dot_R_T_rad} become \pagebreak
\ba
 \partial^{ }_x \R^{ }_v 
 &
 \underset{\rmii{\nr{osc_appro_2},\nr{def_x}}}{
 \overset{\rmii{\nr{dot_R_v_rad}}}{\approx}}
 &
 \alpha \bigl(\, \R^{ }_v - \R^{ }_\T \,\bigr)
 \;, \label{dx_Rv} \\[2mm]
%%%%%%%%
 \partial^{ }_x \R^{ }_\T
 &
 \underset{\rmii{\nr{koaH_out},\nr{def_x}}}{
 \overset{\rmii{\nr{dot_R_T_rad}}}{\approx}}
 &
 \R^{ }_v - \R^{ }_\T 
 \;+\;\frac{\alpha}{3}
 \underbrace{ 
 \biggl(\frac{t}{t_\rmii{out}}\biggr)^{2(1-\alpha)}_{ } 
 }_{ e^{2(1-\alpha)x}_{ }}\;
 \biggl( \frac{k}{a H} \biggr)^2_\rmii{out}
 \, \R^{ }_v
 \;. \label{dx_RT}
\ea

Equations \nr{dx_Rv} and \nr{dx_RT} are transparent enough 
that the qualitative features of their solution can be discussed. 
The last term in \eq\nr{dx_RT} is small at $t \to t^{ }_\rmi{out}$,  
if $k \ll (a H)^{ }_\rmi{out}$. Then we find a stationary 
solution $\R^{ }_v = \R^{ }_\T$, which 
serves as an initial condition. When $x > 0$, the importance of 
the last term in \eq\nr{dx_RT} grows exponentially. 
In spite of this large contribution, the function
$\R^{ }_\T$ cannot grow much, because of the 
damping term $-\R^{ }_\T$. Therefore the magnitude
of $\R^{ }_v$, multiplying the growing coefficient, 
has to decrease exponentially. 
These features are confirmed by a numerical solution
of \eqs\nr{dx_Rv} and \nr{dx_RT}, which is shown for 
$\alpha = \tfr{1}{2}$ 
in \app\ref{app:num_osc}
(cf.\ \fig\ref{fig:acoustic}(left) on p.~\pageref{fig:acoustic}). 
With some more effort,
incorporating a numerical background solution like in 
\fig\ref{fig:bg_thermal}(left) on p.~\pageref{fig:bg_thermal}, 
it can be verified that \fig\ref{fig:acoustic}(left) 
represents a good approximation also to the solution of
the un-approximated
\eqs\nr{dot_R_v_rad} and \nr{dot_R_T_rad}~\cite{alica_copy_2}.

Let us finally discuss the physical meaning of $\R^{ }_\T$.
Knowing the scaling of $H$ and $T$ from 
\eqs\nr{osc_appro_1} and \nr{osc_appro_2}, respectively, 
we rewrite \eq\nr{def_R_T} as 
\be
  \R^{ }_\T
 \;
 \overset{\rmii{\nr{def_R_T}}}{=}
 \; 
 - \biggl( 
  h^{ }_\rmii{D} + \frac{\nabla^2\vartheta}{3} 
   \biggr)
 - \H\, \frac{\delta T}{T'}
  \; = \; 
 - \biggl( 
  h^{ }_\rmii{D} + \frac{\nabla^2\vartheta}{3} 
   \biggr)
 \underbrace{ 
 - \frac{H T}{\dot T}
 }_{ 
    \underset{\rmii{\nr{osc_appro_2}}}
             {\overset{\rmii{\nr{osc_appro_1}}}{\approx}}
    1}
 \, \frac{\delta T}{T}
 \;. \label{R_T_again}
\ee
We see that $\R^{ }_\T$ represents a gauge-invariant
version of a relative temperature fluctuation. 
In addition, given that 
$
 \delta e/\bar{e}\hspace*{0.3mm}' \approx \delta T/T'
$, 
\be
 \R^{ }_\T
 \; \overset{\rmii{\nr{def_R_e}}}{ \approx } \;
 \R^{ }_e
 \; \overset{\rmii{\nr{zeta}}}{=} \;
 \zeta
 \;, \label{zeta_again}
\ee
so it also captures energy density perturbations. 
Furthermore we note that 
another measure of energy density perturbations,
from \eq\nr{Delta_v}, becomes
\ba
 \Delta
 &
 \overset{\rmii{\nr{Delta_v}}}{=}
 &  
  \frac{\delta e}{\bar e} 
 + \frac{\bar{e}\hspace*{0.3mm}'}{\bar{e}}(h-v)
 \; 
 \approx 
 \;
  \frac{\bar{e}^{ }_{,\T}\delta T}{\bar e} 
 + \frac{\bar{e}^{ }_{,\T} T'}{\bar{e}}(h-v)
 \;
 = 
 \;
 \frac{\bar{e}^{ }_{,\T} T'}{\bar{e}}
 \biggl( h - v + \frac{\delta T}{T'} \biggr)
 \nn[3mm]
%%%%%%%
 & = &  
 \frac{\dot{\bar e}}{ \bar e H }
 \bigl(\, \R^{ }_v - \R^{ }_\T \,\bigr)
 \;
  \underset{\rmii{\nr{def_alpha}}}
  {\overset{\rmii{\nr{eq_end0-2}}}{\approx}}
 \;
 \frac{2}{\alpha} 
 \bigl(\, \R^{ }_\T - \R^{ }_v \,\bigr)
 \; \underset{\rmii{\nr{dx_RT}}}
    {\overset{\scriptscriptstyle x\;\gg\; 1 \lift }{\approx}} \; 
 \frac{2}{\alpha}\, \R^{ }_\T
 \;.
 \label{Delta_v_new}
\ea
So, $\R^{ }_\T$ determines this function as well, up to a coefficient
of order unity.

In non-relativistic physics, 
propagating fluctuations 
in the temperature and 
the longitudinal flow velocity are sound waves. Given this analogy,
the oscillations in $\R^{ }_\T$ (and in $\R^{ }_v$)
that we see in \fig\ref{fig:acoustic}(left)
can be called {\em acoustic oscillations}.

%%%%%%%%%%%%%%%%%%%%%%%%%%%%%%%%%%%%%%%%%%%%%%%%%%%%%%%%%%%%%%%%%%%%%%%%%
%
\subsection{What happens to $\R^{ }_\varphi$?}
\label{ss:evol_varphi}

The key variable in the determination of the curvature perturbations
in \ch\ref{se:outside} was $\R^{ }_\varphi$, defined in 
\eq\nr{def_R_varphi}. Recalling the definition of the Hubble time
as $\Delta t \equiv H^{-1}_{ }$, it contains
\be
 \R^{ }_\varphi
 \;
 \overset{\rmii{\nr{def_R_varphi}}}{\supset}
 \; 
 - H \, \frac{\delta\varphi}{\dot{\bar\varphi}}
 \; = \; 
 - \frac{\delta\varphi}{\Delta t \, \dot{\bar\varphi}}
 \; \approx \; 
 - \frac{\delta\varphi}{\Delta \bar\varphi }
 \;. \label{ratio_R_varphi}
\ee
As we enter the regime $\Upsilon \gg H$, 
the background value $\bar\varphi$ 
decreases, but $\delta\varphi$ also decreases, so it is not 
clear how the ratio in \eq\nr{ratio_R_varphi} behaves. 
Physically, for small~$\bar\varphi$, the distinction
between $\bar\varphi$ and $\delta\varphi$ loses its significance, 
and in the end second-order perturbations determine the contribution of 
$\varphi$ to the plasma energy density and pressure
(cf.\ \app\ref{app:thermal_inflaton}). However, 
for the consistency of our framework, 
it is essential that this transition 
is reflected by the equations that we 
are solving, rather than having to be put in by hand. 
The purpose of this section is to demonstrate 
how this happens. 

Let us anticipate the outcome. 
If we solve \eq\nr{dot_R_varphi}, 
the solution $\R^{ }_\varphi$ remains non-trivial 
and large until late times.
However, when $\dot{\bar\varphi}^{\ibit 2}_{ } \ll \bar{e}$, 
$\R^{ }_\varphi$ decouples from the evolution 
of $\R^{ }_v$ and $\R^{ }_\T$, so that 
as far as the latter two are considered, 
it is sufficient to solve \eqs\nr{dot_R_v_rad} and \nr{dot_R_T_rad}.
Therefore, we are free to stop solving for $\R^{ }_\varphi$, 
and simply include its second-order contribution in
our thermodynamic functions if it equilibrates. 

\vspace*{3mm}

In order to demonstrate the decoupling explicitly, we take
the full \eqs\nr{dot_E_v} and \nr{dot_E_T} for $\E^{ }_v$
and $\E^{ }_\T$ as a starting point. 
We then proceed by noting that the background identities from 
\eqs\nr{bg_HH}, \nr{bg_Hprime} and \nr{bg_Tmunu} can be simplified
when $\dot{\bar\varphi}^{\ibit 2}_{ }$ carries little energy density, 
\be
 3 H^2_{ } 
 \; \overset{\rmii{\nr{bg_HH}}}{\approx} \; 
 8 \pi G \bar{e}
 \;, \quad
 \dot{H} 
 \; \overset{\rmii{\nr{bg_Hprime}}}{\approx} \, 
 -\, 4 \pi G (\bar{e} + \bar{p})
 \;, \quad
 \dot{\bar e}  
 \; \overset{\rmii{\nr{bg_Tmunu}}}{\approx} \, 
  -\, 3 H (\bar e + \bar p)
 \;. \label{bg_decouple}
\ee
We also note that 
in this situation there is only one important 
variable, i.e.\ $T$ plays a role for the dynamics 
but $\varphi$ not, so we can write
$
 \bar{p}^{ }_{\der\T}/\bar{e}^{ }_{\der\T}
 \approx \dot{\bar{p}}/\dot{\bar{e}}
$.
Furthermore we can approximate 
$
 4 \pi G \dot{\bar\varphi}^2 / H \ll 
 4 \pi G \bar{e} / H \sim H
$.
Therefore \eq\nr{dot_E_v} becomes
\ba
 0 
 \!\!
 & 
  \overset{\rmii{\nr{dot_E_v}}}{\approx}
 & 
 \!\!
   \bigl(
      \partial^{ }_t + 3 H 
   \bigr)
   \, 
   \overbrace{
   [(\bar e + \bar p)(\R^{ }_v - \R^{ }_\varphi)]
   }^{ 
   \nr{def_E_v}:\;{\rm was}\;
   \E^{ }_v } 
 \, - \, 
   \frac{\dot{\bar{p}}}{\dot{\bar{e}}}
   \, 
   \overbrace{ 
   [\dot{\bar{e}} (\R^{ }_\T - \R^{ }_\varphi) ] 
   }^{ 
   \nr{def_E_T}:\;{\rm was}\;
   \E^{ }_\T }
% \nn[2mm]
%%%%
% & +&   
  \; + \; 
    \bigl(\, \bar e + \bar p \,\bigr) \dot{\R}^{ }_\varphi 
  \, + \, 
 \frac{2 H \nabla^2_{ } \Pi}{3}
 \nn[3mm]
%%%%%%
 \!\!
 & \overset{\rmii{\nr{bg_decouple}}}{=} &
 \!\! 
   \bigl(\, \bar e + \bar p \,\bigr) \dot{\R}^{ }_v 
 \,+\,
 \bigl[\, \cancel{\dot{\bar{e}}} + \dot{\bar{p}} + 
     \cancel{ 3 H (\bar e + \bar p) } \,\bigr]
 (\R^{ }_v - \bcancel{\R^{ }_\varphi} )
 \,+\,
 \dot{\bar{p}}\, ( \bcancel{\R^{ }_\varphi} - \R^{ }_\T)
 \,+\, 
 \frac{2 H \nabla^2_{ } \Pi}{3}
 \nn[3mm]
%%%%%%
 \!\!
 & = & 
 \!\!
    \bigl(\, \bar e + \bar p \,\bigr) \dot{\R}^{ }_v 
 + \dot{\bar{p}}\, (\R^{ }_v - \R^{ }_\T)
 \,+\, 
 \frac{2 H \,\nabla^2_{ } \Pi}{3}
 \;. \hspace*{5mm} \label{dot_E_v_decoupling}
\ea
We see that $\R^{ }_\varphi$ has dropped out, 
and \eq\nr{dot_R_v_rad} is recovered. 

Turning to \eq\nr{dot_E_T}, we can similarly drop explicit
appearances of $\dot{\bar\varphi}$ from the coefficients, 
given that $\dot{\bar\varphi}^2_{ }\ll \bar{e}$. Then we find
\ba
 0 
 & 
 \underset{\rmii{}}
{\overset{\rmii{\nr{dot_E_T}}}{\approx}}
 & 
   \biggl\{\, 
   \partial^{ }_t 
     + 
   \overbrace{
   \frac{4\pi G(\bar e + \bar p) - \dot{H}}{H}
   }^{\nr{bg_decouple}:\; 
  {\rm becomes}\;
   - 2 \dot{H}/H
   }
   \,\biggr\}
   \overbrace{ 
   [-3 H(\bar{e} + \bar{p}) (\R^{ }_\T - \R^{ }_\varphi) ] 
   }^{ 
   \nr{def_E_T},\,\nr{bg_decouple}:\;
   {\rm was}\;\E^{ }_\T }
 \nn[2mm] 
%%%%%%
 &  & \;+\,
   \biggl\{\,
  3 H 
 \biggl( 1 + \frac{\dot{\bar{p}}}{\dot{\bar{e}}}\biggr)
   \,\biggr\} 
   \overbrace{ 
   [\dot{\bar{e}} (\R^{ }_\T - \R^{ }_\varphi) ] 
   }^{ 
   \nr{def_E_T}:\;
   {\rm was}\;\E^{ }_\T }
 \;-\;
 \frac{\nabla^2_{ }}{a^2_{ }} 
 \Bigl\{\, 
 \hspace*{-4mm}
 \overbrace{
   (\bar e + \bar p) \R^{ }_\varphi + \E^{ }_v
 }^{
 \nr{def_E_v}:\;{\rm becomes}\;
 (\bar e + \bar p) \R^{ }_v 
 }
 \hspace*{-4mm}
 \,\Bigr\}
 \nn[2mm] 
%%%%%%
 &  & 
 \,-\,
 \hspace*{-2mm}
 \overbrace{
 \biggl\{\,
 \frac{8\pi G(\bar e + \bar p)\,\bar e}{H}
 \,\biggr\} 
 }^{\nr{bg_decouple}:\;
 {\rm becomes}\;
 3 H (\bar{e} + \bar{p}) 
 }
 \hspace*{-2mm}
 \biggl\{\, 
   \dot{\R}^{ }_\varphi
  - \frac{4\pi G}{H}\,
   \overbrace{
   [(\bar e + \bar p)(\R^{ }_v - \R^{ }_\varphi)]
   }^{ 
   \nr{def_E_v}:\;{\rm was}\;
   \E^{ }_v } 
 \,\biggr\}
 \nn[3mm] 
%%%%%%
 & = &
 \bigl\{\,  
 \cancel{3} - \cancel{3} 
 \,\bigr\}
 \, 
 H (\bar e + \bar p)
 \, 
  \dot{\R}^{ }_\varphi 
 \; - \; 
  \bigl\{\,
  3 H (\bar e + \bar p )
  \,\bigr\}
 \,\dot{\R}^{ }_\T
 \nn[2mm]
%%%%%%
 &  & \;+\, 
  \biggl\{\,
    \bcancel{( 3 - 6) \dot{H}(\bar e + \bar p)}
  + 
  ( \cancel{3} - \cancel{3} ) H (\dot{\bar e} + \dot{\bar p})
  - 
     \bcancel{12\pi G (\bar e + \bar p)^2_{ }}
  \,\biggr\}
 \, \R^{ }_\varphi
 \nn[2mm]
%%%%%%
 &  & \;+\,
  \biggl\{\,
    ( 6 - 3 ) \dot{H}(\bar e + \bar p) 
  + ( \cancel{3} - \cancel{3})  H (\dot{\bar e} + \dot{\bar p})
  \,\biggr\}
 \,\R^{ }_\T
 \nn[2mm]
%%%%%%
 &  & \;+\, 
 \biggl\{\,
    12\pi G (\bar e + \bar p)^2_{ }
  - (\bar e + \bar p) \frac{\nabla^2_{ }}{a^2_{ }}
 \,\biggr\}
 \, \R^{ }_v
 \;. 
 \label{dot_E_T_decoupling}
\ea
Again, $\dot{\R}^{ }_\varphi$ and $\R^{ }_\varphi$ drop out. 
Dividing the remainder 
by $3 H(\bar e + \bar p)$ and using \eq\nr{bg_decouple} once more, 
\eq\nr{dot_R_T_rad} is reproduced. Therefore, we have shown 
that $\R^{ }_\varphi$ has no influence on the evolution of 
$\R^{ }_v$ and $\R^{ }_\T$, 
if $\dot{\bar\varphi}^{\ibit 2}_{ }\ll \bar{e}$.

To summarize, the quantity that we identify as the scalar curvature
perturbation, when determining observable quantities like 
$A^{ }_\scalar$ or $n^{ }_\scalar$ (cf.\ \se\ref{ss:cmb}), is the curvature
perturbation $\R^{ }_\T$ 
(cf.\ \eqs\nr{R_T_again}, \nr{zeta_again}). 
The curvature perturbation~$\R^{ }_\varphi$
is valuable as it determines the constant value of~$\R^{ }_\T$ outside
of the Hubble horizon, from the 
initial inflationary dynamics. However, 
strictly speaking $\R^{ }_\varphi$ is not the
observable that we measure in the end. 

%%%%%%%%%%%%%%%%%%%%%%%%%%%%%%%%%%%%%%%%%%%%%%%%%%%%%%%%%%%%%%%%%%%%%%%%%
%
\subsection{The origin of Jeans instability}
\label{ss:jeans} 

A physical reason for the acoustic oscillations in the CMB spectrum
that we discussed in \se\ref{ss:osc} is sometimes explained
as follows. If we have a region of overdensity, gravity pulls it together,
so that it tends to collapse. But as it does so, its pressure increases.
This prohibits a collapse, and rather leads to oscillations. 

To obtain a realistic picture, 
giving rise to proper structure formation, 
requires the addition of more ingredients. 
The first is that there are at least two matter components which
behave differently: {\em dark matter}, \index{dark matter}
which does not feel pressure
($p^{ }_\dm \ll e^{ }_\dm$), 
and Standard Model matter, which does
($p^{ }_r \approx e^{ }_r/3$). 
The second is that in the physical universe, the nature of
the background evolution changes
as the collapse proceeds, from radiation-dominated to 
matter-dominated. The energy
density of radiation in a late universe, 
$e^{ }_r$, is carried by photons and neutrinos, 
whereas that of matter, $e^{ }_m$,  
resides in dark matter, $e^{ }_\dm$, 
and in baryons, $e^{ }_{b}$. 
At temperatures well below the electron mass, 
$T \ll m^{ }_e$, we have  
\ba
 e^{ }_r 
 & 
 \underset{\rmii{\nr{def_Neff},\nr{p_r}}}
 {\overset{T\;\ll\;m^{ }_e}{\approx}}
 & 
 \frac{\pi^2_{ }T^4_{ }}{30}
 \, 
 \biggl[\,
 \overbrace{
  2
 }^{\mathrm{from}~e^{ }_\gamma}
  + 
 \overbrace{
 \frac{7}{4}\biggl(\frac{4}{11}\biggr)^{4/3}_{ } N^{ }_\rmi{eff}  
 }^{\mathrm{from}~e^{ }_\nu}
 \,\biggr]
 \;, \label{energy_budget1} \\[2mm]
%%%%%% 
 e^{ }_m 
 &
 \overset{T\,\ll\,m^{ }_e}{\approx}
 & 
 e^{ }_\dm + 
 \hspace*{-1mm}
 \underbrace{ 
 m^{ }_p\, n^{ }_\gamma \, \biggl( \frac{n^{ }_\rmii{b}}{n^{ }_\gamma} \biggr)
 }_{\hspace*{11mm} e^{ }_{b} \; (p^{ }_b \, > \, 0)}
 \hspace*{-1mm}
 \;. \label{energy_budget2}
 \index{baryon asymmetry}
 \index{matter-radiation equality}
\ea
Here $m^{ }_p$ is the proton mass, 
$n^{ }_\gamma \equiv 2\int_{\vec{p}} \nB^{ }(p) 
= 2 \zeta(3) T^3_{ }/\pi^2_{ }$
is the photon number density,  
and 
$ n^{ }_{b} / n^{ }_\gamma \approx 6.1 \times 10^{-10}_{ } $, 
referred to as {\em baryon asymmetry}, is among the most important
and mysterious
parameters of post-inflationary cosmology. Inserting a realistic 
number for dark matter, 
$
 e^{ }_\dm \simeq 5.4\, e^{ }_b
$,
{\em matter-radiation equality} is reached
at $T\sim$~eV, whereas CMB decoupling and electron-proton
recombination take place a bit later, at $T \sim 0.3\,$eV.

The ``common wisdom'' for the early stages of gravitational collapse
can now be formulated as follows. Given that dark matter does not feel
radiation pressure, perturbations in dark matter can start 
to grow earlier than in visible matter~\cite{efst}. 
Initially, 
baryons and photons oscillate independently of dark matter. 
At some point, their oscillations start to be  
influenced by dark matter-induced potential wells. 
After recombination, photons decouple and start
streaming freely, constituting the CMB. Baryons stop feeling
radiation pressure, and can therefore now collapse, 
eventually catching up with the dark matter. The 
{\em gravitational collapse} \index{gravitational collapse} 
is sometimes referred to as a {\em Jeans instability}
\index{Jeans instability}
(cf., e.g., refs.~\cite{lss_1,lss_2}).

\vspace*{3mm}

To illustrate some features of this dynamics, we consider an early epoch
in which the universe is still radiation dominated, and we follow 
a perturbation in the subdominant dark matter component
(this assumption is introduced in \eq\nr{dm_small_terms}, 
in order to decouple the evolution of radiation perturbations from 
those of dark matter, so that some qualitative features of the 
solution can be understood analytically). 
We stress that
this is just an ``anticipatory'' phase.
Later on, dark matter is expected to collapse to form
early galaxies during matter domination, 
a process commonly referred to
as {\em halo formation}. 

\index{halo formation}

\vspace*{3mm}

To proceed, let us 
generalize \eqs\nr{dot_R_v_rad} and \nr{dot_R_T_rad} to a situation
in which two different fluids are present (cf.,\ e.g.,\ ref.~\cite{wands}).
Before focussing on dark matter, we keep the equations general, so that
they may apply to other problems as well; the fluids are 
labelled with the indices $i =1,2$. 
For simplicity,
we assume that the fluids interact only via gravity, so that their 
energy-momentum tensors are conserved separately, 
\be
 T^{(i);\mu}_{\mu\nu} \;=\; 0
 \;, \qquad i = 1,2
 \;. \label{dTmunu_i}
\ee
At the background level, this implies
\be
 \bar{e}\hspace*{0.3mm}'_i + 3\H (\bar{e}^{ }_i + \bar{p}^{ }_i ) \; = \; 0 
 \;, \qquad
 i = 1,2
 \;. \label{eom_twofluids_0} 
\ee
At first order in perturbations, 
\eqs\nr{eom_vT_1} and \nr{eom_vT_2} become
\ba
  - \delta e_1' - 3 \H (\delta e^{ }_1 + \delta p^{ }_1)
 + (\bar{e}^{ }_1 + \bar{p}^{ }_1) (\, 3 h_\rmii{D}' + \nabla^2_{ } v^{ }_1 \,)
 % - \H\, \tr\barpPi 
 & \overset{\rmii{\nr{eom_vT_1}}}{=} & 0 
 \;, \label{eom_twofluids_1} \\[2mm]
%%%%%
  \delta p^{ }_1 + (\bar{e}^{ }_1 + \bar{p}^{ }_1) h^{ }_0
 + (\partial^{ }_\tau + 4\H) [(\bar{e}^{ }_1 + \bar{p}^{ }_1)(h - v^{ }_1)]
 + \frac{2}{3} \nabla^2_{ }\barpPi^{ }_1 
 & \overset{\rmii{\nr{eom_vT_2}}}{=} & 0 
 \;, \label{eom_twofluids_2}
\ea
and similarly for $\delta e^{ }_2$ and $\delta p^{ }_2$
(for dark matter, 
$\delta e^{ }_2 \to \delta e^{ }_\dm$
and 
$\delta p^{ }_2 \to \delta p^{ }_\dm \ll \delta e^{ }_\dm$).

In order to align our discussion with the one in \se\ref{ss:evol_v_T}, 
we assume that both fluids can be parametrized by the respective temperatures,
$T^{ }_1$ and $T^{ }_2$. However, this does not need to be the case literally,
as long as there is a unique equation of state in the dark matter fluid, 
which allows us to replace $\delta T^{ }_2$ through the more physical
$\delta e^{ }_2$ in the end.
We return to a further elaboration of this point around 
\eq\nr{one_variable}.

We now express the matter and metric perturbations in terms of curvature
perturbations, defined as in \eqs\nr{def_R_v} and \nr{def_R_T}, 
as well as gauge degrees of freedom, abbreviated as in 
\eq\nr{shorthand}. If we write
$\delta p^{ }_1 = \bar{p}^{ }_{1,\T_{\!\scalebox{0.4}{1}}^{ }} \delta T^{ }_1$,
and then express $\delta T^{ }_1$ like in \eq\nr{vT_deltaT}, 
we can factor out the combination
$
 \bar{p}^{ }_{1,\T_{\!\scalebox{0.4}{1}}^{ }} T_1^{\hspace*{0.3mm}\prime} 
 = \bar{p}_1^{\hspace*{0.3mm}\prime}
$.
Therefore, 
\ba
 h - v^{ }_1 
 & \overset{\rmii{\nr{vT_hv}}}{=} & 
 \frac{\R_{v^{ }_1} + Y}{\H} 
 \;, \quad
 \delta p^{ }_1
 \; \overset{\rmii{\nr{vT_deltaT}}}{=} \; 
 - 
 \frac{\bar{p}_1^{\hspace*{0.3mm}\prime}}{\H}
 \bigl(\, \R^{ }_{\T^{ }_1} + Y \,\bigr)
 \;, \label{dmatter_1}
 \\[2mm]
%%%%%%
 \delta e^{ }_1
 & \overset{\rmii{\nr{vT_deltaT}}}{=} & 
 - 
 \frac{\bar{e}_1^{\hspace*{0.3mm}\prime}}{\H}
 \bigl(\, \R^{ }_{\T^{ }_1} + Y \,\bigr)
 \; \overset{\rmii{\nr{eom_twofluids_0}}}{=} \; 
 3 (\bar{e}^{ }_1 + \bar{p}^{ }_1)
 \bigl(\, \R^{ }_{\T^{ }_1} + Y \,\bigr)
 \;. \label{dmatter_2}
\ea

On the side of the Einstein equations, the right-hand side is
a sum of the two energy-momentum tensors. Therefore, the background
identity from \eq\nr{bg_vT} becomes
\be
 \H^2 - \H'
  \; \overset{\rmii{\nr{bg_vT}}}{=} \; 
 4 \pi G  a^2_{ }(\bar{e}^{ }_1 
 + \bar{e}^{ }_2 
 + \bar{p}^{ }_1 
 + \bar{p}^{ }_2 )
 \;, \label{bg_twofluids}
\ee
whereas $\H''$ can now be expressed as 
\be
 \H'' + 2 \H \H' - 4 \H^3_{ }
 \; \overset{\rmii{\nr{bg_vT_2}}}{=} \; 
 4\pi G a^2_{ }
 \bigl(\, 
    \bar{e}\hspace*{0.3mm}'_1 
  + \bar{e}\hspace*{0.3mm}'_2 
  - \bar{p}\hspace*{0.3mm}'_1 
  - \bar{p}\hspace*{0.3mm}'_2 
 \,\bigr)
 \;.  \label{bg_twofluids_2}
\ee
Turning to metric perturbations, $h^{ }_0$ appearing in 
\eq\nr{eom_twofluids_2} reads
\ba
 h^{ }_0 
 &
  \underset{\rmii{\nr{shorthand}}}
  {\overset{\rmii{\nr{vT_h0}} \lift }{=}} 
 & 
 - \frac{Y'}{\H}
 + \frac{ 4\pi G 
  a^2_{ } }{\H}
 \,\bigl[\,
  ( \bar{e}^{ }_1 + \bar{p}^{ }_1) (v^{ }_1 - h)
 + 
  ( \bar{e}^{ }_2 + \bar{p}^{ }_2) (v^{ }_2 - h)
 \,\bigr] 
 \nn[2mm]
%%%%%
 &
  \underset{\rmii{\nr{bg_twofluids}}}
  {\overset{\rmii{\nr{dmatter_1}} \lift }{=}} 
 & 
 - \frac{Y'}{\H}
 + 
 \biggl(
   \frac{\H'}{\H^2_{ }} - 1
 \biggr)\, Y
 - 
 \frac{ 4\pi G 
  a^2_{ } }{\H^2_{ }}
 \,\bigl[\,
  ( \bar{e}^{ }_1 + \bar{p}^{ }_1) \R_{v^{ }_1}
 + 
  ( \bar{e}^{ }_2 + \bar{p}^{ }_2) \R_{v^{ }_2}
 \,\bigr]  
  \;. \label{twofluids_h0}
\ea
For the metric perturbations appearing in 
\eq\nr{eom_twofluids_1}, we make use of \eq\nr{vT_comb}, 
and write
\ba
 \nabla^2_{ } h
 + 3 h_\rmii{D}' + h_0'
 & \overset{ \rmii{\nr{vT_comb_prime}} }{=} & 
  -  2 \biggl( 2 \H + \frac{\H'}{\H}  \biggr) h^{ }_0
  - \frac{Y'' + 2 \H Y' - \nabla^2_{ }Y}{\H}
 \nn[2mm]
%%%%
 &  & \;+\,
  \frac{ 4\pi G a^2_{ }}{\H}  
 \biggl[  
 \delta p^{ }_1 + \delta p^{ }_2 - \delta e^{ }_1 - \delta e^{ }_2 
   + \frac{2}{3} \nabla^2_{ } \bigl( \barpPi^{ }_1 + \barpPi^{ }_2 \bigr)  
 \biggr]
 \;, \hspace*{7mm} \label{twofluids_comb_prime} 
\ea
as well as 
\ba
 h_0' & \overset{\rmii{\nr{twofluids_h0}}}{=} & 
 -\frac{Y''}{\H}
 + \biggl( \frac{2\H'}{\H^2_{ }} - 1 \biggr) \,Y'
 + \biggl( \frac{\H''}{\H^2_{ }} 
         - \frac{2 \H^{\prime\hspace*{0.3mm}2}_{ }}{\H^3_{ }} \biggr) \,Y
 \nn[2mm]
%%%%%%
 &  & \;+\,
 2 \biggl( \frac{\H'}{\H} - \H \biggr)
 \frac{4\pi G a^2_{ }}{\H^2_{ }}
 \bigl[\, 
 (\bar{e}^{ }_1 + \bar{p}^{ }_1) \R_{v^{ }_1}
 + 
 (\bar{e}^{ }_2 + \bar{p}^{ }_2) \R_{v^{ }_2}
 \,\bigr]
 \label{twofluids_h0_prime} \\[2mm]
%%%%%%
 &  & \;-\, 
 \frac{4\pi G a^2_{ }}{\H^2_{ }}
 \bigl[\, 
  (\bar{e}^{\hspace*{0.3mm}\prime}_1
 + \bar{p}^{\hspace*{0.3mm}\prime}_1) \R_{v^{ }_1}
 + 
    (\bar{e}^{ }_1 + \bar{p}^{ }_1) \R^{\prime}_{v^{ }_1}
 + 
  (\bar{e}^{\hspace*{0.3mm}\prime}_2
 + \bar{p}^{\hspace*{0.3mm}\prime}_2) \R_{v^{ }_2}
 + 
 (\bar{e}^{ }_2 + \bar{p}^{ }_2) \R^{\prime}_{v^{ }_2}
 \,\bigr]
 \;. \hspace*{8mm} \nonumber 
\ea

The task now is to substitute \eqs\nr{dmatter_1}--\nr{twofluids_h0_prime}
into \eqs\nr{eom_twofluids_1} and \nr{eom_twofluids_2}. We start with 
the latter, which is somewhat simpler. Collecting first the terms 
containing $Y$ on the left-hand side (L), we find
\ba
 \mbox{\nr{eom_twofluids_2}}^{ }_\rmii{L} & \supset & 
 \hspace*{-7mm} 
 \overbrace{
  - 
 \frac{\bar{p}_1^{\hspace*{0.3mm}\prime} Y}{\H}
 }^{{\rm from}\;\delta p^{ }_1 \;{\rm via}\;\nr{dmatter_1} }
 \hspace*{-7mm} 
  + 
  (\bar{e}^{ }_1 + \bar{p}^{ }_1)
  \biggl[\,
  \overbrace{
   -\frac{Y'}{\H} +  
  \biggl(
     \frac{\H'}{\H^2_{ }} - 1 
  \biggr) Y
  }^{{\rm from}\; h^{ }_0\;{\rm via}\;\nr{twofluids_h0}}
  \,\biggr]
  + 
  (\partial^{ }_\tau + 4\H)\biggl[\,
  \hspace*{-9mm}
  \overbrace{
  \frac{(\bar{e}^{ }_1 + \bar{p}^{ }_1) Y}{\H}
  }^{{\rm from}\; (\bar{e}^{ }_1 + \bar{p}^{ }_1)(h-v^{ }_1)
     \;{\rm via}\;\nr{dmatter_1} }
  \hspace*{-9mm}
  \,\biggr]
 \hspace*{8mm}
 \nn[2mm]
%%%%%%
 & = & 
 \biggl[\,
  - 
 \cancel{\frac{\bar{p}_1^{\hspace*{0.3mm}\prime}}{\H}}
 + 
  (\bar{e}^{ }_1 + \bar{p}^{ }_1)
  \biggl(
     \bcancel{\frac{\H'}{\H^2_{ }}}
   - 1 
   - \bcancel{\frac{\H'}{\H^2_{ }}}
   + 4 
  \biggr)
 + 
 \frac{\bar{e}_1^{\hspace*{0.3mm}\prime}
     + \cancel{\bar{p}_1^{\hspace*{0.3mm}\prime}} }{\H}
  \,\biggr]\,Y
 \;
  \overset{\rmii{\nr{eom_twofluids_0}}}{=} 
 \; 
 0
 \;. \label{twofluids_canc_1}
\ea
Proceeding to the physical terms, the substitutions yield
\ba
 \H \times 
 \mbox{\nr{eom_twofluids_2}}^{ }_\rmii{L} 
 \hspace*{-5mm}
 & 
 \overset{\scriptscriptstyle Y\;\to\; 0}{=}  
 & 
 \hspace*{-6mm}
 \overbrace{
  \;-\, 
  \bar{p}_1^{\hspace*{0.3mm}\prime} \R_{\T^{ }_1}
 }^{{\rm from}\;\delta p^{ }_1  \;{\rm via}\;\nr{dmatter_1} }
 \hspace*{-6mm}
  - 
  (\bar{e}^{ }_1 + \bar{p}^{ }_1)
  \overbrace{
  \frac{4\pi G a^2_{ } }{\H}
   \,\bigl[\,
  ( \bar{e}^{ }_1 + \bar{p}^{ }_1) \R_{v^{ }_1}
 + 
  ( \bar{e}^{ }_2 + \bar{p}^{ }_2) \R_{v^{ }_2}
 \,\bigr]  
  }^{{\rm from}\; h^{ }_0  \;{\rm via}\;\nr{twofluids_h0} }
 \nn[2mm]
%%%%%%
 &  & \;+\,
  \H\,
  (\partial^{ }_\tau + 4\H)\biggl[\,
  \hspace*{-8mm}
  \underbrace{
  \frac{(\bar{e}^{ }_1 + \bar{p}^{ }_1) \R_{v^{ }_1}}{\H}
  }_{{\rm from}\; (\bar{e}^{ }_1 + \bar{p}^{ }_1)(h-v^{ }_1)
    \;{\rm via}\;\nr{dmatter_1} }
  \hspace*{-8mm}
  \,\biggr]
 + \frac{2\H}{3} \nabla^2_{ }\barpPi^{ }_1 
 \nn[2mm]
%%%%%%
 & 
 \hspace*{-8mm}
 \overset{\scriptscriptstyle 
       \R_{v^{ }_\rmiii{2}} 
 \,=\, \R_{v^{ }_\rmiii{1}} \,+\, 
       \R_{v^{ }_\rmiii{2}} \,-\, 
       \R_{v^{ }_\rmiii{1}} \hspace*{5mm} }{
 \overset{\hphantom{|}}{=}} 
 \hspace*{-8mm}
 &
 \;+\,\bar{p}_1^{\hspace*{0.3mm}\prime} 
 \bigl(\, \R_{v^{ }_1} -  \R_{\T^{ }_1} \,\bigr)
 - 
  \frac{4\pi G a^2_{ }
  (\bar{e}^{ }_1 + \bar{p}^{ }_1)
  (\bar{e}^{ }_2 + \bar{p}^{ }_2) }{\H}
  \bigl(\, \R_{v^{ }_2} - \R_{v^{ }_1} \,\bigr)
 \nn[2mm]
%%%%%%
 &  & \;+\,  
 (\bar{e}^{ }_1 + \bar{p}^{ }_1) \R^{\prime}_{v^{ }_1}
 + \frac{2\H}{3} \nabla^2_{ }\barpPi^{ }_1 
 \label{twofluids_simpl_1} \\[2mm]
%%%%%
 &  & \;+\, 
 \biggl[
 \underbrace{
 \bar{e}^{\hspace*{0.3mm}\prime}_1
 + 
 (\bar{e}^{ }_1 + \bar{p}^{ }_1) 
 \biggl(
 \overbrace{
  - \H 
  + \bcancel{\frac{\H'}{\H}} 
 }^{{\rm via}\;\nr{bg_twofluids}}
  - \bcancel{\frac{\H'}{\H}} 
  + 4\H
 \biggr)}_{ 
  {\rm vanishes~due~to}\;\nr{eom_twofluids_0}
  } \biggr] 
 \R_{v^{ }_1}
 \; \overset{\rmii{\nr{eom_twofluids_2}}}{=} \; 0
 \;. \nonumber 
\ea
Finally, we transform to 
physical time, with $\R^{\prime}_{v^{ }_1} = a \dot{\R}^{ }_{v^{ }_1}$ and
$\H = a H$, and to comoving momentum space. 
Dividing the whole by
$ a( \bar{e}^{ }_1 + \bar{p}^{ }_1 ) $, 
yields the result given in \eq\nr{twofluids_dot_R_v}.

\vspace*{3mm}

Let us then proceed to \eq\nr{eom_twofluids_1}. For the terms 
containing $Y$, we find
\ba
 \mbox{\nr{eom_twofluids_1}}^{ }_\rmii{L} & \supset & 
 \overbrace{
  -  
 \bcancel{
 3 (
 \bar{e}_1^{\hspace*{0.3mm}\prime} 
 + 
 \bar{p}_1^{\hspace*{0.3mm}\prime} 
 )\,Y }
  - 3 (
 \bar{e}_1^{ } 
 + 
 \bar{p}_1^{ } 
 )\,Y' 
 }^{-\delta e^{\prime}_1\;{\rm via}\;\nr{dmatter_2} }
 + 
 \hspace*{-4mm}
 \overbrace{
  \bcancel{
  3 (
 \bar{e}_1^{\hspace*{0.3mm}\prime} 
 + 
 \bar{p}_1^{\hspace*{0.3mm}\prime} 
 )\,Y}
 }^{\delta e^{ }_1 + 
    \delta p^{ }_1\;{\rm via}\;\nr{dmatter_1},\,\nr{dmatter_2} }
 \hspace*{-4mm}
 \nn[2mm]
%%%%%%
 &  & \;+\,
  (\bar{e}^{ }_1 + \bar{p}^{ }_1)
  \biggl\{
    \hspace*{-5mm}
    \overbrace{
      -  \bcancel{\frac{\nabla^2_{ }Y}{\H}}
    }^{\nabla^2_{ }(v_1 - h)\;{\rm via}\;\nr{dmatter_1}}
    \hspace*{-5mm}
  + 
   \overbrace{
   2 \biggl(2\H + \frac{\H'}{\H} \biggr)
   \biggl[
       \frac{Y'}{\H} +  
  \biggl(
     1 - \frac{\H'}{\H^2_{ }} 
  \biggr) Y
   \biggr]
   }^{{\rm from}\;h^{ }_0\;{\rm in}\;\nr{twofluids_comb_prime}
   \;{\rm via}\;\nr{twofluids_h0}}
 \nn[2mm]
%%%%%%
 & & \;-\, 
   \overbrace{
   \frac{\cancel{Y''} + 2 \H Y' - \bcancel{\nabla^2_{ }Y}}{\H}
   }^{ {\rm from}\;\nr{twofluids_comb_prime} }
  + 
  \overbrace{
  \frac{4\pi G a^2_{ }
  ( 
    \bar{e}\hspace*{0.3mm}'_1 
  + \bar{e}\hspace*{0.3mm}'_2 
  - \bar{p}\hspace*{0.3mm}'_1 
  - \bar{p}\hspace*{0.3mm}'_2 
  )
  }{\H^2}\,Y
   }^{{\rm from}\;\delta p^{ }_i,\,\delta e^{ }_i\;{\rm in}
  \;\nr{twofluids_comb_prime}}
 \nn[2mm]
%%%%%%
 &  & \;+\,
  \underbrace{
   \cancel{\frac{Y''}{\H}}
 + \biggl( 1 - \frac{2\H'}{\H^2_{ }}  \biggr) \,Y'
 + \biggl( \frac{2 \H^{\prime\hspace*{0.3mm}2}_{ }}{\H^3_{ }}
         - \frac{\H''}{\H^2_{ }}  \biggr) \,Y
  }_{ 
   {\rm from}\;-h^{\prime}_0\;{\rm via}\;\nr{twofluids_h0_prime}
  }
  \biggr\}
 \nn[2mm]
%%%%%%
 & 
 \overset{\rmii{\nr{bg_twofluids_2}}}{=} 
 & 
 (\bar{e}^{ }_1 + \bar{p}^{ }_1)\,
 \biggl\{\,
 \biggl(\,
 \bcancel{- 3 + 4} + \cancel{\frac{2\H'}{\H^2_{ }}} 
 \bcancel{- 2 + 1} - \cancel{\frac{2\H'}{\H^2_{ }}}
 \,\biggr)\,Y'
 \label{twofluids_canc_2} \\[2mm]
%%%%
 &  & \;+\,
 \biggl(\,
 \cancel{4 \H}
 - \bcancel{\frac{2\H'}{\H}}
 - \cancel{\frac{2 \H^{\prime\hspace*{0.3mm}2}_{ }}{\H^3_{ }}}
 + \frac{ \bcancel{\H''} + \bcancel{2 \H \H'} - \cancel{4 \H^3_{ }} }{\H^2_{ }}
 + \cancel{\frac{2 \H^{\prime\hspace*{0.3mm}2}_{ }}{\H^3_{ }}}
 - \bcancel{\frac{\H''}{\H^2_{ }}}
 \,\biggr)\,Y
 \,\biggr\}
 \; = \; 0
 \;. \hspace*{5mm} \nonumber 
\ea
So all appearances of $Y$  cancel, 
and therefore gauge independence has been verified.

For the physical terms, we find
\ba
 \mbox{\nr{eom_twofluids_1}}^{ }_\rmii{L} 
 \hspace*{-6mm}
 & 
 \overset{\scriptscriptstyle Y\;\to\; 0}{=}
 & 
 \overbrace{
  -  
 \bcancel{
 3 (
 \bar{e}_1^{\hspace*{0.3mm}\prime} 
 + 
 \bar{p}_1^{\hspace*{0.3mm}\prime} 
 )\,\R_{\T^{ }_1} }
  - 3 (
 \bar{e}_1^{ } 
 + 
 \bar{p}_1^{ } 
 )\,\R^{\prime}_{\T^{ }_1} 
 }^{-\delta e^{\prime}_1\;{\rm via}\;\nr{dmatter_2} }
 + 
 \hspace*{-4mm}
 \overbrace{
  \bcancel{
  3 (
 \bar{e}_1^{\hspace*{0.3mm}\prime} 
 + 
 \bar{p}_1^{\hspace*{0.3mm}\prime} 
 )\,\R_{\T^{ }_1}}
 }^{\delta e^{ }_1 + 
    \delta p^{ }_1\;{\rm via}\;\nr{dmatter_1},\,\nr{dmatter_2} }
 \hspace*{-4mm}
 \nn[2mm]
%%%%%%
 &  & \;+\,
  (\bar{e}^{ }_1 + \bar{p}^{ }_1)
  \biggl\{
    \hspace*{-5mm}
    \overbrace{
      -  \frac{\nabla^2_{ } \R_{v^{ }_1}}{\H}
    }^{\nabla^2_{ }(v_1 - h)\;{\rm via}\;\nr{dmatter_1}}
    \hspace*{-5mm}
 \nn[2mm]
%%%%%%
  &  & \;+\,  
   \overbrace{
   2 \biggl(2\H + \cancel{\frac{\H'}{\H}} \biggr)
 \frac{ 4\pi G 
  a^2_{ } }{\H^2_{ }}
 \,\bigl[\,
  ( \bar{e}^{ }_1 + \bar{p}^{ }_1) \R_{v^{ }_1}
 + 
  ( \bar{e}^{ }_2 + \bar{p}^{ }_2) \R_{v^{ }_2}
 \,\bigr]  
   }^{{\rm from}\;h^{ }_0\;{\rm in}\;\nr{twofluids_comb_prime}
   \;{\rm via}\;\nr{twofluids_h0}}
 \nn[2mm]
%%%%%%
 &  & \;+\,
  \frac{4\pi G a^2_{ }  }{\H^2}
  \,\biggl[
  \overbrace{
    (\bar{e}\hspace*{0.3mm}'_1 
  - \bar{p}\hspace*{0.3mm}'_1 )\R_{\T^{ }_1} 
  + (\bar{e}\hspace*{0.3mm}'_2 
  - \bar{p}\hspace*{0.3mm}'_2 ) \R_{\T^{ }_2} 
   }^{{\rm from}\;\delta p^{ }_i,\,\delta e^{ }_i\;{\rm in}
  \;\nr{twofluids_comb_prime}}
  + 
 \overbrace{
 \frac{2\H}{3} \nabla^2_{ } \bigl( \barpPi^{ }_1 + \barpPi^{ }_2 \bigr)
 }^{{\rm from}\;\nr{twofluids_comb_prime}}
  \,\biggr]
 \nn[2mm]
%%%%%%
 &  & \;+\,
  \underbrace{
 \biggl( 2\H - \cancel{\frac{2\H'}{\H}} \biggr)
 \frac{4\pi G a^2_{ }}{\H^2_{ }}
 \bigl[\, 
 (\bar{e}^{ }_1 + \bar{p}^{ }_1) \R_{v^{ }_1}
 + 
 (\bar{e}^{ }_2 + \bar{p}^{ }_2) \R_{v^{ }_2}
 \,\bigr]
  }_{ 
   {\rm from}\;-h^{\prime}_0\;{\rm via}\;\nr{twofluids_h0_prime}
  }
 \nn[2mm]
%%%%%%
 &  & \;+\, 
  \underbrace{
 \frac{4\pi G a^2_{ }}{\H^2_{ }}
 \bigl[\, 
  (\bar{e}^{\hspace*{0.3mm}\prime}_1
 + \bar{p}^{\hspace*{0.3mm}\prime}_1) \R_{v^{ }_1}
 + 
    (\bar{e}^{ }_1 + \bar{p}^{ }_1) \R^{\prime}_{v^{ }_1}
 + 
  (\bar{e}^{\hspace*{0.3mm}\prime}_2
 + \bar{p}^{\hspace*{0.3mm}\prime}_2) \R_{v^{ }_2}
 + 
 (\bar{e}^{ }_2 + \bar{p}^{ }_2) \R^{\prime}_{v^{ }_2}
 \,\bigr]
  }_{ 
   {\rm from}\;-h^{\prime}_0\;{\rm via}\;\nr{twofluids_h0_prime}
  }
  \biggr\}
 \nn[2mm]
%%%%%%%%%%%%%%%%%%%%%%%%%%%%%%%%%%%%%%
 & = & 
  (\bar{e}^{ }_1 + \bar{p}^{ }_1)
  \biggl\{\,
   - 3 \R^{\prime}_{\T^{ }_1}
   -  \frac{\nabla^2_{ } \R_{v^{ }_1}}{\H}
 \nn[2mm]
%%%%%%
 &  & \;+\,
 \frac{ 4\pi G 
  a^2_{ } }{\H^2_{ }}
 \,\biggl[\,
  6\H\, ( \bar{e}^{ }_1 + \bar{p}^{ }_1) \R_{v^{ }_1}
 + 
  6\H\, ( \bar{e}^{ }_2 + \bar{p}^{ }_2) \R_{v^{ }_2}
 \nn[2mm]
%%%%%%
 &  & \;+\, 
     (\bar{e}\hspace*{0.3mm}'_1 
  - \bar{p}\hspace*{0.3mm}'_1 )\R_{\T^{ }_1} 
  + (\bar{e}\hspace*{0.3mm}'_2 
  - \bar{p}\hspace*{0.3mm}'_2 ) \R_{\T^{ }_2} 
  +  \frac{2\H}{3} \nabla^2_{ } \bigl( \barpPi^{ }_1 + \barpPi^{ }_2 \bigr)
 \nn[2mm]
%%%%%%
 &  & \;+\, 
  (\bar{e}^{\hspace*{0.3mm}\prime}_1
 + \bar{p}^{\hspace*{0.3mm}\prime}_1) \R_{v^{ }_1}
 + 
    (\bar{e}^{ }_1 + \bar{p}^{ }_1) \R^{\prime}_{v^{ }_1}
 + 
  (\bar{e}^{\hspace*{0.3mm}\prime}_2
 + \bar{p}^{\hspace*{0.3mm}\prime}_2) \R_{v^{ }_2}
 + 
 (\bar{e}^{ }_2 + \bar{p}^{ }_2) \R^{\prime}_{v^{ }_2}
  \,\biggr]  
  \,\biggr\}
 \nn[2mm]
%%%%%
 & \underset{\rmii{and\;for\;}
  \scriptscriptstyle\R^{\prime}_{v^{ }_\rmiii{2} } }
 {\overset{\rmii{\nr{twofluids_simpl_1}\;for\;}
  \scriptscriptstyle\R^{\prime}_{v^{ }_\rmiii{1} } }{=}} & 
  (\bar{e}^{ }_1 + \bar{p}^{ }_1)
  \biggl\{\,
   - 3 \R^{\prime}_{\T^{ }_1}
   -  \frac{\nabla^2_{ } \R_{v^{ }_1}}{\H}
 \nn[2mm]
%%%%%%
 &  & \;+\,
 \frac{ 4\pi G 
  a^2_{ } }{\H^2_{ }}
 \,\bigl[\,
  6\H\, ( \bar{e}^{ }_1 + \bar{p}^{ }_1) \R_{v^{ }_1}
 + 
  6\H\, ( \bar{e}^{ }_2 + \bar{p}^{ }_2) \R_{v^{ }_2}
 \nn[2mm]
%%%%%%
 &  & \;+\, 
    \bar{e}\hspace*{0.3mm}'_1 ( \R_{\T^{ }_1}  + \R_{v^{ }_1} )
  + \bar{e}\hspace*{0.3mm}'_2 ( \R_{\T^{ }_2}  + \R_{v^{ }_2} ) 
  \,\bigr]  
  \,\biggr\}
 \nn[2mm]
%%%%%
 & 
  \overset{\rmii{\nr{eom_twofluids_0}}}{=}
 & 
  (\bar{e}^{ }_1 + \bar{p}^{ }_1)
  \biggl\{\,
   - 3 \R^{\prime}_{\T^{ }_1}
   -  \frac{\nabla^2_{ } \R_{v^{ }_1}}{\H}
  \label{twofluids_simpl_2} \\[2mm]
%%%%%%
 &  & \;+\,
 \frac{ 12 \pi G 
  a^2_{ } }{\H}
 \,\bigl[\,
  ( \bar{e}^{ }_1 + \bar{p}^{ }_1) ( \R_{v^{ }_1} - \R_{\T^{ }_1} )
 + 
  ( \bar{e}^{ }_2 + \bar{p}^{ }_2) ( \R_{v^{ }_2} - \R_{\T^{ }_2} )
 \,\bigr]  
 \,\biggr\}
 \; \overset{\rmii{\nr{eom_twofluids_1}}}{=} \; 0 
 \;. \hspace*{5mm} \nonumber
\ea
Going subsequently to physical time and to comoving momentum 
space, and dividing by $3 a$, we obtain \eq\nr{twofluids_dot_R_T}.

To summarize, the evolution equations for two fluids read
\begin{empheq}[box=\fbox]{align}
 \;  \vphantom{\Bigg|^b_q} 
 \dot{\R}^{ }_{v^{ }_1} 
 & \overset{\rmii{\nr{twofluids_simpl_1}}}{=} \;
 \frac{\dot{\bar{p}}^{ }_1}{\bar{e}^{ }_1+\bar{p}^{ }_1}
 \,\bigl(\, \R_{\T^{ }_1} - \R_{v^{ }_1} \,\bigr)
 + 
 \frac{2 H\, k^2_{ }\barpPi^{ }_1 }{3 (\bar{e}^{ }_1 + \bar{p}^{ }_1 )}
 + 
  \frac{4\pi G 
  (\bar{e}^{ }_2 + \bar{p}^{ }_2) }{H}
  \bigl(\, \R_{v^{ }_2} - \R_{v^{ }_1} \,\bigr)
 \;,
 \label{twofluids_dot_R_v}
 \\[3mm]
%%%
 \dot{\R}_{\T^{ }_1} & \overset{\rmii{\nr{twofluids_simpl_2}}}{=} \; 
 \frac{\R_{v^{ }_1}}{3 H} \frac{k^2_{ }}{a^2_{ }} 
 \;
 + \;
 \frac{4\pi G}{H}
 \bigl[\,
   (\bar{e}^{ }_1 + \bar{p}^{ }_1)   
  \,\bigl(\, \R_{v^{ }_1} - \R_{\T^{ }_1}  \,\bigr)
  + 
   (\bar{e}^{ }_2 + \bar{p}^{ }_2)   
  \,\bigl(\, \R_{v^{ }_2} - \R_{\T^{ }_2}  \,\bigr)
 \,\bigr]
 \;. 
 \;  \vphantom{\Bigg|^b_q} 
 \label{twofluids_dot_R_T} 
\end{empheq}
The evolution equations for $\R_{v^{ }_2}$ 
and $\R_{\T^{ }_2}$ are obtained through
the interchange $1\leftrightarrow 2$. 

We note that if we look for a stationary solution 
($\dot{\R}_{v^{ }_i} = \dot{\R}_{\T^{ }_i} = 0$) 
and modes outside of the Hubble 
horizon ($k \ll a H$), then the non-vanishing terms
in \eqs\nr{twofluids_dot_R_v} and \nr{twofluids_dot_R_T}
are proportional
to various isocurvature perturbations. There are three 
independent ones, 
$
 \R_{v^{ }_1} - \R_{\T^{ }_1}  
$, 
$
 \R_{v^{ }_2} - \R_{\T^{ }_2}  
$, 
and
$
 \R_{v^{ }_2} - \R_{v^{ }_1} 
$.
The homogeneous system that relates them 
only has a trivial solution. 
In accordance with \se\ref{ss:outside_isocurv}, 
this implies that the isocurvature perturbations
vanish outside of the Hubble horizon.

\vspace*{3mm}

We now return to the physical setting described around
\eqs\nr{energy_budget1} and \nr{energy_budget2}.
We consider a time before matter-radiation equality, so 
that $e^{ }_r \gg e^{ }_m$. 
Fluid~1 is identified as radiation and fluid~2 as dark matter.
Baryonic matter interacts with radiation and should be viewed
as part of fluid~1, however in this epoch
$e^{ }_b \ll e^{ }_r$ and 
$p^{ }_b \ll p^{ }_r$, so it plays no role in practice. 
The dark matter is assumed cold, 
with $p^{ }_\dm \approx 0$. It follows that 
\be 
 \frac{4\pi G(e^{ }_\idm + p^{ }_\idm)}{H}
 \; \approx \; 
 \frac{4\pi G e^{ }_\idm }{H}
 \; \ll \; 
 \frac{4\pi G(e^{ }_r + p^{ }_r)}{H}
 \; \overset{\rmii{\nr{bg_decouple}}}{\approx} \;
 - \frac{\dot H}{H} 
 \;. \label{dm_small_terms}
\ee
Since 
$
 {\dot{\bar{p}}^{ }_1}/({\bar{e}^{ }_1+\bar{p}^{ }_1})
 \approx \dot{T}^{ }_1/T^{ }_1
$, 
and since
$
 \dot{T}^{ }_1/T^{ }_1 \sim \dot{H}^{ }/H^{ }
$
according to \eq\nr{osc_appro_2}, 
% if $\alpha \sim 1$, 
the last terms
of \eqs\nr{twofluids_dot_R_v} and \nr{twofluids_dot_R_T} represent
small corrections as far as the evolutions of 
$\R_{v^{ }_1}$ and $\R_{\T^{ }_1}$ are concerned. Therefore, 
the radiation fluid undergoes acoustic oscillations, 
as described
in \se\ref{ss:osc} and illustrated numerically in 
\app\ref{app:num_osc} 
(cf.\ \fig\ref{fig:acoustic}(left)).

Let us then consider perturbations in the dark matter sector. 
The equations are as in \eqs\nr{twofluids_dot_R_v} and 
\nr{twofluids_dot_R_T}, after interchanging
$1\leftrightarrow 2$.
So far we have used $T^{ }_2$
to parametrize the energy density $e^{ }_2$, 
but we can now undo this parametrization, 
\be
 \frac{\delta T^{ }_2}{T_2^{\ibit\prime} }
 \; \approx \; 
 \frac{\bar{e}^{ }_{2,\T} \delta T^{ }_2} 
     { \bar{e}^{ }_{2,\T} T_2^{\ibit\prime} }
 \; = \; 
 \frac{\delta e^{ }_2}{\bar{e}^{\hspace*{0.3mm}\prime}_2}
 \;. \label{one_variable}
\ee
Therefore, in accordance with \eqs\nr{def_R_T} and \nr{def_R_e},
we can replace 
$
 \R_{\T^{ }_2} \to \R_{e^{ }_2} \equiv \R_{e^{ }_\idm}
$.
This representation is more useful than that
in terms of temperature, 
because in many models, 
dark matter particles become non-relativistic
at low temperatures, and fall out of chemical equilibrium. Then 
their energy density
fluctuations originate dominantly from those in their number
density, rather than in kinetic energy. We remark that if both energy types
are important simultaneously, 
we need to introduce both 
temperature and chemical potential 
fluctuations, and fixing the dynamics
requires a further equation, the conservation law 
for particle number, $J_\mu^{{\scriptscriptstyle (2)};\mu} = 0$. 

Working under the assumption of \eq\nr{one_variable}; 
omitting small terms in accordance
with \eq\nr{dm_small_terms}; 
and denoting 
$
  \R_{v^{ }_2} \to \R_{v^{ }_\idm}
$,
$
  \R_{v^{ }_1} \to \R^{ }_v
$,
and 
$
  \R_{\T^{ }_1} \to \R^{ }_\T
$,
the evaluation of 
\eqs\nr{twofluids_dot_R_v} and \nr{twofluids_dot_R_T}
after the interchange $1\leftrightarrow 2$ yields
\ba
 \dot{\R}_{v^{ }_\idm} 
 & \underset{\scriptscriptstyle 1\,\leftrightarrow\, 2}
   {\overset{\rmii{\nr{twofluids_dot_R_v},\nr{dm_small_terms}}
    \lift  }{\approx}} &
 \frac{2 H\, k^2_{ }\barpPi^{ }_\idm }
      {3 \bar{e}^{ }_\idm }
 + 
  \overbrace{
  \frac{4\pi G 
  (\bar{e}^{ }_r + \bar{p}^{ }_r) }{H}
  }^{{\rm from}\;\nr{bg_decouple}:\;\approx\,-\dot{H}/H }
  \bigl(\, \R^{ }_{v} - \R^{ }_{v^{ }_\idm} \,\bigr)
 \;,
 \label{dot_R_v_dm}
 \\[3mm]
%%%
 \dot{\R}_{e^{ }_\idm} 
  & \underset{\scriptscriptstyle 1\,\leftrightarrow\, 2}
   {\overset{\rmii{\nr{twofluids_dot_R_T},\nr{dm_small_terms}}
    \lift }{\approx}} & 
 \frac{\R_{v^{ }_\idm}}{3 H} \frac{k^2_{ }}{a^2_{ }} 
 \;
 + \;
 \underbrace{
 \frac{4\pi G (\bar{e}^{ }_r + \bar{p}^{ }_r)   }{H}
 }_{{\rm from}\;\nr{bg_decouple}:\;\approx\,-\dot{H}/H}
  \,\bigl(\, \R^{ }_{v} - \R^{ }_{\T}  \,\bigr)
 \;. \label{dot_R_T_dm} 
\ea
We focus on the dynamics at fairly early times, when 
$k/(a H)$ is not yet huge. Then we may 
ignore the anisotropic stress as well. 
Going over to the variables in \eq\nr{def_x}, 
\eqs\nr{dot_R_v_dm} and \nr{dot_R_T_dm} can be converted into
\ba
 \partial^{ }_x \R^{ }_{v^{ }_\idm}
 & 
  \underset{\rmii{\nr{osc_appro_2},\nr{def_x}}}
  {\overset{\rmii{\nr{dot_R_v_dm}}  \ilift  }{\approx}}
 & 
 \R^{ }_{v} - \R^{ }_{v^{ }_\idm}
 \;, \label{dx_R_v_dm} \\[2mm]
%%%%%%%%
 \partial^{ }_x  {\R}^{ }_{e^{ }_\idm} 
  & \underset{\rmii{\nr{koaH_out},\nr{def_x}}}{
     \overset{\rmii{\nr{dot_R_T_dm}}  \ilift  }{\approx}} & 
  \R^{ }_{v} - \R^{ }_{\T}
  \;+\;\frac{\alpha\, e^{2(1-\alpha)x}_{ }}{3}
 \biggl( \frac{k}{a H} \biggr)^2_\rmii{out}
 \, \R^{ }_{v^{ }_\idm}
 \;. \label{dx_R_T_dm}
\ea

How does the solution of \eqs\nr{dx_R_v_dm} and \nr{dx_R_T_dm}
differ from that of \eqs\nr{dx_Rv} and \nr{dx_RT}? From 
\eq\nr{dx_R_v_dm}, we see that 
$\R_{v^{ }_\idm}$
grows if 
$
 \R_{v^{ }_\idm} < \R^{ }_{v}
$,
and decreases if 
$
 \R_{v^{ }_\idm} > \R^{ }_{v}
$.
Therefore, 
$\R_{v^{ }_\idm}$
traces the oscillations of 
$\R^{ }_v$ (one form of matter ``drags'' the other through 
gravity), 
however the amplitude and frequency of its oscillations are smaller. 

In contrast, the evolutions of $ {\R}_{e^{ }_\idm}$
and $\R^{ }_\T$ differ considerably from each other.   
Whereas in \eq\nr{dx_RT} there is a term  
counteracting the growth of $\R^{ }_\T$, 
there is no such term for 
$ {\R}_{e^{ }_\idm} $ in \eq\nr{dx_R_T_dm}.
All the individual 
terms on the right-hand side of \eq\nr{dx_R_T_dm} oscillate. 
Though it is not easy to see this without a numerical solution, 
their sum does not average to zero, but rather oscillates 
around a constant positive value. 
Therefore, $ {\R}_{e^{ }_\idm} $ grows linearly in $x$.
This implies logarithmic growth in $t$ (cf.\ \eq\nr{def_x}), 
and $\sim \ln^2_{ }(t/t^{ }_\rmi{out})$ for the corresponding
power spectrum, $\P^{ }_{\scriptscriptstyle \R_{e_\rmiii{dm}}}$
(the growth is illustrated numerically in 
\app\ref{app:num_osc}, 
cf.\ \fig\ref{fig:acoustic}(right)).
Such a growth means that we soon leave the domain of validity
of linear perturbation theory, and a non-linear analysis 
becomes necessary. 

Let us stress that in 
structure formation, the dark matter energy density  
becomes the dominant component after 
matter-radiation equality 
(cf.\ \eqs\nr{energy_budget1} and \nr{energy_budget2}), 
and then the approximations in \eq\nr{dm_small_terms} 
break down. 
However, \eqs\nr{twofluids_dot_R_v} and 
\nr{twofluids_dot_R_T} and their 
interchange $1\leftrightarrow 2$ 
(with $\R^{ }_{T^{ }_2} \to \R^{ }_{e^{ }_2}$)
remain valid, as long as we stay
in the linear regime. 

%%%%%%%%%%%%%%%%%%%%%%%%%%%%%%%%%%%%%%%%%%%%%%%%%%%%%%%%%%%%%%%%%%%%%%%%%
%
\subsection{Numerical softwares for the physical multicomponent universe}
\label{ss:evol_many} 

\index{evolution equations: physical universe}

In \se\ref{ss:osc}, we have solved for gauge-invariant temperature
and flow-velocity perturbations for a system consisting of just one plasma, 
and in \se\ref{ss:jeans}, we have added a pressureless fluid to mimic 
the presence of dark matter.
However, this is not yet enough to capture the real world. 
As discussed in \ses\ref{ss:cmb} and \ref{ss:history}, 
the modes that are seen in the CMB re-enter inside the Hubble horizon 
only rather late in the history of the universe, around or later than 
the BBN epoch ($T \sim 0.1$~MeV, cf.\ p.~\pageref{bbn}). By this time, 
neutrinos have decoupled from the Standard Model plasma 
(this happens at $T \sim 2$~MeV), and should be represented
as free-streaming particles. Moreover, 
dark matter has probably decoupled from the Standard Model 
plasma much earlier. The precise
moment of its decoupling is model-dependent, and in principle
it is possible that it never entered equilibrium in the first place. 
In any case, simulations of large-scale
structure formation suggest that gravitational collapse in 
the dark sector should start earlier than in the baryonic one, 
and therefore these degrees of freedom need to be handled 
separately~\cite{efst}.

To treat this complex system quantitatively, requires 
that we solve a coupled system of equations, in which the 
various degrees of freedom evolve in a gravitational background, 
and also source new metric perturbations through the Einstein
equations. The equations to be solved depend on the
degree of freedom in question. For particle-like
degrees of freedom out of equilibrium, 
like neutrinos and some models of dark matter, 
it is natural to solve a Boltzmann equation
for a phase space distribution. For very light bosonic
degrees of freedom, for instance some variants of axions, we may 
assume that they can more efficiently be described by a field 
(or a ``condensate''), and solve the corresponding field equation. 
Finally, for species that do equilibrate, like baryons or photons, 
we could consider either hydrodynamic equations as we have done
above (incorporating viscosities to handle shorter wavelengths), or 
Boltzmann equations. It is perhaps appropriate to mention that
hydrodynamic equations are valid in a narrower kinematic range
(only for the smallest momenta and frequencies), but within that
range they offer for a powerful tool, able to  
incorporate radiative corrections from particle interactions. 

\index{transfer function: scalar perturbations}

In any case, this late dynamics can be disentangled from the early one
that yields the solution outside of the Hubble horizon. 
Within the linear regime, the outcome of the late dynamics can be 
represented by a {\em transfer function}, because the 
initial condition depends on one single 
function. For instance, for 
$\R^{ }_\T$, in analogy with \eq\nr{Ps_PRT}, 
we could write
\be
 \R^{ }_\T(t,k) \; \equiv \; 
 X^{ }_{\scriptscriptstyle \R^{ }_\T}(t,t^{ }_\rmi{out},k)
 \,\R^{ }_\T(t^{ }_\rmi{out},k)
 \;, \quad
 \label{def_X}
\ee
with the initial conditions 
\be
 X^{ }_{\scriptscriptstyle \R^{ }_\T}(t,t^{ }_\rmi{out},k)
 \bigr|^{ }_{t\; =\; t^{ }_\rmii{out}}
  \; = \; 1
 \;, \qquad
 \partial^{ }_t
 X^{ }_{\scriptscriptstyle \R^{ }_\T}(t,t^{ }_\rmi{out},k)
 \bigr|^{ }_{t\; =\; t^{ }_\rmii{out}}
  \; = \; 0
 \;.
\ee
Then the power spectrum at photon 
decoupling ($t = t^{ }_\rmi{dec}$) is given by 
\be
 \P^{ }_{\mbit \scriptscriptstyle  \R^{ }_\T }(t^{ }_\rmi{dec},k) 
 \; = \; 
 \bigl[
    X^{ }_{\scriptscriptstyle \R^{ }_\T}(t^{ }_\rmi{dec},t^{ }_\rmi{out},k)
  \,\bigr]^2_{ }\,
 \P^{ }_{\mbit \scriptscriptstyle  \R^{ }_\T }(t^{ }_\rmi{out},k) 
 \;. 
\ee

\index{Einstein-Boltzmann solver}
\index{Sachs-Wolfe effect}

\vspace*{3mm}

In practice, however, there is no need to stop at photon decoupling: we 
rather want to connect the initial curvature perturbations 
to today's CMB observations
(cf.\ \eq\nr{Ps_PRT}). 
Then we need to address also the physics of the last 
Thomson scattering, which induces polarization and  
spectral distortions (cf.\ \se\ref{ss:cmb}). 
The appropriate variable 
for this is not the hydrodynamic temperature, but rather 
the microscopic photon phase-space distribution function, 
$f^{ }_{\gamma}(t,\vec{x},\vec{p})$~\cite{peebles}. 
The photon temperature may be 
viewed as a derived quantity, obtained like in the {\em Stefan-Boltzmann law}, 
from an integral over the phase space distribution, 
\be
 \frac{2\times \pi^2_{ }T^4_\gamma(t,\vec{x})}{30}
 \; 
  \overset{\rmii{\nr{p_r}}}
{\underset{\rmii{\nr{T00_exp}}}{\equiv}} 
 \;
 2\,
  \int \! \frac{{\rm d}^3_{ }\vec{p}}{(2\pi)^3_{ }} \,
 |\vec{p}| \, f^{ }_{\gamma}(t,\vec{x},\vec{p}) 
 \;. \label{stefan-boltzmann} 
 \index{Stefan-Boltzmann law}
\ee
Here the factor 2 counts the photon polarization states, which
could also be treated as separate degrees of freedom.
We remark that the choice of a microscopic Boltzmann approach, 
in terms of $f^{ }_{\gamma}(t,\vec{x},\vec{p})$, comes
with a price, namely that the framework is guaranteed to be 
correct only up to leading order in the electromagnetic 
fine-structure constant, $\alpha^{ }_\rmii{em}$, 
which controls the scattering cross section
(specifically, plasma-induced modifications to 
dispersion relations are not easy to accommodate systematically). 

The desired physical observable
is the deviation of $T^{ }_\gamma(\vec{x})$ from its average value, 
$\bar T^{ }_\gamma \equiv T^{ }_\now$
(cf.\ \se\ref{ss:cmb}), where we have dropped the time 
argument $t = t^{ }_\now$ for simplicity. 
In theoretical considerations, 
the deviation is represented in Fourier space, as
\be
 \delta T^{ }_\gamma(\vec{k})
 \; \overset{\rmii{\nr{fourier_k}}}{=} \; 
 \int_\vec{x} 
 \, \delta T^{ }_\gamma(\vec{x}) 
 \, e^{-i\vec{k}\cdot\vec{x}}_{ }
 \;. \label{deltaT_fourier}
\ee
Observations of $\delta T^{ }_\gamma$ are made as a function 
of a photon propagation direction, $\vec{n}$. \linebreak 
2-point correlators are evaluated by combining observations
in two different directions, $\vec{n}^{ }_i$, $i=1,2$.
In a first step 
we can represent the directional dependence
of $\delta T^{ }_\gamma$ as a function
of the relative angle between $\vec{n}^{ }_i$ and $\vec{k}$, defining 
$
 \cos\theta^{ }_i \equiv \vec{n}^{ }_i\cdot\vec{k} / k
$.
The angular dependence is represented as a series in 
spherical harmonics, 
$
 Y^{ }_{\ell_i m_i}( \theta^{ }_i,\phi^{ }_i )
$.
The coefficients
appearing in \eq\nr{C_l} describe
differences of two observation directions,
$\vec{n}^{ }_1\cdot\vec{n}^{ }_2 \equiv \cos\theta$, 
where the $\theta$-dependence is parametrized through the 
overall multipole, $\ell$, obtained from $\ell^{ }_1$
and $\ell^{ }_2$.
Therefore we are forced to do 
angular momentum additions in 
the domain $\ell \sim 10^3_{ }$, 
which is computationally quite expensive.
Even if this may not be obvious at first sight, 
large comoving momenta $k$ contribute to 
rapid angular variations in~$\theta$, 
and therefore to large angular momenta $\ell$. 

\index{Thomson scattering}

The equation governing the time evolution 
of $f^{ }_{\gamma}(t,\vec{x},\vec{p})$ is 
a Boltzmann equation, 
with {\em Thomson scatterings} appearing as a collision
term~\cite{peebles}. 
The Boltzmann equation is solved in an expanding  
background, in the presence of metric perturbations
(which lead to the {\em Sachs-Wolfe effect} mentioned in \se\ref{ss:cmb}). 
The softwares that have been written for this purpose
are generally referred to as 
{\em Einstein-Boltzmann solvers},  
and include as additional variables
the phase space distribution of neutrinos, 
the density of cold dark matter, 
and a tightly coupled plasma of electrons and baryons.
Perhaps a first popular solver was 
{\tt CMBFAST}~\cite{cmbfast}, 
somewhat more recent frameworks are  
{\tt CAMB}~\cite{camb} and {\tt CLASS}~\cite{class}.
However the topic is not closed, and further packages are being developed. 

The physics that the Einstein-Boltzmann solvers incorporate; 
the approximations that they adopt; as well as the
historical developments that led to the current state of the art, 
are reviewed
in one of the publications that lays out the
foundations for {\tt CAMB}~\cite{covariant}.
This reference employs a gauge independent approach, recalling
the confusions that specific gauge choices may cause.
That said, {\tt CMBFAST} and {\tt CLASS} rely
on gauge-fixed evolution equations, originally 
derived in ref.~\cite{Ma:1995ey_copy}. 
As a technical remark, we note 
that the solvers actually operate in 
conformal time, $\tau$, measured in units of Mpc.

For us, perhaps the most important issue about the solvers 
is to appreciate their choice of initial conditions. To this aim, 
a radiation-dominated era is considered (in terms of 
\eqs\nr{energy_budget1} and \nr{energy_budget2}, 
$T \equiv T^{ }_\gamma \gg \mbox{eV}$).
The initial moment is chosen  
early enough that the modes of interest are deep inside
the Hubble horizon
($|k\tau| \ll 1$,  
cf.\ \fig\ref{fig:history_tau} on p.~\pageref{fig:history_tau}).
In this domain, the evolution equations can be solved analytically, as a 
power series in $k^2_{ }\tau^2_{ }$. 
As we have seen in \se\ref{ss:outside_isocurv}, the solution 
is described by a single function, the power spectrum
related to the curvature perturbation (though, traditionally, 
a Bardeen potential is employed). 
Having fixed the initial condition, the remaining
evolution is uniquely determined. For the photon
phase-space distribution at values of~$k$ relevant for the CMB, 
a solution obtained at linear order in perturbations 
represents a good approximation all the way till today.

Finally, as alluded to at the end of \se\ref{ss:jeans}, if we consider
cold dark matter, density perturbations start to grow rapidly once they are
inside the Hubble horizon. For baryonic matter, the growth starts
after recombination, when baryons no longer feel radiation pressure. 
To treat this system, relevant for large-scale structure formation, 
we need to go beyond the linear order in perturbations. 
Determining the corresponding dynamics 
represents a field of its own, 
traditionally based on so-called 
$N$-body simulations, 
requiring high-performance computing resources. 
We remark that in recent years, 
alternative (partly more analytic) approaches have been 
developed as well. 

\vspace*{3mm}

At this point, it is appropriate to briefly return to the 
assumption that we made at the very beginning, in \se\ref{ss:coordinates}, 
namely that the background solution is homogeneous and isotropic. 
In the end, as is the case in empirical science, 
this assumption cannot be justified {\it a priori}, but needs
rather to be verified from a comparison with data. Even though
tensions always arise and cause excitement, 
at the time of writing the framework
of a homogeneous and isotropic initial background, on top of
which perturbations are solved at linear order as outlined above, 
or beyond it when larger momenta are considered,
provides for a viable and accurate 
description of the observed universe. 

%%%%%%%%%%%%%%%%%%%%%%%%%%% start appendices %%%%%%%%%%%%%%%%%%%%%%%%%%%%%%%%%

\newpage 

%%%%%%%%%%%%%%%%%%%%%%%%%%%%%%%%%%%%%%%%%%%%%%%%%%%%%%%%%%%%%%%%%%%%%%%%%
%
\subsubsection{Numerics for acoustic oscillations and Jeans instability}
\label{app:num_osc}

\addcontentsline{toc}{subsection}{\App\ref{app:num_osc}: 
Numerics for acoustic oscillations and Jeans instability}

%%%%%%%%%%%%%%%%%%%%%%%%%%%% FIGURE %%%%%%%%%%%%%%%%%%%%%%%%%%%%%%%%%
%
\begin{figure}[t]
    \centering
    \includegraphics[width=0.44\linewidth]{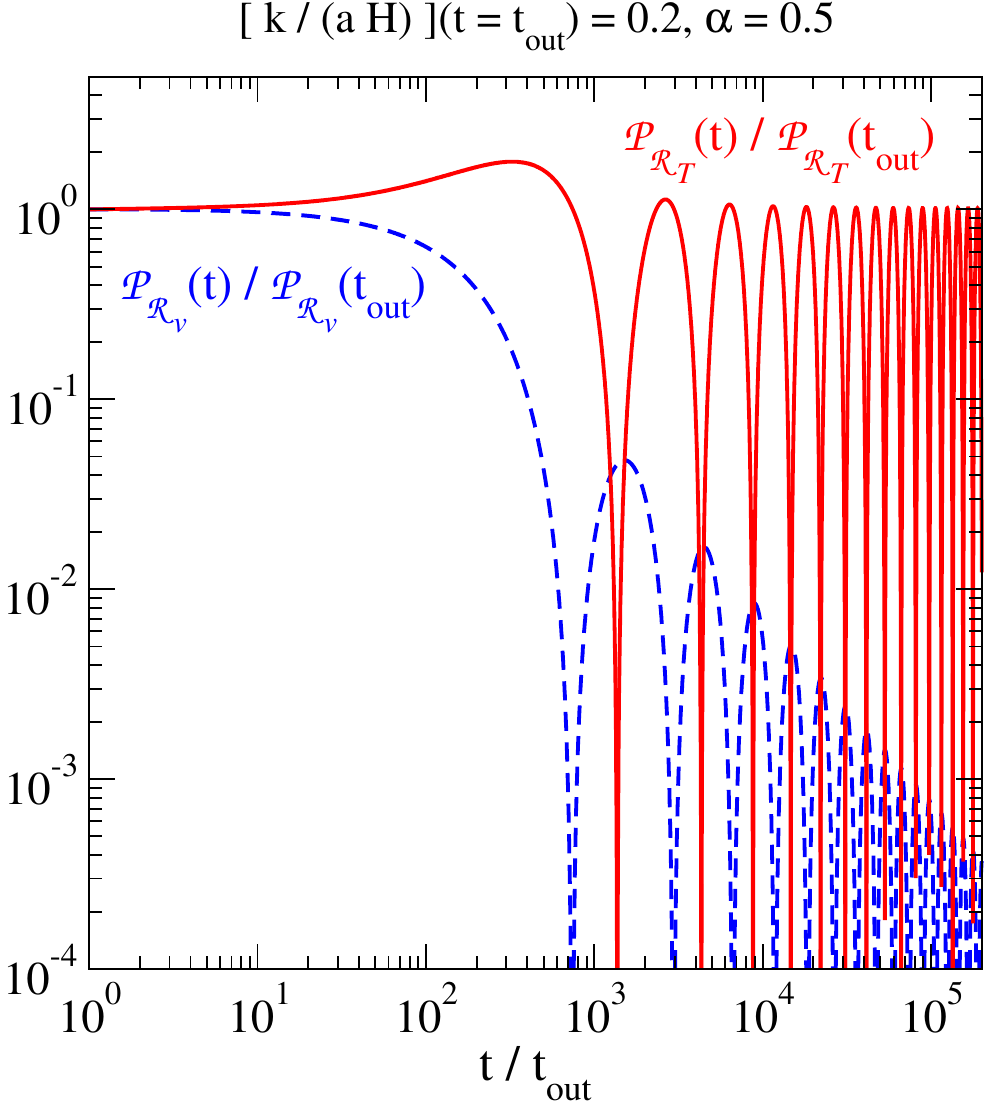}%
    \hspace*{8mm}\vspace*{-2mm}
    \includegraphics[width=0.44\linewidth]{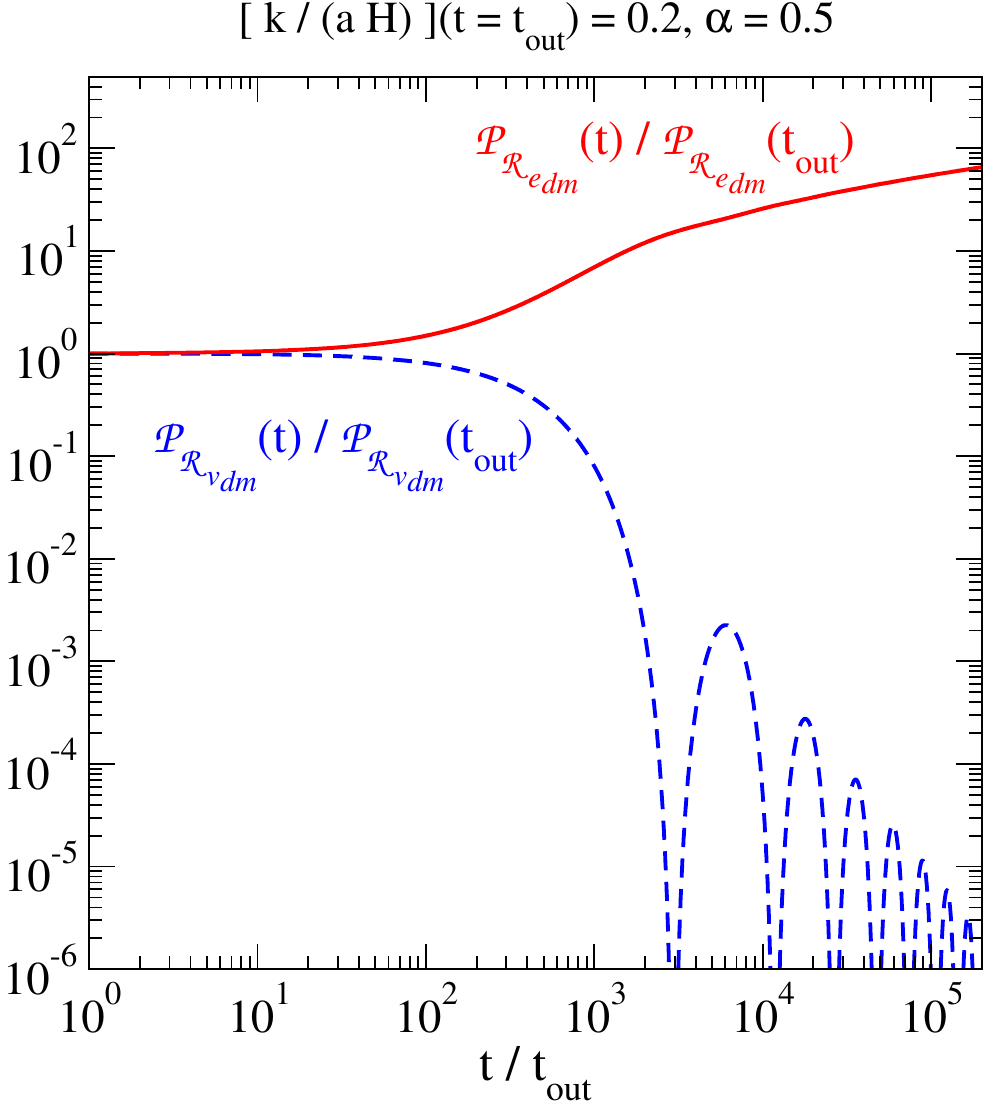}
    \caption{%
    \small
    Left: An example of a numerical solution of 
    \eqs\nr{dx_Rv} and \nr{dx_RT}, for $\alpha = \frac{1}{2}$, 
    corresponding to radiation domination
    (cf.\ \eq\nr{def_alpha}). 
    The solid curve, displaying the power spectrum 
    corresponding to $\R^{ }_\T$, can  
    be interpreted as a determination of a 
    transfer function from \eq\nr{def_X}.
%%%%
    We remark that this  
    is a good approximation to the solution of the full 
    \eqs\nr{dot_R_v_rad} and \nr{dot_R_T_rad}.
%%%%
    Right: Prototypical dark matter perturbations, 
    from \eqs\nr{dx_R_v_dm} and \nr{dx_R_T_dm}. 
    The power spectrum $\P^{ }_{\R_{e_\rmiii{dm}}}$ 
    grows as $\sim \ln^2_{ }(t/t^{ }_\rmi{out})$ in physical time, 
    once the mode is inside the Hubble horizon, however the 
    approximate forms in \eqs\nr{dx_R_v_dm} and \nr{dx_R_T_dm}
    only apply when dark matter is subdominant compared to radiation
    (otherwise the full \eqs\nr{twofluids_dot_R_v}
    and \nr{twofluids_dot_R_T} need to be solved). 
    }
    \index{acoustic oscillations: numerics (figure)} 
    \index{figure: numerics for acoustic oscillations}
    \label{fig:acoustic}
\end{figure}
%
%%%%%%%%%%%%%%%%%%%%%%%%%%%%%%%%%%%%%%%%%%%%%%%%%%%%%%%%%%%%%%%%%%%%

In this appendix, we show how \eqs\nr{dx_Rv} and \nr{dx_RT}
can be solved numerically. This amounts to a determination of
the early stage of a transfer function for scalar 
perturbations, describing the effect of acoustic oscillations. 
Simultaneously, we solve \eqs\nr{dx_R_v_dm} and \nr{dx_R_T_dm},
as an example of a dark matter component, illustrating
the origin of a Jeans instability (unattenuated growth). 
The results are illustrated in \fig\ref{fig:acoustic}.
A {\tt python} script producing these solutions is given below. 

%%%%%%%%%%%%%%%%%%%%%%%%%%%%%%%%%%%%%%%%%%%%%%%%%%%%%%%%%%%

\index{code: numerics for acoustic oscillations}

{\fontsize{8pt}{10pt}\selectfont
\begin{verbatim}
# acoustic oscillations and Jeans instability after re-entry [numerics_acoustic_jeans.py]
#
# import basic tools and integration routines
import numpy as np
from scipy.integrate import solve_ivp

# parameters 
koaH = 0.2   # initial momentum over Hubble rate
alpha = 0.5  # rate of background expansion (radiation dominated)

# sources for [Rv, RT, Rvdm, Redm]
# derivatives are taken with respect to x = ln(t/t_out)
def derivatives(x,y):
    Rv, RT, Rvdm, Redm = y
    dRv_dx = alpha*(Rv - RT)
    dRT_dx = Rv - RT + alpha/3*np.exp(2*(1-alpha)*x)*koaH*koaH*Rv
    dRvdm_dx = Rv - Rvdm
    dRedm_dx = Rv - RT + alpha/3*np.exp(2*(1-alpha)*x)*koaH*koaH*Rvdm
    dy_dx = [dRv_dx, dRT_dx, dRvdm_dx, dRedm_dx]
    return dy_dx

# initial conditions 
Rv_0 = 1.0; RT_0 = 1.0; Rvdm_0 = 1.0; Redm_0 = 1.0

# integrate up to chosen point 
sol = solve_ivp(derivatives,
                [0, 15/2/(1-alpha)],
                [Rv_0, RT_0, Rvdm_0, Redm_0],method='DOP853',
                t_eval=np.linspace(0, 15/2/(1-alpha),3000)) 
x_grid = sol.t
Rv_sol = sol.y[0]; RT_sol = sol.y[1]; Rvdm_sol = sol.y[2]; Redm_sol = sol.y[3]

# print to file
np.savetxt('numerics_acoustic_jeans.dat',
           np.c_[np.exp(x_grid),Rv_sol**2, RT_sol**2, Rvdm_sol**2, Redm_sol**2], 
           fmt='%.6e',newline='\n',
           header='columns: t/t_out, Rv^2, RT^2, Rvdm^2, Redm^2')

\end{verbatim}
}

%%%%%%%%%%%%%%%%%%%%%%% end appendices %%%%%%%%%%%%%%%%%%%%%%%%%%%

%%%%%%%%%%%%%%%%%%%%%%%%% BIBLIO %%%%%%%%%%%%%%%%%%%%%%%%%%%%%%%%
%

%%%%%%%%%%%%%%%%%%%%%%%%%%%% SECTION %%%%%%%%%%%%%%%%%%%%%%%%%%%%%%%%%%
\newpage 

\section{Gravitational-wave probes of the inflationary and reheating epochs}
\label{se:gw}

\paragraph{Abstract:}

While the motivation for inflationary cosmology comes from
scalar perturbations 
(the source of anisotropies in the CMB and of 
structure formation, cf.\ \se\ref{ss:cmb}), 
a future probe of this period might 
be offered by tensor 
perturbations, manifesting themselves as gravitational waves
(cf.\ \eq\nr{einstein_ij_t}). 
A key property of gravitational waves is that they propagate almost 
freely until present time, 
i.e.\ that their transfer function (cf.\ \eq\nr{eq_Ogwdef}) is simple. 
Apart from the inflationary epoch, we discuss 
how gravitational waves could originate at or after reheating, 
from hydrodynamic fluctuations, second-order scalar perturbations, 
or elementary particle decays and scatterings.
Many other sources have been proposed
in the literature, such as preheating, 
topological defects, or phase transitions, but these are 
strongly model-dependent, and not discussed here. 
As gravitational-wave science will grow in 
importance in the next decades, we end the book with a summary
of the various frequency domains that can 
hopefully be empirically investigated one day. 

\paragraph{Keywords:} 
gravitational-wave energy density, 
tensor perturbations, 
tensor-to-scalar ratio, 
anisotropic stress, 
scalar-induced gravitational waves, 
gravitational waves from particle scatterings and decays, 
transfer function in the tensor channel, 
gravitational strain.

%%%%%%%%%%%%%%%%%%%%%%%%%%%%%%%%%%%%%%%%%%%%%%%%%%%%%%%%%%%%%%%%%%%%%%%%%
%
\subsection{What is gravitational-wave energy density?}
\label{ss:gw_energy}

\index{gravitational-wave energy density} 

A physical manifestation of gravitational waves is that
they carry energy density. When we discuss various sources of 
gravitational waves from the early universe, 
we normalize the {\em gravitational-wave energy density}, $e^{ }_\gw$,  
\index{$e^{ }_\rmii{gw}$ (gravitational-wave energy density)}
to the current critical energy density, ${e}^{ }_\rmi{crit}$
(cf.\ \eq\nr{eq_Ogwdef}). The critical energy density is the one  
which would yield 
the {\em current Hubble rate}, $H^{ }_\now$, if $\kappa = 0$, 
\be
 H_\now^2 
 \;  
 \underset{\rmii{\nr{mpl}}}{
 \overset{\rmii{\nr{eq_end0-1}}}{\equiv}}
 \; 
 \frac{8\pi}{3} \frac{{e}^{ }_\rmii{crit}}{\mpl^2}
 \;. \label{ecrit_again}
 \index{$H^{ }_\inow$ (current Hubble rate)}
\ee
Measuring $H^{ }_\now$ is not easy, and therefore the result is often
parametrized as % (cf.\ \eq\nr{eq_Ogwdef})
\be
 H^{ }_\now 
 \;=\;  
 h \times \bar H^{ }_\now 
 \;, \quad
 \bar H^{ }_\now 
 \;\equiv\;
 100\,\tfr{\rm km}{\rm s\bit{}Mpc}
 \;, \label{def_h_red}
\ee
where the observed value now appears as 
the {\em reduced Hubble rate}, $h$.  \index{$h$ (reduced Hubble rate)}
The fractional gravitational-wave energy density is expressed as
\be
 \Omega^{ }_{\gw,\inow} 
 \;
 \equiv
 \; 
 \frac{e^{ }_{\igw,\inow}}{e^{ }_\rmii{crit}}
 \;
 \overset{\rmii{\nr{ecrit_again}}}{=}
 \; 
 \frac{8\pi e^{ }_{\igw,\inow} }{3 \mpl^2 H^{2}_\inow} 
 \;. \label{def_Omega_gw}
 \index{$\Omega^{ }_\rmii{gw}$ (gravitational-wave energy fraction)}
\ee
Given \eq\nr{def_h_red}, the quantity normally plotted
is $h^2_{ } \Omega^{ }_{\gw,\inow}$, 
or the spectrum, 
    $
     h^2_{ } {\rm d}
     \Omega^{ }_{\igw,\inow}
    / 
    {\rm d} \ln\! f^{ }_\inow
    $. 

Obtaining a formal expression for $e^{ }_\rmi{gw}$ is non-trivial. 
We first determine the gravitational contribution
to the Einstein-Hilbert action, in \eq\nr{einstein_hilbert}. Then, from 
the action, we identify the energy density. In order to arrive at this 
result, we need to expand the Ricci scalar, $R$, and the determinant
of the metric, $\sqrt{-g}$, 
to {\em second order in perturbations}. \index{second-order perturbations} 
As a simplification, we may adopt the role of an observer
rather than a source. Then we may restrict ourselves 
to modes well within the horizon, and consider a locally 
Minkowskian space-time, so that the scale factor $a$
can be taken to be constant (i.e.\ its time derivative is omitted). 
In the notation of \eq\nr{eq_metric-pert}, we therefore consider
\be
 g^{ }_{\mu\nu}
 \;
 \equiv
 \;
 \hspace*{-3mm}
 \underbrace{
 a^2_{ }
 }_{\rm constant}
 \hspace*{-1mm}
 (\eta^{ }_{\mu\nu} + h^{ }_{\mu\nu})
 \;, \quad
 \eta^{ }_{\mu\nu} \; \equiv \;
 \mathop{\rm diag}(\mbox{$-$$+$$+$$+$})
 %\mbox{diag($-$$+$$+$$+$)}
 \;. \label{g_munu_2}
\ee
The inverse metric is given by \eq\nr{inv_met}, {\it viz.}
\be
 g^{\mu\nu}_{ }
 \;
 \overset{\rmii{\nr{inv_met}}}{=}
 \; 
 a^{-2}_{ }
 \, \bigl(\, 
 \eta^{\mu\nu}_{ } - h^{\mu\nu}_{ } % + h^{\mu\lambda}_{}h^\nu_\lambda 
 \,\bigr)
% +  \ord(\delta^2_{ })
 + \ord(h_{\mu\nu}^2) 
 \;. 
 \label{inv_g_mink}
\ee

\index{Levi-Civita symbol: 4-dimensional}

Next, we compute the determinant of the metric. It can be represented
with the help of the antisymmetric {\em Levi-Civita symbol}, 
$
 \epsilon^{\mu\nu\rho\sigma}_{ }
$, 
as 
\be
 g
 \; = \;
 \det g^{ }_{\mu\nu}
 \; = \;  
 \frac{1}{24} 
 \epsilon^{\mu\nu\rho\sigma}_{ }
 \epsilon^{\alpha\beta\kappa\lambda}_{ }
 g^{ }_{\mu\alpha}\,
 g^{ }_{\nu\beta}\,
 g^{ }_{\rho\kappa}\,
 g^{ }_{\sigma\lambda}\,
 \;. \label{det_gmunu}
\ee
As discussed around \eq\nr{R_h1}, for constant $a$ the Ricci scalar
starts at first order in~$h^{ }_{\mu\nu}$, and the Einstein-Hilbert 
action contains $\sqrt{-g}\,R$, 
so we only need $g$ up to first order in~$h^{ }_{\mu\nu}$. Then 
at least three of the metric tensors are represented by $\eta^{ }_{\mu\nu}$, 
and we encounter
\ba
% \eta^{ }_{\mu\alpha}\,
% \eta^{ }_{\nu\beta}\,
% \epsilon^{\mu\nu\rho\sigma}_{ }\,
% \epsilon^{\alpha\beta\kappa\lambda}_{ }
% & = & 
% -\,2\, \bigl(  \eta^{\rho\kappa}_{ }\eta^{\sigma\lambda}_{ }
%            - \eta^{\rho\lambda}_{ }\eta^{\sigma\kappa}_{ }
%      \bigr)
% \;, \\[2mm]
%%%
 \epsilon^{\mu\nu\rho\sigma}_{ }\,
 \epsilon^{\alpha\beta\kappa\lambda}_{ }\,
 \eta^{ }_{\mu\alpha}\,
 \eta^{ }_{\nu\beta}\,
 \eta^{ }_{\rho\kappa}
 & = & 
 -\,6\, \eta^{\sigma\lambda}_{ }
 \;, \label{leci_3} \\[2mm]
%%%
 \epsilon^{\mu\nu\rho\sigma}_{ }\,
 \epsilon^{\alpha\beta\kappa\lambda}_{ }\,
 \eta^{ }_{\mu\alpha}\,
 \eta^{ }_{\nu\beta}\,
 \eta^{ }_{\rho\kappa}\,
 \eta^{ }_{\sigma\lambda}
 & = & 
 -\,24
 \;. \label{leci_4}
\ea
It follows that 
\ba
 g
 &
 \underset{\rmii{\nr{det_gmunu}}}{
 \overset{\rmii{\nr{g_munu_2}} \lift }{=}}
 & 
 \frac{a^8_{ }}{24} 
 \epsilon^{\mu\nu\rho\sigma}_{ }
 \epsilon^{\alpha\beta\kappa\lambda}_{ }
 \bigl(\, \eta^{ }_{\mu\alpha} + h^{ }_{\mu\alpha} \,\bigr)
 \bigl(\, \eta^{ }_{\nu\beta} + h^{ }_{\nu\beta} \,\bigr)
 \bigl(\, \eta^{ }_{\rho\kappa} + h^{ }_{\rho\kappa} \,\bigr)
 \bigl(\, \eta^{ }_{\sigma\lambda} + h^{ }_{\sigma\lambda} \,\bigr)
 \nn[2mm]
%%%%%%%%%%%%%%%
 &
 \underset{\rmii{\nr{leci_4}}}{
 \overset{\rmii{\nr{leci_3}} \lift }{=}}
 & 
 \frac{a^8_{ }}{24} 
 \bigl(\, 
  -24
  - 4 \times 6\, \eta^{\sigma\lambda}_{ } h^{ }_{\sigma\lambda}
 % - 6 \times 2\,
 %     \bigl(  \eta^{\rho\kappa}_{ }\eta^{\sigma\lambda}_{ }
 %           - \eta^{\rho\lambda}_{ }\eta^{\sigma\kappa}_{ }
 %     \bigr) h^{ }_{\rho\kappa} h^{ }_{\sigma\lambda}
 \,\bigr) + \ord(h^2_{\mu\nu})
 \; = \; 
 - a^8_{ }\,\bigl(\, 
 1 + h^\alpha_\alpha
% + \tfr{1}{2}
% \bigl(\, h^\alpha_\alpha h^\beta_\beta
%   - h^\alpha_\beta h^\beta_\alpha \,\bigr)
 \,\bigr)
 + \ord(h^2_{\mu\nu})
 \;. \label{det_gmunu_exp1}
\ea
The square root reads
\ba
 \sqrt{-g}
% & = & 
% a^4_{ }\,\Bigl[\, 
% 1 + h^\alpha_\alpha
% + \tfr{1}{2}
% \bigl(\, h^\alpha_\alpha h^\beta_\beta
%   - h^\alpha_\beta h^\beta_\alpha \,\bigr)
% \,\Bigr]^{\frac{1}{2}}_{ }
% + \ord(h^3_{\mu\nu}) 
% \nn[2mm]
%%%%
 &
 \overset{\rmii{\nr{det_gmunu_exp1}}}{=}
 & 
 a^4_{ }\,\bigl(\, 
 1 
 + \tfr{1}{2} h^\alpha_\alpha
% - \tfr{1}{8} h^\alpha_\alpha h^\beta_\beta
% + \tfr{1}{4}
% \bigl(\, h^\alpha_\alpha h^\beta_\beta
%   - h^\alpha_\beta h^\beta_\alpha \,\bigr)
 \,\bigr)
 + \ord(h^2_{\mu\nu})
 \;. \label{sqr_g}
\ea

Next, we compute 
the {\em Christoffel symbols} from  \index{Christoffel symbols} 
\eq\nr{eq_ChSy-a}, 
\ba
 {\Gamma}^\rho_{\mu\nu}
 &
 \overset{\rmii{\nr{eq_ChSy-a}}}{=}
 &
 \frac{1}{2} 
 {g}^{\rho\sigma}_{ } 
 \bigl(\, 
    {g}_{\sigma\mu,\nu}^{ }
  + {g}_{\sigma\nu,\mu}^{ }
  - {g}_{\mu\nu,\sigma}^{ } 
 \,\bigr)
 \nn[2mm] 
%%%%%%
 &
 \underset{\rmii{\nr{inv_g_mink}}}{
 \overset{\rmii{\nr{g_munu_2}} \lift }{=}}
 & 
 \frac{1}{2}
 \bigl(\,
  \eta^{\rho\sigma}_{ } - h^{\rho\sigma}_{ } 
 \,\bigr)
 \bigl(\,
    {h}_{\sigma\mu,\nu}^{ }
  + {h}_{\sigma\nu,\mu}^{ }
  - {h}_{\mu\nu,\sigma}^{ } 
 \,\bigr)
 + \ord(h_{\mu\nu}^3)
 \;. 
 \label{chr_h2}
\ea
With their help, the {\em Ricci scalar}  \index{Ricci scalar}
follows from 
\eqs\nr{eq_rt-a} and \nr{bg_ricci_scalar}, 
\ba
 R 
 & \underset{\rmii{\nr{bg_ricci_scalar}}}
  {\overset{\rmii{\nr{eq_rt-a}} \lift }{=}} & 
 g^{\mu\nu}_{ }
 \bigl(\,  {\Gamma}^\alpha_{\mu\nu,\alpha}
 -  {\Gamma}^\alpha_{\mu\alpha,\nu}
 +  {\Gamma}^\beta_{\mu\nu} {\Gamma}^\alpha_{\beta\alpha}
 -  {\Gamma}^\beta_{\mu\alpha} {\Gamma}^\alpha_{\beta\nu}
 \,\bigr)
 \nn[2mm]
%%%%
 & \overset{\rmii{\nr{inv_g_mink}}}{=} & 
 a^{-2}_{ }
 \bigl(\,
  \eta^{\mu\nu}_{ } - h^{\mu\nu}_{ } 
 \,\bigr)
 \bigl(\,  {\Gamma}^\alpha_{\mu\nu,\alpha}
 -  {\Gamma}^\alpha_{\mu\alpha,\nu}
 +  {\Gamma}^\beta_{\mu\nu} {\Gamma}^\alpha_{\beta\alpha}
 -  {\Gamma}^\beta_{\mu\alpha} {\Gamma}^\alpha_{\beta\nu}
 \,\bigr)
 + \ord(h_{\mu\nu}^3)
 \;. \label{R_h2} \hspace*{6mm}
\ea

Let us start at $\ord(h^{ }_{\mu\nu})$. The Christoffel symbols become
\be
 {\Gamma}^\rho_{\mu\nu}
 \; \overset{\rmii{\nr{chr_h2}}}{=} \;
 \frac{1}{2}
 \bigl(\,
    {h}_{\mu,\nu}^{\rho}
  + {h}_{\nu,\mu}^{\rho}
  - {h}_{\mu\nu}^{,\rho } 
 \,\bigr)
 + \ord(h_{\mu\nu}^2)
 \;, \label{G_h1}
\ee
and then 
\ba
 a^2_{ } R 
 &
 \overset{\rmii{\nr{R_h2}}}{=}
 &
  \eta^{\mu\nu}_{ } 
 \bigl(\,  {\Gamma}^\alpha_{\mu\nu,\alpha}
 -  {\Gamma}^\alpha_{\mu\alpha,\nu}
 \,\bigr)
 + \ord(h_{\mu\nu}^2)
 \label{pre_R_h1} \\[2mm]
%%%%
 &
 \overset{\rmii{\nr{G_h1}}}{=}
 & 
 \frac{1}{2}
   \eta^{\mu\nu}_{ } 
 \bigl(\,
    \cancel{{h}_{\mu,\nu\alpha}^{\alpha}}
  + {h}_{\nu,\mu\alpha}^{\alpha}
  - {h}_{\mu\nu,\alpha}^{,\alpha} 
  - \cancel{{h}_{\mu,\alpha\nu}^{\alpha}}
  - {h}_{\alpha,\mu\nu}^{\alpha}
  + {h}_{\mu\alpha,\nu}^{,\alpha} 
 \,\bigr) 
 + \ord(h_{\mu\nu}^2)
 \nn[2mm]
%%%%
 & = & 
 \frac{1}{2}
 \bigl(\,
    {h}_{\nu,\alpha}^{\alpha,\nu}
  - {h}_{\mu,\alpha}^{\mu,\alpha} 
  - {h}_{\alpha,\mu}^{\alpha,\mu}
  + {h}_{\alpha,\nu}^{\nu,\alpha} 
 \,\bigr) 
 + \ord(h_{\mu\nu}^2)
 \nn[2mm]
%%%%
 & = & 
    {h}_{\nu,\alpha}^{\alpha,\nu}
  - {h}_{\mu,\alpha}^{\mu,\alpha} 
 + \ord(h_{\mu\nu}^2)
 \;.  \label{R_h1}
\ea
When we multiply this with the square root from \eq\nr{sqr_g},
we find 
\be
 \sqrt{-g}\,R 
 \; \underset{\rmii{\nr{R_h1}}}{\overset{\rmii{\nr{sqr_g}}}{=}} \;  
 a^2_{ } \bigl(\,  {h}_{\nu,\alpha}^{\alpha,\nu}
  - {h}_{\mu,\alpha}^{\mu,\alpha} \,\bigr)
 + \ord(h_{\mu\nu}^2)
 \;. \label{S_E-H_1}
\ee
However, for constant $a$, 
this is a total derivative, 
and plays no physical role in \eq\nr{einstein_hilbert}. 

We then proceed to $\ord(h_{\mu\nu}^2)$. The first non-trivial 
term originates when we multiply \eq\nr{R_h1} with the 
first-order contribution from \eq\nr{sqr_g}, yielding
\ba
 \sqrt{-g}\,R 
 & \underset{\rmii{\nr{R_h1}}}{\overset{\rmii{\nr{sqr_g}}}{\supset}} &  
 \frac{1}{2}
  a^2_{ }\, h^\beta_\beta \, 
 \bigl(\,  {h}_{\nu,\alpha}^{\alpha,\nu}
  - {h}_{\mu,\alpha}^{\mu,\alpha} \,\bigr)
 \nn[2mm]
 & \overset{\rmii{IBP}}{=} & 
 \frac{1}{2}
  a^2_{ }\, 
 \bigl(\,  
   {h}^{\beta}_{\beta,\alpha}  {h}_{\mu}^{\mu,\alpha} 
  -  {h}^{\beta,\nu}_{\beta} {h}_{\nu,\alpha}^{\alpha}
 \,\bigr)
 \;. \label{gR_1}
\ea
Here IBP stands 
for {\em integration by parts}
(i.e.\ the omission of total derivatives, 
leading to boundary terms in \eq\nr{einstein_hilbert}), 
\index{IBP (integration by parts)}
which we employ in order to set the
expression in a form in which only single derivatives appear. 
The second term 
comes by replacing $\eta^{\mu\nu}_{}\to -h^{\mu\nu}_{}$
in \eq\nr{pre_R_h1}, producing 
\ba
 \sqrt{-g}\, R 
 & \underset{\rmii{\nr{pre_R_h1}}}
  {\overset{\rmii{\nr{inv_g_mink}}}{\supset}} &
 - \frac{1}{2} a^2_{ }
   h^{\mu\nu}_{ } 
 \bigl(\,
    {h}_{\nu,\mu\alpha}^{\alpha}
  - {h}_{\mu\nu,\alpha}^{,\alpha} 
  - {h}_{\alpha,\mu\nu}^{\alpha}
  + {h}_{\mu\alpha,\nu}^{,\alpha} 
 \,\bigr) 
 \nn[2mm]
%%%%
 & = & 
 - \frac{1}{2} a^2_{ }
   h^{\nu}_{\mu} 
 \bigl(\,
    {h}_{\nu,\alpha}^{\alpha,\mu}
  - {h}_{\nu,\alpha}^{\mu,\alpha} 
  - {h}_{\alpha,\nu}^{\alpha,\mu}
  + {h}_{\alpha,\nu}^{\mu,\alpha} 
 \,\bigr) 
 \nn[2mm]
%%%%
 & \overset{\rmii{IBP}}{=} & 
 \frac{1}{2} a^2_{ }
 \bigl(\,
    2 h^{\nu,\mu}_{\mu} 
    {h}^{\alpha}_{\nu,\alpha}
  -
    h^{\nu,\alpha}_{\mu} 
    {h}^{\mu}_{\nu,\alpha} 
  -
    h^{\nu,\mu}_{\mu} 
    {h}^{\alpha}_{\alpha,\nu}
 \,\bigr) 
 \;,  \label{gR_2}
\ea
where in the last step we have renamed indices, 
in order to combine identical structures. 
The third term originates from 
the contributions quadratic in Christoffel symbols
in \eq\nr{R_h2}, which after the insertion of \eq\nr{G_h1} give
\ba
 \sqrt{-g}\, R 
 & \underset{\rmii{\nr{G_h1}}}{\overset{\rmii{\nr{R_h2}}}{\supset}} &
 \frac{1}{4} a^2_{ } \eta^{\mu\nu}_{ }
 \Bigl[ 
 \bigl(\,
    {h}_{\mu,\nu}^{\beta}
  + {h}_{\nu,\mu}^{\beta}
  - {h}_{\mu\nu}^{,\beta} 
 \,\bigr)
 \bigl(\,
    \cancel{ {h}_{\beta,\alpha}^{\alpha} }
  + {h}_{\alpha,\beta}^{\alpha}
  - \cancel{ {h}_{\beta\alpha}^{,\alpha} }
 \,\bigr)
 \nn[2mm]
 & & \hspace*{8mm}
 -
 \bigl(\,
    {h}_{\mu,\alpha}^{\beta}
  + {h}_{\alpha,\mu}^{\beta}
  - {h}_{\mu\alpha}^{,\beta} 
 \,\bigr)
 \bigl(\,
    {h}_{\beta,\nu}^{\alpha}
  + {h}_{\nu,\beta}^{\alpha}
  - {h}_{\beta\nu}^{,\alpha} 
 \,\bigr)
 \Bigr]
 \nn[2mm]
 & = & 
 \frac{1}{4} a^2_{ } 
 \Bigl[ 
 \bigl(\,
    2 {h}_{\mu}^{\beta,\mu}
  - {h}_{\mu}^{\mu,\beta} 
 \,\bigr)
  {h}_{\alpha,\beta}^{\alpha}
% \nn[2mm]
% & & \hspace*{8mm}
 -
 \bigl(\,
    {h}_{\mu,\alpha}^{\beta}
  + {h}_{\alpha,\mu}^{\beta}
  - {h}_{\mu\alpha}^{,\beta} 
 \,\bigr)
 \bigl(\,
    {h}_{\beta}^{\alpha,\mu}
  + {h}_{,\beta}^{\alpha\mu}
  - {h}_{\beta}^{\mu,\alpha} 
 \,\bigr)
 \Bigr]
 \nn[2mm]
 & \overset{\rmii{IBP}}{=} & 
 \frac{1}{4} a^2_{ } \Bigl[\,
 h^{\beta,\mu}_{\mu} h^{\alpha}_{\alpha,\beta}
 \bigl( +2 \bigr)
 + 
 h^{\mu,\beta}_{\mu} h^{\alpha}_{\alpha,\beta}
 \bigl( -1 \bigr)
 +
 h^{\beta,\mu}_{\mu} h^{\alpha}_{\beta,\alpha}
 \bigl( -1-\cancel{1}-\cancel{1}+\cancel{1}+\cancel{1}-1\bigr)
 \nn[2mm]
%%%%%
 & & \hspace*{6mm}
 +
 h^{\beta}_{\mu,\alpha} h^{\mu,\alpha}_{\beta}
 \bigl( +\cancel{1} -\cancel{1} \bigr)
 + 
 h^{,\beta}_{\mu\alpha} h^{\mu\alpha}_{,\beta}
 \bigl( +1 \bigr)
 \,\Bigr]
 \;. \label{gR_3}
\ea
In principle there is also a fourth term, originating
by inserting the $\ord(h_{\mu\nu}^2)$ expressions 
of $\Gamma^\alpha_{\mu\nu}$ and $\Gamma^\alpha_{\mu\alpha}$ 
in \eq\nr{pre_R_h1}, however it is a total derivative, and can 
therefore be omitted. 

Now we can sum together \eqs\nr{gR_1}--\nr{gR_3}. 
Relabelling indices, and lowering and raising them where it makes
the appearance more elegant (which also permits for us to see that
the last two structures in \eq\nr{gR_3} are equivalent), 
but retaining the order of appearances from \eq\nr{gR_3}, 
we get~\cite{isa,choi} 
\ba
 \sqrt{-g}\, R 
 &
 \overset{\rmii{\nr{gR_1}--\nr{gR_3}} \lift }{=}
 &
 \frac{1}{4} a^2_{ } \Bigl[\,
 h^{\mu,\beta}_{\beta} h^{\alpha}_{\alpha,\mu}
 \overbrace{\bigl( -\,2-2+2 \bigr)}^{-2}
 + 
 h^{\beta,\mu}_{\beta} h^{\alpha}_{\alpha,\mu}
 \overbrace{\bigl( +\,2+0-1 \bigr)}^{+1}
 \label{gR_sum}  \\[2mm]
%%%%%
 & & \hspace*{6mm}
 + \,
 h^{\mu,\beta}_{\beta} h^{\alpha}_{\mu,\alpha}
 \underbrace{\bigl( +\,0+4-2 \bigr)}_{+2}
 +
 h^{\beta}_{\mu,\alpha} h^{\mu,\alpha}_{\beta}
 \underbrace{\bigl( +\,0-2+1 \bigr)}_{-1}
 \,\Bigr]  
 + \ord(h_{\mu\nu}^3)
 \;. \hspace*{6mm} \nonumber
\ea
Here the contributions originate from 
\eqs\nr{gR_1}, \nr{gR_2}, and \nr{gR_3}, respectively.

Next we would like to extract the contribution of tensor modes
to \eq\nr{gR_sum}. 
We denote the 
{\em tensor perturbations} by $ h^\tensor_{ij} $, 
whereby it is good to keep in mind the relation to our previous notation
from \eq\nr{g_munu}, 
\be
 h^\tensor_{ij}
 \;
 \underset{\rmii{\nr{g_munu_2}}}{
 \overset{\rmii{\nr{g_munu}}}{\equiv}}
 \;
 2 \vartheta^\tensor_{ij}
 \;. \label{h_t_relation}
 \index{$h^\tensor_{ij},\,\vartheta^\tensor_{ij}$ (tensor perturbation)}
\ee
Recalling from 
\eq\nr{eq_metric-pert3} that 
$
 \delta^{ij}_{ } h^\rmi{t}_{ij} = 0 
 = \partial^{i}_{ } h^\rmi{t}_{ij}
$, 
trace parts, $h^\alpha_\alpha$, or divergences, 
$h^\alpha_{\mu,\alpha}$, contain no tensor part. 
Consequently, only the last term of \eq\nr{gR_sum}  
has a tensor part. The {\em Einstein-Hilbert action} 
from \eq\nr{einstein_hilbert} therefore contains
\ba
 S 
 &
    \underset{\rmii{\nr{gR_sum}}}
    {\overset{\rmii{\nr{einstein_hilbert}}}{\supset}}
 &
 \frac{1}{ 32 \pi G } \int_\X \! a^2_{ } \,
 \biggl(\, -\frac{1}{2} \partial^{ }_\alpha h^\rmi{t}_{ij}
   \, \partial^\alpha_{ } h^\rmi{t}_{ij} \,\biggr)
 \nn[2mm]
%%%%
 &
 \underset{\rmii{\nr{rescale_htij}}}{
 \overset{\rmii{\nr{inv_g_mink}} \lift }{=}}
 & 
 \int_\X \! a^4_{ } \,
 \biggl(\, -\frac{g^{\mu\nu}_{ }}{2}
      \partial^{ }_\mu \varh^\rmi{t}_{ij}
   \, \partial^{ }_\nu \varh^\rmi{t}_{ij} \,\biggr)
 \;+\; \ord(h_{\mu\nu}^3)
 \;. \label{einstein_hilbert_2}
 \index{Einstein-Hilbert action}
\ea 
In the second step we have rescaled
\be
 h^\rmi{t}_{ij} 
 \; \equiv \; 
 \sqrt{32\pi G} \, 
 \varh^\rmi{t}_{ij} 
 \;,
 \label{rescale_htij}
 \index{$\varh^\rmi{t}_{ij}$ (rescaled tensor perturbation)}
\ee
where $\varh^\rmi{t}_{ij}$ 
has the canonical dimension of a bosonic field, 
for us measured in GeV. 

We observe that 
$ \varh^\rmi{t}_{ij} $ appears in exactly the same way
in \eq\nr{einstein_hilbert_2}
as 
$ \varphi $ appears in \eq\nr{einstein_hilbert}, provided
that the potential $V(\varphi)$ is omitted. 
Therefore, we can directly take over familiar results from 
a massless minimally coupled scalar field, 
and apply them to $ \varh^\rmi{t}_{ij} $, 
being only careful with the counting of the 
independent degrees of freedom. In particular, the scalar
field energy density is given by $T^{ }_\rmii{$00$}$ 
in \eq\nr{T_00_2nd}, 
so that 
\be
 \boxed{
 \quad
 e^{ }_\gw 
 \;
 \underset{\rmii{\nr{T_00_2nd}}}{
 \overset{\rmii{\nr{einstein_hilbert_2}} \ilift }{\approx}}
 \;  
 \sum^{ }_{ij}
 \frac{1}{2 a^2_{ }}  
 \bigl[\,
   ( \varh^{\rmi{t}\,\prime}_{ij} )^2_{ } 
  + 
   | \nabla \varh^{\rmi{t}}_{ij} |^2_{ }
 \,\bigr]
 \;
 \underset{\underset{\rmii{\raise-0.5ex\hbox{average}}}{\rmii{oscillation}}}{
 \overset{\rmii{\nr{rescale_htij}} \ilift }{\simeq}}
 \; 
 \sum^{ }_{ij}
 \frac{
 \langle\,
 ( {h}^{\rmi{t}\,\prime}_{ij} )^2_{ }
 \,\rangle
 }{32\pi G a^2_{ }}   
 \;. 
 \quad \vphantom{\Bigg|^b_q }
 }
 \label{e_gw}
\ee
It is customary to take the second step, but it requires
additional assumptions, namely that $h^\tensor_{ij}$ has 
the form of a plane wave, where spatial gradients
and time derivatives contribute equally on average; 
and that we do take that average. 
Instead, the first form requires no average, 
given that oscillatory terms cancel for waves: 
$
 [\partial^{ }_\tau \sin (k\tau)]^2 + k^2_{ } \sin^2_{ }(k\tau)
 = k^2_{ }
$.
 
%%%%%%%%%%%%%%%%%%%%%%%%%%%%%%%%%%%%%%%%%%%%%%%%%%%%%%%%%%%%%%%%%%%%%%%%%
%
\subsection{Gravitational waves from vacuum fluctuations}
\label{ss:gw_vac}

\index{gravitational waves: vacuum fluctuations}

We have seen in \se\ref{ss:gw_energy} that in local Minkowskian 
coordinates, tensor perturbations have, up to overall normalization, 
the same action as a massless minimally coupled scalar field
(cf.\ \eq\nr{einstein_hilbert_2}). 
Moreover, the Einstein
equation satisfied by tensor perturbations, in \eq\nr{einstein_ij_t_eta}, 
has the same form as that for scalar perturbations, 
in \eq\nr{eq_field-eq-pert}, provided that we remove 
the potential~$V$ and the metric 
perturbations $h^{ }_0$, $h^{ }_\rmii{D}$ and $h$ from the latter. 
This analogy implies that we can take over parts of the 
discussion from \ch\ref{se:dS}, and derive the spectrum of 
tensor perturbations produced by vacuum fluctuations. 

Before we proceed, there is however one difference to scalar
perturbations that is worth anticipating. While for the 
scalar perturbations that affect the CMB and large-scale 
structure, we are interested in very small momenta or frequencies, 
which did exit the Hubble horizon at some point, and
re-entered only relatively late in the history of the
universe (cf.\ \se\ref{ss:history}), for tensor perturbations, 
future experiments give access also 
to much larger momenta and frequencies
(cf.\ \se\ref{ss:overview_f_0}). Such modes may have never 
crossed outside of the Hubble horizon. In this case, 
the way we treat them differs from that for scalar
perturbations (cf.\ \se\ref{ss:gw_transfer}).
In the remainder of the present section, we assume that the
modes did exit the Hubble horizon, sooner or later. 

\index{helicity: gravitational waves}
\index{projector: traceless transverse tensor}

Let us next clarify the role of polarization (or {\em helicity}) 
in tensor perturbations. The {\em projector to tensor perturbations} 
has been given in \eq\nr{T_ijmn}, and we now rewrite it in 
momentum space, 
\be
 \mathbbm{T}^{mn}_{ij}(\vec{k})
 \; 
 \underset{\scriptscriptstyle \vec x \;\to\; \vec k}{
 \overset{\rmii{\nr{T_ijmn}} \lift }{\equiv}} 
 \; 
 \frac{1}{2}
 \Bigl(
   \projK^{m}_{i}\projK^{n}_{j}
 + \projK^{n}_{i}\projK^{m}_{j}
 - \projK^{ }_{ij}\projK^{mn}_{ } 
 \Bigr)
 \;, \quad
 \projK^{ }_{ij} 
 \; \equiv \; \delta^{ }_{ij} - \frac{k^{ }_i k^{ }_j}{\vec{k}^2_{ }}
 \;. \label{T_ijmn_k}
 \index{$\projP^{i}_{m}, \projK^{i}_{m}$ (transverse projector)}
\ee
If we consider the vector space of symmetric $3\times 3$ matrices, 
its projection by \eq\nr{T_ijmn_k} defines a 2-dimensional invariant 
subspace. We can imagine finding an orthonormal basis in this subspace,
spanned by 
the {\em polarization vectors} 
\index{polarization vectors: gravitational waves}
$\epsilon^\lambda_{ij}(\vec{k})$, 
$\lambda \in \{+,\times\}$. It is not necessary for us to give 
an explicit representation of these vectors, 
but we need their orthonormality and completeness relations.
The basis vectors can be complex, and it is convenient to 
reverse the index positions for complex conjugates, so that  
\be 
 \sum_{ij}
  \epsilon^{\lambda}_{ij}(\vec{k})\,
  [\epsilon^{ij}_{\lambda'}(\vec{k})]^*_{ }
 \; = \; 
 \delta^{\lambda}_{\lambda'} 
 \;, \quad
 \sum_{\lambda}
  \epsilon^{\lambda}_{ij}(\vec{k})\,
  [\epsilon^{mn}_{\lambda}(\vec{k})]^*_{ }
 \; = \; 
 \mathbbm{T}^{mn}_{ij}(\vec{k}) 
 \;. 
 \label{basis} 
\ee
This implies that a general tensor perturbation can be written as
% (we omit $\vec{k}$ for brevity)
\be
 h^\rmi{t}_{ij}
 \;
 = 
 \;
 \sum_{mn}
 \mathbbm{T}^{mn}_{ij}
 h^\rmi{t}_{mn} 
 \;
 \overset{\rmii{\nr{basis}} \lift }{=} 
 \;
 \sum_{\lambda}
  \epsilon^{\lambda}_{ij}\,
 \sum_{mn}
  (\epsilon^{mn}_{\lambda})^*_{ }
  h^\rmi{t}_{mn} 
 \; \equiv \; 
 \sum_{\lambda}
  \epsilon^{\lambda}_{ij}\,
  h^\rmi{t}_\lambda 
 \;. \label{complete}
\ee 
A sum over $i,j$ like in \eqs\nr{einstein_hilbert_2}
and \nr{e_gw} is, after going to momentum space, \linebreak
equivalent to a sum over $\lambda$, 
\be
 \sum^{ }_{ij} 
 h^\rmi{t}_{ij}
 h^{ij*}_\rmi{t}
 \; \underset{\rmii{\nr{complete}}}
    {\overset{\rmii{\nr{basis}} \lift }{=}}
 \; 
 \sum^{ }_\lambda
 h_\lambda^\rmi{t}
 h_\rmi{t}^{\lambda*}
 \;. \label{h_basis_trafo}
\ee
We have included complex conjugates because relations such 
as \eq\nr{eq_angular_average} imply that, 
in the presence of translational 
invariance or volume average, 
a local product of real quantities turns into a product of 
a Fourier transform and its complex conjugate. 

The {\em power spectrum of tensor perturbations} is now defined as 
\be
 \P^{ }_\tensor(k)
 \; \equiv \; 
 \sum_\lambda
 \P^{ }_{ h^\rmii{t}_\lambda}(k)
 \;. \label{def_P_T}
\ee
Given that $h_\lambda^\rmi{t}$ satisfies the same equation, 
the power spectrum 
$
 \P_{ h^\rmii{t}_\lambda}
$
is related to the one 
of a massless scalar field, 
$
 \P^{ }_{\mbit\scriptscriptstyle\Q^{ }_\varphi}
$, 
which can be obtained from \eq\nr{P_stoch}, 
\be
 \P^{ }_{\mbit\scriptscriptstyle\Q^{ }_\varphi}(k) 
 \;
 \overset{\rmii{\nr{P_stoch}}}{\approx}
 \;
 \biggl( \frac{H^{ }_*}{2\pi} \biggr)^2_{ }
 \biggr|^{ }_{H^{ }_* \,=\, k/a^{ }_*}
 \;. \label{P_Q_final}
\ee
The subscript $(...)^{ }_*$ denotes the time at which 
the horizon-crossing condition, $ H^{ }_* = k/a^{ }_* $, is satisfied.
We also recall that a massless scalar field 
corresponds to the rescaled $\varh^\rmi{t}_{ij}$, related to 
$h^\rmi{t}_{ij}$ via \eq\nr{rescale_htij}. 
Putting all these ingredients together yields
\be
 \P^{ }_\tensor(k) 
 \; \approx \; 
 \underbrace{ 
 2
 }_{\sum_\lambda}
 \times 
 \underbrace{
 32\pi G
 }_{\nr{rescale_htij}}
 \times 
 \underbrace{
 \biggl( \frac{H^{ }_*}{2\pi} \biggr)^2_{ }
 \biggr|^{ }_{H^{ }_* \,=\, k/a^{ }_*}
 }_{\nr{P_Q_final}}
 \; = \; 
 \frac{16}{\pi}
 \biggl( \frac{H^{ }_*}{\mpl^{ }} \biggr)^2_{ }
 \biggr|^{ }_{H^{ }_* \,=\, k/a^{ }_*}
 \;. \label{P_T_final}
 \index{power spectrum: tensor perturbations} 
\ee
This applies to times during which the modes
are outside of the Hubble horizon
($t^{ }_\rmi{out} > t^{ }_*$). 
Remarkably, 
the amplitude of the tensor power spectrum directly
probes the magnitude of the Hubble 
rate in Planck units~\cite{gwr1,gwr2,gwr3,gwr4}. 

At small frequencies, 
the tensor power spectrum can be tested by the effect 
that it has on CMB polarization
(cf.\ \se\ref{ss:probes_gw}). 
No definite sign of primordial $B$-mode polarization 
has been observed, \index{$B$-mode polarization}
which sets an upper bound on the 
magnitude of the tensor power spectrum, as has been
discussed around \eq\nr{r_obs}. A theoretical prediction
for the {\em tensor-to-scalar ratio}, $r$, 
can be obtained
by combining \eqs\nr{P_R_final} and \nr{P_T_final}, 
\ba
%%%
 r^{ }_\rmii{0.002}
 &
 \underset{\scriptscriptstyle \R^{ }_\rmiii{\it T} 
  \, \approx \, \R^{ }_{\varphi}}{
 \overset{\rmii{\nr{r_obs}} \lift }{\equiv}}
 & 
 \frac{ \P^{ }_\tensor(0.04\,k^{ }_*) }
      { \P^{ }_{\mbit \scriptscriptstyle \R_\varphi}(0.04\,k^{ }_*) }
 \;
 \underset{\rmii{\nr{P_R_final}}}{
 \overset{\rmii{\nr{P_T_final}} \lift }{\approx}}
 \; 
 \pi G\, \biggl( 
           \frac{8\dot{\bar{\varphi}}^{ }_* }{H^{ }_*}
         \biggr)^2_{ }
 \;
 \overset{\rmii{\nr{r_obs}} \lift }{<}
 \;
 0.038
 \;. \label{r} 
 \index{$r$ (tensor-to-scalar ratio)}
\ea
In terms of the slow-roll parameters, this becomes
\ba 
%%%
 r^{ }_\rmii{0.002} 
 & 
 \overset{\rmii{\nr{r}} \lift }{\approx} 
 & 
 \frac{ 64\pi G \dot{\bar\varphi}^2_*}{H^2_*}
 \; 
 \overset{\rmii{\nr{assume}} \lift}{\approx}
 \; 
 \frac{ 64\pi G V_{,\varphi}^2}{9 H^4_*}
 \; 
 \overset{\rmii{\nr{slowroll_params}} \lift }{\approx}
 \; 
 16 \biggl( \frac{ 8 \pi G V}{3 H^2_*} \biggr)^2_{ }
 \, \epsilon^{ }_\rmii{$V$}
 \; 
 \overset{\rmii{\nr{slowroll_HH}} \lift }{\approx}
 \; 
 16 \epsilon^{ }_\rmii{$V$} 
 \;. \label{r_slowroll} 
\ea
A comparison with the numerical value from \eq\nr{r}
then indicates that the slow-roll parameter $\epsilon^{ }_\rmii{$V$}$  
needs to be small indeed, $ \epsilon^{ }_\rmii{$V$} \lsim 0.003$.
For the {\em tensor tilt}, defined according to \eq\nr{P_tensor}, 
a computation like in 
\eqs\nr{n_s_x1}--\nr{ns} yields
\ba
 n^{ }_\tensor 
 \; \underset{\rmii{\nr{ns},\nr{P_T_final}}}
   {\overset{\rmii{\nr{P_tensor}} \lift }
   {\approx}} \; 
   \frac{2H^{ }_*}{H^2_{*} + \dot{H}^{ }_*}
  \frac{\dot{H}^{ }_*}{H^{ }_*} 
 \; \overset{\rmii{\nr{slowroll_Hdot}} \lift }{\approx} \; 
  - 2 \epsilon^{ }_\rmii{$V$}
 \;. \label{nt}
 \index{$n^{ }_\tensor$ (tensor tilt)} 
\ea
The small value of $r^{ }_\rmii{0.002}$ 
thus guarantees the smallness of the tensor tilt.

%%%%%%%%%%%%%%%%%%%%%%%%%%%%%%%%%%%%%%%%%%%%%%%%%%%%%%%%%%%%%%%%%%%%%%%%%
%
\subsection{Gravitational waves from a matter energy-momentum tensor}
\label{ss:gw_Tij}

\index{gravitational waves: from matter $T^{ }_{\mu\nu}$}

Apart from vacuum fluctuations, gravitational waves can also 
be generated later on, for instance from hydrodynamic fluctuations
according to the evolution equation \nr{einstein_ij_t_eta}. 
Re-phrasing the latter in terms
of $h^\rmi{t}_{ij}$, via \eq\nr{h_t_relation}, we are faced with 
\be
   \bigl[\,
   \partial_\tau^2 + 
  \bigl(\, 2 \H
          + 16 \pi G a \eta 
  \,\bigr) \partial_\tau^{ } - \nabla^2_{ }
  \,\bigr]
   h^\rmi{t}_{ij}
 \; \underset{\rmii{\nr{h_t_relation}}}
   {\overset{\rmii{\nr{einstein_ij_t_eta}} \lift }{=}} \; 
 16\pi G a^2_{ } S^\rmi{t}_{ij} 
 \;, 
 \label{einstein_ij_t_eta_again}
\ee
where $\eta$ is the shear viscosity. 
This is an inhomogeneous linear partial differential equation. 
Even if the physical interpretation is different, 
formally the problem is similar to that considered
for scalar perturbations 
in \se\ref{sec_varrho}, with the role of the thermal noise, $\varrho$,
now taken over by hydrodynamic fluctuations, $S^\rmi{t}_{ij}$
(cf.\ \eq\nr{delta_S}). 
The method of solution, with a Green's function, is also analogous.

The friction term in \eq\nr{einstein_ij_t_eta_again}, 
$16\pi G a \eta$, 
was estimated in \eq\nr{shear_estimate}, 
and argued to be usually 
small compared with the Hubble dilution term, $2\H$. 
In order to permit for analytic manipulations, the friction
will be omitted in the following. Nevertheless, 
it turns out that there is a particular epoch, 
at low temperatures in a late universe, 
where the matter-induced damping of gravitational waves 
does play a significant role, and we return to this below \eq\nr{eq_X_T}.

With the reservations mentioned, 
we reinstate a general energy-momentum tensor on the right-hand side
of the gravitational wave equation, 
considering then 
\be
 (\partial_\tau^2 + 2 \H \partial_\tau^{ } - \nabla^2_{ })
  h^\tensor_{ij}
 \;
 \overset{\rmii{\nr{einstein_ij_t}} \lift }{
 \underset{\rmii{\nr{h_t_relation}}}{=}}
 \; 
 16\pi G a^2_{ } \barpPi^\rmi{t}_{ij} 
 \;. 
 \label{einstein_ij_t_again}
\ee
We stress that 
the right-hand side can depend on the unknown
function~$ h^\rmi{t}_{ij} $: 
the movement of the matter
that generates anisotropic stress is influenced by the 
gravitational background in which it propagates. When this 
is ignored, we are working at zeroth order in the sense of 
linear response theory (cf.\ \se\ref{sec_eom}). 

\vspace*{3mm}

In order to solve for a power spectrum, we go to momentum
space in spatial directions, 
and introduce a {\em retarded Green's function}
(cf.\ \eqs\nr{eq_phiGrho} and \nr{J_expansion_2}). 
The Green's function satisfies 
\ba
  (\partial_\tau^2 + 2 \H \partial_\tau^{ } +k^2_{ })
  G^{ }_\iR(\tau,\tau^{ }_1,k) & \equiv &  
  \delta(\tau - \tau^{ }_1)
 \;, \label{greens_ht} \\[2mm]
%%%%
  G^{ }_\iR(\tau,\tau^{ }_1,k) & \equiv & 0
 \quad \mbox{for} \quad \tau < \tau^{ }_1
 \quad \mbox{and} \quad \tau \to \tau_1^+
 \;, \label{greens_retard} \\[2mm]
%%%
 \lim_{\tau \to \tau_1^+} \partial^{ }_\tau
  G^{ }_\iR(\tau,\tau^{ }_1,k) & = & 1
 \;, \label{greens_deriv} 
 \index{retarded Green's function: tensor}
 \index{$G^{ }_\iR$ (retarded Green's function)}
\ea
where \eqs\nr{greens_ht} and \nr{greens_retard} serve as 
definitions, and \eq\nr{greens_deriv} follows from them, 
by integrating over the Dirac-$\delta$. 
The formal solution of \eq\nr{einstein_ij_t_again} then 
reads 
\be
 h^\rmi{t}_{ij}(\tau,k)
 \; = \; 
 h^\rmi{t}_{ij}(\tau,k) |^{ }_\vac 
 + 16\pi G 
 \int_{-\infty}^{\tau} \! \dd \tau^{ }_1 \, 
 G^{ }_\iR(\tau,\tau^{ }_1,k) \,
 a^2_{ }(\tau^{ }_1) \hspace*{0.3mm}
 \barpPi^\rmi{t}_{ij}(\tau^{ }_1,k) 
 \;, \label{ht_soln}
\ee
where 
$
  h^\rmi{t}_{ij}(\tau,k) |^{ }_\vac 
$
is a general solution of the homogeneous equation.
We consider the integration constants to be fixed like
for quantum-mechanical mode functions,
so that the power spectrum from
the vacuum part corresponds to that in \eq\nr{P_T_final}.

To make contact with the literature, 
we recall that, by rescaling variables, 
the basic equations can be set in different forms. 
For instance, if we define 
$
 \widehat{h}^\rmi{t}_{ij} \equiv a h^\rmi{t}_{ij}
$, 
then we can read off from 
\eqs\nr{rsc_1}--\nr{rescaled_pert_scalar}
that \eq\nr{einstein_ij_t_again} turns into
\be
  \biggl(
  \partial_\tau^2  - \nabla^2_{ } - \frac{a''}{a}
  \biggr)
   \widehat{h}^\rmi{t}_{ij}
 \;
 \underset{\rmii{\nr{rsc_1}--\nr{rescaled_pert_scalar}}}{
 \overset{\rmii{\nr{einstein_ij_t_again}} \lift }{=}}
 \; 
 16\pi G a^3_{ } \barpPi^\rmi{t}_{ij} 
 \;. \label{hijt_conformal}
\ee
As explained around \eq\nr{greens_2}, this form
has the benefit that the Green's function can be expressed in 
a simple way in terms of normalized solutions 
of the homogeneous equation. Moreover, as will be discussed 
around \eqs\nr{trace} and \nr{trace_concrete}, \eq\nr{hijt_conformal}
gets simplified in a radiation-dominated universe, given that 
$a''/a$ vanishes in the limit that the plasma can be approximated
as a collection of non-interacting massless particles.   
On the other hand, 
the un-rescaled form in \eq\nr{greens_ht} has its own advantages, 
notably that the small-$k$ asymptotics of its solution is transparent. 
This will be important for us in the following,  
so we stick to \eqs\nr{greens_ht} and \nr{ht_soln}.

With \eq\nr{ht_soln}, we can write down an expression for the 
tensor power spectrum, generalizing on \eq\nr{P_T_final}. 
The vacuum part remains intact, as long as the modes are outside
of the Hubble horizon in a certain time window, 
from which we choose a representative, denoted by $\tau^{ }_\rmi{out}$. 
Manipulating the Green's function part like in 
\eq\nr{noise_average}
(and omitting mixed terms since they average to zero), 
the late-time tensor power spectrum becomes 
\ba
 \P^{ }_\tensor(\tau,k) 
 &
 \underset{\rmii{\nr{P_T_final},\,\nr{def_X_T}}}{
 \overset{\rmii{\nr{noise_average},\nr{ht_soln}} \lift }{\approx}}
 & 
 X^2_\tensor(\tau,\tau^{ }_\rmi{out},k)
 \, 
 \frac{16}{\pi}
 \biggl( \frac{H^{ }_*}{\mpl^{ }} \biggr)^2_{ }
 \biggr|^{ }_{H^{ }_* = k/a^{ }_*} 
 \nn[2mm]
%%%%%
 &  & \;+\,
 \frac{k^3_{ }}{2\pi^2_{ }}
 \,(16\pi G)^2_{ }
 \int_{-\infty}^{\tau} \! \dd \tau^{ }_1 \, 
 \int_{-\infty}^{\tau} \! \dd \tau^{ }_2 \, 
 G^{ }_\iR(\tau,\tau^{ }_1,k) \,
 G^{ }_\iR(\tau,\tau^{ }_2,k) \,
 \nn[2mm]
%%%%%
 &  & \hspace*{26mm} \; \times \,  
 a^2_{ }(\tau^{ }_1) 
 a^2_{ }(\tau^{ }_2)
 \,
 P^{ }_\rmii{$\Pi$}(\tau^{ }_1,\tau^{ }_2,\vec{k})
 \;, \label{P_T_final_2} 
 \index{power spectrum: tensor perturbations} 
%%%%%%%%
\ea
where 
$
 X^{2}_\tensor(\tau,\tau^{ }_\rmi{out},k)
$
from \eq\nr{def_X_T}
is a transfer function in the tensor channel 
(this is discussed at some length in \se\ref{ss:gw_transfer}).
The energy-momentum correlator reads
\ba
 P^{ }_\rmii{$\Pi$}(\tau^{ }_1,\tau^{ }_2,\vec{k})
 & \overset{\rmii{\nr{eq_PQdef}}}{\equiv} & 
 \int \! \dd^3_{ } \vec{x} \, e^{-i\vec k \cdot \vec x}_{ }
 \bigl\langle\,  
 \barpPi^\rmi{t}_{ij}(\tau^{ }_1,\vec x) 
 \barpPi^{\rmi{t}}_{ij}(\tau^{ }_2,\vec 0) 
 \,\bigr\rangle
 \nn[2mm]
%%%%%%
 & \overset{\rmii{\nr{T_ijmn_k}}}{=} & 
 \mathbbm{T}^{mn}_{ij}(\vec{k})
 \int \! \dd^3_{ } \vec{x} \, e^{-i\vec k \cdot \vec x}_{ }
 \bigl\langle\,  
 \barpPi^{ }_{ij}(\tau^{ }_1,\vec x) 
 \barpPi^{ }_{mn}(\tau^{ }_2,\vec 0) 
 \,\bigr\rangle 
 \;. \label{P_Pi}
\ea
Usually this correlator has some symmetry properties.
For instance, assuming parity and translation invariance of the 
ensemble average, we can write
\ba
 \bigl\langle\,
   \barpPi^\rmi{t}_{ij}(\tau^{ }_1,\vec x) 
   \barpPi^{\rmi{t}}_{ij}(\tau^{ }_2,\vec 0) 
 \,\bigr\rangle
 & \underset{\rmii{ }}{\overset{\rmii{parity} \lift }{=}} & 
 \bigl\langle\,
   \barpPi^\rmi{t}_{ij}(\tau^{ }_1,-\vec x) 
   \barpPi^{\rmi{t}}_{ij}(\tau^{ }_2,\vec 0) 
 \,\bigr\rangle
 \nn[2mm]
%%%%%
 & \underset{\rmii{ }}{\overset{\rmii{translation} \lift }{=}} &  
 \bigl\langle\,
   \barpPi^\rmi{t}_{ij}(\tau^{ }_1,\vec y -\vec x) 
   \barpPi^{\rmi{t}}_{ij}(\tau^{ }_2,\vec y) 
 \,\bigr\rangle
 \nn[2mm]
%%%%%
 & \overset{\scriptscriptstyle \vec{y} \;\to\; \vec{x}
  \lift }{=} & 
 \bigl\langle\,
   \barpPi^\rmi{t}_{ij}(\tau^{ }_1,\vec 0) 
   \barpPi^{\rmi{t}}_{ij}(\tau^{ }_2,\vec x) 
 \,\bigr\rangle
 \;. \label{parity} 
\ea
In the classical approximation, operators commute, implying that 
\be
 P^{ }_\rmii{$\Pi$}(\tau^{ }_2,\tau^{ }_1,\vec{k})
 \;
 \underset{\rmii{classical~limit}}{
 \overset{\rmii{\nr{P_Pi},\nr{parity}} \lift }{\approx}} 
 \;
 P^{ }_\rmii{$\Pi$}(\tau^{ }_1,\tau^{ }_2,\vec{k})
 \;. \label{time_reversal}
\ee

Let us now ask how \eq\nr{P_T_final_2} behaves as a function 
of $k$. In particular, we can investigate 
the small-$k$ asymptotics (the large-$k$ limit will be discussed
in \se\ref{ss:gw_scat}). 
As far as $P^{ }_\rmii{$\Pi$}$ from \eq\nr{P_Pi}
goes, if we consider very large distances, we are in a spacelike
domain, and then causality dictates that correlations  
decay fast
(in the tensor channel, there is no 
disconnected term like \eq\nr{discon}). 
Therefore we can assume that the Fourier transform 
remains finite for $\vec{k} \to \vec{0}$. 

The key is then to 
determine what happens with the Green's function. 
If we set $k\to 0$ in \eqs\nr{greens_ht}--\nr{greens_deriv}, 
an explicit solution can be found, 
\be
 \lim_{k\to 0}
 G^{ }_\iR(\tau,\tau^{ }_i,k)
 \; 
 \overset{\scriptscriptstyle \tau \,\ge\, \tau^{ }_i \vphantom{ \big | } }
 {\underset{ }{ = }}
 \; 
 a^2_{ }(\tau^{ }_i)
 \int_{\tau^{ }_i}^{\tau} \! \frac{\dd \tau'}{a^2_{ }(\tau')}
 \;. 
\ee
This demonstrates that the small-$k$ limit of the Green's function is finite. 

Inspecting now \eq\nr{P_T_final_2}, we can draw an interesting conclusion. 
The first part, the vacuum contribution, is to a good approximation 
independent of $k$. 
Instead, the second part grows as $k^3_{ }$ at small $k$
according to the discussion above, because we can set $k\to 0$
within the integrand (in $G^{ }_\iR$ and $P^{ }_\rmii{$\Pi$}$) 
without meeting a singularity. 
So, if there is any anisotropic stress in the universe,
this produces a 
{\em growing spectrum of gravitational waves}.
Of course, the growth cannot continue
forever, and we return to this in \se\ref{ss:gw_scat}.

Finally, let us make \eq\nr{P_T_final_2} more concrete, by assuming
that after it has heated up, the universe
contains a thermalized plasma (cf.\ \ch\ref{se:thermal}). 
In such a plasma, 
{\em hydrodynamic fluctuations} \index{hydrodynamic fluctuations}
take place, as has been discussed
in \app\ref{app:viscous}. 
In particular, 
the contribution to $P^{ }_\rmii{$\Pi$}$ originates from the fluctuations
in \eq\nr{delta_S}, as shown by \eq\nr{einstein_ij_t_eta_again}.
We also need to recall the relation between momentum space
correlators from \eq\nr{eq_PQdef}, which eliminates 
$
 (2\pi)^3_{ }\delta^{(3)}_{ }(\vec k + \vec q)
$ 
from \eq\nr{delta_S}, yielding then 
\ba 
 P^{ }_\rmii{$\Pi$}(\tau_1^{ },\tau^{ }_2,\vec{k})
 &
 \overset{\rmii{\nr{einstein_ij_t_eta_again},\,\nr{P_Pi}}
          \lift }{
 \underset{\rmii{\nr{delta_S},\,\nr{eq_PQdef}}}{\approx}}
 & 
 \mathbbm{T}^{mn}_{ij}(\vec{k})\, 
 \frac{ 
 2 T \, \delta(\tau^{ }_1 - \tau^{ }_2)
 }{a^2_{ }(\tau^{ }_1) a^2_{ }(\tau^{ }_2)}
%  \nn[2mm]
% & \times &  
 \,\Bigl[
 \eta \, \bigl( 
                 \delta^{ }_{im} \delta^{ }_{jn}
               + \delta^{ }_{in} \delta^{ }_{jm}
         \bigr)
 +       \Bigl( 
                 \zeta - \frac{2\eta}{3}
         \Bigr) \, 
                 \delta^{ }_{ij} \delta^{ }_{mn} 
 \Bigr] 
 \nn[3mm]
%%%%%
 & \overset{\rmii{\nr{T_ijmn_k}} \lift }{=} & 
 \frac{ 
 8 T \eta\, \delta(\tau^{ }_1 - \tau^{ }_2)
 }{a^2_{ }(\tau^{ }_1)\, a^2_{ }(\tau^{ }_2)}
 \;, \label{noise_t}
\ea
where $\eta$ is the shear viscosity. We have shown time
arguments for $a$ (suppressing them for $T$ and $\eta$), 
to make it clear that, 
when the expression is inserted in \eq\nr{P_T_final_2}, 
the factor $a^2_{ }(\tau^{ }_1)\, a^2_{ }(\tau^{ }_2)$ 
cancels, and we get
\begin{empheq}[box=\fbox]{align}
 \quad \vphantom{\Bigg|^b_q} 
 \P^{ }_\tensor(\tau,k) 
 &
 \underset{\rmii{\nr{def_X_T}}}{
 \overset{\rmii{\nr{P_T_final_2},\nr{noise_t}}}{\supset}}
 \;
 X^2_\tensor(\tau,\tau^{ }_\rmi{out},k)
 \, 
 \frac{16}{\pi}
 \biggl( \frac{H^{ }_*}{\mpl^{ }} \biggr)^2_{ }
 \biggr|^{ }_{H^{ }_*\; = \; k/a^{ }_*} 
 \nn[2mm]
%%%%%
 & \hspace*{1.9cm}
 \; + \, 
 \frac{32^2_{ }k^3_{ }}{\mpl^4}
 \int_{-\infty}^{\tau} \! \dd \tau^{ }_1 \, 
 G^{2}_\iR(\tau,\tau^{ }_1,k) \,
 T(\tau^{ }_1)\hspace*{0.3mm}\eta(\tau^{ }_1)
 \;.
 \quad  \vphantom{\Bigg|^b_q} 
 \label{P_T_final_3}
\end{empheq}
We note that 
$H_*^2 \approx 8\pi e^{ }_*/(3\mpl^2)$, 
cf.\ \eq\nr{eq_end0-1}, so both the vacuum contribution
on the first line of \eq\nr{P_T_final_3}, and 
the hydrodynamic contribution on the second line,   
are formally of $\ord(1/\mpl^4)$, 
even if they display different dependences on $k$.
We return to \eq\nr{P_T_final_3} around \eq\nr{hydro_fluct_5},  
re-phrasing the result in terms of measurable quantities.

%%%%%%%%%%%%%%%%%%%%%%%%%%%%%%%%%%%%%%%%%%%%%%%%%%%%%%%%%%%%%%%%%%%%%%%%%
%
\subsection{Scalar-induced gravitational waves}
\label{ss:gw_sigw}

\index{gravitational waves: scalar-induced}

\index{scalar-induced gravitational waves}

In \eq\nr{P_T_final_3}, we have given a prediction for 
the gravitational-wave power spectrum 
originating from vacuum and hydrodynamic fluctuations. 
However, the most significant contributions might 
not originate from such expected
``equilibrium'' sources, but rather from more peculiar phenomena, 
active perhaps only for a short period of time, but undergoing violent 
``non-equilibrium'' dynamics during that instance
(cf.,\ e.g.,\ ref.~\cite{gw-review}).  
These phenomena are in general model-dependent, and thus speculative. 
In the present section, we discuss one generic source, which
originates from the curvature perturbations that we have discussed
in the previous chapters, and is therefore quite likely to be present, 
even if the amplitude of curvature perturbations at momenta larger
than probed by CMB and structure formation, is 
not strongly constrained at the time of writing. 

Let us consider \eq\nr{Tmunu_mixed}, 
complemented by the first-order-in-gradients 
viscous corrections from \eq\nr{Tmunu_hydro}, 
\be
 T^{ }_{\mu\nu}
 \; 
  \underset{\rmii{\nr{Tmunu_hydro}}}{
  \overset{\rmii{\nr{Tmunu_mixed}}}{\equiv}} 
 \; 
 \varphi^{ }_{,\ibit\mu} \varphi^{ }_{,\ibit\nu}
 \;-\; \frac{ g^{ }_{\mu\nu}\, 
 \varphi^{ }_{,\alpha} \varphi^{,\alpha}_{ } }{2} 
 \;+\; ( e + p ) \,  u^{ }_{\mu}u^{ }_{\nu}
 \;+\; p \, g^{ }_{\mu\nu}
 \;+\;
 \hspace*{-6mm}
 \overbrace{
 a^2_{ } \Pi^{ }_{\mu\nu}
 }^{\rm involves~gradients}
 \hspace*{-6mm}
 \;. \label{Tmunu_mixed_again}
\ee
At {\em linear order in perturbations}, 
the spatial part $T^{ }_{ij}$ 
has two tensor components. One originates from 
$g^{ }_{ij} \supset 2 a^2_{ }\vartheta^\rmi{t}_{ij}$
(cf.\ \eq\nr{g_munu}), 
which multiplies the second and fourth terms, 
but this cancels 
after the use of background identities, 
cf.\ the discussion leading up to \eq\nr{einstein_ij_t}.
The other originates from $\Pi^{ }_{ij}$ 
(cf.\ \eqs\nr{decomposition} and \nr{delta_S}). 
The latter led to \eq\nr{P_T_final_3}.
However, we can talk about viscous hydrodynamics only under 
special conditions (cf.\ \se\ref{sec_T}).
This prompts us to ask how else 
anisotropic stress could arise?

One definite source for anisotropic stress is found by going 
up to {\em second order in perturbations}. Restricting to scalar
perturbations on the matter side
(although second-order effects from 
vector and tensor perturbations can also be considered), 
let us insert
\be
  \varphi 
  \; \overset{\rmii{\nr{delta_varphi}}}{=} \; 
  \bar\varphi + \delta\varphi
  \;, \quad
  u_\mu^{ }   
  \;
 \underset{\rmii{\nr{eq_v-v-s}}}
 {\overset{\rmii{\nr{eq_u_down}} \ilift }{=}}
 \;
  a(-1-h_0,h^{ }_{,i} - v^{ }_{,i}) + \ord(\delta^2_{ })  
 \label{2nd_ord_perts}
 \index{second-order perturbations}
\ee
into the various structures in \eq\nr{Tmunu_mixed_again}. 
Recalling the metric from \eqs\nr{g_munu} and \nr{g^munu}, 
and omitting the parts proportional to $\delta^{ }_{ij}$, 
which have no tensor component, this yields
\ba
 \varphi^{ }_{,\ibit i}\, \varphi^{ }_{,\ibit j}
 & 
 \overset{\rmii{\nr{2nd_ord_perts}}}{=} 
 & 
 \delta\varphi^{ }_{,\ibit i}\, \delta\varphi^{ }_{,\ibit j}
 \;, \label{phi_prime_i_j} \\[2mm]
%%%%%%
 - \frac{ g^{ }_{ij}\, 
 \varphi^{ }_{,\alpha} \varphi^{,\alpha}_{ } }{2} 
 & \underset{\rmii{\nr{delta_Tij_varphi}}}
  {\overset{\rmii{\nr{g_munu}}}{\supset}} & 
  \vartheta^{ }_{ij} 
  \bigl[\, 
  (1 - 2 h^{ }_0) (\bar{\varphi}\ibit')^2 
  + 2 {\bar\varphi}\ibit'\delta\varphi'
  \,\bigr]
 + \ord(\delta^3_{ })
 \;, \label{phi_t_diag} \\[2mm]
%%%%%%
 ( e + p ) \,  u^{ }_{i}u^{ }_{j}
 & 
  \overset{\rmii{\nr{2nd_ord_perts}}}{=} 
 &
  a^2_{ }(\bar{e} + \bar{p})
 (h-v)^{ }_{,\ibit i}
 (h-v)^{ }_{,\ibit j}
 + \ord(\delta^3_{ })
 \;, \label{uu_t} \\[2mm]
%%%%%%
 p \, g^{ }_{ij}
 & \underset{\rmii{\nr{delta_e_p}}}
  {\overset{\rmii{\nr{g_munu}}}{\supset}} & 
 2 a^2_{ }\vartheta^{ }_{ij} 
 \bigl( \bar{p} + \delta p \bigr)
 \;. \label{p_t_diag}
%%%%%%
\ea
We note that, within our definitions of $\vartheta^{ }_{ij}$, 
$\delta p$ and $\delta \varphi$
in 
\eqs\nr{g_munu}, 
\nr{delta_e_p}, and \nr{delta_varphi},
respectively, no perturbative expansion has been
made in \eqs\nr{phi_prime_i_j} and \nr{p_t_diag}.

Now, an important point is that second-order perturbations, 
notably those in \eqs\nr{phi_prime_i_j} and \nr{uu_t}, 
by themselves {\em generate} anisotropic stress, as we will see
in the remainder of this section. 
So, it can be dangerous to include simultaneously 
the first-order $a^2\Pi^{ }_{ij}$
from \eq\nr{Tmunu_mixed_again}, 
as this could lead to double counting. This issue will
manifest itself concretely below, however for the moment
we keep both contributions in the equations. 

To proceed, we slightly deviate from our previous
philosophy of working in a general gauge, and instead do
a {\em partial} gauge fixing. This simplifies the derivation
of the Einstein tensor, where we have to go up to second 
order in perturbations. A complete derivation, without
any gauge fixing and with multiple fluids, 
can be found in ref.~\cite{sigw_theory}.

Concretely, in the language of \eqs\nr{gauge1}--\nr{gauge7}, we choose
a partially Newtonian gauge, 
\be
 \xi \;=\; -\vartheta
 \; \overset{\rmii{\nr{gauge4}} \lift }{\Rightarrow} \; 
 \tilde\vartheta \;=\; 0 
 \;. \label{gauge_fixed}
\ee
The gauge parameter
$\xi^0_{ }$ is left free, permitting for a check of gauge dependence. 
The gauge-fixed quantities are denoted by 
$\tilde h^{ }_0$, $\tilde h^{ }_\rmii{D}$, $\tilde h$, $\tilde v$, 
and $\delta{\tilde\varphi}$. 
%
% (according to \eqs\nr{gauge2} and \nr{gauge3}, 
% $\tilde h = X$ and $\tilde h^{ }_\rmii{D} = Y$, 
% where $X$ and $Y$ are from \eq\nr{shorthand}).
%
% only true for $\xi^0_{ }=0$
%
With this choice, the metric from \eq\nr{g_munu} becomes
\be
  g^{ }_{\mu\nu}
 \; 
  \underset{\rmii{\nr{gauge_fixed}}}
 {\overset{\rmii{\nr{g_munu}} \lift }{=}} 
 \; 
  a^2_{ } \begin{pmatrix}
 -1-2\tilde h^{ }_0 & \tilde h^{ }_{\der j} \\[2mm]
 \tilde h^{ }_{\der i}    & (1-2\tilde h^{ }_\rmii{D})\,\delta^{ }_{ij}
 +2\vartheta^\rmi{t}_{ij}
 \end{pmatrix} 
 \ . \label{g_munu_fixed}
 \index{metric: perturbed}
\ee
Here vector perturbations have been omitted.

The first task is to invert the metric. For a matrix that is of the form
\be
 g^{ }_{\mu\nu} 
 \;=\;
 a^2_{ }
 \,\bigl(\, \eta + A \,\bigr)
 \;, \quad
 \eta \;\equiv\;
 \biggl( \begin{array}{cc} -1 & 0 \\ 0 & \mathbbm{1} \end{array} \biggr) 
 \;,
\ee
the inverse reads 
\be
 g^{\mu\nu}_{ } \;=\; \frac{1}{a^2_{ }}
 \,\bigl(\, 
 \eta - \eta A\hspace*{0.3mm} \eta 
+ \eta A\hspace*{0.3mm} \eta A\hspace*{0.3mm} \eta + \ldots
 \,\bigr)
 \;. \label{general_inverse}
\ee
This yields 
\be
  g^{\mu\nu}_{ } 
  \;
  \underset{\rmii{\nr{general_inverse}}}
 {\overset{\rmii{\nr{g_munu_fixed}}  \lift  }{=}}
 \; 
 \frac{1}{a^{2}_{ }} \begin{pmatrix}
 -1+2\tilde h^{ }_0
  - 4 \tilde h_0^2 + \tilde h^{ }_{\!\der k} \tilde h^{ }_{\!\der k} & 
 \tilde h^{ }_{\!\der j}\,[1 + 2 (\tilde h^{ }_\rmii{D} - \tilde h^{ }_0)]
 \\[2mm]
%%%%%
 \tilde h^{ }_{\!\der i}\,[1 + 2 (\tilde h^{ }_\rmii{D} - \tilde h^{ }_0)]  & 
 (1+2\tilde h^{ }_\rmii{D} + 4 \tilde h^2_\rmii{D})\,\delta^{ }_{ij}
 - \tilde h^{ }_{\!\der \ibit i} \ibit \tilde h^{ }_{\!\der \ibit j}
 -2\vartheta^\rmi{t}_{ij}
 \end{pmatrix} 
 + \ord(\delta^3_{ })
 \ , \label{g^munu_fixed}
\ee
where $\vartheta^\rmi{t}_{ij}$ has been counted
as being of $\rmO(\delta^2_{ })$, so that only its
first-order appearance 
(without $\tilde h^{ }_0$, $\tilde h^{ }_\rmii{D}$, $\tilde h$) 
has been retained. 

Subsequently, we insert \eqs\nr{g_munu_fixed} and 
\nr{g^munu_fixed}
in \eq\nr{eq_ChSy-a}, 
in order to 
determine the Christoffel symbols, $\Gamma^{\rho}_{\mu\nu}$;
then in \eq\nr{eq_rt-a}, 
for the Ricci tensor, $R^{ }_{\mu\nu}$;
in \eq\nr{bg_ricci_scalar},
for the Ricci scalar, $R$; 
and in \eq\nr{bg_einsteindowndown}, 
for the Einstein tensor, $G^{ }_{\mu\nu}$. Given that
the equations become lengthy, we implement them with
a symbolic script, as illustrated 
in \app\ref{app:sigw}.
For the spatial components $i\neq j$, 
relevant for the tensor channel, we find
\ba
  G^\rmi{ }_{ij}
 & 
 \overset{\scriptscriptstyle i\;\neq\; j  \lift  }{\supset}
 &
 \overbrace{
 (\partial_\tau^2 + 2 \H \partial_\tau^{ } - \nabla^2_{ }) 
 \vartheta^\rmi{t}_{ij}
 -2 (2\H' + \H^2_{ }) \vartheta^\rmi{t}_{ij}
 }^{{\rm from}\;\nr{tensor_full}} 
 \; - \; 
 \overbrace{
 \bigl(\, \tilde h^{ }_0 - \tilde h^{ }_\rmii{D}
          + \tilde h' + 2 \H \tilde h \bigr)^{ }_{\der ij}
 }^{{\rm from~the}\;i \neq j\;{\rm part~of}\;\nr{delta_Gij_s} }
 \nn[2mm]
%%%%
 &  & \;+\,
 \tilde h^{ }_{0\der i} \ibit
 \tilde h^{ }_{0\der j}
 + 
 3 \tilde h^{ }_{\rmii{D}\der i} \ibit
   \tilde h^{ }_{\rmii{D}\der j}
 + 
 2 ( \tilde h^{ }_0 \ibit
     \tilde h^{ }_{0\der ij} 
   + \tilde h^{ }_\rmii{D} \ibit
     \tilde h^{ }_{\rmii{D}\der ij}
   )
 - (  
 \tilde h^{ }_{0\der i} \ibit
 \tilde h^{ }_{\rmii{D}\der j}
 + 
 \tilde h^{ }_{0\der j} \ibit
 \tilde h^{ }_{\rmii{D}\der i}
 ) 
 \nn[2mm]
%%%%
 &  & \;+\, 
 (\, \tilde h^\prime_0 + \tilde h^\prime_\rmii{D}
 + 4 \H \tilde h^{ }_0 \,) \ibit
   \tilde h^{ }_{\der ij}
 + 2 \tilde h^{ }_0 \ibit
     \tilde h^{\prime}_{\der ij}
 \nn[2mm]
%%%%
 &  & \;-\,
  (\partial^{ }_\tau + 2 \H)
  ( \tilde h^{ }_{\rmii{D}\der i} \ibit
    \tilde h^{ }_{\der j}
 +  \tilde h^{ }_{\rmii{D}\der j} \ibit
    \tilde h^{ }_{\der i} )
 \; + \; 
 \tilde h^{ }_{\der kk} \ibit
 \tilde h^{ }_{\der ij}
 - 
 \tilde h^{ }_{\der ki} \ibit
 \tilde h^{ }_{\der kj}
 + 
 \ord(\delta^3_{ })
 \;. \label{Gij_t_1}
\ea

Next, we go over to gauge-invariant variables. The Bardeen potentials
from \eqs\nr{def_phi} and \nr{def_psi} become
\ba
 \phi 
 &
  \underset{\rmii{\nr{gauge_fixed}}}
 {\overset{\rmii{\nr{def_phi}} \lift }{=}} 
 & 
 \tilde h^{ }_0 + \tilde h^\prime_{ }+ \H \ibit \tilde h
 \;, \label{phi_fixed}
 \\[2mm]
%%%%
 \psi
 &
  \underset{\rmii{\nr{gauge_fixed}}}
 {\overset{\rmii{\nr{def_psi}} \lift }{=}} 
 & 
 \tilde h^{ }_\rmii{D} - \H \ibit \tilde h
 \;. \label{psi_fixed} 
\ea
Therefore we can substitute
\begin{align}
 \tilde h^{ }_0 
 & \overset{\rmii{\nr{phi_fixed}}}{=} 
 \phi - \tilde h' - \H \ibit \tilde h
 &
 &\Rightarrow 
 &
 \tilde h^{\prime}_0 
 &\; = \; 
 \phi' - \tilde h'' - \H' \tilde h - \H \ibit \tilde h'
 \;, \label{tilde_h0_subst} 
 \\[2mm]
%%%%%
 \tilde h^{ }_\rmii{D} 
 & \overset{\rmii{\nr{psi_fixed}}}{=}  
 \psi + \H  \ibit \tilde h
 &
 &\Rightarrow 
 &
 \tilde h^{\prime}_\rmii{D} 
 &\; = \; 
 \psi' + \H' \tilde h + \H \ibit \tilde h'
 \;. \label{tilde_hD_subst} 
\end{align}
Inserting these into \eq\nr{Gij_t_1}, we get
\ba
  G^\rmi{ }_{ij} 
 & 
  \underset{\rmii{\nr{tilde_h0_subst},\,\nr{tilde_hD_subst}}}
  {\overset{\rmii{\nr{Gij_t_1}}}{\supset}} 
 &
 (\partial_\tau^2 + 2 \H \partial_\tau^{ } - \nabla^2_{ }) 
 \vartheta^\rmi{t}_{ij}
 -2 (2\H' + \H^2_{ }) \vartheta^\rmi{t}_{ij} 
 \; + \; 
 (\, \psi - \phi \,)^{ }_{\der ij}
 \nn[2mm]
%%%%
 &  & \;+\,
 \phi^{ }_{\der i} \ibit
 \phi^{ }_{\der j}
 + 
 3 \psi_{\der i} \ibit
   \psi_{\der j}
 + 
 2 (
 \phi \, \phi^{ }_{\der ij}
 + 
 \psi \, \psi^{ }_{\der ij} )
 - (  
 \phi_{\der i}\,
 \psi_{\der j}
 + 
 \phi_{\der j}\,
 \psi_{\der i}
 ) 
 \hspace*{8mm} \nn[2mm]
%%%%
 &  & \;-\,
   (\phi^{ }_{\der i} \ibit
    \tilde h'
  + \psi' \ibit
   \tilde h^{ }_{\der i} )^{ }_{\der j}
 - (\phi^{ }_{\der j} \ibit
    \tilde h'
  + \psi' \ibit
    \tilde h^{ }_{\der j})^{ }_{\der i}
 \;
 + 
 \;
 ( \phi' + 3 \psi' )\, \tilde h^{ }_{\der ij}
 \nn[2mm]
%%%%
 &  & \;+\, 
   2 \H \,[\,
   2 \phi\, \tilde h^{ }_{\der ij}
   + (\psi\, \tilde h - \phi\, \tilde h)^{ }_{\der ij}
 \,]
 \;
 +
 \;
 2 (\H^2_{ }-\H') \ibit 
 \tilde h^{ }_{\der i} \ibit 
 \tilde h^{ }_{\der j}
 \nn[2mm]
%%%%
 &  & \;+\, 
 ( \tilde h^{ }_{\der kk} - 2 \H\, \tilde h' - \tilde h'')\,
 \tilde h^{ }_{\der ij}
 + 
 \tilde h^{\prime}_{\der i} \ibit
 \tilde h^{\prime}_{\der j}
 - 
 \tilde h^{ }_{\der ki} \ibit
 \tilde h^{ }_{\der kj}
 \;+\; 
 \ord(\delta^3_{ })
 \;. \label{Gij_t_2}
\ea

For gravitational waves, we need to determine the tensor part 
of \eq\nr{Gij_t_2}, denoting the result by~$G^\tensor_{ij}$.
We recall that the tensor projector from \eq\nr{T_ijmn} is transverse
with respect to spatial derivatives (or, in momentum space, spatial momenta), 
\be
       \mathbbm{T}^{mn}_{ij} (...)^{ }_{\der mn} 
 \;=\; \mathbbm{T}^{mn}_{ij} (...)^{ }_{\der m}
 \;=\; 0
 \;. 
 \label{tmn_props}
\ee
This implies that we can effectively carry out {\em integrations by parts}
(IBP) in the tensor channel, \index{IBP (integration by parts)}
\be
 \mathbbm{T}^{mn}_{ij} \phi\, \phi^{ }_{\der mn}
 \; = \; 
 \mathbbm{T}^{mn}_{ij}
 \,[\, (\phi\, \phi^{ }_{\der n})^{ }_{\der m} - 
       \phi^{ }_{\der m}\, \phi^{ }_{\der n} \,]
 \;
 \overset{\rmii{\nr{tmn_props}}}{=}
 \; 
 - 
 \mathbbm{T}^{mn}_{ij} \phi^{ }_{\der m}\, \phi^{ }_{\der n}
 \;. \label{ibp_spatial}
\ee
Furthermore, recalling that we are considering waves, we can 
envisage taking a time average over an oscillation period, or
a spatial average over a wavelength. Given that the
background is time-dependent, this has to be implemented in 
a covariant way, as 
\ba
 \langle \, f' \, g' \,\rangle
 & \equiv &
 \frac{1}{\N} 
 \int_{\X} \, 
 \overbrace{
 \sqrt{-\bar{g}} 
 }^{a^4_{ }}
 \, 
 \overbrace{
 \bar{g}^{00}_{ }
 }^{-1/a^2_{ } }
 \, ( - \, f' \, g' )
 \; = \; 
 \frac{1}{\N} 
 \int_{\X} \, 
 a^2_{ }\, \partial^{ }_\tau f \, \partial^{ }_\tau g
 \nn[2mm]
%%%%%
 & = &  
 \frac{1}{\N} 
 \int_{\X} \, 
 \bigl[\, 
  \overbrace{
  \partial^{ }_\tau (a^2_{ } f \, \partial^{ }_\tau g)
  }^{{\rm boundary~terms}}
  - 2 a a '\, f \, \partial^{ }_\tau g
  - a^2_{ } \, f \, \partial^{2}_\tau g
 \,\bigr]
 \nn[2mm]
%%%%%
 & = &  
 \frac{1}{\N} 
 \int_{\X} \, 
 \sqrt{-\bar{g}} 
 \, \bar{g}^{00}_{ }
 \, f \, \bigl(\, 2 \H g' + g'' \,\bigr)
 \; + \; 
 \mbox{(boundary~terms)}
 \nn[2mm]
%%%%%
 & = &  
 \bigl\langle \, f \, \bigl(\, - 2 \H g' - g'' \,\bigr) \,\bigr\rangle 
 \; + \; 
 \mbox{(boundary~terms)}
 \;, \label{ibp_time}
 \hspace*{6mm}
\ea
where $\N$ is a normalization factor. For spatial derivatives, 
the same logic gives
\ba
 \langle \, f^{ }_{\der k} \, g^{ }_{\der k} \,\rangle
 & \equiv &
 \frac{1}{\N} 
 \int_{\X} \, 
 \overbrace{
 \sqrt{-\bar{g}} 
 }^{a^4_{ }}
 \, 
 \overbrace{
 \bar{g}^{kl}_{ }
 }^{\delta^{ }_{kl}/a^2_{ } }
 \, ( \, f^{ }_{\der k} \, g^{ }_{\der l} )
 \; = \; 
 \frac{1}{\N} 
 \int_{\X} \, 
 a^2_{ }\, \partial^{ }_k f \, \partial^{ }_k g
 \nn[2mm]
%%%%%
 & = &  
 \bigl\langle \, f \, \bigl(\, - g^{ }_{\der kk} \,\bigr) \,\bigr\rangle 
 \; + \; 
 \mbox{(boundary~terms)}
 \;. \label{ibp_space}
 \hspace*{6mm}
\ea
Omitting the boundary terms, 
as would be viable for periodic motion, this implies that 
in the first term on the last line of \eq\nr{Gij_t_2}, we can write 
\begin{align}
 \bigl\langle\, 
 (\, - 2 \H \tilde h' - \tilde h'' \,)\, \tilde h^{ }_{\der ij}
 \, \bigr\rangle 
 &\;
 \underset{\scriptscriptstyle
           g\,\to\,\tilde h,
         \,f\,\to\,\tilde h^{ }_{\der ij}}{
 \overset{\rmii{\nr{ibp_time}}  \lift  }{=}} 
 \;
 \bigl\langle\, 
 \tilde h'  
 \, \tilde h^{\prime}_{\der ij}
 \,\bigr\rangle
 &&\hspace*{-7mm} 
 \xrightarrow[\scriptscriptstyle
           \mathbbm{T}^{mn}_{ij}]
             {\rmii{\nr{ibp_spatial}}  \lift  }
 \;\hspace*{2mm} 
 \bigl\langle\, 
 - \, 
 \tilde h^{\prime}_{\der i} \tilde h^{\prime}_{\der j}
 \,\bigr\rangle
 \;,  \label{ibp_temporal}
 \\[2mm]
%%%%%%%%%%%%%%%%%%%%%%%%%
 \bigl\langle\, 
 \tilde h^{ }_{\der kk} \, \tilde h^{ }_{\der ij}
 \, \bigr\rangle 
 &\;
 \underset{\scriptscriptstyle
           g\,\to\,\tilde h,
         \,f\,\to\,\tilde h^{ }_{\der ij}}{
 \overset{\rmii{\nr{ibp_space}} \lift }{=}} 
 \;
 \bigl\langle\, 
 - \tilde h^{ }_{\der k} \, \tilde h^{ }_{\der kij}
 \, \bigr\rangle  
 &&\hspace*{-7mm} 
 \xrightarrow[\scriptscriptstyle
           \mathbbm{T}^{mn}_{ij}]
             {\rmii{\nr{ibp_spatial}}  \lift  }
 \;\hspace*{2mm}
 \bigl\langle\, 
 \tilde h^{ }_{\der ki} \, \tilde h^{ }_{\der kj}
 \, \bigr\rangle  
 \;. \label{ibp_spatial_2}
\end{align}
With these, the last line of \eq\nr{Gij_t_2} drops out. 
In the following, we denote
\be
 \{ (...)^{ }_{ij} \}^\rmi{t}_{ } 
 \; \equiv \; 
 \mathbbm{T}^{mn}_{ij} \langle\, ... \,\rangle^{ }_{mn} 
 \;. \label{proj_t}
\ee
Thereby \eq\nr{Gij_t_2} can be converted into 
\ba
  G^\rmi{t}_{ij} 
 & 
  \underset{\rmii{\nr{ibp_temporal}--\nr{proj_t}}}
  {\overset{\rmii{\nr{Gij_t_2}} \lift }{=}} 
 &
 (\partial_\tau^2 + 2 \H \partial_\tau^{ } - \nabla^2_{ }) 
 \vartheta^\rmi{t}_{ij}
 -2 (2\H' + \H^2_{ }) \vartheta^\rmi{t}_{ij} 
 \nn[2mm]
%%%%
 &  & \;+\,\bigl\{\,
 \;-\,
 \phi^{ }_{\der i}\,
 \phi^{ }_{\der j}
 + 
   \psi_{\der i}\,
   \psi_{\der j}
 - (  
 \phi_{\der i}\,
 \psi_{\der j}
 + 
 \phi_{\der j}\,
 \psi_{\der i}
 ) 
 \nn[2mm]
%%%%
 &  & \hspace*{4mm} \;+\, 
 (  \phi' + 3 \psi' + 4\H\phi )\, \tilde h^{ }_{\der ij}
 + 2 (\H^2_{ }-\H') \ibit 
  \tilde h^{ }_{\der i} \ibit 
  \tilde h^{ }_{\der j}
 \,\bigr\}^\rmi{t}_{ }
  \;+\; 
 \ord(\delta^3_{ })
 \;. \hspace*{7mm} \label{Gij_t_3}
\ea

\vspace*{3mm}

Turning to the right-hand side of the Einstein equations, we express
it in terms of curvature perturbations, defined 
in \eqs\nr{def_R_varphi} and \nr{def_R_v}, 
\ba
 \R^{ }_\varphi
 &
  \underset{\rmii{\nr{gauge_fixed}}}
 {\overset{\rmii{\nr{def_R_varphi}} \lift }{=}} 
 & 
 - 
  \tilde h^{ }_\rmii{D} 
  - \H\,\frac{\delta\tilde\varphi}{\bar\varphi\hspace*{0.3mm}{}'}
 \label{R_varphi_fixed}
 \;, \\[2mm] 
%%%
 \R^{ }_v
 &
  \underset{\rmii{\nr{gauge_fixed}}}
 {\overset{\rmii{\nr{def_R_v}} \lift }{=}} 
 & 
 - 
  \tilde h^{ }_\rmii{D} 
 + \H \, (\tilde h - \tilde v)
 \;. \label{R_v_fixed}  
\ea
In \eqs\nr{phi_prime_i_j} and \nr{uu_t}, 
we substitute 
\ba
 \delta\tilde\varphi^{ }_{\der i}
 \,\delta\tilde\varphi^{ }_{\der j}
 & 
  \overset{\rmii{\nr{R_varphi_fixed}}}{=}  
 & 
 \frac{(\bar\varphi\hspace*{0.3mm}')^2_{ }}{\H^2}
 (\R^{ }_\varphi + \tilde h^{ }_\rmii{D})^{ }_{\der i}
 (\R^{ }_\varphi + \tilde h^{ }_\rmii{D})^{ }_{\der j}
 \;, \label{Tij_varphi_fixed} 
 \\[2mm] 
%%%%
 (\tilde h - \tilde v)^{ }_{\der i}
 (\tilde h - \tilde v)^{ }_{\der j}
 & 
  \overset{\rmii{\nr{R_v_fixed}}}{=}  
 & 
 \frac{1}{\H^2}
 (\R^{ }_v + \tilde h^{ }_\rmii{D})^{ }_{\der i}
 (\R^{ }_v + \tilde h^{ }_\rmii{D})^{ }_{\der j}
 \;. \label{Tij_v_fixed} 
\ea
Then the right-hand side of the Einstein equations becomes
\ba
 8\pi G\, T^\rmi{t}_{ij} 
 \!\!
 & 
  \underset{\rmii{\nr{Tij_varphi_fixed},\nr{Tij_v_fixed}}}
 {\overset{\rmii{\nr{Tmunu_mixed_again}--\nr{p_t_diag}}
           \lift }{=}} 
 & 
 \!\!
 8 \pi G \, \bigl\{ \, 
 \,\bigl[\,
  (\bar{\varphi}\hspace*{0.3mm}')^2_{ } 
 + 2 a^2_{ } \bar{p} 
 \,\bigr]\,\vartheta^\rmi{t}_{ij}
 \; + \; 
 a^2_{ }\Pi^\rmi{t}_{ij}
 \,\bigr\} 
 \nn[2mm]
%%%%%
 &  & \!\! + \,
 \frac{8\pi G}{\H^2_{ }}
 \biggl\{ 
   (\bar{\varphi}\hspace*{0.3mm}')^2_{ } 
   \R^{ }_{\varphi\der i}
   \R^{ }_{\varphi\der j}
  + 
   a^2_{ }(\bar e + \bar p)
   \R^{ }_{v\der i}
   \R^{ }_{v\der j}
 \nn[2mm]
%%%%%
 &  & \;+\, 
 [ (\bar{\varphi}\hspace*{0.3mm}')^2_{ } \R^{ }_\varphi 
  + a^2_{ }(\bar e + \bar p) \R^{ }_v ]^{ }_{\der i}
 \, \tilde{h}^{ }_{\rmii{D}\der j} 
 + 
 [ (\bar{\varphi}\hspace*{0.3mm}')^2_{ } \R^{ }_\varphi 
  + a^2_{ }(\bar e + \bar p) \R^{ }_v ]^{ }_{\der j}
 \, \tilde{h}^{ }_{\rmii{D}\der i} 
 \nn[2mm]
%%%%
 &  & \;+\, 
 [\, (\bar{\varphi}\hspace*{0.3mm}')^2_{ } 
 + a^2_{ }(\bar e + \bar p) \,]
 \, \tilde{h}^{ }_{\rmii{D}\der i} 
 \, \tilde{h}^{ }_{\rmii{D}\der j} 
 \biggr\}^\rmi{t}_{ }
  \;+\; 
 \ord(\delta^3_{ })
 \;. \label{Tij_t_fixed_1}
\ea
Here, we insert the background identities from \eq\nr{bg_again}, as 
\ba
 8\pi G \bigl[\, (\bar{\varphi}\hspace*{0.3mm}')^2_{ } 
 + 2 a^2_{ }\bar{p} \,\bigr]
 &
 \overset{\rmii{\nr{bg_again}}}{=}
 & 
 -2(2\H' + \H^2_{ })
 \;, \label{bg_yet_again_1}
 \\[2mm]
%%%%
 8\pi G \bigl[\, 
 (\bar{\varphi}\hspace*{0.3mm}')^2_{ }
 +
 a^2_{ }(\bar{e} + \bar{p})
 \,\bigr]
 &
 \overset{\rmii{\nr{bg_again}}}{=}
 & 
 2(\H^2_{ } - \H') 
 \;. \label{bg_yet_again_2}
\ea
For the third line of \eq\nr{Tij_t_fixed_1}, we need 
the first-order equation from \eq\nr{delta_einstein_0i}, 
\ba
    \H \tilde h^{ }_0 
 +   
   \tilde h_\rmii{D}' 
 \!\!
 & 
  \underset{\rmii{\nr{gauge_fixed}}}
  {\overset{\rmii{\nr{delta_einstein_0i}} \lift }{=}} 
 &
 \!\!
 4\pi G \bigl[\, 
 \overbrace{
  a^2_{ }( \bar{e} + \bar{p}) (\tilde v - \tilde h)
  + {\bar\varphi}' \delta\tilde\varphi
 }^{{\rm insert}\;\nr{R_varphi_fixed},\nr{R_v_fixed}}
  \,\bigr]
 \nn[2mm]
%%%%%
 \!\!
 & 
  \underset{\rmii{}}
  {\overset{\rmii{}}{=}} 
 &
 \hspace*{-3mm}
 -\frac{4\pi G}{\H}
 \biggl\{\,
  (\bar{\varphi}\hspace*{0.3mm}')^2_{ } \R^{ }_\varphi 
  + a^2_{ }(\bar e + \bar p) \R^{ }_v
 + 
 \bigl[\,
 \hspace*{-3mm}
 \underbrace{ 
 (\bar{\varphi}\hspace*{0.3mm}')^2_{ }
 +
 a^2_{ }(\bar{e} + \bar{p})
 }_{{\rm via}\;\nr{bg_yet_again_2}:\ (\H^2_{ } - \H')/(4\pi G)}
 \hspace*{-3mm}
 \,\bigr]\,\tilde h^{ }_\rmii{D} 
 \,\biggr\}
 \;, \hspace*{10mm}
 \label{delta_einstein_0i_again}
\ea 
which implies 
\be
 - 2 
 \,\biggl[\,
 \tilde h^{ }_0 + \frac{\tilde h^\prime_\rmii{D}}{\H}
 + \biggl( 1 - \frac{\H'}{\H^2_{ }} \biggr) \tilde h^{ }_\rmii{D}
 \,\biggr]
 \; 
  \underset{\rmii{\nr{delta_einstein_0i_again}}}
 {\overset{\rmii{\nr{bg_yet_again_2}} \lift }{=}}
 \; 
 \frac{8\pi G}{\H^2_{ }}
 \bigl[\,
  (\bar{\varphi}\hspace*{0.3mm}')^2_{ } \R^{ }_\varphi 
  + a^2_{ }(\bar e + \bar p) \R^{ }_v
 \,\bigr]
 \;. \label{eom_0i_fixed}
\ee
With these substitutions, \eq\nr{Tij_t_fixed_1} becomes 
\ba
 8\pi G\, T^\rmi{t}_{ij} 
 & 
  \underset{\rmii{\nr{eom_0i_fixed}}}
 {\overset{\rmii{\nr{Tij_t_fixed_1}} \lift }{=}} 
 & 
 - \overbrace{ 
    2 (2\H' + \H^2_{ })
    }^{{\rm via}\;\nr{bg_yet_again_1} }
 \,\vartheta^\rmi{t}_{ij}
 + 
 8 \pi G \,  a^2_{ }\Pi^\rmi{t}_{ij}
 \nn[2mm]
%%%%%
 &  & \;+\,
 \biggl\{\,
 \frac{8\pi G}{\H^2_{ }}
 \bigl[\, 
   (\bar{\varphi}\hspace*{0.3mm}')^2_{ } 
   \R^{ }_{\varphi\der i}
   \R^{ }_{\varphi\der j}
  + 
   a^2_{ }(\bar e + \bar p)
   \R^{ }_{v\der i}
   \R^{ }_{v\der j}
  \,\bigr]
 \nn[2mm]
%%%%%
 &  & \hspace*{4mm} \;-\,
 2\,\bigl(\,
  \tilde h^{ }_{0\der i} \ibit 
  \tilde h^{ }_{\rmii{D}\der j}
 + 
  \tilde h^{ }_{0\der j} \ibit
  \tilde h^{ }_{\rmii{D}\der i}
 \,\bigr) 
 - \frac{2}{\H}\,\bigl(\,
  \tilde h^{ }_{\rmii{D}\der i} \ibit
  \tilde h^{\prime}_{\rmii{D}\der j}
 + 
  \tilde h^{ }_{\rmii{D}\der j} \ibit
  \tilde h^{\prime}_{\rmii{D}\der i}
 \,\bigr)
 \nn[2mm]
%%%%
 &  & \hspace*{4mm} \;+\, 
 \hspace*{-9mm}
 \underbrace{(-\hspace*{0.5mm}4 + 2)}_{
 {\rm 3rd~and~4th~line~of}\;\nr{Tij_t_fixed_1}
 }
 \hspace*{-6mm}
 \biggl( 1 - \frac{\H'}{\H^2_{ }} \biggr)
 \, \tilde{h}^{ }_{\rmii{D}\der i} 
 \, \tilde{h}^{ }_{\rmii{D}\der j} 
 \;\biggr\}^\rmi{t}_{ }
  \;+\; 
 \ord(\delta^3_{ })
 \;. \label{Tij_t_fixed_2}
\ea
Finally, we express $\tilde h^{ }_0$ and $\tilde h^{ }_\rmii{D}$
via $\phi$ and $\psi$, through 
\eqs\nr{tilde_h0_subst} and \nr{tilde_hD_subst}, 
and make use of IBP relations, from 
\eq\nr{ibp_spatial}. 
Thereby \eq\nr{Tij_t_fixed_2} turns into
\ba
 8\pi G\, T^\rmi{t}_{ij} 
 & 
  \overset{\rmii{\nr{Tij_t_fixed_2},\nr{ibp_spatial}} \lift }
 {\underset{\rmii{\nr{tilde_h0_subst},\nr{tilde_hD_subst}}}{=}} 
 & 
 -  2 (2\H' + \H^2_{ })
 \,\vartheta^\rmi{t}_{ij}
 + 
 8 \pi G \,  a^2_{ }\Pi^\rmi{t}_{ij}
 \nn[2mm]
%%%%%
 &  & \;+\, 
 \biggl\{\;
 \frac{8\pi G}{\H^2_{ }}
 \bigl[\, 
   (\bar{\varphi}\hspace*{0.3mm}')^2_{ } 
   \R^{ }_{\varphi\der i}
   \R^{ }_{\varphi\der j}
  + 
   a^2_{ }(\bar e + \bar p)
   \R^{ }_{v\der i}
   \R^{ }_{v\der j}
  \,\bigr]
 \nn[2mm]
%%%%%
 &  & \hspace*{4mm} \;-\,
 2\,\bigl(\,
  \phi^{ }_{\der i}\, \psi^{ }_{\der j}
 + 
  \phi^{ }_{\der j}\, \psi^{ }_{\der i}
 \,\bigr) 
 + 
  2 \biggl( \frac{\H'}{\H^2_{ }} - 1 \biggr)
 \, \psi^{ }_{\der i} 
 \, \psi^{ }_{\der j} 
 - \frac{2}{\H}
  \,\bigl( \psi^{ }_{\der i}\, \psi^{ }_{\der j} \bigr)'
 \nn[2mm]
%%%%
 &  & \hspace*{4mm} \;+\, 
 4 ( \psi' +  \H\phi  )\, \tilde h^{ }_{\der ij}
 + 2 (\H^2_{ } - \H')\, \tilde h^{ }_{\der i}\, \tilde h^{ }_{\der j}
 \;\biggr\}^\rmi{t}_{ } 
 \;+\; 
 \ord(\delta^3_{ })
 \;. \label{Tij_t_fixed_3}
\ea

The remaining step is to equate \eqs\nr{Gij_t_3} and \nr{Tij_t_fixed_3}.
Transporting the source terms to the right-hand side, we obtain 
\ba
 \boxed{
 \begin{array}{rcl} 
 \hphantom{\hspace*{0.6cm}}
 && \hspace*{-1.5cm} \displaystyle \vphantom{\Bigg|^b_q}
 (\partial_\tau^2 + 2 \H \partial_\tau^{ } - \nabla^2_{ }) 
 \vartheta^\rmi{t}_{ij}
 \; 
  \underset{\rmii{\nr{Tij_t_fixed_3}}}
  {\overset{\rmii{\nr{Gij_t_3}} \lift }{=}} 
 \; 
 8 \pi G \,  a^2_{ }\Pi^\rmi{t}_{ij}
 \nn[4mm]
%%%%%
 & + & \displaystyle \biggl\{\,
 \frac{8\pi G}{\H^2_{ }}
 \bigl[\, 
   (\bar{\varphi}\hspace*{0.3mm}')^2_{ } 
   \R^{ }_{\varphi\der i}
   \R^{ }_{\varphi\der j}
  + 
   a^2_{ }(\bar e + \bar p)
   \R^{ }_{v\der i}
   \R^{ }_{v\der j}
  \,\bigr]
 \nn[5mm]
%%%%%
 &  & \;+\, \displaystyle
 \phi^{ }_{\der i}\,\phi^{ }_{\der j}
 - (  \phi^{ }_{\der i}\,\psi^{ }_{\der j} 
    + \phi^{ }_{\der j}\,\psi^{ }_{\der i})
 + \biggl(\frac{2\H'}{\H^2_{ }} -3 \biggr) 
 \psi^{ }_{\der i}\,\psi^{ }_{\der j}
 - \frac{2}{\H}\bigl( \psi^{ }_{\der i}\,\psi^{ }_{\der j} \bigr)' 
 \quad
 \nn[5mm]
%%%%%
 &  & \;+\, \displaystyle
 (\psi - \phi)'\, \tilde h^{ }_{\der ij}
 \,\biggr\}^\rmi{t}_{ }
 + \ord(\delta^3_{ })
 \;. \vphantom{\Bigg|^b_q}
 \end{array} 
 }
 \nn[-26mm]
 \label{sigw}
 \\[14mm]
 \nonumber
\ea
A significant, but nevertheless incomplete, cancellation
took place 
in the contributions 
involving $\tilde h$ (cf.\ \eq\nr{g_munu_fixed}).
In order to obtain physical predictions from \eq\nr{sigw}, 
usually simplified versions thereof are solved with the Green's function
method that we will illustrate in \se\ref{ss:gw_reheat}. 
Let us elaborate on the key points
related to \eq\nr{sigw}: 

\vspace*{-3mm}

%%%%%%%%%%%%%%%%%%%%%%%%%%%%%%%%%%%%%%%%%%%%%%%%%%%%%%%%%%%%%%%%%

\paragraph{(i) Gauge dependence.}

In much of the literature, a gauge has been fixed
($\tilde h = 0$). If this is done from the outset,  
then one misses the information that the right-hand side of \eq\nr{sigw}
is actually gauge dependent~\cite{gauge}. 
Gauge dependence disappears, and the 
result is unambiguous, only if $\psi = \phi$, 
which according to \eq\nr{dot_phi} requires the 
absence of the scalar part of anisotropic stress, $\Pi$. 
The absence of $\Pi$ 
alleviates the concerns about
double counting that we express at the beginning
of this section. If $\Pi = 0$, then 
\eq\nr{sigw} can be written in terms of gauge-invariant variables, 
\ba
 && 
 \hspace*{-0.8cm}
 (\partial_\tau^2 + 2 \H \partial_\tau^{ } - \nabla^2_{ }) 
 \vartheta^\rmi{t}_{ij}
 \;
  \underset{\scriptscriptstyle \Pi^{ }_{ }\;=\;0}
  {\overset{\rmii{\nr{sigw},\nr{dot_phi}} \lift }{=}} 
 \; 
 8 \pi G \,  a^2_{ }\Pi^\rmi{t}_{ij}
 \label{sigw_2} \\[1mm]
%%%%%
 & + &
 \biggl\{\,
 \frac{8\pi G}{\H^2_{ }}
 \bigl[\, 
   (\bar{\varphi}\hspace*{0.3mm}')^2_{ } 
   \R^{ }_{\varphi\der i}
   \R^{ }_{\varphi\der j}
  + 
   a^2_{ }(\bar e + \bar p)
   \R^{ }_{v\der i}
   \R^{ }_{v\der j}
  \,\bigr]
 \, + \, 
 2\,\biggl(\frac{\H'}{\H^2_{ }} - 2 - \frac{\partial^{ }_\tau}{\H} \biggr) 
 \psi^{ }_{\der i}\,\psi^{ }_{\der j}
 \,\biggr\}^\rmi{t}_{ }
 + \ord(\delta^3_{ })
 \;. 
 \nonumber 
\ea

\paragraph{(ii) Index placement.}

Often, instead of $G^{ }_{ij}$,
the component ${G^{\,i}_{ }}^{ }_j$ of the Einstein tensor
is considered (cf.,\ e.g.,\ ref.~\cite{matarrese}). 
At the second order, given the metric
in \eq\nr{g_munu_fixed}, the relation to our expression reads
\ba
 G^{\scriptscriptstyle (2)}_{ij}
 \; = \; 
 [ g^{ }_{i\mu} {G^{\,\mu}_{ }}^{ }_j ]^{\scriptscriptstyle (2)}_{ }
 \!\!& = &\!\! 
 g^{\scriptscriptstyle (0)}_{i\mu} 
   [{G^{\,\mu}_{ }}^{ }_j]^{\scriptscriptstyle (2)}_{ }
 \;+\; 
 g^{\scriptscriptstyle (1)}_{i\mu}
   [{G^{\,\mu}_{ }}^{ }_j]^{\scriptscriptstyle (1)}_{ }
 \;+\;
 \hspace*{-5mm} 
 \overbrace{
 g^{\scriptscriptstyle (2)}_{i\mu}  
 }^{{\rm via}\;\nr{g_munu_fixed}\;\to\;0}
 \hspace*{-5mm} 
 [{G^{\,\mu}_{ }}^{ }_j]^{\scriptscriptstyle (0)}_{ }
 \nn[2mm]
%%%%
 \!\!&
 \overset{\rmii{\nr{g_munu_fixed}}}{=} 
 &\!\! 
 a^2_{ }\,\Bigl\{\, 
 [{G^{\,i}_{ }}^{ }_j]^{\scriptscriptstyle (2)}_{ }
 \;-\;
 2 \tilde h^{ }_\rmii{D}\,
 [{G^{\,i}_{ }}^{ }_j]^{\scriptscriptstyle (1)}_{ }
 \;+\;
 \tilde h^{ }_{\der i}\,
 [{G^{\,0}_{ }}^{ }_j]^{\scriptscriptstyle (1)}_{ }
 \,\Bigr\}
 \;. \hspace*{6mm} \label{G_ij_2}
\ea
Here, in turn,  we can write
\ba
 a^2_{ } [{G^{\,i}_{ }}^{ }_j]^{\scriptscriptstyle (1)}_{ }
 & = & 
 a^2_{ } [g^{i\mu}_{ } {G^{ }_{\mu j }}]^{\scriptscriptstyle (1)}_{ }
 \; = \; 
 a^2_{ } 
 \hspace*{-6mm}
 \overbrace{
 [g^{i\mu}_{ }]^{\scriptscriptstyle (0)}_{ }
 }^{{\rm via}\;\nr{g^munu_fixed}:\;\delta^{ }_{i\mu}/a^2_{ } }
 \hspace*{-7mm}
 \,  G^{\scriptscriptstyle (1)}_{\mu j }
 \; + \; 
 a^2_{ }  
 [g^{i\mu}_{ }]^{\scriptscriptstyle (1)}_{ }
 \hspace*{-13mm}
 \overbrace{
 \, G^{\scriptscriptstyle (0)}_{\mu j }
 }^{{\rm via}\;\nr{G_ij}:\; - (2 \H' + \H^2_{ } ) \delta^{ }_{\mu j}}
 \hspace*{-13mm}
 \nn[2mm]
%%%%%
 & = & 
 \hspace*{-5mm}
 \overbrace{
 G^{\scriptscriptstyle (1)}_{i j }
 }^{{\rm insert}\,\nr{delta_Gij_s},\,\nr{gauge_fixed} }
 \hspace*{-5mm}
 - \;
 (2 \H' + \H^2_{ } )
 \hspace*{-3mm}
 \overbrace{
  a^2_{ }  
 [g^{ij}_{ }]^{\scriptscriptstyle (1)}_{ }
 }^{{\rm via}\;\nr{g^munu_fixed}:\;2 \tilde h^{ }_\rmiii{D} \delta^{ }_{ij}  }
 \hspace*{-3mm}
 \nn[2mm]
%%%%%
 & \overset{\scriptscriptstyle i\;\neq\;j
             \lift }{\supset} & 
 (\tilde h^{ }_\rmii{D} - \tilde h^{ }_0 - \tilde h'
 - 2 \H\,\tilde h )^{ }_{\der ij}
 \nn[2mm]
%%%%% 
 & 
 \underset{\rmii{\nr{tilde_hD_subst}}}
{\overset{\rmii{\nr{tilde_h0_subst}} \lift }{=}} 
 & 
 (\psi - \phi)^{ }_{\der ij}
 \;, \label{G_ij_1} \\[3mm]
%%%%%%%%%%%%%%%%%%%%
 a^2_{ } [{G^{\,0}_{ }}^{ }_j]^{\scriptscriptstyle (1)}_{ }
 & = & 
 a^2_{ } [g^{0\mu}_{ } {G^{ }_{\mu j }}]^{\scriptscriptstyle (1)}_{ }
 \; = \; 
 a^2_{ } 
 \hspace*{-6mm}
 \overbrace{
 [g^{0\mu}_{ }]^{\scriptscriptstyle (0)}_{ }
 }^{{\rm via}\;\nr{g^munu_fixed}:\;-\,\delta^{ }_{0\mu}/a^2_{ } }
 \hspace*{-7mm}
 \,  G^{\scriptscriptstyle (1)}_{\mu j }
 \; + \; 
 a^2_{ }  
 [g^{0\mu}_{ }]^{\scriptscriptstyle (1)}_{ }
 \hspace*{-13mm}
 \overbrace{
 \, G^{\scriptscriptstyle (0)}_{\mu j }
 }^{{\rm via}\;\nr{G_ij}:\; - (2 \H' + \H^2_{ } ) \delta^{\mu j}_{ }}
 \hspace*{-13mm}
 \nn[2mm]
%%%%%
 & = & 
 \overbrace{
 - \, G^{\scriptscriptstyle (1)}_{0 j }
 }^{{\rm insert}\,\nr{delta_G0i_s},\,\nr{gauge_fixed} }
 - \;
 (2 \H' + \H^2_{ } )
 \hspace*{-2mm}
 \overbrace{
  a^2_{ }  
 [g^{0j}_{ }]^{\scriptscriptstyle (1)}_{ }
 }^{{\rm via}\;\nr{g^munu_fixed}:\; \tilde h^{ }_{\der j}  }
 \hspace*{-3mm}
 \nn[2mm]
%%%%%
 & = & 
 -2 (\tilde h^{\prime}_\rmii{D} + \H \tilde h^{ }_0 )^{ }_{\der j}
 + \,
 \bcancel{(2 \H' + \H^2_{ } ) \tilde h^{ }_{\der j} }
 - \,
 \bcancel{(2 \H' + \H^2_{ } ) \tilde h^{ }_{\der j} }
 \nn[2mm]
%%%%%%
 &
 \underset{\rmii{\nr{tilde_hD_subst}}}
{\overset{\rmii{\nr{tilde_h0_subst}} \lift }{=}} 
 & 
 -2\, [\, \psi' + \H \phi + (\H' - \H^2_{ })\tilde h \,]^{ }_{\der j}
 \;. \label{G_0j_1}
\ea
Substituting \eqs\nr{G_ij_1} and \nr{G_0j_1} in \eq\nr{G_ij_2}, 
where we write $\tilde h^{ }_\rmii{D} = \psi + \H\,\tilde h$
from \eq\nr{tilde_hD_subst}, we get 
\ba
 a^2_{ }[{G^{\,i}_{ }}^{ }_j]^{\scriptscriptstyle (2)}_{ }
 & 
 \overset{\rmii{\nr{G_ij_2}} \lift }{
 \underset{\scriptscriptstyle 
   i\; \neq\; j 
  \vphantom{|} }{=}}
 & 
 G^{\scriptscriptstyle (2)}_{ij} + 2
 \overbrace{ 
 (\psi + \H\,\tilde h)
 }^{{\rm from}\;\nr{tilde_hD_subst}}
 \;
 \overbrace{
 (\psi - \phi)^{ }_{\der ij} \vphantom{\tilde h} 
 }^{{\rm from}\;\nr{G_ij_1}}
 \; + \; 
 2 \tilde h^{ }_{\der i}\,
 \overbrace{
 [\,\psi' + \H \phi + (\H' - \H^2_{ })\tilde h \,]^{ }_{\der j}
 }^{{\rm from}\;\nr{G_0j_1}}
 \nn[2mm]
%%%%
 & = & 
 G^{\scriptscriptstyle (2)}_{ij}
 + 2 \psi \ibit (\psi - \phi)^{ }_{\der ij}
 \nn[2mm]
%%%%
 &  & \; + \,
  2 \H\ibit \tilde h\ibit (\psi - \phi )^{ }_{\der ij}
 + 
 2 \tilde{h}^{ }_{\der i} \ibit ( \psi'_{\der j} + \H \phi^{ }_{\der j} )
 + 
 2 (\H' - \H^2_{ }) \ibit 
  \tilde{h}^{ }_{\der i} \ibit
  \tilde{h}^{ }_{\der j}
 \;. \hspace*{6mm} \label{Gij_updown}
\ea
Adding together the terms on the first line of \eq\nr{Gij_updown} 
and on the second line of \eq\nr{Gij_t_2}, 
the result agrees with the corresponding terms given
in \eq(4) of ref.~\cite{sigw}, where the gauge 
$\tilde h = 0$ is chosen, so that 
the second line of \eq\nr{Gij_updown} is absent.
We remark that 
if we want to write down the Einstein equations with this
index placement, the energy-momentum tensor 
$a^2_{ } [{T^i_{ }}^{ }_j]^{\scriptscriptstyle (2)}_{ }$ 
needs to be similarly
worked out, however the final equation
is equivalent to \eq\nr{sigw}, 
if the same gauge choice is employed. 

\vspace*{-3mm}

\paragraph{(iii) Simplifications if one enthalpy component dominates.}

Depending on the physical setting, the background enthalpy
density might be dominated either by the inflaton field, i.e.\ 
$ 
 (\bar{\varphi}\hspace*{0.3mm}')^2_{ } 
 \gg
 a^2_{ }(\bar e + \bar p)
$, 
or the plasma, i.e.\ 
$ 
 (\bar{\varphi}\hspace*{0.3mm}')^2_{ } 
 \ll 
 a^2_{ }(\bar e + \bar p)
$. 
In these cases, 
we can express the corresponding curvature
perturbation, $\R^{ }_\varphi$ or $\R^{ }_v$, 
in terms of the Bardeen potentials, 
through \eqs\nr{eom_0i_fixed}, \nr{tilde_h0_subst}, 
and \nr{tilde_hD_subst}. Then only Bardeen potentials 
(and gauge artifacts) appear on the right-hand 
side of \eq\nr{sigw}. 

\vspace*{-3mm}

\paragraph{(iv) Remarks on the anisotropic stress, $\Pi^\tensor_{ij}$.}

At the beginning of this section, 
we express concerns about anisotropic stress, 
however in the literature it is sometimes kept and re-expressed
in a peculiar way. Suppose that we write 
$\Pi^{ }_{ij} \equiv p\, \pi^{ }_{ij}$, 
and consider a scalar contribution to the latter part, 
$
 \pi^\text{s}_{ij} \equiv \pi^{ }_{\der ij} 
 - \delta^{ }_{ij} \nabla^2_{ }\pi/3
$. 
Since $\pi$ starts at first order, 
we may write 
$
 \Pi^{\scriptscriptstyle (2)}_{ij} \supset (\bar p + \delta p )
 \, \pi^{\scriptscriptstyle (1)}_{\der ij}
$.
The term $\bar p\, \pi^{\scriptscriptstyle (1)}_{ }$ is identified as 
the scalar anisotropic stress $\Pi$ in \eq\nr{dot_phi},  
whereas the pressure perturbation is written
as $\delta p \simeq c_s^2\, \delta e$, where $c_s^2$ is 
the speed of sound squared. Then, formally,  
$\{ \delta e\, \pi^{\scriptscriptstyle (1)}_{\der ij} \}^\rmi{t}_{ } \neq 0$.
Concretely, if $\delta e$ is solved for
from \eq\nr{delta_einstein_00} (in the absence of~$\varphi$),
and $\bar p\, \pi^{\scriptscriptstyle (1)}_{ } = \Pi$ 
is solved from \eq\nr{dot_phi}, 
this yields yet another 
structure expressed in terms of Bardeen potentials. 
However, $\Pi \neq 0$ implies that the result
is gauge dependent (see point~(i)).  
It is also unclear to what extent this second-order effect
is independent of the other second-order terms in \eq\nr{sigw}.

\vspace*{-3mm}

\paragraph{(v) Example of how double counting can be avoided.}

If the inflaton field thermalizes to a temperature~$T$, 
then at small $k/a \ll T$, the contribution from 
$\delta\varphi^{ }_{\der \ibit i} \ibit \delta\varphi^{ }_{\der \ibit j}$
in \eq\nr{phi_prime_i_j}, 
which has turned into 
$
   (\bar{\varphi}\hspace*{0.3mm}')^2_{ } 
   \R^{ }_{\varphi\der i}
   \R^{ }_{\varphi\der j}
$ in \eq\nr{sigw}, 
generates a shear viscosity,~$\eta$.
The shear viscosity parametrizes anisotropic stress 
(cf.\ \app\ref{app:viscous}), 
and contributes to the  
gravitational-wave power spectrum according to \eq\nr{P_T_final_3}.
However, as discussed at the end of 
\app\ref{rayleigh-jeans}, the typical
fluctuations of a thermal inflaton are so ``hard'', with physical
momenta $p\sim T$ (or $p\sim \sqrt{2 m T}$ if the inflaton
has a mass $m \gg T$), 
that they
cannot be described by classical field theory,  
because it suffers from a Rayleigh-Jeans divergence.
To compute the contribution 
of~$\delta\varphi^{ }_{\der \ibit i} \ibit \delta\varphi^{ }_{\der \ibit j}$ 
to $\eta$ requires tools 
from thermal field theory~\cite{reheat}
(the problem is even more delicate if $\varphi$ does not interact with 
a separate plasma, but the plasma is rather generated by its
self-interactions~\cite{jeon}).   
Once this task has been accomplished, 
we can address the issue
of double counting: it would {\em not} be consistent
to include both $\Pi^\rmi{t}_{ij}$ with $\eta$
(cf.\ \eqs\nr{decomposition} and \nr{delta_S}), 
and the {\em full} second-order contribution from \eq\nr{phi_prime_i_j}, 
in a single computation.
Rather, the fluctuations in $\delta\varphi$ should be divided
into soft and hard momentum ranges, like in effective field theories.
The contribution from hard momenta could be incorporated 
in the shear viscosity, whereas in \eq\nr{phi_prime_i_j} 
(and subsequently in \eq\nr{sigw}), we should include only 
momenta that are soft enough to be treated classically.  
For completeness, we note that the second-order 
contribution can contain divergences even if 
no temperature is present, and therefore in general
requires some form of renormalization or matching~\cite{brems}.

\label{double_count}

%
%%%%%%%%%%%%%%%%%%%%%%%%%%%%%%%%%%%%%%%%%%%%%%%%%%%%%%%%%%%%%%%%%

% \vspace*{3mm}

To summarize, scalar perturbations induce tensor 
perturbations at second order, as shown by \eq\nr{sigw}.
However, we should not double-count 
contributions from first-order anisotropic stress and 
explicit second-order terms. This is underlined
by the fact that \eq\nr{sigw} is gauge dependent if 
$\psi\neq \phi$, which according to \eq\nr{dot_phi} 
happens if $\Pi \neq 0$.

%%%%%%%%%%%%%%%%%%%%%%%%%%%%%%%%%%%%%%%%%%%%%%%%%%%%%%%%%%%%%%%%%%%%%%%%%
%
\subsection{From anisotropic stress to gravitational-wave energy density}
\label{ss:gw_reheat}

The gravitational-wave source that we have obtained
in \eq\nr{sigw} does not rely on the presence 
of an equilibrated plasma (even though it permits for this possibility), 
and could be active during different epochs, such as 
inflation, 
the early matter-domination era that may follow it, 
a reheating stage, 
or the subsequent radiation-dominated expansion. 
In the present section, 
we specialize on 
{\em phenomena which take place after reheating}, 
\index{gravitational waves: after reheating}
when the universe may be radiation-dominated.
Furthermore, the processes are assumed to be   
localized within the Hubble horizon ($k \gg \H$), which 
according to \fig\ref{fig:history_tau} on p.~\pageref{fig:history_tau} 
is true after some time. 
Local out-of-equilibrium physics could be involved, 
such as that induced by a first-order phase transition. 
Under these circumstances, we can take 
further steps with an expression like \eq\nr{sigw}, converting
an ``abstract'' wave equation into a concrete formula for the
energy density that gravitational waves carry. 

As a starting point, we consider 
the second term of \eq\nr{P_T_final_2}.
The anisotropic stress $\Pi^\tensor_{ij}$ there 
can be interpreted as any source term, 
whereby the result can also be used in the context of \eq\nr{sigw}.
In \eq\nr{P_T_final_2}, the Green's function $G^{ }_\rmii{R}$, 
defined in \eq\nr{greens_ht}, determines how gravitational waves
propagate from the source until observation, 
and we therefore need to discuss how $G^{ }_\rmii{R}$ behaves. 
Specifically, \eq\nr{greens_ht} is the
harmonic oscillator equation of motion, with a time-dependent
damping coefficient. When $k \gg \H$, 
\eq\nr{greens_ht} has to be solved in the {\em underdamped regime}. 
\index{underdamped regime}
 
To appreciate the challenge of determining $G^{ }_\rmii{R}$, 
let us first take the ansatz
\be
 G^{ }_\iR(\tau,\tau^{ }_i,k)
 \;\overset{\rmii{?}}{\sim}\; 
 A^{ }_\iR(\tau,\tau^{ }_i,k)
 \; \equiv \; 
 \alpha \sin\bigl( k (\tau - \tau^{ }_i) \bigr)
 \;. \label{A_R}
\ee
With this ansatz, 
$
 (\partial^2_\tau + k^2_{ })
 A^{ }_\iR = 0
$, 
so we are correctly accounting for the largest
terms in \eq\nr{greens_ht}. 
But if we want to know $G^{ }_\rmii{R}$ for a long time interval,
functional dependences like $f(\H\tau)$ could emerge. 
They lead to terms like 
$\partial^2_\tau f = \H' f' + \H^2_{ } f''$.
In comparison, the damping from \eq\nr{greens_ht} yields 
$
 \H \partial^{ }_\tau A^{ }_\iR \sim k \H A^{ }_\iR
 \gg \H^2_{ } A^{ }_\iR
$.
So, if $|f''|\sim |f'| \sim |f|$, the damping 
$\H \partial^{ }_\tau A^{ }_\iR $ 
is more important than $\partial_\tau^2 f$. 
However, \eq\nr{A_R} omits the damping.

Fortunately, a better approximation
% than in \eq\nr{A_R}  
can be found. Consider the ansatz
\be
 G^{ }_\iR(\tau,\tau^{ }_i,k)
 \;\overset{\rmii{?}}{\sim}\; 
 B^{ }_\iR(\tau,\tau^{ }_i,k)
 \; \equiv \; 
 \frac{\alpha\, a(\tau^{ }_i)}{a(\tau)}\,
 \sin\bigl( k (\tau - \tau^{ }_i) \bigr)
 \;. \label{B_R}
\ee
It follows that 
\ba
 \partial^{ }_\tau B^{ }_\iR 
 &
 \overset{\rmii{\nr{B_R}}}{=}
 & 
 \frac{\alpha\, a(\tau^{ }_i)}{a(\tau)}\,
 \biggl\{
  - \frac{a'}{a}\, 
    \sin\bigl( k (\tau - \tau^{ }_i) \bigr)
  + k\, \cos\bigl( k (\tau - \tau^{ }_i) \bigr)
 \biggr\}
 \;, \label{dB_R} \\[2mm]
%%%% 
 \partial^{2}_\tau B^{ }_\iR 
 &
 \overset{\rmii{\nr{dB_R}}}{=}
 & 
 \frac{\alpha\, a(\tau^{ }_i)}{a(\tau)}\,
 \biggl\{
 \biggl[\,
  - \frac{a''}{a}
  + \frac{2(a')^2_{ }}{a^2}
  - k^2_{ }
 \,\biggr] 
    \sin\bigl( k (\tau - \tau^{ }_i) \bigr)
  -\frac{2 a' k}{a}\,   
    \cos\bigl( k (\tau - \tau^{ }_i) \bigr)
 \biggr\}
 \nn[2mm]
%%%%
 & \Rightarrow & 
 \bigl(\, \partial_\tau^2 + 2 \H \partial^{ }_\tau + k^2_{ } \,\bigr) 
 B^{ }_\iR 
 \; = \; 
 -\frac{a''}{a}
 B^{ }_\iR 
 \; 
 \overset{\rmii{\nr{eq_hubble}}}{=}
 \; 
 - \bigl(\, \H' + \H^2_{ } \,\bigr)
 B^{ }_\iR
 \; 
 \overset{?}{\sim}
 \;  
 0
 \;. \label{eom_B_R}
\ea
Now the remainder is smaller than with $A^{ }_\iR$, because
it contains no term of $\ord(k\H)$. Furthermore, 
for 
$
 (\bar{\varphi}\hspace*{0.3mm}')^2_{ }
 \ll
 a^2_{ }(\bar{e} - 3 \bar{p})
$, 
\eqs\nr{bg_yet_again_1} and 
\nr{bg_yet_again_2}
imply that
\be
 \frac{a''}{a} 
 \;
 \overset{\rmii{\nr{eq_hubble}}}{=}
 \;
 \H' + \H^2_{ } 
 \;
 \underset{\scriptscriptstyle
   (\bar{\varphi}\hspace*{0.3mm}')^\rmiii{2}_{ }
   \;\ll\;
   a^\rmiii{2}_{ }(\bar{e} - 3 \bar{p})
 }{
 \overset{\rmii{\nr{bg_yet_again_1},\nr{bg_yet_again_2}}
          \lift }{\approx}}
 \;
 \frac{4\pi G a^2_{ }}{3}
   \bigl(\, \bar{e} - 3 \bar{p} \,\bigr)
 \;. \label{trace} 
 \index{trace anomaly}
\ee
The left-hand side of \eq\nr{trace} is proportional to 
the Ricci scalar (cf.\ \eq\nr{bg_ricci_scalar}), and is often 
referred to as {\em curvature} in general relativity. 
\index{curvature ($R$): definition}
The combination $ \bar{e} - 3 \bar{p} $ appearing 
on the right-hand side of \eq\nr{trace} is known
as the {\em trace anomaly} in thermodynamics, 
with the ``trace'' referring to that of the 
energy-momentum tensor. To give the trace anomaly 
a concrete meaning, we note that
with the parametrization from \eq\nr{p_r}, 
\be
 e^{ }_r - 3 p^{ }_r
 \;
 \overset{\rmii{\nr{p_r}}}{=} 
 \;
 \frac{2 (g^{ }_* - h^{ }_*) \pi^2_{ }T^4_{ }}{15}
 \;. \label{trace_concrete}
\ee
This vanishes in the massless non-interacting limit, 
where $g^{ }_* = h^{ }_*$, perhaps justifying calling it an ``anomaly''.  
Exceptions to its smallness are special points in the phase diagram,
such as phase transitions,
where interactions are strong (cf.,\ e.g.,\ ref.~\cite{gubser}), 
or temperatures at which some thermalized particle species 
``crosses a mass threshold'', i.e.\ has a mass of the order of
the temperature. However, the latter happens for one particle species
at a time, so it is usually a relatively small effect.

Given \eqs\nr{eom_B_R}--\nr{trace_concrete}, 
we conclude that we can normally adopt 
$B^{ }_\iR$ as
a good approximation for $G^{ }_\iR$ from \eq\nr{greens_ht}. 
There are some exceptions, and we return to them 
in the last bullet point
on p.~\pageref{trace_anomaly_exceptions}. 
Fixing the integration constant, $\alpha$, through 
\eqs\nr{greens_retard} and \nr{greens_deriv}, we thus find
\be
 G^{ }_\iR(\tau,\tau^{ }_i,k)
 \;\; 
 \overset{\scriptscriptstyle \tau \,\ge\, \tau^{ }_i \lift }
         {\scriptscriptstyle \underset{k \,\gg\, \H}{\approx}}
 \;\; 
 \frac{a(\tau^{ }_i)\, \sin(k(\tau - \tau^{ }_i)) }{a(\tau)\,k}
 \;, \qquad
 i
 \overset{\rmii{\nr{P_T_final_2}}}{\in}
 \{1,2\}
 \;. \label{greens_ht_BR}
\ee
Making use of 
$
 \sin c^{ }_1 \sin c^{ }_2 = \frac{1}{2}
 [\,
  \cos( c^{ }_1 - c^{ }_2 ) - \cos( c^{ }_1 + c^{ }_2 ) 
 \,]
$, 
this yields
\be
 G^{ }_\iR(\tau,\tau^{ }_1,k) \,
 G^{ }_\iR(\tau,\tau^{ }_2,k)  
 \;
 \underset{\scriptscriptstyle k\,\gg\,\H}{
 \overset{\rmii{\nr{greens_ht_BR}} \lift }{\approx}}
 \; 
 \frac{a(\tau^{ }_1)\,a(\tau^{ }_2)}{2k^2_{ } a^2_{ }(\tau)} 
 \bigl[\,
  \cos\bigl(k(\tau^{ }_2 - \tau^{ }_1)\bigr)
 - 
  \cos\bigl(k(2\tau - \tau^{ }_1 - \tau^{ }_2)\bigr)
 \,\bigr]
 \;. \label{trigo} 
\ee
If we instead consider time derivatives of $h^\tensor_{ij}$,
which play a role in the energy density (cf.\ \eq\nr{e_gw}),
and insert 
$
 \cos c^{ }_1 \cos c^{ }_2 = \frac{1}{2}
 [\,
  \cos( c^{ }_1 - c^{ }_2 ) + \cos( c^{ }_1 + c^{ }_2 ) 
 \,]
$,
then we find
\be
 G^{\hspace*{0.3mm}\prime}_\iR(\tau,\tau^{ }_1,k) \,
 G^{\hspace*{0.3mm}\prime}_\iR(\tau,\tau^{ }_2,k)  
 \;
 \underset{\scriptscriptstyle k\,\gg\,\H}{
 \overset{\rmii{\nr{greens_ht_BR}} \lift }{\approx}}
 \; 
 \frac{a(\tau^{ }_1)\,a(\tau^{ }_2)}{2 a^2_{ }(\tau)} 
 \bigl[\,
  \cos\bigl(k(\tau^{ }_2 - \tau^{ }_1)\bigr)
 + 
  \cos\bigl(k(2\tau - \tau^{ }_1 - \tau^{ }_2)\bigr)
 \,\bigr]
 \;. \label{trigo_2} 
\ee
 
Now, we may recall from electrodynamics that when a source emits 
radiation, there is a ``near zone'' in which the exact solution 
is not yet wavelike, and a ``far zone'' in which it settles to   
a propagating wave. Similarly, for gravitational waves, 
their physical characteristics can be unambiguously defined
only once the wave motion has had time to take shape. 
Concretely, this requires that $k(\tau - \tau^{ }_i) \gg 1$.
In this domain, the second terms in \eqs\nr{trigo} and \nr{trigo_2}
are rapidly oscillating, and average to zero, implying that 
\be
 \bigl\langle\,
 G^{ }_\iR % (\tau,\tau^{ }_1,k)
 \,
 G^{ }_\iR % (\tau,\tau^{ }_2,k)  
 \,\bigr\rangle
 \;
 \underset{\scriptscriptstyle k(\tau-\tau^{ }_i)\,\gg\,1}{
 \overset{\rmii{\nr{trigo}} \lift }{\approx}}
 \; 
 \frac{a(\tau^{ }_1)a(\tau^{ }_2) 
      \cos\bigl(k(\tau^{ }_2 - \tau^{ }_1)\bigr)
      }{2k^2_{ } a^2_{ }(\tau)} 
 \;
 \underset{% k\,\gg\,\H\,,\,
           \scriptscriptstyle 
           k(\tau-\tau^{ }_i)\,\gg\,1}{
 \overset{\rmii{\nr{trigo_2}} \lift }{\approx}}
 \; 
 \frac{
 \bigl\langle\,
 G^{\hspace*{0.3mm} \prime}_\iR % (\tau,\tau^{ }_1,k)
 \,
 G^{\hspace*{0.3mm} \prime}_\iR % (\tau,\tau^{ }_2,k)  
 \,\bigr\rangle
 }{ k^2_{ }}
 \;.  \label{trigo_appro}
\ee 
On the other hand, the gravitational
energy density from the first variant of \eq\nr{e_gw} is
proportional to the quadratic expression
$
 G^{\prime}_\iR
 G^{\prime}_\iR
 + 
 k^2_{ }
 G^{ }_\iR
 G^{ }_\iR
$.
With this form, 
fast oscillations cancel between 
\eqs\nr{trigo} and \nr{trigo_2}
at {\em any time}, and no averaging is required. 

Inserting $\langle\, G^{ }_\iR G^{ }_\iR \,\rangle$ from 
\eq\nr{trigo_appro} into \eq\nr{P_T_final_2}, which means
that we are assuming $k(\tau - \tau^{ }_i) \gg 1$ 
and therefore averaging over oscillations, we find
\ba
 && \hspace*{-1.9cm}
 \P^{ }_\tensor(\tau,k) 
 \;
 \underset{\rmii{\nr{trigo_appro}}}{
 \overset{\rmii{\nr{P_T_final_2}} \lift }{\supset}}
 \;
 \frac{64 k}{\mpl^4\, a^2_{ }(\tau) }
 \int_{-\infty}^{\tau} \! \dd \tau^{ }_1 \, 
 \int_{-\infty}^{\tau} \! \dd \tau^{ }_2 \, 
      \cos\bigl(k(\tau^{ }_2 - \tau^{ }_1)\bigr)
 a^3_{  }(\tau^{ }_1) a^3_{  }(\tau^{ }_2)
 P^{ }_\rmii{$\Pi$}(\tau^{ }_1,\tau^{ }_2,\vec{k})
 \nn[2mm]
%%%%%
 & \!\! \overset{ \scriptscriptstyle 
                  \tau^{ }_\rmiii{1} \leftrightarrow \tau^{ }_\rmiii{2}
                  \lift }
        {\underset{\rmii{\nr{def_S_Pi}}}{=}}\!\! & 
 \frac{128 k}{\mpl^4\, a^2_{ }(\tau)}
 \int_{-\infty}^{\tau} \! \dd \tau^{ }_1 \, 
 \int_{-\infty}^{\tau^{ }_1} \! \dd \tau^{ }_2 \, 
      \cos\bigl(k(\tau^{ }_2 - \tau^{ }_1)\bigr)
 a^3_{  }(\tau^{ }_1) a^3_{  }(\tau^{ }_2)
 S^{ }_\rmii{$\Pi$}(\tau^{ }_1,\tau^{ }_2,\vec{k})
 \;. \hspace*{8mm} \label{P_T_final_4} 
\ea
For a source active only for a finite period of time, 
until a moment~$\tau^{ }_{\sfin}$, 
we may replace $\tau \to \tau^{ }_{\sfin}$ 
in the upper bounds of the $\tau^{ }_1$ and $\tau^{ }_2$-integrals, 
and the residual time dependence
of the spectrum lies in the prefactor $1/a^2_{ }(\tau)$.
In the second step, we have divided the integration into the domains
$\tau^{ }_1 > \tau^{ }_2$ and $\tau^{ }_1 < \tau^{ }_2$; 
renamed $\tau^{ }_1\leftrightarrow \tau^{ }_2$ in the latter; 
and defined a time-symmetrized correlator as 
\be
 S^{ }_\rmii{$\Pi$}(\tau^{ }_1,\tau^{ }_2,\vec{k})
 \; \equiv \; 
 \frac{1}{2} 
 \bigl[\, 
   P^{ }_\rmii{$\Pi$}(\tau^{ }_1,\tau^{ }_2,\vec{k}) + 
   P^{ }_\rmii{$\Pi$}(\tau^{ }_2,\tau^{ }_1,\vec{k})
 \,\bigr]
 \;. \label{def_S_Pi}
\ee
In the classical limit, $P^{ }_\rmii{$\Pi$}$ is automatically
time-symmetric (cf.\ \eq\nr{time_reversal}), however 
as this is not the case in quantum mechanics, 
it is instructive to avoid the assumption.
Writing the cosine in terms of exponentials 
then yields
\ba
 && \hspace*{-1.8cm}
 \P^{ }_\tensor(\tau,k) 
 \; 
 \overset{\rmii{\nr{P_T_final_4}} \lift }{\supset} 
 \; 
 \frac{128 k}{\mpl^4\,  a^2_{ }(\tau)}
 \int_{-\infty}^{\tau} \! \dd \tau^{ }_1 \, 
 \int_{-\infty}^{\tau^{ }_1} \! \dd \tau^{ }_2 \, 
 \frac{ e^{ik(\tau^{ }_1 - \tau^{ }_2) }_{ } +
        e^{ik(\tau^{ }_2 - \tau^{ }_1) }_{ } }{2} \,
 a^3_{  }(\tau^{ }_1) a^3_{  }(\tau^{ }_2)
 \bit S^{ }_\rmii{$\Pi$}(\tau^{ }_1,\tau^{ }_2,\vec{k})
 \nn[2mm]
%%%%
 & \overset{\scriptscriptstyle 
            \tau^{ }_\rmiii{2}\,=\, \tau^{ }_\rmiii{1} + \tilde\tau 
            \lift }{=} & 
 \frac{64 k}{\mpl^4\,  a^2_{ }(\tau)}
 \int_{-\infty}^{\tau} \! \dd \tau^{ }_1 \, 
 \int_{-\infty}^{0} \! \dd \tilde\tau \, 
      \bigl( 
        \underbrace{
        e^{- ik \tilde\tau }_{ }
        }_{
          \tilde\tau\;\to\;-\tilde\tau
        }
         +
        e^{ik \tilde\tau }_{ } \bigr) \,
 a^3_{  }(\tau^{ }_1) a^3_{  }(\tau^{ }_1 + \tilde\tau)
 \bit S^{ }_\rmii{$\Pi$}(\tau^{ }_1,\tau^{ }_1 + \tilde\tau,\vec{k})
 \nn[2mm]
%%%%%%
 & 
 \underset{\rmii{}}{
 \overset{}{=}}
 & 
 \frac{64 k}{\mpl^4\, a^2_{ }(\tau)}
 \int_{-\infty}^{\tau} \! \dd \tau^{ }_1 \, 
 a^3_{  }(\tau^{ }_1) 
 \int_{-\infty}^{\infty} \! \dd \tilde\tau \, 
 e^{ik \tilde\tau }_{ } \,
  a^3_{  }(\tau^{ }_1 - |\tilde\tau|)
  \bit S^{ }_\rmii{$\Pi$}(\tau^{ }_1,\tau^{ }_1 - |\tilde\tau|,\vec{k})
 \;. \hspace*{8mm} 
 \label{P_T_final_5} 
\ea

Before taking the final step, 
let us relate $\P^{ }_\tensor$ to 
the gravitational-wave energy density, 
as it appears in \eq\nr{eq_Ogwdef}. 
The tensor power spectrum can be expressed as 
\be
 \bigl\langle\, 
 h^\rmi{t}_{ij}(\tau,\vec{x})
 h^\rmi{t}_{ij}(\tau,\vec{x})
 \,\rangle 
 \; 
 \overset{\rmii{\nr{eq_angular_average}}}{=}
 \;
 \int_{-\infty}^{\infty} \! \dd \ln k \,
 {\textstyle \sum_{i,j}} 
 \P^{ }_{h^\rmiii{t}_{ij}}(\tau,k)
 \; 
 \underset{\rmii{\nr{def_P_T}}}{
 \overset{\rmii{\nr{h_basis_trafo}} \lift}{=}} 
 \; 
 \int_{-\infty}^{\infty} \! \dd \ln k \,
 \P^{ }_\tensor(\tau,k)
 \;.  
 \label{def_P_T_again}
\ee
From \eq\nr{e_gw}, adopting now 
the second variant, the energy density reads
\be
 e^{ }_\gw(\tau) 
% &
 \;
 \overset{\rmii{\nr{e_gw}}}{\simeq}
 \;
% &
 \frac{1}{32\pi G a^2_{ }(\tau)}  
 \bigl\langle\, 
    {h}^{\rmi{t}\,\prime}_{ij} (\tau,\vec{x}) 
    {h}^{\rmi{t}\,\prime}_{ij} (\tau,\vec{x}) 
 \,\rangle
% \nn[2mm] 
%%%%%%
% & 
 \;
  \overset{\rmii{\nr{trigo_appro}} \lift }
 {\underset{\rmii{\nr{def_P_T_again}}}{\approx}}
 \;
% & 
 \frac{1}{32\pi G a^2_{ }(\tau)}  
 \int_{-\infty}^{\infty} \! \dd \ln k \, k^2_{ }\,
 \P^{ }_\tensor(\tau,k)
 \;. \label{e_gw_again}
\ee
This yields the differential energy density
\ba
 && \hspace*{-1.8cm}
 \frac{{\rm d} e^{ }_\rmii{gw}(\tau,k)}{{\rm d}\ln k}
 \;
   \underset{\scriptscriptstyle G\, =\, m_\rmiii{pl}^{-\rmiii{2}}}
   {\overset{\rmii{\nr{e_gw_again}} \lift }{\simeq}}  
 \;
 \frac{\mpl^2 k^2_{ }}{32\pi a^2{ }(\tau)}
 \, 
 \P^{ }_\tensor(\tau,k)
 \nn[2mm]
%%%%%%
 \hspace*{-3mm}
 &
 \overset{\rmii{\nr{P_T_final_5}}}{\supset}
 & 
 \hspace*{-3mm}
 \frac{2 k^3_{ }}{\pi \mpl^2\,  a^4_{ }(\tau)}
 \int_{-\infty}^{\tau} \! \dd \tau^{ }_1 \, 
 a^3_{  }(\tau^{ }_1) 
 \int_{-\infty}^{\infty} \! \dd \tilde\tau \, 
 e^{ik \tilde\tau }_{ } \,
  a^3_{  }(\tau^{ }_1 - |\tilde\tau|)
  \bit S^{ }_\rmii{$\Pi$}(\tau^{ }_1,\tau^{ }_1 - |\tilde\tau|,\vec{k})
 \;. \hspace*{9mm}
 \label{dk_e_gw} 
\ea
We remark that \eq\nr{dk_e_gw} can also be obtained
{\em without} an oscillation average, if we do not make 
use of $\P^{ }_\tensor$, but instead take the first
variant of \eq\nr{e_gw} as a starting point. 

Let us then turn to the production process and 
inspect the evolution {\em rate} of the \linebreak
gravitational-wave energy density. 
There are two different physical phenomena present in
\eq\nr{dk_e_gw}: overall redshift, represented by the 
prefactor $1/a^4_{ }(\tau)$, and sourced production, 
represented by the integral over $\tau^{ }_1$.
It is convenient to handle both at the same time,
as is also suggested by our notation in which  
$\tau$ and  $\tau^{ }_{\sfin}$ are not distinguished.  
Taking a derivative with respect to this common value, we get
\be
 \bigl( \partial^{ }_\tau + 4 \H \bigr)
 \frac{{\rm d} e^{ }_\rmii{gw}(\tau,k)}{{\rm d}\ln k}
 \;
 \overset{\rmii{\nr{dk_e_gw}} \lift }
{\underset{\scriptscriptstyle k\,\gg\,\H}{\supset}} 
 \;
 \frac{2 k^3_{ }}{\pi \mpl^2\, a(\tau) }
 \int_{-\infty}^{\infty} \! \dd \tilde\tau \, 
 e^{i k \tilde\tau}_{ }
 \, 
  a^3_{  }(\tau^{ } - |\tilde\tau|)
  \bit S^{ }_\rmii{$\Pi$}(\tau^{ },\tau^{ } - |\tilde\tau|,\vec{k})
  \label{dt_e_gw_1}
 \;. 
\ee
Let us anticipate that 
below \eq\nr{dt_e_gw_6}, we show how an equation
of the form in \eq\nr{dt_e_gw_1} can be integrated in $\tau$, 
if the right-hand side is a known function.
We also remark that 
computing rates like in \eq\nr{dt_e_gw_1}
is often safer than computing time-integrated densities, 
because so-called {\em secular terms} \index{secular terms}
(which grow rapidly with time and spoil
perturbative treatments) are avoided. 

We can make further use of the assumption $k \gg \H$.
The Fourier transform in \eq\nr{dt_e_gw_1} probes
variations of the energy-momentum tensor, with the typical
contribution given by $\tilde \tau \sim 1/k \ll \H^{-1}_{ }$. 
On such time scales, 
$a^3_{ }(\tau \pm \tilde \tau) \approx a^3_{ }(\tau)$.
Furthermore, at short time intervals,
we may assume physics to be time-translationally
invariant. Then we find
\ba
 && \hspace*{-2.2cm}
 S^{ }_\rmii{$\Pi$}(\tau^{ },\tau^{ } - |\tilde\tau|,\vec{k})
% \nn[2mm]
%%%%
% & = &
 \; = \; 
 \theta(-\tilde\tau) 
  \,
  \overbrace{
  S^{ }_\rmii{$\Pi$}(\tau^{ },\tau^{ } + \tilde\tau,\vec{k})
  }^{{\rm symmetry}\; \tau^{ } \,\leftrightarrow\, \tau^{ } + \tilde\tau }
  \; + \; 
  \theta(\tilde\tau) 
  \,
  \overbrace{
  S^{ }_\rmii{$\Pi$}(\tau^{ },\tau^{ } - \tilde\tau,\vec{k})
 }^{{\rm translation}\;\tau\,\to\, \tau^{ } + \tilde\tau }
 \nn[2mm]
%%%%
 & \approx & 
 \theta(-\tilde\tau) 
 \, 
  S^{ }_\rmii{$\Pi$}(\tau^{ } + \tilde\tau,\tau^{ },\vec{k})
  \; + \; 
  \theta(\tilde\tau) 
 \, 
  S^{ }_\rmii{$\Pi$}(\tau^{ } + \tilde\tau,\tau^{ },\vec{k})
% \nn[2mm]
%%%%
% & = & 
 \; = \; 
  S^{ }_\rmii{$\Pi$}(\tau^{ } + \tilde\tau,\tau^{ },\vec{k})
 \nn[2mm]
%%%%
 &  
 \underset{\rmii{\nr{def_S_Pi}}}{
 \overset{\rmii{\nr{P_Pi}} \lift }{=}} 
 & 
 \int \! \dd^3_{ } \vec{x} \, 
 e^{- i \vec k \cdot \vec x }_{ }
 \, \frac{1}{2}
  \bigl\langle\,
  \Pi^\tensor_{ij}(\tau^{ }  + \tilde{\tau} ,\vec x) \,
  \Pi^\tensor_{ij}(\tau^{ },\vec 0) 
  + 
  \Pi^\tensor_{ij}(\tau^{ },\vec x) \,
  \Pi^\tensor_{ij}(\tau^{ } + \tilde{\tau},\vec 0) 
 \,\bigr\rangle
 \nn[2mm]
%%%%
 &  
 \overset{\rmii{\nr{parity}} \lift }{=}
 & 
 \int \! \dd^3_{ } \vec{x} \, 
 e^{- i \vec k \cdot \vec x }_{ }
 \, \frac{1}{2}
  \bigl\langle\,
  \Pi^\tensor_{ij}(\tau^{ }  + \tilde{\tau} ,\vec x) \,
  \Pi^\tensor_{ij}(\tau^{ },\vec 0) 
  + 
  \Pi^\tensor_{ij}(\tau^{ },\vec 0) \,
  \Pi^\tensor_{ij}(\tau^{ } + \tilde{\tau},\vec x) 
 \,\bigr\rangle
 \;. \hspace*{7mm} \label{symms}
\ea
The parity invariance that was employed as a part of \eq\nr{parity}
guarantees that the value of this integral is real. 
Given that our derivation relied on classical
field theory, we can now also invoke the corresponding assumption
from \eq\nr{time_reversal}, according to which 
operator ordering plays no role. 
Simplifying the notation as 
$
 \X \equiv (\tilde\tau,\vec x)
$ 
and
$
 \K\cdot \X \equiv - k \tilde\tau + \vec k \cdot \vec x
$,  
and abbreviating $\tau \equiv (\tau,\vec{0})$ as a space-time location, 
we finally get 
\ba
 \boxed{
 \;
 \bigl( \partial^{ }_\tau + 4 \H \bigr)
 \frac{{\rm d} e^{ }_\rmii{gw}(\tau,k)}{{\rm d}\ln k}
 \;
 \overset{
           \overset{
           \rmii{\nr{dt_e_gw_1},\nr{symms}}}{
           \rmii{\nr{time_reversal}}}
          }{
 \underset{\scriptscriptstyle k\;\gg\;\H % \,,\,{\rm oscillation~average}
                      }{\supset}} 
 \;
 \frac{2 k^3_{ } a^2(\tau) }{\pi \mpl^2 }
 \int_\X 
 e^{ - i \K\cdot \X }_{ }
 \bigl\langle\,
  \Pi^\rmi{t}_{ij}(\tau) \,
  \Pi^\rmi{t}_{ij}(\tau + \X) 
 \,\bigr\rangle
 \;.
 \; \vphantom{\Bigg|^b_q}
 }
 \nn[2mm] 
 \label{dt_e_gw_2}
 \index{gravitational-wave production rate}
\ea

To summarize, the main result of this section is 
\eq\nr{dt_e_gw_2}. It shows that
the production rate of the energy density carried 
by gravitational waves 
(after oscillation-averaging, if we adopt the 
second variant of \eq\nr{e_gw} as a definition),
gets a contribution from 
the Fourier transform of the 2-point correlation function of
anisotropic stress, assuming
that this originates from subhorizon 
physics, with $k \gg \H$. 

%%%%%%%%%%%%%%%%%%%%%%%%%%%%%%%%%%%%%%%%%%%%%%%%%%%%%%%%%%%%%%%%%%%%%%%%%
%
\subsection{Graviton production from particle decays and scatterings}
\label{ss:gw_scat}

\index{graviton}
\index{gravitational waves: from particles}

As we increase the momentum $k$ of the gravitational waves considered,
we see from the coefficient of the right-hand side 
of \eq\nr{dt_e_gw_2} that the energy-density production rate
increases. Simultaneously, a larger $k$ implies that we are
probing the properties of the energy-momentum tensor 
with an increasing resolution. At sufficiently large $k$, we cannot 
describe the plasma part of $\Pi^\tensor_{ij}$ with hydrodynamic degrees
of freedom any more, but instead need to use 
elementary quantum fields in $\Pi^\tensor_{ij}$, 
as was already done with 
$\varphi$ in \eq\nr{sigw}. 
Then, the processes responsible for generating
gravitational waves are particle decays and scatterings. 
This is interesting at the level of 
fundamental physics, as it may offer a window
to particles and interactions beyond the Standard Model. 

Like in \se\ref{ss:gw_reheat}, we consider here an epoch late enough
that a given mode is inside the Hubble horizon, $k \gg \H$
(or more conservatively, 
since $\H$ has a maximal value according to 
\fig\ref{fig:history_tau}(right), 
momenta with $k \gg \H^{ }_\rmii{max}$).
In terms of a physical momentum,  
$p \equiv k/a$, this corresponds to $ p \gg H$.
We take the statistical average of the 2-point correlator of 
$\Pi^\tensor_{ij}$, assuming that the universe has reheated. 
Then the result also depends on the temperature, $T$.
Given that $H\sim T^2_{} /\mpl^{ }$ (cf.\ \eq\nr{Hubble}) 
and that we normally
have $T \ll \mpl^{ }$, we expect $ H \ll T$. 
We can work in a local Minkowskian frame, and 
there are two regimes in which $p$ can lie, 
$H \ll p < T$ and $H \ll T < p$, separated by what we 
call the {\em thermal energy scale}, at $p\sim T$
(for simplicity, we focus here on ultrarelativistic particles, 
with masses $m \ll T$). 
\index{thermal energy scale}
We will see in this section that the gravitational-wave energy spectrum 
originating from  
particle scatterings {\em peaks at the thermal energy scale}. 
 
As discussed in \se\ref{ss:history}, the temperature
evolves  approximately as 
$T \sim T^{ }_\rmi{reheat} ( a^{ }_\rmi{reheat} /a)$ after reheating, 
which has the same functional form as $p = k/a$ with respect to $a$.  
So, the relation of $p$ to $T$ does not change much
in the history of the universe. In terms of today's 
frequency, the thermal energy scale corresponds to 
$f^{ }_\now |^{ }_{p\,\sim\, T} \sim 10^{11}_{ }$\hspace*{0.3mm}Hz
(cf.\ \eq\nr{p0_f0}), 
similar to the peak frequency of the CMB Planck spectrum.  

Let us compare the frequency 
$f^{ }_\now  \sim 10^{11}_{ }$\hspace*{0.3mm}Hz
with the interferometric measurements mentioned
in \se\ref{ss:probes_gw}. Most of them probe frequencies
$f^{ }_\now  \ll 10^{11}_{ }$\hspace*{0.3mm}Hz, and therefore
momenta $p \ll T$. In ultrarelativistic plasma physics, 
we identify the domain $p \ll T$ \index{hydrodynamic regime}
as a {\em hydrodynamic regime}, provided that the corresponding 
wavelength is much larger than the mean free path of particle
scatterings. In contrast, 
the so-called 
ultra-high frequency (UHF) detector concepts,
mentioned at the end of \se\ref{ss:probes_gw},
may offer for an opportunity
to observe the physics of momenta $p\sim T$, 
where particle-gravity interactions play a role. 

When we consider large momenta, 
we ultimately enter a domain
in which the classical approximation breaks down. For thermal
systems, this happens when $p \sim T$. For non-thermal
systems, the breakdown happens when the states of the Fock space
that contribute to the correlation function have occupancies
of order unity (rather than being $\gg 1$, which would justify
the classical approximation). 
As an example, this implies that fermions,
which obey the Pauli principle, can be treated 
classically only in the hydrodynamic domain, $p \ll T$, where their
contribution is fully captured by their overall effect on the 
energy density, pressure, and viscosities. 

\index{operator ordering}

In quantum mechanics and quantum field theory, 
{\em operator ordering}
plays a role. For instance, standard
scattering computations make use of time-ordered
perturbation theory (reflected by the $i\epsilon$
prescription of a Feynman propagator).
If time ordering is important, we cannot arbitrarily 
change it, as we did when going from \eq\nr{symms} 
to \eq\nr{dt_e_gw_2}. This implies that the production rate of
the energy density carried by gravitational waves needs to 
be re-derived, by treating the perturbed Einstein-Hilbert
action as a quantum field theory. 
In this case it is natural to refer to 
gravitational waves as {\em gravitons}.

We now explain three ways of obtaining the correct quantum-mechanical
generalization of \eq\nr{dt_e_gw_2}.
The methods
offer complementary benefits, namely that:
\bi

\item[(i)] {\em Boltzmann equations} are a practical tool 
for leading-order computations in particle cosmology
(such as 
those related to dark matter production, or neutrino or photon decoupling), 
and can be adapted to graviton production as well. 

\item[(ii)] {\em Thermal particle production} 
refers to a way to derive the 
graviton production rate directly from quantum field
theory, and can thus be used for establishing the proper operator ordering
of the 2-point correlator of the energy-momentum tensor. 

\item[(iii)] {\em Linear response theory} sets graviton
production in a broader context, relating it to the equilibration or
damping phenomena that we met in \se\ref{sec_eom}, and providing for
a physical explanation of an important Bose distribution, $\nB^{ }$, 
in \eq\nr{dt_e_gw_4}.

\ei
To keep the expressions simple, we work with local Minkowskian
coordinates in the following ($a = \mbox{constant}$), 
denoting $\vec{p} \equiv \vec{k}/a$,
$\vec{r} \equiv a \hspace*{0.3mm}\vec{x}$, and $\R = a \hspace*{0.3mm}\X$. 
 
%%%%%%%%%%%%%%%%%%%%%%%%%%%%%%%%%%%%%%%%%%%%%%%%%%%%%%%%%%%%%%%%
%
\paragraph{(i) Boltzmann equations.}

\index{Boltzmann equations}

{}From the practical point of view, the most 
straightforward avenue to a 
quantum treatment of gravitons
might be to abandon classical fields altogether, and to re-start
from scratch, from a particle picture. 
The tool to be used is that 
of Boltzmann equations, and the variable of interest
is the graviton phase-space distribution, $f^{ }_\gw$.
For simplicity, we assume that both polarization states are 
equally populated, and that $f^{ }_\gw$ denotes the common value. 
The differential energy density in gravitons reads
\be
 \dd e^{ }_\gw 
 \;
 \supset
 \;
 2 p\, f^{ }_\gw 
 \, \frac{{\rm d}^3\vec{p}}{(2\pi)^3_{ }}
 \;, \label{e_vs_f}
\ee
where 2 counts the polarization states and $p$ 
the energy that they carry. Assuming that the production 
is isotropic, so that 
$
 {\rm d}^3_{ }\vec{p} = 4 \pi p^2_{ }\dd p
 = 4 \pi p^3_{ }\, \dd \ln p
$, 
the production rate then becomes
\be
 \frac{\dd e^{ }_\rmii{gw}}{\dd t \, \dd \ln p}
 \;
 \underset{\rmii{isotropy}}{
 \overset{\rmii{\nr{e_vs_f}} \lift }{\supset}}
 \; 
 \frac{p^4_{ } \dot{f}^{ }_\rmii{gw}}{\pi^2_{ }}
 \;. \label{dt_e_gw_3}
\ee

%%%%%%%%%%%%%%%%%%%%%%%%%%%% FIGURE %%%%%%%%%%%%%%%%%%%%%%%%%%%%%%%%%
%
\begin{figure}[t]
    \centering
    \includegraphics[width=0.97\linewidth]{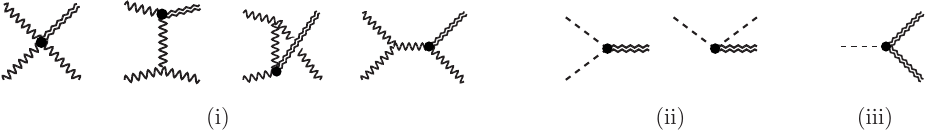}
  
    \vspace*{-2mm}

    \caption{%
    \small
    Examples of {\em Feynman diagrams for graviton production}.
    Gravitons are denoted by doubled curly lines. 
    (i)~$2\to 2$ scatterings, as described by \eq\nr{gain}.
    The wiggly lines denote gauge bosons, for instance gluons, 
    which give the dominant contribution in the Standard Model
    at high temperatures, 
    due to their large multiplicity as well as a logarithmic
    enhancement of the so-called $t$-channel process~\cite{full_lo}.
    The filled circle denotes the anisotropic stress tensor, to which 
    gravitons couple. 
    (ii)~$2\to 1$ and $1\to 2$ processes, for which we can draw
    simple diagrams, however these processes are  
    {\em kinematically forbidden} in local Minkowskian 
    spacetime, due to energy conservation. 
    The dashed line denotes a scalar particle, for instance the inflaton. 
    (iii) $1\to 2$ decays. The vertex needed for this process is not
    present for a minimally coupled scalar field, like in 
    the Einstein-Hilbert action~\nr{einstein_hilbert}, however 
    it can be induced by non-minimal couplings. 
    Processes with two gravitons also originate  
    at second order in anisotropic stress~\cite{hh_1,hh_2}.
    }
    \index{Feynman diagrams}
    \index{figure: diagrams for graviton production}
    \label{fig:diagrams}
\end{figure}
%
%%%%%%%%%%%%%%%%%%%%%%%%%%%%%%%%%%%%%%%%%%%%%%%%%%%%%%%%%%%%%%%%%%%%

The time derivative of the phase space distribution
in \eq\nr{dt_e_gw_3} is given by
what is referred to as a 
{\em collision term} \index{collision term}
of a Boltzmann equation. 
In our case it consists of ``inelastic'' processes, with at least one
graviton in the final state. Given that gravitons do not thermalize, 
only ``gain'' terms need to be included. Under these conditions, 
a typical gain process 
(examples are shown in \fig\ref{fig:diagrams}(i))
has the structure
\ba
 \dot{f}^{ }_\gw 
 & \supset & 
 \frac{\varc}{2p}
 \int
 \frac{{\rm d}^3_{ }\vec{p}^{ }_{i^{ }_1}}
   {(2\pi)^3_{ }\,2\epsilon^{ }_{i^{ }_1}} 
 \frac{{\rm d}^3_{ }\vec{p}^{ }_{i^{ }_2}}
   {(2\pi)^3_{ }\,2\epsilon^{ }_{i^{ }_2}} 
 \frac{{\rm d}^3_{ }\vec{p}^{ }_{e^{ }_1}}
   {(2\pi)^3_{ }\,2\epsilon^{ }_{e^{ }_1}} 
 (2\pi)^4_{ }\delta^{(4)}_{ }(\P^{ }_{i^{ }_1} + \P^{ }_{i^{ }_2}
                            - \P^{ }_{e^{ }_1} - \P )
 \nn[2mm]
%%%%
 & & \hphantom{ \frac{\varc}{2p} \int }
 \times\, 
 f^{ }_{i^{ }_1} f^{ }_{i^{ }_2\vphantom{1}} (1 \pm f^{ }_{e^{ }_1} )
 \sum |\M|^2_{ }
 \;. \label{gain}
\ea
In the sum over matrix elements squared, 
$\sum |\M|^2_{ }$, it is convenient to include both 
graviton polarization states, as well as both particle 
and antiparticle states of {\em each} momentum. 
The combinatorial factor, $\varc \equiv 1/4$, is chosen
to cancel overcountings from the sums and integrals: 
from polarization states (factor~1/2); 
from non-identical initial-state particles
(factor~1/2, because we should have assigned them
to separate momenta); 
or from identical initial-state particles
(factor~1/2, because we should have integrated only 
over half of the momentum space).
The functions 
$f^{ }_{i^{ }_n}$ are the phase-space distributions of
initial-state particles, 
$f^{ }_{e^{ }_1}$ is that of an end-state particle produced
in association with the graviton, and $\pm$ refer
to bosons and fermions, respectively. 
For the matrix elements, we need a vertex with which
a graviton interacts with matter fields, and \eq\nr{gain}
assumes that for its determination the graviton field has
been canonically normalized, 
like in \eq\nr{einstein_hilbert_2}. 

Equation~\nr{gain} corresponds to a $2\to 2$ scattering, 
but in addition gravitons can be produced from $1\to 3$ decays of
a massive particle, such as the inflaton
(cf.,\ e.g.,\ refs.~\cite{brems0,brems1,brems2,brems3}). 
We remark in passing that 
with so-called minimal couplings, gravitons attach to 
the energy-momentum tensors
of matter fields, which are quadratic in the fields. 
Therefore, we can also 
draw diagrams for $2\to 1$ or $1\to 2$ processes
(cf.\ \fig\ref{fig:diagrams}(ii)). 
However, they have two
lines of equal mass and one massless line, and are thus
not kinematically allowed in locally Minkowskian spacetime. If
there are non-minimal vertices, which lead to kinematically allowed
$2\to 1$ or $1\to 2$ processes (e.g.\ one massive particle
decaying into two massless ones, 
cf.\ \fig\ref{fig:diagrams}(iii)), 
they should be included (cf.,\ e.g.,\ refs.~\cite{bremsX,bremsY}).

One strength of Boltzmann equations is that they can also be used for
non-equilibrium situations. In other words, the distribution functions
$f^{ }_{i^{ }_n}$ and $f^{ }_{e^{ }_1}$ do not necessarily need to be 
equilibrium ones. If the particles are in equilibrium, 
we should use the Bose and Fermi distributions for bosons and 
fermions, respectively. 

Even if it is not easy to show this on an abstract level, 
an explicit computation demonstrates that, 
at leading order in coupling constants~\cite{full_lo}, 
\eqs\nr{dt_e_gw_3} and \nr{gain}
yield the same result as a perturbative
evaluation of the quantum-field-theory  
formula in \eq\nr{dt_e_gw_4}, to which we now turn.

%%%%%%%%%%%%%%%%%%%%%%%%%%%%%%%%%%%%%%%%%%%%%%%%%%%%%%%%%%%%%%%%
%
\paragraph{(ii) Thermal particle production.}

\index{thermal particle production}

A second possibility for a quantum treatment
of graviton production is to adopt the framework of 
{\em thermal particle production}.
Via this formalism, two statements can be proven. 
First, the correct time ordering happens to be precisely the one 
shown in \eq\nr{dt_e_gw_2}, where the space-time coordinates $\X$ 
appear on the right in the Fourier transform. This time ordering
is known as one of the {\em Wightman correlators}, 
\index{Wightman function}
and in 
physical momentum space, it is denoted by $G^\rmi{\,t}_{<}(\P)$.
Second, in thermal equilibrium, the Wightman correlator can 
be related to the corresponding {\em retarded Green's function} 
(cf.\  \se\ref{sec_eom}), through 
\be 
 G^\rmi{\,t}_{<}(\P)
 \;
  =
 \;
  2\hspace*{0.3mm} \nB^{ }(p) \im G^\rmi{\,t}_\iR(\P)
 \;, \quad 
 \P
 \;
  \equiv
 \;
 (p,\vec{p})
 \;. \label{kms}
 \index{retarded Green's function: quantum}
\ee
Therefore, employing 
$\partial^{ }_\tau = a \partial^{ }_t$, 
$k = p\hspace*{0.3mm} a$, and 
% $k^3_{ } = p^3_{ } a^3_{ }$, 
$\int^{ }_\X a^4_{ } = \int^{ }_\R$,  
\eq\nr{dt_e_gw_2} becomes
\be
 \boxed{
 \quad
 \frac{{\rm d} e^{ }_\rmii{gw}}{{\rm d}t\, {\rm d}\ln p}
 \;
 \underset{\scriptscriptstyle p \; \sim \;  T\;\gg\;H}
 {\overset{\rmii{\nr{dt_e_gw_2},\nr{kms}} \lift }{\supset}} 
 \;
 \frac{4 p^3_{ } }{\pi \mpl^2 } \,
 \nB^{ }(p) \,  
 \im G^\rmi{\,t}_\iR (\P)
 \;.
 \quad \vphantom{\Bigg|}
 }
 \label{dt_e_gw_4}
\ee
We now rederive \eq\nr{dt_e_gw_4} from quantum field theory
(after re-interpreting $e^{ }_\rmii{gw}$ as an expectation
value of an operator, 
i.e.\ $e^{ }_\rmii{gw} \to \langle \hat{e}^{ }_\rmii{gw} \rangle$), 
verifying the time ordering. 
The proof of \eq\nr{kms}, 
via a straightforward evaluation of the 2-point 
correlators in the energy eigenbasis,  
can be found in thermal field theory text books
(cf.,\ e.g.,\ ref.~\cite[sec.~8.1]{basics_copy}).

To get started we note that, 
after the rescaling in \eq\nr{rescale_htij}, and going over
to local Minkowskian coordinates, 
we can set \eq\nr{einstein_ij_t_again} in a form similar 
to \eq\nr{eqEoM}, 
\be
   \ddot{\,\varh}^\rmi{t}_{ij}
   - \nabla^2_{\vec r}\, \varh^\rmi{t}_{ij}
 \;
 \underset{\rmii{Minkowskian}}{
 \overset{\rmii{\nr{einstein_ij_t_again},\nr{rescale_htij}}
  \lift }{=}}
 \; 
 \underbrace{ 
 \frac{16\pi G}{\sqrt{32\pi G}} }_{ \sqrt{8\pi}/\mpl^{ }
 }\, \barpPi^\rmi{t}_{ij} 
 \;. \label{eom_ht_mink}
\ee
In order to streamline the notation, it is helpful to go over
to the polarization basis, with $\lambda \in \{+,\times\}$, 
in accordance with \eqs\nr{basis}--\nr{h_basis_trafo},  
and omit the superscript $(...)^\rmi{t}_{ }$ in $\varh$ and $\Pi$. 
Furthermore, we denote 
\be
 \frac{ \sqrt{8\pi}\, \Pi^{ }_\lambda }{ \mpl^{ } }
 \;
 \equiv
 \;
 - J^{ }_\lambda
 \;, \label{J_lambda}
\ee
where the (inconsequential) minus sign has been inserted in order to
maintain an analogy with \eqs\nr{eqEoM} and \nr{eq_Hamilton}. 

Next, we quantize the graviton field and the operator to which it 
couples, 
$
 \varh^{ }_{\lambda}\to\hat{\,\varh}^{ }_{\lambda}
$ 
and 
$
 J^{ }_{\lambda}\to \hat J^{ }_{\lambda}
$. 
The observable that we are interested in is related to the graviton part
of the Hamiltonian, normalized by a spatial volume, $L^3_{ }$, which will
drop out in the final result, 
\be
 \hat{e}^{ }_\gw(t)
% \; \equiv \;
% \frac{ \hat H^{ }_\gw(t) }{L^3_{ }}
 \; \overset{\rmii{\nr{e_gw}}}{\equiv} \;  
 \frac{1}{L^3_{ }}
 \int^{ }_{\vec{r}} \, 
 \sum_\lambda
 \frac{1}{2} 
 \Bigl(\,
   \dot{\hat{\,\varh}}^{ }_{\lambda} 
   \dot{\hat{\,\varh}}^{ }_{\lambda} 
  + 
   \nabla^{ }_\vec{r}{\hat{\,\varh}}^{ }_{\lambda} 
    \cdot
   \nabla^{ }_\vec{r}{\hat{\,\varh}}^{ }_{\lambda} 
 \,\Bigr) 
 \;. \label{H_gw}
\ee 

We represent the graviton field 
in terms of a mode expansion, similarly to  
\eqs\nr{mode_expansion}--\nr{commutators}, 
but now in local Minkowskian space-time.
Formally, this means that we go over to 
an {\em interaction picture}, \index{interaction picture}
in which the time evolution of the operators is determined
by the free Hamiltonian 
(cf.\ \eq\nr{H_gw}), 
whereas that of states originates from a perturbation
(cf.\ \eq\nr{H_int}). Being pedantic about which objects
are operators, we write the mode expansion as 
\be
 \hat{\,\varh}^{ }_{\lambda}(\R)
 \; = \; 
 \int \! \frac{{\rm d}^3_{ }\vec p}{\sqrt{(2\pi)^3_{ } 2 p}}\,
 \Bigl(\,
   \hat{w}^{\lambda}_\vec{p}\, 
   e^{i \P \cdot \R}_{ }
 + 
   \hat{w}^{\lambda\,\dagger}_\vec{p}\, 
   e^{-i \P \cdot \R}_{ }
 \,\Bigr)
 \;, \label{modes_h}
\ee
where 
the free field equation requires that 
the energy is on-shell, $\P = (p,\vec{p})$, 
and we are using the signature
$
 \P\cdot \R = - p\hspace*{0.3mm} t + \vec p \cdot \vec r
$. 
We note that, strictly speaking, if we go to a finite volume,
like in \eq\nr{H_gw}, 
momentum integrals should be replaced with momentum sums. 
However, the finite volume is only an intermediate regulator, 
and in the end we take the limit $L\to\infty$. We anticipate
this by maintaining the integral notation.  

Inserting \eq\nr{modes_h} in \eq\nr{H_gw}, integrating over $\vec{r}$, 
and making use of the commutators from \eq\nr{commutators} to 
``normal-order'' the appearance of the creation and annihilation
operators (which simply means that the vacuum contribution 
from \se\ref{ss:gw_vac} is omitted), 
we find that our observable contains the part 
\be
 \hat{e}^{ }_\gw(t) 
 \;
 \underset{\rmii{\nr{modes_h}}}{
 \overset{\rmii{\nr{H_gw}}}{\supset}}
 \; 
 \frac{1}{L^3_{ }}
 \int {\rm d}^3_{ }\vec{p} \, |\vec{p}| \, \sum_\lambda 
 \hat{w}^{\lambda\,\dagger}_\vec{p} 
 \hat{w}^{\lambda}_\vec{p}
 \;. \label{e_gw_quantized}
\ee
We need to compute the expectation value of this operator. 
If we assume that originally there are no gravitons
(or, more physically, that the phase-space density of gravitons
of momentum~$\vec p$ is far
below equilibrium), then we start from
the vacuum state, $|0\rangle$. In the interaction picture, states
evolve with a non-trivial {\em time-evolution operator}, 
\be
 \hat U^{ }_{\iI}(t;-\infty)
 \; = \; \hat T 
 \exp \biggl[\, {-i \int_{-\infty}^t \! {\rm d}t' \, \hat H^{ }_\iI(t') }_{ }
      \,\biggr]
 \;, \label{U_I}
 \index{time-evolution operator} 
\ee
where $\hat T$ is the time-ordering operator, and 
$\hat H^{ }_{\iI}$ denotes 
the interaction Hamiltonian (cf.\ \eq\nr{eq_Hamilton}), 
taken in the interaction picture (cf.\ \eq\nr{I_pict}),  
\be
 \hat H^{ }_\iI (t)
 \; 
 \underset{\rmii{\nr{I_pict}}}{
 \overset{\rmii{\nr{eq_Hamilton}} \lift }{=}} 
 \; 
 \int_\vec{r}
 \sum_\lambda 
 \hat{\,\varh}^{ }_{\lambda}(\R)\, 
 \hat J^{ }_{\lambda}(\R)
 \;, \quad
 \R \; \equiv \; (t,\vec r)
 \;. \label{H_int}
\ee
We note that unlike in \se\ref{sec_eom}, 
in \eq\nr{U_I} it is convenient to set the initial
Minkowskian time to $-\infty$.
The initial vacuum state evolves into
\be
 | 0(t) \rangle
 \;
 \equiv
 \;
 \hat U^{ }_{\iI}(t;-\infty) | 0 \rangle
 \;. \label{vacuum}
\ee
In this state, 
the expectation value of \eq\nr{e_gw_quantized} is non-zero, 
\ba
 \langle\, 0(t) \,|\, 
 \hat{w}^{\lambda\,\dagger}_\vec{p} 
 \hat{w}^{\lambda}_\vec{p}
 \,|\, 0(t) \,\rangle 
 & 
 \overset{\rmii{\nr{vacuum}} \lift }{=} 
 & 
 \langle\, 0\, |
 \, 
 \hat U^{\dagger}_{\iI}(t;-\infty)
 \, 
 \hat{w}^{\lambda\,\dagger}_\vec{p} 
 \hat{w}^{\lambda}_\vec{p}
 \, 
 \hat U^{ }_{\iI}(t;-\infty)
 \, 
 |\, 0 \,\rangle 
 \nn[2mm]
%%%%%%%
 & 
 \underset{\scriptscriptstyle \rmO(\hat H^\rmiii{2}_\rmiii{\it I} )}{
 \overset{\rmii{\nr{U_I}} \lift }{\supset}} 
 & 
 \langle\, 0 \,|
 \, 
 \int_{-\infty}^t \! {\rm d}t' \, \hat H^{\dagger}_\iI(t')
 \, 
 \hat{w}^{\lambda\,\dagger}_\vec{p} 
 \hspace*{-3mm} 
 \underbrace{
 \mathbbm{1}
 }_{\sum_n |n\rangle\langle n | }
 \hspace*{-3mm}
 \hat{w}^{\lambda}_\vec{p}
 \, 
 \underbrace{
 \int_{-\infty}^t \! {\rm d}t' \, \hat H^{ }_\iI(t')
 }_{{\rm linear~in}\; 
 \hat{w}^{\lambda}_\vec{p}\,,\,
 \hat{w}^{\lambda\,\dagger}_\vec{p} 
 }
 \, 
 |\, 0 \,\rangle 
 \hspace*{8mm}
 \nn[2mm]
%%%%%%%
 & = & \hspace*{-4mm}
 \langle\, 0 \,|
 \, 
 \int_{-\infty}^t \! {\rm d}t' \, \hat H^{\dagger}_\iI(t')
 \, 
 |\vec{p};\lambda \rangle
 \, 
 \langle \vec{p};\lambda |
 \, 
 \int_{-\infty}^t \! {\rm d}t' \, \hat H^{ }_\iI(t')
 \, 
 |\, 0 \,\rangle 
 \;, 
 \hspace*{8mm}
 \label{in-in}
\ea
where in the last step we have realized that only 
the vacuum state contributes in the sum 
$
 \sum_n |n\rangle\langle n |
$, 
and subsequently denoted 
$
 | \vec{p};\lambda \rangle
 \equiv 
 \hat{w}^{\lambda\,\dagger}_\vec{p}
 | 0 \rangle
$.
We remark that expectation values like in 
\eq\nr{in-in} are referred to as the 
{\em ``in-in'' formalism}, because constraints are 
imposed only on states at the initial time
(final states originate from the unit operator).  \index{in-in formalism}

To proceed, we need to include the heat bath in the computation. 
We thus consider an initial and final state of the type
\be
 |I\,\rangle \;\equiv\; |\ini\rangle \otimes 
 | 0 \rangle 
 \;, \quad
 |F\,\rangle \;\equiv\; |\fin\rangle \otimes 
 | \vec{p};\lambda \rangle
 \;, \label{states_prod}
\ee
where $|\ini\rangle$ and $|\fin\rangle$ 
are states in the Fock space of the heat bath.  
Generalizing on \eq\nr{in-in}, 
the {\em transition matrix element} \index{transition matrix element}
and its absolute value squared read
\ba
 T^{ }_\rmi{$F$$I$}(\vec{p};\lambda)
 \hspace*{-7mm}
 &
 \equiv
 & 
 \hspace*{-7mm}
 \langle F\, | \int_{\tmin}^{t} \! {\rm d}t' \, \hat H^{ }_\iI(t') \, 
 | I\,\rangle
 \nn[2mm]
%%%%%
 &
 \underset{\rmii{\nr{H_int},\nr{modes_h}}}
 {\overset{\rmii{\nr{states_prod}} \lift }{=}}
 & 
 \int_{\R'} \frac{e^{-i\P\cdot\R' }_{ }}
                          {\sqrt{(2\pi)^3_{ } 2 p}}
 \, 
 \langle \fin | \hat{J}^{ }_{\lambda}(\R') | \ini \rangle
 \;, \label{transition_matrix}
 \\[2mm]
%%%%%
 |T^{ }_\rmi{$F$$I$}|^2_{ } 
 \hspace*{-7mm}
 &
 \underset{\rmii{ }}{
 \overset{\rmii{\nr{transition_matrix}} \lift }{=}}
 & 
 \hspace*{-7mm}
 \frac{1}{(2\pi)^3_{ }2 p}
 \int_{\R',\R''}
 e^{i\P\cdot(\R'' - \R')}_{ }
 \langle \ini | \hat{J}^\dagger_\lambda (\R'') | \fin\rangle
 \langle \fin | \hat{J}^{ }_{\lambda}(\R') | \ini \rangle
 \;, 
 \hspace*{9mm}
 \label{TIF_squared}
\ea
where $\R' \equiv (t',\vec{r}')$.
Like in \eq\nr{in-in}, 
the final states, $|\fin\rangle$,
are a representation of the unit operator, 
$
 \sum^{ }_{\sfin} | \fin \rangle \langle \fin | = \mathbbm{1}
$, 
and can be removed. 
The initial states of the heat bath are summed over, with
the assumption that they are thermally distributed, with the weights
$e^{ -E^{ }_\sini/T}_{ } / \Z^{ }_\rmii{bath}$, 
where 
$
 \Z^{ }_\rmii{bath} \equiv 
 \sum_\sini e^{ -E^{ }_\sini/T}_{ }
$ 
is the {\em canonical partition function}.
\index{$\Z$ (canonical partition function)} 
A quantum-statistical average is defined as 
\be
 \langle ... \rangle^{ }
 \; \equiv \; 
 \frac{1}{\Z^{ }_\rmii{bath}}
 \sum_{\sini} e^{ -E^{ }_\sini/T}_{ }
 \langle \ini | ... | \ini \rangle
 \;. \label{therm_ave}
\ee
Then, from \eqs\nr{e_gw_quantized}, \nr{in-in}, 
\nr{TIF_squared}, and \nr{therm_ave}, 
the thermally averaged production rate of the 
gravitational-wave energy density becomes  
\ba
 \frac{{\rm d}\langle \hat{e}^{ }_\rmii{gw} \rangle}
      {{\rm d}t\, {\rm d}^3_{ }\vec{p}}
 & 
 \underset{\rmii{\nr{in-in}}}{
  \overset{\rmii{\nr{e_gw_quantized}} \lift }{\supset}} 
 &  
 \lim_{L\to\infty}
 \frac{|\vec{p}|}{L^3_{ }}
 \sum_{\lambda,\sfin,\sini}
 \frac{ {\rm d}
 |T^{ }_\rmi{$F$$I$}|^2_{ } }{{\rm d}  t } 
 \frac{ e^{ -E^{ }_\sini/T}_{ } }{\Z^{ }_\rmii{bath}}
 \nn[2mm]
%%%%%%
 &
 \underset{\rmii{\nr{therm_ave}}}{
 \overset{\rmii{\nr{TIF_squared}} \lift }{=}}
 & 
 \frac{1}{16 \pi^3_{ } }
 \lim_{L\to\infty} \frac{1}{L^3_{ }}
 \sum_{\lambda}
 \frac{{\rm d}}{{\rm d}t}
 \int_{\R',\R''}
 e^{i\P\cdot(\R'' - \R')}_{ }
 \bigl\langle\,
         \hat{J}^\dagger_\lambda (\R'') 
         \hat{J}^{ }_{\lambda}(\R') 
 \,\bigr\rangle
 \;. \hspace*{8mm}
 \label{d_e_gw_qm_1}
\ea
The time dependence in \eq\nr{d_e_gw_qm_1} follows the pattern
\ba
 & & \hspace*{-2.0cm}
 \frac{{\rm d}}{{\rm d}t}
 \int_{\tmin}^t \! {\rm d}t' 
 \int_{\tmin}^t \! {\rm d}t''
 \, e^{ip(t'-t'')}_{ }
 \, f(t'',t')
 \nn[2mm]
%%%%
 & = & 
 \int_{\tmin}^t \! {\rm d}t''
 \,
 \underbrace{ 
   e^{ip(t -t'')}_{ }
 \, f(t'',t)}_{
              t'' \;=\; t \,-\, \tilde t} 
 \; + \; 
 \int_{\tmin}^t \! {\rm d}t'
 \,
 \underbrace{ 
   e^{ip(t' -t)}_{ }
 \, f(t,t'\,)}_{
              t' \;=\; t \,+\, \tilde t} 
 \nn[2mm]
%%%%
 & = & 
 \int_0^{\tmax} \! {\rm d}\tilde t
 \,
 e^{ip\hspace*{0.3mm} \tilde t }_{ }
 \hspace*{-7mm}
 \underbrace{ 
 \, f(t - \tilde t ,t)}_{
              \approx\;{\rm time\hspace*{0.2mm}\mbox{-}translation~invariant}} 
 \hspace*{-7mm}
 \; + \; 
 \int_{\tmin}^0 \! {\rm d}\tilde t
 \,
   e^{ip\hspace*{0.3mm} \tilde t}_{ }
 \, f(t,t + \tilde t\,)
 \nn[2mm]
%%%%
 & \approx & 
 \int_{\tmin}^{\tmax} \! {\rm d}\tilde t
 \,
   e^{ip\hspace*{0.3mm} \tilde t}_{ }
 \, f(t,t + \tilde t\,)
 \;. \label{time_ints}
\ea
For the spatial integrals in \eq\nr{d_e_gw_qm_1}, we analogously get
\ba
 && \hspace*{-2.0cm}
 \lim_{L\to\infty}
 \frac{1}{L^3_{ }}
 \int_{\vec r',\vec r''}
 \underbrace{
 e^{i\vec p\cdot (\vec r'' - \vec r') }_{ }
 \, g(\vec r'',\vec r')
 }_{\vec r' \; = \; \vec r'' \,+\, \tilde{\vec r} }
 \nn[2mm] 
%%%%
 & = & 
\lim_{L\to\infty}
 \frac{1}{L^3_{ }}
 \int_{\tilde{\vec r},\vec r''}
 \, 
 e^{ - i\vec p\cdot \tilde{\vec r} }_{ }
 \hspace*{-7mm}
 \underbrace{
 \, g(\vec r'',\vec r'' + \tilde{\vec r} )
 }_{\rm spatial\hspace*{0.2mm}\mbox{-}translation~invariant}
 \nn[2mm] 
%%%%
 & = & 
 \int_{\tilde{\vec r}}
 \, 
 e^{ - i\vec p\cdot \tilde{\vec r} }_{ }
 \, 
 g(\vec 0,\tilde{\vec r})
 \;. \label{spat_ints}
\ea
The final steps are to write 
$ 
 {\rm d}^3_{ }\vec{p} = 4 \pi p^3_{ }\, 
 {\rm d} \hspace*{-0.3mm} \ln p
$; 
insert the Hermitean interaction term
$
 \hat J^{ }_\lambda = 
  - \sqrt{8\pi}\, \hat \Pi^{ }_\lambda / \mpl^{ } 
$
from \eq\nr{J_lambda}; 
and define 
\be
 G^\rmi{t}_{<}(\P) 
 \; \equiv \; \sum_{\lambda} 
 \int_{\tilde\R}
 e^{-i\P\cdot\tilde\R }_{}
 \bigl\langle\,
  \hat \Pi^\rmi{t}_\lambda (t)\, \hat \Pi^\rmi{t}_\lambda(t+\tilde\R) 
 \,\bigr\rangle
 \;, \label{G_wight}
\ee
where now $t \equiv (t,\vec 0)$, 
and the integral over $\tilde\R \equiv (\tilde t,\tilde{\vec r})$ 
runs over the full space-time
according to \eqs\nr{time_ints} and \nr{spat_ints}.
Thereby we finally get
\be
 \frac{{\rm d} \langle \hat{e}^{ }_\rmii{gw} \rangle}
      {{\rm d}t\, {\rm d}\ln p}
 \;
 \overset{\rmii{\nr{d_e_gw_qm_1}--\nr{G_wight}} \lift }{\supset}
 \;
 \frac{2 p^3_{ } }{\pi \mpl^2 } \,
 G^\rmi{\,t}_{ < } (\P)
 \;.
 \label{dt_e_gw_4_pre}
\ee
This is also valid for non-thermal systems, 
provided that we adjust the expectation value
in \eq\nr{therm_ave} accordingly. 
Inserting 
$G^\rmi{\,t}_{<}(\P) = 2 \nB^{ }(p) \im G^\rmi{\,t}_\iR(\P)$
from \eq\nr{kms}, which applies to an equilibrated plasma
($\nB^{ }$ contains $T$),
\eq\nr{dt_e_gw_4_pre} verifies \eq\nr{dt_e_gw_4}.

%%%%%%%%%%%%%%%%%%%%%%%%%%%%%%%%%%%%%%%%%%%%%%%%%%%%%%%%%%%%%%%%
%
\paragraph{(iii) Linear response argument.}

\index{linear response}

The third possibility for treating graviton production
at the thermal scale, $p\sim T$, or above it, 
is that we adapt our 
quantum-mechanical linear-response treatment of the friction coefficient
$\Upsilon$, from \se\ref{sec_eom}, to the present context.
Combining with a 
{\em detailed balance} argument, we
can again obtain the production rate. 
We should stress that this last logic 
is conceptually non-trivial,
because physically speaking gravitons are
not close to equilibrium, which is often considered a necessary
requirement for the validity of linear-response theory. The argument
turns out to be viable for a different reason, namely that gravitons
are weakly coupled to the equilibrium plasma. 

Let us start again from \eq\nr{eom_ht_mink}, 
but now in momentum and helicity space, 
\be
   \ddot{\,\varh}^{ }_{\lambda}
   + p^2_{ } \varh^{ }_{\lambda}
 \;
 \overset{\rmii{\nr{eom_ht_mink}} \lift }{=}
 \; 
 \frac{ \sqrt{8\pi} }{ \mpl^{ } }\,
 \barpPi^{ }_{\lambda} 
 \;
 \overset{\rmii{\nr{J_lambda}} \lift }{=}
 \; 
 - J^{ }_\lambda
 \;. \label{eom_ht_mink_p}
\ee
From \eq\nr{im_GR}, we get a friction coefficient associated
with the gravitational field, 
\be
 \Upsilon^{ }_{\!\lambda}(p) 
 \;
 \overset{\rmii{\nr{im_GR}} \lift }{=}
 \; 
 \frac{\im G^{ }_\iR[\,J^{ }_{\lambda}\,]}{p}
 \;
 \overset{\rmii{\nr{eom_ht_mink_p}} \lift }{=}
 \; 
 \frac{8\pi}{\mpl^2}
 \frac{\im G^{ }_\iR[\,\Pi^{ }_{\lambda}\,]}{p}
 \;.  \label{Ups_lambda}
\ee

With the friction coefficient known, we need
to review its role in the evolution equations.
The principle of detailed balance asserts that 
every interacting particle species in principle
approaches equilibrium sooner or later, implying that 
\be
 \dot{f}^{ }_{\lambda} \; = \; - \Upsilon^{ }_{\!\lambda}(p) 
 \bigl(\, f^{ }_{\lambda} - \nB^{ }(p) \,\bigr)
 + 
 \ord\biggl( \frac{1}{\mpl^4}\biggr)
 \;. \label{lin_rep} 
\ee
Close to equilibrium, this functional form could 
be postulated as a leading term in an expansion in 
$ f^{ }_{\lambda} - \nB^{ }(p) $. For us, more important
is that the same form also applies far from equilibrium, 
at leading order in 
the interaction strength between the graviton and the plasma, 
i.e.\ $1/\mpl^2$, 
as can be established through 
a non-trivial analysis~\cite{sangel}. 

In the early universe, 
gravitons are far from reaching equilibrium, so that  
$ 
 f^{ }_{\lambda} \ll \nB^{ }(p)
$
in most momentum bins (there could be exceptions
at specific momenta, if gravitational waves 
are produced by non-equilibrium processes such as phase transitions).
Then \eq\nr{lin_rep} implies 
$
 \dot{f}^{ }_{\lambda} \approx \Upsilon^{ }_{\lambda}(p)\,\nB^{ }(p)
$.
For single polarization modes, \eq\nr{e_vs_f} becomes
\be
 \dd e^{ }_{\lambda}
 \;
 \overset{\rmii{\nr{e_vs_f}} \lift }{\supset}
 \;
 p\, f^{ }_{\lambda } 
 \, \frac{{\rm d}^3\vec{p}}{(2\pi)^3_{ }}
 \;
 \overset{\rmii{isotropy} \lift }{=}
 \; 
 f^{ }_\lambda \, \frac{p^4_{ }}{2\pi^2_{ }}\, {\rm d}\ln p
 \;, \label{prod_lam}
\ee
which then yields
\ba
 \frac{\dd e^{ }_\rmii{gw}}{\dd t \, \dd \ln p}
 & 
 \overset{\rmii{\nr{prod_lam}} \lift }{\supset}
 & 
 \frac{p^4_{ }}{2\pi^2_{ }} 
 \sum_\lambda \dot{f}^{ }_{\lambda}
 \;
 \underset{\scriptscriptstyle 
            f^{ }_{\lambda}\;\ll\; n^{ }_\rmiii{B}}{
 \overset{\rmii{\nr{lin_rep}} \lift }{\approx}}
 \; 
 \frac{p^4_{ }}{2\pi^2_{ }} 
 \sum_\lambda \Upsilon^{ }_{\!\lambda}(p)\,\nB^{ }(p)
 \nn[2mm]
%%%% 
 & \overset{\rmii{\nr{Ups_lambda}} \lift }{=} & 
 \frac{4 p^3_{ }}{\pi \mpl^2}
 \, \nB^{ }(p) \, \im G^{\tensor}_\iR(\P)
 \;, \quad \P \; = \; (p,\vec p)
 \;. 
 \label{dt_e_gw_5}
\ea
This once again agrees with \eq\nr{dt_e_gw_4}.

%%%%%%%%%%%%%%%%%%%%%%%%%%%%%%%%%%%%%%%%%%%%%%%%%%%%%%%%%%%%%%%%

\vspace*{3mm}

Let us summarize the findings of this section. 
If we consider physical momenta $p > T$, the growth of the 
gravitational-wave spectrum that we had observed at smaller
momenta (cf.\ \eq\nr{P_T_final_2}), 
finally turns around into an exponential falloff, 
because of the Bose factor, 
$\nB^{ }(p)$, in \eq\nr{dt_e_gw_5}. 
The physical reason is that, to produce a graviton
of momentum~$p$, the energy needs to be extracted
from fluctuations, which in a thermal plasma 
are Boltzmann-suppressed 
at high energies. 
Non-trivial contributions at energies higher than $p\sim T$ 
can only originate from non-equilibrium processes, 
for instance if the gravitons
are produced from the decays of heavy 
inflaton particles, whose abundance can be much above
equilibrium during an early matter-dominated epoch 
(cf.\ \app\ref{app:num_bg_vac}). 

%%%%%%%%%%%%%%%%%%%%%%%%%%%%%%%%%%%%%%%%%%%%%%%%%%%%%%%%%%%%%%%%%%%%%%%%%
%
\subsection{Transfer function for tensor modes}
\label{ss:gw_transfer}

The generic result for the tensor power spectrum 
(cf.\ \eq\nr{P_T_final_2}), which in turn 
determines the gravitational-wave
energy density (cf.\ \eq\nr{e_gw_again}), 
originates from two kinds of sources. 
First of all, there are the perturbations that were produced 
by vacuum fluctuations and then extended beyond the Hubble horizon by
a period of inflationary expansion. This component has the
special property that the corresponding power spectrum ``froze out''
for a long period of time, while the modes were outside of the 
Hubble horizon. And second, there are the gravitational waves that
are continuously produced by physical phenomena within the Hubble
horizon, notably during the period of reheating and afterwards. 
Gravitational waves can also get damped by the plasma
that they traverse, via the anisotropic stress induced by 
the shear viscosity (cf.\ \eq\nr{einstein_ij_t_eta_again}) or,
more generally, by a damping coefficient that can be
computed from linear response theory (cf.\ \eq\nr{lin_rep}).
The purpose of the present section is 
to explain how the primordial power spectra 
get converted to the gravitational-wave energy density that is 
measurable today. 

Given that a distinction is made between modes that did or
did not exit the Hubble horizon, it is important to know which
today's frequency, $f^{ }_\now$, is the maximal one that did so, 
$f^\rmii{(max)}_\now$.
The answer is not unique, but depends on 
the energy scale of inflation, and on the expansion history
that followed inflation (for instance, on whether there was 
an extended matter-domination period after inflation, 
cf.\ \app\ref{app:num_bg_vac}). 
As can be deduced from \fig\ref{fig:bg_num} on 
p.~\pageref{fig:bg_num}, modes
crossed outside of the Hubble horizon until very briefly after
inflation ended. If we {\em assume} that the 
universe is radiation-dominated immediately when the modes re-enter 
inside the Hubble horizon, then the subsequent
scaling is as discussed in \se\ref{ss:history}.
For instance, if 
$T^\rmii{(max)}_\rmi{re-entry} = 10^{15}_{ }$\hspace*{0.3mm}GeV, 
then 
$
 f^\rmii{(max)}_\now = 
 2.6\times 10^7_{ }\,
$Hz
(cf.\ \eq\nr{max_gw_vac}).
If, instead, there is a matter-dominated period after
inflation, then the very 
same initial momentum mode, $(k/a^{ }_i)^\rmii{(max)}_{ }$, 
corresponds to a smaller wave vector and frequency today
(cf.\ \fig\ref{fig:bg_thermal}(right) on p.~\pageref{fig:bg_thermal}).

We now turn to a concrete evaluation of the transfer function
for the two cases mentioned above.  
The discussion applies to any realization 
of the post-inflationary history, however if 
a perturbation re-enters inside the Hubble
horizon during radiation domination, 
then the final result can be simplified quite a bit
(cf.\ the discussion around \eq\nr{res_T_T_final}).

%%%%%%%%%%%%%%%%%%%%%%%%%%%%%%%%%%%%%%%%%%%%%%%%%%%%%%%%%%%%%%%%%%%
%
%\subsubsection*
\paragraph{(i) Vacuum-generated tensor perturbations.}

The present-day 
($\tau \to \tau^{ }_\inow$, $a\to a^{ }_\inow$)
gravitational-wave energy density from \eq\nr{e_gw}, 
now written in terms of helicity states, reads
\be
 e^{ }_{\rmi{gw},\inow} 
 \; 
 \overset{\rmii{\nr{e_gw}}}{\simeq}
 \; 
 \frac{1}{32\pi G a_\inow^2}
 \sum_{\lambda} 
 \bigl\langle\,
   h^{\prime}_{\lambda} (\tau^{ }_\inow,\vec{x})\,
   h^{\prime}_{\lambda} (\tau^{ }_\inow,\vec{x})
 \,\bigr\rangle
 \;, \label{e_gw_0}
\ee
where the left-hand side does not depend on $\vec x$, 
because of translational invariance 
(cf.\ \eq\nr{eq_angular_average}), and
the angular brackets denote an average over an oscillation period
and a wavelength.  
Recalling that
$h^{ }_\lambda(\tau^{ }_\rmi{out},k)$ is constant
if $\tau^{ }_\rmi{out}$ is a moment at which the mode
is outside of the Hubble horizon, 
we express the subsequent
evolution in comoving momentum space as 
a functional of this initial value, 
\ba
 h^{ }_\lambda(\tau^{ }_\inow,k)
 & \equiv & 
 X^{ }_\tensor(\tau^{ }_\inow,\tau^{ }_\rmi{out},k) \, 
 h^{ }_\lambda(\tau^{ }_\rmi{out},k)
 \;, \label{def_X_T} \\[2mm]
%%%
 X^{ }_\tensor(\tau^{ }_\rmi{out},\tau^{ }_\rmi{out},k)
 & \equiv & 1
 \;, \qquad
 \partial^{ }_\tau 
 X^{ }_\tensor(\tau,\tau^{ }_\rmi{out},k)
 \overset{\scriptscriptstyle \;\;\;\tau\, =\, \tau^{ }_\rmiii{out}
           \lift }{=} \; 
 0
 \;. \label{initial_X_T} 
\ea
Then, from \eqs\nr{def_Omega_gw}, \nr{e_gw_0}, \nr{def_X_T},  
and \nr{eq_angular_average}, 
\be
 \frac{{\rm d}
 \Omega^{ }_{\igw,\inow} 
 }{{\rm d}\ln k} 
 \;\;
    \overset{\rmii{\nr{def_Omega_gw}} \lift }{=} \;\; 
 \frac{8\pi}{3 \mpl^2 H^{2}_\inow } \,
 \frac{{\rm d} e^{ }_{\rmii{gw,0}}}{{\rm d}\ln k}
 \;\; 
    \underset{\rmii{\nr{def_X_T}}}
    {\overset{\rmii{\nr{e_gw_0}} \lift }{\simeq}} 
 \;\; 
 \underbrace{ 
 \frac{
 [\partial^{ }_{\tau^{ }_\inow} X^{ }_\tensor
 (\tau^{ }_\inow,\tau^{ }_\rmii{out},k)]^2_{ }
 }{12 a_\inow^2 H^2_\inow}
 }_{\;\equiv\; \mathcal{T}^{ }_\itensor(k)} 
 \; 
 \underbrace{
 \, \P^{ }_\tensor(\tau^{ }_\rmi{out},k)
 }_{{\rm like~}\nr{eq_angular_average}}
 \;. \label{def_T_T}
 \index{transfer function: tensor perturbations}
\ee   % 
Here 
$
 \mathcal{T}^{ }_\tensor
$
denotes the {\em transfer function} in the tensor channel. 
We made use of the fact that, with real initial conditions,
the solution of \eq\nr{eq_X_T} is real, 
$X^{ }_\tensor \in \mathbbm{R}$.

The remaining task is to determine the time evolution 
of $X^{ }_\tensor$ from \eq\nr{def_X_T}. 
It satisfies the same equation
as the tensor perturbations, 
\be
  (\partial_\tau^2 + 2 \H \partial_\tau^{ } + k^2_{ })
   h^{ }_{\lambda}
 \;  
 \overset{\rmii{\nr{einstein_ij_t_again}} \lift }{=} 
 \; 
 16\pi G a^2_{ } \barpPi^{ }_{\lambda} 
 \;. 
\ee
As we have accounted for the right-hand side 
via the second term in \eq\nr{P_T_final_2}, it will be 
omitted from the transfer function of the vacuum contribution, 
and we therefore consider
\be 
 (\partial_\tau^2 + 2 \H \partial_\tau^{ } + k^2_{ })
 \, X^{ }_\tensor
 \;
 \underset{\rmii{see~below}}{
 \overset{\rmii{for~assumptions} \lift }{\approx}}
 \; 
 0 
 \;. \label{eq_X_T}
\ee
As for the assumptions behind this approximation, 
we recall that, apart from being produced,
tensor modes can also be damped by the anisotropic stress 
(cf.\ \eq\nr{einstein_ij_t_eta_again}). 
Indeed it has been realized that 
decoupled particle species,
which are free-streaming in a late universe, 
can have an influence
on the transfer function of the tensor modes that re-enter inside
the Hubble horizon at late times~\cite{sw}. For instance,
{\em neutrinos decouple} \index{decoupling: neutrinos}
from the electromagnetic plasma \index{neutrino decoupling}
at $T\sim 2$~MeV. So, according to \eq\nr{f0_MeV}, we expect
them to have an influence at frequencies
$f^{ }_\now < 4\times 10^{-11}_{ }\,$Hz.  
A concrete computation shows an effect 
at $f^{ }_\now <  10^{-9}_{ }/(2\pi)\,$Hz~\cite{eos1}. 
If we have in mind an interferometric observation of 
gravitational waves at larger frequencies, 
then this effect can probably be ignored, 
and we will do this in the following. 
Explorations of the influence 
of viscous damping at larger 
frequencies can be found, e.g.,\ 
in refs.~\cite[fig.~1]{nmg}, \cite[fig.~1]{jena},
and \cite[fig.~4]{bari}.

Despite its simple structure, the solution of \eq\nr{eq_X_T}, 
with initial conditions from \eq\nr{initial_X_T},  
is quite non-trivial.
In fact, the solution is not unlike that shown in 
\fig\ref{fig:acoustic} on p.~\pageref{fig:acoustic}: 
initially, it is constant, 
and then it starts oscillating fast. 

The strategy that is normally adopted for the solution 
is as follows~\cite{eos2}.
In general, we may integrate the equation numerically, similarly to what
was done in \app\ref{app:num_osc}, until a time when
the oscillations become fast. For the latter domain, 
it is helpful to work out 
an analytical approximation, because integrating over
many fast oscillations is numerically expensive. 
The numerical and analytical solutions are
matched onto each other in a regime where both are valid; 
let us call this moment $\tau^{ }_\rmii{match}$.
Subsequently, the analytic approximation is used for extrapolating
to present day. 

As far as the analytic approximation goes, 
it is discussed around \eqs\nr{B_R}--\nr{trace}. 
Taking over the form from \eq\nr{B_R}, we find that 
\be
 X^{ }_\tensor(\tau,\tau^{ }_\rmi{out},k)
 \; \underset{\rmii{\nr{B_R}}}
             {\overset{\scriptscriptstyle 
              \tau\,\gg\,\tau^{ }_\rmiii{out}
               \lift }{\approx}} \; 
 \frac{\alpha\, a(\tau^{ }_\rmii{out})}{a(\tau)}\,
 \sin\bigl(\, k (\tau - \tau^{ }_\rmi{out}) + \beta \,\bigr)
 \;, \label{X_T_asymptotic}
\ee
where $\alpha$ and $\beta$ are two integration constants. 
The integration constants can be fixed by matching the 
numerical solution and its time derivative to 
\eq\nr{X_T_asymptotic},
\ba
 \overbrace{
 X^{ }_\tensor(\tau^{ }_\rmi{match},\tau^{ }_\rmi{out},k)
 }^{\rm determined~numerically}
 &
 \overset{\rmii{\nr{X_T_asymptotic}} \lift }{=}
 & 
 \frac{\alpha\, a(\tau^{ }_\rmii{out})}{a(\tau^{ }_\rmii{match})}\,
 \sin\bigl(\, k (\tau^{ }_\rmi{match} - \tau^{ }_\rmi{out}) + \beta \,\bigr)
 \;, \label{match_1} 
 \\[2mm]
%%%%
 \underbrace{
 ( \partial^{ }_{\tau} + \H)
 X^{ }_\tensor(\tau^{ }_\rmi{match},\tau^{ }_\rmi{out},k)
 }_{\rm determined~numerically}
 & 
 \overset{\rmii{\nr{X_T_asymptotic}} \lift }{=}
 & 
 \frac{\alpha\, k\, a(\tau^{ }_\rmii{out})}{a(\tau^{ }_\rmii{match})}\,
 \cos\bigl(\, k (\tau^{ }_\rmi{match} - \tau^{ }_\rmi{out}) + \beta \,\bigr)
 \;. \hspace*{8mm} \label{match_2}
\ea
Subsequently, making use of the hierarchy 
$k \gg \H$ valid at $\tau \ge \tau^{ }_\rmii{match}$, 
the transfer function from \eq\nr{def_T_T} 
can be approximated as 
\be
 \boxed{
 \quad
 \mathcal{T}^{ }_\tensor(k)
 \; \underset{\rmii{\nr{X_T_asymptotic}}}
    {\overset{\rmii{\nr{def_T_T}} \ilift }{\approx}} \; 
 \frac{\alpha^2_{ }}{12}
 \frac{k^2_{ } a^2_{ }(\tau^{ }_\rmii{out}) }{a_\inow^4 H^2_\inow}
 \cos^2_{ }\bigl(\, k (\tau^{ }_\now - \tau^{ }_\rmi{out}) + \beta \,\bigr)
 \;, 
 \quad \vphantom{\Bigg|}
 }
\label{res_T_T}
\ee
where $\alpha$ and $\beta$ are fixed through
\eqs\nr{match_1} and \nr{match_2}. We remark that, 
as discussed below \eq\nr{dk_e_gw}, we could
eliminate the rapid oscillations from \eq\nr{res_T_T} by taking 
the first variant of \eq\nr{e_gw} as the definition of the
gravitational-wave energy density or, alternatively, 
we could replace the oscillations by their average, 1/2.  

There is now a remarkable point, following 
from \eqs\nr{eom_B_R} and \nr{trace}. Namely, if $\bar{e} - 3\bar{p} = 0$
during the time at which modes re-enter inside the Hubble horizon, then 
\eq\nr{X_T_asymptotic} is an {\em exact} solution of \eq\nr{eq_X_T},
even when $k \sim \H$ or $k \ll \H$. In this case, we just need to
fix the integration constants. 
Imposing the boundary conditions of \eq\nr{initial_X_T} at the 
time $\tau = \tau^{ }_\rmi{out}$ yields
\ba
 1 & 
  \underset{\rmii{\nr{match_1}}}
  {\overset{\rmii{\nr{initial_X_T}} \lift }{=}}
   & 
 \alpha\,
 \sin \beta 
 \;, \label{ini_1} \\[2mm]
%%%%%
 0 & 
  \underset{\rmii{\nr{match_2}}}
 {\overset{\rmii{\nr{initial_X_T}} \lift }{=}}
   & -  
 \alpha \H^{ }_\rmi{out}
 \bit \sin \beta 
 + 
 \alpha\, k \,
 \cos \beta 
 \;. \label{ini_2}
\ea
Consequently, we find that 
\be
 \tan\beta 
 \;
 \overset{\rmii{\nr{ini_2}} \lift }{=}
 \; 
 \frac{k}{\H^{ }_\rmii{out}}
 \;\; 
 \overset{\scriptscriptstyle
           k\;\ll\; \H^{ }_\rmiii{out}
            \lift  }{\Rightarrow}
 \;\; 
 \beta \; \approx \; \frac{k}{\H^{ }_\rmii{out}}
 \;\;
 \overset{\rmii{\nr{ini_1}}  \lift  }{\Rightarrow}
 \;\;
 \alpha \; \approx \; 
 \frac{\H^{ }_\rmii{out}}{k}
 \;
 \overset{\rmii{\nr{eq_HH'}}  \lift  }{=}
 \; 
 \frac{a^{ }_\rmii{out} H^{ }_\rmii{out}}{k}
 \;. \hspace*{4mm}
 \label{alpha_beta}
\ee
In \eq\nr{res_T_T}, 
the $k$-dependence 
$\alpha^2_{ }\sim 1/k^2_{ }$ 
cancels against the explicit factor $k^2_{ }$, 
so that
\be
 \mathcal{T}^{ }_\tensor(k)
 \; 
 \underset{\rmii{\nr{res_T_T},\nr{alpha_beta}}}{
 \overset{\vphantom{\big | }
          \scriptscriptstyle
          \bar e\, -\, 3 \bar p \; \approx \; 0
           \lift  }{\approx}}
 \; 
 \frac{1}{12}
 \biggl( \frac{a^{ }_\rmii{out}}{a^{ }_\inow} \biggr)^4_{ }
 \biggl( \frac{H^{ }_\rmii{out}}{H^{ }_\inow} \biggr)^2_{ }
 \cos^2_{ }\bigl(\, k (\tau^{ }_\now - \tau^{ }_\rmi{out}) \,\bigr)
 \;. \label{res_T_T_final}
\ee
The same comments as below \eq\nr{res_T_T} 
can be made about the oscillations. % in \eq\nr{res_T_T_final}.

Let us summarize how the transfer function from \eq\nr{res_T_T}, 
or its approximate form from \eq\nr{res_T_T_final},  
affects the spectral shape and amplitude of the energy density
carried by vacuum-generated gravitational waves:
\bi

\item
As for the shape, 
\eq\nr{res_T_T_final} is to a good approximation  
independent of~$k$, just like the original power spectrum
from \eq\nr{P_T_final}
(the latter corresponds to the tensor spectral tilt being very
small, cf.\ \eq\nr{nt}). 
Therefore, for
frequencies small enough to cross outside of the Hubble horizon
during inflation (cf.\ the discussion in the 
paragraph above \eq\nr{e_gw_0}), 
and to re-enter it during an epoch where 
the trace anomaly is small, 
the current energy density is almost independent of the frequency. 

\item
As for the overall amplitude of the 
gravitational-wave energy density, its order of magnitude 
can be approximated from \eq\nr{res_T_T_final}.
Inserting the evolution of the scale factor 
(cf.\ \eq\nr{evolution_a}) 
and the Hubble rate 
(cf.\ \eq\nr{Hubble}) 
as well as the critical energy density 
(cf.\ \eq\nr{ecrit_again}), 
we obtain
\be
 \biggl( \frac{a^{ }_\rmii{out}}{a^{ }_\inow} \biggr)^4_{ }
 \biggl( \frac{H^{ }_\rmii{out}}{H^{ }_\inow} \biggr)^2_{ }
 \; 
 \overset{\rmii{\nr{Hubble},\nr{evolution_a}} \lift }
 {\underset{\rmii{\nr{ecrit_again}}}{=}}
 \;
    \biggl( \frac{\bar s^{ }_\inow / T_\inow^3 }
                 {\bar s^{ }_\rmii{out} / T_\rmii{out}^3}
                  \biggr)^{4/3}_{ }
    \biggl( 
    \frac{\bar e^{ }_\rmii{out} / T_\rmii{out}^4}
         {\bar e^{ }_{\gamma,\inow} / T_\inow^4 }       
    \biggr)
    \overbrace{
    \biggl( 
    \frac{\bar e^{ }_{\gamma,\inow}}{ e^{ }_\rmii{crit}}
    \biggr)
    }^{\;\equiv\; \Omega^{ }_{\gamma,\inow} }
 \;,  \label{T_T_factors}
\ee
where $ \bar e^{ }_{\gamma,\rmi{0}} $ is the current energy
density of CMB photons. The last factor in \linebreak
\eq\nr{T_T_factors} is the numerically most 
significant one: even if thermal photons 
carry a large fraction of the 
current entropy density, they contribute very little 
to the current
energy density, $\Omega^{ }_{\gamma,\now}\sim 5\times 10^{-5}_{ }$, 
which is instead dominated by dark energy and dark matter. 
As concerns the other two factors,
they do differ from unity, 
because the effective numbers of light degrees of freedom, as 
defined in \eq\nr{p_r}, change with the temperature. 
However, this dependence is moderate, and the two factors also partly
compensate against each other. As a consequence,
the choice of $\tau^{ }_\rmi{out}$ has little influence. 
Including the factor $1/12$,
the numerical magnitude of the transfer
function at $f^{ }_\now > 10^{-8}_{ }\,$Hz is then 
$h^2_{ } \mathcal{T}^{ }_\tensor \sim 10^{-6}_{ }$.

\begin{figure}[t]
    \centering
    \includegraphics[width=0.98\linewidth]{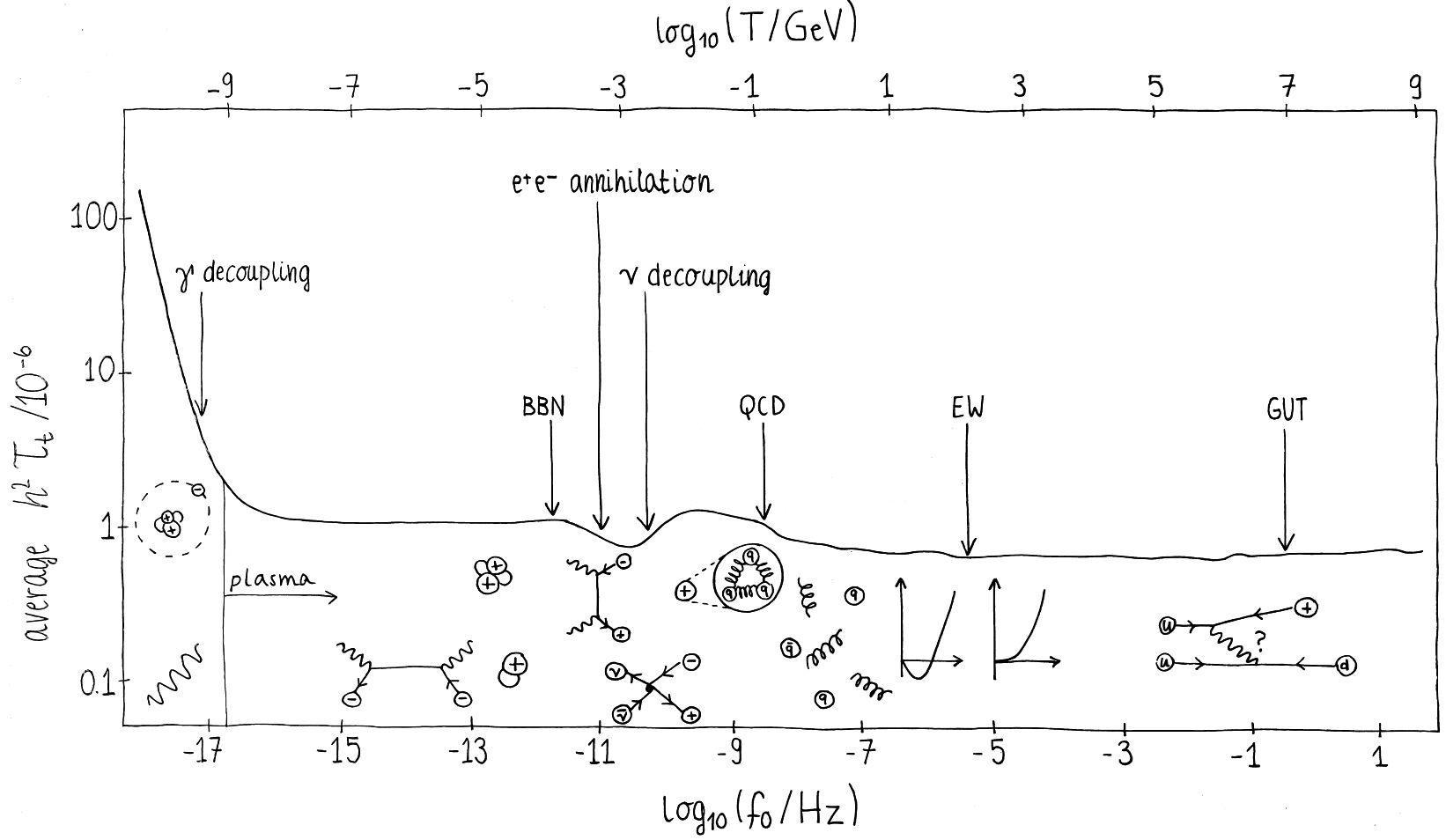}

    \caption{%
    \small
    A sketch of the transfer 
    function for the tensor channel,
    $ h^2_{ } \mathcal{T}^{ }_\tensor $,  
    from \eq\nr{res_T_T}.
    The lower horizontal axis displays the current 
    frequency, $f^{ }_\inow$/Hz, 
    and the upper one the temperature at which 
    the corresponding mode 
    re-enters the Hubble horizon, 
    cf.\ \eqs\nr{max_gw_vac} and \nr{f0_MeV}.
    The small features correspond
    to mass thresholds, or the QCD and electroweak
    (or higher-scale) 
    phase transitions,
    which induce a trace anomaly in the thermodynamics
    of the radiation plasma
    (cf., e.g., ref.~\cite{gubser}). 
    Quantitative evaluations of $ h^2_{ } \mathcal{T}^{ }_\tensor $
    can be found, 
    e.g., in refs.~\cite{eos2,scan}, with the former 
    displaying also the rapid oscillations in \eq\nr{res_T_T}.
    In the sketch above,
    we only show the physically meaningful average,
    in line with the discussion 
    below \eq\nr{res_T_T}.
    }
    \index{tensor transfer function: sketch (figure)}
    \index{figure: sketch of tensor transfer function}
    \label{fig:T_T}
\end{figure}

\item
We arrived at \eq\nr{res_T_T_final} by assuming the absence of 
a trace anomaly, i.e.\ that $\bar e - 3\bar p = 0$. 
However, certain frequencies re-enter inside the
Hubble horizon when 
the trace anomaly is substantial~\cite{gubser}, for instance
during the QCD crossover at $T \sim 150$~MeV. For these frequencies, 
$\mathcal{T}^{ }_\tensor$ has to be determined numerically, as 
specified by \eqs\nr{match_1} and \nr{match_2}. 
This produces small but non-trivial ``features'' in the 
current gravitational-wave energy density, interpolating 
between the flat parts existing at other frequencies.
We illustrate these patterns in \fig\ref{fig:T_T}
on p.~\pageref{fig:T_T}. 
If a sufficient level of precision can be reached one day, 
such features offer for a direct way 
to see the thermal history of the early universe
imprinted on the gravitational-wave background~\cite{djs}.

\label{trace_anomaly_exceptions}

\ei

%%%%%%%%%%%%%%%%%%%%%%%%%%%% FIGURE %%%%%%%%%%%%%%%%%%%%%%%%%%%%%%%%%
%
%
%%%%%%%%%%%%%%%%%%%%%%%%%%%%%%%%%%%%%%%%%%%%%%%%%%%%%%%%%%%%%%%%%%%%

%%%%%%%%%%%%%%%%%%%%%%%%%%%%%%%%%%%%%%%%%%%%%%%%%%%%%%%%%%%%%%%%%%%
%
%\subsubsection*
\paragraph{(ii)~Tensor perturbations from reheating and 
post-inflationary dynamics.}

Between \eqs\nr{def_T_T} and \nr{T_T_factors}, we have determined the 
transfer function for tensor perturbations produced by vacuum fluctuations
that were extended beyond the Hubble horizon during inflationary expansion. 
However, as discussed in 
\ses\ref{ss:gw_Tij}--\ref{ss:gw_scat}, 
later moments in the history of the universe
can produce additional
contributions to the 
gravitational-wave background. 
Here we discuss how these get redshifted to the current day. 

As a starting point, we re-express the gravitational 
energy density production rate from \eq\nr{dt_e_gw_2}
with microscopic information  
according to \eq\nr{dt_e_gw_4}, 
\be
 \bigl( \partial^{ }_\tau + 4 \H \bigr)
 \frac{{\rm d} e^{ }_\rmii{gw}(\tau,k)}{{\rm d}\ln k}
 \;
 \underset{\scriptscriptstyle k/a,\,T\;\gg\;H}
 {\overset{\rmii{\nr{dt_e_gw_2},\nr{dt_e_gw_4}} \lift }{\approx}} 
 \;
 \frac{4 k^3_{ } }{\pi \mpl^2\, a^2_{ } } \,
 \nB^{ }\bigl(\tfr{k}{a}\bigr) \,  
 \im G^\rmi{t}_\iR \bigl(\tfr{\K}{a}\bigr)
 \;.
 \label{dt_e_gw_6}
\ee
Here 
$
 G^\rmi{t}_\iR
$
is the retarded 2-point correlator associated with anisotropic stress, 
and the Fourier transform is taken in local Minkowskian coordinates 
($\int^{ }_\R e^{-i\P\cdot\R}_{ } = 
a^4_{ }\int^{ }_\X e^{-i\K\cdot\X}_{ }$). 
We transform \eq\nr{dt_e_gw_6} into physical time, which 
according to \eqs\nr{eq_HH'} and \nr{eq_HH} goes via 
$
 \partial^{ }_\tau + 4 \H = a (\partial^{ }_t + 4 H)
$.
Then we define 
\be
 \mathcal{E}(t,k)
 \; \equiv \; 
 \frac{a^4_{ }}{a_\inow^4} 
 \frac{{\rm d} e^{ }_\rmii{gw}(t,k)}{{\rm d}\ln k}
 \qquad
 \Rightarrow
 \qquad
 \frac{{\rm d} e^{ }_\rmii{gw}(t,k)}{{\rm d}\ln k}
 \; = \; 
 \frac{a_\inow^4}{a^4_{ }} \,
 \mathcal{E}(t,k)
 \;. \label{def_E}
\ee
It follows that 
\be
 \frac{4 k^3_{ } }{\pi \mpl^2\, a^3_{ } } \,
 \nB^{ }\bigl(\tfr{k}{a}\bigr) \,  
 \im G^\rmi{t}_\iR \bigl(\tfr{\K}{a}\bigr)
 \;
 \underset{\rmii{\nr{def_E}}}
 {\overset{\rmii{\nr{dt_e_gw_6}} \lift }{=}} 
 \;
 (\partial^{ }_t + 4 H) 
 \biggl[\,
   \frac{a_\inow^4}{a^4_{ }} \,
   \mathcal{E}(t,k) 
 \,\biggr] 
% \nn[2mm]
%%%%
 \;
  = 
 \; 
 \frac{a_\inow^4}{a^4_{ }} \,
 \partial^{ }_t\hspace*{0.3mm} \mathcal{E}(t,k)
 \;, \hspace*{7mm}
\ee
which can be integrated into
\ba
 \mathcal{E}(t^{ }_\inow,k) - \mathcal{E}(t^{ }_e,k)
 & = & 
 \int_{t_e^{ }}^{t^{ }_\inow}
 \! {\rm d}t \, 
 \frac{4 k^3_{ } a }{\pi \mpl^2\, a_\inow^4 } \,
 \nB^{ }\bigl(\tfr{k}{a}\bigr) \,  
 \im G^\rmi{t}_\iR \bigl(\tfr{\K}{a}\bigr)
 \nn[2mm]
%%%%%%
 \; \overset{\rmii{\raise1ex\hbox{\nr{def_E}}}}{\Leftrightarrow} \;
 \frac{{\rm d} e^{ }_\rmii{gw}(t^{ }_\inow,k)}{{\rm d}\ln k}
 & = & 
 \frac{a_e^4}{a_\inow^4}
 \frac{{\rm d} e^{ }_\rmii{gw}(t^{ }_e,k)}{{\rm d}\ln k}
 \label{int_e_gw}
 \\[2mm]
%%%%%%
 &  &  \; + \,
 \int_{t_e^{ }}^{t^{ }_\inow}
 \! {\rm d}t \, 
 \frac{ a^4_{ } }{ a_\inow^4 }
 \frac{4 }{\pi \mpl^2 } \,
 \biggl(\frac{k}{a}\biggr)^3_{ }\,
 \nB^{ }\bigl(\tfr{k}{a}\bigr) \,  
 \im G^\rmi{t}_\iR \bigl(\tfr{\K}{a}\bigr)
 \;. \hspace*{8mm}
 \nonumber 
\ea
Here the initial moment, $t^{ }_e$, 
where the subscript refers to the beginning of emission
(which may coincide with the end of inflation), 
should be chosen late enough
that the production takes place within the Hubble horizon 
(otherwise we need to return to a transfer function
like in \eq\nr{res_T_T}).

The physical interpretation of 
\eq\nr{int_e_gw} can be as follows. 
The first term on the right-hand side indicates that, 
if there is an initial inside-horizon spectrum at time~$t^{ }_e$, 
it redshifts until today with the factor
$a_e^4 / a_\inow^4$. The second term tells that, 
in addition, we need to integrate over the 
production taking place between $t^{ }_e$ and $t^{ }_\now$, 
redshifting the momenta at which the production
happens ($k/a$), so that we consider a fixed frequency today.
Moreover,  
we need to redshift the energy densities produced
at a given time by~$a^4_{ }/a_\inow^4$.

\vspace*{3mm}

Let us now illustrate the use of \eq\nr{int_e_gw} with a practical example. 
In order to have a simple right-hand side, we consider the signal sourced
by {\em hydrodynamic fluctuations}, \index{hydrodynamic fluctuations} 
previously considered in \eq\nr{P_T_final_3}. Given that hydrodynamic 
fluctuations appear in the classical domain, $H \ll k/a \ll T$, we rewrite
the retarded Green's function in \eq\nr{int_e_gw} in terms of the 
Wightman function from \eq\nr{kms}, which has a direct classical limit, 
\be
 \frac{{\rm d} e^{ }_\rmii{gw}(t^{ }_\inow,k)}{{\rm d}\ln k}
 \; 
 \underset{\rmii{\nr{kms}}}{
 \overset{\rmii{\nr{int_e_gw}}  \lift  }{\supset}} 
 \;  
 \int_{t_e^{ }}^{t^{ }_\inow}
 \! {\rm d}t \, 
 \frac{ a^4_{ } }{ a_\inow^4 }
 \frac{2 }{\pi \mpl^2 } \,
 \biggl(\frac{k}{a}\biggr)^3_{ }\,
 \int_\R 
 e^{ - i \frac{\K}{a} \cdot \R }_{ }
 \bigl\langle\,
  \Pi^\rmi{t}_{ij}(t) \,
  \Pi^\rmi{t}_{ij}(t + \R) 
 \,\bigr\rangle
 \;. \label{hydro_fluct_0}
\ee 
The 2-point correlator of the anisotropic stress can be determined
with \eqs\nr{Pi_splitup} and \nr{delta_S}.
Replacing $\Pi^{ }_{ij}$ by its noise part $S^{ }_{ij}$
according to \eq\nr{Pi_splitup}, 
the correlator can be expressed in 
comoving momentum space as 
\ba
 \int_\R 
 e^{ - i \frac{\K}{a} \cdot \R }_{ }
 \bigl\langle\,
  \Pi^\rmi{t}_{ij}(t) \,
  \Pi^\rmi{t}_{ij}(t + \R) 
 \,\bigr\rangle
 & 
 \underset{\scriptscriptstyle a\,\approx\,\rm constant}{
 \overset{\scriptscriptstyle \R\,=\,a\X  \lift  }{\supset}} 
 &
 a^4_{ }
 \int_\X 
 e^{ - i \K \cdot \X }
 \bigl\langle\,
  S^\rmi{t}_{ij}(\tau) \,
  S^\rmi{t}_{ij}(\tau + \X) 
 \,\bigr\rangle
 \label{S_X_to_K} \\[2mm]
%%%%%
 & 
 \overset{\rmii{\nr{fourier_x}} \lift }{
 \underset{\rmii{\nr{fourier_k}}}{=}} 
 & 
 a^4_{ }
 \int_{\tau'} e^{i k \tau' }_{ }
 \int_\vec{q} 
 \underbrace{
 \bigl\langle\,
  S^\rmi{t}_{ij}(\tau,\vec{q}) \,
  S^\rmi{t}_{ij}(\tau + \tau',\vec{k}) 
 \,\bigr\rangle
 }_{{\rm insert}\;\nr{delta_S} }
 \;. \hspace*{8mm} \nonumber 
\ea
The traceless and transverse index contraction 
($
 S^\rmi{t}_{ij} S^\rmi{t}_{ij} 
 = 
 \mathbbm{T}_{ij}^{kl}
 S^\rmi{ }_{ij} S^\rmi{ }_{kl} 
$)
yields
\ba
 & & \hspace*{-1.5cm}
 \overbrace{ 
  \frac{1}{2}
 \Bigl(
   \projK^{k}_{i}\projK^{l}_{j}
 + \projK^{l}_{i}\projK^{k}_{j}
 - \projK^{ }_{ij}\projK^{kl}_{ } 
 \Bigr)
 }^{ 
 \mathbbm{T}_{ij}^{kl}\;{\rm from}\;\nr{T_ijmn_k}
 }
 \overbrace{
 \biggl[
 \eta \, \bigl( 
                 \delta^{ }_{ik} \delta^{ }_{jl}
               + \delta^{ }_{il} \delta^{ }_{jk}
         \bigr)
 +       \biggl( 
                 \zeta - \frac{2\eta}{3}
         \biggr) \, 
                 \delta^{ }_{ij} \delta^{ }_{kl} 
 \biggr] 
 }^{{\rm part~of}\;\langle S^\rmi{ }_{ij} S^\rmi{ }_{kl} 
 \rangle\;{\rm from}\;\nr{delta_S}}
 \label{T_contraction}
 \\[2mm]
%%%%%%
 &
 \overset{\rmii{}}{=}
 & 
 \frac{\eta}{2}
 \bigl(
  2 \projK^{i}_{i}\projK^{j}_{j}
 + \cancel{ 2 \projK^{j}_{i}\projK^{i}_{j} }
 - \cancel{ 2 \projK^{j}_{i}\projK^{i}_{j} }
 \bigr)
 \; + \;
 \frac{1}{2}
 \biggl( 
                 \zeta - \frac{2\eta}{3}
         \biggr) \, 
 \bigl( 
  \bcancel{ 2 \projK^{j}_{i}\projK^{i}_{j} }
 - \bcancel{ \projK^{i}_{i}\projK^{j}_{j} }
 \bigr)
 \; = \; 4 \eta
 \;. 
 \hspace*{8mm}
 \nonumber 
\ea
Putting everything together, we find
\be
 \int_\R 
 e^{ - i \frac{\K}{a} \cdot \R }_{ }
 \bigl\langle\,
  \Pi^\rmi{t}_{ij}(t) \,
  \Pi^\rmi{t}_{ij}(t + \R) 
 \,\bigr\rangle
 \; 
 \underset{\rmii{\nr{delta_S}}}{
 \overset{\rmii{\nr{S_X_to_K},\nr{T_contraction}}\vphantom{\big |}}{\supset}} 
 \;
 8 T \eta
 \; 
 = 
 \; 
 8 T^4_{ } \hat\eta
 \;, \label{hydro_fluct_1}
\ee  
where the rescaled $\hat\eta \equiv \eta / T^3_{ }$ 
is dimensionless.
We plug this into \eq\nr{hydro_fluct_0}, and convert
the time integral to a temperature one with the 
help of \eq\nr{dt_dT}, 
\ba
 \frac{{\rm d} e^{ }_\rmii{gw}(t^{ }_\inow,k)}{{\rm d}\ln k}
 & 
 \underset{\rmii{\nr{hydro_fluct_1}}}{
 \overset{\rmii{\nr{hydro_fluct_0}} \lift }{\supset}} 
 &  
 \int_{t_e^{ }}^{t^{ }_\iinow}
 \! {\rm d}t \, 
 \frac{ a^4_{ } }{ a_\inow^4 }
 \frac{2 }{\pi \mpl^2 } \,
 \biggl(\frac{k}{a}\biggr)^3_{ }\,
 8 T^4_{ } \hat\eta
 \nn[2mm]
%%%%%
 &
 \overset{\rmii{\nr{dt_dT}}}{\approx}
 & 
 \frac{3\sqrt{5}}{2\sqrt{\pi^3_{ }}}
 \frac{16}{\pi\mpl^{ }}
 \biggl(\frac{k}{a_\inow}\biggr)^3_{ }\,
 \int_{T_\iinow^{ }}^{T_e} \!
 {\rm d}T
 \frac{\hat\eta}{\sqrt{g^{ }_*}}
 \frac{a T}{a^{ }_\inow} 
 \;. \label{hydro_fluct_2}
\ea
The integrand is slowly varying, so for $T^{ }_e \gg T^{ }_\inow$ 
the value of the integral is dominated by the region close to 
the upper boundary, and the result can 
be approximated as 
\be
 \int_{T_\iinow^{ }}^{T_e} \!
 {\rm d}T
 \frac{\hat\eta}{\sqrt{g^{ }_* \vphantom{ t }}}
 \frac{a T}{a^{ }_\inow} 
 \;\; 
 \overset{\scriptscriptstyle 
          T^{ }_e \; \gg \; T^{ }_\iinow
           \lift  }{
 \approx}
 \;\;  
 T^{ }_e  
 \frac{\hat\eta^{ }_e}{\sqrt{g^{ }_{*,e} \vphantom{ t } }}
 \frac{a^{ }_e T^{ }_e}{a^{ }_\inow} 
 \;\; 
 \overset{\rmii{\nr{evolution_a}}}{
 \underset{\rmii{\nr{p_r}}}{=}} 
 \;\; 
 T^{ }_e T^{ }_\inow  
 \frac{\hat\eta^{ }_e}{\sqrt{g^{ }_{*,e} \vphantom{ t } }}
 \biggl(
 \frac{h^{ }_{*,\inow}}{h^{ }_{*,e}} 
 \biggr)^{1/3}_{ }
 \;. \label{hydro_fluct_3}
\ee
Furthermore, we write $k = a^{ }_\inow\hspace*{0.3mm} p^{ }_\inow$
(cf.\ \eq\nr{k_aeHe})
and 
$e^{ }_{\gamma,\inow} = g^{ }_{\gamma,\inow} \pi^2_{ } T_\inow^4 /30$
(cf.\ \eq\nr{p_r}). 
With these, the current spectrum takes the form
\be
 \frac{{\rm d} e^{ }_\rmii{gw}(t^{ }_\inow,k)}{{\rm d}\ln k}
 \;\;
 \underset{\rmii{\nr{hydro_fluct_3}}}{
 \overset{\rmii{\nr{hydro_fluct_2}}}{\supset}} 
 \;\; 
 \frac{3\sqrt{5}}{2\sqrt{\pi^3_{ }}}
 \frac{16 T^{ }_e}{\pi\mpl^{ }}
 \biggl(\frac{p^{ }_\inow}{T_\inow}\biggr)^3_{ }\,
 \frac{\hat\eta^{ }_e}{\sqrt{g^{ }_{*,e} \vphantom{ t } }}
 \biggl(
 \frac{h^{ }_{*,\inow}}{h^{ }_{*,e}} 
 \biggr)^{1/3}_{ }
 \; 
 \overbrace{
 \frac{30\, e^{ }_{\gamma,\inow}}{\pi^2_{ } g^{ }_{\gamma,\inow} }
 }^{\scriptstyle T_\iinow^4 \lift }
 \;. \label{hydro_fluct_4}
\ee
The final step is to normalize the result 
to the critical energy density
(cf.\ \eq\nr{def_Omega_gw}),
\ba
 \frac{ h^2_{ } 
          {\rm d} \Omega^{ }_{\igw,\inow} }
         {{\rm d}\ln\! f^{ }_\inow} 
 &
 \underset{\rmii{\nr{def_Omega_gw}}}{
 \overset{\rmii{\nr{hydro_fluct_4}}}{\supset}}
 & 
 \frac{720\sqrt{5}}{\pi^3_{ }\sqrt{\pi^3_{ }}}
 \biggl(
 \overbrace{
 \frac{p^{ }_\inow}{T_\inow}
 }^{ 
 \nr{p0_f0}
 }
 \biggr)^3_{ }
 \,
 \overbrace{
 \frac{1}{g^{ }_{\gamma,\inow} \, \sqrt{g^{ }_{*,e} \vphantom{ t } } }
 }^{ 
 \approx\,\frac{1}{2.0\,\sqrt{106.75 \vphantom{t^a} }}
 }
 \,
 \biggl(
 \overbrace{
 \frac{h^{ }_{*,\inow}}{h^{ }_{*,e}} 
 }^{ 
 \approx\,\frac{3.93}{106.75 \vphantom{t^a} }
 }
 \biggr)^{1/3}_{ }
 \,
 \hspace*{-2mm}
 \overbrace{
 \frac{h^2_{ }e^{ }_{\gamma,\inow}}{e^{ }_\rmii{crit} }
 }^{ 
 \approx\; 2.47\,\times\,10^{-5}_{ }
 }
 \hspace*{-2mm}
 \,
 \frac{\hat\eta^{ }_e T^{ }_e}{\mpl^{ }}
 \hspace*{8mm}
 \nn[2mm]
%%%%%
 & 
 \approx
 & 
 2.02 \times 10^{-29}_{ }
 \, 
 \frac{\hat\eta^{ }_e T^{ }_e}{\mpl^{ }}
 \biggl( 
 \frac{f^{ }_\inow}{\mbox{kHz}}
 \biggr)^3_{ }
 \;, \quad
 f^{ }_\now \; \ll \; 10^{11}_{ }\,\mbox{Hz}
 \;. \label{hydro_fluct_5}
\ea
This shows that the maximal temperature (and shear viscosity) of 
the radiation-dominated epoch is in principle visible, as the coefficient
of $f_\now^3$ in the gravitational-wave energy density in the 
frequency domain of the planned ET and CE interferometers
(cf.\ \se\ref{ss:probes_gw} and \fig\ref{fig:overview}). 
However, an exquisite resolution might be needed 
for observing the signal.

%%%%%%%%%%%%%%%%%%%%%%%%%%%%%%%%%%%%%%%%%%%%%%%%%%%%%%%%%%%%%%%%%%%%%%%%%
%
\subsection{Overview of frequency domains}
\label{ss:overview_f_0}

\index{gravitational waves: frequency domains}

Let us summarize the results of this chapter. 
Tensor perturbations, or gravitational waves, carry 
energy density (cf.\ \se\ref{ss:gw_energy}), 
and can be produced from 
vacuum fluctuations by the same mechanism as curvature perturbations
(cf.\ \se\ref{ss:gw_vac}). After these modes re-enter inside the Hubble
horizon, their frequency redshifts, however the spectrum remains
almost flat, with small features
superimposed on it by special epochs during the subsequent
expansion (cf.\ \se\ref{ss:gw_transfer}). 

There are also new
sources for gravitational waves. A model-independent 
contribution originates from hydrodynamic fluctuations
of the reheated plasma that fills the universe
(cf.\ \se\ref{ss:gw_Tij} and \eq\nr{hydro_fluct_5}). 
Moreover, the curvature perturbations that we have discussed in 
previous chapters, produce a gravitational-wave background
at second order 
(cf.\ \se\ref{ss:gw_sigw}). 
In contrast to vacuum fluctuations, 
the later sources generate a spectrum which shows overall growth 
with frequency, modified by the 2-point correlation function
of anisotropic stress (cf.\ \se\ref{ss:gw_reheat}). 
Ultimately, going to very high frequencies, 
gravitational waves are produced by quantum-mechanical elementary particle 
decays and scatterings (cf.\ \se\ref{ss:gw_scat}). However, the
energy carried by them must come out of 
somewhere. Above the scale of thermal motion, $p\sim T$, there are
no guaranteed sources present, and the amplitude of the primordial 
spectrum will likely turn down. 

As has been reviewed in \se\ref{ss:probes_gw}, these primordial
sources can conceivably be probed with future observations, in 
particular by combing interferometric searches sensitive to different
frequency domains. Of course, the sensitivity of each
individual experiment is limited by
various ``backgrounds'', ranging from instrumental noise
to astrophysical foregrounds. At the time of writing, 
the conceptual and technical development
of the instruments is work in progress, and it is difficult to judge
what their sensitivities will be. However, even if 
the observed gravitational-wave signal will be dominated 
by astrophysical sources, it is clear that we are facing the advent 
of a new probe for the physics of the early universe. 

Complementary to interferometers,
there is an important indirect constraint on gravitational 
waves in any frequency range,  
and that is their contribution to 
the {\em effective number of neutrinos}, 
\index{$N^{ }_\rmii{eff}$ (effective number of neutrinos)}
$N^{ }_\rmi{eff}\,$, 
as explained around \eq\nr{eq_Neff}. 
Traditionally, this is referred to as a BBN constraint, because
big bang nucleosynthesis works successfully only if the overall
expansion rate, and therefore the energy density at that time 
($T^{ }_\bbn \sim 0.1\,$MeV), 
is very close to that predicted by the Standard Model. 
In addition, the overall energy density is constrained
by the physics of photon decoupling
($T^{ }_\cmb \sim 0.3\,$eV). 
Even though these two constraints, 
$[\Delta N^{ }_\rmi{eff}]{}^{ }_\bbn$ and
$[\Delta N^{ }_\rmi{eff}]{}^{ }_\cmb$,
are independent, their  
numerical relationship to today's $\Omega^{ }_{\gw,\inow}$
is the same, once we note that the BBN constraint
is by convention evaluated at 
$T \ll T^{ }_\bbn$, so that the electron mass can be approximated as heavy, 
$\exp(-m^{ }_e/T) \ll 1$. 
In phenomenological determinations, the BBN and CMB data
sets are often combined, in order to obtain a more precise
estimate of the observed value of  
$N_\rmi{eff}^\rmi{ }$. 
Therefore, the additional 
subscript from $\Delta N^{ }_\rmi{eff}$ is usually dropped. 
Below we illustrate how the bound is imposed,  
focussing on the CMB case 
for concreteness  (cf.\ \eq\nr{Neff}).  

Following \eq\nr{eq_Neff}, we write the gravitational-wave energy density 
as an undetermined radiation contribution allowed by 
observational uncertainties around CMB decoupling,  
\be
 e^{ }_{\gw,\cmb} 
 \;\overset{\rmii{\nr{eq_Neff}} \lift }{\equiv}\; 
 \Delta N^{ }_\rmi{eff}\,
 \frac{7}{8} \biggl( \frac{4}{11} \biggr)^{4/3}_{ } \bar e^{ }_{\gamma,\cmb}
 \;. \label{e_gw_bbn}
\ee
This can be redshifted to today according to the first term on 
the right-hand side of \eq\nr{int_e_gw}, and subsequently normalized
to the critical energy density
(cf.\ \eq\nr{def_Omega_gw}), 
\be
 \Omega^{ }_{\gw,\inow} 
 \;
  \underset{\rmii{\nr{int_e_gw}}}{
  \overset{\rmii{\nr{def_Omega_gw}} \lift }{=}}
 \; 
  \frac{e^{ }_{\igw,\icmb}}{e^{ }_\rmii{crit}} 
  \frac{a^4_\icmb}{a_\inow^4}
 \;. \label{Omega_gw_bbn}
\ee
Then we insert the evolution of the scale factor
from \eq\nr{evolution_a}, and also divide and multiply with 
the energy density in CMB photons, like in \eq\nr{T_T_factors},  
\ba
 \Omega^{ }_{\gw,\inow} 
 & 
  \underset{\rmii{\nr{e_gw_bbn},\nr{Omega_gw_bbn}}} 
           {\overset{\rmii{\nr{evolution_a}} \lift }{=}} 
 & 
 \Delta N^{ }_\rmi{eff}\,
 \frac{7}{8}\biggl( \frac{4}{11} \biggr)^{4/3}_{ }
 \biggl( \frac{\bar s^{ }_\inow / T_\inow^3}
              {\bar s^{ }_\icmb / T^3_\icmb} \biggr)^{4/3}_{ }
 \biggl( 
 \frac{\bar e^{ }_{\gamma,\icmb} / T^4_\icmb }
      {\bar e^{ }_{\gamma,\inow} / T_\inow^4  }
 \biggr)
 \frac{\bar e^{ }_{\gamma,\inow} }{ e^{ }_\rmii{crit}  }
 \hspace*{8mm}
%%%%%%
 \nn[3mm] 
 & \overset{\rmii{\nr{p_r}} \lift }{=} & 
 \Delta N^{ }_\rmi{eff}\,
 \underbrace{ 
 \frac{7}{8}\biggl( \frac{4}{11} \biggr)^{4/3}_{ }
 \biggl( \frac{h^{ }_{*,\inow}}{h^{ }_{*,\icmb}} \biggr)^{4/3}_{ }
 \biggl( 
 \frac{g^{ }_{\gamma,\icmb}  }
      {g^{ }_{\gamma,\inow}  }
 \biggr)
 \hspace*{-6mm}
 \overbrace{
 \frac{\bar e^{ }_{\gamma,\inow} }{ e^{ }_\rmii{crit}  }
 }^{ 
   \approx\, 2.47 \times 10^{-5}_{ } / h^2_{ }
 }
 }_{
   \approx\, 5.62 \times 10^{-6}_{ } / h^2_{ }
 }
 \;. \label{Neff}
\ea
We recall from the discussion below \eq\nr{evolution_a} that, 
by convention, the current entropy density encompasses the effect of 
the decoupled neutrinos, whereas in the energy density, only the
thermalized electromagnetic plasma is included. 
For the numerical estimates, 
the photon energy densities have been approximated via
$ g^{ }_{\gamma,\cmb} \approx g^{ }_{\gamma,\inow} \approx 2.0$, 
and we have also set $h^{ }_{*,\cmb} \approx h^{ }_{*,\inow}$, 
given that neutrino decoupling is 
completed well before we reach $T^{ }_\cmb$.
{}From the observational upper bound on 
$ \Delta N^{ }_\rmi{eff} $ from BBN and CMB physics, 
we therefore obtain an empirical upper bound on 
gravitational-wave energy density, 
\be
 h^2_{ }\Omega^{ }_{\rmi{gw},\rmii{0}} 
 \;
 \overset{\rmii{\nr{Neff}}}{<}
 \; 
 5.62\times 10^{-6}_{ } 
 \, \Delta N^\rmi{obs}_\rmi{eff} 
 \;, \label{Omega_vs_Neff}
\ee
where 
$
 \Delta N^\rmi{obs}_\rmi{eff} 
 \simeq 0.15\pm 0.11
$
according to \eqs\nr{Neff_SM} and \nr{Neff_obs}.

\begin{figure}[p]
    \centering
    \includegraphics[width=0.99\linewidth]{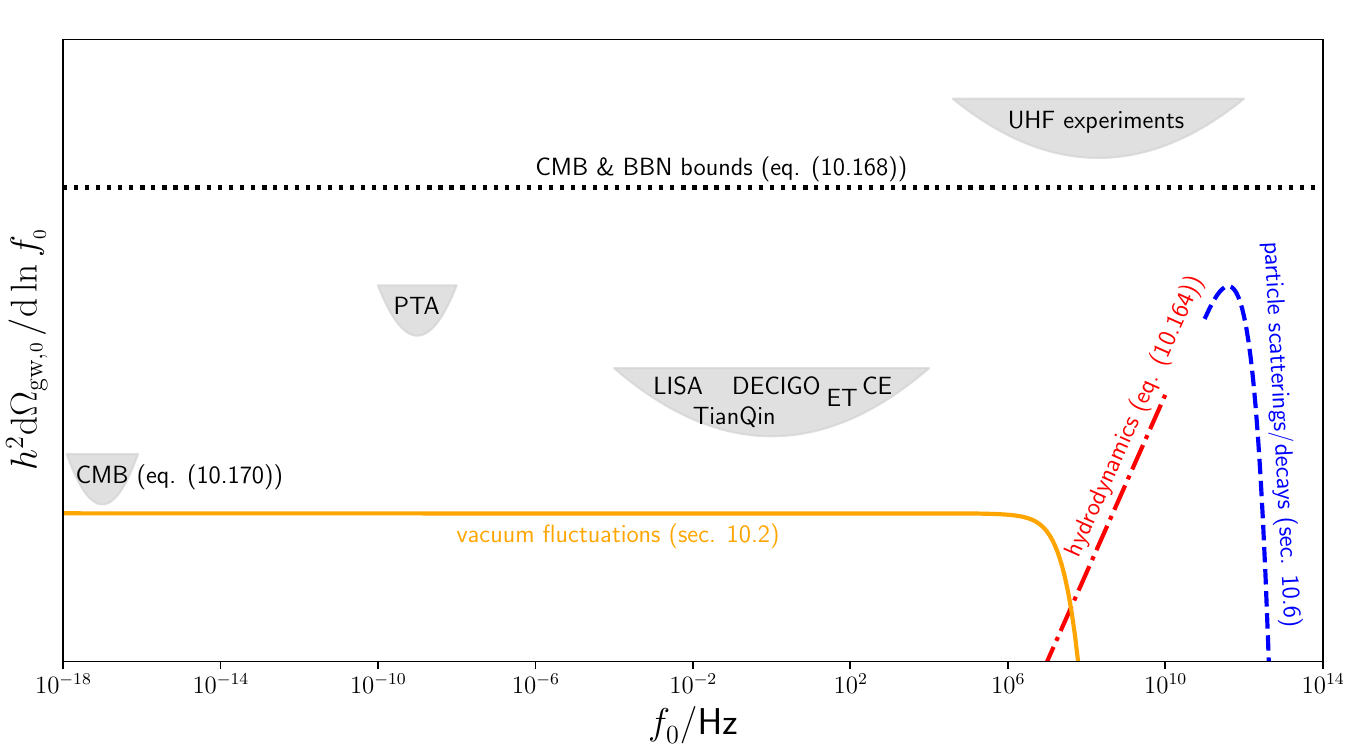}%

    \caption{%
    \small
    An illustration of the contributions of inflation and 
    post-reheating physics 
    to the current gravitational-wave energy density. 
    Only features that are guaranteed to be present are displayed, 
    even though their amplitudes are still unknown 
    (they depend, respectively, on the energy scale of inflation, 
    and on the maximal temperature after reheating).
    The curves that we show are oversimplified; 
    an example of a more accurate determination of the vacuum part, 
    from \eqs\nr{P_T_final} and \nr{def_T_T}, can be found in  
    fig.~8 of ref.~\cite{eos2}, whereas model realizations 
    of the high-frequency spectrum 
    can be found, e.g.,\ in refs.~\cite{qualitative,ringwald,xu}.
    The ``CMB \& BBN bounds'' from \eq\nr{Omega_vs_Neff},
    describing the gravitational-wave contribution 
    to the effective number of 
    neutrinos ($N^{ }_\rmii{eff}$), 
    applies to the integral over the spectrum
    rather than differentially. 
    The planned interferometers, 
    described in \se\ref{ss:probes_gw}, 
    are only indicated schematically, 
    given that their  
    ultimate resolutions  
    are difficult to anticipate, due to 
    astrophysical foregrounds and instrumental challenges.
    We remark that in the literature, 
    the $y$-axis of this type of plots is frequently labelled by
    $ h^2_{ } \Omega^{ }_{\igw,\inow} $, even though what is meant is 
    the differential spectrum, 
    $
     h^2_{ } {\rm d}
     \Omega^{ }_{\igw,\inow}
    / 
    {\rm d} \ln\! f^{ }_\inow
    $. 
    }
    \index{gravitational-wave spectrum (figure)}
    \index{figure: overview of $h^2{\rm d}\Omega^{ }_{\igw,\inow}/{\rm d}\ln f^{ }_\inow$}
    \label{fig:overview}
\end{figure}

It should be stressed that \eq\nr{Omega_vs_Neff} 
represents a constraint on the {\em integral} over 
the gravitational-wave energy density spectrum,
\be
 \boxed{
 \quad
 h^2_{ }\Omega^{ }_{\gw,\inow} 
 \; = \; 
 \int_{-\infty}^{\infty} \! {\rm d} \ln\! f^{ }_\now 
 \, \frac{ h^2_{ } 
          {\rm d} \Omega^{ }_{\igw,\inow} }
         {{\rm d}\ln\! f^{ }_\inow} 
 \;. 
 \quad
 }
 \label{O_vs_dO}
\ee 
Quite often
in the literature, the result from \eq\nr{Omega_vs_Neff}
is displayed as a frequency-independent
upper bound, and we repeat this in 
\fig\ref{fig:overview}. Unfortunately, there is also 
a widespread convention of denoting the {\em differential}
spectrum by $ \Omega^{ }_{\gw,\inow} $, having no separate
notation for the integrated gravitational-wave energy density,  
which increases the likelihood of misunderstanding.
In any case, if there is a peak in the spectrum, 
the maximal amplitude of
$
   { h^2_{ } 
          {\rm d} \Omega^{ }_{\igw,\inow} } / 
         {{\rm d}\ln\! f^{ }_\inow}
$
could well violate \eq\nr{Omega_vs_Neff}, 
as long as the peak is narrow, so that the 
overall gravitational-wave energy density satisfies it. 
Conversely, if the spectrum is nearly flat, the would-be constraint
is stronger than \eq\nr{Omega_vs_Neff} --- for example, 
a constant of {\em any} magnitude is excluded, as the 
integral in \eq\nr{O_vs_dO} would diverge.

\vspace*{3mm}

Let us also briefly quantify the constraint on 
$
   { h^2_{ } 
          {\rm d} \Omega^{ }_{\igw,\inow} } / 
         {{\rm d}\ln\! f^{ }_\inow}
$
from CMB polarization, discussed around \eq\nr{r_obs}.
Adapting \eq\nr{def_T_T}, we can express it as 
\be
   \frac{ h^2_{ } 
          {\rm d} \Omega^{ }_{\igw,\inow} } 
         {{\rm d}\ln\! f^{ }_\inow}
  \; 
  \overset{\rmii{\nr{def_T_T}} \lift }{\simeq}
  \; 
  \hspace*{-2mm}
  \underbrace{
  h^2_{ } \mathcal{T}^{ }_\tensor(k)
  }_{
  \rmii{fig.\hspace*{0.3mm}\ref{fig:T_T}}:\;\sim\;10^{-6}_{ }
  }
  \hspace*{0mm}
  \underbrace{
 \,
  \frac{ \P^{ }_\tensor(\tau^{ }_\rmii{out},k) }
       { \P^{ }_\scalar(\tau^{ }_\rmii{out},k) }
 }_{\nr{r_obs}:\;<\; 4\times 10^{-2}_{ }}
 \;
 \underbrace{
 \, \P^{ }_\scalar(\tau^{ }_\rmi{out},k)
 }_{\nr{ex_A_s}:\;\sim\; 2\times 10^{-9}_{ }}
 \;\, 
 \lsim % \lesssim 
 \;\, 
 10^{-16}_{ }
 \;. \label{cmb_constraint}
\ee 
This is much stronger than \eq\nr{Omega_vs_Neff}, however
only applicable at very small frequencies: the momentum 
scale  
$
 k/a^{ }_\now = 0.002 \, \mbox{Mpc}^{-1}_{ } = 2\pi/\lambda^{ }_\now
$ 
from \eq\nr{r_obs}
corresponds to $f^{ }_\now \approx 3 \times 10^{-18}_{ }\,\mbox{Hz}$ 
according to \eq\nr{f0_lam0}. 

\vspace*{3mm}

\index{gravitational strain} 
\index{$\hstrain^{ }_\rmi{t}$ (gravitational strain)}

Finally, we would like to explain the relation of the gravitational-wave
energy density to another quantity relevant for the observational side. 
It is meant as an instrumental tool, 
as it gives a direct feeling about the relative magnitude of a 
space-time distortion that is caused 
by a gravitational wave entering an interferometer.

As a starting point, let us inspect \eq\nr{e_gw}, normalized
to the critical energy density like in \eq\nr{def_Omega_gw}, 
but now in physical time, 
\be 
 \Omega^{ }_{\rmi{gw},\rmii{0}}
  \;\;
  \underset{\rmii{\nr{e_gw}}}{
  \overset{\rmii{\nr{def_Omega_gw}}  }{\simeq}}
 \;\; 
 \frac{1}{12 H_\inow^2}
 \sum_{ij} 
 \bigl\langle\, (\, \dot{h}^\rmi{t}_{ij} \,)^2_{ }\,\bigr\rangle\,
 \;. \label{Omega_gw_again}
\ee
The idea is to re-parametrize 
the differential spectrum, 
$
 {\rm d}
 \Omega^{ }_{\igw,\inow}
 / 
 {\rm d} \ln\! f^{ }_\inow
$,
through a new quantity, $\hstrain^{ }_\rmi{t}(f^{ }_\now)$, 
called the {\em gravitational strain}. 
A recipe for this is to 
replace  $h^\rmi{t}_{ij}$ in \eq\nr{Omega_gw_again}
by a plane wave of angular frequency 
$\omega^{ }_\now = 2 \pi f^{ }_\now$, 
which is ``unpolarized'', so that the sum $\sum_{ij}$ is replaced
by a factor~$\Sigma \equiv 4$. 
In addition, as explained below \eq\nr{res_T_T}, 
we should take an oscillation average, 
multiplying the result by $1/2$, 
even if this is not always implemented in the literature;
we indicate this with $\langle \Sigma \rangle$. 
After these steps, we write the differential spectrum as
\be
 \frac{ 
 {\rm d}
 \Omega^{ }_{\igw,\inow}
 }{
 {\rm d} \ln\! f^{ }_\inow
 } 
 \; \equiv \; 
 \frac{\langle \Sigma \rangle \hspace*{0.3mm} 
        \pi^2_{ } f_\inow^2 \hstrain_\rmi{t}^2 (f^{ }_\inow)}{3 H_\inow^2}
 \; \equiv \; 
 \frac{\langle \Sigma \rangle \hspace*{0.3mm}
       \pi^2_{ } f_\inow^3}{3 H_\inow^2} 
 \, S^{ }_{\hstrain^{ }_\rmii{t}} (f^{ }_\now)
 \;, \label{strain}
 \index{strain sensitivity}
 \index{$S^{ }_{\hstrain^{ }_\rmii{t}}$ (strain sensitivity)}
\ee
where 
$ 
 S^{ }_{\hstrain^{ }_\rmii{t}} \equiv 
 \hstrain_\rmi{t}^2 (f^{ }_\now) / f^{ }_\now
$ 
is the {\em strain sensitivity} 
(often $ S^{1/2}_{\hstrain^{ }_\rmii{t}} [ \mbox{Hz}^{-1/2}_{ } ] $
is plotted). 
The factor $f_\now^3$ 
in \eq\nr{strain}
matches the $k^3_{ }$ 
in \eq\nr{dt_e_gw_2}.
Assuming that technology allows to probe similar
strain sensitivities over many frequencies, 
this suggests opportunities for mapping the physics
induced by anisotropic stress across 
a broad spectrum of phenomena. 

%%%%%%%%%%%%%%%%%%%%%%%%%%%% FIGURE %%%%%%%%%%%%%%%%%%%%%%%%%%%%%%%%%
%
%
%%%%%%%%%%%%%%%%%%%%%%%%%%%%%%%%%%%%%%%%%%%%%%%%%%%%%%%%%%%%%%%%%%%%

A schematic summary of various sources and detection 
opportunities is shown in \fig\ref{fig:overview}
on p.~\pageref{fig:overview}. 
Here it is appropriate
to repeat the warning from below \eq\nr{O_vs_dO}: in the literature, 
the $y$-axis of this type of plots is frequently labelled by
$ h^2_{ } \Omega^{ }_{\igw,\inow} $, even though what is meant is 
the differential spectrum, 
$
 h^2_{ } {\rm d}
 \Omega^{ }_{\igw,\inow}
  / 
 {\rm d} \ln\! f^{ }_\inow
$. 

%%%%%%%%%%%%%%%%%%%%%%%%%%%%%%%%%%%%%%%%%%%%%%%%%%%%%%%%%%%

\newpage

%%%%%%%%%%%%%%%%%%%%%%%%%%%% start appendices %%%%%%%%%%%%%%%%%%%%%%%%%%%%%%%

%%%%%%%%%%%%%%%%%%%%%%%%%%%%%%%%%%%%%%%%%%%%%%%%%%%%%%%%%%%%%%%%%
%
\subsubsection{Computer algebra for Einstein tensor at second order}
\label{app:sigw}

\addcontentsline{toc}{subsection}{\App\ref{app:sigw}: 
Computer algebra for Einstein tensor at second order}

\index{computer-algebraic methods}
\index{symbolic manipulation}

In the script below, we carry out the main computations that were 
reported in \se\ref{ss:gw_sigw}. First, \eq\nr{Gij_t_1}
is checked. Then its scalar-induced part is moved to the right-hand
side of the Einstein equation, where it is combined with the relevant
parts of \eq\nr{Tij_t_fixed_1}. The variables are expressed in terms
of Bardeen potentials via \eqs\nr{tilde_h0_subst} 
and \nr{tilde_hD_subst}. What is {\em not} implemented are the
IBP relations from \eqs\nr{ibp_spatial}, \nr{ibp_temporal}, 
and \nr{ibp_spatial_2}.
Therefore, the final result is a bit more complicated 
than \eq\nr{sigw}, however the expressions are equivalent
if IBP relations are subsequently employed. 

\index{code: algebra for $G^{ }_{\mu\nu}$ at second order}

{\fontsize{8pt}{10pt}\selectfont
\begin{verbatim}
# computer-algebraic determination of the source for scalar-induced GWs [symbolic_sigw.py]
#
# basic coordinates and functions appearing
from sympy import *
i, j, k, l = symbols('i j k l',cls=Idx); x = IndexedBase('x'); eps = symbols('eps')
a = Function('a')(x[0]); H = Function('H')(x[0])
h = Function('h')(x[0],x[1],x[2],x[3])
h0 = Function('h0')(x[0],x[1],x[2],x[3]); hD = Function('hD')(x[0],x[1],x[2],x[3])
phi = Function('phi')(x[0],x[1],x[2],x[3]); psi = Function('psi')(x[0],x[1],x[2],x[3])

# partial derivative with respect to x[i]
def d(h,i): return diff(h,x[i],evaluate=True)
def dd(h,i,j): return diff(h,x[i],x[j],evaluate=True)            # doesn't symmetrize as should :(
def dds(h,i,j): return (dd(h,i,j)+dd(h,j,i))/2                   # symmetrized
def ddd(h,i,j,k): return diff(h,x[i],x[j],x[k],evaluate=True)    # doesn't symmetrize as should :(
def ddds(h,i,j,k): return (ddd(h,i,j,k)+ddd(h,i,k,j))/2          # symmetrized in last two args

# restrict to eps^order for possibly more efficient execution
def trunc2(h): return (h.expand()).subs([(eps**5,0),(eps**4,0),(eps**3,0)])
def trunc1(h): return (h.expand()).subs([(eps**2,0)])
order = 2

# input metric with both indices down
a00 = a**2*(-1-2*eps*h0)
a01 = a**2*(eps*d(h,1));  a02 = a**2*(eps*d(h,2)); a03 = a**2*(eps*d(h,3))
a10 = a01; a20 = a02; a30 = a03
a11 = a**2*(1-2*eps*hD); a22=a11; a33=a11
a12 = 0; a13 = 0; a23 = 0; a21 = a12; a31 = a13; a32 = a23
gdown=Matrix([[a00,a01,a02,a03],[a10,a11,a12,a13],[a20,a21,a22,a23],[a30,a31,a32,a33]])

# input metric with both indices up
b00 = a**(-2)*(-1+2*eps*h0 + eps**2*( -4*h0**2 + d(h,1)**2 + d(h,2)**2 + d(h,3)**2 ))
b01 = a**(-2)*d(h,1)*(eps +  eps**2*(  2*hD - 2*h0  ))
b02 = a**(-2)*d(h,2)*(eps +  eps**2*(  2*hD - 2*h0  ))
b03 = a**(-2)*d(h,3)*(eps +  eps**2*(  2*hD - 2*h0  ))
b10 = b01; b20 = b02; b30 = b03
b11 = a**(-2)*(1 + 2*eps*hD + eps**2*( 4*hD**2 - d(h,1)**2 ))
b22 = a**(-2)*(1 + 2*eps*hD + eps**2*( 4*hD**2 - d(h,2)**2 ))
b33 = a**(-2)*(1 + 2*eps*hD + eps**2*( 4*hD**2 - d(h,3)**2 ))
b12 = a**(-2)*(0 + eps**2*( - d(h,1)*d(h,2) ))
b13 = a**(-2)*(0 + eps**2*( - d(h,1)*d(h,3) ))
b23 = a**(-2)*(0 + eps**2*( - d(h,2)*d(h,3) ))
b21 = b12; b31 = b13; b32 = b23
gup=Matrix([[b00,b01,b02,b03],[b10,b11,b12,b13],[b20,b21,b22,b23],[b30,b31,b32,b33]])
if order==1: gup = trunc1(gup) 

# check inversion
test=(gdown*gup) 
if order==1: print(trunc1(test)) 
else: print(trunc2(test)) 
print("inversion checked")

# christoffel symbols
def dgdown(m,n,i): return d(gdown[m,n],i)
def gamma(r,m,n):
    return Rational(1/2)*Sum( gup[r,i]*(dgdown(i,m,n)+dgdown(i,n,m)-dgdown(m,n,i)),(i,0,3)).doit()

# ricci tensor with both indices down
def prericcidown(m,n):
    return ( Sum( d(gamma(k,m,n),k) - d(gamma(k,m,k),n),(k,0,3) ).doit() +
        Sum( gamma(l,m,n)*gamma(k,l,k) - gamma(l,m,k)*gamma(k,l,n),(k,0,3),(l,0,3) ).doit() )
riccidown=trunc2(Matrix(4,4,prericcidown))
if order==1: riccidown=trunc1(riccidown)

# ricci tensor with indices up and down
ricciupdown = trunc2(gup*riccidown)
if order==1: ricciupdown=trunc1(ricciupdown)

# ricci scalar
ricciscalar = ricciupdown.trace()

# einstein tensor with both indices down
einstein = trunc2(riccidown - ricciscalar*gdown/2)
if order==1: einstein=trunc1(einstein)
    
# replace derivatives of the scale factor by the hubble rate
einstein = einstein.subs([(Derivative(a,x[0]),a*H),(Derivative(a,x[0],2),a*(d(H,0)+H**2))])

# adjust expression until the difference subtracts to zero 
testeinstein12 = (0 + eps*( - dds(h0,1,2) + dds(hD,1,2) - ddds(h,0,1,2) - 2*H*dds(h,1,2) )
                  + eps**2*(
                      + d(h0,1)*d(h0,2) + 3*d(hD,1)*d(hD,2)
                      + 2*( h0*dds(h0,1,2) + hD*dds(hD,1,2) )
                      - d(h0,1)*d(hD,2) - d(hD,1)*d(h0,2)
                      + ( d(h0,0) + d(hD,0) )*dds(h,1,2)
                      + 4*H*h0*dds(h,1,2) + 2*h0*ddds(h,0,1,2)
                      - d(hD,1)*dd(h,0,2) - d(hD,2)*dd(h,0,1)
                      - dd(hD,0,1)*d(h,2) - dd(hD,0,2)*d(h,1)
                      - 2*H*( d(hD,1)*d(h,2) + d(hD,2)*d(h,1) )
                      + ( dd(h,1,1) + dd(h,2,2) + dd(h,3,3))*dds(h,1,2)
                      - ( dd(h,1,1)*dds(h,1,2) + dds(h,2,1)*dd(h,2,2) + dd(h,3,1)*dd(h,3,2) )
    #### remnant of non-symmetrized derivatives
                      + (dd(h,1,1)-dd(h,2,2))*(dd(h,2,1)-dd(h,1,2))/2   ) ) 
if order==1: testeinstein12=trunc1(testeinstein12)
print(expand((einstein[1,2]+einstein[2,1])/2 - testeinstein12))
print("einstein 12 checked to 2nd order")

# construct effective energy-momentum tensor by moving metric perturbations to right-hand side
rhs12=-testeinstein12 + eps**2*(
    -2*( d(h0,1)*d(hD,2) + d(hD,1)*d(h0,2) ) 
    -2/H*( dd(hD,0,1)*d(hD,2) + dd(hD,0,2)*d(hD,1) )
    +2*(d(H,0)/H**2 - 1)*d(hD,1)*d(hD,2) )
if order==1: rhs12=trunc1(rhs12)

# substitute bardeen potentials
rhs12=rhs12.subs([(h0,phi-d(h,0)-H*h),(hD,psi+H*h)]).doit()
if order==1: rhs12=trunc1(rhs12)

# adjust expressions until the difference subtracts to zero
testrhs12 = (0 + eps*( dds(phi,1,2)-dds(psi,1,2) )
             + eps**2*(
                 -d(phi,1)*d(phi,2) - 2*phi*dds(phi,1,2)
                 -( d(phi,1)*d(psi,2) + d(phi,2)*d(psi,1) ) - 2*psi*dds(psi,1,2)
                 +( 2*d(H,0)/H**2 - 5 )*d(psi,1)*d(psi,2)
                 -2/H*( dd(psi,0,1)*d(psi,2) + dd(psi,0,2)*d(psi,1) )
                 +2*H*( h*dds(phi,1,2) - phi*dds(h,1,2) )
                 -d(phi,0)*dds(h,1,2) + 2*dds(phi,1,2)*d(h,0) + d(phi,1)*dd(h,0,2) + d(phi,2)*dd(h,0,1)
                 -2*H*( d(psi,1)*d(h,2) + d(psi,2)*d(h,1) + psi*dds(h,1,2) + h*dds(psi,1,2) )
                 -d(psi,0)*dds(h,1,2) - dd(psi,0,1)*d(h,2) - dd(psi,0,2)*d(h,1)
                 +dd(h,1,1)*dds(h,1,2) + dds(h,2,1)*dd(h,2,2) + dd(h,3,1)*dd(h,3,2)
                 -( dd(h,1,1) + dd(h,2,2) + dd(h,3,3))*dds(h,1,2)
                 +dd(h,0,0)*dds(h,1,2) - dd(h,0,1)*dd(h,0,2) + 2*H*d(h,0)*dds(h,1,2)
    #### remnant of non-symmetrized derivatives
                 - (dd(h,1,1)-dd(h,2,2))*(dd(h,2,1)-dd(h,1,2))/2  ) )
if order==1: testrhs12=trunc1(testrhs12)
print(expand(rhs12-testrhs12))
print("sigw 12 checked to 2nd order")
\end{verbatim}
}

%%%%%%%%%%%%%%%%%%%%%%% end appendices %%%%%%%%%%%%%%%%%%%%%%%%%%%

%%%%%%%%%%%%%%%%%%%%%%%%% BIBLIO %%%%%%%%%%%%%%%%%%%%%%%%%%%%%%%%
%

%%%%%%%%%%%%%%%%%%%%%%%%%%%%%%%%%%%%%%%%%%%%%%%%%%%%%%%%%%%%%%%%%%%%%
\newpage 

%%%%%%%%%%%%%%%%%%%% INDEX %%%%%%%%%%%%%%%%%%%%%%%%%%%%%%%%%%%%%%%%%%
%
\clearpage
\addcontentsline{toc}{section}{Index}
\small
\printindex

\end{document}